\documentclass[a4paper,twoside,fleqn]{article}
\usepackage[figuresright]{rotating}
\usepackage{graphicx}
\usepackage{multicol}
\usepackage{enumerate}
\usepackage{url}
\usepackage[font=small]{caption}
\usepackage[authoryear]{natbib}
\bibpunct{(}{)}{;}{a}{}{,}
\usepackage[hyperindex,breaklinks]{hyperref}
\usepackage{breakurl}
\date{}
\title{\large\bf\flushleft A beacon of new physics: the Pioneer anomaly modelled as a path
based speed loss driven by the externalisation of aggregate non-inertial QM energy}
\author{\parbox{\textwidth}{\flushleft
\vspace{-0.5cm}
{\it \quad \, Paul G. ten Boom}\\
\vspace{0.4cm}
{\small \quad \quad School of Physics, University of New South Wales, Sydney NSW 2052, Australia}\\
{\small \quad \quad Email: ptb@phys.unsw.edu.au,
paul.tenboom@unswalumni.com}}}
\begin{document}
\begin{changemargin}{0cm}{0cm}
\begin{minipage}{.99\textwidth}
\vspace{-1cm} \maketitle
\end{minipage}
\end{changemargin}
%\twocolumn[
%Original format (pre toc lot lof)
\begin{changemargin}{.8cm}{.5cm}
\begin{minipage}{.88\textwidth}
%\vspace{-1cm} \maketitle
%
%
%%%%%%%%%%%%%     ABSTRACT    %%%%%%%%%%%%%
%Abstract of no more than 1920 characters (approx. 260 words) goes here.
\small{\bf Abstract:} This treatise outlines how a non-systematic
based Pioneer anomaly, with its implied violation (re: `low' mass
bodies only) of both general relativity's weak equivalence
principle and the Newtonian inverse-square law, can be
successfully modelled. These theoretical hurdles and various
awkward observational constraints, such as the low value of
Pioneer 11's anomaly pre-Saturn encounter, have (to date) not been
convincingly modelled. Notwithstanding the recent trend to embrace
a non-constant Sun/Earth-directed heat based explanation of this
anomalous deceleration, the actual: nature, direction, and
temporal and spatial variation of the Pioneer anomaly remain an
open arena of research. Working backwards from the observational
evidence, and rethinking: time, mass, quantum entanglement and
non-locality, we hypothesise a mechanism involving a quantum
mechanical energy source and a new type of `gravitational' field;
neither of which lie within general relativity's domain of
formulation/application. By way of a systemic conservation of
energy principle, an internally inexpressible (aggregate)
non-inertial energy discrepancy/uncertainty --- involving a myriad
of quantum (lunar/third-body residing) atomic and molecular
systems moving in analog curved spacetime --- is (non-locally)
re-expressed externally as a (rotating) non-Euclidean spatial
geometry perturbation. At a moving body each ``rotating
space-warp" induces sinusoidal proper acceleration and speed
perturbations, as well as a path-based constant (per cycle) rate
of speed shortfall relative to predictions that omit the
additional effect. `Solutions' of the new model may extend to: the
Earth flyby anomaly, solar system related large-scale anomalies in
the CMB radiation data, the nature of dark energy, and how a
theory of everything unification agenda is inadvertently impeding
a deeper understanding of physical reality and quantum
entanglement.

%%%%%%%%%%%%%     KEYWORDS    %%%%%%%%%%%%%
\medskip{\bf Keywords:} Pioneer anomaly --- gravitation --- time ---
Earth flyby anomaly --- dark energy
% Please write all keywords in lower case. PASA uses the
% standard list of subject headings adopted by The Astrophysical Journal
% and available from http://www.journals.uchicago.edu/ApJ/keywords_text.html.
% Keywords are separated by em-dashes, i.e. ---

%%%%%%%%DO NOT EDIT%%%%%%%%%%%%
\medskip
\medskip
\end{minipage}
\end{changemargin}
%]
%end of previous two column 1st page
%
%%%%%%%%EDIT FROM HERE%%%%%%%%%%%%
\small

\sloppy
% OPTIONAL USAGE - leave off for the writing?

\tableofcontents

\listoffigures

\listoftables

\twocolumn
%-------------------------------------------------------------------------------------------------------------
\section{Introduction}
\label{Section:introduction} The Pioneer 10 and 11 spacecraft,
launched in 1972 and 1973 respectively, represent an ideal system
to perform precision celestial mechanics experiments
\citep*{Anderson_02a}. In the quiescent outer solar system (and
beyond), where the effects of solar radiation pressure are
minimal, it was strikingly counter to expectations that the
radio-metric Doppler tracking data of these spacecraft
unambiguously indicated the presence of a small, anomalous,
blue-shifted frequency drift --- `uniformly' changing with a rate
of $\approx6\times10^{-9}$ Hz/s. This drift (applicable to both
spacecraft) has generally been interpreted as a `constant' (mean)
`inward' deceleration of magnitude
$a_P=(8.74\pm1.33)\times10^{-10}\rm~m/s^2$. This deviation away
from expected/predicted spacecraft navigational behaviour has
become known as the Pioneer anomaly.

Although the anomaly had been apparent in the data from as early
as 1980, the first publication of this well deliberated ``Pioneer
anomaly" was \citet*{Anderson_98a}. Subsequently, there has been
ongoing debate as to whether this anomaly is a harbinger of `new'
physics or merely an overlooked or unappreciated systematic
effect. Throughout this debate, the latter and more conservative
position has understandably (and rightfully) been favoured; with
recent publications from: \citet*{Francisco_11},
\citet*{Rievers_11}, \citet*{Turyshev_11a}, and
\citet*{Turyshev_12a} increasingly confident that the `solution'
lies in rectifying the modelling of onboard (anisotropic) heat
emissions, particularly radiative momentum transfer. Over the
years, this confidence has been bolstered by the failure of all
conceivable alternative hypotheses to convincingly explain the
anomaly.

In Section 2, as well as presenting a fuller overview of the
Pioneer anomaly and its primary observational features, concerns
with this conventional anisotropic heat based explanation are
presented, not the least of which is the minor qualitative
inconsistencies of the various models with each other. John D.
Anderson, first author of the initial comprehensive study of the
anomaly, remains sceptical of this approach \citep*{Anderson_11a}.
Supporting his stance is the disparity of these
thermal-radiation/heat based hypotheses with both the general
pre-2010 consensus of a long-term \emph{constant} anomaly, and the
minor role ascribed to anisotropic/non-isotropic heat effects in
the initial comprehensive investigation \citep{Anderson_02a}.

The main purpose of this paper is to construct a model that is
able to fully describe a non-systematic based Pioneer anomaly.
This is achieved by way of utilising: a process of elimination, a
process of abductive reasoning (i.e. guessing/hypothesising), and
inference to the best conceivable/possible explanation --- that is
free of preconceptions and primarily based upon \emph{all} facets
of the (awkward) observational evidence. As such, and in light of
the fact that motion in the solar system has historically provided
fertile `ground' for scientific advancement, the (outer solar
system) Pioneer anomaly is conceivably a bellwether of new physics
--- as the inner solar system anomalous precession of Mercury's
perihelion once was. Herein, it is \emph{not} the case that
general relativity (GR) is (in any way) quantitatively wrong or in
need of modification; rather, GR's domain of application is seen
as incomplete in that a unique supplementary source-type of
non-Euclidean geometry or gravitation (in its widest sense) ---
that is not describable by way of GR's formalism (and conceptual
basis) --- is hypothesised. Importantly, the unusual observational
constraint of long-term constancy in the Pioneer anomaly
contributes to this new supplementary acceleration/gravitational
field being compatible with the invariance requirements of
relativity. This additional/secondary `gravitational' field type
is seen to (perturb and) ``piggyback" upon the standard
gravitational field. The Pioneer anomaly arises from a `low' mass
moving body encountering several of these supplementary
(effectively oscillatory) acceleration/gravitational fields ---
that coexist in superposition. Above a maximum ``cut-off mass"
value --- noting that there are several different, distance
dependent, and coexistent ``cut-off mass" values --- `high' mass
bodies such as: large asteroids and comets, moons, and planets are
not additionally/anomalously affected.

In broad terms, a Heisenberg uncertainty principle (``wiggle
room") based extremely low energy situation is hypothesised as a
root cause. This hypothesis pertains to the role of the intrinsic
angular momentum of every elementary fermion particle within (the
prodigious number of) atoms/molecules constituting/comprising a
lunar \emph{(third) celestial body} --- that itself is part of a
celestial global system. Furthermore, we take issue (at a
`fractional' quantum mechanical level) with general relativity's
stance that: celestial \emph{geodesic} motion is (without
exception) \emph{always} inertial motion. The paper builds,
Section by Section, a conceptually intricate physical model that
mathematically is fairly straightforward. Five distinct
`sizes'/levels of physical `particles' and/or systems are
important to the model; these are: (1) elementary fermion
particles; (2) atoms and molecules; (3) (bulk matter) moons; (4)
Sun-planet-moon systems; and (5) the universe. A hybrid mechanism,
\emph{arising} from both (a many-particle and spin based) quantum
mechanical energy and curved spacetime \emph{together}, is implied
by the observational data/evidence. Thus, the conceptualisation
and mathematical formalisms of general relativity and/or quantum
mechanics in isolation are insufficient to apprehend and
appreciate the mechanism proposed herein. Mathematically, the use
of a single system Lagrangian or Hamiltonian is not viable in this
`lunar' \emph{geodesic motion} based situation, that encompasses
and `spans' both the microscopic \emph{and} macroscopic realms;
nor is a metric (tensor) based approach appropriate/viable. A
($\Delta t>>0$) \emph{process} based externalisation of
(fractional and `internally' inexpressible) \emph{non-inertial}
quantum mechanical spin/intrinsic angular momentum rate (i.e.
energy) is involved, with this (externalisation process) dependent
upon a new type of (non-local) `quantum' entanglement between: the
`overall' non-inertial (and virtual) \emph{quantum mechanical}
spin angular momentum (per process cycle), and the
(acceleration/gravitational) amplitude of the new/secondary
\emph{macroscopic} non-Euclidean geometry (i.e. space-time
curvature) field perturbation effect.

The model's development is best described as a three-stage
process. Primary model development specifics and the violation of
GR's equivalence principles are addressed in the second and third
stages respectively. In Section 3 (the first/inceptive stage), the
awkward observational constraints of: a constant long-term mean
value for the (outer solar system and beyond) anomalous
acceleration, as well as the temporal variation about this
constant value, drive the model's origination. By way of a process
of elimination, first order \emph{constant amplitude} sinusoidal
acceleration/gravitational field perturbations (in one dimension)
upon the pre-existing gravitational field, that each induce a type
of (perturbation based) ``celestial `simple' harmonic motion" in
the spacecraft's `translational' motion, are implied.
Interestingly, the phases associated with the (four)
\emph{sinusoidal perturbations} (around their non-stationary
`equilibrium' values) --- i.e. the gravitational field and a
body's: (proper) distance, speed, and acceleration (responses)
--- are all different and offset from each other by either 90 or
180 degrees (see subsections \ref{subsubsection:CSHPM} and
\ref{subsubsection:real_relative}). Furthermore, additional
observational evidence implies the acceleration/gravitational
fields are actually (in two and three dimensions) rotating
curvature/deformations of spacetime, i.e. rotating ``space-warps"
(RSWs) of cosmological extent, and that their rotation and energy
source is related to Sun-planet-moon (three-body) celestial
motion. The axis of rotation of the (`thick' planar or
cylindrical-like) space-warp is coexistent with the spin axis of
the moon with/to which it is associated/affiliated. Both this
axis, and each plane of space-warping orthogonal to it, extend to
`infinity'. Note that one side of the rotating space-warp is
above, and the opposite side is below, the equilibrium curvature
of spacetime produced by general relativistic gravitation, and as
such the (overall) ``net spacetime curvature" remains unchanged.
Close examination of the Pioneer spacecrafts' diurnal and annual
(post-fit) residuals facilitates and enlightens this modelling
process. The units/dimensions (of the physical `quantities')
involved necessitates that these RSWs have a conjoint mass aspect
(discussed in Sections 6 and 7, the third and `formulated' stage)
that accounts for: the \emph{dispersion of energy} at increasing
distances from a RSW's source `region', and the restriction of the
oscillatory perturbation motion effect to ``low mass" bodies ---
i.e. bodies whose mass is below at least one of the individual
``cut-off mass" values associated with the numerous rotating
space-warps. The unsteady/oscillatory speed (and acceleration)
components induced by the coexistence of several of these
(superpositioned) oscillatory/perturbation fields account for: (1)
the (Pioneer) anomalous deceleration as a constant rate of speed
shortfall, relative to predicted/expected spacecraft motion, i.e.
relative to `steady' motion in the absence of these
superpositioned supplementary (or secondary) fields; (2) the
quasi-stochastic temporal variation of (both) the model's
(supplementary) `gravitational' field strength, and (Pioneer 10)
spacecraft speed (and deceleration) about/around their respective
mean values, as well as how these temporal variations relate to
measurement noise and overall observational variation in the data;
and (3) a lower value of the Pioneer 11 based anomaly's magnitude
pre-Saturn encounter (cf. Pioneer 10's value at $>$20\,AU), and
its `rapid' increase post-Saturn encounter.

Sections 4 and 5 (the model's second, formative and innovative
stage) addresses: (1) the apparent incompatibility and coexistence
of these rotating field deformations (or RSWs) with (special and
general) relativity theory; as well as (2) how this secondary
(type of) gravitational field (or non-Euclidean geometry) is
`generated'; and (3) how each rotating space-warp necessarily
coexists with an appropriate initial ``non-local mass" value
[$m^*(r_1)$], and a non-local mass distribution (NMD) of
cosmological extent that determines the (distance/spherical-radius
dependent) ``cut-off mass" (particular to its associated RSW).
``Non-local mass", a new concept and new physical quantity
demanded by the model's use of quantum (spin) entanglement and
quantum non-locality, is enacted in order to provide a ``best/only
fit" to the observational evidence. The Pioneer anomaly is mainly
accounted for by way of five main RSWs, emanating `from' (or
rather, by way of) Jupiter's four Galilean moons and Saturn's
large moon Titan. Note that the amplitude/strength of a rotating
space-warp is dependent upon (both) a moon's mass, and the orbital
progression angle (around the Sun) of the moon's (host) planet per
lunar orbital cycle.

In the spirit of the historian and philosopher of science Thomas
Kuhn \citep{Kuhn_70}, the requirement of some kind of ``new
physics" (in response to a genuine anomaly) necessitates Section
\ref{Section:PhiloTheory}'s `non-destructive' \emph{conceptual}
(both physical and philosophical based) re-think of certain
foundational features of physics, particularly: mass, space and
time, energy, and a theory of everything unification agenda.
Concerns of philosophers of science (and physics) regarding the
physical status of general covariance in general relativity, and
the non-relativistic aspects of rotation and acceleration, are
explored. Combining these concerns with the issue of time's
treatment in quantum mechanics as compared to time's treatment in
special (and general) relativity \citep{Albert_09}, we promote a
Vladimir Fock-like stance/attitude towards relativity
\citep{Fock_59}; in the sense that special relativity is more
appropriately viewed as a theory of invariance, and general
relativity as a relativistic theory of gravitation. Responding to
the absence of a (classical) global/systemic perspective in
relativistic theorisation, and the existence (and global/systemic
implications) of non-local entanglement in quantum mechanics,
leads to the proposal of a (``noumenal" cf.
`phenomenal'/observational) \emph{supplementary} and
background/hidden ontological\footnote{``Ontology" (a
philosophical term and branch of philosophy) is concerned with the
nature of: being, existence, and reality. Herein its usage is
largely synonymous with ``physical reality", but its usage
additionally implies a depth of enquiry --- into physical
existence and reality --- that reaches beyond (current and
standard) `scientific' means/methods.} stance (or perspective)
regarding space and time. By virtue of the
``noumenal--phenomenal\footnote{Use of the adjective `phenomenal'
as a noun is intentional, so as to indicate its juxtaposition
alongside `noumenal' (as defined/modified herein, see subsection
\ref{subsubsection:noumena}), which has traditionally been used
(in philosophy) as either an adjective or a noun.}" `bifurcation'
arising from this ontological/physical supplementation it is shown
that: this paper's proposed (unique) second/new `type' of
gravitation (i.e. a second means to non-Euclidean geometry) can
coexist with general relativity's (standard) `coverage' of
gravitation. Further, in order to appease quantum non-locality and
entanglement, this `complementary' (and background/hidden)
ontological stance features a (universe-encompassing)
background/\emph{hidden} systemic and digital (cf. analog)
process-based approach to the ``passing of time" --- in the sense
of a sequential arrangement of events. This allows the
conservation of energy principle to be applicable on a
cosmological scale, but only in the context of the model's
entanglement and non-local based circumstances.

In Section 5 (and early in Section 6) the specifics of this
(`boutique') model/mechanism are further developed and the
emphasis is decidedly quantum mechanical, specifically involving
(lunar-based) \emph{discrete} quantum \emph{systems} (i.e.
individual spin-orbit coupled atoms and molecules dominated by
electromagnetic forces) --- moving (along a geodesic) in
\emph{continuous/analog} curved spacetime --- acquiring a
non-inertial frame status \emph{relative} to a systemic/global
background inertial frame (and other unaffected solar system
bodies). It is proposed that under appropriate circumstances
(composite/whole) atoms and molecules in the (spin-orbit resonant)
\emph{lunar/third body} of a macroscopic three-body
Sun-planet-moon celestial system --- by way of their constituent
fermion elementary particles and a (closed loop and path based)
relative (spin-based) \emph{geometric phase} offset --- attain (or
acquire) an extraordinarily small ($\sim 10^{-34} \times 10^{-6}$
via $\leq\frac{1}{2}\hbar \Delta t^{-1}$) \emph{virtual} rate of
intrinsic (spin) angular momentum offset relative to inertial
(\emph{spin}) frame conditions; i.e. $\sim 10^{-40}$ J of energy
per atom or molecule, which is an \emph{extraordinarily small
amount of} atomic/molecular based \emph{energy}.

Importantly, for conceptual explanatory/heuristic reasons, this
virtual/`fictitious' (quantum mechanical) geometric spin phase
offset --- that entails a virtual/fractional intrinsic angular
momentum offset --- is hypothetically conceptualised (i.e.
conceived of) as a relative spin phase \emph{precession} `induced'
by the closed-\emph{orbital} path of a moon's \emph{geodesic}
motion; with this motion being `governed' by its host planet's
curved spacetime, that (in turn) is \emph{nested} `within' the
solar system's dominant central Sun based spacetime curvature.
Furthermore, (lunar based) intra-atomic/molecular \emph{orbital}
(phase and) angular momentum are \emph{not} affected, and
electromagnetic spin-orbit coupling thwarts/denies any possible
actual/real spin-based response; thus ensuring a non-inertial spin
configuration is acquired/attained. This virtual spin offset's
magnitude and (\emph{externally} `imposed') projection plane is
common to all the elementary fermion particles within
(effectively) all $\sim 10^{49}$ atoms/molecules in a large
non-rigid `solid' moon --- with this being an
\emph{extraordinarily large number} of atoms and molecules. In
order to ensure and maintain global/universal and systemic
\emph{conservation of energy}, the (additive) total amount of this
(non-decohered) internally inexpressible `fractional' and
`\emph{fictitious}' quantum \emph{energy} --- which (along with
its underlying geometric phase offset) can \emph{not} be `carried
over' into the following closed loop cycle --- is necessarily
expressed \emph{externally} as both: a rotating
acceleration/gravitational field space-perturbation/deformation or
`rotating space-warp', and a (conjointly existing) non-local mass
distribution. Facilitating this (cyclic) externalisation process
there is --- for each individual Sun-planet-moon (or
barycentre-planet-moon) system --- a (newly proposed) non-local
entanglement relationship between (the \emph{en masse} total value
of shared/common) unresolved and non-inertial spin/intrinsic
angular momentum and the perturbation amplitude of the
(`gravito-quantum') rotating space-warp (RSW).

Scientific knowledge and concepts related to this aforementioned
physical mechanism/effect are: (1) a non-locality associated with
fermion quantum waves (as is similarly the case with the
Aharonov-Bohm effect); (2) ``a [closed loop/orbit based relative]
geometric phase [also known as \emph{Berry's phase}] associated
with the motion [i.e. kinematics] of the state of the quantum
system and \emph{not} with the motion of the Hamiltonian [as was
the case with the approach of Sir Michael Berry] \citep[
p.1864]{Anandan_88}"; (3) the inertial circumstances associated
with the `intrinsic spin' of elementary fermion particles (and
atomic/molecular systems) within a lunar/third-body, specifically
a geometric phase based exception to: ``\ldots the tendency of
intrinsic spin to keep its aspect with respect to a global
background inertial frame \citep{Mashhoon_06}"; (4) a new
(`reversed') instantiation of the energy-time version of
Heisenberg's uncertainty principle (HUP), noting that $\Delta t$
in this version of HUP is not/never an operator belonging to a
particle --- and in the model's case it is the closed (lunar
spin-orbital) loop/cycle duration ($\sim10^6$ s); and (5) because
(intrinsic angular momentum) entanglements are generated between
the many-atomed/moleculed (lunar) `quantum system' and its
environment (i.e. the surrounding universe), quantum mechanical
\emph{energy} (cf. information) is effectively transferred to (or
re-expressed in) the surroundings/environment. Finally, (6) the
proposed mechanism --- which also involves fermion wave
self-interference --- requires \emph{quantum decoherence} to be
averted; this requires/involves (model based) spin geometric phase
offsets that are below a tiny ($2\pi$ radian) half fermion
wavelength decoherence onset (and spinor sign change) threshold.
This final concept/feature is (indirectly) associated with a
denial of the existence of the graviton particle --- as a
(quantum) decoherence `agent' at least. To ensure (whole/entire
moon) decoherence is not `triggered', a number of rather stringent
pre-conditions must be met, including: that moons be predominantly
(non-rigid) solid bodies, and celestial moon-planet 1:1 (or
synchronous) \emph{spin-orbit resonance} --- with the latter also
referred to as: (lunar) synchronous rotation, tidal locking, and
`phase' lock.

Section 6 returns to a largely (conventional) physics based
approach; it begins with an overview of the model's major
features\footnote{As such, and due to the length of this paper,
the first time reader may wish to jump from the end of Section 3
to the beginning of Section 6.} and a brief discussion of errors.
A variety of implications of the hypothesised (mathematical and
conceptual) model are then outlined, primarily within the solar
system but also further afield, e.g. an ecliptic plane based
signature in the cosmic microwave background radiation. A
surprisingly simple (qualitative) resolution of the \emph{Earth
flyby anomaly} is proposed, involving (geocentric inbound and
outbound) trajectory based geometric angles in cooperation with
observational ramifications arising from the influence of the
model's RSWs upon spacecraft motion --- with each RSW's plane of
rotation (`near to' Earth) having been `refracted' by the Earth's
($\sim$10 billion times stronger) gravitational field so as to be
parallel with the Earth's equatorial plane. This proposal is
consistent with the (quantitative) findings/model proposed by
\citet*{Anderson_08}. Subsequent to this, the (Pioneer anomaly
based) model is fully quantified. Particularly relevant is an
energy equality --- specific to each participating Sun-planet-moon
(three-body) system --- involving an exact value of total
(virtual) non-inertial quantum mechanical (spin) energy and the
`re-expression' of this energy magnitude in the external
environment as two scalar fields: (1) a \emph{constant} (across
all space) `gravitational' \emph{amplitude} (gravito-quantum)
rotating (non-Euclidean geometry) space-warp, and a
\emph{conjointly} existing (2) `non-local mass' distribution. The
geometrical/orbital configuration and kinematics of each
Sun-planet-moon system determines (in a quasi-empirical manner)
the relative geometric phase offset (per closed loop/cycle), which
then quantifies the efficiency with which the minimum real (i.e.
maximum virtual) quantum mechanical intrinsic spin
($\frac{1}{2}\hbar$) --- of all elementary (fermion) particles
within a composite (lunar-based) atom or molecule --- is affected
per loop/cycle. Note that (for geometric reasons) the Earth's moon
is not a RSW `generator'\footnote{The `excessive' (i.e. $>2\pi$
rad) geometric spin phase offset associated with the
Sun-Earth-Moon system's (collision-originated) orbital
configuration triggers a (lunar) decoherence effect/event. This
event, that encompasses every one of its atoms/molecules, renders
the (whole) Moon a non-quantum/classical body. We note that the
decoherence of large-scale macroscopic (celestial) bodies is
qualitative, and not (purely) size/quantitatively based --- as is
mistakenly assumed/envisaged by some physicists (subsection
\ref{subsubsection:Vlatko Vedral}).}. Also determined are: the
various (individual) constant space-warp/space-deformation
`gravitational-acceleration' amplitudes $\Delta a$ [which are
collectively/non-singularly represented as $(\Delta a)_i$] and
their (attendant/conjoint) variable non-local mass distributions;
the latter by way of a constant/invariance relationship involving
the product of non-local mass and enclosed volume. We show/argue
that each RSW's specific energy is proportional to
$\frac{1}{2}(\Delta a)^2$, and that a root sum of squares (RSS)
approach determines the \emph{model's} overall (Pioneer) anomalous
rate of speed shortfall ($a_p$). This value provides a fit well
within the error range of the experimental/\emph{observational}
based value ($a_P$). Finally (in section \ref{Subsection:EquivPr
Comment}), the model's appeasement of the apparent violation of
general relativity's equivalence principles (especially the weak
equivalence principle) and the general principle of relativity are
examined and discussed in considerable detail.

Section 7 utilises and applies aspects of the preceding
model/mechanism to argue that the currently favoured
interpretation of observations pertaining to (distant) type 1a
supernovae and baryon acoustic oscillations, which implies
accelerated cosmological expansion and (hence) \emph{dark energy},
is quite conceivably misguided. It is proposed that the model's
(cosmological scale) non-local mass scalar field distributions,
that have been active for approximately four and a half billion
years --- i.e. from when the large moons (of Jupiter and Saturn)
attained spin-orbit resonance `with' their (respective) host
planet --- subjects (incoming) propagating electromagnetic (EM)
radiation to an increasing energy field over the duration of its
journey. Upon recognising/correcting for this additional
(distance/radius dependent and \emph{spherical} volume based)
foreground `effect' --- that slightly increases (i.e. blueshifts)
the frequency of all (`incoming' and) received EM radiation ---
the resultant relationship between the `observed' redshifts and
(cosmological) comoving spatial coordinates now includes a
signature that matches/mimics the (additional) effect/effects
attributable to dark energy's (`distance' dependent) presence.
Subsequently, an \emph{accelerating} expansion of the universe
(interpretation) is no longer implied by the data/evidence. This
conclusion is supported by the (presumed) onset of accelerated
expansion (at $z_t\approx0.44$) being remarkably consistent with
the initial attainment of (gas giant) planet-moon spin-orbit
resonance approximately 4.56 billion (`lookback') years ago. The
new model's implications for the (cosmological) ``flatness
problem" and other features of the `concordance model' are
(necessarily) also discussed.

Section 8 summarises and discusses the paper's major findings,
both qualitative and quantitative. Major quantitative results,
predictions, applications, and equations of the proposed model and
mechanism are presented; as well as its pertinent (primary)
conceptual features, and circumstantial simplifications that
greatly benefited the model's formulation. Broader implications of
the model/mechanism are also reviewed and discussed. As is the
case with this introductory Section, Section 8 is quite extensive.
This has been deemed necessary because an investigation into an
anomalous physical phenomenon (and physical reality in general)
that seeks to be progressive --- thereby questioning the existing
and generally accepted state of affairs (within physics) --- is
easily misconstrued.
%------------------------------------------------------------------------------------------------------------------------
\section{Background and discussion of the Pioneer anomaly}
\label{Section:Status} In this section a brief background of the
Pioneer anomaly, and how it is to be addressed in this paper, is
outlined. Additionally, the primary observational evidence and
what constitutes a ``new physics" is discussed. Other solar system
motion concerns are mentioned, followed by comments on the
anomaly's status.

\subsection{Stance taken in this paper: real anomalous spacecraft motion}
The Pioneer anomaly is the \emph{difference} between the predicted
behaviour of the Pioneer spacecraft and their observed or measured
behaviour. Radiometric tracking and navigation of the Pioneer 10
and 11 spacecraft was performed using (electromagnetic wave based)
Doppler observations. By way of a raw measurement involving a
phase count\footnote{The Doppler phase difference between
transmitted and received (S-band frequency) phases is counted
\citep*[ p.58]{Turyshev_10b}.} divided by the count time duration,
the anomaly is indicated by a steady anomalous Doppler
frequency\footnote{The Doppler data (in and of itself) is
\emph{not} a frequency measurement, but this data does share the
same unit as frequency which is cycles per second or Hertz [Hz].}
(blue-shift) drift (of magnitude $\approx
6.0\times10^{-9}~\rm{Hz~s^{-1}}$), which implies that the
spacecraft speed is \emph{less} than the speed theoretically
predicted (i.e. expected). This speed offset/anomaly occurs at a
constant/steady rate, which may be interpreted as a constant
anomalous acceleration. Thus, as the Pioneer 10 and 11 spacecraft
travel away from the inner solar system and into deep space, their
speed retardation is greater than that theoretically predicted; or
in other words, the speed of the spacecraft (and distance covered)
is less than the predicted values.

In the Introduction the direction of the Pioneer anomaly was
(somewhat) ambiguously referred to as `inward'. The word ``inward"
was used to highlight that analyses of the observational data do
\emph{not} conclusively favour a Sun-directed anomalous
deceleration, to the extent of precluding either (or both) an
Earth-directed or/and a (spacecraft) \emph{path}-directed anomaly
(i.e. directed along the spacecraft's ``velocity vector").

The stance taken in this paper is that this anomalous measurement
is not an observational or systematic artifact; rather, the
anomaly is considered to be an indication of real anomalous
spacecraft (S/C) motion and, on the whole, the observational
measurements are considered to be reliable\footnote{Certainly, the
anomaly's: error, variance, and temporal evolution remain somewhat
ambiguous. This is not surprising considering the navigational
accuracy of the \textit{in situ} observations of the Pioneers is
unsurpassed. The Voyager spacecraft with three-axis stabilization
are much less precise navigators than the spin stabilised Pioneer
10 and 11. For a discussion of a heat basis to the anomaly see
\mbox{section \ref{Subsection:Heat}.}}. This stance is based upon
an appraisal of the observational evidence that yields an ``open
verdict", and the underdevelopment of models acknowledging and
addressing a non-systematic based (`real') Pioneer anomaly.

The anomaly is well described in the literature, with the
comprehensive \citet{Turyshev_10b} review paper and the brief
\citet*{Nieto_0702} review paper recommended. The Pioneer
spacecraft yield a unique test/measurement, of a very small
effect, spanning a long period of time (i.e. years) cf. tests of
(general relativistic) time delays involving electromagnetic
radiation propagation.

Herein it is accepted that the Pioneer anomaly represents a
strange and significant non-conformity with the usual
Newtonian/Einsteinian gravitational behaviour of solar system
bodies. Note that the Pioneers' \textit{predicted} motion (and
measurement thereof) incorporates general relativity and very
sophisticated modelling of the spacecrafts' behaviour, with the
anomaly itself being in excess of five orders of magnitude greater
than the corrections to Newtonian gravitation associated with
General Relativity at 50AU \citep[ p.808]{Rathke&Izzo_06}.

At 20AU the Pioneer anomaly is approximately $0.006\%$ of the
Sun's Newtonian gravitational value; at 40 AU it is $0.024\%$, and
at 80AU it is $0.094\%$. By way of comparison, Mercury's (inner
solar system) observed precession of the perihelion is $\approx
5600$ seconds of arc per century, with $43$ seconds of arc per
century attributed to general relativity, representing
$\approx0.774\%$ of the ($5557$ seconds of arc) non-relativistic
value. Considering Mercury's anomalous precession was first
reported by Urbain Le Verrier in 1859, and that observational
accuracy and precision has come a long way since then, we begin to
appreciate the `significance' of the Pioneer anomalous
measurement\footnote{As the story goes: in 1994 physicist Michael
Martin Nieto, who had a good knowledge of the observational
accuracy of gravitational theory, almost fell off his chair when
JPL's John Anderson quantified the discrepancy in the Pioneer
spacecrafts' position and speed.}.

\subsection{On heat emission as the cause of the Pioneer anomaly}
\label{Subsection:Heat} The remaining alternative
hypothesis\footnote{Over time all other conceivable hypotheses
have been gradually eliminated.} to a real Pioneer anomaly is a
thermal (radiation) recoil force hypothesis, arising from an
anisotropic emission of waste heat generated on board the
spacecraft. This waste heat is dominated by the radioisotopic
thermoelectric generators (RTGs), but other internal heat sources
are also relevant --- particularly the electrical equipment
located inside the spacecraft compartments.

The RTGs of non-Pioneer spacecraft are mounted much closer to
their spacecraft's mainframe, e.g. Cassini, and also the New
Horizons space probe (\emph{en route} to Pluto). For all
spacecraft with radioisotopic thermoelectric generators the heat
generated by the RTGs overwhelms the tiny (Pioneer) anomalous
acceleration. Although the magnitude of the thermal radiation (or
heat) from the Pioneer S/C RTGs is much larger than the anomaly
itself [approximately 2200W vs. 63W \citep*{List_08}\footnote{In
so much as a 63 Watt collimated beam of photons will produce an
acceleration upon the S/C equivalent to the Pioneer anomaly.
\citet[ p.75]{Turyshev_10b} determine the value to be 65W.}], the
symmetric and perpendicular to the spin axis nature of the
released heat \citep*[ p.7]{Toth_09b}, is herein considered to
negate this factor as the primary cause --- although this
continues to be a subject of ongoing debate.

A web post by \emph{Symmetry Magazine} --- a joint Fermilab/SLAC
publication, concerning a presentation at the American Physical
Society's \mbox{April 2008} conference --- reported that no more
than one third of the anomaly (21W) could conceivably be heat
related. In contrast, \citet*{Bertolami_08} argued that up to two
thirds (67\%) of the anomaly's magnitude may be heat related; even
though the general consensus (in 2008) of a long-term constant
anomaly was in conflict with their stance, because over time there
is RTG radioactive decay and a (pronounced) reduction in available
electrical power. Prior to 2011, the standard response to this
(constancy) concern was to cite \citet{Markwardt_02}, who found
that a temporal variation of the Pioneer anomaly could not (at
that stage) be ruled out. As of 2011, one may now also cite
\citet{Toth_09a} and the long-awaited paper of
\citet{Turyshev_11a}: ``Support for temporally varying behaviour
of the Pioneer anomaly from the extended Pioneer 10 and 11 Doppler
data sets".

By way of their numerical analysis and discussion, \citet*[
Section 3]{Levy_09a} provide a strong counter argument to
accepting a non-constant (decaying/diminishing) anomaly; thus
supporting the stance of \citet[ Section VIII D]{Anderson_02a} ---
reaffirmed in a recent interview \citep{Anderson_11a} --- and our
stance herein, that anisotropic heat effects upon the Pioneer
spacecraft are not considered to be of great significance.

Counter to these sentiments, (most recently) the three independent
and mutually supportive models of: \citet{Francisco_11},
\citet{Rievers_11}, and \citet{Turyshev_12a} have claimed or
strongly suggested that (thermal) radiative momentum transfer
effects can \emph{fully} account for the Pioneer anomaly.
Regarding the first of these three models/papers, the comments of
Rob
Cook\footnote{\url{http://www.iop.org/careers/workinglife/profiles/page_50796.html}}
in a Physics arXiv
blog\footnote{\url{http://www.technologyreview.com/blog/arxiv/26589/}}
highlight some (potentially grave) concerns regarding the authors'
modelling of diffusive and specular reflection effects. The other
two models/papers --- simply by assuming the veracity of a ``known
physics" approach --- may well have fallen victim to the
``liberties of parametrisation" afforded to the `maker' of such a
many-faceted model, with this complexity evident in the abstract
of \citet{Rievers_11}. Subsequently, and regardless of the fact
that one or more of these models may be `correct', we shall pursue
our hypothesis that the Pioneer anomaly indicates a `real' (and
non-systematic based) phenomenon associated with some kind of
`new' (or overlooked) physics.

\subsection{Why a temporally diminishing Pioneer anomaly is not
assured}\label{Subsection:Diminishing} One would assume that the
major shift in stance from a long accepted/established (time
`variable' but) temporally long-term-mean \emph{constant} anomaly
to a temporally \emph{diminishing} (or decaying) anomaly ---
implied by \citet{Turyshev_11a} --- would be based upon a clear
and decisive scientific case for embracing the new stance over the
former stance. This aim of this section is to argue --- by citing
specific remarks in \citet{Turyshev_11a} --- that this latter
stance has not been decisively attained; nor has the previous
stance been found wanting. Indeed, ``[the] batched stochastic
[method or] model [as used by \citet{Anderson_02a}], \ldots as the
model with the most estimated parameters, it is [most] likely to
produce the best possible fit (p.2)."

Bearing in mind that: ``the [post-fit] residuals show significant
[temporal] structure (p.2)" --- which is a well credentialed
feature of the Pioneer anomaly --- and that: the ``gradually
decreasing linear and exponential decay models [and the stochastic
model] yield [only] \emph{marginally} [italics added] improved
fits when compared to the [steadily] constant acceleration model
(pp.3-4)"; this finding is only significant in that the extended
data set challenges the acceptance level of the anomaly's
long-term mean constancy. Noise and errors inherent in the data
ensure that neither case prevails.

Subsequent to \citet{Anderson_02a}, it was common sense to expect
that an extended Doppler data set would improve our understanding
of the Pioneer anomaly. However, ``the addition of earlier data
arcs [see footnote 1, p.2], with greater occurrences of [spin axis
orientation] maneuvers [and greater solar radiation pressure], did
not help as much as desired (p.4)." Additionally, the
standardisation of the heroically saved/retrieved older Doppler
data into a common format was laden with challenges and complexity
(e.g. see \url{http://www.newscientist.com/article/dn11304}\,).

Significantly, by way of the extended Pioneer 11 deep space (P11
DS) data, \citet{Turyshev_11a} declare that: ``We can [now]
exclude an anomaly directed along the spacecraft [path or]
velocity vector (p.4)", even though the quality of the additional
data is (somewhat) inferior to the latter/original data. Further,
``for Pioneer 11, the rms residuals [in 3D cf. 1D] improve when
considering an \emph{unknown} [italics added] constant force,
perpendicular to the spacecraft-Earth direction (p.3)"; and with
regard to the possible onset (or ramping up) of the anomaly
post-Saturn encounter, results for Pioneer 11's ``Saturn approach"
were disappointing:
$a_{P11}=(4.58\pm11.80)\times10^{-10}\rm~m/s^2$. Consequently, the
possibility of a path-directed anomaly is \emph{retained}.

\subsection{Primary observational evidence}\label{Subsection:Primary observational}
Any viable model, that assumes the Pioneer anomaly is real, must
satisfy the following three primary observational constraints.
\begin{enumerate}
\item\label{Observ1}{Only low mass bodies are affected. Planets,
moons, larger comets\footnote{Assuming a real anomaly,
\citet*{Whitmire_03} assert that cometary bodies of mass
$\geq10^{14}\rm kg$ ($\approx 7\rm km$ diameter, assuming a comet
density of $0.5\times 10^{3}{\rm \,kg \, m^{-3}}$) are \emph{not}
affected --- at least between 20 and 70 AU.} and larger asteroids
(whose mass is known) appear to not display the
anomaly\footnote{The 21st century NASA ephemerides, overseen by E.
Myles Standish, includes effects from over 300 `larger' asteroids.
Successful modelling of the motion of larger celestial bodies does
\textit{not} require the inclusion of the Pioneer anomaly
correction. At present, the masses of `smaller' asteroids and
comets have yet to be determined with good accuracy.} --- see
\citet[ p.41]{Iorio_07,Whitmire_03,Anderson_02a}}; and \citet*[
p.11]{Wallin_07}. \item\label{Observ2}{General constancy (of the
\emph{mean} anomalous `deceleration') and isotropy of the anomaly,
at larger heliocentric distances, i.e. $r>15$AU. The magnitude of
the anomaly\footnote{It has been common practice to express the
Pioneer anomaly (motion \emph{shortfall} rate) in terms of a
positive acceleration magnitude, and to use centimetres in the
units.} is usually expressed as:
${a_P}=(8.74\pm1.33)\times10^{-8}~\rm{cm~s^{-2}}$ which equals
${a_P}=(8.74\pm1.33)\times10^{-10}~\rm{m~s^{-2}}$.}
\item\label{Observ3}{The pre-Saturn flyby values of the anomaly
($a_{P}$), associated with the Pioneer 11 spacecraft, are much
less than the ``headline" constant value --- see Figure 7 in
\citet*{Anderson_02a}.}
\end{enumerate}

While a variety of solutions have been proposed, no model has to
date successfully addressed all three of these constraints
together. Modified general relativity, e.g.
\citet{Brownstein&Moffat_07} omits constraint \ref{Observ1},
whereas an electromagnetic wave/photon propagation effect
\citep*{Mbelek_07} omits constraint \ref{Observ3}. Other
approaches are similarly restricted; e.g. cosmological stretching
of spacetime, clock acceleration, dark matter, and
MOND\footnote{Modified Newtonian dynamics \citep{Milgrom_09}.}, to
name a few.

A surprising number of proposals restrict themselves simply to
constraint \ref{Observ2}. Such disregard for the subtleties of the
observational evidence, by way of only addressing the headline
result, is quite alarming --- in that the observational evidence
is paramount to any explanation\footnote{When the observational
evidence is unprecedented, as in constraint \ref{Observ1}, and
awkward as in constraint \ref{Observ3}, disbelief is somewhat
understandable, but not strictly an \mbox{objective/scientific}
approach --- notwithstanding the success of general relativity.}.

Assuming the observations were reliably obtained, and also not a result
of systematics\footnote{Whose causal basis is either onboard or
external to the spacecraft.}, the appeasement of these three
\textit{inconvenient} constraints (together) necessitates the
introduction of some form of ``new physics".

\subsection{Veracity and impediments to new physics.}
\textbf{\,\,\,\,\,\,\,\,What is new physics?} In this paper it is
considered to be the application of known physical principles and
concepts to a particular system and/or set of circumstances in a
previously unforseen manner.

\textbf{Why new physics?} The Pioneer 10 and 11 spacecraft are
relatively simple\footnote{They were based somewhat upon the
amazingly successful and reliable Pioneer 6, 7, 8 and 9 space
probes, although Pioneer 10 and 11 are more complicated.}, and the
quiescent\footnote{As far as radiative forces acting upon the
spacecraft are concerned, e.g. direct solar radiation pressure.
Other non-gravitational influences such as the:
Poynting-Robertson, Yarkovsky, and YORP effects do not affect the
Pioneer S/C.} outer solar system is an ideal place to test the
`truth' of gravitational theory. With the permutations of the more
likely explanations conceivably exhausted, by process of
elimination new physics becomes a distinct option. This option may
indeed be misguided, but it is the essence of science to attempt
such things, especially when ``the jury is out" on an anomalous
phenomenon.

The alternative (approach) of seeking a missing systematic merely
seeks to maintain the ``status quo". It is a conservative
approach, albeit reasonable, but not at present supported by a
convincing and progressive\footnote{Progressive in the sense of
fruitful, in that other issues/anomalies are positively affected;
i.e. benefits and advancement arise --- e.g. insight into the
Earth flyby anomaly --- as compared to just ``putting the fire
out".} explanation.

\textbf{Was it too soon (circa 2005-10) to attempt to model the
anomaly using new physics?} Certainly the detailed re-analysis of
the full Pioneer 10 and 11 data sets (then underway) was to be
insightful, especially regarding the Saturn flyby; but the
formulation and presentation of a hypothesis/model need not wait
for the final results of this analysis to be published. The
analysis covering 1987-1998 was rigorous, and it utilised the
highest quality Doppler data set.

\textbf{Does existing gravitational experimental evidence deny a
``real" Pioneer anomaly?} Authorities such as Clifford M. Will and
NASA's E. Myles Standish\footnote{Private communication.} are
inclined to be sceptical. This is understandable because General
Relativity (GR) has passed all tests to date\footnote{Including
impressively accurate tests within the solar system. For example:
Shapiro electromagnetic wave propagation delay (Cassini
spacecraft), and lunar laser ranging experiments that test the
strong equivalence principle.}. Further, the failure of modified
GR, with its flexibility of parametrization, to appease the
anomaly has led many to take refuge in a sceptical or
non-committal stance. Nevertheless, the onus lies upon the sceptic
to find a (quantifiable) chink in the observational evidence, once
a purportedly reliable scientific analysis is completed/presented.

Additionally, very accurate assessment of low mass bodies such as
spacecraft travelling at speed\footnote{The Pioneers exceed
$10~\rm{km~s^{-1}}$.} in deep space, over substantial periods of
time, is a situation \emph{not} previously (or subsequently)
encompassed by other gravitational experiments.

\textbf{Two impediments to new physics.} (1) Inadequate
appreciation of the observational evidence, and (2) conceptual
rigidity regarding the theoretical \textit{interpretation} of
physical observations. Regarding the latter, \citet[
p.808]{Rathke&Izzo_06} correctly cite that an additional
gravitational \emph{force} is in contradiction with the planetary
ephemerides\footnote{Further, and more recently, supported clearly
at Jupiter and most likely beyond in an article by E. M.
\citet*{Standish_08} titled: ``Planetary and Lunar Ephemerides:
testing alternate gravitational theories".} and the weak
equivalence principle; but to then exclude an unforseen
contributor to spacetime curvature assumes a complete and final
understanding of gravitation, \textit{and} its interaction with
quantum mechanical energy. Although apparently unlikely, something
may still be missing from our understanding that intervenes in an
unexpected manner. This paper seeks the reader's open-mindedness
to illustrate how such an oversight may (and can) exist.

\subsection{Further concerns regarding the solar system}
\label{subsection:further concerns} A brief list of concerns is
presented. These may or may not be relevant to a non-systematic
(Pioneer) anomaly. Nevertheless, they should be of at least
peripheral interest to anyone investigating the Pioneer anomaly.
\begin{enumerate}
\item{The Earth flyby anomaly \citep*{Anderson_07,Anderson_08},
involving an anomalous increase (and decrease) in kinetic energy
(from a geocentric perspective.).} \item{The timescale problem,
involving the too rapid formation of the ice giants Uranus and
Neptune --- assuming the core accretion hypothesis. See
\citet{Boss_02} amongst numerous sources. Planetary migration is
considered to appease this concern.} \item{A less than expected
number of small comets
--- i.e. less than $1\, \rm{km}$ in diameter. See \citet[ Figures 1 \&
2]{Kuzmitcheva_02}, \citet{Francis_05} and \citet*{Zahnle_04}.}
\item{The fading problem, involving a major shortfall in the
number of returning comets predicted by dynamical models of the
solar system. See \citet*{Levison_02} and \citet{Rickman_05}.}
\item{The (counter to expectations) drastic drop-off in the number
of large objects beyond 50 AU \citep*[ p.76]{Malhotra_01,Luu_02}
--- known as the ``Kuiper cliff".} \end{enumerate} Finally, note
that the extrapolation of Newtonian gravitation to galaxies
assumes that the solar system's mechanics is well understood. A
real and non-systematic based Pioneer anomaly, i.e. anomalous
motion for low mass bodies, denies this assumption and complicates
the appeasing `invocation' of dark matter --- which has not yet
been directly detected.

\subsection{Author's comments on the anomaly's status}
A model that is inconsistent with any aspect of the observational
evidence is a model that is falsified by the observations
themselves. Assuming a real Pioneer anomaly and the invalidity of
a thermal-radiation/heat based explanation, the anomaly
(conceivably) remains unexplained and in need of a suitable model.
As such, the solar system effectively becomes (in a sense) a new
``terra incognita" \citep*{Turyshev_07}.

Sceptics of a non-systematic based and unresolved Pioneer anomaly
are asked to consider the (new) model presented herein (simply) as
a \emph{contingency plan}.

The extraordinary observational evidence necessitates an
extraordinary model\footnote{Certainly, as the saying goes:
``extraordinary claims require extraordinary evidence".
Unfortunately, there is a short to medium-term inability to
improve S/C navigational accuracy and precision in the outer solar
system. Nevertheless, the current observational evidence is not
insufficient for attempts at an original model to be pursued.
Acceptance of such a model is another thing entirely.}, that will
need to be: consistent with \textit{all} the observational
evidence, predictive, progressive, and not in conflict with
physical principles, theory, and other astrophysical observations.
An alternative to the anomaly's `preferred' explanation as a
conjectured hidden (or ``dark") systematic is sought. Explaining
the apparent violation of the (weak) Equivalence Principle (in the
case of low mass bodies) is crucial.
%*************************************************************************************************************
\section{First stage modelling of the Pioneer anomaly}\label{Section:PrelimModel}
This Section begins the piecemeal task of modelling a real Pioneer
anomaly ($a_p$). A lowercase `p' subscript is used when dealing
with the model's acceleration determination(s), as distinct from
the observation based Pioneer anomaly's standard nomenclature
($a_P$).

The aim of this Section is restricted to establishing a basic (S/C
path-based, i.e. velocity vector based) model that has the ability
to match the full suite of observational evidence, with a number
of lesser observational aspects of the Pioneer analysis being
introduced. A new domain of application of \emph{known}
mathematical physics (Rayleigh's energy theorem\footnote{Also
known as Parseval's theorem.}) is introduced.

Partial theoretical appeasement of the observationally implied
model is the basis of the Section following this one (i.e. Section
\ref{Section:PhiloTheory}), which then paves the way for further
development of the model in \mbox{Section \ref{Section:general
model}.} Full quantification, and some further conceptualisation,
of the model is given in Section \ref{Section:Quantif Model}. This
Section, and the three Sections to follow, alter in stages what is
in need of explanation. The complexity of the model (to be
presented) demands such a piecemeal approach. This Section begins
with an (assumed) real $a_P$ devoid of \emph{any} suitable
explanation, and incompatible with present theoretical
constraints\footnote{Thermal recoil force may play a minor role,
but (herein) its influence is considered to comprise no more than
20\% of the anomalous acceleration ($a_P$).}.

\subsection{Loosening the subtle shackles of General Relativity}
\label{Subsection:Shackles}In this section (i.e.
\ref{Subsection:Shackles}) the initial conceptual landscape of the
model is outlined in a rough preliminary/outline form. The detail
of the model/picture shall then be added slowly, step by step.
Subsequently, the reader should not expect full clarity of the
final product at this early/formative stage; this is because the
model covers vast ground, extending into a number of different
academic territories.

At this stage we simply need to appreciate that the relationship
between General Relativity (GR) and Quantum Mechanics (QMs) has
not been finalised. Herein, General Relativity is accepted as the
correct theory of gravitation --- at least for planets, moons, and
larger asteroids and comets. The existence of a `real' Pioneer
anomaly implies that GR may be \emph{incomplete} as a theory
describing celestial motion; i.e. a minor supplementation of
spacetime curvature\footnote{In GR ``spacetime curvature" is
loosely used to describe non-Euclidean geometry.}, beyond the
scope of GR (alone), may exist --- for example involving
(celestial) three-body systems and (celestial) spin-orbit
resonance. To proceed, such a limitation needs to be provisionally
accepted. Section \ref{Section:PhiloTheory} shall more fully
investigate this assertion of limitation, and how the GR
`fortress' is structured to deny any such limitation and/or
supplementation. By process of elimination, only a \emph{system}
involving some subtle aspect of quantum mechanical [internal
(spin-orbit coupled) motion/momentum-based] energy, in curved
spacetime, can conceivably overcome this restriction.

The invariance and general covariance\footnote{Also known as
diffeomorphism covariance or general invariance.} associated with
Special and General Relativity (respectively) must not be violated
--- at least not beyond a minimum amount associated with
Heisenberg's uncertainty principle, nor above a physically
relevant quantum mechanical \emph{first} energy level. If such a
(`fractional' quantum mechanical) effect is shared by a vast
number of atoms/molecules\footnote{Of the order of $10^{50}$, in
the case of large solar system moons --- indeed a ``large
number".} the sub-quantum or \emph{virtual} energy involved is not
insignificant. Later, we shall see that a significant amount of
quantum mechanical (QM) excess (virtual) internal energy is
necessarily expressed externally (and singularly) as a large-scale
rotating warp-like curvature of space. One side of this
``space-warp\footnote{In the sense of aeroplane wing warping,
which involves twisting in opposite directions, but involving
space `deformation'.}" (or space-fold\footnote{In the geological
sense of the word `fold', which implicitly assumes an initial
surface that is essentially level.}) is above, and the opposite
side is below, the equilibrium curvature of spacetime, so that on
average (and `overall') there is no deviation from the background
gravitational field's strength.

It is well known that in GR gravitational energy cannot be
localised. We shall exploit the logical loophole that when QMs is
involved, there is no impediment to a (supplementary) energy
associated with spacetime curvature being non-localised in some
manner\footnote{As Paul Davies says: ``\ldots QMs is `non-local'
because the state of a quantum system is spread throughout a
region of space \citep[ p.20]{Davies_04}."}. In what follows, we
need to clearly distinguish (and `divide out') mass from
gravitational acceleration and (then) mass from \emph{specific}
energy. By way of quantum mechanics (QMs), the notion of non-local
(or distributed\footnote{As is the case with force fields, energy
fields, and wave phenomena.}) `mass' demands
consideration\footnote{As such, this non-local mass (occupying
space) is not considered `matter' \emph{per se}.}.

In these unusual and highly restrictive supplementary
circumstances, we may: (1) (fortunately) neglect high-speed
\emph{special} relativistic effects; and shall (2) give credence
to the notion of ``the total energy of space curvature" --- but
only as regards the supplementary space curvature being pursued.

\subsection{Pioneer Anomaly as a shortfall relative to \emph{predicted}
motion}\label{Subsection:Shortfall}
\subsubsection{Introductory remarks} Observational evidence
rules out any additional direct \emph{force}, but the conceptual
notion of (intrinsic) curved spacetime remains viable. Awkwardly,
the three observational constraints presented in section
\ref{Subsection:Primary observational} effectively shrink any
`conceptual modelling space' to zero, i.e. no conceivable model.
To proceed with a (non-systematic based) ``real anomaly"
hypothesis, at least one unconventional extra or missing
ingredient is required.

One, and possibly the only, viable way forward is to conceive of
the anomaly as a (constant rate of) shortfall in predicted motion,
associated with an additional and unforseen perturbation of
spacetime curvature --- whose influence is restricted to lesser
masses. How can this be achieved in our solar system?

Initially, we shall employ the idealisation of spacecraft (S/C)
motion as a radially (i.e. outwardly) directed point mass, in
conjunction with a systemic reference frame centered at the solar
system barycentre\footnote{Note that the \emph{rate} of time (for
the system as a whole) can be either as per the Earth's surface
(Terrestrial time), or as per the barycentre (Barycentric
Coordinate Time).}. Subsequently, and in conjunction with the fact
that supplementary curvature is to be determined by way of an
energy magnitude\footnote{Note that atomic/molecular mass and (a
minimum) angular momentum (over a cyclic process time) are
involved in this energy magnitude. The reduced Planck constant
($\hbar$), i.e. Dirac's constant, shall also play a key role.}, we
restrict the visual conceptualisation of the analysis to curved
space cf. curved spacetime.

We examine the difference between a `steadily' moving mass and one
additionally subjected to a supplementary (and global) undulatory
``gravitational" field\footnote{The term ``gravitational" is being
used in its broadest sense; thus it includes hypothesised and
unforseen additional space curvature --- which may not necessarily
be representable by GR's `manifestation' of non-Euclidean
geometry. Later, to avoid confusion, the expression
``acceleration/gravitational field" is usually preferred. The
expressions ``supplementary gravitational field" and
``acceleration field" are occasionally used.}. It shall be seen
that a difference in translation motion arises between the two
cases --- as ``observed" from/at (or near) the solar system
barycentre. Actually, the Pioneer anomaly is modelled as a simple
linear superposition of a (multiple) number of these effects
(subsection \ref{subsubsection:a_p and mass comment}), but the
bulk of this section is concerned with examining a \emph{single}
instantiation of (such) an undulatory acceleration/gravitational
field.

It shall become evident that the covariant formalism of GR is
unsuited to describing this situation, and cannot attain the
ensuing (systemic and) barycentic result. The alternative, and
simpler, point-based formalism of Eulerian (or Continuum)
Mechanics, as used in fluid mechanics for example, is required. We
shall now examine the kinetic energy of, and the \emph{specific}
gravitational energy affecting, a moving point mass (\emph{at} the
point mass) over a \emph{single} oscillatory cycle.

\subsubsection{Conceptual aspects of a shortfall in motion relative to
predicted motion} For a moving body, an unsteady (i.e. undulating
around a mean value) acceleration/gravitational field, on top of
the standard gravitational field, necessarily induces a component
of unsteady motion. These newly postulated field undulations are
seen to lie or ``piggy-back" \emph{upon} the pre-existing
gravitational field, much in the manner of ocean waves upon an
otherwise calm sea. Clearly, these undulations are
\emph{completely unrelated to GR's gravitational waves}, and their
influence relates especially to the long-term behaviour of (low
mass) moving bodies. In what (immediately) follows, it is (a
\emph{single} idealised instantiation of) this new/supplementary
field perturbation that shall only concern us --- not the
underlying pre-existing gravitational field.

The following shall show that from a barycentric/systemic
perspective a supplementary (oscillatory)
acceleration/gravitational field leads to a \emph{redistribution}
of some steady translational energy, into a \emph{longitudinal}
oscillatory motion (about a mean speed); which then detracts from
the (maximum possible) translational speed of the moving body.
Such a scenario results in a shortfall (per oscillation cycle),
when \emph{compared to} predictions that omit or overlook the
unsteady `acceleration' field aspect. Consideration is initially
restricted to a moving point mass subject to a (single and pure)
sinusoidal field influence. The shortfall directly \emph{opposes
the} (linear outward) \emph{path} and/or velocity vector of the
motion.

We shall see that consistency with GR's invariance requirements
necessitates the field (acceleration) undulations have constant
amplitude throughout space. Subsequently, energy dispersion, with
increasing radius from the undulation source, necessarily arises
from a `new' variable and distributed/`non-condensed' (non-local)
mass aspect that has a precise value at all places/points in the
field\footnote{We shall see later (in section
\ref{Subsection:warp's mass}) that the value of the supplementary
field's ``non-local mass", whose existence is related to quantum
mechanical \emph{non-locality}, is inversely proportional to the
spherical volume associated with a radius from the energy source.
Also of interest shall be the conditional response of ``inertial
spacecraft mass" --- which incidentally always equals (passive and
active) gravitational mass --- to the field's acceleration
undulation magnitude \emph{and} the non-local mass value `at' the
spacecraft. Subsequently, only bodies whose mass is below the
field's (non-local) ``cut-off mass" will experience the
retardation effect.}. If multiple sources of such undulations
exist, then their (linear) superposition may explain the Pioneer
spacecraft observations --- although their sum will need to mimic
a roughly constant (i.e. low variance) effect, and account for
Pioneer 11's low $a_P$ value around 6AU (pre-Saturn flyby). These
characteristics markedly constrain any prospective model.

To validate this scenario, firstly the spacecraft's (specific)
kinetic energy `shortfall' needs to be quantified; and secondly,
the (until recently) generally accepted ``pure-constancy" of the
anomaly shall need to be found wanting. The former follows and the
latter is clarified in sections \ref{Subsection:Varying},
\ref{Subsection:Approx annual} and \ref{Subsection:Molding}.

The global or systemic distribution of the additional space
curvature/undulation, and the nature of the driving mechanism that
generates the undulation, are established later in the paper. The
reader is asked to consider that both of these are physically
viable at this stage (see subsection
\ref{subsubsection:Forward_reach} for further comment).

\subsubsection{General preliminaries, idealisations, and undulation
features of the model} \label{subsubsection:intra-undulation} The
following \emph{idealisation} suits the Pioneer 10 and 11
spacecraft's motion quite well. Initially, a simple linear
(one-dimensional) situation is considered. The approach is
decidedly classical, even though it utilises the notion of a
(deformable) space continuum. Note that the (deeper) nature of the
field strength's sinusoidal variation is elucidated in subsection
\ref{subsubsection:Space-warp}.

\begin{enumerate}
\item{The Pioneer 10 and 11 spacecraft are considered to be in
purely linear-radial (outward, one dimensional) motion from the
Sun/Barycentre\footnote{Later we shall see that S/C (and ``low
mass" bodies) in non-radial motions are equally affected, but
line-of-sight \emph{observations} of these non-radial motions are
less than (or equal to) the actual magnitude of the (path-based)
effect.}.} \item{Barycentric: radius ($x$), speed ($v$), and
Sun-dominated gravitational acceleration ($g$) are very much
greater ($>>$) than undulation-based perturbation magnitudes
($\Delta x,~\Delta v,~\Delta g$) around the mean $x$,~$v$, and $g$
--- over cycle time ($\Delta t$).} \item{Undulation time ($\Delta
t$) is considered to be (celestially) small/short. It is of the
order of a week (7 days).} \item{Change in $g$ over $\Delta t$ is
considered negligible --- indicative of a very low gradient
gravitational field/well (in the S/C's immediate vicinity).}
\item{Thus, barycentric speed ($v$) is considered effectively
constant over $\Delta t$ --- in the absence of a ($\Delta g$)
undulation. It is of the order of $10~\rm{km~s^{-1}}$, and as such
$v<<c$ (the speed of light).} \item{The (supplementary)
acceleration/gravitational field undulations are assumed to exist
on a large scale, i.e. the order of or greater than solar system
size.} \item{Subsequently, an idealised \emph{equivalence} can be
envisaged. Over $\Delta t$ the field's undulation at the
\emph{moving} spacecraft approaches the acceleration/gravitational
field's undulation at a position \emph{fixed} with respect to the
barycentre.} \item{Thus, our idealisation permits an equivalence
between (supplementary) gravitational (field) undulations and an
objects's physical (or proper) acceleration\footnote{I.e. (an
object's) measurable acceleration --- achievable by way of an
onboard accelerometer.}, such that: $\Delta g=\Delta a$. From here
on we use $\Delta a$ to signify both.} \item{Let us also
distinguish between two types of speed and acceleration
perturbation: a \emph{local} variation relative to the
\emph{moving} body's mean value, and a perturbation as measured
from the (global/systemic) \emph{inertial} and \emph{fixed}
barycentre.}
\end{enumerate}

Of primary interest is whether the spacecraft's barycentric mean
speed ($v$) remains unchanged over $\Delta t$, in the presence of
a ($\Delta a$) field undulation/sinusoid --- that induces a
`measurable' speed sinusoid (magnitude $\Delta v$) at the
spacecraft. In other words, by way of a cyclic field perturbation,
has a barycentric speed shortfall ($\delta v$) occurred (after
and) during $\Delta t$?

\subsubsection{Briefly introducing celestial simple harmonic
(proper) motion}\label{subsubsection:CSHPM} We now briefly
consider the qualitative effect of the introduction of a
\emph{single} (sinusoidal) field undulation upon the motion of a
mass moving in otherwise uniform motion. In general, the sign of
the \emph{field} acceleration undulation/perturbation (relative to
a mean value) opposes the sign of the \emph{spacecraft's} speed
perturbation\footnote{The variation of space curvature, i.e.
(supplementary) acceleration/gravitational field strength, may
also be thought of as a sinusoidal variation in a `tilt' angle
(i.e. small inclination) --- around its mean or flat equilibrium
condition --- \emph{encountered} by the moving spacecraft. Maximum
tilt angle (representing maximum field strength) is associated
with minimum spacecraft speed and vice versa. Alternatively,
imagine a fast moving ice skater on a frictionless (ice) surface
comprising a series of sinusoidal (small height) undulations. As
they reach the top of each hill (representing maximum field
strength and tilt angle) their speed is minimised, whereas at the
bottom (\emph{representing} minimum field strength and extreme
negative tilt angle) their speed is maximised. Stripping out the
overall mean speed leaves a sinusoidal speed
perturbation/undulation, in the same manner that we have removed
the mean \mbox{speed ($v$)} of the moving spacecraft to only
consider the speed undulations/sinusoids (amplitude $\Delta v$) of
the spacecraft.}. The acceleration/gravitational field
perturbation/sinusoid
\mbox{$a_{\rm{field}}=a_{\rm{field}}(t)=\Delta a \sin(\omega
t-\varphi)$} induces a (proper) speed perturbation
\mbox{$v_{\rm{proper}}=-\Delta v \sin(\omega t-\varphi)$}; which
in turn can be associated with a (proper) position perturbation
\mbox{$x_{\rm{proper}}=\Delta x \cos(\omega t-\varphi)$} and an
acceleration undulation \mbox{$a_{\rm{proper}}=-\Delta a
\cos(\omega t-\varphi)$.} We note that the phase of
$a_{\rm{proper}}$ lags $a_{\rm{field}}$ by $90^o$ or $\pi/2$
radians, and that the phase of $v_{\rm{proper}}$ lags
$a_{\rm{field}}$ by $180^o$ or $\pi$ radians.

The phase of each of the \emph{four} sinusoids (at a given time)
is offset by $90^o$ relative to their `neighbouring' sinusoids ---
which (with $360^o$ in a full cycle) is arguably an elegant `set'
of circumstances. Upon correcting for non-sinusoidal motions also
influencing the spacecraft\footnote{Primarily the spacecraft's
motion along its trajectory, and (later) also the `real' monotonic
(Pioneer) anomalous acceleration (i.e. anomalous rate of speed
change).}, the spacecraft moves `to and fro' about a \emph{mean
position} (and mean speed and acceleration). From the perspective
of a barycentric coordinate system, this mean position is in
(translational) motion i.e. non-stationary. Similarly, Earth-based
(inertial frame-based) simple harmonic motion neglects extraneous
motions associated with its location, e.g. motion associated with
the spin of the Earth on its axis.

Note that there is no restoring \emph{force} (directed towards the
mean position) associated with the (spacecraft's speed or
position) undulation itself as is the case with the position
undulation of a spring, pendulum or ocean wave for example.
Nevertheless, the spacecraft's (position, speed, and acceleration)
undulations are considered to be driven, by way of the
supplementary gravitational field undulation --- which also
possesses an associated total energy. The (immediate) basis,
rather than the (ultimate) source or nature, of this
acceleration/gravitational field undulation is the ``rotating
space-warps" (superficially) introduced in section
\ref{Subsection:Shackles}.

\subsubsection{Agenda and (relative) undulation maximum amplitude
relationships}\label{subsubsection:rel undu ampl} The mathematical
relationships between: the \emph{hypothesised}
acceleration/gravitational field undulation amplitude ($\Delta
a$), speed undulation amplitude ($\Delta v$), translational speed
shortfall per cycle ($\delta v$), and the \emph{rate} of
translational speed shortfall ($\delta a$) are to be established
in a systemic reference frame. It will be eventually shown (noting
that $\delta a_{\rm{spacecraft}}=\delta a_{\rm{s/c}}$) that:
\begin{displaymath}
\Delta a_{\rm field}^2=\delta a_{\rm proper}^2 \,
({\rm{and~in~idealised~1D~}} \Delta a_{{\rm field}}^2=\delta
a_{\rm s/c}^2)
\end{displaymath} and thus --- assuming we are dealing with a
\emph{linear} system --- superposition of the newly proposed
field's wave-like nature\footnote{Note that wave `height' is
somewhat analogous to the field's (spatially-constant)
`acceleration' amplitude/magnitude.} and effect upon spacecraft
motion is conceptually non-problematic\footnote{The physical
validity of this superposition is further discussed in subsection
\ref{subsubsection:field to S/C} and elsewhere.}.

To proceed we shall need to make use of the following two
(\emph{magnitude only}) amplitude relationships, that are based
upon $v_{\rm{proper}}=v_{\rm{proper}}(t)=-\Delta v \sin(\omega
t-\varphi)$ and (celestial) simple harmonic (proper) motion:
\begin{equation}\label{eq:amp1}
\Delta a=\omega\Delta v
\end{equation}
\begin{equation}\label{eq:amp2}
\Delta v=\omega\Delta x
\end{equation}
A constant $\Delta a$ amplitude and constant $\omega$ ensure that
the $\Delta v$ and $\Delta x$ amplitudes are also constant, and
(as such) they are not functions of time.

Equation \ref{eq:amp1} has been discussed in \citet[ p.37 Section
IX, Part C]{Anderson_02a}. Its form relates to Equations 1, 13 and
50 in \citet{Anderson_02a}. Note that \citet[ Equation
50]{Anderson_02a} uses $A_0$ (cf. $\Delta a$) to denote (apparent
angular) acceleration amplitude, and $\Delta v$ represents the
(amplitude of) change/variation in the \emph{line-of-sight}
Doppler velocity data. \emph{Constant} angular frequency
($\omega$) was indicative of either an Earth day based
($\omega_{d.t.}$) or an Earth year based ($\omega_{a.t.}$) angular
velocity. For \citet[ p.38]{Anderson_02a} \mbox{Equation
\ref{eq:amp1}} was representative of ``errors in any of the
parameters of the spacecraft orientation with respect to the
chosen reference frame", e.g. Earth equator orientation; and as
such, these \emph{Earth-based} errors `generate' undulatory
(idealised to sinusoidal) diurnal and annual residuals in the
Doppler data. In section \ref{Subsection:Approx annual} we
re-access this account of the annual residual, and discuss a major
concern with the \emph{magnitude} ascribed to it.

\subsubsection{Relative undulation amplitude relationships:
for two different cases}\label{subsubsection:real_relative} A
crucial aspect of our hypothesis of a real Pioneer anomaly is that
these relationships (Equations \ref{eq:amp1} and \ref{eq:amp2})
are seen to also apply to: real Pioneer spacecraft (i.e. physical
or proper) motion fluctuations ($\Delta v$) --- \emph{at the
spacecraft}. The (constant) angular frequency ($\omega$) is the
magnitude of a (vector) angular velocity associated with an (as
yet to be clearly discussed) rotational (space-warp) mechanism
(see subsection \ref{subsubsection:Space-warp}). This mechanism
produces a sinusoidal variation in the acceleration/gravitational
field strength (amplitude $\Delta a$); and subsequently a
sinusoidal undulation in proper motion speed ($\Delta v$),
position/`range' ($\Delta x$) and proper acceleration (also
$\Delta a$). Recall that these are undulatory/oscillatory
\emph{amplitude-only} expressions.

Indeed, the kinematics of the moving body's sinusoidal
variation/perturbation in position, speed, and acceleration ---
relative to a `comoving' (equilibrium) reference
frame/point\footnote{Alternatively, a frame not influenced by the
undulatory gravitational/accelerational field effect.} --- is
essentially that of (one dimensional) simple harmonic
motion\footnote{Particularly the fact that the undulatory proper
acceleration of the spacecraft is proportional to, and oppositely
directed to, its displacement from a (non-stationary) ``mean
position".}; but the kinetics and dynamics involved in the
situation presented are quite different. At this stage, not even
the $\Delta v$ value can be unambiguously related to
barycentre-based, or (with appropriate corrections) Earth-based,
observations. We shall examine two different cases.

Firstly, for the no speed shortfall case ($\delta v=0$) ---
designated ``case 1" --- we accept that in addition to:
\begin{equation}\label{eq:equal a's} \Delta g=\Delta
a_{\rm{field}}=\Delta a=\Delta a_{\rm{proper}}
\end{equation} we also have --- with the provision/caveat that mean
speed $v$ is removed and $\Delta t$ is relatively short --- that:
\begin{equation}\label{eq:caveat} \Delta a_{\rm{proper}}=
\Delta a_{\rm{spacecraft}}\,\,(=\Delta a_{\rm{s/c}})\end{equation}
Thus, we have declared that \emph{observed} spacecraft
acceleration (perturbation) amplitude $\Delta a_{\rm{s/c}}$,
\emph{after} correcting for actual spacecraft (S/C) motion, is
also equal in magnitude to proper acceleration amplitude $\Delta
a_{\rm{proper}}$\,. The sinusoids associated with $\Delta
a_{\rm{field}}$ and $\Delta v_{\rm{s/c}}$ are 180 degrees out of
phase, with this being the primary \emph{physical} relationship;
whereas the lesser physical relationships involving sinusoidal
variations/amplitudes of $\Delta x_{\rm{s/c}}$ and $\Delta
a_{\rm{s/c}}$, relative to $\Delta v_{\rm{s/c}}$, are both 90
degrees out of phase (with respect to the latter) --- as is the
case with standard simple harmonic motion.

In terms of `causation' of events/effects we have: $\Delta
a_{\rm{field}}$ $\Rightarrow$ $\Delta v_{\rm{s/c}}$ $\Rightarrow$
$\Delta a_{\rm{s/c}}$ and $\Delta x_{\rm{s/c}}$; and in terms of
$90^o$ ($\pi/2$) phase offset jumps, the sequence (reference to
lagging) is `signified' by: $a_{\rm{field}}$, $a_{\rm{s/c}}$,
$v_{\rm{s/c}}$, then $x_{\rm{s/c}}$ (noting/recalling the four
interrelated time dependent sinusoidal expressions given in
subsection \ref{subsubsection:CSHPM}).

Secondly, in the case of (our hypothesised) monotonic spacecraft
speed shortfall $(\delta v\neq0)$, these sinusoidal
amplitude/magnitude equalities still apply, but only if the $\delta
v$ monotonic effect has \emph{also} been removed; i.e. compensated
for in observed $\Delta v_{\rm{s/c}}$ and $\Delta a_{\rm{s/c}}$ ---
in addition to the (previous) removal of the mean speed of the
spacecraft (over suitably short time scales). Later, we shall refer
to this (speed shortfall) case as ``case 2".  We note that $\delta
v<<v$.

By way of the adjustments associated with \mbox{case 1} and case 2
(alternatively situation-1 and situation-2), we've ensured that we
are simply dealing with $\Delta a_{\rm{field}}$ and three
(celestial) simple harmonic motion amplitudes. With this
restriction in place, we have: $\Delta v_{\rm{proper}}=\Delta
v_{\rm{s/c}}$ and $\Delta x_{\rm{proper}}=\Delta x_{\rm{s/c}}$, as
well as: $\Delta a_{\rm{field}}=\Delta a=\Delta
a_{\rm{proper}}=\Delta a_{\rm{s/c}}$. Subsequently, the various
subscripts may be dropped and we may simply use the nomenclature
of: $\Delta a$, $\Delta v$, $\Delta x$, and $\delta v$ from here
on --- unless we seek to highlight a particular feature of the
model.

\subsubsection{Celestial (simple) harmonic motion and its influence
upon average speed} We (once again) enquire: does the
\emph{average} translational speed ($v$) of a body change when
subjected to this undulatory acceleration field effect? In
(\emph{symmetric}) GR the default answer is `no'. The
Eulerian-like analysis of subsection \ref{subsubsection:Rayleigh
Power}, which examines conditions at a moving point mass, implies
the purely relative (barycentre-to-S/C) relationship of
\mbox{Equation \ref{eq:amp1}}, as employed in \citet{Anderson_02a}
and case 1, has neglected a (system-based) shortfall in motion
($\delta v$) --- as mentioned in case 2. Note that this shortfall
is quantified by way of a comparison to circumstances in the
\emph{absence} of an acceleration/gravitational field undulation;
i.e. it is quantified relative to predicted/expected
(non-anomalous) spacecraft motion.

We shall assume that it is the spacecraft --- by way of a new
supplementary field --- and \emph{not} an Earth-based error [as
assumed by \citet{Anderson_02a}], that is responsible for both the
$\Delta v$ oscillation signal \emph{and} the (rate of) speed
shortfall ($\delta a=\delta v/\Delta t$). Thus, we (need to)
appreciate that Equation \ref{eq:amp1} only describes the
oscillation/undulatory aspect of what is actually a two pronged
(overall) speed variation --- as measured by an Earth-based
observer or (in our idealised case) a barycentric-based observer.

In the next subsection (\ref{subsubsection:Rayleigh Power}) a second
(and new) relationship, between field undulation amplitude ($\Delta
a$) and spacecraft speed variation is proposed; with the latter
involving the monotonic $\delta v$ shortfall as compared to the
sinusoidal/undulatory $\Delta v$ variation. This (monotonic aspect)
requires the S/C be considered as moving \emph{relative to the
barycentre}, which is the fixed central point of a solar-systemic
reference frame. Importantly, recalling subsection
\ref{subsubsection:intra-undulation}, our point of analysis
`attached' to the moving body, effectively experiences the same
undulation of acceleration/gravitational field as a fixed point in
the vicinity --- over the (short duration) cyclic undulation time
$\Delta t$.

\subsubsection{Rayleigh's Power Theorem in curved space (at a point
mass)}\label{subsubsection:Rayleigh Power} Rayleigh's Energy
Theorem is sometimes known as Parseval's theorem, and \emph{vice
versa}. This theorem is an energy conservation theorem and (for a
continuous function) may be written as:
\begin{displaymath}
\int_{-\infty}^{\infty}x^{2}(t)\,dt=\int_{-\infty}^{\infty}|X(f)|^{2}\,df
\end{displaymath}
It says that a signal\footnote{The terminology describing Fourier
transforms is biased towards electrical engineering interests.}
contains the same amount of energy regardless of whether that
energy is computed in the time domain or in the frequency
domain\footnote{An alternative interpretation is that: the total
energy contained in a waveform $x(t)$ summed across all of time
$t$ is equal to the total energy of the waveform's Fourier
transform $|X(f)|$ summed across all its frequency
\mbox{components $f$.}}.

\citet[ p.120]{Bracewell_00} points out that: ``\ldots the theorem
is true if [only] one of the integrals exists." and he reassures
us that Fourier transforms of physically real, yet finite
sinusoidal waves, do mathematically exist.

Rayleigh's theorem, by way of ``the power theorem", is also
applicable to the rate of \emph{energy transfer}, or power, of a
`signal'~\citep[ p.121]{Bracewell_00}.

Applying this mathematical equality to a space curved into
sinusoidal undulations\footnote{By way of (for example) the
(constant angular velocity) ``rotating space-warps" briefly
introduced in section \ref{Subsection:Shackles}.}, and the
resultant motion of a moving point mass\footnote{We shall need to
examine the role of mass more closely at a latter stage; but for
now, only specific energy, acceleration and speed are considered.}
(with a \mbox{systemic/global} reference frame implicit), implies:
\begin{displaymath}
\int_{-\infty}^{\infty}{a^{2}}(t)\,dt=\int_{-\infty}^{\infty}|{v}(f)|^{2}\,df
\end{displaymath}
One can distinguish two \emph{physical} aspects: the moving point
mass and the (sinusoidal) field acting upon it. Ignoring the
steady (average/mean) power of the ``signal" and restricting
ourselves to the `power' in the undulations, this mathematical
coexistence relationship appears to also describe a physically
relevant (coexistence) relationship --- which we can alternatively
think of as a (one-way) \emph{`transfer'/causal} effect. Note that
the (disregarded) mean acceleration field and mean speed terms are
indicative of conditions arising from standard gravitational
theory, i.e. general relativity.

It is hypothesised that: when a body, moving with a
constant/steady translational speed, is subjected to (i.e.
affected by) a \emph{single} (constant amplitude) sinusoidal
acceleration/gravitational field undulation ($\Delta a=\Delta
a_{\rm{field}}$), over a \emph{solitary} cycle (time $\Delta t$),
the power theorem implies:
\begin{equation}\label{eq:undulate}
\frac{1}{2}\Delta a^{2}\Delta t=\frac{1}{2}\delta v^{2}f
\end{equation}
where $f=\Delta t^{-1}=\omega\,(2\pi)^{-1}$ (a single exact
value), and the unit in (the per cycle relationship that is)
\mbox{Equation \ref{eq:undulate}} is $[\rm{m^2}/{s^3}]$ i.e.
\emph{specific} power. Recall that: the point of analysis is in
motion within a much larger system. Note that: $\frac{1}{2}\delta
v^{2}$ is a fixed magnitude quantity, and because negative
frequency ($-f$) is non-physical it has been neglected ---
effectively halving the (mathematical) value of the
frequency-based `integral'.

Physically, we consider $\frac{1}{2}\delta v^{2}$ (i.e.
$\frac{1}{2}\delta v^{2}_{\rm{s/c}}$) to be indicative of the
specific (kinetic) energy associated with the body/spacecraft's
\emph{unsteady} motion energy component; and $\frac{1}{2} \delta
v^2$ is also seen to be indicative of a (barycentric referenced)
kinetic energy `shortfall' (per $\Delta t$ cycle) relative to a
fully-steady motion case. Thus, $\delta v$ is seen to represent a
speed shortfall (per $\Delta t$ cycle), relative to (predicted)
fully-steady motion --- which assumes the absence of an
undulatory/sinusoidal gravitational/accelerational field effect.
Note that the body/spacecraft's total kinetic energy is
predominantly in ``steady" (i.e. non-oscillatory) motion.
Rearranging Equation \ref{eq:undulate} gives:
\begin{equation}\label{eq:specific energy}
\Delta e=\frac{1}{2}\Delta a^{2}\Delta t^{2}=\frac{1}{2}\delta
v^{2}
\end{equation}
where this common specific energy aspect ($\Delta e$) of the point
mass motion (is seen to) also physically quantify the
\emph{specific} (`potential') energy of the supplementary
oscillatory acceleration/gravitational field (\emph{per cycle}),
i.e. the `rotating space-warp' mentioned previously\footnote{Note
the similar expressions for specific (field energy) $\Delta
e=\frac{1}{2}\Delta t^{2}\Delta a^{2}$ and the (specific)
\emph{total mechanical} energy of (spring-based) simple harmonic
motion:
$e=\frac{E}{m}=\frac{1}{2}\frac{k}{m}A^2=\frac{1}{2}\omega^2A^2$
--- where $A$ is the amplitude (i.e. maximum \emph{displacement} from
the equilibrium position).}.

The compound motion\footnote{Steady \emph{and} unsteady motion,
arising from standard GR based celestial motion and the new
supplementary field undulations respectively.} involved and the
ensuing asymmetric resultant motion effect ($\delta v$) is beyond
both: GR's conceptual and representational scope, and Newtonian
gravitation's perceived content.

\subsubsection{Physical interpretation of the Power Theorem
(over a single cycle)}\label{subsubsection:Interpret power}
Physically, introducing the field undulation redirects some
``given" steady kinetic energy\footnote{Arising from rocket
propulsion and a gravitational assist manoeuvre at Jupiter --- in
the case of Pioneer 10 cf. Jupiter \emph{and} Saturn in the case
of Pioneer 11.} to unsteady energy, and thus the \emph{initial}
steady motion is incrementally decreased \emph{each cycle} (by
$\delta v$). The field has effectively transformed, by way of a
redistribution, a small proportion of the body's (pre-cycle)
energy of motion.

\begin{figure}[h!]
\centerline{\includegraphics[height=13.5cm,
angle=0]{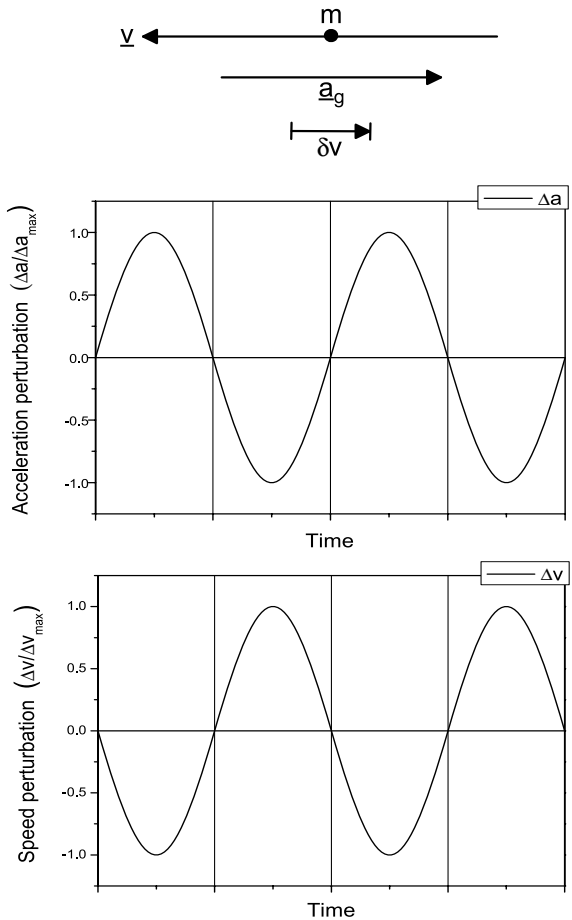}}
\begin{center}
\caption[The vector schematic diagram (very top) shows the
relative magnitude and direction of a body's motion and the speed
loss ($\delta v$) (per cycle). The (lower) two diagrams illustrate
the fact that oscillatory acceleration/gravitational \emph{field}
perturbation amplitude ($\Delta a$) and the coexisting speed
perturbation amplitude ($\Delta v$) of a moving body are 180
degrees out of phase.]{The vector schematic diagram (very top)
shows the relative magnitude and direction of a body's motion and
the speed loss ($\delta v$) (per cycle), with the (approximate)
direction of the Sun's gravitational acceleration also shown ---
assuming the body is moving outwards from the barycentre. The
(lower) two diagrams illustrate the fact that oscillatory
acceleration/gravitational \emph{field} perturbation amplitude
($\Delta a$) and the coexisting speed perturbation amplitude
($\Delta v$) of a moving body are 180 degrees out of phase. Thus,
a stronger acceleration/gravitational field (relative to its
average value) results in a reduced speed for the moving body ---
and \emph{vice versa}.} \label{Fig:VelocityAccel}
\end{center}
\end{figure}

Another way of saying this is: since the equilibrium `point' of
the longitudinal motion oscillation is stationary (relative to the
steady motion), the unsteady energy cannot contribute to directed
translational energy. Note that \emph{within} each
(undulatory/oscillatory) cycle a partial analogy to pendulum
motion is apparent, in that increasing gravitational potential
energy detracts from kinetic energy and \emph{vice versa}
--- with the (constant) speed oscillation amplitude ($\Delta v$)
being quite distinct from the asymmetric (and monotonic)
retardation effect ($\delta v$) attributed to the existence of the
\emph{full} cycle (see \mbox{Figure \ref{Fig:VelocityAccel}).}
Note that, the same loss of (steady) translation speed occurs for
each and \emph{every} cycle. In brief, constant amplitude $\Delta
a$ gives a constant $\delta v$ shortfall per ($\Delta t$) cycle
(i.e. a constant $\delta a$).

%********************************** Fig:VelocityAccel ******************************
%********************************** end of Fig:VelocityAccel  *************************

It is also worth noting that for a frictionless pendulum or spring
the system's \emph{total} energy remains constant, whereas for a
spacecraft in undulatory curved space, a (simple) constant total
energy notion (K.E.+P.E.) is no longer physically meaningful. From
a barycentric perspective, \emph{the system} (i.e. frictionless
S/C motion) \emph{is} (effectively) \emph{dissipative}, with
`given' (predominantly steady) K.E. being incrementally eroded
(per cycle) by the unsteady (energy) component of the motion.
Further, Equation \ref{eq:specific energy} indicates that (the
perturbation based) `specific gravitational field energy'
($\frac{1}{2}\Delta a^{2}\Delta t^{2}$) and specific kinetic
energy loss per cycle ($\frac{1}{2}\delta v^{2}$) have
\emph{equal} magnitudes.

\subsubsection{Two different ways in which celestial simple
harmonic motion occurs}\label{subsubsection:2 ways to CSHM} Before
we establish a link between systemic (translational) speed
reduction per cycle (i.e. $\delta v$) and (constant) speed
undulation amplitude ($\Delta v$) in subsection
\ref{subsubsection:amplitude to shortfall}, we shall need to
distinguish two quite distinct physical situations associated with
this unsteady motion; which is essentially a (perturbation-based)
``celestial simple harmonic motion"\footnote{Note that the orbital
motion of planets around the Sun, and moons around a planet can be
considered as a means of inducing a sinusoidal motion perturbation
upon a freely moving body in deep space. This \emph{distance
dependent} amplitude scenario is negligible cf. the (similar)
\emph{constant amplitude} celestial \emph{simple} harmonic motion
being proposed.} (CSHM).

Firstly, we can have a \emph{pseudo effect} arising from the
(one-dimensional) projection of (errors associated with) uniform
(Earth-based) `circular' motion\footnote{For example: ``The
transformation from a Earth fixed coordinate system to the
International Earth Rotation Service (IERS) Celestial System is a
complex series of rotations that includes precession, nutation,
variations in the Earth's rotation (UT1-UTC) and polar motion
\citep[ p.57]{Turyshev_10b}."} onto the observed/apparent position
(and motion) of a spacecraft. This results in a (pseudo)
sinusoidal position variation, that upon differentiation gives the
associated speed and acceleration `sinusoidal' oscillations. The
Pioneer's diurnal and annual variations are generally seen to be
based upon such a scenario.

Secondly, we have a \emph{real effect} arising from the influence
of the (constant amplitude) sinusoidally varying
acceleration/gravitational field associated with one of a number
of (constant angular frequency) rotating space-warps (RSWs). Both
situations, but more so the first situation, are largely analogous
to the relationships that describe spring-based simple harmonic
motion; notwithstanding the fact that Newton's second law and
Hooke's law play no direct role. In Einstein's gravity there is no
\emph{force} of gravity (\emph{per se}), only spacetime curvature.
In both \emph{situations} we retain the `truth' that: the
(perturbation) acceleration of an object is proportional to its
(perturbation) displacement and is oppositely directed --- which
is a defining characteristic of simple harmonic motion (SHM).

In the first (pseudo) situation the trigonometric differentiation
process (to speed and then acceleration `sinusoids') has its basis
in a position/displacement-based sinusoid, whereas in the (second)
situation/scenario a differentiation from a (proper) speed
sinusoid to (proper) acceleration sinusoid is paramount --- with
this (further) relying on the equivalence of $\Delta
a_{\rm{field}}$ and $\Delta a_{\rm{proper}}$ magnitudes. In the
second situation, which is pivotal to our explanation/modelling of
the Pioneer anomaly, the integration from a speed sinusoid to a
displacement `sinusoid' (around a `moving' mean position) is of
lesser importance in the model.

With particular emphasis upon the second (RSW-based) situation, a
subtlety involving (specific) energy arises --- regarding its
analogy to (spring-based) SHM; particularly concerning the role of
terms/quantities involving rotation. Firstly, for springs, we note
that: `$\rm{A}$' designates the amplitude (i.e. the maximum
\emph{displacement} from the equilibrium position),
$E_{\rm{tot}}={\rm{constant}}={\rm{P.E.+K.E.}}=\frac{1}{2}kx^2+\frac{1}{2}mv^2$,
$\omega^2=\frac{k}{m}$, \mbox{$x(t)={\rm{A}}\cos(\omega
t-\varphi)$,} and $v(t)={\rm{-A}}\omega\sin(\omega t-\varphi)$;
with the \emph{total mechanical} energy of (spring-based) SHM (per
unit mass) represented by:
$e=E_{\rm{tot}}/m=\frac{1}{2}\frac{k}{m}
\rm{A}^2=\frac{1}{2}\omega^2 \rm{A}^2$. Secondly, from Equation
\ref{eq:specific energy} we have the expression for the specific
(field energy) of a RSW: $\Delta e=\frac{1}{2}\Delta t^{2}\Delta
a^{2}$. These two energy expressions, representing different types
of SHM, are somewhat similar. In the first specific energy
expression, $\omega$ actually has the unit of
$[\frac{\rm{rad}}{\rm{s}}]$ and thus the energy per unit mass of
SHM has a non-standard (energy) unit of
$\left[\frac{\rm{m^2\,rad^2}}{\rm{s^2}}\right]$ --- where m is
metres, s is seconds, and rad is radians.

By way of comparison, (from Equation \ref{eq:undulate}) specific
power \emph{per cycle} has `units' (alternatively meaning ``the
unit of") $\left[\frac{\rm{m}^2}{\rm{s}^3\,\rm{cycle}}\right]$;
and thus the (specific) energy of a rotating space-warp (itself)
has (model/theoretical) units of
$\left[\frac{\rm{m^2}}{\rm{s^2\,\rm{cycle^2}}}\right]$, with
duration $\Delta t$ having the (non-standard) unit of
$[\rm{s/\rm{cycle}}]$ which is the inverse of `unit' frequency
$[\rm{cycle}\rm{/s}]$\footnote{The National Institute of Standards
and Technology (NIST) recognises the radian [rad] as a derived
unit, but \emph{not} the (unit of) cycle [cycle]. Herein, the
model's introduction and discussion of celestial simple harmonic
motion, in conjunction with a cycle-based (rotating space-warp)
specific energy, necessitates the introduction of this (new)
derived unit; which is also (loosely) indicative of, and similar
to: a (full) `turn', full circle, (single) revolution, (completed)
rotation, or (a single and completed) cycle.}. Further, we need to
appreciate a distinction between: the (constant/fixed) total
specific energy of the physical phenomenon that is an entire
rotating space-warp, and the temporal sinusoidal influence of this
RSW upon a moving mass/body. The former retains the same specific
energy throughout time, whereas in the latter case the cyclic
nature of the influence results in a ($\delta v$) speed shortfall
\emph{per} full ($2\pi$ radian) cycle, i.e. per revolution of the
rotating space-warp --- with $\delta v$ having units of
$\left[\frac{\rm{m}}{\rm{s}} \frac{1}{\rm{cycle}}\right]$.

This subtlety, regarding the not specifically stated radian unit
in angular frequency --- occurring in SHM, and our (projection of
Earth-based measurement error) situation-1, by way of the
trigonometric differentiation process --- necessitates that when
Equations \ref{eq:amp1} and \ref{eq:amp2} are applied in
situation-2 --- which concerns the (real/non-pseudo) celestial
simple harmonic motion (of a body/spacecraft) --- a unit-based
modification of Equations \ref{eq:amp1} and \ref{eq:amp2} is
required. In a manner of speaking, we do this so as to reconcile
the (radian-based) mathematics with the (cyclic-based) physics
that both describe aspects of the phenomenon.

In effect, when dealing with spacecraft (or proper) speed
amplitude (i.e. $\Delta v=\Delta a/\omega$), the angular frequency
needs to have a cyclic basis rather than a radian basis. Thus, we
need to convert the units of speed amplitude, i.e.
$[\frac{\rm{m}}{\rm{s^2}}\frac{\rm{s}}{\rm{rad}}]=[\frac{\rm{m}}{\rm{s.rad}}]$
to $[\frac{\rm{m}}{\rm{s.cycle}}]$. This is achieved by
multiplying the magnitude of $\Delta v$ (per radian) by $2\pi$.
Think of converting from a number in metres per second to a number
in metres per minute; the longer time involves a numerically
larger quantity, and similarly a larger angular measure (cycles
cf. radians) requires a larger numerical value for the quantity.

Thus, $\Delta v=(2\pi) \Delta a/\omega_{\rm{rad}}=\Delta
a/\omega_{\rm{cyc}}=\Delta v_{\rm cyc}$. Situation-2 (or case 2)
also requires a similar correction to the simple harmonic motion
and situation-1 relationship: $\Delta x=\Delta v/\omega$. This
change in numerical \emph{magnitude} by way of a conversion of
units, yields `per cycle', rather than `per radian', values of
$\Delta v$ and $\Delta x$ (as regards a moving spacecraft for
example). Subsequently, these `adjusted' units make the
situation-2 values of $\Delta v$ and $\Delta x$ consistent with:
the Rayleigh power theorem based determination of the physically
real RSW specific energy ($\Delta e$), and particularly the
associated velocity shortfall \emph{per cycle} ($\delta v$)
induced by the rotating space-warp field strength variation at a
`point'.

Note that with $\Delta v$ being necessarily a sinusoid-based
`wave' amplitude, and \emph{actual} spacecraft motion (cf. a
theoretical description) being described by speed $\rm[m/s]$, the
use of units $\rm[m/s]$ for $\Delta v$, when dealing with the
relationship $\Delta x=\Delta v/\omega$, is considered appropriate
(and indeed necessary); whereas in the case of
(non-sinusoidal/monotonic) $\delta v$, with units of $[\rm
\frac{m}{s}\frac{1}{cycle}]$, such an alteration would be
inappropriate (and inaccurate at best). Similarly,
sinusoidal-based $\Delta e=\frac{1}{2}\Delta a^2 \Delta t^2$, when
describing a rotating space-warp (itself), can be thought of
(simply) as a specific energy $[\rm \frac{m^2}{s^2}]$.

With $\Delta a$ effectively equal to $f \Delta v$ (and $\Delta v$
effectively equal to $f \Delta x$) it is (somewhat) tempting to
re-write the sinusoids as $a_{\rm{s/c}}(t)=-\Delta a \cos(f
t-\varphi)$ and $v_{\rm{s/c}}(t)=-\Delta v \sin(f t-\varphi)$ (for
example); but radians play a unique role in trigonometric
differentiation and in the form taken by a solution of the
spring-based simple harmonic motion (second-order linear)
differential equation: $m\,\frac{d^2 x}{dt^2}=ma=-kx$. In short,
simple trigonometric derivatives and integrals only `work' if the
argument of sine and cosine functions are expressed in
\emph{radians}. Consequently, with regard to quantities that vary
sinusoidally, this frequency-based notation, and possible
consequences thereof, is neither promoted nor pursued.

Finally, we recall and note that the analogy to SHM involving the
(\emph{pseudo} cf. real) first situation --- which involves
Earth-based diurnal-spinning\footnote{With the observer on the
Earth's surface at a fixed radius from its centre.} and
annual-orbiting `projected' motion (errors) --- does \emph{not}
require this (radians to cycles) unit correction.

To proceed we summarise the above by noting that: in (pseudo)
situation-1: $\omega=\omega_{\rm{rad}}$, whereas in (real)
situation-2: $\omega=\omega_{\rm{cyc}}=f=\Delta t^{-1}$. Also note
that: $\omega_{\rm{rad}}=2 \pi \omega_{\rm{cyc}}=2\pi f$, and that
$\omega$ (with no subscript) is ambiguous. The relationship of
acceleration amplitude to velocity amplitude is best encapsulated
by:$$\Delta a_{\rm field}=\Delta a_{\rm proper}=\omega \Delta
v=\omega_{\rm rad}\Delta v_{\rm rad}=\omega_{\rm cyc}\Delta v_{\rm
cyc}$$

\subsubsection{Amplitude to shortfall relationships and an
equality of specific energies}\label{subsubsection:amplitude to
shortfall} We are now in a position to establish a link between
systemic translational speed reduction per cycle ($\delta v$) and
speed undulation amplitude ($\Delta v$), when it is a rotating
space-warp that is causing the celestial simple harmonic motion
(i.e. situation-2 as discussed in subsection \ref{subsubsection:2
ways to CSHM}).

As a preliminary we need to appreciate that by way of $\Delta
a_{\rm{field}}=\Delta a_{\rm{proper}}=\Delta a_{\rm{s/c}}$ (recall
Equations \ref{eq:equal a's} and \ref{eq:caveat} in subsection
\ref{subsubsection:real_relative}) Equation \ref{eq:amp1}, i.e.
$\Delta a=\omega \Delta v$, can also be used to describe a
relationship between (relative) speed amplitude (per cycle) and
acceleration perturbation/wave amplitude. Furthermore, Equations
\ref{eq:undulate} and \ref{eq:specific energy} describe a
relationship \emph{at} the moving point mass. This appreciation,
allows us to substitute (\mbox{Equation \ref{eq:amp1}}) into
either \mbox{Equation \ref{eq:undulate}} or \ref{eq:specific
energy}, and vice versa.

Using the situation-2 result obtained in subsection
\ref{subsubsection:2 ways to CSHM}, i.e. $\Delta
a=\omega_{\rm{cyc}} \Delta v=f \Delta v=\Delta v/\Delta t$, which
is an altered form of Equation \ref{eq:amp1}, and upon
substituting $\Delta a=\Delta v/\Delta t$ into Equation
\ref{eq:specific energy}, i.e. $\Delta e=\frac{1}{2}\Delta
a^{2}\Delta t^{2}=\frac{1}{2}\delta v^{2}$, we obtain:
\begin{equation}\label{eq:the v's}
\Delta v =|\delta v| \quad (\rm{per~cycle})
\end{equation}
Note that if we had simply substituted Equation \ref{eq:amp1}
(situation-1) into Equation \ref{eq:undulate} or \ref{eq:specific
energy} the result would have been: $2\pi \Delta v=|\delta v|$;
with the units of $\Delta v$ (or rather $\Delta v_{\rm{rad}}$)
being $\left[\frac{\rm{m}}{\rm{s}} \frac{\rm{1}}{\rm{rad}}\right]$
rather than $\left[\frac{\rm{m}}{\rm{s}}
\frac{1}{\rm{cyc}}\right]$ as required by the model --- which
necessarily employs a (per cycle) situation-2 based definition of
$\Delta v$. An alternative way of expressing this is: $2\pi \Delta
v_{\rm{rad}}=\Delta v_{\rm{cyc}}$ with $\Delta v_{\rm cyc}=|\delta
v|$ (per cycle).

Also note that we have not assumed $\delta v$ is positive, and
from subsection \ref{subsubsection:real_relative} we recall that
(with some caveats --- particularly one dimensional radial
motion): $\Delta v=\Delta v_{\rm{proper}}=\Delta v_{\rm{s/c}}$.
Further, by way of defining:
\begin{equation}\label{eq:a to change in v}
\delta a=\frac{\delta v}{\Delta t}
\end{equation}
it follows that:
\begin{equation}\label{eq:deltas_a}
\Delta a=|\delta a|
\end{equation}
Note that the (model's) unit for $\Delta v$ and $\delta v$ is
$[\rm \frac{m}{s}\frac{1}{cycle}]$, whereas the unit for $\Delta
a$ and $\delta a$ is $[\rm \frac{m}{s^2}]$. Next, by way of
Equations \ref{eq:the v's}, \ref{eq:a to change in v} and
\ref{eq:deltas_a} we confirm (re: situation-2):
\begin{equation}\label{eq:a to delta v}
\Delta a=\frac{\Delta v}{\Delta t}=f \Delta v=\omega_{\rm{cyc}}\,
\Delta v=\frac{\omega_{\rm{rad}}\Delta v}{2\pi}
\end{equation}
A discussion of this acceleration result is presented in
subsection \ref{subsubsection:field to S/C}. The terminology of
Equations \ref{eq:the v's} and \ref{eq:deltas_a} is a bit
confronting, i.e. two `deltas' (upper and lower case) in each of
the two equalities, but they are indicative/representative of two
different types of \emph{change}: a sinusoidal perturbation
(maximum) amplitude and a scalar (monotonic) difference.

Equation \ref{eq:the v's} indicates that: the total loss of
(solar-systemic) steady speed in a \em{single cycle} \rm ($\delta
v$), relative to a case where only non-undulatory steady speed
($v$) persisted throughout the cycle time, equals the amplitude of
the sinusoidal speed variation ($\Delta v$). Importantly,
\emph{the speed shortfall always opposes the direction of motion}.
The idealised actual situation is:
\mbox{$v_{\rm{final}}-v_{\rm{initial}}=\delta v$ with $\delta
v<0$}; and additionally $|\delta v| << v$ such that:
$v_{\rm{final}} \approx v_{\rm{initial}}$.

By way of Equation \ref{eq:the v's} it follows that per cycle:
$$\frac{1}{2}\Delta v^2=\frac{1}{2}\delta v^2$$
This result is significant in that: the \emph{unsteady} specific
energy associated with the celestial harmonic motion of an
otherwise steadily moving body is \emph{equal in magnitude} to the
(monotonic) \emph{shortfall} (cf. predictions) in the specific
energy of its motion. In other words, the energy of the body's
oscillatory variation in motion is equal to the (per cycle) loss
in kinetic energy of the body; with this loss being relative to
the kinetic energy the body would otherwise have had --- in the
absence of the sinusoidal field's influence. In short, and in
agreement with conservation of energy, a moving body's loss of
steady (translational) kinetic energy equals its gain in unsteady
(oscillatory) kinetic energy. Furthermore, the two (kinetic)
energy magnitudes also equal the (specific) `potential' energy of
the rotating space-warp phenomenon (itself); that is:
$$\Delta e=\frac{1}{2}\Delta a^{2}\Delta t^{2}=\frac{1}{2}\Delta
v^{2}=\frac{1}{2}\delta v^{2}$$ Thus, the model requires that
there be \emph{three} different (and equal) specific energy
effects `in play' at the same time (and over the same cyclic
duration), as compared to spring-based simple harmonic motion with
its solitary total energy equation. Finally, it is worth noting:
firstly, the \emph{constancy} of the specific energies, and the
spring-based total mechanical energy per unit mass, involved in
the preceding discussion; and secondly, that two very different
\emph{transfer} of energy \emph{mechanisms} are `active' in (this)
celestial simple harmonic motion (scenario).

Beyond this section's largely accurate idealisation, in reality, a
\emph{non-constant} equilibrium speed ($v$) value exists because
both: the Sun-based gravitational field ($g$), and kinetic energy
redistribution ($\delta v$ per cycle) act to alter the S/C's
steady translational speed ($v$).

\subsubsection{Acceleration field strength to spacecraft acceleration
relationship}\label{subsubsection:field to S/C} In subsection
\ref{subsubsection:amplitude to shortfall} it was established
(Equation \ref{eq:deltas_a}) that: $\Delta a=|\delta a|$.

This result can \emph{also} be established from \emph{situation-1
conditions}: i.e. the (angular frequency) undulatory relationship
\mbox{$\Delta a=\omega\Delta v$} (i.e. Equation \ref{eq:amp1}),
and $2\pi\Delta v=|\delta v|$ --- as discussed in subsection
\ref{subsubsection:amplitude to shortfall}. Subsequently:
\begin{displaymath}
\Delta a=\omega \frac{|\delta v|}{2\pi}=f |\delta v|=\frac{|\delta
v|}{\Delta t}=|\delta a|
\end{displaymath}

With the preceding Rayleigh (Energy and Power) Theorem based
relationships being physically illucidated/explained in terms of
\emph{energy}-based equalities, and because the total energy in
simple harmonic motion is proportional to amplitude squared, it is
(physically) more appropriate to state this equality in terms of
the square of each quantity, such that:
\begin{equation}\label{eq:accelerations}
\Delta a_{\rm field}^{2}=\delta a_{\rm proper}^{2}\,(=\delta
a_{\rm s/c}^{2} \rm{~~if~idealised~1D~radial})
\end{equation}
Note that the cycle time ($\Delta t$) is common to both
quantities, and by using Equation \ref{eq:accelerations} the
specific energy equality (Equation \ref{eq:specific energy}) is
upheld.
\begin{displaymath}
\Delta e=\frac{1}{2}\Delta a^{2}\Delta t^2=\frac{1}{2}\delta
a^{2}\Delta t^2=\frac{1}{2}\delta v^{2}
\end{displaymath}
This final equality cross-check gives us confidence in the
approach we have pursued, although it should be noted that the
physical origin and nature of the sinusoidal $\Delta a$ field
undulations, that causes the spacecraft's $\Delta v$ (sinusoidal)
oscillations, has yet to be adequately discussed. Finally, we note
(via Equation \ref{eq:the v's}) that: $\Delta e \propto \Delta
v^2$, in addition to $\Delta e \propto \Delta a^2$.

\subsubsection{The Pioneer anomaly \& commenting upon the model's
use of `mass'}\label{subsubsection:a_p and mass comment} The model
postulates that when multiple sources for this type of
supplementary (acceleration) field effect occur and coexist, the
fields act together simultaneously and continuously upon a moving
mass. Assuming an (additive) linear system, the superposition
principle applies to both the $(\Delta a)_i$ and $(\delta a)_i$
effects\footnote{Such that the net response (at a given time and
place) caused by multiple `stimuli' is the sum of the `responses'
which would have been caused by each stimulus individually. The
\emph{`i' subscript} indicates that multiple instantiations of the
(particular) variable exist/coexist in superposition}.

For the model to fit the (Pioneer anomaly) observational evidence,
a (linear) superposition of squared (shortfall) accelerations,
that are representative of a superposition of energies, is
(necessarily) proposed\footnote{Certainly an unorthodox quantity,
but in order to relate to both the sinusoidal
gravitational/accelerational field \emph{energy} affecting a body,
and the (unsteady) kinetic \emph{energy} of that moving body ---
over the same `time' period ($\Delta t$) --- a squared quantity is
appropriate (recall Equation \ref{eq:specific energy}).}. This
leads to a total ``shortfall" in expected motion over a given
period of time; which is conceivably the Pioneer 10 and 11
spacecraft's (long-term average) anomalous acceleration ($a_P$).
Thus, drawing upon Equations \ref{eq:specific energy} and
\ref{eq:accelerations}, the model proposes that:
\begin{displaymath}\sum \frac{1}{2}(\Delta a_{\rm field})^{2}_i
=\sum \frac{1}{2}(\delta a_{\rm proper})^{2}_i=\frac{1}{2}
\overline{a_p(t)^2}=\frac{1}{2} a_p^2
\end{displaymath} which is re-expressed as:
\begin{equation}\label{eq:SumA}
a_p=\overline{a_p(t)}=\sqrt{\sum (\delta a_{\rm
proper})^{2}_i}=\sqrt{\sum (\Delta a_{\rm field})^{2}_i}
\end{equation}

Note that due to the clarity of context we've omitted the index of
summation ($i$) and the finite upper bound of summation ($n$) from
the sigma (summation) notation, such that $\sum_{i=1}^{n} x_{i}^2$
(for example) is simply expressed as $\sum x_i^2$. \emph{Note}:
use of the `i' subscript outside of a summation emphasises a
variable's plurality.

Equation \ref{eq:SumA} is \emph{the crucial} result/output of the
model; it asserts that (theoretically) the magnitude of the
Pioneer anomaly may be conceived as a simple square root of the
\emph{summation} of the individual \emph{field} undulation
amplitudes $(\Delta a)$ squared. Fortuitously, the
\emph{different} period times, of each summed field element, are
included in the individual (spacecraft) acceleration terms
$(\delta a=\delta v/\Delta t)$. This is further discussed in
subsection \ref{subsubsection:multiple warps}. We interpret the
quantity $\frac{1}{2} (\Delta a)^2$ or $\frac{1}{2} \Delta a^2$,
with units $[\frac{L^2}{T^4}]$, as being representative of the
rate of specific energy \emph{transferred} per cycle duration
($\Delta t$) for a given single rotating
space-warp\footnote{Alternatively, $\frac{1}{2} \Delta a^2$ may be
thought of as a specific power per cycle. It (also) equals the
magnitude of the `rate' of specific (steady) kinetic energy
\emph{loss} per cycle $[\frac{1}{2} (\delta v/ \Delta t)^2]$ ---
from a barycentric perspective and concerning a given (single)
rotating space-warp.}.

Correct modelling would mean this instantiation of the model's $a_p$
value is our best match to observational $a_P$. This is largely the
case, although further modelling of this (idealised) `theoretical'
$a_p$ value is required. For example, a reduction arising from an
offset angle between Pioneer 10's path direction, and the
(observational) line-of-sight, is required (see
\ref{subsubsection:corrected accel}). Later, in subsection
\ref{subsubsection:multiple warps}, we obtain a `raw' value for the
Pioneer anomaly of $a_p=8.65\times 10^{-8}~{\rm cm~s^{-2}}$, which
upon correction (for line-of-sight offset angle) gives a final/`best'
value of: $a_p=8.52\pm 0.66 \times 10^{-8}~{\rm cm~s^{-2}}$.

Obviously, not all bodies experience this supplementary effect,
because `high mass'/larger bodies, such as planets and moons, are
immune to it. Note that, at this stage, only an equality (Equation
\ref{eq:specific energy}) concerning \emph{specific}
(gravitational/accelerational) field energy and a specific kinetic
energy redistribution has been considered. The mass aspect of the
body's kinetic energy is obvious, but the field's (distributed or
non-condensed) mass distribution and hence energy interaction with
the `condensed' mass\footnote{In the sense of `compact' mass, i.e.
solids and liquids.} body/spacecraft is yet to be fully outlined.

Clearly, the nature of mass in the model --- particularly its
association with the acceleration field --- needs to be further
considered. Later we shall see that a physical link between a
quantum mechanical based energy and this new type of
acceleration/`gravitational' field (expressed as an) energy is
achievable. This link shall require the introduction of a new
type/class of mass at the macroscopic level: ``non-local mass" ---
see subsection \ref{subsubsection:non-local inertial mass} for a
proper introduction. Note that the equivalence of \emph{inertial}
mass with (active and passive) gravitational mass
\mbox{($m_i=m_g$)} shall remain valid; but we further note that
this equality, at the macroscopic/celestial level, has implicitly
involved (i.e. been restricted to) \emph{local} bulk matter.

\subsubsection{Summary of: celestial simple harmonic motion, and the
Pioneer anomaly as a motion shortfall rate}
\label{subsubsection:Summary Rayleigh} In summary, this subsection
(\ref{Subsection:Shortfall}) has examined the retardation effect
of a hypothesised supplementary (acceleration/gravitational)
undulatory/sinusoidal field phenomenon upon a radially (outward)
directed \emph{point} mass --- at the point mass, within a
systemic (barycentric) reference frame. The retardation direction
opposes the motion, mimicking a (flight) path-based drag. Note
that this undulatory acceleration field has nothing to do with
general relativity's gravitational waves, whose source are large
\emph{macroscopic} masses in accelerated motion, e.g. binary star
systems.

By way of a (\emph{single} case) `constant' amplitude field
(acceleration/gravitational) sinusoid, expressions for spacecraft:
position, speed and acceleration sinusoids were determined, as
well as relationships between their (time and position
independent) fixed/constant amplitudes. By way of Rayleigh's power
theorem, an equality involving the (gravitational/accelerational)
sinusoid's energy ($\Delta e$) and an unsteady specific kinetic
energy is established, with the latter expression containing a
motion shortfall term ($\delta v$). This motion shortfall (per
sinusoidal cycle) is defined relative to predictions that overlook
an unsteady component within the spacecraft's kinetic energy ---
where steady (non-anomalous) motion is (exclusively) assumed. A
concern with the units of angular frequency in celestial simple
harmonic motion, as compared to `standard' simple harmonic motion,
was also discussed; this affects the magnitude of the
speed/velocity amplitude ($\Delta v$). This result is especially
relevant to section \ref{subsection:attenuation}.

This section culminated (subsection \ref{subsubsection:a_p and
mass comment}) in a hypothesis/declaration that: the Pioneer
anomalous acceleration is determined from a simple linear
superposition of motion shortfall rates arising from (and via)
\emph{multiple} field sinusoids, (but) with the summed amplitudes
having an energy-basis, rather than merely being a simple
summation of the sinusoidal/undulation acceleration amplitudes
$(\delta a=\delta v/\Delta t)$. The contents of this section are
pivotal to the explanation and model of the Pioneer anomaly
pursued herein.

\subsubsection{Interlude: what's still to come}
\label{subsubsection:Forward_reach} The previous subsection
(\ref{subsubsection:Summary Rayleigh}) gave a brief (overall)
summary of this section. In this final subsection, as something of
an interlude, we briefly describe aspects of the model that,
although not thoroughly discussed as yet, are of relevance --- in
so much as they provide some degree of wider context to the rather
specific (or narrow) aspects of the model `presently' being
discussed and formulated.

The (quantum mechanical) energy source of a single
(acceleration/gravitational) field undulation remains inadequately
explained, and the nature and global distribution of this new
field (type) are yet to be established. The field's
three-dimensional (3-D) distribution might alter the conditions at
the examined point mass from the preceding discussion's `linear'
scenario --- especially if the steady motion involved is not
linear-radial, i.e. elliptical, circular, parabolic or hyperbolic
motion. Fortunately, the speed retardation \emph{phenomenon} is
ubiquitous (i.e. path/direction and speed independent in the outer
solar system and beyond); as compared to its line-of-sight
\emph{observation} which will vary with the (object's) velocity
vector to (observer's) line-of-sight angle. Thus, in three
dimensions: $a_{\rm{proper}}\neq (a_{\rm {s/c}})_{\rm
{observed}}$\,.

We shall see that further observational evidence implies that a
(2-D planar/3-D cylindrical) rotating space-warp, centered
`toward' the far distant barycentre\footnote{In the case of the
Pioneer 10 spacecraft beyond 40AU (approximately Pluto's orbital
radius); i.e. \emph{far} beyond the orbital radii of Jupiter and
Saturn.}, describes the supplementary \emph{global} field
distribution; thus causing a sinusoidal/undulatory field at the
moving point mass. Subsequently, the type and/or direction of a
mass's motion, has no affect upon the (path or velocity vector
based) shortfall result established previously; although the same
cannot be said for line-of-sight measurements (thereof). The
author previously envisaged spherical divergent undulations, but
this has been superseded in order to be consistent with the
observational evidence (see subsection
\ref{subsubsection:Space-warp}).

Getting a fair bit ahead of ourselves, note that: the energy of
each field is seen to not exceed some `wiggle' room associated
with an atomic/molecular-based quantum mechanical energy
uncertainty that is shared concurrently by numerous ($\sim
10^{50}$) \mbox{atoms/molecules} in a celestial ``third-body" ---
this being a (`geometrically' suitable) spin-orbit `coupled' moon.
Physical concepts such as: entanglement, geometric phase,
self-interference, and decoherence (or lack thereof, regarding a
fractional/virtual geometric phase offset) are also
involved\footnote{The reader who would balk at, or straight out
deny, the ability of quantum mechanical entanglement to persist in
(special circumstance) macroscopic physical systems is in need of
``updating their world-view" --- by way of consulting the
(Scientific American) summary paper of \citet*{Vlatko_11}. A
proposal for entanglement (to be) acting on a scale of $10^{20}$
atoms has gained general acceptance, as has recognition of
``macroscopic entanglement in materials such as copper carboxylate
at room temperature and higher \citep[ pp.41-42]{Vlatko_11}" ---
``even though molecular jiggling might be expected to disrupt
entanglement (p.39)."}. The total \emph{virtual} excess internal
energy (below a minimum change in discrete energy levels), is
expressed non-locally (and externally) in the form of the
hypothesised \emph{real} supplementary acceleration/gravitational
field. This quantum mechanical (virtual) energy discrepancy is
indirectly induced by curved spacetime effects in a
global/systemic reference frame, and it is this (yet to be fully
explained) energy source that `drives' each of the (real)
``rotating space-warps" --- that coexist (together) in
superposition.
%*************************************************************************************
\subsection{Variation in the $a_P$
observations}\label{Subsection:Varying} In this section (and
section \ref{Subsection:Approx annual}) the periodic
\emph{temporal variation} of the Pioneer 10 anomalous acceleration
values ($a_P$) is our central focus. We shall step back from the
forward reaching aspects of subsection
\ref{subsubsection:Forward_reach}, to simply examine what the
temporal variation of the observation data of the Pioneer anomaly
($a_P$) is indicating/telling us. As such, we seek to approach
this aspect of the observational data with a fairly uncommitted
mind, (thus) essentially putting to one side aspects of the model
introduced and argued for in sections \ref{Subsection:Shackles}
and \ref{Subsection:Shortfall}, and `previewed' in subsection
\ref{subsubsection:Forward_reach}. From this (largely) unbiased
perspective (and somewhat of a `new beginning') we seek to support
features of the model that have already been `introduced'.

Upon completion of this (initial) investigation of temporal
variation in the $a_P$ data, section \ref{Subsection:Molding}
furthers the model's development --- or rather this Section's
preliminary version of the model. This (then) paves the way for
section \ref{subsection:attenuation}, in which the model's
relationship to various aspects of the Pioneer observational
evidence --- i.e. the anomaly itself, its temporal variation, and
the magnitude of the (post-fit) residuals in the data --- are more
fully appreciated.

\subsubsection{Introduction}
\citet*[ Section 2.1]{Turyshev_06} state that there is a spatial
and temporal variation of the order of 10\% for each spacecraft.
\citet{Turyshev_06}, \citet{Olsen_07}, and \citet[ Section
3]{Levy_09a} all stress the \emph{constancy} of the anomaly; (with
the first two of these latter citations) implicitly assuming that
without the noise of observation, which includes Earth-based
diurnal and annual residuals, the data would be free of spatial
and (short-term) temporal variance. This is a reasonable
assumption, but nevertheless it is an interpretation of the
observational data. In this, and the following subsection
(\ref{Subsection:Approx annual}), the validity of this ``latent
constancy" (or ``latent steadiness") interpretation is scrutinised
and found wanting --- a view somewhat supported by
\citet{Levy_09a} and now endorsed by \citet{Turyshev_10b}.

\subsubsection{Spatial variation of $a_P$}\label{subsubsection:Spatial}
The different average values for Pioneer 10 and 11 may or may not
indicate spatial variation of $a_P$. At the end of Section VI in
\citet[ p.25 and reiterated p.34]{Anderson_02a} the experimental
(excluding total bias) magnitudes are given as \mbox{$(7.84$ and
$8.55)\times10^{-8}~\rm{cm~s^{-2}}$} respectively. Recalling the
total error for the anomaly is $\pm1.33 \, \times10^{-8}~\rm{cm
~s^{-2}}$, the spatial variation of $a_P$ is (inherently)
ambiguous.

\subsubsection{A comment on observational errors}
\label{subsubsection:A comment on}
The unprecedented navigational accuracy delivered by the Pioneer
spacecraft Doppler data, and the lack of other similarly precise
S/C, makes interpretation involving all, or part, of the error
data a ``non-exact science". \citet[ Section X]{Anderson_02a}
understandably lump all the experimental and systematic errors
together in a least squares uncorrelated manner to obtain the
total error.

\citet[ Table II]{Markwardt_02} finds the RMS residuals of all the
non-extreme measurements is approximately $8~\rm{mHz}$
\mbox{($\sim0.5~\rm{mm~s^{-1}}$),} and states that
\citet{Anderson_02a} assigned a standard error of $15~\rm{mHz}$ to
their Doppler data processing. Note that for the Pioneer
spacecraft's S-band Doppler, \mbox{1 cycle = 0.0652m}
\citep*{Null_81}, so $10~\rm{mHz}\Rightarrow0.652~\rm{mm~s^{-1}}$;
and Markwardt's analysis spans \mbox{1987-1994} whereas the
\citet{Anderson_02a} rigorous analysis spans 1987-1998.

Some distinction between error sources and types is useful,
especially considering the magnitude of any real temporal
variation in $a_P$, and the total error, are of similar
magnitudes.

\subsubsection{Temporal variation of $a_P$}
Five types of temporal variation may be distinguished.
\begin{enumerate}
\item{Pure bias (i.e. no temporal variation) --- e.g. radio beam
reaction force. Note that RTG heat reflected off the craft is a
non-pure bias.} \item{Monotonic. The \citet{Markwardt_02} and
\citet{Olsen_07} analyses are inconclusive. Monotonic variation is
not considered from here on.} \item{Effectively stochastic (or
random). } \item{Effectively discontinuous --- e.g. at manoeuvres,
and for Pioneer 11's Saturn flyby \citep*{Turyshev_06}.}
\item{Periodic. This includes solar plasma effects, and the
diurnal and $\sim$annual\footnote{Markwardt's terminology for a
residual of not quite one year's duration \citep{Markwardt_02}.}
residuals.}
\end{enumerate}
The annual residual's source has been deemed ``undetermined" by
\citet[ p.11]{Markwardt_02}, and ``[somewhat] unmodeled" by
\citet[ p.397]{Olsen_07}\footnote{To be fair, we note that Olsen
does not believe there is anything troublesome or irregular in the
existence of the $\sim$annual residual, attributing it primarily
to ``an artifact of the maneuver estimation algorithm".}.
Regarding (temporal) variation of the anomaly, and attempts at
modelling the anomaly, a good understanding of the `annual'
residual is (in all likelihood) crucial.

\subsubsection{Non-specific variation in
$a_P$}\label{subsubsection:Non-specific} 
\begin{figure}[h!]
\centerline{\includegraphics[height=5.0cm, angle=0]{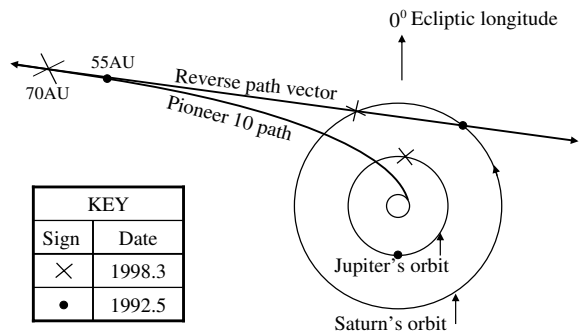}}
\begin{center}
\caption[Schematic view down from the north ecliptic pole
illustrating where a tangent from the Pioneer 10's path (at latter
times) is in relation to the orbits of Jupiter and
Saturn.]{Schematic view down from the north ecliptic pole
illustrating where a tangent from the Pioneer 10's path (at latter
times) is in relation to the orbits of Jupiter and Saturn.
Saturn's orbit crosses this line in 1992 and 1998. Jupiter's orbit
lies nearest this (negative) path vector in 1998, but in 1992 this
is not the case. Coincidentally, early 1998 produced an all-time
maximum in the Pioneer anomaly's (measured) magnitude.}
\label{Fig:RevTang}
\end{center}
\end{figure}

In the following case
both spatial \emph{and} temporal variation may exist. In early
1998 a fairly steeply peaked maximum acceleration value of over
$10\times10^{-8}~\rm{cm~s^{-2}}$ for Pioneer 10 \citep[ Figure
14]{Anderson_02a} occurred. Coincidentally, at this time the
(negative) path vector of the S/C is aligned with Saturn, and
Jupiter makes its closest approach to the path vector --- see
Figure \ref{Fig:RevTang}. Thus, the solar system's planets (and/or
moons) are \emph{possibly} implicated in a physical model.

%********************************** Fig:RevTang  ********************************
%********************************** end of Fig:RevTang  *************************

\subsubsection{The diurnal residual}\label{subsubsection:diurnal}
The diurnal residual is a good example of a periodic residual
arising from inaccurate parameter constants, rather than
observation error or a modelling inadequacy\footnote{These are the
three possible causes of residuals, arising from a method of least
squares analysis \citep[ Ch. 8]{BrouwerClemence_61}.}. It
\emph{is} an Earth-based effect. Note that the Doppler data only
measures the relative Earth to spacecraft speed, and that numerous
effects --- including the Earth's movement relative to the solar
system's barycentre --- are incorporated in the analysis. Thus, in
\emph{raw} data format a significant annual variation necessarily
exists.

It appears the periodic nature of the diurnal residual arises
primarily from \emph{both} Earth Orientation Parameter (EOP) error
and Earth Ephemeris error acting together in the Orbital
Determination Program\footnote{This interpretation is based on
correspondence (circa 2003) with E. Myles Standish of JPL, and is
analogous to how celestial pole offsets arise. Celestial pole
offsets are required because the model (of the Earth's
orientation), that \emph{combines} precession and nutation, relies
on fixed parameters for the Earth's shape (geodesy) and internal
structure; but since these are not fixed the offsets inevitably
arise. Other (lesser) effects exist beyond this somewhat idealised
account.}. \citet[ p.38]{Anderson_02a} comment that the period is:
``approximately equal to the Earth's sidereal [spin] rotation
period", which is unfortunately vague.

The diurnal residual's magnitude is crucial to understanding the
$\sim$annual residual. The noise of the diurnal residual is
greatest around solar conjunction, but at solar opposition, near a
minimum in the solar cycle\footnote{There was a broad minimum in
the solar cycle around May 1996. The Pioneer 10 S/C was beyond 60
AU.}, exceptionally good data is available: see~\citet[ Figure 18,
p.38\,]{Anderson_02a}. Non-periodic noise is evident, arising
primarily from interplanetary plasma and the Earth's ionosphere.

\citet{Markwardt_02} gives a figure of $10~\rm{mHz}$ (i.e.
$0.65~\rm{mm~s^{-1}}$) for the amplitude (on average) of the
Pioneers' diurnal residuals. \citet[ Figure 18,
p.38\,]{Anderson_02a} indicates a Nov--Dec 1996 solar opposition
amplitude of $0.1373~\rm{mm~s^{-1}}$ that may be obtained from:
the amplitude relationship $\Delta a=\omega \Delta v$ (Equation
\ref{eq:amp1}), the diurnal period
$\omega_{d.t.}=7.292\times10^{-5}\rm~rad~s^{-1}$, and $\Delta
a_{d.t.}=(100.1\pm7.9)\times10^{-10}~\rm{m~s^{-1}}$ given
by~\citet[ p.38]{Anderson_02a}. This gives, via $\Delta v=\omega
\Delta x$ (i.e. Equation \ref{eq:amp2}, subsection
\ref{subsubsection:rel undu ampl}), a \emph{best}-case cyclic
diurnal position offset amplitude of approximately $\pm\,1.9\rm \,
metres$ (i.e. \mbox{$\Delta x\approx1.9\,\rm{m}$)}, and about 9
metres on average. This is consistent with the NEAR\footnote{Near
Earth asteroid rendezvous mission. Note that this spacecraft uses
(the later generation) X-band carrier frequencies whereas the
Pioneer 10/11 used S-band.} spacecraft's range accuracy of 1
metre, and average position error of 25 metres
\citep*{ElliotHellings_97}.

Such a small periodic position error promotes confidence in our
belief that the Pioneer spacecraft exhibit exceptional
navigational accuracy and precision. The individual ``beads on a
string" (or catena) placement of most observational points in
Figure 18 \citep{Anderson_02a}\footnote{Note that only one of
NASA's three Deep Space Network tracking stations is used
throughout.} involves an accuracy of approximately $0.01
\rm~mm~s^{-1}$, which is ultimately related to the precise and
stable frequency produced by a ground-based hydrogen maser atomic
clock. With improved clock precision a best-case Doppler S/C
accuracy of $0.0001 \rm~mm~s^{-1}$ is now potentially achievable
\citep*{Asmar_05} --- although this is not achievable in practice.
%********************************************************************************
\subsection{The $\sim$\,annual residual}\label{Subsection:Approx annual}
As with the diurnal residual it is the Earth, and not the
spacecraft, that is generally accepted as the source of the
``annual" residual. This assertion is now challenged by way of the
period and amplitude observational data. This subsection shares
the sentiments of footnote 125 in \citet{Anderson_02a}.
\begin{quote} We thank E. Myles Standish of JPL, who encouraged us
to address in greater detail the nature of the annual-diurnal
terms seen in the Pioneer Doppler residuals. \ldots
\end{quote}
The fact that the Pioneer anomaly ($a_P$) itself is inconsistent
with Standish's non-problematic ephemerides
generation\footnote{For the larger (`high mass') mass bodies of
the solar system at least.} results in an inevitable and
understandable tension between the Ephemeris and the (real)
Pioneer anomaly camps\footnote{E.g. \citet[ p.178]{Standish_05}.}
--- necessitating the restriction of a hypothesised ``real" anomaly
to low mass bodies.

\subsubsection{Period mismatch}\label{subsubsection:Period}
The period as stated is not $365.25$ days. \citet[ Section IX,
Part C]{Anderson_02a} gives the angular velocity (over interval
III --- 1992.5 to 1998.5) as $\omega_{a.t.}=(0.0177\pm0.0001)$
rad/day; this value equates to a cyclic period of $355\pm2$ days.

This stance is supported by Figure 2 in \citet*{Scherer_97}
showing the real part of the autocorrelation function of the later
Pioneer 10 data (1987--1995). By averaging (from the graph) the
clear maximum and minimum range of values, at the half and full
year, a period of $\sim355$~d~(i.e days) is confirmed. The shape
of the real part of the autocorrelation function indicates a
solitary sinusoidal-like oscillation dominates the spectral aspect
of a time series representing the (long-term) Pioneer Doppler
data.

The difference in period is too large to be attributed to
spacecraft position drift in the celestial sphere over a year.

The only coincidental period to match this is an obscure 356 day
\emph{lunar} orbital-based beat duration\footnote{A beat, in this
case, is the interference between two (lunar-based) rotating
space-warps of slightly different frequencies, `registered' as a
periodic variation in the superposition amplitude of the two
$[\Delta a \sin(\omega t-\varphi)]_i$ inputs. Its rate is the
difference between the two input frequencies.} (and associated
$\Delta a_p$ amplitude modulation\footnote{See subsection
\ref{subsubsection:summations} for more information on $\Delta
a_p$\,.}) from the perspective of the Pioneer 10 S/C --- see
Doppler residuals (in Hz units) \citet[ Figures 9,
13]{Anderson_02a}, rather than the more processed 1-day and 5-day
batch-sequential (acceleration) data of Figures 17 and 14
respectively. The heliocentric-based orbital periods of Jupiter's
moon Callisto and Saturn's Titan are respectively: $16.689018$ and
$15.945421$ days. Gravitational tidal effects ensure that both
these moons are in spin-orbit resonance `around' their respective
host planets. \emph{Together}, these moons also exhibit a joint
spin (or rotational) resonance, in so much that their rotational
(and orbital) phases --- and their associated rotating space-warp
phases --- will coincide/resonate every 357.9~d ($m=n+1$ where
$n\approx21.445$). As observed \emph{from onboard} (and thus at)
the Pioneer 10 spacecraft, this period (for years spanning 1992.5
and 1998.5) goes to approximately 356.1\,d --- by way of making a
(synodic-like) correction for the motions of the host planets of
these moons\footnote{In the 6 years of interval III (1992.5 to
1998.5) Jupiter tracks, with respect to Pioneer 10's location and
trajectory, approximately $+11^{o}$ prograde, whereas Saturn's
position remains essentially unchanged. Callisto's prograde
progression is thus~$\approx1.8^{o}$ (relative to Titan) per
357.9~d resonant cycle. This yields a shortening of the
($360^{o}$) resonance cycle of \mbox{$\approx1.8$~d. Note: `d'
indicates days.}}. Easy access to spacecraft positions may be
found at: \url{http://cohoweb.gsfc.nasa.gov/helios/heli.html}.
Alternatively, the more accurate ``JPL's Horizons" system may be
consulted online.

Thus, if a real Pioneer anomaly exists it may possibly be related
to the (larger) moons of the solar system.

\subsubsection{Amplitude non-compliance}
The existence of an Earth-based (exactly) annual residual is
\emph{undeniable}. Note that this amplitude is independent of the
Earth's diurnal residual amplitude.

In \citet[ p.38]{Anderson_02a} the amplitude of the periodic
Doppler frequency/speed variation ($0.1053~\rm{mm~s^{-1}}$
--- in interval III) is at a relative \emph{minimum} compared with
its average value between 1987 and 1998 \citep[ Figure
1]{Turyshev_99}. This amplitude is very similar to the diurnal
residual's (minimum) amplitude ($0.1373~\rm{mm~s^{-1}}$), but the
position `error' of approximately $\pm530$
metres\footnote{Determined by way of the ``scalable set of annual
oscillation amplitude values" given in the following paragraph.}
is quite different due to the different angular velocities, with
\mbox{$(\omega_{a.t.})_{\rm{Earth}}=1.991\times10^{-7}\rm~rad~s^{-1}$}
corresponding a 365.256 day sidereal orbit period.

An easily scalable set of annual oscillation amplitude values is:
$\Delta x=1~\rm{km}$, $\Delta v=0.2~\rm{mm~s^{-1}}$ and $\Delta
a=0.4\times10^{-8}~\rm{cm~s^{-2}}$ --- via \mbox{Equations
\ref{eq:amp1} and \ref{eq:amp2}} in subsection
\ref{subsubsection:rel undu ampl}, and the (observed) angular
velocity value $\omega_{\rm{a.t.}}=0.0177$ rad/day given in
subsection \ref{subsubsection:Period}. Note that:
$\omega_{\rm{a.t.}}=\omega_{\sim\rm{annual}} \approx 2.0
\times10^{-7}$ rad/sec.

In the literature on the Pioneer anomaly, the $\sim$annual
residual is usually vaguely linked to Earth orientation and
position error, and/or spacecraft pointing error. To estimate the
magnitude of the $\sim$annual residual directly is an uncertain
exercise; (and) we are basically restricted to a comparative
analysis involving pulsar timing measurements and ephemerides
accuracy.

If the annual residual's amplitude, arising from Earth and/or
spacecraft position error, is below 100 metres (corresponding to
$0.02~\rm{mm~s^{-1}}$), then it is effectively invisible to the
analysis; and the observed larger amplitude $\sim$annual residual
may \emph{additionally} exist, and be spacecraft based. This is
distinctly likely because the Earth's heliocentric distance, at
the time of DE405\footnote{Developmental Ephemeris 405 is used for
the Pioneer analysis 1987-1998.} was known to \mbox{$<20$
metres}\footnote{Private communication with E. Myles Standish. The
reference cited then adds support to this number. Heliocentric
accuracy is `now' i.e. \emph{circa} 2006 (via DE410 and EPM2004)
approximately 2~m \citep[ p.183]{Pitjeva_05}.}
\citep{Standish_05}. We note that Earth orbit orientation
uncertainties of about $\pm$1~km\footnote{Standish: further email
correspondence in 2004.}, implying extremely small angular
uncertainty ($\sim\rm{nrad}$) in the Pioneer line-of-sight Doppler
observations, are of negligible significance.

Millisecond Pulsar timing experiments rely on, and conceivably
feedback to ensure, an accurate Earth ephemerides. Their annual
residuals act as a check upon the Earth's orbit orientation
accuracy. With long-term millisecond Pulsar timing measurements
known to better than $300\rm~nanoseconds$ ($0.3\rm~\mu s$)
\citep*[ Section 5.6]{Splaver_05}, the Earth's ephemeris (DE405)
is known to better than $\pm$90 metres --- a view predated by
\citet{Chandler_96}.

If the aforementioned amplitude residual \mbox{($\pm$530 m)} is
Earth-based, then it is inconsistent with a lack of concern,
regarding the magnitude of annual residuals, in Pulsar timing
experiments. Actually, \emph{long-term} Doppler data and
millisecond Pulsar timing are not accurate enough to feedback data
to improve ephemerides accuracy. It is CCD astrometric
observations, spacecraft ranging data, and $\Delta$VLBI data of
orbiting spacecraft that dominate ephemeris establishment today
\citep{Standish_04}.

A comparison of DE405 with the earlier DE200 further supports this
stance. Firstly, Pulsar timing data employing DE200 \citep*[
Figure 5 and Section 6.1]{Kaspi_94} had accuracy to $2~\rm{\mu s}$
(i.e. 600m), which is nearly an order of magnitude less accurate
than the use of DE405 gives. This order of magnitude improvement
is confirmed by Standish's comparison of DE200 with the later
DE405 \citep{Standish_04}. Further, \citet[ Section IX,
p.36]{Anderson_02a} discounts both Earth orientation parameters
and the planetary ephemeris as possible causes of error or
significant residuals. Regarding the ephemeris they say:
``post-fit residuals to DE405 were virtually unchanged from those
using DE200."

We may conclude: firstly, that the existence of a true-annual
residual, arising from parameter inaccuracies similar to those
causing the diurnal residual does exist, but its amplitude is
\emph{below} the level of the Doppler data's accuracy. This view
is supported by the spectral analyses of \citet[ Section
4]{Levy_09a} and \citet[ Section 4.4]{Toth_09a}. Secondly, (we
conclude) that the $\sim$annual residual's unaccountably large
amplitude, by process of elimination, appears to be
spacecraft-based and real. Just what this ambiguous amplitude
actually represents is becoming a pressing concern.

This second conclusion need not necessarily be ruled out by the
aforementioned two references involving spectral analysis, because
the sampling intervals are not uniform \citep[ p.20]{Toth_09a},
and the temporal duration of the ($\sim$annual) signature is quite
long cf. total sample time. Further, \citet*[ Section 5]{Levy_09b}
declare that: ``[their] fit\footnote{Albeit involving a
true-annual periodicity.} indicates the presence of a significant
annual term though this periodicity was not detected by the
spectral analysis." Interestingly, \citet[ Section 4]{Levy_09b}
indicate that: ``[the] multiple period searching method from Van
Dongen et al. as well as the SparSpec method indicate the presence
of a semi-annual period which was not detected by the first three
methods considered; \ldots" They declare ``a periodicity at
(177$\pm$3.7) days", which indicates an $\sim$annual period
(roughly) within the range of 350 to 358 days.

Without our (real) model taking its first small steps, the
ambiguity surrounding an explanation of the `annual' signature
would be inevitably sustained.

\subsubsection{Average amplitude consensus, and discussion of amplitude ambiguity}
\label{subsubsection:amplitude ambiguity} Removing the constant
$a_P$ value from the \mbox{Pioneer 10} analysis data leaves an
$\sim$annual signature, which unfortunately is of the same order
of magnitude as the noise/error of the data; ambiguity abounds.
Figures 13 and 17 of \citet{Anderson_02a} appear to favour a
``stationary" amplitude, within the stochastic noise; whereas
\citet[ Figure 1B]{Turyshev_99} imply a non-stationary (dampened)
value. One recent analysis, \citet[ Figures 1,4,5]{Olsen_07},
favours neither and further complicates the situation by also
finding apparent stochastic behaviour in the signature. Further,
Olsen highlights the amplitude's sensitivity to the (200 day)
correlation time. The stochastic-like appearance of the Olsen
residuals loosely supports a hypothesis of (linear) superposition
of periodic effects --- the dominant one being the $\sim$annual
residual ($\approx 355$ days, recalling subsection
\ref{subsubsection:Period}).

Significantly, \citet[ p.396]{Olsen_07} says there is no clear
annual variation for Pioneer 11, which is at odds with \citet[
Section 5]{Nieto_05} and \citet{Anderson_02a}. Surely, an
Earth-based residual should not exhibit such vagaries between
analyses?  This reinforces our leaning towards multiple periodic
contributors, of different amplitudes, in the temporal data.
Certainly, other non-periodic factors influence the data, thus
``clouding" the veracity of this stance.

Fortunately, a consensus upon the \emph{average} amplitude of the
$\sim$annual residual, in the Pioneer 10 data, can be established.

In the month of data given by~\citet[ Figure 18,
p.38]{Anderson_02a} for the diurnal residual, covering about
$30^{o}$ of the annual anomaly's cyclic period, a smooth amplitude
increase of about $0.1~\rm{mm~s^{-1}}$ (to a maximum) over the 30
days is evident. If this implies \mbox{[via $(1-\cos
30^{o})^{-1}=(0.134)^{-1}\approx7.5$]} a speed sinusoid amplitude
of $\sim0.75~\rm{mm~s^{-1}}$ then this roughly agrees with the
\citet{Markwardt_02} $\sim$annual value of 10mHz or
$0.65~\rm{mm~s^{-1}}$. \citet[ Equation 13]{Olsen_07} concurs with
a stated value of $1.5\times10^{-8}~\rm{cm~s^{-2}}$ (corresponding
to $0.75~\rm{mm~s^{-1}}$).

Of some concern is that the aforementioned data, especially the
1996 diurnal-data based value occurring during interval III
(1992-1998), exposes an ambiguity with the significantly smaller
average value of \mbox{interval III,} i.e.
$0.1053~\rm{mm~s^{-1}}$. No single reason can be given with
confidence\footnote{Possibly, this amplitude mitigation arises
from alterations to the Kalman filter estimation of the data, in
conjunction with occasional upgrades to the Deep Space Network
System that monitors spacecraft. Why? In order to appease the
awkwardly large amplitude of a (wrongly) perceived Earth-based
residual.}. A further mission to investigate and clarify such
ambiguities is sorely needed.

Nevertheless, an average value of about $0.7~\rm{mm~s^{-1}}$,
corresponding to undulatory/oscillatory amplitudes of
$1.4\times10^{-8}~\rm{cm~s^{-2}}$ and 3.5~km, seems to be the
common ground between the various analyses and the figures
referred to above. Note that these last two specific values only
apply in situation-1 circumstances (recall subsection
\ref{subsubsection:2 ways to CSHM}); i.e. an oscillatory
Earth-based Doppler residual --- or (alternatively) the
mismodelling of solar plasma effects upon the spacecraft's `radio'
signal \citep[ p.125]{Turyshev_10b}. The (situation-1) derived
annual acceleration and position variations (around their
equilibrium values) are \emph{not} quantitatively representative
of (situation-2) real Pioneer 10 spacecraft motion --- i.e. after
the anomalous shortfall effect ($a_P$) has been taken out. This
real scenario is examined in subsection
\ref{subsubsection:Callisto-Titan}.

\subsubsection{Concluding remarks on the $\sim$annual residual}
Taken on their own, the period and amplitude analyses are not
galvanising in their discrediting of an Earth-based residual ---
so as to be in favour of a spacecraft-based $\sim$annual residual.
But taken together, and in conjunction with the accuracy of the
diurnal residuals \mbox{($\pm~1.9$ metres)}, they demand a rethink
of the (reasonably well supported) $\sim$annual residual and the
nature of temporal $a_P$ variation. A superposition of various
undulation effects is conceivably implicated. Further discussion
regarding the quantification, and the physical basis, of the
$\sim$annual residual is presented in subsection
\ref{subsubsection:Callisto-Titan}.

\subsection{Molding the Model}\label{Subsection:Molding}
In contrast to \citet{Turyshev_06}, who (among others) promoted a
constant/static (magnitude) $a_P$ behind the noise of
observations, the stance of this article is to accept variance of
the Pioneer anomaly data around its long-term \emph{constant} mean
as real and indicative of the mechanism underlying the anomaly. It
is reassuring that this feature is now supported by \citet[
p.121]{Turyshev_10b}. Additionally, (herein) \emph{long-term}
(anomalous acceleration) constancy --- as a best fit term/result
for the data --- is seen to make heat merely a secondary/non-major
aspect of an explanation (\mbox{recall} section
\ref{Subsection:Heat}).

\subsubsection{Aspects of the mechanism}
\label{subsubsection:aspects} So far in this paper a number of
features of a mechanism have been implied; in point form these
are: \begin{enumerate} \item{Compound (i.e. steady \emph{and}
unsteady) motion in a systemic reference frame, and an Eulerian
(point-based\footnote{Recalling that the solar system's scale is
so vast that from a systemic (barycentric) perspective the
spacecrafts' \emph{change} in speed and position, over time
$\Delta t$, is effectively negligible.}) analysis, was
instigated.} \item{Sinusoidal field undulations (amplitude $\Delta
a$) lead to (sinusoidal) speed undulations ($\Delta v$) in the
motion of the Pioneer 10 and 11 spacecraft.} \item{The individual
field undulations are (each) required to have constant amplitude.
To appease GR's need for no preferred location (i.e. positional
invariance), this amplitude constancy extends out to
``infinity"\footnote{Infinity in the sense of \emph{mathematical}
infinity; physically, this means an extension to the `end of the
universe' --- which itself is a somewhat ambiguous concept.}.}
\item{For each undulation a shortfall in predicted speed occurs
($\delta v$), in conjunction with an oscillatory motion effect
($\Delta v$) around a mean value.} \item{The moons of the solar
system are implicated; by way of subsections:
\ref{subsubsection:Non-specific} (position),
\ref{subsubsection:Period} (period `resonance'), and also
\ref{subsubsection:Space-warp} ($\sim$annual phase difference
between Pioneer 10 and 11).} \item{Physically, the Pioneer anomaly
is considered to be associated with the summation of (specific)
kinetic energy (`loss' or) shortfalls [$\frac{1}{2}(\delta v)^2$
or $(\frac{1}{2}\delta v^2)$\,] over their various cycle durations
$(\Delta t)$.} \item{$a_p=\overline{a_p(t)}=\sqrt{\sum (\delta
a_{\rm proper})^{2}_i}=\sqrt{\sum (\Delta a_{\rm field})^{2}_i}$
where $\delta a=\delta v/\Delta t$, and $a_p$ (as compared to
$a_P$) refers to the \emph{model's} (theoretical) estimation of
the (observed) Pioneer anomaly ($a_P$).} \item{Multiple amplitudes
and periods of the field undulations --- by virtue of different
moons producing different field undulations --- gives the
impression of both periodic and stochastic behaviour in the
$a_P(t)$ data\footnote{Albeit, with a fairly small mean deviation
from the (constant) mean (rate of speed shortfall) value.}.}
\end{enumerate}

Further aspects and ramifications, of the mechanism being
formulated, are developed and discussed in the following
subsections.

%********************************** Fig:P10vsP11 ********************************
\begin{figure}[h!]
\centerline{\includegraphics[height=11.8cm,
angle=0]{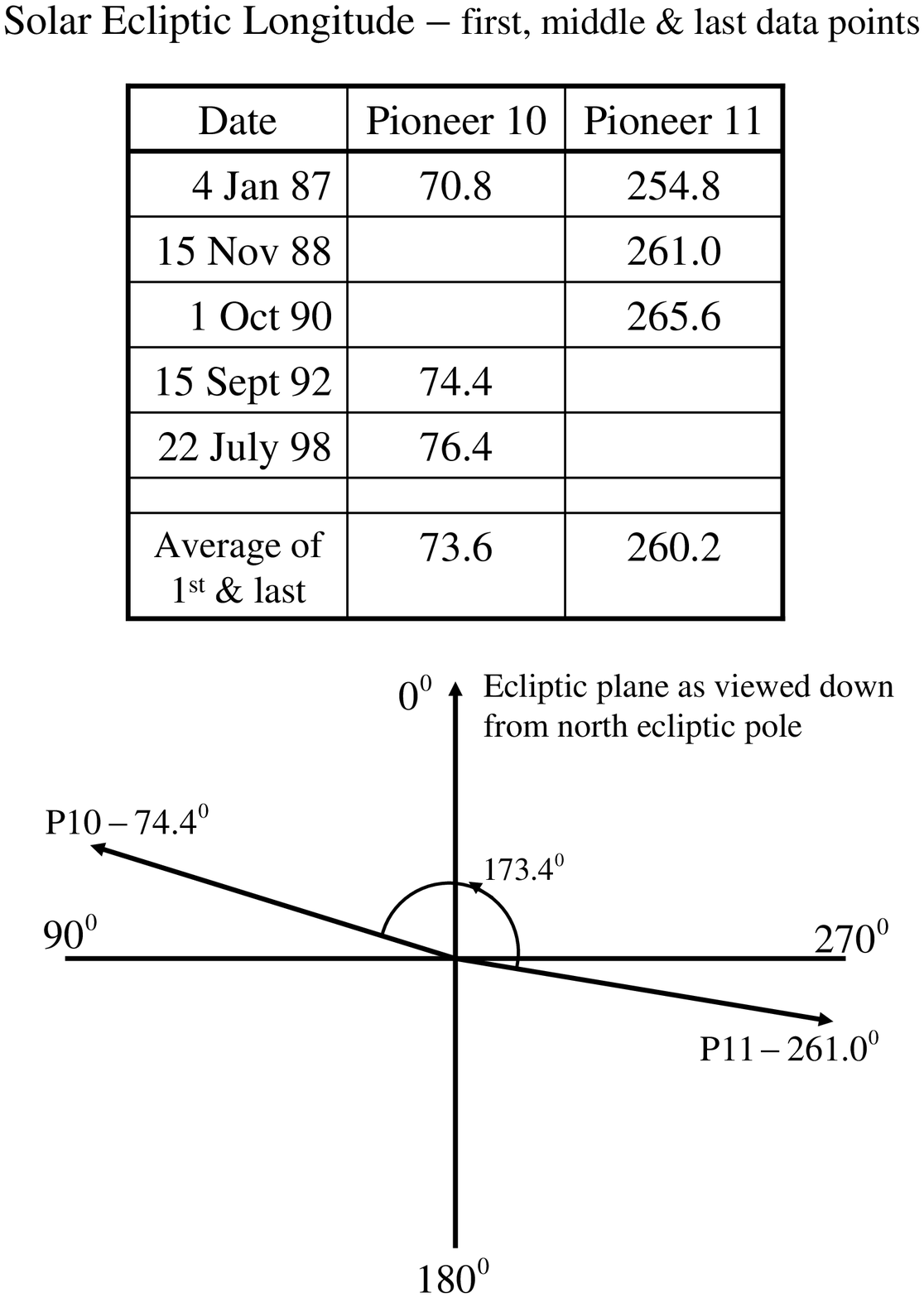}}
\begin{center}
\caption[Schematic diagram and table showing the direction (in the
ecliptic plane, relative to the barycentre) of the Pioneer 10 and
11 spacecraft for the mid-time of the Anderson \mbox{et. al.} data
set (1987-1998).]{Schematic diagram and table showing the
direction (in the ecliptic plane, relative to the barycentre) of
the Pioneer 10 and 11 spacecraft for the mid-time of the Anderson
\mbox{et. al.} data set (1987-1998). Note that the Pioneer 11 data
spans only 3.75 years. The diagram's vectors only indicate
\textit{direction} (i.e. \mbox{solar} ecliptic longitude), and not
position. The angle of 173.4 degrees is maintained when the
average of the first and last data values are used. Ecliptic
longitude data was acquired from
http://cohoweb.gsfc.nasa.gov/helios/heli.html}
\label{Fig:P10vsP11}
\end{center}
\end{figure}
%********************************** end of Fig:P10vsP11  *************************

\subsubsection{Field undulations as planar rotating space-warps, and
lunar comment}\label{subsubsection:Space-warp}
$a_p=\overline{a_p(t)}=\sqrt{\sum (\delta a_{\rm
proper})^{2}_i}=\sqrt{\sum (\Delta a_{\rm field})^{2}_i}$ (i.e.
Equation \ref{eq:SumA}) is applicable for durations much longer
than the various $(\Delta t)$ oscillation periods [or (i.e.) the
$(\Delta t)_i$ values]. This equation indicates that
acceleration/gravitational field undulations coexist with their
respective rates of (spacecraft proper) motion shortfall. The
nature of the field undulations, and resonances thereof, are
constrained by observational evidence.
\begin{quote}  The difference in phase between the
Pioneer 10 and 11 [$\sim$annual] waves is $173.2^{o}$, similar to
the angular separation of the two spacecraft in ecliptic longitude
\citep[ Section 5]{Nieto_05}.\end{quote}

\begin{figure}[h!]
\centerline{\includegraphics[height=5.3cm, angle=0]{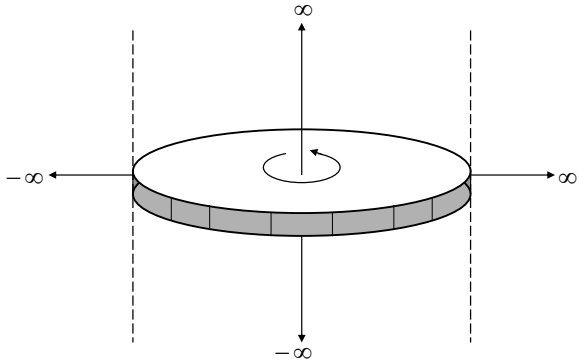}}
\begin{center}
\caption[Schematic representation of a \textit{non-warped} thin
disk `of' space, that can conceivably be curved or warped;
further, this warp (or perturbation) can conceivably rotate. By
way of contrast, Figures \ref{Fig:R3SideAbove},
\ref{Fig:R3SideElevNew} and \ref{Fig:R3FrontElev} illustrate the
(curved space) `shape' of the (gravito-quantum) rotating
space-warps proposed in this paper. Note that space itself does
not rotate.]{Schematic representation of a \textit{non-warped}
thin disk `of' space, that can conceivably be curved or warped;
further, this warp (or perturbation) can conceivably rotate. This
planar disk extends to the `end' of the universe, with this
feature represented by the `infinity' symbol. By way of contrast,
Figures \ref{Fig:R3SideAbove}, \ref{Fig:R3SideElevNew} and
\ref{Fig:R3FrontElev} shall illustrate the (curved space) `shape'
of the (gravito-quantum) rotating space-warps proposed in this
paper, for which the acceleration/gravitational perturbations
($\Delta a$) deviate from zero --- i.e. deviate from the flat
space scenario illustrated above. Note that space itself does not
rotate; rather our concern lies with the rotation of a curvature
of space --- similar (in part) to both: the rotation of a warped
vinyl record, and wave propagation upon an ocean.}
\label{Fig:FlatDisk}
\end{center}
\end{figure}

This evidence confines the (constant acceleration amplitude) field
undulations to rotating space-warps; rotating at the rate, and in
the directional sense, of their associated moon. The angle of
$173.2^{o}$ is prograde\footnote{So as to be compatible with
Callisto and Titan's prograde spin motion/direction. These spin
directions are/were established by way of: i) the planets'
prograde (direction of) orbit, and ii) the formation process that
(gravitationally) `bound' these large moons to their host
planets.}, stretching from Pioneer 11 to Pioneer 10 (see
\mbox{Figure \ref{Fig:P10vsP11}).} We hypothesise that this
implies the (lunar `centered') individual rotating space-warps are
disk-like and aligned with (or near to) the plane of the moon's
orbit around its host planet\footnote{The rotating space-warp is
similar to a warped rigid (vinyl) record, in that it retains its
`shape' and orientation throughout a rotation.} (see \mbox{Figure
\ref{Fig:FlatDisk}} to Figure \ref{Fig:R3FrontElev}). Further,
spacecraft motion inclined to the space-warp's
plane-of-orientation is seen to \emph{also} receive the full
motion shortfall (per moon, per cycle). For spacecraft motion,
geometrically offset/inclined to the Doppler line-of-sight
\emph{observations}, a cosine angle correction is required. For
Pioneer 10 and 11 (1987-1998) this correction is very minor (see
subsection \ref{subsubsection:three dimensions}).

%********************************** Fig:FlatDisk *********************************
%********************************** end of Fig:FlatDisk  *************************
%********************************** Fig:R3SideAbove ******************************
\begin{figure}[h!]
\centerline{\includegraphics[height=7.8cm,
angle=0]{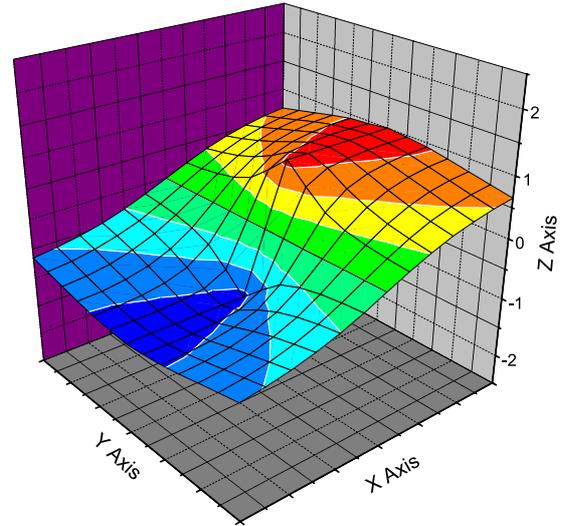}}
\begin{center}
\caption[The first of three (`snapshot') views of a rotating
space-warp shown in a three-dimensional grid-like representation.
The Z axis represents acceleration perturbation (away) from
`standard' conditions. Note that (beyond the first grid line)
points at a constant radius from the central point have a Z (i.e.
acceleration/curvature) value that varies sinusoidally around the
`circumference'.]{The first of three (`snapshot') views of a
rotating space-warp shown in a three-dimensional grid-like
representation. The Z axis represents acceleration perturbation
(away) from `standard' conditions, i.e. additionally curved space.
Front and side elevation views (Figures \ref{Fig:R3SideElevNew}
and \ref{Fig:R3FrontElev}) are preceded by this elevated view
lying roughly midway between the front elevation and side
elevation views. Note that (beyond the first grid line) points at
a constant radius from the central point have a Z (i.e.
acceleration/curvature) value that varies sinusoidally around the
`circumference'; furthermore, the rate of (space-warp) rotation
coincides with the rate of rotation and duration ($\Delta t$) of a
(suitable) moon in spin-orbit `resonance' around its host planet.}
\label{Fig:R3SideAbove}
\end{center}
\end{figure}
%********************************** end of Fig:R3SideAbove *************************
%********************************** Fig:R3SideElevNew ******************************
\begin{figure}[h!]
\centerline{\includegraphics[height=7.7cm,
angle=0]{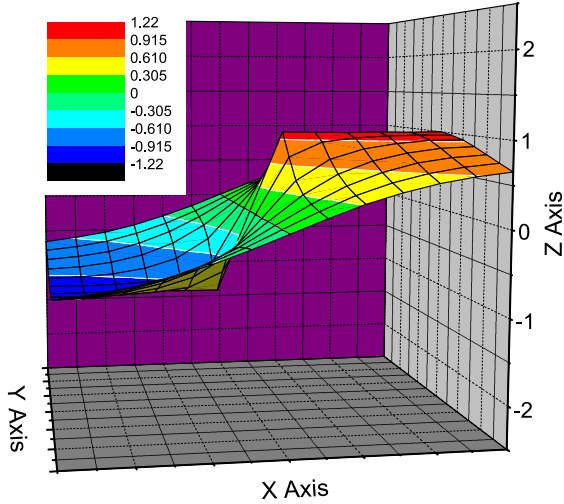}}
\begin{center}
\caption[Side elevation view of a (gravito-quantum) rotating
space-warp.]{Side elevation view of a (gravito-quantum) rotating
space-warp. Note that the colour scale merely illustrates changes
in the contour colours on the diagram, with maximum and minimum
amplitudes occurring at $1.00$ and $-1.00$ respectively.}
\label{Fig:R3SideElevNew}
\end{center}
\end{figure}
%********************************** end of Fig:R3SideElevNew  *************************
%********************************** Fig:R3FrontElev ***********************************
\begin{figure}[h!]
\centerline{\includegraphics[height=7.4cm,
angle=0]{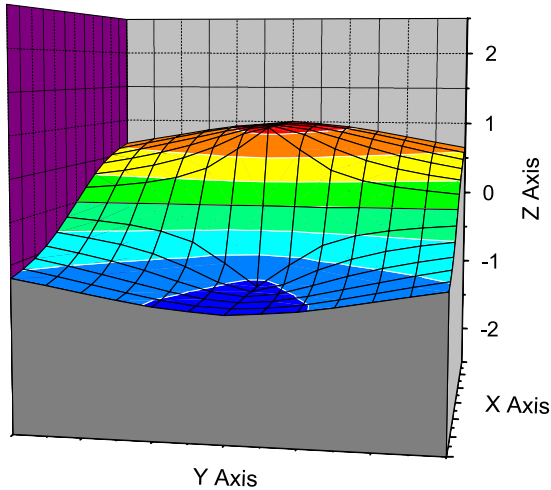}}
\begin{center}
\caption[Front elevation view of a (gravito-quantum) rotating
space-warp.]{Front elevation view of a (gravito-quantum) rotating
space-warp (RSW). Unfortunately, the actual (planar) rotational
motion around the central point cannot be shown in this
fixed/still image.} \label{Fig:R3FrontElev}
\end{center}
\end{figure}
%********************************** end of Fig:R3FrontElev *************************

The point based nature of the Eulerian analysis, and our
conceptualisation of gravitation as curved space, favours a purely
two-dimensional circumstance --- which then needs to be modified to
suit three-dimensional space (see subsection \ref{subsubsection:three
spatial}). Previously, a (non-rotating) spherical undulation was
envisaged by the author \citep{ten_Boom_05}, but a perturbed
(inherently) planar or disk-like (rotating) phenomenon is demanded by
the observational evidence and simple harmonic motion's
($\omega$-based) mathematical relationships. In three dimensions the
phenomenon/mechanism is cylinder-like --- although open and of
`infinite' extent.

Not all moon-planet-sun systems are suitably configured to
``generate" a rotating space-warp\footnote{Suitability involves:
i) lunar 1:1 spin-orbit resonance `around' its host planet, and
ii) a certain relationship between: lunar angular progression
around the planet, and the (host) planet's progression angle
around the Sun.}. In Section \ref{Section:Quantif Model} we shall
see that Jupiter's four Galilean moons and Saturn's Titan dominate
the undulated curved space landscape\footnote{The smallest of
these, Europa, is ``\ldots more massive than all known moons in
the Solar System smaller than itself combined [Wikipedia:
\emph{Europa} (moon), 2011-12\,]."}, with Earth's moon not being a
contributor/`generator'.

Importantly, with Io, Europa, and Ganymede being in a 4:2:1
orbital resonance (with one another), their contribution to the
`variation' of $a_p(t)$ about its mean value $\overline{a_p(t)}$
[designated/signified as: $\Delta a_p(t)$], is conceivably one of
(additional) attenuation. Hence the (356 day) Callisto-Titan
cyclic/`beat' repetition, evidenced by the $\sim$annual residual
and `sensitive' to the use of a 200-day correlation time (see
subsections \ref{subsubsection:Callisto-Titan} and
\ref{subsubsection:Facade}), is \emph{the} significant
characteristic of the temporal variation of $a_P$ (i.e. the
Pioneer anomaly).

\subsubsection{Anomalous CMB results and lunar spin plane rotating space-warps}
\label{subsubsection:WMAP} Interestingly, anomalous cosmic
microwave background (CMB) results --- by way of the Wilkinson
Microwave Anisotropy Probe (WMAP) --- implicate a solar system
\emph{ecliptic plane} effect as possibly influencing the otherwise
isotropic CMB radiation data
\citep*{Copi_07,de_Oliveira-Costa_04,Huterer_06}. For a recent
review article see \citet*{Copi_10}. At present, there is no good
reason why maps of temperature anisotropies in the CMB should be
correlated with any \emph{geometric} features belonging to our
local (solar) system.

In the anisotropic power spectrum the low multipole vectors,
quadrupole ($l=2$) and octopole ($l=3$), have an unexplained
alignment to: each other, and the geometry and direction of motion
of the solar system\footnote{Additionally, the quadrupole's
magnitude is significantly less than expected, and the third
octopole alignment correlation involves the supergalactic plane
(SGP).}. Note that the simple dipole ($l=1$) is interpreted as
resulting from the Doppler shift caused by the solar system's
\emph{motion} relative to the nearly isotropic (cosmic microwave)
background field.

The physical nature of the (cosmological size) rotating
space-warps (RSWs) proposed herein influence all low mass
bodies\footnote{Below a mass cut-off threshold.}, and conceivably
all photons, \emph{equally}. These (gravito-quantum) RSWs are
axisymmetric (and rotationally symmetric), `bound' to, and based
in the solar system. Jupiter's Galilean moons dominate the new
mechanism, and their inclination angles are offset less than $2^o$
from the \emph{ecliptic} plane.

Interestingly, for each rotating space-warp, the wave-like
(specific) energy ($\Delta e$) involved in the new mechanism may
be expressed (recalling end of subsection \ref{subsubsection:field
to S/C}) as the \emph{square} of either: a speed undulation
($\Delta v$), or an oscillatory acceleration/gravitational field
($\Delta a$). A common signature in the octopole and quadrupole,
that supplements the dipole ($l=1$), is
conceivable\footnote{Possibly some vector cross product effect is
involved.}. With CMB photons arriving (at Earth) from all
different directions, it may well be that the presence of the
(foreground) influence of a superposition of RSWs affects the
\emph{measurement} of CMB photons in a non-isotropic
manner\footnote{Three of the four planes determined by the
quadrupole and octopole are orthogonal to the ecliptic plane.}.

Section \ref{section:Type1a} discusses a very different type of
foreground influence that (also) may arise from the model.

\subsubsection{Steep increase of Pioneer 11 anomaly `around'
Saturn encounter discussed}\label{subsubsection:Saturn jump}
Observations, concerning Pioneer 11 in particular rather than
Pioneer 10, indicate that the (aforementioned) root sum of squares
(RSS) approach to $a_P(t)$ appears to \emph{not} apply for Pioneer
11 `\emph{between}' Jupiter and Saturn. Subsequently, the RSS
approach appears to be restricted to the case of all acceleration
field undulations, in the form of rotating space-warps, having the
same direction of rotation with respect to the spacecraft's
direction of motion. This is certainly the case for the Pioneer 10
analysis spanning 1987 to 1998 (40 to 70.5 AU), where Pioneer 10
is beyond the orbital radius of all planets in the solar system.

In the case of the Pioneer 11 S/C between Jupiter and Saturn, the
space-warp rotations associated with Saturn and Jupiter's
(prograde) moons are opposed at the S/C --- see \citet[ p.10
Figure 2]{Nieto_0702} and \citet[ Figure 3]{Anderson_02a} for S/C
trajectories. At this stage, we shall assume the spacecrafts's
\emph{path}-based response (to the different individual rotating
space-warps) is one of simple superposition --- as is the case
with the rigorous Pioneer 10 analysis (beyond the outer solar
system).

We treat the individual and different (simple harmonic) speed
variations/oscillations acting on the spacecraft
\mbox{$[v=v(t)=-\Delta v \sin(\omega t-\varphi)]$\,,} in response
to individual acceleration/gravitational field
variations/undulations $[a_{\rm{field}}=(\Delta a_{\rm{field}})
\sin(\omega t-\varphi)]$\,, in the manner of interfering ocean
waves --- with the different $\Delta a_{\rm{field}}$ amplitudes
replacing wave height amplitudes. The unequal period times (or
angular frequencies) of the (sinusoidal) field undulations results
in periods of constructive and destructive interference at the
spacecraft. Thus, the superposition of spacecraft speed
undulations/oscillations, and consequently (also) the real time
(effective/overall rate of) monotonic speed shortfall [$a_p(t)$],
can be quite variable --- with this occurring over extended
periods of time (of the order of days, weeks and months). Later
(in Section \ref{Section:Quantif Model}) we shall see that moon
size is (indirectly and roughly) indicative of undulation
amplitude, and thus the effects of \emph{Saturn's Titan and
Jupiter's four Galilean moons dominate the model}.

With Pioneer 11 redirected from Jupiter \emph{across} the solar
system to Saturn [see diagram \citep[ Figure 2]{Nieto_0702}], the
line-of-sight correction to $a_P$ is significant --- persisting
(even) after Saturn encounter (but to a much lesser extent).
Subsequently, as Pioneer 11 approaches Saturn, an (assumed)
Sun-directed Pioneer anomaly would `appear' smaller; and after the
spacecraft has encountered Saturn the anomaly steadily rises to
its standard value \citep{Turyshev_06}. Also see \citet[ Figure
7]{Anderson_02a}, noting the exceedingly large error bar ``around"
Saturn encounter.

The \emph{shape} of the best fit line is consistent with the
notion of a path to line-of-sight (cosine angle based) correction.
Awkwardly, the Doppler-based late 1977 (10-day sample interval)
data point (at $\sim$ 6AU) \citep[ Figure 7]{Anderson_02a} of
$<2\times10^{-8}~ \rm{cm~s^{-2}}$ implies a near removal of
respective shortfall rates, which is a concern for the model and
our (initial approach) to explaining this Pioneer 11
`discrepancy'. Even with a near $40\%$ ($\sim 52^o$) line-of-sight
to path angle reduction\footnote{By way of using
$a_P=8.74\pm1.33\times10^{-8}~ \rm{cm~s^{-2}}$ as a reference
value.} in $a_P$, and a large variation in effective/overall $a_P$
around its mean value at the time of measurement\footnote{The
short 10-day sample interval for the data point \citep[
pp.17-18]{Anderson_02a} needs to be at a low point in the
(superposition-based) temporal variation of $a_P(t)$.} --- i.e. a
\emph{reduction} in $a_P$ equal to the amplitude of the
$\sim$annual residual ($1.4\times10^{-8}~ \rm{cm~s^{-2}}$) --- the
model predicts measured $a_p$ of barely less than $4.0\pm1.33
\times10^{-8}~ \rm{cm~s^{-2}}$.

The mitigating circumstance of this concern is that the data
point's magnitude is contentious (in this author's opinion) for a
number of reasons. Firstly, solar radiation pressure is
significantly greater than $a_P$ [see \citet[ Figure
6]{Anderson_02a}]. Secondly, Doppler tracking data and the orbit
determination program's accuracy are somewhat compromised when the
direction of travel is at a large angle away from the
line-of-sight. Thirdly, the data point is for the not so rigorous
earlier analysis.

Alternatively, and somewhat necessarily, it may be the case that
(in late 1977) the \emph{opposing} (rotational) \emph{direction}
of Titan's rotating space-warp (later\footnote{See Table
\ref{Table:acceleration} in section \ref{Subsection:Model
quantifies external}.} found to have amplitude/magnitude
$3.35\times10^{-8}~ \rm{cm~s^{-2}}$), which opposes the `spin'
direction of Jupiter's four `Galilean' rotating space-warps
(RSWs), introduces a further type of destructive
interference\footnote{In addition to the \emph{scalar}
constructive and destructive (amplitude) interference that arises
from the superposition of multiple sinusoidal `signals'.} (not
previously appreciated). This (other) interference affects the
acceleration/gravitation \emph{field} (amplitude $\Delta
a_{\rm{field}}$) superposition \emph{prior to} its affect upon
spacecraft proper motion and acceleration. This subsection's
previous discussion of superposition assumed all the individual
acceleration/gravitation fields \emph{fully} participate (i.e.
standard linear system addition/superposition by way of their
coexistence) when they influence the path-based variation of
spacecraft speed around the S/C's mean value ($v$)
--- with this implicitly being the case in the outer solar system
(and thus/also throughout the rest of this paper). Consequently,
an anomalous acceleration $a_P(t)$ (near 6 AU in late 1977) of
around $0.7\times10^{-8}~ \rm{cm~s^{-2}}$ is conceivable ---
especially with Jupiter and Saturn on opposing sides of the
Pioneer 11 spacecraft at this `position' in space and time, see
\citet[ Figure 2]{Nieto_0702}.

If this alternative/`richer' second scenario is indeed the valid
case, the model then conceivably \emph{predicts} that the
noise/variation of the $a_p(t)$ data `between' Jupiter and Saturn
[i.e. $(\Delta a_p)_{\rm{mean}}$] should be at least somewhat
muted when compared to the mean variation of $a_p(t)$ in the outer
solar system --- all other things being equal (\emph{ceteris
paribus}).

\subsubsection{Spin rate changes may support a real Pioneer
anomaly}\label{subsubsection:spin rate changes} There were
\emph{opposing} variations of spin rate changes for the two
Pioneers \emph{between} manoeuvres, Pioneer 10 down and Pioneer 11
up \citep[ Figures 11,12]{Anderson_02a}. A real basis to these
quite precise variations is almost unavoidable.
\begin{quote}For the Pioneers there were anomalous spin-rate
changes that could be correlated with changes of the exact values
of the short-term $a_{P}$. The correlations between the spin-rate
changes and $a_{P}$ were good to $0.2\times10^{-8}~\rm{cm~s^{-2}}$
and better \citep[ p.4019]{Nieto_04}.\end{quote}

The Pioneer spacecrafts' solar system motion is hyperbolic orbits
in roughly \emph{opposing} directions. The not solely
linear-radial (outward) motion of the S/C, in conjunction with the
proposed (prograde) rotating space-warp mechanism, may have some
bearing upon an explanation of this unusually precise correlation.
%*********************************************************************************
\subsection{The model's relationship to the Pioneer observational
evidence}\label{subsection:attenuation} In section
\ref{Subsection:Shortfall} we investigated single/solitary
acceleration field undulations and their associated: (proper)
speed, proper acceleration, and position undulations --- about
their respective mean values. We then established/deciphered the
overall motion shortfall rate arising from the coexistence of
multiple field undulations, with this being the \emph{model's}
value for the Pioneer anomalous acceleration $(a_p)$. In this
section we turn our attention to further effects arising from
\emph{multiple} acceleration/gravitational field undulations
affecting the motion of a (`low mass') body or spacecraft.

\subsubsection{A two component Doppler residual \& the main objective
of this section}In addition to residuals arising from measurement
noise, the (sinusoidal) motion/speed and acceleration variations
arising from the model's (multiple) rotating space-warps shall
also be classed as \emph{residuals}. In the absence of this
further effect, measurement noise is (simply) the difference
between the calculated/expected\footnote{As `determined' by (a
particular) orbit determination program.} and observed/measured
Doppler values\footnote{Utilising the NASA Deep Space Network.};
but herein this difference shall include the model's additional
(residual/variation) component.

Unlike the monotonic anomalous acceleration (speed shortfall rate)
previously discussed, the mean speed and mean acceleration arising
from the model's undulatory/sinusoidal effects --- once the
monotonic anomalous acceleration is removed (or corrected for) ---
is zero. Thus, the proper motion and proper acceleration
(amplitude) variations around their `equilibrium' condition shall
be referred to as ``residual proper motion" and ``residual proper
acceleration" respectively.

In this section we firstly examine the nature of the model's
temporal amplitude `variation'\footnote{As compared to `variance',
as used in probability theory and statistics.} of residual proper
acceleration, taking note of its relation to the variation of
$a_P$ through time [i.e. $a_P(t)$]; and secondly, variation in the
residual proper motion/speed --- (in both cases) \emph{after}
adjustments/corrections for: the spacecraft's predicted
trajectory, systematics, and the constant (per unit time) Pioneer
anomalous speed shortfall, have been made. Initially, we are
interested in short-term temporal variations, of the order of days
and weeks; whereas later in this section, medium and long-term
variations are considered, particularly the $\sim$annual residual.

\subsubsection{Superposition of (sinusoidal) proper acceleration and
proper velocity}\label{subsubsection:summations} Multiple
acceleration/gravitational field `undulations' ($a_{\rm
field})_i=[\Delta a_{\rm{field}}\, \sin(\omega t-\varphi)]_i$\,,
and/or the spacecraft proper acceleration undulations (represented
by) $(a_{\rm proper})_i=[-\Delta a_{\rm proper}\, \cos(\omega
t-\varphi)]_i$\,, are assumed to coexist in, and be described by,
a simple superposition. This superposition results in the maximum
(possible) amplitude of (residual) proper acceleration being:
\begin{equation}\label{eq:SumB} (\Delta a_p)_{\rm max}=\sum
(\Delta a_{\rm proper})_i\end{equation} Although involving space
curvature/deformation, this superposition is (assumed to be)
conceptually and mathematically analogous to a superposition
involving ocean wave heights (for example). In subsection
\ref{subsubsection:stochastic}, we discuss why this
(superposition-based) acceleration summation, from a
reconciliation with the observational evidence perspective, is
considered secondary to a (superposition-based) summation of
\emph{speed} amplitudes:
\begin{equation}\label{eq:SumC} (\Delta v_p)_{\rm max}=\sum
(\Delta v_{\rm proper})_i\end{equation} where each $v=v(t)=-\Delta
v_{\rm proper}\, \sin(\omega t-\varphi)$, with each of the speed
and acceleration undulatory amplitudes related by (a number of
alternative expressions):$$\Delta a=(\omega_{\rm rad}/2\pi)\Delta
v=\omega_{\rm{cyc}} \Delta v=f \Delta v=\Delta v/\Delta t$$ This
amplitude relationship is indicative of situation-2 circumstances
applying\footnote{Recalling Equation \ref{eq:a to delta v} and the
distinction between situation-1 and situation-2 established in
subsection \ref{subsubsection:2 ways to CSHM}.} --- i.e.
real/non-pseudo celestial simple harmonic motion (arising from
rotating space-warps). Subsequently, $\Delta v_{\rm cyc}$ with
units $[\frac{\rm m}{\rm s}\frac{1}{\rm cycle}]$
--- as compared to $\Delta v_{\rm rad}$ with units $[\frac{\rm
m}{\rm s}\frac{1}{\rm rad}]$ --- is implied by the use of $\Delta
v$ [or $(\Delta v)_i$ `component'] values.

Previously (Equation \ref{eq:SumA} in subsection
\ref{subsubsection:a_p and mass comment}), the model quantified
(and defined) the Pioneer anomaly's
magnitude\footnote{Fortuitously, because $a_p(t)$ has a long-term
\emph{constant} mean --- primarily by way of constant rotating
space-warp $(\Delta a_{\rm field})_i$ amplitude values and
constant $(\omega)_i$ values --- long-term (mean) $a_p(t)$ may be
written (simply) as $a_p$\,. We note that each (Sun-planet-moon
system based) set of: $\delta a$, $\Delta a_{\rm proper}$, $\Delta
v$, and $\delta v$ values also have fixed/constant magnitudes ---
both temporally and spatially.} as:
$$a_p=\overline{a_p(t)}=\sqrt{\sum (\delta a_{\rm
proper})^{2}_i}=\sqrt{\sum (\Delta a_{\rm field})^{2}_i}$$ This
root sum squared approach, based upon the various rate of speed
shortfalls (per cycle) $(\delta a=\delta v/\Delta t)$, is quite
different to the maximum amplitude expressions of superposed
acceleration and speed presented above (Equations \ref{eq:SumB}
and \ref{eq:SumC} respectively) --- even though each $|\delta
a|=\Delta a$. This distinction is due to \mbox{$(\delta
a)^2=(\delta v/\Delta t)^2$} being proportional to the rate of
(unsteady) \emph{energy} (per cycle), whereas the superposition of
individual motion/speed (perturbation) sinusoids and the proper
acceleration sinusoids (derived from them) are based upon
sinusoidal \emph{motion}. In this section
(\ref{subsection:attenuation}), it is the undulatory/sinusoidal
(residual) proper \emph{motion} (or speed) arising from a
superposition of multiple (rotating space-warp based)
acceleration/gravitational fields that is being examined.

\subsubsection{Background information concerning the (discrete) Pioneer
observations}\label{subsubsection:Background to observ}Both Jet
Propulsion Laboratory (JPL) and The Aerospace Corporation (Aerospace)
assign a standard (1-$\sigma$) uncertainty of 1 mm/s over a 60 second
count time for the S-band Doppler data \citep[ p.10]{Anderson_02a}
--- after calibration for transmission media effects\footnote{The
dominant systematic error that can affect S-band tracking data is
ionospheric transmission delays \citep[ p.60]{Turyshev_10b}.}. It
takes nearly 2 weeks for the anomalous acceleration ($8.74 \times
10^{-10}~\rm{m~s^{-2}}$) to achieve a \mbox{1 mm/s} motion
anomaly/shortfall. This is some 20,000 times longer than the 60
second count time, and some $13\frac{1}{2}$ times greater than the
total anomalous change in speed over a 1-day time scale. Over the
substantially longer time scale of \mbox{$\sim$ 3 months}, the
uncertainty circumstance (for the Doppler data) is similar in
that: the Doppler residuals (from both JPL's and Aerospace's
analyses) --- with the anomalous acceleration taken out --- are
distributed about zero Doppler velocity with a systematic
variation of $\sim$ 3.0 mm/s \citep[ p.19 \& p.37]{Anderson_02a}.

The noise of the raw data, especially (thermal or) plasma
noise\footnote{Notwithstanding the implementation of a
batch-sequential filtering and smoothing algorithm used in
conjunction with JPL's Orbit Determination Program \citep[
p.16]{Anderson_02a}.}, together with the use of just 5 data points
per day on average\footnote{Based on 20055 data points for Pioneer
10 over an $11\frac{1}{2}$ year period, cf. Pioneer 11 with 10616
data points over its $3\frac{3}{4}$ year analysis period \citep[
p.20]{Anderson_02a}; i.e. approximately 5 and 8 data points per
day, respectively.}, effectively masks any clear, continuous and
unambiguous signature in the $a_P(t)$ observational data --- other
than the constant `frequency'/velocity drift (cf. predictions)
that is the ``Pioneer anomaly" and (to a lesser extent) the
$\sim$annual residual. Upon removing the anomaly, the `noise' in
the data and the manner of its determination, makes an accurate
`real time' \emph{continuous} assessment of the very short-term
(i.e. minutes or hours) temporal behaviour of $v_p(t)=\sum
[v(t)]_i$ and $a_p(t)=\sum [a(t)]_i$ an impractical/unrealistic
task\footnote{These comments are in no way meant to disparage the
data analysis performed by: JPL, the Aerospace Corporation, or
other investigators. The intricacy and comprehensiveness of the
orbit determination programs are especially awe inspiring. Indeed,
the model presented herein, by way of merely dealing with the
\emph{anomalous} acceleration and residuals, is decidedly less
intricate (in this regard).}.

\subsubsection{The difficulty associated with assessing the
temporal variation in $a_P(t)$}\label{subsubsection:stochastic}
The major difficulty associated with relating the temporal
\emph{variation} in observational anomalous acceleration $a_P(t)$
to the model's temporal variation in $a_p(t)$ is that
observational $a_P(t)$ is derived from non-continuous Doppler
velocity data `points'. Even though the raw Doppler data is
received \emph{continuously}, this data is (then) integrated over
a time interval to give an average for the interval.

With the raw observations directly addressing spacecraft speed,
rather then acceleration, and in light of observational and data
processing circumstances that impinge upon the Doppler data, our
main focus shall be upon a comparison involving how well the
variation/range in observed Doppler velocity data corresponds to
the quasi-periodic (or alternatively quasi-stochastic) variation
of speed implicated by the model\footnote{With its superposition
of multiple (constant and very small amplitude) sinusoidal
gravitational/accelerational field effects, all of which have
different frequencies, and the associated superposition of
(multiple) sinusoidal speed variations/perturbations.}, especially
$(\Delta v_p)_{\rm{max}}$ (as defined by Equation \ref{eq:SumC} in
subsection \ref{subsubsection:summations}). This `correspondence'
is quantitatively addressed in subsection
\ref{subsubsection:Velocity_vs_Doppler}.

In subsection \ref{subsubsection:accel attenuation} we shall give
a brief overview of the model's $\Delta a_p$ variation/range,
noting that there is a \emph{disconnect} between this variation
and the variation apparent in the $a_P$ data points, that are
determined over the course of the 11$\frac{1}{2}$ year long-term
(Pioneer 10) rigorous data analysis period (3 Jan 1987 to 22 July
1998) --- see \citet[ Figures 13, 14 and 17]{Anderson_02a}. This
is because the model's \emph{derived} proper acceleration sinusoid
amplitudes: $\Delta a=\Delta v/\Delta t$\,, do not have a common
temporal basis, in that the cycle/period duration ($\Delta t$) is
different for each $\Delta a$. For a summation (rather than a
superposition) of the monotonic motion shortfall rates $(\delta
a=\delta v/\Delta t)$, this ``timing issue" is not a concern. Note
that these two preceding (and similar) equalities involve the
following relationship of units: $[\frac{\rm m}{\rm
s^2}]=[\frac{\rm m}{\rm s}\frac{1}{\rm cycle}][\frac{\rm
cycle}{\rm s}]$. Neither is this timing issue a concern for the
model's $\Delta v$ values (recall subsection \ref{subsubsection:2
ways to CSHM}), because these are actual values accessible to
`real time' observations --- irrespective of their (pedantic) `per
cycle' unit basis in the model.

Somewhat restating the above: with discrete (line-of-sight) Doppler
velocity data as the primary ``means of observational contact", there
is a tension/disconnect between the model's approach to
\emph{sinusoidal} acceleration terms, and \emph{observational}
acceleration, with the latter \emph{derived} via a rate of change of
discrete velocity measurements (with respect to \emph{time})
--- i.e. observation time, which is based upon the standard unit
of seconds. In short, the conversion from observational discrete
speed/velocity `data points' to acceleration values overlooks the
`time span' adjustments required to correctly ascertain the
(sinusoid-based) proper acceleration \emph{component} values, and
hence the actual (\emph{superposed}) physical variation of
(anomalous/residual) proper acceleration [i.e. $a_p(t)$] proposed
and formulated by the model. Recall that the spacecraft's
\emph{overall} `proper acceleration' is the acceleration that
would be displayed by a (sufficiently sensitive) onboard
accelerometer, and thus in principle the `residual' proper
acceleration is measurable/determinable --- after correcting for
the effects of standard (Newtonian and) general relativistic
gravitation.

Subsequently, with the actual observational variation in $a_P$ not
being indicative/representative of the model's $a_p(t)$ variation
(around a mean value), the model's relationship to the (Doppler
speed/velocity) observational data evidence is restricted to a
purely velocity basis. As such, the model's value of $(\Delta
v_p)_{\rm max}$, i.e. the maximum range of (the superposed
resultant) $v_p(t)$, is of vital importance; it supplements the
measurement noise/residuals inherent to Doppler data based
spacecraft navigation. The model's $(\Delta v_p)_{\rm max}$ value
is quantified in subsection
\ref{subsubsection:Velocity_vs_Doppler}.

\subsubsection{The model's $(\Delta a_p)_{\rm max}$ value and the
attenuation of maximum amplitude}\label{subsubsection:accel
attenuation} By way of Table \ref{Table:acceleration} in
subsection \ref{subsubsection:geometric acceler} and Equation
\ref{eq:SumB}, $(\Delta
a_p)_{\rm{max}}\approx18.3\times10^{-10}~\rm{m~s^{-2}}$, which is
approximately twice $a_P$, as compared to the observational
evidence \citep[ Figures 13, 14 and 17]{Anderson_02a} which
implies $(\Delta a_p)_{\rm{max}}< a_P$. For convenience the
individual acceleration amplitudes --- that are actually derived
much later in the paper --- have been included/previewed in Table
\ref{Table:Velocity}. As previously mentioned (see discussion in
subsections \ref{subsubsection:Background to observ} and
\ref{subsubsection:stochastic}), the variation/range of $a_P(t)$
is not symptomatic of the model's range/variation in $a_p(t)$.
Furthermore, subsection \ref{subsubsection:summations} illustrated
the quite different \emph{summation} formulations for the motion
shortfall rates $(\delta a)$, as compared to the sinusoidal motion
amplitudes $(\Delta v)$ and their rate of change $\Delta
a\,\,(=\Delta v/\Delta t)$. Consequently, we shall (from here on)
largely restrict our (temporal variation based) discussion to the
validity of the relationship between: the model's $(\Delta
v_p)_{\rm max}$ value, and the range/variation of the
observational Doppler velocity/speed data residuals. We also need
to take into consideration the typical level/magnitude of noise in
the Doppler measurement `process'.

Attenuation of the maximum possible amplitude variation of the
\emph{model's} $v_p(t)$ values [i.e. $(\Delta v_p)_{\rm max}$]
conceivably arises in two main ways. Firstly, with five major
moons involved (six including the much lesser role played by
Neptune's Triton) and their fairly \emph{short} period durations
ranging from Io's 1.77 days to Callisto's 16.69 days, any
maximum/minimum value is short lived
--- being of the order of hours (and minutes) rather then days.
With a longest data integration time of 1980 seconds (33 min) and
count times (and data integration times) of 10 and 60 seconds
quite typical \citep[ p.58]{Turyshev_10b} --- although `Aerospace'
had a preference for the longer count times of 600 and 1980
seconds \citep[ p.10]{Anderson_02a} --- the attenuation in
$(\Delta v_p)_{\rm max}$ (arising from this short-term feature of
the model) is almost negligible\footnote{By way of contrast, the
$\sim$annual (356 day) beat duration of the Titan-Callisto
superposition is a relatively much longer-term variation. Indeed,
it is the model's longest-term variation, primarily evidenced by
its (long-term) amplitude modulation.}, and we shall assume it to
be $<\,$3\,\%. In other words, the observational data's sampling
rate is considered sufficiently dense so as to `appreciate' (i.e.
largely not miss) the shortest-term temporal variation(s) in
$v_p(t)$ propounded by the model.

Secondly, attenuation conceivably arises by way of the orbital
resonance of Jupiter's Galilean moons: Io, Europa and Ganymede, in
that they are in a 4:2:1 resonance; and additionally by way of
Ganymede and Callisto which effectively display a (very near to,
and subtle) 7:3 resonance. As such, all four (spin-orbit
coupled/resonant) moons are never `aligned' (on the same side of
Jupiter), and thus their `related' rotating space-warps (RSWs)
possibly never attain maximum (constructive) or minimum
(destructive) interference. This preceding line of thought assumes
that the (four) `initial' phases $[(\varphi)_i]$ of the rotating
space-warps (associated with the \emph{four} Galilean moons) are
dependent upon their orbital location, but (it shall become
apparent later that) there is no basis for this assumption.
Notwithstanding this qualifying remark, Galilean lunar orbital
resonance will influence periodicity in the $v_p(t)$ data, but it
does \emph{not} follow that their resonant motion involves
significant/major attenuation in the range/variation of the
Doppler data\footnote{Later, we shall see that these initial
phases --- although `knowable' in principle (but not currently
determinable in practice) --- have (herein) not been
ascertained.}. We argue (below) that the attenuation is no greater
than 12 percent, by way of one (i.e. the least) of the five major
moons being unable to contribute to a maximum superposition
magnitude/amplitude.

In determining the Doppler data, a `batch-sequential'
method\footnote{``Though the name may imply otherwise,
batch-sequential processing does not involve processing the data
in batches. Instead, in this approach any small anomalous forces
may be treated as stochastic parameters affecting the spacecraft
trajectory. As such, these parameters are also responsible for the
stochastic noise in the observational data. To characterize these
noise sources, we split the data interval into a number of
constant or variable size batches with respect to the stochastic
parameters, and make assumptions on the possible statistical
properties of the noise factors. We then estimate the mean values
of the unknown parameters within the batch and their second
statistical moments \citep[ p.81]{Turyshev_10b}."} was implemented
by JPL (utilising a 1-day, 5-day, or 200-day batch duration), and
the data is filtered, weighted and smoothed. Although this (data
processing) would appear to possibly result in attenuation of the
$v_p(t)$ [and $a_p(t)$] data, it is a technique associated with
orbit determination; and as such, it \emph{reduces} measurement
noise. Thus, the range/variation of the model's $v_p(t)$ [and
$a_p(t)$] (residual) values do not incur attenuation as a result
of its use.

%*********************************** Table 0 *******************************
\begin{table*}[t]
\small \caption{Lunar orbital periods \& frequencies, and
component proper acceleration and speed amplitudes.
\label{Table:Velocity}}
\begin{center}
\begin{tabular}[t]{lcccccccc} \hline
MOON$^a$              &\textit{Units}       &Luna$^{b}$ &Io
&Europa &G'mede$^{c}$   &Callisto &Titan &Triton$^{d}$
\\\hline
Moon's orbital period   &(days)               &27.32     &1.769   &3.552        &7.155   &16.69      &15.95     &5.877  \\[0.5ex]
Moon orbit frequency ($f$)  &($10^{-6}~{\rm s^{-1}}$) &0.424 &6.542  &3.259  &1.618  &0.694     &0.726   &1.969 \\[0.5ex]
Accel. amplitude$^e$ ($\Delta a$)  &($10^{-8}~{\rm cm/s^{2}}$) &0.00 &5.139 &2.190   &5.354    &1.884    &3.348    &0.416 \\[0.5ex]
Speed amplitude$^f$ ($\Delta v$)  &($10^{-2}~{\rm mm/s}$) &0.00 &7.855 &6.719   &33.10    &27.17    &46.12    &2.112 \\[0.5ex]
Speed amplitude$^g$ ($\Delta v$)  &(${\rm mHz}$) &0.00 &1.205 &1.031   &5.076    &4.167    &7.074    &0.324 \\[0.5ex]\hline
\end{tabular}
\end{center}
\begin{center}
$^a$The lunar data is taken from Table \ref{Table:angular
progression} in Section \ref{Section:Quantif Model}; with lunar
orbital period and frequency rounded off to 3 or 4 significant
figures. $^b$Earth's moon. $^c$Ganymede is abbreviated to G'mede.
$^d$Retrograde motion. $^e$Acceleration amplitude data is taken
from Table \ref{Table:acceleration} (subsection
\ref{subsubsection:geometric acceler}), where it is called
``weighted acceleration" ($\Delta a_{w}$). Note that the square
root of the summation of squared (weighted) acceleration is
$8.64\times 10^{-8}~{\rm cm~s^{-2}}$ without Triton, and
$8.65\times 10^{-8}~{\rm cm~s^{-2}}$ with Triton's contribution;
as compared to the Pioneer anomaly's quoted magnitude of $a_{P} =
8.74\pm 1.33 \times 10^{-8}~{\rm cm\,s^{-2}}$. $^f$Based on
$\Delta v=\Delta a/ \omega_{\rm{cyc}}=\Delta a/f$. Note that
$\Delta v_{\rm{max}}=\sum (\Delta v)_i\approx1.23\,{\rm
mm~s^{-1}}$. $^g$Via conversion factor of $10\, \rm{mHz}=0.652\,
\rm{mm\,s^{-1}}$ (recalling subsection \ref{subsubsection:A
comment on}), with $\Delta v_{\rm{max}}=\sum (\Delta
v)_i\approx18.9\,{\rm mHz}$.
\end{center}
\end{table*}
%*********************************** End Table 0 *****************************

Consequently, neither of these two possible (model-based)
attenuation factors, nor the implementation of batch-sequential
processing, are seen to cause any major attenuation (i.e. $>15\%$)
of the model's proposed variation/range in the $v_p(t)$ data and
(specifically) the magnitude of $(\Delta v_p)_{\rm max}$. By way
of Table \ref{Table:Velocity} we see that the contributions from
different moons are not equal. Of the five moons that dominate the
model, Europa is the smallest contributor to maximum \emph{speed}
amplitude $[(\Delta v_p)_{\rm max}]$\footnote{Whereas Callisto is
the smallest (and Europa the second smallest) contributor to the
(theoretical) monotonic anomalous \emph{acceleration} (i.e. speed
shortfall rate) --- remembering that $a_p$ is based upon a root
sum squared (RSS) derivation --- and Callisto (necessarily also)
makes the smallest contribution to proper acceleration amplitude
variation.}. With regard to Galilean moon orbital resonance, the
removal of Europa's contribution leads to an attenuation effect
upon $(\Delta v_p)_{\rm max}$ of about $5\frac{1}{2}\%$ of the
`overall' total; and $11\%$ if (directly) opposed/negative (which
is unlikely). Indeed, it is more likely that the attenuation in
maximum speed/velocity amplitude (from Galilean moon orbital
resonance) is closer to $5\frac{1}{2}\%$.

Importantly, the model will be rendered invalid if the `standard'
long-term maximum amplitude variation of the observed Doppler
velocity data --- independent of large variations (and spikes)
associated with manoeuvres and other spurious effects --- fails to
be consistent (i.e. a good `match') with: expected measurement
noise, and more importantly, the model's estimated value of
$(\Delta v_p)_{\rm max}=\sum (\Delta v)_i$ (see subsection
\ref{subsubsection:Velocity_vs_Doppler}). Finally, we note that in
addition to (mean longer-term) $a_P$ being constant, the model's
$(\Delta v_p)_{\rm max}$ and $(\Delta a_p)_{\rm max}$ values are
also constant --- at least as regards the 3 Jan 1987 to 22 July
1998 data period of Pioneer 10 when/where the spacecraft is far
beyond the orbits of Jupiter and Saturn.

\subsubsection{Comparing the model's velocity range to the
Doppler observations}\label{subsubsection:Velocity_vs_Doppler} As
previously mentioned, Table \ref{Table:Velocity} \emph{previews}
results (relevant to this section, i.e.
\ref{subsection:attenuation}), that are delivered much later in
the paper. From the acceleration/gravitational field amplitude
values, and lunar (spin and) orbital `frequencies', we establish
the individual (contributing) speed amplitudes\footnote{Note that
we have adjusted the model's (pedantic) speed amplitude and
frequency units of $[\frac{\rm m}{\rm s}\frac{1}{\rm cycle}]$ and
$[\frac{\rm cycle}{\rm s}]$ (respectively), to their observational
(and more standard) format. As such, the [cycle] `unit' has been
left out.} $(\Delta v)$. By way of: Equation \ref{eq:SumC} (in
subsection \ref{subsubsection:summations}), noting that: $0.652\,
\rm{mm/s}=10\, \rm{mHz}$ (recall subsection \ref{subsubsection:A
comment on}), and neglecting the (minor) influence of attenuation
effects, we have:
$$(\Delta v_p)_{\rm max}=\sum (\Delta v)_i=1.23~\rm mm/s=18.9~\rm mHz$$
With 10\% attenuation, this maximum value (or upper bound)
reduces to about 17 mHz, and with (an estimated maximum) 15\%
attenuation, this upper bound reduces to about 16 mHz.

It is reassuring that the (constant) amplitude range/variation
\emph{bound}, established by the model, is consistent (and
somewhat less than) the residuals `generated' by: \citet[ Figure
7]{Levy_09b}, and \citet[ Figure 2 top right diagram]{Toth_09a}
--- with effectively all of their (non-spurious) residuals spread
between $\pm$20 mHz and $\pm$25 mHz, respectively. \citet[ Section
5]{Levy_09b} also found that the standard deviation of their
residuals was 9.8 mHz --- i.e. about half the model's maximum
variation.

We necessarily conjecture/hypothesise that the measured (speed)
residuals are the additive `resultant' of \emph{both} measurement
noise, \emph{and} the variation inherent in the model's instantiation
of multiple and superposed sinusoidal speed variations --- around a
mean value.

This $\pm \,$20 to 25 mHz residual range (derived from the
literature) represents a median value (in the literature) in that:
\citet[ Figure 13]{Anderson_02a} has slightly smaller
residuals\footnote{Possibly because it is based upon a weighted
least-squares (WLS) approach cf. a batch-sequential filter (BSF)
algorithmic approach \citep[ p.23]{Anderson_02a}.} (in general),
whereas the (later data Pioneer 10) residuals in \citet[ Figure
2]{Olsen_07} are bigger (about $\pm$30 mHz). \citet[ Figure
2]{Olsen_07} also displays an annual (or $\sim$annual) amplitude
modulation, which is supportive of the Callisto-Titan beat
behaviour discussed in subsection \ref{subsubsection:Space-warp}
and elaborated upon in subsection
\ref{subsubsection:Callisto-Titan}. Interestingly, the (shorter
sample duration) Pioneer 11 residuals in \citet[ Figure
6]{Olsen_07} have a (maximum) range of about 20 mHz, which is also
(roughly) the maximum amplitude of the (non-spurious)
CHASMP-based\footnote{Compact High Accuracy Satellite Motion
Program. Note that CHASMP employs a weighted least squares
estimation approach.} residuals displayed in \citet[ Figure
9]{Anderson_02a}.

\emph{Importantly}, \citet[ Section 4]{Levy_09b} note that: ``It
should be emphasised that the level of the residuals in [their]
\mbox{Fig. 7} is higher than the measurement noise. It is also
clear from the figure that the postfit residuals do not correspond
to white gaussian noise." Further, \citet[ Section 4.6,
p.80]{Turyshev_10b} declare that: ``Several, independent efforts
to analyze the trajectories of the two [Pioneer] spacecraft have
demonstrated that this [i.e. providing a model of as large a
segment of the spacecraft's trajectory as possible, using a
consistent set of parameters and minimizing the model residual]
can be accomplished using multi-year spans of data with a
root-mean-square \emph{model residual of no more than a few mHz}
[italics added]."

This quantitative circumstantial evidence, although not directly
supporting the model, is strongly indirectly supportive of the
proposed/hypothesised model. Indeed, \citet[ Section 5]{Levy_09b}
found that they could almost \emph{halve} the standard deviation
of their model's residuals (down to 5.5 mHz from 9.8 mHz) simply
by introducing (into their model) periodic terms --- relating to
the Earth's diurnal and annual periods.

\subsubsection{The Callisto-Titan `beat' amplitude dominates
the Doppler variation}\label{subsubsection:Callisto-Titan} From
Table \ref{Table:Velocity}'s previewed results, we see that the
amplitude of the 356 day ($\sim$annual) Callisto-Titan beat period
(recall subsection \ref{subsubsection:Period}) is 0.733 mm/s or
11.24 mHz, via $(\Delta v)_{\rm{Callisto}}+(\Delta
v)_{\rm{Titan}}=73.3\times 10^{-2}~{\rm mm/s}$. With this value
being sizeable, at about 60\% of the maximum speed amplitude
predicted by the model [$\Delta v_{\rm max}=(\Delta v_p)_{\rm
max}=1.231~\rm{mm/s}$\,], and because this `amplitude modulation'
has (by far) the \emph{longest period} of any of the periodic
amplitude variations within the model, it is necessarily the
dominant long-term amplitude effect in the observed/measured
Doppler residuals (predicted by the model). This
($\sim$annual/356-day period) amplitude modulation is clearly
evident in \citet[ Figure 2]{Olsen_07}.

From the model's perspective, this `beat' duration/frequency and
associated amplitude modulation has been (in the literature)
mistakenly understood as an Earth-based annual residual.
\citet{Markwardt_02} was sufficiently concerned with its period
that he referred to it as the ``$\sim$annual" residual. This
subsection's (model-based) quantitative result --- i.e. the 0.73
mm/s ($\Delta v$) magnitude of the `annual' residual --- is fully
supportive of (and compatible with): the observational evidence
for the amplitude of the $\sim$annual residual; as well as the
line of argument, and concluding remarks presented in section
\ref{Subsection:Approx annual}'s fairly extensive discussion of
this residual. We recall that in subsection
\ref{subsubsection:amplitude ambiguity} the observational evidence
implied the $\sim$annual residual had an amplitude of roughly 0.7
mm/s.

\subsubsection{On the variation of $a_P$ data around its
long-term constant mean value}\label{subsubsection:Facade} In this
subsection we present a (further) brief discussion relating to the
observed acceleration (cf. speed) residuals, and amplitude
variation/modulation in the long-term $a_P$ data.

From \citet[ p.37 and p.24]{Anderson_02a}, we can compare the
1-day batch-sequential acceleration residuals (Figure 17) to the
5-day batch-sequential acceleration (Figure 14) data --- after
subtracting the $a_P$ value from Figure 14, and upon adjusting for
the $[\rm{km/s^2}]$ units of Figure 17. The 5-day residuals appear
to be roughly \emph{one third} the level/amplitude of the 1-day
residuals. Further, the much longer 200-day acceleration residuals
(obtained using Aerospace's CHASMP software and included in Figure
14) are in good agreement with the 5-day ODP/SIGMA
results\footnote{``The latest results from JPL are based on an
upgrade, SIGMA, to JPL's ODP software \citep[
p.22]{Anderson_02a}."} --- with the latter using batch-sequential
filtering with a 200-day correlation time. Further, we note that
the amplitude range of the 200-day CHASMP data is closely related
to (about three-quarters) the range/variation of the much shorter
5-day sample averages.

Assuming the information described by these: two figures, three
batch-sequential durations, and two orbit determination programs
(ODPs) are comparable\footnote{Noting that: ``The CHASMP 200-day
averages suppress the solar conjunction bias inherent in the ODP
5-day averages \ldots \citep[ p.24]{Anderson_02a}."}, the
implication is that most of the measurement (or observational)
`noise' influencing the ODPs is short-term (i.e. less than five
days). Of interest, is the amplitude variation (around the
long-term constant mean value) that \emph{remains} evident in the
$a_P$ data [e.g. \citet[ Figure 14]{Anderson_02a}], or equally,
the variation of residual acceleration upon removal of $a_P$. This
situation is consistent with the model (being proposed), in that
in addition to the long-term $\sim$annual residual and short-term
amplitude variations, there will be two medium-term amplitude
variations (cf. modulations) that are associated with the pairings
of: Ganymede with Callisto, and Ganymede with Titan. We note that
these medium-term (spacecraft-based) amplitude variations are
`induced' by (superposed pairs of) acceleration/gravitational
fields, with these fields (in turn) having `emanated' from two
\emph{lunar}-based rotating space-warps. Finally, from Table
\ref{Table:Velocity} we see that these two medium-term variations
have (superposed) amplitudes comparable to that of the (Titan with
Callisto) $\sim$annual residual, because Titan, Ganymede, and
Callisto (in that order) are the dominant proper motion/speed
residuals.

\emph{Interestingly}, \citet[ Section 3.1]{Iorio_07} notes that
(with particular regard to the annual residual): ``\ldots changes
in the [200-day] correlation time lead to quite different results
with the Kalman filter." The model is able to propose an
explanation for this circumstance in that: with Ganymede and
Callisto in a 7:3 orbital resonance relationship, that spans 50.1
days --- and Io, Europa and Ganymede in a 4:2:1 orbital resonance
--- the use of a 200-day correlation time is in remarkable
synchronisation with four full 50-day cycles of this
Ganymede-Callisto orbital duration relationship. Subsequently, any
deviation from a 200-day correlation time will lessen the
distinctiveness of the $\sim$annual residual, by way of
introducing phase offsets and hence additional amplitude effects,
that are otherwise (fortuitously) suppressed.

\subsubsection{Concluding remarks on the model's consistency with
observational data} The model (at this preliminary stage) is
qualitatively and quantitatively consistent with the Pioneer 10
(and Pioneer 11) observational evidence; which in addition to the
Pioneer anomalous acceleration ($a_P$), implies (a number of)
short and medium-term periodic amplitude variations, and a
(long-term) $\sim$annual amplitude modulation\footnote{Noting that
this (beat-based) amplitude modulation `contains' numerous
short-term amplitude variations.} of the (post-fit Doppler) speed
and acceleration residuals. As such, in the $a_P$ data there is a
noticeable temporal variation (i.e. `wandering') around the (very)
long-term constant mean (Pioneer anomaly) value --- see \citet[
Figure 14]{Anderson_02a}. The model's approach allows this
(apparently) stochastic temporal variation in $a_P$ around its
mean value, and the (overall and non-spurious data based)
amplitude of the Doppler residuals, to `exist' at a
level/magnitude that exceeds the (respective) amplitude variations
that can simply (and solely) be attributed to measurement noise.

In response to the favourable relationship the model has towards
the Pioneer observational evidence, we shall further pursue and
build upon this Section's ``first stage modelling of the Pioneer
anomaly" --- so as to eventually establish (amongst other things)
the results that have been partially \emph{previewed} in this
section (\ref{subsection:attenuation}) --- in particular Table
\ref{Table:Velocity}.
%**********************************************************************************
\subsection{Summary of the model's first steps and what's still to be
done}\label{Subsection:Summary 1st steps} The
hypothesis/stipulation of a rotating warp (or rotating `fold') of
space curvature (or space deformation) --- first mentioned in
subsection \ref{Subsection:Shackles} and implied by observational
results discussed in section \ref{subsubsection:Space-warp}
--- is a promising, multi-faceted and provocative first step
towards a model of the Pioneer anomalous observations [$a_P(t)$].
Later we shall see that this new mechanism is related to a clash
involving a (common) exceedingly small geometric phase offset
within (a great many, i.e. $\sim10^{50}$) \emph{discrete}/quantum
mechanical systems (i.e. atoms/molecules) by way of their (``bulk
solid" body) motion in \emph{analog} curved spacetime.

Utilising a barycentric systemic reference frame, and applying (a
power version of) Rayleigh's Energy Theorem (also known as
Parseval's theorem) to compound motion (i.e. steady \emph{and}
unsteady motion) in deep space (and curved spacetime), allows the
Pioneer anomaly to be conceived as a (path-based, i.e. velocity
vector based) shortfall in predicted spacecraft motion ---
irrespective of the spacecraft's direction of motion (and speed).
Standard general relativistic gravitation determines the
(temporally) `steady' motion component, whereas the unsteady
component embodies the new/supplementary physical mechanism
presented (and espoused) herein.

Each `constant' (i.e. the same everywhere) amplitude ``rotating
space-warp" (RSW) leads to a low mass body (such as a spacecraft)
experiencing an (unsteady) sinusoidal variation in
acceleration/gravitational field strength, which in turn causes a
sinusoidal variation in speed around an `equilibrium' speed ---
all other things being equal. From a barycentric (or Earth-based)
perspective, this oscillatory variation in speed (and kinetic
energy) results in a constant rate of motion shortfall (per
cycle), cf. the (predicted) motion in the absence of the unsteady
motion. The individual RSWs and their effects coexist in a
wave-like (linear) superposition, although the effect of this RSW
superposition upon spacecraft proper motion and proper
acceleration, as compared to the (monotonic) rate of motion
shortfall ($a_p$), are quite distinct (see subsection
\ref{subsubsection:summations}).

Later we shall see that five major rotating space-warps account
(almost completely) for the Pioneer anomaly, specifically
involving Jupiter's four Galilean moons and Saturn's large moon
Titan --- i.e. large moons that are part of a spin-orbit coupled
three-body (Sun-planet-moon) celestial system. For geometric
suitability reasons (explained in subsection
\ref{subsubsection:three-body phase}), the Earth's moon is
\emph{not} a RSW `generator'. It should be noted that direct
variations in gravitational field strength arising simply from
lunar motion is \emph{not} (at all) what is being proposed!
Furthermore, subsection \ref{subsubsection:WMAP} briefly discussed
how these RSWs could be non-isotropically affecting the
measurement of CMB (radiation) photons, e.g. `producing' a
`foreground' solar ecliptic plane signature upon the CMB data.

Significantly outstanding at this stage, is an understanding of
the restriction of the new mechanism's effects to low mass bodies;
(in so much that) the affected (spacecraft) mass is below the
`non-local' cut-off masses associated with each (individual)
rotating space-warp. This was the first of subsection
\ref{Subsection:Primary observational}'s three primary
observational constraints. Only by going beyond the scope of
General Relativity's explanatory domain and standard contemporary
physics (a process begun in \mbox{Section
\ref{Section:PhiloTheory}}) can this crucial constraint be
circumvented.

We have adhered to sound scientific methodology, in that the
establishment of the model's `form' is being dictated by the
Pioneer anomaly's observational evidence --- especially its more
subtle (and awkward) aspects. Section \ref{subsubsection:aspects}
outlined the proposed mechanism's \emph{essential
characteristics}. Compatibility with the fine detail of the
observations is both a primary motivator, and the distinguishing
feature, of the model; e.g. the low $a_P$ reading of Pioneer 11
\emph{en route} to and `around' Saturn encounter (subsection
\ref{subsubsection:Saturn jump}). Additionally, the temporal
variation of the $a_P$ data (\emph{around} its constant long-term
mean) --- which is quite distinct from the (anomalous) monotonic
shortfall aspect --- and the (rotating space-warp `driven')
temporal variation of the model's (proper) speed/motion and
acceleration `residuals' were investigated (section
\ref{subsection:attenuation}). Prior to this, the diurnal and
annual residuals were closely examined (subsection
\ref{subsubsection:diurnal} and section \ref{Subsection:Approx
annual} respectively), and the (sinusoidal) `annual' residual has
been recast as an $\sim$annual (356 day) non-Earth-based amplitude
modulation --- that arises from a long-term `beat' duration
involving two RSW's of similar (spin and) orbital
frequencies/periods (i.e. Jupiter's Callisto and Saturn's Titan).
Further, the model's amplitude (of \,0.73 mm/s) for this
Titan-Callisto beat frequency is a good match to the observed
($\sim$\,0.7 mm/s) amplitude of the $\sim$annual residual --- see
subsection \ref{subsubsection:Callisto-Titan}.

The intractability of comparing the amplitude variation of the
model's $a_p(t)$ proper acceleration value with the observational
$a_P$ data (through time) is discussed (subsection
\ref{subsubsection:accel attenuation}). Subsequently, the Doppler
speed/velocity residuals become a primary indicator of the model's
veracity. These observable residuals are comprised of (i.e.
indicative of) \emph{both} measurement noise and the model's
proposed sinusoidal/undulatory proper motion effects (around an
equilibrium speed). Additionally, we discuss the quite minor
attenuation effect (upon the model's proper motion/speed
residuals) arising from the (1:2:4) orbital resonance of Ganymede,
Europa and Io, and the (3:7) orbital resonance of Callisto and
Ganymede (subsection \ref{subsubsection:accel attenuation}).

By way of previewing results determined later in the paper (Table
\ref{Table:Velocity}), it was shown (firstly) that the model's
long-term (constant and monotonic) \emph{mean} rate of motion
shortfall value $a_p$ is consistent with the observation-based
Pioneer anomalous acceleration value $a_P$ (subsection
\ref{subsubsection:a_p and mass comment}). Secondly, the combined
magnitude/amplitude of a (superposition-based) maximum amplitude
summation of the model's proper motion/speed residuals, on top of
typical (non-spurious) measurement noise-based residuals inherent
in the Doppler method, is consistent with the observed/measured
spacecraft Doppler residuals determined by various investigators
(subsection \ref{subsubsection:Velocity_vs_Doppler}).

Certainly, there is still much more that a robust explanation
needs to cover and examine, including:
\begin{enumerate}
\item{Ascertaining how the model can theoretically coexist with
General Relativity.} \item{Establishing the amplitude and energy
of the (individual) acceleration/gravitational field undulations
$(\Delta a)$, which equal and correspond to a rate of spacecraft
speed shortfall $(\delta a)$. These $\delta a$ (components)
together (in superposition) then largely quantify the model's
Pioneer anomaly estimation ($a_p$) --- recall subsection
\ref{subsubsection:a_p and mass comment}.} \item{Understanding how
a \emph{virtual} spin phase offset --- relative to the spin phase
(of all elementary fermion/matter particles) within numerous
individual lunar atoms and molecules (exhibiting spin-orbit
coupling) --- that is below a minimum (quantum) energy level,
yields a non-inertial (excess) spin energy. This virtual energy,
is recognised globally (i.e. systemically) but it is locally (i.e.
at the QM `level') inexpressible. Thus, this (\emph{total} QM)
energy --- pertaining to a great number of atoms/molecules --- is
necessarily released externally as a: `single', \emph{real},
supplementary acceleration/gravitational field --- so as to
preserve conservation of energy \emph{globally}.}
\item{Appreciating the non-local nature of the mass aspect of the
(total) virtual quantum mechanical \emph{energy} --- with this
energy necessarily having the same magnitude as the externally
expressed \emph{supplementary} (`acceleration undulation/warp' and
`mass' based) `gravitational energy'. This (latter) real energy is
comprised of both a macroscopic ``rotating space-warp" (RSW) and
its (associated) non-local mass `distribution'.}
\item{Establishing the qualitative and quantitative distribution
of \emph{non-local} mass (cf. standard inertial or gravitational
mass) in the field.}
\end{enumerate}

A useful, yet overly simplified, \emph{encapsulation} of the new
model is that: in standard general relativity, mass and
(\emph{macroscopic}) motion tell space how to curve, which then
tells mass how to move --- and let us herein consider the ensuing
motion to be \emph{steady}. In contrast, and supplementary to
this, is the (as yet uncorroborated) proposal that: in special
circumstances yet to be fully outlined, (an angular momentum-based
energy associated with) \emph{microscopic} mass `motion' (`within'
a great many atomic and molecular `systems') can also tell space
how to curve (i.e. sinusoidal-like undulations upon the
gravitational field at a point mass, or RSWs more generally),
which then tells (a moving) mass how to \emph{vary} its movement.
The latter --- with regard to a single rotating space-warp ---
involves an oscillatory variation of speed about a mean
motion/speed i.e. \emph{unsteady} motion, that leads to (or rather
coexists with) a very small (incremental) loss/decay of a body's
`given' kinetic energy (per cycle). In this supplementary
`gravitational' case a global/systemic basis to energy
(quantification and `transfer') is necessary, incorporating both
the macroscopic and microscopic `domains' (or realms) together.
Once again we note that only the motion of `low mass' bodies are
affected.

%******************************************************************
%******************************************************************
\section{Theoretical concerns and philosophical aspects of the proposed mechanism}
\label{Section:PhiloTheory} The awkward observational constraints
discussed in \mbox{subsection \ref{Subsection:Primary
observational}} have been addressed and largely satisfied (recall
section \ref{Subsection:Summary 1st steps}) by the model
preliminaries introduced in Section \ref{Section:PrelimModel}.
Unfortunately, giving the reliable observational evidence full
priority, so as to make the Pioneer anomaly real, has lead to a
quite bizarre preliminary model. The challenge of this Section is
to show that this model, although restricted and supplementary to
current theory, is not incompatible with accepted theory --- once
certain assumptions (metaphysical in some cases) influencing
physicists' current understanding are appreciated, and certain
ontological\footnote{Philosophy may be divided into three main
branches: ethics/morality, the theory of knowledge (epistemology),
and ontology; the latter being concerned with the nature of being,
existence, and reality.} ramifications of quantum non-locality are
appreciated.

This Section effectively forms a conceptual bridge between the
awkward implications of the observational evidence and the model's
mathematical quantification (given partly in Section
\ref{Section:general model} and mainly in Section
\ref{Section:Quantif Model}). After a series of preliminary
remarks, the model's tensions in relation to assumptions
underlying contemporary physics are addressed; e.g.
\emph{theoretical} reductionism\footnote{Theoretical reduction is
the reduction of one explanation or theory to another --- that is,
it is the absorption of one of our ideas about a particular thing
into another idea; for example Newtonian gravitation into general
relativity.}. Further conceptual features of the
model\footnote{Especially quantum mechanical aspects.} are
presented in Section \ref{Section:general model}, and prior to
that where it is considered appropriate.

This Section is necessary because the conceptual, ontological, and
mathematical foundations of: QMs, `gravitation', mass, space,
time, and energy are being touched upon. Subsequently, the
explanation to follow needs to address `philosophical' issues
beyond the usual scope of today's ``physics".

\subsection{Preliminary remarks}\label{subsection:preliminary remarks}
In order to enrich the model being developed, a number of
preliminary remarks need to be presented. The remarks are wide
ranging involving: philosophy, philosophy of science, limitations
of GR, a (decoherence) interpretation of QMs; and issues such as:
Ehrenfest's theorem, quantum-to-classical transition, quantum
condensate behaviour, and angular momentum addition. Together,
they can be thought of as building blocks for the new model, and
the write-up that follows.

\subsubsection{Societal consensus in physics}
Unlike philosophy which (to the best of its ability) pursues and
holds alternative interpretations with rigour and relish, the
mathematical nature of physics, with its singular solutions,
appears (in practice) to also extend to explanations of anomalous
physical phenomena --- e.g. dark matter, dark energy. Generally, a
dominant conceptual ``party line" is favoured; and subsequently,
alternatives face a daunting task, compounded by the interrelated
nature of conventional physical modelling (e.g. cosmology's
`concordance model'). Due to the interrelated nature of this
(Pioneer anomaly motivated) unconventional model, the reader is
asked to refrain from drawing firm conclusions until the
\emph{end} of this Section (i.e. Section
\ref{Section:PhiloTheory}) is reached.

\subsubsection{Philosophy of science comment}
Borrowing terminology from the philosopher of science Imre Lakatos
(1922-1974), it may be said that an alternative ``research
programme" (RP) is to be presented; and thus alternatives to
beliefs in the ``hard core" of the current research programme
shall be considered\footnote{An increase in the number of
`anomalies' is a feature of a ``degenerative" research program.
For example, see: ``Do we really understand the solar system?
\citep*{Dittus_06}"}. The combined alternative views to be
presented, assemble to form, in the author's opinion, a
`progressive' research program.

Interestingly, some of the alternatives to be pursued are external
to the curriculum of today's physicists; and yet well-known to
philosophers of science and physics. One need only remember the
surprise with which an expanding universe [post Vesto Slipher
(1917) and Hubble's observations] was greeted to appreciate that:
a lack of doubt may be a physicist's worst enemy\footnote{Before
Hubble's Law (1929), the \emph{assumption} of a `static' universe
was the generally accepted view.}. The assurance with which both:
GR fully/correctly describes the solar system's (curved spacetime
based) mechanics, and that theoretical ``reductionism" (e.g. GR to
QMs) is currently favoured; are the primary accepted beliefs to be
confronted herein --- at least in the low energy, mass era of the
universe.

\subsubsection{Physics as possibly one step ontologically removed
from ``deep-reality"}\label{subsubsection:noumena} Physics relies
ultimately on observations to understand the world. It is worth
mentioning at the outset that mass, spacetime\footnote{Of which we
shall restrict ourselves to just three dimensions of space, and
one of time, interlinked \emph{observationally} by the demands of
Special Relativity (SR), in conjunction with electromagnetism, to
form ``spacetime".}, and the meaning of intrinsic curvature remain
somewhat mysterious beyond their role in physical law. These are
philosophical (rather then scientific) concerns, that may be a
waste of time to some physicists; but while anomalous phenomena
exist, philosophy of physics issues should not lie beyond
consideration.

The observations of physicists, regarding gravitation beyond
Earth, utilise electromagnetic (EM) radiation. It is not clear
whether ``measurements exhaust reality" or whether a physical
Kantian-like\footnote{Immanuel Kant (1724-1804), a highly
influential German philosopher.} distinction between phenomena
(measurement) and noumena (a reality beyond direct measurements)
is valid. Note that we have stated ``a Kantian-like distinction"
as compared to explicitly employing Kant's version of the
distinction. Nevertheless, the terminology shall be retained
herein with the \emph{new} (or rather modified) meanings of
`phenomena' and `noumena' gradually outlined.

If this distinction is valid, then local measurements might be one
step removed from a further reality, i.e. a reality with non-local
(i.e. global or systemic) hidden variables. What is certain is
that physicists generally reject such a Kantian
distinction\footnote{Were it not for the Pioneer and other
anomalies, Ockham's razor would strongly support their view.}, and
that \emph{local} hidden variables cannot exist. Yet, physics
concerns what we can say about the world or nature, not what
nature is\footnote{These sentiments are often accredited to Niels
Bohr and Werner Heisenberg.}; and the distinction cannot be easily
resolved, nor easily removed.

\subsubsection{A wider sense to use of the word ``gravitation"}
Herein the expression/word ``gravitation" is used in a wider sense
than normal, since we are entertaining the fact that
(predominantly macroscopic) mass, momentum, and energy (comprising
GR's stress-energy-momentum tensor) may not be the only
contributors to non-Euclidean geometry. Microscopic-based energy,
by way of a virtual\footnote{`Virtual' as in below the first
energy level or a minimum change in energy levels, and necessarily
implying non-local systemic (i.e. global) hidden variables.}
intrinsic angular momentum offset per specified cycle time (i.e. a
virtual spin energy), is seen to also contribute to non-Euclidean
geometry --- in the very special circumstances to be described in
Section \ref{Section:general model}. Later, it shall become clear
why this virtual/hidden energy contribution may not be included in
the stress-energy tensor\footnote{This is due to the non-local
role (quantum mechanics based) mass plays in the energy
concerned.}. Rather, we need to treat it as an independent and
supplementary contribution to non-Euclidean geometry; necessarily
expressed as a rotating space-warp, and arising from the
`experience' of quantum mechanical (QM) `\emph{systems}',
specifically atoms and molecules, moving in curved spacetime.
Importantly, this virtual effect/experience, arising via
\emph{analog} curved spacetime, causes no measurable change in the
(\emph{digital}/quantum) atomic and molecular (systems) themselves
--- by way of EM forces being dominant within an atom/molecule.

The term ``gravitation" continues to be coextensive with
``non-Euclidean geometry"; and the path of any mass in `unforced'
motion (in a non-Euclidean geometry) continues to be a `geodesic',
in the sense that it is governed by something akin to a
``principle of least action". Although, it can no longer be said
that: ``matter responds to the geometry of spacetime, as based
upon a stress-energy tensor, and nothing else." Thus, the
statement: ``\emph{only} a metric theory can describe
gravitation," is necessarily being cast into doubt.

\subsubsection{Beyond the scope of GR's formalism}\label{subsubsection:Beyond GR}
The model being developed requires the existence of moon-planet
\emph{macroscopic} spin-orbit resonance, also known as: synchronous
rotation, tidal locking, and phase-lock. This arises from tidal
effects acting upon a (non-rigid-body) planet-moon system. Later, we
see that a moon-planet-sun (three body) system is important in the
new model. It is worth noting that GR is `set-up' to see neither of
these circumstances as in any way significant --- and certainly not
related to (a supplementary) non-Euclidean geometry. Note that the
expression ``non-Euclidean geometry" is not meant to (in any way)
imply the existence of absolute space.

General Relativity itself has issues. It is problematic at small
distances and very high energies/temperatures; additionally, the
reality of spacetime singularities is an ongoing issue. GR's
foundation stone, the principle of equivalence (Pr. of Eq.), is
restricted to local effects, i.e. very small volumes and/or small
bodies. For example, in large bodies, tidal effects lead to
non-uniformity in field strength, which means the Pr. of Eq.
cannot be applied throughout. The point being made is that,
solutions to the equations of GR are restricted to quite simple
circumstances; but the model herein is based upon a (quite
different) complicated set of circumstances, particularly
involving three-body celestial motion.

\subsubsection{The equivalence principle and the uncertainty
principle}\label{Subsubsection:Equiv and Uncert} Concerning the
relationship between GR and QMs, Raymond Chiao dispels any concern
regarding the conceptual tension between Heisenberg's uncertainty
principle and the locality of the equivalence principle.
\begin{quote} \ldots, whenever the correspondence principle holds, the
\emph{centre of mass} of a quantum wave packet (for a single particle
or for an entire quantum object) moves according to Ehrenfest's theorem
along a classical trajectory, and \emph{then} it is possible to
reconcile the two principles \citep*[ p.267]{Chiao_04}. \end{quote} The
tracks of subatomic particles in a bubble chamber illustrate this
circumstance.

In Section \ref{Section:general model} we shall examine the subtle
ramifications of: three-body macroscopic/celestial orbital motion
in curved spacetime affecting the geometric/Berry phase of
atomic/molecular intrinsic spin --- especially the relationship
between orbital phase and spin phase in curved spacetime.

\subsubsection{Interpretative and ontological stance for quantum mechanics}
The idealisation to a classical (and thus simple) approach
utilised in this write-up, favours the acceptance of both: quantum
non-locality and a realist stance on QM atomic and/or molecular
`particle' position --- \emph{but only as regards their centre of
mass} (recall subsection \ref{Subsubsection:Equiv and Uncert}).
The closest interpretation of QMs to this is the \mbox{de
Broglie-Bohm} (pilot wave) theory, which incorporates non-local
hidden variables. This philosophical stance is \emph{not} being
explicitly promoted; rather its ontological preferences overlap
somewhat with the model that is to be presented. Unlike Bohmian
Mechanics the model denies (QM) particles having definite
positions at all times. Additionally, we embrace Decoherence
Theory --- albeit applied with somewhat non-standard background
philosophical assumptions.

\subsubsection{Transition from quantum to `classical' behaviour
--- via decoherence}\label{subsubsection:decoherence} ``Quantum
decoherence" is the mechanism whereby quantum systems interact
(irreversibly) with their environment, to give the
\emph{appearance} of wave function collapse, and hence classical
behaviour. Herein we deviate from a strict (reductive) decoherence
agenda that seeks to:
\begin{quote} \ldots eliminate primary classical concepts, thus neither
relying on an axiomatic concept of observables nor on a probability
interpretation of the wave function in terms of classical concepts
\citep*[ Preface to 2nd ed.]{Joos_03}
\end{quote} We shall retain the classical conception of the physical
world as a `complementary adjunct' to post-decoherence behaviour.
A classical formalism, although a secondary approach, is seen as
an indispensable aspect of physical description\footnote{Can the
concept of a non-Euclidean geometry exist without the existence,
conceptually at least, of a Euclidean geometry?}. Thus,
`decoherence' herein is primarily associated with a disappearance
of \emph{observable} quantum (superposition and entanglement)
effects, and (importantly) a suppression of interference effects.
\begin{quote} In this [decoherence] process, the quantum superposition
is turned into a statistical mixture, for which all the information on
the system can be described in classical terms, so our usual perception
of the world is recovered \citep*[ p.1295]{Davidovich_96}. \end{quote}

For everyday macroscopic objects the ``decoherence time" is
extremely short. Thus, bulk matter is considered to be (always)
free of (directly measurable at least) quantum superposition (and
entanglement) effects. Herein, we shall explore an exception to
this rule arising from analog curved spacetime effects upon a QM
system --- with the effect being \emph{below} a lowest energy
level and/or increment.

In the spirit of the decoherence (many-worlds) interpretation of
QMs, the QM aspects of macroscopic bodies upon measurement are
considered to \emph{not} be obliterated out of existence --- as
compared to the case with the wave function collapse of the
Copenhagen interpretation\footnote{Obviously, experiments do
disrupt a system's wave function but `reality', the `subject' of
the measurements, necessarily persists --- albeit quite
differently post-measurement or post-decoherence.}.

\subsubsection{On quantum condensate behaviour}
Herein, we need to entertain the idea that there is a second way
to attain quantum condensate behaviour. In standard QMs,
`macroscopic' quantum condensate behaviour is restricted to very
low temperature circumstances\footnote{There is also decoherence
by thermal emission of radiation.}. Alternatively, very low
\emph{energy} circumstances (below a minimum or first energy
level), although \emph{hidden} from any direct observation, will
starve off decoherence; and may well exhibit coherence over a
large number of such atomic/molecular (QM) systems --- i.e. a bulk
mass. Subsequently, this situation could facilitate a (new)
non-standard macroscopic condensate behaviour. Quantum
entanglement is an important auxiliary condition for attaining
this behaviour.

The origin of such a common (virtual) QM energy, and its relation
to the rotating gravitational field perturbations (of Section
\ref{Section:PrelimModel}) is the major concern of Sections
\ref{Section:general model} and \ref{Section:Quantif Model}. Later
we see that this energy is indicative of a relative (and virtual)
angular momentum offset achieved (through self-interference) over
a certain (finite) `process' time interval ($\Delta t$).

\subsubsection{On the addition of (rate of change of) angular momentum}
\citet[ p.80]{Feynman_65} ruminates upon the fact that, even
though (classically) angular momentum depends upon a `projection'
angle, quantum mechanical angular momentum always involves integer
multiples of some quantity --- regardless of the axis about which
we measure. Certainly, within a quantum \emph{system} we have an
inability to `count up' angular momentum, as with (say) electric
charge; but this additive restraint need not be extended to
duplicate or multiple quantum systems (i.e. atoms and molecules)
exhibiting a common (i.e. equal and shared) effect. For the model,
the effect in question is a common \emph{rate} of (virtual)
angular momentum offset (effectively) extending over an entire
macroscopic body --- in our case a moon.

\subsection{Tensions of the model regarding reduction, unification, and GR}
\label{Subsection:Tensions}
As was the case in section \ref{subsection:preliminary remarks}, this
section constructs further building blocks for the new model, but more
importantly it begins the process of `deconstructing' certain
presumptions that stand in the way of the new model.

\subsubsection{Celestial ``gravitational" motion as a dissipative
process}\label{subsubsection:Dissipation} General Relativity is
based upon a perceived conservative or non-dissipative process;
and thus the useable (kinetic) energy of a body in motion cannot
be `lost' in the manner argued herein for the Pioneer spacecraft.
Loosely speaking, this paper has argued that cyclic variation in
acceleration/gravitational field strength (at a point), arising
from (new) acceleration undulations, detract from `useable'
kinetic energy (recall subsection \ref{subsubsection:Interpret
power}). Thus, by way of this supplementary
space-warping/curvature, the motion (of low mass bodies) is seen
(from an Earth-based or barycentric perspective) to be depreciated
or `dissipated', and hence the motion is time-asymmetric. Our
common sense notion of time-irreversibility for macroscopic
phenomena is reinforced; but \emph{not} the standard belief that
all gravitational theorisation is inherently time-symmetric.

Section \ref{Section:Status} and \ref{Section:PrelimModel}'s
hypothesising of a non-systematic based (real) Pioneer anomaly
implies that: the astrophysical/celestial motion of low mass
bodies in the solar system joins the ranks of other macroscopic
phenomena \emph{always} acted upon by some type of dissipative
effect, such as: air resistance, friction, and increasing entropy
for example\footnote{Note that this comment/declaration ignores
the well known effects of (various) celestial radiative forces,
and is thus restricted to our newly proposed non-Euclidean
space-time effect.}. It should be noted that there is a
distinction between microscopic/quantum and macroscopic phenomena
regarding the basis for dissipation. The uncertainty principle
ensures time-asymmetry at quantum scales\footnote{For example,
quantum vacuum fluctuations provide disturbances that may result
in radioactive decay.}, whereas a variety of processes, including
the second law of thermodynamics, are responsible at `classical'
scales. We shall see that the uncertainty principle (applied to
energy, expressed as a rate of angular momentum) is the
`facilitator' of the new supplementary space curvature being
proposed.

\subsubsection{Linking a very large number and a very small (energy)
number}\label{subsubsection:Large number} Herein (loosely
speaking), the quantum mechanical energy-time uncertainty
principle ($\Delta E \Delta t\geq \hbar/2$) is given a new
interpretative twist. We exploit the fact that (unlike $\Delta x$,
$\Delta p$, and $\Delta E$) $\Delta t$ is not an operator
belonging to a particle (see subsection \ref{subsubsection:maximum
virtual} for further discussion). Energy-time uncertainty can also
be seen to give an \emph{upper} bound to the ``no change"
condition of an \emph{external} effect imposing itself upon a QM
system's internal (spin) energy over a given (cyclic/process) time
(i.e. its intrinsic angular momentum). If \mbox{$\Delta E \Delta t
< \hbar/2$,} then \emph{no} quantum mechanical influence upon the
system can be expressed within the (microscopic) system itself ---
with such effects considered to be ``below the first (or minimum)
energy level", and hence below a conceivable decoherence
threshold. Thus, the state of the (microscopic) quantum system is
\emph{not changed}. Note that regarding this new externally driven
effect --- dependent upon motion in curved spacetime --- an exact
QM energy (albeit virtual) is assumed to exist in this
interpretation of the energy-time uncertainty relation.

A (virtual) energy `difference' ($\Delta E$) conceivably exists.
For the moment let us simply say that this energy is only
meaningful at a `universal systemic' level. We assume that by way
of curved spacetime, the global inertial conditions for a QM
system can differ from the local QM equilibrium conditions ---
when a QM system is taken around a closed loop (see Section
\ref{Section:general model}). Later, we argue that this (virtual)
energy difference quantifies the QM system's non-inertial status
--- from a universal systemic perspective.

With specific regard to the model, the (maximum) non-inertial
energy ``not internally incorporated"\footnote{Note that we drop
the term ``energy uncertainty".}, that can be associated with
Dirac's constant\footnote{More commonly referred to as the
``reduced Planck constant".} (minimum angular momentum) per
(typical) lunar spin-orbital period: $\Delta E\sim \frac{1}{2}\,
\hbar \, \Delta t^{-1}$, is of the order of $10^{-40}$ Joules.
With typically $10^{50}$ atoms/molecules in a large (solar system)
moon\footnote{See Table \ref{Table:energy} at the end of section
\ref{Subsection:Model quantified internal}.}, an energy effect
\emph{commonly} shared by all atoms/molecules could (conceivably)
involve (up to) approximately 10 gigajoules of energy.

Later we shall see that this commonality, i.e. space curvature
affecting neighbouring lunar QM (atomic/molecular) systems in
effectively the same way, is possible (in part) because of the
small diameter of a (celestial) moon, relative to the distance
from a planet to its moon. In the model, it is the \emph{total}
(virtual) energy that gets re-expressed \emph{externally} to
appease a universal (systemic) conservation of energy constraint
--- arguably as (one of) the rotating space-warps discussed in Section
\ref{Section:PrelimModel}. The manner of the external curved
spacetime (input) effect that influences QM internal energy is
discussed very briefly in the following two subsections
\ref{subsubsection:basis for condensate} and
\ref{subsubsection:basis for energy uncertainty}.

\subsubsection{The basis for (quantum) condensate behaviour, and QM
spin}\label{subsubsection:basis for condensate} Landau and
Lifshitz use the example of a dielectric, where a non conductor
gets a displacement charge, to argue that:
\begin{quote} the separation of the first energy levels of a
macroscopic body may even be independent of the size of a body, as
for example in the electronic spectrum of a dielectric \citep[
p.14 fn.]{Landau_58}.
\end{quote} This independence from the size of the body is an
important concept in this paper.

Moving away from electromagnetic phenomena, we seek to show that
condensate behaviour may arise in relation to the same (i.e. a
common) effect, below the first (or a minimum) energy level,
acting upon every atom/molecule's (elementary fermion/matter
particle-based) intrinsic angular momentum (over a given process
time), i.e. the spin energy. This common effect is simply
additive, yielding a total virtual energy.

Note that in QMs, total energy is (also) a function of QM spin.

Historically, the application of an external magnetic field to
electrically neutral particles (or atoms), led to the `discovery'
of the quantised nature of intrinsic/spin angular
momentum\footnote{In retrospect, the first direct experimental
evidence of particle spin (e.g. the electron spin) was the
Stern-Gerlach experiment of 1922, named after Otto Stern and
Walther Gerlach.}. Herein, we seek to show that the presence of
curved spacetime acts upon QM (spin) angular momentum, resulting
in something new --- albeit (only) if (and when) non-decoherence
behaviour of a many atomed/moleculed body is `active'.

\subsubsection{The basis for the energy uncertainty}
\label{subsubsection:basis for energy uncertainty} Assuming a
systemic reference frame datum and incorporating moon-planet
orbital resonance, looped motion in curved spacetime is seen to
slightly affect (i.e. $\leq $1/2 a
wavelength\footnote{Circumstances where the effect is $> $1/2 a
wavelength exist but they are not relevant to this discussion.})
the wave phase of \emph{spin} angular momentum (per cycle time)
but \emph{not} orbital angular momentum `within' an atom or
molecule. To maintain a multitude of electromagnetic internal/QM
angular momentum coupling relationships as well as conservation of
energy, as they would be in a flat spacetime, the QM system(s)
involved simply export any imbalance arising from motion in curved
spacetime (S/T). This scenario is restricted to a \emph{virtual}
imbalance, i.e. below a first or minimum change in energy level.

Importantly, this virtuality permits a common spin ``precession"
\emph{direction} and magnitude, \emph{relative} to inertial frame
conditions, to (conceivably) apply to \emph{all} elementary
fermion/matter particles within \emph{all} atoms/molecules
regardless of their `internal' spin orientation. With a great many
atoms/molecules similarly involved, the (real) rotating
space-warps previously hypothesised are (conceivably) the
\emph{singular} (condensate) expression of this energy
exportation/deferral process. Thus, internally/microscopically it
is as if this (new) curved spacetime based effect is not there;
but from a `universal' systemic perspective things are quite
different. The process illustrates nature ensuring stability in a
simple, elegant and economic manner.

\subsubsection{Questioning physicists' objective --- regarding
reduction and unification}  Upon removal of dissipative effects
the laws of mechanics are considered to be time reversal
invariant. `Conservative' systems are described by a Lagrangian
and/or Hamiltonian formulation, e.g. Hilbert's variational
approach to GR's field equations. Time symmetry (in this sense)
supports the reductive inclinations of contemporary physicists ---
and draws confirmation from the previously successful reduction of
thermodynamics to statistical mechanics. The generalisation of
reduction to all macroscopic phenomena needs to be recognised as a
goal or agenda and not `a given'.

The current unification agenda/goal of physicists; i.e. weak,
strong, electromagnetic, and gravitational `forces', is ostensibly
`force' based and involves the exchange of particles. Other
macroscopic phenomena such as: elasticity/deformation of
materials, springs (Hooke's Law), and rotor/propeller behaviour in
fluid mechanics\footnote{Which are described ``exactly" by (an
idealised) continuum mechanics approach, or (alternatively)
approximately by finite element methods.} are assumed to `fall in
line'. Is this (to date unsuccessful\footnote{The failure of
string theory and quantum loop gravity to deliver a successful
unification, or a falsification criteria, makes them currently
non-rigidly scientific. This may change, but the recent advent of
a perceived accelerating universe expansion was not even upon the
radar of these ``cutting-edge" approaches. Both of these
approaches are particle based, and hence assume reduction is
viable. For more information see \citet*{Cartwright_07}, and
George Ellis' review \citep*{Ellis_06} of Lee Smolin's ``The
Trouble with Physics" \citep*{Smolin_06}.})
metaphysical\footnote{In the sense of: based on speculative or
abstract reasoning.} reductive agenda too simplistic? Anecdotal
evidence suggests that both Niels Bohr and Erwin Schr\"{o}dinger
doubted a grand unified theory was possible.

In this paper the emphasis is upon an interaction \emph{between}
the microscopic and macroscopic realms --- in the form of an
energy \emph{re-expression}. The word `re-expression' is preferred
to transfer, because the latter has particle connotations. The
nature of their \emph{coexistence}, from a global/systemic
perspective, is considered to lead a new (``emergent") phenomenon,
i.e. (gravito-quantum) rotating space-warps\footnote{Together with
a ``non-local mass" distribution.} --- this being the external
energy expression of unaccounted-for (virtual) (and non-inertial)
QM energy. Possibly ``non-reductive" is a better word than
`emergent'. We cite quantum mechanical non-local entanglement as
the mechanism/`vehicle' by which this re-expression is achieved.

Alternatively, if we assume the correctness of GR and QMs, and
that the current reduction and unification agendas/objectives are
correct; then a `real' Pioneer anomaly is at odds with accepted
theory and these related agendas. Subsequently, a systematic
explanation (involving heat) is experiencing favour within the
Pioneer anomaly community; essentially to make the anomaly ``go
away".

\subsubsection{The differences that divide microscopic and macroscopic
physics}\label{Subsubsection:MM Differences} The existence of a
divide between the microscopic (i.e. up to atoms/molecules) and
macroscopic/classical realms, and the laws describing them, is
undeniable. Aspects of this circumstance include:
\begin{enumerate} \item{The inherent incompatibility of GR and QMs.}
\item{The absence of electron radiation within atoms.} \item{The
nature of dissipation (recall \ref{subsubsection:Dissipation}).}
\item{The `exact' symmetries (and hence quantised gauge theories)
displayed by microscopic systems cf. the approximate symmetries
displayed by macroscopic systems.} \item{The role of special
relativity is primary in QMs, whereas in a variety of celestial
circumstances, e.g. the solar system and galaxies, SR plays a
minor role.} \end{enumerate} On the other hand, an aspect of
commonality between the micro and the macro realms is waves and
wave superposition; simply because wave behaviour is ubiquitous.

To proceed with a real explanation of the Pioneer anomaly, the
\emph{divide} between micro- and macroscopic physics needs to be
emphasised. In the following subsection
(\ref{subsubsection:Prop/HAWT}) we examine if a force and particle
platform, as used in microscopic physics, is sufficient for the
description of a significant (non-gravitational) macroscopic
physical phenomena.

\subsubsection{The role of energy and power in the Vortex Theory of Propellers}
\label{subsubsection:Prop/HAWT} In general physics, energy is
considered to be the capacity (of a system) to do work; with work
defined as force applied over a distance. This implicitly assumes that
force (usually in conjunction with energy) is \emph{necessarily} a
primary quantity in the analysis of major physical interactions.

Theodore Theodersen's Theory of (Aircraft) Propellers
\citep{Theodorsen_48} primarily examines the physics of a
`frictionless' propeller/airscrew, i.e. drag effects are ignored.
Minor corrections may then be made for drag effects. A
vortex/circulation\footnote{``Circulation" is a fluid mechanical
term. It is essentially a macroscopic continuum mechanics
equivalent to angular momentum, and closely related to the
(velocity) potential in a ``potential flow" analysis. In fluid
dynamics, potential flow describes the velocity field as the
gradient of a (scalar) velocity potential --- if the flow is
irrotational.} approach is employed, with this approach being
uniquely applicable to (low solidity\footnote{Solidity refers to
the amount of (swept) disc area occupied by the (solid) material
blades themselves.}) airscrews and (horizontal axis) wind turbines
cf. other types of (high solidity) \emph{open}
turbomachinery\footnote{Examples of \emph{closed} turbomachinery
are numerous, including: centrifugal pumps, water turbines (e.g.
Pelton wheel and Kaplan turbine, steam turbines and (jet aircraft)
gas turbines.)} such as ship propellers and (household) fans.  It
is worth noting that:
\begin{quote} It was not until 1929 that [Sydney] Goldstein solved
the potential flow problem completely for a lightly loaded
single-rotation [aircraft] propeller of small advance ratio. This was
unquestionably the greatest single step in the evolution of the
propeller theory \citep[ p.1]{Theodorsen_48}. \end{quote} Equally
significant, Theodorsen extended the theory, with its ability to
determine optimum (i.e. minimised energy loss) circulation
distributions, to incorporate non-light/heavy propeller blade
loadings and pragmatic advance ratios.

Utilising a control volume, dimensionless expressions for: shaft
power ($P$), blade thrust ($T$), and energy lost to the wake ($E$)
are derived (pp.28-29), which may then be converted back to their
dimensional quantities\footnote{Theodorsen's theory remains (in a
sense) the pinnacle of classical propeller theory. Its development
during the late 1930s/early 1940s coincided with the dawn of the
jet age. The theory is just as effectively applied to
propeller-style wind turbines. Some iteration is required for
optimisation, and it is necessary to optimise for a single
`design' airspeed.}. The point being made is that a force basis
fails to \emph{accurately} represent (i.e. model) the less than
100\% efficiency of converting (mechanical) rotational
energy/power into linear (aerodynamic) thrust work. A power based
expression $P=TV+dE/dt$ is the equality describing the system's
behaviour, where V is (freestream) airspeed. In physics, rather
than aeronautical (or aerospace) engineering, a force basis to the
power expressions is preferred, in conjunction with an efficiency
($\eta$) factor\footnote{Similarly, for an airfoil's lift, a
simple action-reaction force basis is given preference by
physicists. The circulatory airflow, where the `work' is done, is
given either secondary emphasis or no emphasis.}; usually written
as $\frac{1}{2}\rho A V^{3} \eta$ --- the emphasis being the
kinetic energy of the airflow relative to an aircraft's
propellers. This is an inferior explanatory approach because the
efficiency factor is not theoretically determinable as is the case
with the vortex method\footnote{Finite elements methods are
preferred today because the effects of: engine nacelles, aircraft
fuselage, etc., may be incorporated. Vortex theory treats the
rigid vortex sheet, with its surface of discontinuity, in
isolation from \emph{all} interference effects --- including drag
arising from fluid viscosity.}.

In the theory \citep[ Ch.4]{Theodorsen_48} the dimensionless power
coefficient \mbox{$c_{p}=2 \chi \bar{w} (1+\bar{w})
(1+\frac{\epsilon}{\chi}\bar{w})$} [Equation 37, p.29\,] can be (with
some algebraic effort) re-expressed as a \emph{dimensional} power
equation \mbox{$P=\rho A V n \Gamma_{e}$} which then may be reduced
to a new type of (potential) energy equation/expression:
\begin{displaymath}
E_p=\rho A L \Gamma_{e} n=m \Gamma_{e} n \qquad {\rm{where}} \quad
\Gamma_{e}=B \bar{\Gamma} (1+\frac{\epsilon}{\chi}\bar{w})
\end{displaymath} or alternatively (and more appropriately herein)
\begin{equation}\label{eq:Vortex}
E_p=m \Gamma_{e} f \end{equation} where: $m$ (`air' mass) is the
fluid density ($\rho$) multiplied by cylindrical volume (blade
swept area `times' the axial length/distance --- associated with
one revolution), \mbox{$n=f=(\omega)(2\pi)^{-1}$} is the number of
(complete) revolutions per second, and $\Gamma_{e}$ is (newly)
defined as the effective circulation around the blades.
$\Gamma_{e}$ (with `dimensions' $[\frac{L^2}{T}]$\,) is
essentially the number of blades times the average circulation
(across each blade) multiplied by the (Theodorsen) efficiency
factor of an \emph{optimum} propeller --- i.e. one where the
trailing flow, in a frictionless fluid, is shed into the wake as a
``rigid" vortex sheet, i.e. a (three-dimensional) `solid' helix or
helicoid\footnote{At optimisation the (dimensional) displacement
velocity ($w=\bar{w} V$) of the airflow, for all radial locations
along the blade span, are equal; it is positive for propellers,
and negative for wind turbines. In the nomenclature of Theodorsen
$\Gamma_{e}=w H \chi (1+\frac{\epsilon}{\chi} \bar{w})$ where $w H
\chi=\bar{\Gamma} B$ with $\bar{\Gamma}$ being the average
circulation `over' each blade, $B$ is number of blades,
\mbox{$\epsilon$ is} the axial energy loss factor, and $\chi$ is
the (all important) ``mass coefficient"
--- see \citet{Theodorsen_48} for further details. \mbox{$H$ is} the
pitch of the rigid helix, with $H n=V+w$. A further (minor)
correction for propeller wake contraction, or wind turbine wake
expansion, has not been included.} --- which together with a
(two-dimensional) plane is the only other surface of revolution
that is also a \emph{ruled} minimal surface\footnote{Such that:
for any point on the surface there is a line on the surface
passing through it; and with a zero mean, the surface area is
minimised. Note that a `catenoid' only exhibits the latter
characteristic.}.

Of interest here is firstly, the fact that this new (fluid
mechanical circulation based) energy expression is proportional to
the rate of (cyclic) rotation ($n$ or $f$), so there is (also) a
\emph{unit energy} corresponding to a \emph{single rotation} of
the blades; and secondly, that a solely force and/or kinetic
energy basis fails to be sufficient to (theoretically) best
represent this linear \emph{and} rotational system --- which can
only be realistically modelled in \emph{three} spatial
`dimensions'\footnote{The far wake (also) plays a pivotal role in
the model.}.

Thus, we have a further difference between macro- and microscopic
systems; in terms of the most dimensionally ``rich" physical quantity
(i.e. power $[\frac{ML^2}{T^3}]$ cf. energy $[\frac{ML^2}{T^2}]$)
needed to `best' model the essence of a system's physical behaviour ---
albeit that the case of `vortex' propeller theory is somewhat idealised
and ``old school" in its approach.

\subsubsection{Energy to frequency proportionality}
An alternative physical expression of energy, other than work (a
force applied over a distance), is wave energy; where the wave's
energy is proportional to wave amplitude squared. A further
alternative route to energy is the representation of electromagnetic
wave/radiation energy as proportional to a frequency; i.e. Planck's
formula\footnote{Also called the Planck relation or the
Planck-Einstein equation, and alternatively written as: $E=h\nu$.}
$E=hf$. In general, frequency can be conceived as either (and
sometimes both) indicating: a rotational process that occurs over a
finite time\footnote{Philosophically, time in-itself, can be
conceived as process based; this being an unorthodox (although well
established) stance.}, or simply the number of (wave) oscillations
over a given time --- with the role of a time standard, i.e. the
second, essential. Note that precision clocks are frequency based.

The vortex theory energy (Equation \ref{eq:Vortex}) discussed in
subsection \ref{subsubsection:Prop/HAWT}, with mass as fluid
density multiplied by a cylindrical volume (and thus non-solid),
is a macroscopic instantiation of an energy to `frequency'
proportionality. When energy is proportional to frequency (i.e.
revolutions per second or cycles per second), the importance of
angular momentum is paramount; whether (expressed) as a
macroscopic fluid circulation or by way of Planck's constant (re:
electromagnetic radiation/waves).

Later, we see that the energy of interest herein is a geometric
phase based intrinsic/spin angular momentum (offset\footnote{The
description of what is involved in this offset, and what this
offset is relative to, is far from trivial --- see subsection
\ref{subsubsection:Two faces}.}) \emph{per cycle time}, or angular
momentum multiplied by frequency; i.e. spin energy --- with this
energy being below the first and/or minimum energy level, or below
a minimum energy uncertainty `level', in/of an atomic/molecular
`quantum' system. This energy, and the energy of vortex propeller
theory, may be contrasted with the standard \emph{position}-based
potential energy (P.E.) at a given time (of Classical Mechanics).
Conventional P.E. has no role to play here, although for our
purposes we may still speak of the supplementary field energy
\emph{over a given unit cycle time} $\Delta t$, at a given
distance from the (rotating space-warp's) energy
source\footnote{Noting we are (at present) restricting ourselves
to position in a plane when discussing (gravito-quantum) rotating
space-warps (GQ-RSWs).} --- and similarly, when a moving body's
positional change over $\Delta t$ is considered insignificant
(when compared to the scale of the system).

The aim of this subsection has been to show that requiring potential
energy to be exclusively position (i.e. location) based is somewhat
misguided; because classical vortex theory (of a
\emph{three}-dimensional propeller\footnote{The trailing flow far
downstream of the blades is fundamental to the theory.}) and to a
lesser extent Planck's formula in electromagnetism provide counter
examples. Non-kinetic energy, i.e. `potential' energy\footnote{With
`potential' now used in a wider sense, in that it is not exclusively
a function of position.}, is conceivably also frequency-based, and/or
alternatively cyclic process time-based. Recall $\Delta
e=\frac{1}{2}\Delta a^{2}\Delta t^{2}=\frac{1}{2}\delta v^{2}$ from
subsection \ref{subsubsection:Rayleigh Power}; proposing that the
\emph{specific} `potential' energy of the proposed supplementary
oscillatory gravitational field (per cycle\footnote{Per cycle, or
alternatively, over the course of a full cycle.}) equals the
\emph{unsteady} specific kinetic energy component of a body in
otherwise steady (i.e. non-oscillatory) motion (per cycle).

\subsubsection{Is macroscopic gravitation mediated by the graviton particle?}
\label{subsubsection:graviton} Returning to a discussion of four
force unification, we accept that the three micro forces are
particle based\footnote{As evidenced by the success of the
Standard Model of particle physics.}. On the contrary, GR is
primarily a geometric theory that is considered to be mediated by
the (to date undetected) graviton particle. Our proposed
supplementary space curvature, on the other hand, cannot be a
standard particle based phenomenon, because it is (partly)
QM-based and exhibits a non-local `influence' aspect
--- although changes to the strength/amplitude of undulations
conceivably propagate at the speed of light. Awkwardly and
inevitably, some form of very-fast hidden variable theory appears
to be required in the conjectured model --- to convey the
impression of either a non-local `correlation' or `action'. The
later is inconsistent with the particle based (reductive) theory
agenda of modern physicists that employs the speed of light (c)
upper limit (to information exchange); whereas a correlation-based
understanding is \emph{not} inconsistent with experiments
illustrating the presence of quantum non-locality. In this sense,
non-local QM (spin) effects only \emph{appear} to involve an
instantaneous causal `action'. Contemporary physics cites
entanglement as a precedent (condition) for non-locality.

The explanatory waters are muddied (so to speak) by the fact that the
existence of the graviton particle remains an open issue. The
non-detection of a graviton particle means that: possibly GR's
spacetime geometry is `enforced' by a non-particle (non-graviton)
mechanism --- albeit with changes propagating at the speed of light.

\subsubsection{Hypothesising a non-local mechanism}\label{subsubsection:non-locality}
The following proposal of a `hidden' supplemental mechanism for
`explaining' non-local correlations is non-rigorous. Time (and mass)
even today retain a certain mystery, and particle based interactions
in (macroscopic) `spacetime' \emph{may} not exhaust reality.
Conceivably, the universe's evolution might only \emph{appear}
analog; such that from a wider (temporal\footnote{Immediately, upon
their use in (general) discussion, the meaning of the words ``time"
or ``temporal" becomes inexact and somewhat ambiguous.}) perspective
it may well evolve in a staccato manner, that is inherently
imperceptible to physical observations
--- (conceivably) by way of the limits imposed by Heisenberg's uncertainty
principle (HUP). Thus, physical reality would effectively `occur'
in a digital (go-pause-go-pause-...) manner, with a new type of
``process" occurring between the instants/moments (or states) of
measurable time, i.e. in the `pause'; thus complementing reality
as is currently understood.

Since this process lies outside of (or beyond) observational time,
and within limits imposed by HUP, it is effectively invisible to
(i.e. \emph{hidden} from) measured reality. Some type of (unknown)
cosmological size/scale \emph{wave} effect(s) is one conceivable,
yet vague, option for what this process involves\footnote{If
rotational, the direction of (possibly two orthogonal) rotation(s)
could leave a signature that results in the laws of physics being
slightly different for matter and anti-matter.}. This cyclic wave
process facilitates a form of subliminal whole-universe
communication and interaction, that is quite conceivably geometric
and topological\footnote{Note that there is no
\emph{observational} difference between the digital `reality'
moments occurring ``all at once", or sequentially `around' the
universe --- by way of an observationally subliminal `scanning'
process. The latter scenario seems more appropriate and viable.}.
Subsequently, non-locality can regain a (new type of) `cause and
effect' explanatory narrative. Additionally, invariance and
conservation demands may be appeased; particularly regarding
(non-local) QM intrinsic angular momentum. This hypothesis is
consistent with the phenomena--noumena distinction
introduced/discussed in subsection \ref{subsubsection:noumena}.

Finally, this cosmological process allows the universe to evolve
in a temporally harmonious/coordinated (and stable) manner. Due to
the imperceptible brevity between (digital) `moments', we
effectively have a hidden mechanism that ensures a universal
\emph{background} time simultaneity --- even though clocks may run
at different rates within the observable universe. \mbox{Section
\ref{subsection:SR's ontology}} shall elaborate upon this
(admittedly sketchy) hypothesis.

\subsubsection{Interim conclusion and subtleties of the model}
Neither the non-particle, nor the non-reductive aspects, of our
proposed supplementation to non-Euclidean geometry, is sufficient
(as yet) to scuttle the model being pursued. Further, the
differences between the micro- and macroscopic domains (and their
theoretical descriptions) has been emphasised. The
non-simultaneity of special and general relativity (SR and GR)
presents a far more challenging denial of the model, and this
shall begin to be addressed in subsection \ref{Subsection:GR
fortress}.

The model's need to externalise a (coherent) virtual quantum spin
energy offset as a large scale (rotating) acceleration/gravitational
field perturbation implies a global nature to the model, and hence
global coordination between the microscopic (QM) and macroscopic
(curved spacetime) domains. This in turn seems to require a
coordination that effectively exhibits temporal simultaneity,
reminiscent of (historical) classical mechanics; and thus the model
would benefit from interpreting non-locality as implying some form of
\emph{apparent} instantaneous `causation' (recall subsection
\ref{subsubsection:non-locality}). Non-locality regarding QM spin is
undeniable, and remains open to interpretation. Appropriately, QM
(intrinsic) angular momentum is central to the `global' model being
hypothesised. It is insufficient to simply cite GR's v\,$<$\,c
information speed limit and then reject the model, because the model
involves \emph{both} curved spacetime and QMs acting in concert.

From an (idealised) classical conception, we can (with some
imagination) visualise a \emph{constant amplitude} rotating
space-warp, with the acceleration amplitude representable in a
field based manner\footnote{Varying cyclically (and sinusoidally)
at a `fixed' point in a barycentric reference frame.} (recall
Figures \ref{Fig:R3SideAbove} to \ref{Fig:R3FrontElev}). As a
first stage idealised approach, the use of classical mechanics
involving: (a QM process based) energy, a continuum of space,
rotating acceleration perturbations, and utilising (classical)
time simultaneity, appears feasible. Further, if we regard the
accelerations (equating to gravitational field strength) proposed
in Section \ref{Section:PrelimModel} as `true' accelerations, in
the sense of reference frame independent `proper' accelerations;
then (qualitatively) any \emph{ensuing} speed perturbations (of
low mass bodies) also need not be frame dependent. Note that even
though the rotating space-warps (RSWs) rotate about a central
point/region, and the associated acceleration is considered a
`true' (cyclically) varying acceleration, the motion of any body
`receiving' this influence is inevitably described using a
reference frame --- e.g. the Pioneer spacecraft relative to the
solar system's barycentre.
%************************************************************************************
\subsection{Tackling fortress General Relativity and its
non-simultaneity}\label{Subsection:GR fortress} The aim of this
and the next section (\ref{subsection:SR's ontology}) is to cast
doubt upon the ontological commitments of GR. By assuming a `real'
(i.e. non-systematic based) Pioneer anomaly we necessarily take
issue with GR's confident assertion that: there are no (indeed
never) global reference frames, by way of an underlying acceptance
of relativism. This leads into \mbox{subsection
\ref{subsection:SR's ontology}}, where an alternative
interpretation of: time and length dilation is examined, so as to
illustrate that space and time's coexistence or `union' may be
something other than \emph{only} SR's and GR's `spacetime'. It
would be so much easier to simply declare the model
non-special-relativistic, and indeed the sceptical reader should
keep this in mind; but the model appears to demand simultaneity of
time (throughout the field), by way of quantum non-locality being
involved. This is compatible with a solar system (systemic) time
standard; one of whose roles is to quantify (a RSW's) rotation
duration ($\Delta t$).

\subsubsection{Physicists and philosophical thought} Science involves
both concepts and mathematics, and usually physics has a
mathematical emphasis because the concepts involved are stable.
Here the emphasis is strongly conceptual, because we are
attempting to establish \emph{new} physics, in the sense of a
supplement involving an alternative conceptual structure, or
``disciplinary matrix" as philosophers of science would call it.
We do this not in order to unnecessarily attack physics, but
merely to resolve at least one (Pioneer) anomaly, and possibly
further our understanding of how curved S/T and QMs coexist in a
celestial material \emph{system}.

The huge success of physics from about 1930, when Paul Dirac's
\emph{Principles of Quantum Mechanics} was published, has led to
today's theoretical physicists under-valuing conceptual
discussion. About 1975 the standard model was ensconced, and 1975
is when Vera Rubin and Kent Ford's observations of spiral galaxy
flat rotation curves appeared, marking the beginning of
significant ``unfinished physics business"; e.g. dark matter and
dark energy. From a (Thomas) Kuhnian perspective \citep{Kuhn_70},
the years 1930 to 1975 typifies a period of \emph{normal science},
and 1975 (possibly) heralds the beginning of a rise in conspicuous
scientific issues and anomalies --- which may require rethinking
our conceptualisation of the physical world.
\begin{quote} [Indeed at] \ldots the beginning of the 20th century ---
Einstein, Bohr, Mach, Boltzmann, Poincare, Schr\"{o}dinger,
Heisenberg --- thought of theoretical physics as a philosophical
endeavor. They were motivated by philosophical problems, and they
often discussed their scientific problems in the light of a
philosophical tradition in which they were all at home. For them,
calculations were secondary to a deepening of their conceptual
understanding of nature \citep*{Smolin_05}. \end{quote} Assuming a
real Pioneer anomaly implies something of a (minor) crisis in
physics, and necessitates a return to this early 20th century
conceptual approach. If the model went straight to the
mathematics, we would be implicitly assuming that any associated
conceptualisation was non-problematic --- i.e. normal science or
``business as usual".

\subsubsection{Looking for cracks in the General Relativity fortress}
It is worth mentioning that herein we accept without question the
standard model's unification of the three (microscopic)
forces/interactions (weak, strong, electromagnetic\footnote{Note
that electromagnetism has both microscopic and macroscopic
aspects.}) that incorporates SR. It is the extension of this to
(large-scale macroscopic) gravitational effects that is our
concern.

Loosely speaking, GR describes the space between \emph{points}.
Aspects of GR include: \begin{quote}[The Relativity Principle:]
The laws of nature are merely statements about spacetime
coincidences; they therefore find their only natural expression in
generally covariant equations \citep{Barbour_95}; [and] the
distance between adjacent points is expressed by a metric equation
\ldots The metric coefficients in the metric equation depend on
the arbitrarily coordinates chosen and geometric properties of the
space are expressed in the form of differential equations showing
how the metric coefficients vary from place to place \citep[
p.198]{Harrison_00}.\end{quote} Since GR is a generalisation of
SR, the theory necessarily builds upon the mathematical structure
of SR, more so than SR's conceptual structure. Central to GR are
Einstein's equations, which link the geometry of a
four-dimensional semi-Riemannian manifold representing spacetime
with the energy-momentum contained in that spacetime.

Further, since GR is a generalisation of SR, it has an inherently
localised (theoretical) starting-`point', providing no conceivable
avenue with which one could express the (as yet to be derived)
excess QM \emph{energy} arising from the closed loop (finite time)
motion of (bulk) QM matter in curved spacetime upon a geodesic.
The supplementary space curvature (proposed herein), although not
explicitly expressed as a metric theory, could possibly be
supplemented to GR's formalism as a pure energy supplementation
--- but the need for an apparent simultaneity/non-locality feature
in the new model remains in direct conflict with the
\emph{pairing} of: SR's non-simultaneity and GR's general
principle of relativity\footnote{Not forgetting GR's use of the
(Einstein) Equivalence Principle as effectively a `foundation
stone' for generalizing special relativistic physics to include
gravitational effects.}. These two aspects of relativity ensure
that GR does not permit/contain any invariant geometric background
structures. A solution of the Einstein equations consists of a
semi-Riemannian manifold (usually defined by giving the metric in
specific coordinates) on which are defined matter fields.

Each solution of Einstein's equations describes a whole history of
a universe; i.e. we have a ``blocktime" basis to the physical
explanation. A prominent and persistent philosophical concern of
``blocktime" is that it effectively denies an individual's
freewill. This inconsistency with our own perceptions is a central
issue in philosophy, and is reason enough to examine the bedrock
upon which GR is constructed. Additionally, GR's non-linear
partial differential equations are very difficult to solve, and
only a few exact solutions have direct physical
application\footnote{Parts of this subsection paraphrase
Wikipedia: \emph{General Relativity}; as modified on 16 May 08, at
02.24.}.

\subsubsection{The `principle' of general covariance: the logic behind
it and its generality}\label{subsubsection:logic of GCoV} Special
relativity eliminated absolute space and absolute motion. The
theory of general relativity is an extraordinary theory of
gravitation, and has been an unmitigated success. With GR,
Einstein sought to extend SR's \emph{relativity} of motion to
acceleration, so as to generalise relativity; and with the
possible exception of the Pioneer anomaly, this conclusion appears
quite reasonable. If we accept that acceleration is (also)
relative, then a number of implications follow:
\begin{enumerate}
\item{Most importantly, an acceptance of a \emph{principle} of
general covariance.} \item{Matter, momentum and energy
distribution are the sole determiners of spacetime curvature.}
\item{Gravitational theorisation must utilise a metric, because
the structure of spacetime is encoded in the metric field tensor,
with the curvature encoding `gravity' at the same time.} \item{The
principle of equivalence applies to all forms of accelerated
motion.} \item{There can be no privileged reference frames.}
\end{enumerate} Accepting the strict relativism of Einstein's GR
then strongly supports a reductive stance to physics, especially
four force unification.

The first of these listed points is the most important, with the
other points logically dependent upon it. Our concern is not with
GR as concerns celestial matter, rather it is with a quantum
mechanical (QM) system moving in curved spacetime\footnote{Support
is possibly lent to this by way of the black hole information
paradox (BHIP). Stephen Hawking concludes the paradox implies the
usual rules of QMs cannot apply in \emph{all} situations. Note,
this situation (also, like the BHIP) involves QMs \emph{and}
gravitation.}.

In the spirit of (Sir) Arthur Stanley Eddington we see (he saw)
(QMs and) GR ``\ldots as fundamentally `epistemological' in
character, meaning that they provided insight into \emph{how} we
see the world, rather than \emph{what} the world is \citep*[
p.37]{Stanley_05}." In this subsection we shall argue, in line
with certain philosophers of physics, that general covariance
(GCoV) should be regarded as simply a new mathematical technique,
and not necessarily an expression of physical content (i.e. an
ontological aspect) --- especially since the physical implications
of SR's invariance of the interval (under Lorentz transformations)
remain far from obvious (or fully settled).

Only if GCoV looses its (general) `principle' status, i.e. beyond
GR's formalism, can our model be viable; because it has features
such as: acceleration/gravitation encoding curvature (cf. \emph{vice
versa}); no need for a metric nor an equation of (`point' mass)
geodesic motion; and (crucially) a situation that effectively
violates GR's (weak) principle of equivalence --- with this last
feature not fully addressed until subsection
\ref{subsubsection:Violation of Eq.Pr}.

In the philosophy of physics, and for some physicists, the following is
accepted. \begin{quote} Einstein offered the principle of general
covariance as the fundamental physical principle of his general theory
of relativity and as responsible for extending the principle of
relativity to accelerated motion. This view was disputed almost
immediately with the counterclaim that the principle was no relativity
principle and was physically vacuous. The disagreement persists today
\citep[ Abstract]{Norton_93}. \end{quote}

Certainly, GR is a generalisation of SR and subsequently, this
greatly restricts any supplementary alternative (rather than
modified) gravitational theorisation; but the mathematical
technique of general covariance is not necessarily `active' beyond
GR (i.e. gaining ``principle" status). Thus, the various
restrictions to gravitational theorisation (listed above) that
(logically) follow from a \emph{principle} of GCoV are not validly
deduced. Admittedly, it is one thing to say the principle of GCoV
is logically restricted, and another (far more difficult) thing to
explain why. Section \ref{Subsection:EquivPr Comment} addresses
this issue in some depth.

Recalling the exclusions and idealised circumstances associated
with general relativity outlined in subsection
\ref{subsubsection:Beyond GR}, GR's `completeness' credentials are
clearly non-`water-tight'\footnote{For example, concerning GR's
\emph{restricted} coverage to non-complicated physical phenomena;
in that exact solutions to N-body systems elude (NMs and) GR.
Admittedly, numerical methods are promising in this regard.}. To
simply use GR's relativism (in particular) to deny the
supplementation of non-Euclidean geometry proposed herein is
totally inappropriate (although understandable) --- especially
since both curved spacetime \emph{and} (non-inertial aspects of)
QM atomic/molecular systems are involved in our model. Note that
QM atomic/molecular systems are physically dominated by
electromagnetic concerns cf. the inertial (and \emph{mass}-based)
concerns of the model.

\subsubsection{Non-relativistic aspects of gravitation in the model}
\label{subsubsection:Non-rel aspect of g} Gravitational
theorisation involves a clear distinction between accelerating and
non-accelerating motion (i.e. non-inertial and inertial motion).
The model utilises both: the concept of a rotating space-warp, and
a (constant amplitude) \emph{acceleration} (perturbation) that is
related to (and induces) the anomalous motion of the Pioneer
spacecraft; but the (rotating) acceleration/gravitational field is
only indirectly related to (the driving energy of)
`\emph{motion/movement}' within (a great many) atoms and
molecules. The \emph{specific} potential energy of the proposed
supplementary oscillatory gravitational field (per cycle)
\mbox{$\Delta e=\frac{1}{2}\Delta a^{2}\Delta t^{2}$} (recall
subsection \ref{subsubsection:Rayleigh Power}) is a surprisingly
simple law-like expression, with $\Delta t$ representing the
duration time for a full rotation. The need for a systemic time
standard is clear.

Regarding gravitation, philosophers of science have argued that it
is (\emph{only}) acceleration and rotation that (in at least one
sense) effectively resist a completely relativistic
interpretation\footnote{The issues involved herein are very deep
and are necessarily muted in their presentation. This and other
parts of section \ref{Subsection:GR fortress} have drawn somewhat
upon an article called ``Absolute and Relational Theories of Space
and Motion" in the (online) Stanford Encyclopedia of Philosophy.};
e.g. (hypothetical) whole universe rotation. Hence they retain,
not an absolute aspect, but rather in the case of acceleration a
`true' aspect (independent of relative motion); and in the case of
rotation a sense of being global (or systemic). Fortuitously, only
these two concepts are involved in the model's quantification of
$\Delta e$, and the proposed existence of RSWs (of cosmological
extent). Further, with the model's cyclic acceleration amplitude
(necessarily) constant `everywhere', the simplification of the
model's quantification could not be greater\footnote{No metric, no
equation of motion, no stress-energy tensor, etc.}; providing a
compelling reason to (further) advance the model. In section
\ref{subsection:SR's ontology} we do this by way of a continued
re-evaluation of SR's and GR's ontological commitments.
%*******************************************************************************
\subsection{Evaluating SR's and GR's ontological commitment}
\label{subsection:SR's ontology} In this section we seek to
further expose the not so rigid ontological underbelly upon which
GR is founded. Indirectly we are examining the scope of GR and
questioning ontological assumptions behind SR, that are subsumed
into GR. We also outline a variety of `invariance' aspects
required for the supplementary ontological perspective to be
presented in section \ref{subsection:Reversal}, which (amongst
other things) are consistent with the model's use of a single time
variable to describe the whole system. Eventually, it shall be
seen that this secondary perspective is not in defiance of special
and general relativity (SR and GR), rather it is supplementary;
suiting the very different ingredients associated with the model
--- e.g. virtual QM energy, QM non-locality, and a three-body
celestial system. Specifics of GR (on its own), such as the
relativistic energy-momentum tensor are incompatible with the
model's `ingredients'. This and the next section are
philosophical/conceptual rather then mathematically physical, and
concentrate upon re-examining space, time, their coexistence; as
well as mass, motion and energy. Our motivation is to eventually
explain how a new contribution to gravitation (in its widest
sense) can coexist with the standard approach of general
relativity --- with the latter (by far) the dominant approach to
gravitation.

\subsubsection{Concerns regarding an ontological commitment to spacetime}
\label{subsubsection:SR's ontological commit} Special Relativity
demands that space and time be fused together into spacetime. This
arises in part from the `operational' approach employed in
Einstein's SR\footnote{Operationalization is the process of
defining a concept as the operations that will measure the concept
(variables) through specific observations. That even the most
basic concepts in science, like ``length," are defined solely
through the operations by which we measure them, is the discovery
of Percy Williams Bridgman, whose methodological position is
called operationalism (source: Wikipedia, 2011).}; this being
almost always associated with an additional ontological commitment
to the idea that: measurements dictate reality (without any
remainder); indeed this is the essence of the scientific approach.
Spacetime is seen by physicists as the correct understanding of
(now obsolete) classical space and time; successfully rebutting
all alternative (in the sense of different) interpretations of the
Lorentz transformations that have been proposed. The existence of
the latter is indicative of a suspicion that SR's spacetime and
GR's (physical) ontology may be somewhat restricted, i.e. not
quite the whole `story'.

Supporting this broad concern are a number of specific issues or
concerns. Firstly, locality was assumed `universal' in the days
prior to QMs' full establishment in the 1920s and beyond.
Historically, SR resolved the tension between electromagnetism
(EM) and classical mechanics, that had became apparent at the end
of the nineteenth century. Quantum mechanics' incorporation of
special-relativistic effects has been amazingly successful, but
non-locality remains a vexing issue. Secondly, in the
\emph{formalism} used to describe physics' two realms, the macro-
and microscopic, the relationship between space and time is quite
different. A locally classical/flat space \emph{background} is
(almost always) implicit in (micro) QMs, whereas curved spacetime
is the essence of (macro) GR; and further: \begin{quote} \ldots in
quantum mechanics, time retains its Newtonian aloofness, providing
the stage against which matter dances but never being affected by
its presence. These two conceptions of time don't gel \citep*[
p.18]{Merali_09}. \end{quote} Thirdly, time still remains
mysterious outside of physics' (possibly incomplete) accounts of
`reality' in its widest sense\footnote{See Martin Heidegger's
``Being and Time" for example.}. Finally, quoting from a
philosophical article:
\begin{quote} \ldots the conventional view of space-time as the
ground of physics is [for some people] increasingly being called
into question \citep[ p.53]{Clarke_93}.
\end{quote}

By way of an assumed unification of reality, `spacetime' is
standardly seen to be common to the micro- and the macroscopic
realms. The new approach pursued herein pursues the coexistence of
a secondary \emph{global} ``neo-classical" notion of time (and
space), that cannot be directly `grasped' by local observations.
This supplementation, by way of two different idealisations, is
necessary; so as to physically link (in the model) an energy
(magnitude) between both realms seamlessly (and simply). From this
perspective, we begin to conjecture that GR's need for general
covariance is a direct consequence of (further) adding
gravitational effects to the (already additional) conformal
structure demanded by SR's Lorentz symmetry.

\subsubsection{The failure of alternative approaches to special relativity}
\label{subsubsection:failure} S. J. Prokhovnik's ``The Logic of
Special Relativity" \citep{Prokhovnik_78} sought to highlight
(p.84) that at different levels of description, and/or describing
different sets of phenomena, SR's Lorentz transformations can be
associated with a secondary (rather then alternative)
interpretation, and hence further ontological aspects. We shall
not pursue a detailed discussion here. The reader is simply asked
to recognise a historical pursuit of (unsuccessful) alternative
interpretations of SR's Lorentz transformations, pursued by
scientists and philosophers who perceive logical contradictions in
SR and `spacetime', as (possibly indicative of) an incomplete
`grasp' of physical reality --- in particular the twins
paradox\footnote{For example: consider a two observer, figure
eight symmetric version of the twins paradox, involving geodesic
motion in a suitably configured asteroid belt. Who's clock runs
(relatively) slow when the two S/C, moving in opposing rotational
directions, \emph{periodically} observe each other's clocks at the
nexus of the `eight'? Note that asymmetrical acceleration is
\emph{not} present in this curved spacetime (\emph{geodesic}
motion, i.e. non-force) scenario.}. Indeed, Prokhovnik and others
have struggled to justify their alternative interpretations; and
we shall not champion any of these approaches.

\subsubsection{Other indicators that GR may be incomplete and/or
restricted}\label{subsubsection:GR incomplete indi} A real Pioneer
anomaly demands something additional to GR rather than a simple
modification. The elegance of GR tends to imply that it completes
gravitational theorisation. There is no assurance of GR's
completeness, and subsequently we cannot assume that unification
of GR and QMs is the next (and possibly final) major step in
theoretical physics. Alternatively, the elegance of GR may be
associated with restrictiveness in some sense.

The extrapolation of GR's local gravitational curvature to the
global/cosmological case is neither supported, nor denied, by
observations indicating that (cosmologically) space is
flat/uncurved ($k=0$). Thus, we may question whether whole
universe/global curvature \mbox{($k=\pm1$),} implied by GR's
perceived \emph{exclusive} role in describing gravitation, is
capable of being physically `realised'. In other words, global
(whole universe) curvature remains an unproven conjecture,
supported by an assumed `totality' of GR's scope. It is quite
conceivable that GR's domain of application might always be
restricted to a subset of the whole universe, applying to the
largest of galactic clusters but not necessarily to the universe
as a `whole'. The model needs to enforce this (sub-global to
global) distinction.

\subsubsection{Three levels or scales of mass by way of the amount of binding
energy}\label{subsubsection:Three mass levels} To establish the
model it is necessary to elaborate upon our previous distinction
between micro- and macroscopic matter/mass (subsection
\ref{Subsubsection:MM Differences}), and distinguish \emph{three}
`levels' (or scales) of matter/mass.
\begin{enumerate}
\item{Sub-atomic particles in which binding energy, as a
proportion of total mass/energy, can play a \emph{major} role.}
\item{Atomic and molecular matter in which binding energy, as a
proportion of total mass/energy, plays a \emph{minor} role.}
\item{Macroscopic matter in which binding energy, as a proportion
of total mass/energy, plays a \emph{negligible} role.}
\end{enumerate}
Note that the expression ``microscopic matter" (or mass) shall
imply either or both levels one and two. Also note that vast
quantities of gas or plasma, e.g. vast celestial gas clouds, are
considered under this criteria to be matter on a `macroscopic'
\emph{scale}.

The value of this distinction is that standard gravitation (and
the use of GR) pertains (in practice) to macroscopic matter as
defined above; whereas the model involves (non-stationary) mass at
(and `below') the level of atoms and molecules --- notwithstanding
the fact that these atoms and molecules are encased in a very
large macroscopic (lunar) body. In particular, the model involves
a virtual energy discrepancy\footnote{Later we see that this
atomic/molecular virtual energy, from a global/systemic (whole
universe) perspective, is actually a real inertial energy
discrepancy.} arising from (macroscopic) celestial (geodesic)
motion \emph{of} (second level) atomic/molecular matter; with the
energy, or rather a rate of angular momentum, necessarily
expressed externally as a rotating acceleration/gravitational
field perturbation. Thus, the second type of non-Euclidean
geometry proposed herein clearly has its basis (in part) at a
different (non-macroscopic) level of matter (in motion).

The new model's acceleration/gravitational field is a global
(whole universe) scenario, and it also exhibits global homogeneity
(re: its sinusoidal amplitude). If GR is seen as restricted to
sub-universe macroscopic scenarios, then these two types of
gravitational influence are clearly distinct and non-overlapping.

\subsubsection{Reintroducing the notion of relativistic mass}
\label{subsubsection:relativistic mass} Observations indicate
contraction or dilation effects for length, time and
relativistic/apparent mass, but not electric charge. ``The concept
of relativistic mass has gradually fallen into disuse in physics
since 1950, when particle physics showed the relevance of
invariant [or rest] mass''\footnote{Wikipedia: \emph{Mass in
special relativity}, August 2007.}. Unsaid in this quote is the
fact that: invariance in physical laws, a principle based approach
to physics, and gauge theories have been on the rise since about
then; with this emphasis harbouring a particle-based unification
agenda, along with a sense that ontological interpretation of SR's
Lorentz transformation \emph{is} complete, or even unnecessary. In
line with our distinction between the micro- and macroscopic
realms (subsections \ref{Subsubsection:MM Differences} and
\ref{subsubsection:Three mass levels}) we shall distinguish and
utilise two approaches to mass in SR: microscopically we embrace
the emphasis given to (particle) energy and momentum, but at the
macroscopic/cosmological level, and for the discussion of
ontological concerns, we shall revert back to the
\emph{simplicity} of relativistic/apparent mass \citep{Sandin_91}.

\subsubsection{Extending the scope of invariance, and
`time'}\label{subsubsection:Scope} The distinction between
relativistic mass and invariant (or rest) mass outlined in
subsection \ref{subsubsection:relativistic mass} shall be
supplemented in the following two subsections by arguments for an
existence of invariant aspects of time and length/space in
idealised, yet ultimately practical, global circumstances. The
motive for doing this is to eventually get around the restraint
that SR's invariance of the interval has upon the
conceptualisation of gravitation; (and we do this) by way of
seeking out new and further conceivable instances of invariance.
In section \ref{subsection:Reversal} we shall see that (in
idealised circumstances) the distinction between relativistic mass
and invariant mass can be extended to length and time; but prior
to this extension certain idealised physical circumstances need to
be established by way of conjecture.

Often in physics the expression ``invariance" means: unaffected by
a transformation of coordinates. In this section invariance shall
also be used (somewhat trivially) in the sense of: a homogeneity or uniformity
throughout the global distribution, or occurrence, of a particular
quantity.

The consideration of a less simple approach to the `time' of
physical observations appears to be required by way of the ability
of QM systems to `act' non-locally i.e. effectively
instantaneously, in terms of \emph{observational} time (recall
subsection \ref{subsubsection:non-locality}) --- at least as
regards correlations between properties of distant systems, and
(we shall conjecture) how these correlations are
\emph{maintained}. This alone is reason enough to elaborate upon
time's ontology/reality\footnote{Paul Davies, from personal
observation, strongly objects to more than one time being
propounded; but this (completely appropriate) \emph{scientific}
stance unavoidably involves a philosophical stance upon what time
\emph{isn't}.}; otherwise we are somewhat in denial of a startling
physical observation --- albeit with non-locality (probably)
restricted to the QM \emph{spin} of particles/atoms/molecules, and
collections thereof. \citet*{Albert_09} provide a good overview of
this recently rejuvenated concern.

\subsubsection{Idealised invariant cosmological time --- the universe's
fastest ticking clock} \label{subsubsection:fastest time} Atomic
clock based Terrestrial time (TT) differs from (atomic clock
based) Geocentric Coordinate Time (TCG) by a constant
rate\footnote{See, for example, IERS (International Earth rotation
and Reference systems Service) Technical Note No. 32, Section 10:
General Relativistic Models for Space-time Coordinates and
Equations of Motion.}. The unit of time interval in TT is defined
as the SI second at mean sea level. The reference frame for TCG is
not rotating with the surface of the Earth and not in the
gravitational potential of the Earth; and thus TCG ticks slightly
faster than TT. Barycentric Coordinate Time (TCB) is the
equivalent of TCG for calculations relating to the solar system
beyond Earth orbit. TCB ticks slightly quicker than TCG. The
difference between TCB and TCG involves a four-dimensional
spacetime transformation. We note that all these times: TT, TCG
and TCB are theoretical ideals; but their `constancy' allows them
to have practical applications.

We may conceivably extend this idealisation of time rates to a
galactic centre coordinate time, then the centre of the local
galactic group, and then the local supercluster centre; with the
rate of clock ticking being slightly faster for each
four-dimensional spacetime transformation. Assuming that we could
observe the whole universe, and know the positions, motions, and
gravitational potentials of other superclusters; there is a
conceivable and final conceptual extrapolation that results in an
(idealised) `Local' Cosmological Coordinate Time (TCC) located at
the local supercluster centre. This would be a clock for practical
applications involving the \emph{whole} universe; and it is a
quicker/faster (ticking) clock than all conceivable others.

\subsubsection{Non-locality and an idealised invariant cosmological time (rate)}
\label{subsubsection:universal time} Time can be understood as
either the interval between two events or as a sequential
arrangement of events. To account for QM non-locality, subsection
\ref{subsubsection:non-locality} raised the possibility of the
universe's evolution having a step-like aspect, which is
`separated' and coordinated/harmonised by a hidden
cosmological/global (cyclical) \emph{process} of non-measurable
duration. Observational reality, which is the basis of the
scientific method, is necessarily oblivious to this supplementary
ontological process. This cyclic process ensures that a background
coordination of sequential (cosmological) ``moments" is
conceivable, which introduces a major shift in our
conceptualisation of what observational time may involve (and be
coexistent with). Briefly entertaining this broader temporal
attitude/perspective, it is conceivable that these (available to
observation) `reality moments' --- when compared to the `duration'
of an accompanying hidden non-measurable (background `scanning')
process --- (would most likely) have an extremely
short/`infinitesimal' duration; i.e. the hidden process is a
\emph{comparatively} much `longer process'. This
(moment-to-moment) situation would (in all likelihood) also
involve incremental local\footnote{Where ``local" also implies
extremely small `regions' of the universe.} changes throughout the
universe.

We may conceive that this cyclic (go-pause-go-pause-... or
pivot-pause-pivot-pause-...) universal occurrence can be broken up
into fractional parts. By definition: any measurement of time is
ultimately based on counting the cycles of some regularly
recurring phenomenon and accurately measuring fractions of that
cycle. Although this hidden (i.e. noumenal) and cyclic process is
beyond `measurement', hypothesising it allows observational (i.e.
relativistic) time to latently possess a supplementary hidden
aspect that is effectively (temporally) \emph{digital} and
cosmologically uniform in `nature' --- in addition to the usual
(solely) \emph{analog} perspective of observational (and
relativistic) time. Note that the scientific merit of this
hypothesised scenario is based upon the \emph{observation} of QM
non-locality, as well as discontent with simply citing
entanglement when `explaining' the `occurrence' of non-locality in
experiments.

In subsection \ref{subsubsection:fastest time} we hypothesised an
idealised cosmological (coordinate) time (TCC). Such a `global'
approach to time is fully consistent with a universe that evolves
in a (digital and hence) \emph{globally `coordinated'} manner.
Note that an idealised cosmological time \emph{rate} is \emph{not}
a reprise of classical or absolute time in an absolute space
reference frame, although it is well suited to a systemic
approach. Clearly there is a distinction between idealised
``cosmological time" and ``observational time", with the latter
incorporating the relativity and temporal variances (i.e.
non-simultaneity) associated with observations of a real, dynamic
and non-homogeneous universe. It is worth noting that
historically, before international atomic time was given
preference, the observational time of an `Earth day' or `Earth
year' was a sufficiently exact (reference) of \emph{cyclic}
duration.

The important feature of this discussion is the \emph{invariance}
of the TCG, TCB, and (the new) TCC (idealised) `time rates'; and
their applicability throughout a given system --- something that
clearly eludes GR. Satellite and spacecraft navigation illustrate
the necessity of establishing an idealised systemic time rate in
practical and scientific astro-dynamical experiments. There will
be a vast finite (indefinite) number of `noumenal'
cycles/increments occurring within any SI second. The existence of
the model proposed (herein), to explain a real Pioneer anomaly,
appears to demand this global coordination of events. Indeed, from
a universal/global perspective, a coordinating/harmonising
(noumenal or hidden background) process appears to be a very good
way to ensure that all parts (of a vast system) can be maintained
in stable `coordination'.

\subsubsection{Benefits of appreciating a broader scope to the concept of invariance}
\label{subsubsection:two timing} The concept of time dilation depends
upon one's stance upon what time is. Ontologically speaking, time and
time dilation are considered to be analog (cf. digital), such that
time, relative to a stationary observer, simply runs \emph{slow} at
relativistic speeds. Alternatively (and hypothetically), if we accept
the introduction of a second/further fixed (yet hidden) aspect of
cosmological time, then some \emph{measured} time could simply be
\emph{lost} at the level of observations, as a result of motion --- in
the manner that relativistic/apparent mass is \emph{gained}. This
notion is the major outcome to be established in section
\ref{subsection:Reversal}; it is a subtle concept and requires the
preliminaries discussed throughout this section (\ref{subsection:SR's
ontology}).

Regarding mass, we inevitably examine the \emph{same}
rest/intrinsic mass, but as (relative) speed increases from zero
the \emph{magnitude} of mass (in physical formulae) is modified.
Could the same be true with time (and length)? Does motion alter
the \emph{measurement} of `time', thus giving the impression, via
\emph{time lost}, of time dilation; but not altering a deeper
(non-observable) true \emph{intrinsic} invariant aspect of time
(recall subsection \ref{subsubsection:universal time})? Or should
we refuse to seek progress by way of new hypotheses; trusting in
general principles and invariance, and accepting that mass is
(always and only reducible to) \emph{just} energy. Thus,
(effectively) dispensing with questions that seek a (potentially)
richer explanation of: (observed) time dilation, energy's increase
with momentum, the origin of mass, and so forth. Accepting a real
Pioneer anomaly demands we must strive towards something
different.

Our alternative view results in a distinction between: the
standard relativistic appreciation of time\footnote{Figuratively,
time (or rather the rate of time) is elastic, stretching and
shrinking \emph{with} the clock that measures it.}; and a (new)
scenario that has \emph{both} a globally fixed (and hidden) time
rate\footnote{In the sense of a globally coordinated/harmonious
reality, i.e. state of the universe, that (unbeknownst to
observational physics) pauses before re-expressing the next
cosmological state of the universe; one of a long sequence of
observable (cosmological scale) `reality' states (or steps). Both
perceptually and observationally, (our and scientific) reality
appears `to be' continuous; with the additional (noumenal) aspect
always hidden from experienced and observational (i.e.
`phenomenal') reality.}, \emph{and} a non-fixed time rate for
relative motion, various physical systems, etc. --- with only the
latter being ``measurable". In section \ref{subsection:Reversal}
we see that this new scenario allows an appreciation of observed
time (and length) similar to (old-fashioned) relativistic mass; in
that there is both a rest/intrinsic value (albeit in idealised
circumstances) \emph{and} a relativistic/apparent (measured)
value.

\subsubsection{Substantivalism without a cosmologically global reference frame}
\label{subsubsection:substantivalism} We now turn our attention to the
notion of universal background space. Physicists such as: E. A. Milne,
H. W. McCrea, V. Fock, and S. J. Prokhovnik, have all entertained the
idea of a cosmological substratum; i.e. a fundamental/global reference
frame acting as a background (or stage) to physical interactions,
events, and processes. Not conflicting with this notion is:
\begin{enumerate} \item{space's (vacuum) impedance of 377 ohms,}
\item{the astronomical aberration of light, for observations from
an Earth that orbits the Sun, and} \item{a dipole effect
associated with Earth's motion through the microwave background
radiation.}
\end{enumerate}

The idea of a cosmological substratum is certainly not a consensus
(nor popular) view. Indeed, all explanations of SR and GR
incorporating absolute motion and/or absolute space \emph{should}
be rejected. Opposition to such proposals have been strongly
stated, e.g. Clifford Will; and quite rightly, since nearly all
experimental evidence, \emph{excluding a real Pioneer anomaly} and
non-locality, supports an implementation of Ockham's razor ---
thus, excluding any `hidden' (metaphysical) aspects beyond direct
observational evidence.

A question that remains is: must all discussion of a \emph{global}
perspective to space and/or time \emph{always} be strongly
rejected? Certainly, regarding observational physics employing
light rays, the answer is yes; but a cosmological substratum need
\emph{not} be always localised (nor `real' cf. virtual), so as to
necessitate an accompanying absolute reference frame.

The (or an) exception to this rule would be a globally homogeneous
cosmological substratum, with all points/places the same; thus, it
would exhibit spatial invariance by way of global homogeneity.
This substratum could only exist in the \emph{absence} of any
mass, motion, momentum or physical field energy; and thus it does
not `exist' \emph{per se}; rather it is necessarily a conceptual
idealisation. This includes `taking out' the motion of the solar
system barycentre with respect to cosmic microwave background
(CMB) radiation (i.e. the CMB dipole); and then `taking out' the
energy field associated with the CMB's (very nearly) homogeneous
temperature. Effectively establishing an `absolute rest
frame\footnote{The dipole is a frame-dependent quantity, and one
can thus determine the `absolute rest frame' as that in which the
CMB dipole would be zero \citep*[ section 23.2.2]{Scott_10}.}' at
a temperature of absolute zero. One interpretation would be to see
this (idealised) globally homogeneous substratum as effectively
the empty mold (form, or die) upon which physical reality (all
that exists) is `cast'. If so, it has a latent physical relevance.

In philosophical terms we are skirting around the debate between
relativism and substantivalism. The substantial `something' being
proposed is not SR's spacetime \emph{per se}; rather it is an
idealised (beyond physical reality) \emph{background} continuum of
uncurved (i.e. flat) space involving three dimensions. Only
conceptually can this empty universe coexist with a non-empty
universe\footnote{Note that dark (or vacuum) energy is not being
addressed in this dichotomy. Actually, the RSWs (rotating
space-warps) proposed by the model, conceivably affect redshift
measurements of EM radiation (cf. alter its speed); and thus, dark
energy may not exist, once the existence of RSWs is recognised ---
see subsection \ref{subsubsection:Brief list} and Section
\ref{section:Type1a}.}; yielding a relativistic substantivalism.
Additionally, this idealised background also has an ability to
coexist with the (idealised) form of (staccato) time
`simultaneity' discussed in subsection
\ref{subsubsection:universal time} (also recall subsection
\ref{subsubsection:non-locality}).

Clearly, an ontological open frame of mind is required to
entertain this ontological `supplementation'; involving a
\emph{combination} (or coexistence) of measured reality exhibiting
relativity, together with the additional conceptual `existence' of
an (idealised and hidden) background/platform of space (and time).

\subsubsection{On idealised invariant cosmological empty space}
What is the purpose of proposing an (idealised) `substantial
nothing' background in addition to the measured/observed world?
How can the existence of an (ongoing) homogeneous nothingness
(throughout and) of space, in the absence of gravitational (and
all other) effects, possibly be related to the relativism of
special and general relativity? Although special and general
relativity do not explicitly involve/require this supplementary
(physical) ontology --- notwithstanding Ockham's razor --- they
cannot deny it.

The use of a (flat) space background, together with the
(noumenal/background) time simultaneity discussed in subsections
\ref{subsubsection:universal time} and \ref{subsubsection:two
timing}, indirectly supports the additional (perturbative)
non-Euclidean geometry of the model to be mathematically
formalised by way of (simple) `neo-classical'
mechanics\footnote{The attendant conceptualisation, that allows
this simple ``neo-classical" formalism, is anything but
`classical' and far from simple.}. The conceptual arguments
permitting such a simple approach, in defiance of special and
general relativity's usual implications, are far from trivial and
require further discussion, as well as the subtle conceptual move
outlined in section \ref{subsection:Reversal}.

\subsubsection{Supplementary invariant ontology \& the model's invariant attributes}
The hypothesising of: a hidden cosmological evolution process, a
latent universal time coordination, and a latent homogeneous space
background, are all anathema to GR with its metric based approach.
With quantification in the new model only requiring (several
instantiations of): a systemic (i.e. barycentric) cyclic
time\footnote{Note that this (sidereal) rotation period/time takes
into consideration all (minor) general relativistic effects such
as geodetic precession.} ($\Delta t$), a (field based) cyclic
variation of a `true' acceleration/gravitation ($\Delta a$) that
is of \emph{fixed} magnitude everywhere\footnote{Later we see that
an (inner) central `hole' is required because the rotating
space-warp is the externalisation of an inexpressible QM energy
arising from atomic/molecular mass in a non-inertial
configuration.}, and a \emph{constant} (non-local) mass ($m^*$) to
(celestial space) enclosed volume ($V$)
relationship\footnote{Non-local mass is a new concept unique to
the model. It is more elaborately `defined'/introduced in
subsection \ref{subsubsection:non-local inertial mass}.}; a simple
neo-classical \emph{formalism} is sufficient to account for the
Pioneer anomaly. Note the \emph{global invariance} (i.e.
homogeneity) associated with the latter two quantities: $\Delta a$
and $(m^*V)$.

\subsubsection{Another kind of `uniform' acceleration that can be visualised}
Other than the product of non-local mass and volume enclosed, the
model has only one (type of) field value (sinusoidal $\Delta a$)
(which is) of fixed value/amplitude throughout the universal
system. If we allow the acceleration/gravitation to be represented
by space curvature (cf. spacetime curvature), it becomes possible
to visualise each (gravito-quantum) rotating space-warp (in two
dimensions) as a rotating `rigid' warped disk (recall Figures
\ref{Fig:FlatDisk} to \ref{Fig:R3FrontElev}). This ability to
visualise is more in keeping with a (neo-classical) fluid
mechanical treatment incorporating a vast \emph{space} continuum
and time simultaneity; rather than anything remotely resembling
general relativity's formalism. Expressing the model's features as
a supplementation of GR, in the language of GR, is not feasible;
nor is a modified Newtonian mechanics (MOND) viable\footnote{MOND
tends to alter the mathematics of gravitation without any
supportive conceptual justification.}.

It is important to note that with: the $\Delta a$ amplitude (at
any given point) being \emph{non-uniform}, in the sense of
sinusoidally or \emph{cyclically} variable; and uniform or
homogeneous, in the sense of being the same magnitude/amplitude
everywhere; compatibility with GR (by way of the latter sense) is
trivially assured. Supporting this new type of field is the fact
that GR's conceptual foundations are restricted to \emph{uniform}
(i.e. locally homogeneous) gravitational fields. In GR
non-uniformity is both expressed as, and restricted to, curved
spacetime cf. curved space in time. Pragmatically, each
instantiation of the model's (very small) supplementary
acceleration/gravitational field is seen to sit (evenly) on top of
any existing curved spacetime field\footnote{Over extended periods
of time, i.e. a fair bit longer than the longest $\Delta t$.}; but
conceptually and mathematically this coexistence is an uneasy
alliance (see \mbox{subsection \ref{subsubsection:model's
separate}).}

\subsubsection{The model's separate gravitational relationship cf.
general relativity}\label{subsubsection:model's separate} The
implicit contention herein is that general relativity is ``not all
of (gravitational) reality"; and that a (solely) relativistic
approach is restrictive in some negative sense. Indirectly
supporting a notion of incompletion and restrictiveness is GR's
inability to incorporate/ascertain total energy (and total mass).
Primarily, SR and GR are restrictive in a positive sense, in that
the \emph{laws} of physics are invariant with respect to
coordinate transformations.

Whereas SR utilises local frame-to-frame relative relationships in
conjunction with `spacetime interval' invariance, our global
approach has highlighted a system-upon-idealised background
relationship in conjunction with the amplitude invariance of
sinusoidal $\Delta a$. These relationships are distinctly
different. This further confirms that the systemic/global nature
of the model is completely foreign to what a relativistic theory
can encompass.

Fortunately, the three levels/scales of mass systems, outlined in
subsection \ref{subsubsection:Three mass levels}, provides some
clarity in distinguishing the basis of GR's predominantly
(macroscopic) gravitation from the model's (ultimate) basis in
atomic/molecular `mass'. More specifically, this basis is an
atomic/molecular (process-based) rate of angular momentum;
relating to an asymmetrical relationship between the QM spin and
orbital values arising from closed loop celestial (geodesic)
motion of the atoms/molecules in curved spacetime, and also
involving the \emph{self}-interference\footnote{By way of a
difference between pre- and post-orbital-cycle wave phase
conditions.} and QM fermion `wave' aspects of these
atoms/molecules over the course of each complete orbit (of their
macroscopic `host' body).

Furthermore, \emph{scalar} magnitudes such as: total and specific
energy, (constant amplitude) acceleration, and (non-local) mass
(albeit referenced to an enclosed volume), are all well suited to
the model's global perspective. Later (section
\ref{Subsection:Model quantified internal}) we see that the energy
of each rotating space-warp --- together with its initial
non-local mass ($m_{1}^*$) --- is \mbox{$\Delta E_w=\frac{1}{2}
m_{1}^* \Delta a_w^{2}\Delta t^{2}$.}

\subsubsection{Invariant light speed and the limits of observational physics}
\label{subsubsection:light speed} Electromagnetic radiation's
constancy of propagation speed is pivotal to relativity's
mathematical formalism; and subsequently how space and time are
seen to go together (at least observationally). Further:
\begin{quote} It is a striking fact that \emph{all} the
established departures from the Newtonian picture have been, in
some way, associated with the behaviour of \emph{light} \citep[
p.285, Note 1 Chapter 5]{Penrose_90}.
\end{quote}

Responding to this quote of Roger Penrose regarding light, a sceptic
of pure-relativism can hypothesise upon additional ontological
aspects excluded by SR and GR. Particulary, whether one's
observational (ontological) framework in some way \emph{guarantees}
the (measured) constancy of light (EM radiation) speed in a vacuum
($c$) in all relative motion circumstances. This implies that the
outcome of an act or process of observation utilising electromagnetic
radiation, and reality itself, may be deeply entwined
--- but not in a way that conflicts with the SR view that $c$ (the
speed of light) is a fundamental feature of the way space and time
are unified as spacetime. Once again the finality of (direct)
observational evidence (alone) is being questioned and deeper
ontological circumstances entertained.

Associated with this EM radiation speed invariance is whether (and
if so, why) the specific energy ($E/m$) of all phenomena is
limited by a $c^2$ upper limit? This is further examined in
section \ref{subsection:Reversal}.

Heisenberg's uncertainty principle and the Planck (or reduced
Planck) constant support both: the notion that observational
precision is limited in a quantifiable manner, and that hidden
features (cf. variables) of the physical universe may exist,
especially where (QM) uncertainty prevails --- e.g. below a
minimum (quantum) energy level, or change in \emph{energy} levels,
in an atomic or molecular \emph{system}.

\subsubsection{Section overview and summary} \label{subsubsection:summerview}
Although a real Pioneer anomaly is a very minor
gravitational/accelerational effect\footnote{Note that the
proposed mechanism to explain the Pioneer anomaly is of
cosmological scale; and its influence may (conceivably) become
more and more significant as the periods of time involved are
extended.}, its existence (and associated constancy) requires
significant modification of our accepted understanding of reality.
Its existence is in defiance of GR's (assumed) `completion' of
gravitational theorisation.

In reply to the strong dependence of SR upon invariance (of the
interval) this section (\ref{subsection:SR's ontology}) has sought
to establish further invariances --- in the widest (less
technical) sense of the term, i.e. including homogeneity,
constancy and uniformity\footnote{In physics, invariance of
angular momentum and energy are directly related to conservation
\emph{laws}.}. We may divide these into three types.

\begin{enumerate}
\item{Special relativity has the invariance of the interval under
Lorentz transformations (Lorentz invariance) and the constancy of
the velocity of light (in a vacuum).} \item{In the model to be
presented there are two major types of (observational) invariants:
the amplitude of the cyclic acceleration in the field ($\Delta
a$), and the product of non-local mass and volume enclosed
($m^*V$).} \item{Two new idealised `invariances' have been
introduced. By way of non-locality we conjectured a hidden
(background) cosmological coordination of time, i.e. a (type of)
frequency invariance. Also, by way of space being (observed as)
cosmologically `flat' ($k=0$), we inferred that empty space ---
i.e. in the absence of matter and energy --- is homogeneous; and
thus this (idealised background) empty space displays spatial
invariance.} \end{enumerate}

What is the point of this proliferation of invariance, both overt
and covert? Firstly, \emph{hidden} background aspects `support'
the classical-like formalism in the model's account of the Pioneer
anomaly\footnote{Note that the model requires standard gravitation
to be conceptualised as curved spacetime, rather than as a
gravitational \emph{force}.}; especially since the field
perturbation/curvature arises from a \emph{virtual} asymmetrical
energy offset, that although inexpressible at/within the
atomic/molecular level of matter\footnote{`Virtual' energy;
because the dominant electromagnetic forces, at this level of
matter, overwhelm any non-inertial energy effects that could arise
from atomic/molecular (i.e. mass) motion in (celestial scale)
\emph{curved} spacetime. Actually, it is the intrinsic spin of
\emph{all} elementary fermion/matter particles `within' each
(composite/whole) atom/molecule that is `influenced' in a virtual
manner.}, it is witnessed/registered at the global background
`level'. Secondly, we note that the four new invariances
introduced, all have a global or systemic application.

More importantly, we may distinguish two distinct types of
`relative' effects (especially involving motion): the standard
object-to-object (`phenomenal') relativity, and a \emph{new}
relationship between the observable world and the empty space and
coordinated/harmonious time of a (hidden background) `noumenal'
idealised `world'. This latter, local-phenomenal to global
background/noumenal physical relationship, argues for the
coexistence of relativity's invariance of spacetime with both
homogeneity of (empty) space, and uniformity of time in this empty
space. Thus, mass (for example) can be seen to curve an otherwise
flat space, \emph{and} cause a deviation from `the' (empty space)
rate of time. Whereas, in SR's \emph{formalising} of space and
time, it is the invariance of the spacetime interval that is
paramount.

The reader should be aware that we are \emph{not} seeking to
downplay the importance and exclusivity of the Minkowski metric,
nor are we promoting a ``disguised" Euclidean metric. Although
similar, but with a subtle difference, we shall promote the notion
of global reality as a ``distortion" of an idealised Euclidean
metric --- that would exist in a universe bereft of (mass,
momentum and) energy.

By way of highlighting: certain shortcomings of special and
general relativity; the limits of observational physics; and
conceivable restrictions upon the capacities of the process of
observation; this subsection has sought to undermine confidence in
the \emph{completeness} of (GR's) existing gravitational
theorisation. Note that within GR's scope of application, i.e.
effectively everything except the model's specific/unique case,
GR's correctness and/or accuracy is \emph{not} doubted.

A primary aim of this section has been to ``set the scene" for the
decisive conceptual and theoretical move to follow (section
\ref{subsection:Reversal}); whereby SR's Lorentz transformation is
given a new/further and physically supplementary/different
interpretation --- involving conditions in a real/measured
universe \emph{relative} to inertial circumstances, with the
latter being associated with a new (cosmological scale) specific
energy based \emph{invariance}. This new conceptual landscape
results from our assumption/hypothesis that the Pioneer anomaly is
a real and new physical effect.
%****************************************************************************************
\subsection{Local relativistic effects within a globally invariant universe}
\label{subsection:Reversal} Section \ref{subsection:SR's ontology}
supplemented the local and (and as measured) invariances of SR,
i.e. light speed and the laws of physics in constant relative
motion, with the establishment of \emph{idealised global}
invariances --- i.e. homogeneous space and time on the largest
scale possible. Regarding the latter, note that globally (and
beyond observations) space and time are treated \emph{separately}.
This, together with the weaknesses in the exclusivity of special
and general relativity previously outlined, forms the platform for
this section. This section (\ref{subsection:Reversal}) discusses
the specifics of how the new model, with its noumenal-time
simultaneity and (further) incorporation of (measurable) global
(invariant) quantities, can \emph{coexist} with special and
general relativity's account of gravitation.

\subsubsection{Preliminaries}\label{subsubsection:preliminaries}
Loosely speaking, we are countenancing two aspects of reality: one
measured; and one beyond measurement, yet also (in a certain sense) the
platform `behind' measurement\footnote{Note that this stance is based
upon anomalous \emph{observational} evidence in need of an
explanation.}. For observations/measurements we retain standard or
`proper' physics; e.g. the relativistic energy-momentum equation
$E^2-(pc)^2=(mc^2)^2$ with its relativistic (observer dependent) values
of energy ($E$) and momentum ($p$). In what follows, it is necessary to
move away from a purely observational basis that gives light rays and
coordinate systems (i.e. operationalism) exclusive priority.

Moving from the standard ``phenomenal-to-phenomenal"
[\emph{sic}\footnote{We shall intentionally use the adjective
`phenomenal' also as a noun, in order to highlight its
new/specific meaning herein; as well as to indicate its
juxtaposition with/alongside `noumenal', which has usage as either
an adjective or a noun.}] approach of relativity, to a
phenomenal-to-background (i.e. a phenomenal-to-noumenal) approach,
we recognise a (flat) background space continuum with a (latent)
global time simultaneity; but not a universal time
\emph{coordinate}, nor a background reference frame that could be
used with \emph{observations}. Only this latter approach can
facilitate the additional global gravitational/accelerational
effect proposed herein. Additionally, GR's domain of application
is perceived as always a subset of the whole universe, applying up
to the largest of galactic clusters. Relativity's curved spacetime
is systemic, but herein we need to deny its (seamless)
extrapolation to universally (i.e. globally) systemic conditions.

The digital global sequenced temporality discussed in section
\ref{subsection:SR's ontology}, allows us to appreciate one
(idealised) global `now' existing with (or in amongst) many
different local time rates. The temporal symmetry of most laws of
physics tends to support this picture of reality (i.e. hidden
simultaneity); but note that the model/mechanism to be presented
is unlike this and relies upon an irreversible asymmetry (and/or
offset) of QM \emph{energy} `achieved' (and externally
re-expressed) over a (system-based) cyclic time period.

\subsubsection{Inertial frames and: nothing, something, and
everything}\label{subsubsection:nothing,something} In Newtonian
Mechanics inertial frame circumstances apply to bodies in uniform
motion or at rest (in an attendant inertial frame). The concept of an
idealised global/universal empty space is (herein) considered to be a
deeper (i.e. more fundamental) inertial circumstance than uniform
\emph{motion} conditions --- albeit not `realistically' (i.e.
measurably or phenomenally) viable.

If `the phenomenal' is seen to always involve `something', then
the idealised (noumenal) stance pursued herein involves both:
`nothing' as in truly empty space; but somewhat by way of
invariance, the idealised (noumenal) stance also involves
`everything' (spatial) --- albeit in another sense. Pursuing this
idea, a response to the philosophical question: ``Why is there
something rather than nothing?" should necessarily \emph{also}
include/involve the notion of `everything'.

When dealing with finite quantitative circumstances, we are
necessarily dealing with `something'. A (coexistent)
nothing-everything dichotomy lies (either at, or) beyond the
bounds of (the something within) a minimum-maximum dichotomy.
Quantitatively, any minimum `amount' would appear to also require
a conceivable maximum as the other `book-end'. For example,
regarding velocity we have rest and light speed.

In subsection \ref{subsubsection:Copernican} the something of
physical reality is considered to lie within the coexistent
(quantitative) limiting `book-ends' of the empty and the full ---
corresponding to the (aforementioned) nothing-everything
dichotomy. This coexistent dichotomy has an attendant global
invariance; i.e. one existing everywhere (and) at all `times'. Our
objective is to look at (hitherto non-phenomenal or noumenal)
`empty space', i.e. energy-less space, in a second coexistent way;
as energy-full --- with this different/reversed (i.e.
\emph{noumenal}) perspective being then understood as the more
appropriate noumenal perspective.

Initially, we restrict ourselves to a discussion of SR's uniform
relative motion circumstances. Later in this section
(\ref{subsection:Reversal}) gravitational circumstances are examined.

\subsubsection{A Copernican-like reversal regarding inertia and
specific energy invariance} \label{subsubsection:Copernican} By
way of: envisaging restriction in one's own observational capacity
(subsection \ref{subsubsection:light speed}); homogeneous
background length/space and time, light speed invariance in SR;
and $E_{\rm{rest}}=m_0 c^2$; it is feasible that a `richer'
invariant (background/noumenal) global quantity may exist. All
that we can be sure of is that the background must be invariant,
at all places and times in some sense; and if a magnitude is
involved, it most probably involves $c$ or $c^2$ --- with the
latter having the dimensions/units of specific energy ($E/m$) i.e.
$[L^2/T^2]$. Thus, $c^2$ conceivably quantifies an invariant
physical (or phenomenal) maximum in some way.

To appease this situation, it appears that we need to also
conjecture a Copernican-like reversal regarding inertial motion
(in SR) and (specific) energy; where zero energy (and hence zero
force, zero angular momentum, etc.) at a `point' is
\emph{actually}, from a reversed perspective, maximum possible
specific energy ($c^2$). From this reversed perspective, i.e. a
noumenal supplementation (or `complementation') to the
phenomenal/observational, an inertial frame or inertial motion is
`full' of \emph{latent} `potential' (specific) energy.
Subsequently and additionally, what appears (observationally) to
be an increase in specific energy (from zero) is, from this
\emph{reversed perspective}, actually a \emph{drawing down} from
an (everywhere and always) constant universal-wide
latent-\emph{potential} specific-energy \emph{source}. Left
completely alone this (background) potential would be spatially
homogeneous at $c^2$, at all times --- but the phenomenal universe
is anything but fully empty.

As was the case in subsection
\ref{subsubsection:nothing,something} regarding inertial
circumstances, this reversed (noumenal) perspective, that coexists
with an observational/phenomenal perspective, is seen to be the
deeper (or more fundamental) of the two perspectives. Some
discussion of mass is required before we examine the ramifications
of this reversed (and supplementary/complementary) perspective (in
subsection \ref{subsubsection:reversal ramifications} and beyond).

In reality, mass cannot be said to ever truly exist at a point,
because this implies an infinite density. Further, stable
macroscopic mass, and QM mass especially, always occupy a volume
of space, and thus their (self) energy is distributed over a
finite region --- with the electron being the most `point-like' of
all masses. Observationally, it is considered that: ``atoms are
largely comprised of `empty' space". Herein, we also need to
recognise the volume of space that an object's mass encompasses,
with various mass-to-volume relationships existing for different
types of mass --- i.e. particles, atoms, molecules, and bulk
matter.

Consider the tremendous expansion the universe has undergone
subsequent to the highly `dense'/energetic conditions existing
just after the ``big bang". Thus, when mass is in atomic and
molecular form, the mass to volume ratio is negligible compared
with what was (and is) capable of being achieved. In other words,
the existence of an actual atom/molecule in space is seen to
involve only a very minor proportion of the maximum (noumenal)
`specific' energy possible/available, cf. the case where all of an
atom's space (i.e. every point in its volume) has the maximum
specific energy possible (i.e. $c^2$) --- notwithstanding the huge
energies associated with nuclear fission or fusion effects. Thus,
the existence of a \emph{single} atom/molecule is seen to only
involve a very minor ``drawing-down" from the maximum possible
specific energy available/possible throughout space.

\subsubsection{Ramifications of this reversal regarding specific
energy}\label{subsubsection:reversal ramifications} There are a
number of ramifications of utilising the `reverse' stance
conjectured in subsection \ref{subsubsection:Copernican}. Firstly,
we hypothesise that (observations made from a frame in uniform)
relative motion involves a `drawing-down' from this
(reversed/noumenal) `potential hill' (as compared to a `potential
well'). In subsection \ref{subsubsection:relative motion} we shall
argue that this (also) gives rise to the same type of Lorentz
transformations that occur in SR. Significantly, gravitational
effects also draw down, in a \emph{different} way, from the
\emph{same} `hill'. It is the double-draw, (for example) of a
body's mass and its motion in a field, that is seen to necessitate
the use of general covariance (GCoV) in the mathematical formalism
of GR; thus giving GR its uniqueness amongst physical theories.

Secondly, and on a lesser note, regarding high speed relativistic
motion; no rocket-ship (of non-zero mass) could ever get to the
speed of light ($c$) because the (noumenal) latent specific energy
is just `not available' for this (combination of mass and motion
at light speed) to be achievable --- the noumenal potential hill
is ``only so high". By way of contrast, the `compulsion' of
\emph{massless} photons to always be observed as moving at light
speed makes them unique amongst particles\footnote{Assuming
neutrinos have some mass, they (then) travel at very \emph{close}
to the speed of light.}.

Finally, in order to proceed with this discussion, it is proposed
that the maximum specific (latent) `potential' energy (in a
non-point-like region of spatial volume) $c^2$ is `complex',
literally in the sense of being able to possess real and imaginary
components.

\subsubsection{An alternative magnitude for the invariant specific
energy source/`hill'} Regarding \emph{relative motion} it appears
that $\frac{1}{2}c^2$ is a better choice (cf. $c^2$) for maximum
latent specific energy available, i.e. the amount that physical
reality can be `drawn upon'. Four reasons are given to account for
this preferred magnitude (or `potential hilltop' value); none
entirely compelling (individually or collectively).
\begin{enumerate} \item{Neo-classically this is maximum conceivable
specific kinetic energy, in that $\frac{1}{2}v^2 \rightarrow
\frac{1}{2}c^2$ at the limit.} \item{With relative motion
involving \emph{\emph{two}} bodies, the specific energy of an
`examined' body can only (in some sense) be half of the total
maximum value.} \item{By way of time symmetry in electromagnetic
theory, we might embrace Huw Prices's philosophical argument that
a (secondary) non-observational (--ve) time may exist, permitting
subliminal `backwards' causation\footnote{The negative frequency
(--$f$) of subsection \ref{subsubsection:Rayleigh Power}, that was
deemed non-physical, could be indicative of a (--ve) time
(direction).}; thus halving any latent-potential for real (+ve)
time motion \citep{Price_96}.} \item{The virial theorem, for
rotating gravitational systems, yields $\frac{1}{2}P.E.=K.E.$ and
a cosmological instantiation of it may exist for maximum specific
latent-potential energy and maximum specific kinetic energy.}
\end{enumerate}

In what follows, the choice of $\frac{1}{2}c^2$ or $c^2$ as an
invariant maximum specific energy has no major impact upon the
discussion. Indeed, depending on the circumstances both of these
maximums may have a different role to play.

\subsubsection{Relative motion and background \mbox{invariant}
specific energy}\label{subsubsection:relative motion} This subsection
begins the process of showing how SR's Lorentz transformations can
also be related to the phenomenal-to-background/noumenal distinction
previously discussed (subsections \ref{subsubsection:preliminaries}
and \ref{subsubsection:nothing,something}). Accepting the reversed
perspective on `reality' argued for in subsection
\ref{subsubsection:Copernican}, the Lorentz transformations appear to
(also) quantify the proportion of \emph{globally} invariant specific
energy that is no longer available to (observable) physical reality
--- when two bodies in uniform relative motion observe each other.
Compared to inertial circumstances, there is an equal loss in the
(square of) ``available to measurement" time rate and length (in the
direction of motion). We shall see how this alters our standard
`phenomenal' understanding of: a clock's \emph{local} `time dilation'
relative to an inertial frame, and (especially) relative to other
clocks in motion.

If we consider the maximum latent specific energy \emph{available}
to relative motion as a complex `quantity' (i.e. magnitude and
unit), then $\frac{1}{2}c^2$ is the invariant square of the
hypotenuse of a right triangle, and this allows a break down of
total (invariant) specific energy into real and imaginary
components. Further, if we consider relative motion at $v$ as
contributing (or involving) a $\frac{1}{2}v^2$ imaginary (number
based) component of specific energy, then the real component of
specific energy remaining is $\frac{1}{2}c^2 - \frac{1}{2}v^2$
with `dimensions' $[L^2/T^2]$. As a \emph{proportion} of maximum
available specific energy this is $(c^2-v^2)/c^2$ or $1-v^2/c^2$.
Note that the same result is obtained if $c^2$ and $v^2$ are
(alternatively) taken as the maximum specific energy, and total
kinetic energy respectively. Thus, for motion, which has
dimensions $[L/T]$; a (non-dimensional) proportionality factor
$\sqrt{1-v^2/c^2}$ is required to represent the proportion of time
and length availability \emph{lost}; cf. time dilation and length
contraction.

Alternatively, $\sqrt{1-v^2/c^2}$ is simply $a/c$, if $a$ is the
third side of a (complex) right triangle with sides: imaginary
$v$, real $a$ and complex hypotenuse $c$. The same would be true
for a similar triangle with sides: $v/ \sqrt{2}$, $a/ \sqrt{2}$
and $c/ \sqrt{2}$. Subsequently, an inertial (rest) frame has an
\emph{idealised} background specific energy `potential' (i.e.
availability) that is `actually' a fully \emph{non-imaginary}
quantity. Importantly, we note that, from an observer's
perspective, cf. a universe-based (noumenal) perspective on
things, motion (at $v$) is considered `real' --- whereas from the
universe's perspective motion (at $v$) is considered `imaginary'.

From observational evidence, and an observational perspective, we
make the (new) interpretation that real physical \emph{phenomena},
and hence measurements thereof, may only obtain the real component
of specific energy. By way of the reversal we have considered, we
may say that: relative motion `draws' down from a latent and
invariant specific energy `potential hill', with relative motion
giving rise to an imaginary component that can no longer
contribute to `reality', in the sense of ``be made available" to
reality. Further, to satisfy dimensional constraints, \emph{both}
the `units' of length and time (together) must be equally reduced
from their `inertial' (fully real) magnitudes. In SR, this speed
based influence is quantified as the time dilation, and length
contraction, involving $\gamma=(\sqrt{1-v^2/c^2}\,)^{-1}$. Mass
dilation is discussed in subsection \ref{subsubsection:mass
dilation}.

\subsubsection{A different interpretation of SR's time dilation and length contraction}
In special relativity, relative to an \emph{observer} in an
inertial or stationary system, a clock ($t'$) moving with velocity
$v$ appears to run slow according to the relation:
\mbox{$(t'_2-t'_1)=(t_2-t_1) \sqrt{1-v^2/c^2}$}. In addition to
this, the length contraction of a rod, in a rest frame, as
observed from a moving frame is: \mbox{$(x'_2-x'_1)=(x_2-x_1)
\sqrt{1-v^2/c^2}$}.

The relationships of SR's description of time dilation and length
contraction are not the same; they are conceptual opposites, i.e.
a dilation vs. a contraction. In contrast, from our `universal'
perspective, they are seen to describe the loss in time (rate) and
loss in (measurable) length separation respectively. The
difference lies in one's observational `perspective', with SR's
results using light ray propagation and coordinate differences. SR
also has a different (and simpler) ontological stance upon what
time `is' and how time `passes', with this grounded in an
`operationalism' approach. In SR time itself slows, whereas from
the new perspective, we say that: in a non-inertial frame temporal
\emph{processes} simply take `longer' than in the local inertial
frame. Note that the ticking of a clock is a `process'.

To maintain the spirit of (uniform relative motion) special
relativity, the time loss/dilation scenario discussed above must
only apply (relatively) between two observers. Thus, all relative
inertial frames are seen to involve maximum invariant specific
energy\footnote{There are issues that need to be pursued,
especially regarding the relationship between a global background
and nested reference frames moving relative to each other --- e.g.
Earth in solar system, in galaxy, in galactic cluster, etc. SR
demands that this hierarchy is irrelevant to local relative
observations, which further mystifies just what reality is `in
itself' when it is not being observed.}.

\subsubsection{Further discussion upon relativistically altered time and length}
If we accept that (from a beyond just a measurement-based
perspective) a digital universal pivot-pause-pivot-pause...
temporal evolution exists\footnote{Without an associated universal
(absolute) time coordinate (recall subsection
\ref{subsubsection:universal time}).}, and subsequently that the
universe retains a form of global coordination as it `evolves';
then the non-inertial frame is seen to miss a bit/proportion of
each (inertial) time increment. Total time loss/dilation is
therefore an accumulation of very small time losses. Length
loss/contraction is `slaved' to this time loss so that: in the new
ontology a (material) rod \emph{appears} contacted because the
full length of space containing it cannot be fully perceived by a
viewer in relative motion. A given process in a stationary frame
requires fewer `sequential moments', i.e. digital instances of
(phenomenal/measured) reality, than the same process in a moving
frame.

A second (non-standard) interpretation of the same measurements is
possible. Consider: clocks upon an aircraft, moving
\emph{relative} to the Earth's (effectively/essentially) inertial
frame --- noting that the Earth's motion relative to other bodies
is irrelevant. Interestingly, the time \emph{lost} off the clock
on the aircraft during its flight remains after the experiment,
whereas the length lost (off a rod) is reversed once relative
motion ceases. This arises because the \emph{rate} of time was
only slower during the flight. This highlights a primacy of time
loss over `ensuing' length loss. Length contraction is `real' only
in the sense of being an apparent observational artifact.

As a final step in this new (and complementary) interpretation let
us declare that: uniform (inertial) relative motion leads to an
alteration in the magnitude of the standard \emph{unit} or scale
of measurement of length and/or time. This unit contraction
results in a change to the \emph{magnitude} of measured length or
time --- time actually slows and length appears contracted. By way
of background ontological aspects of the universe, our standards
of measurement are adjusted or compromised, rather than time
(itself) dilating and an object's length (itself) contracting.

Finally, we note that at a deeper ontological level, uniform
relative motion is (conceptually) seen as a less than optimal
idealised-`inertial' system. For a fully
\emph{idealised}-`inertial' system we need to have an empty
universe devoid of all relative motion. When considering
gravitational effects (see subsection
\ref{subsubsection:conceptual ramifications}) this notion is
conceptually useful, although in practice/reality (due to the
relativity of observational reality) it is not required in special
relativity (SR).

\subsubsection{Briefly on the ``clock paradox"} In SR, the twins clock
paradox is resolved by using a Minkowski spacetime diagram and
appreciating the asymmetry of the situation\footnote{Once and for
all, in the minds of most physicists. This `party line' is
championed by Paul Davies in: ``About Time: Einstein's unfinished
revolution" \citep{Davies_95}.}. By way of the ontological
supplementation pursed herein, Minkowski spacetime is not the only
ontological perspective of space and time. In the model's
new/complementary (background and global) perspective two clocks
in the same motion (relative to an inertial frame) will `slow' by
an equal amount/proportion.

This is made particularly evident in a figure eight symmetric
version of the twins paradox, involving geodesic motion, discussed
in a footnote within subsection \ref{subsubsection:failure}. The
paradoxical aspect of SR's (dual) time dilation dissolves, once
and for all; and the dreaded third observer (at rest), so easily
conceptualised, need not be derided --- as it must be in SR. The
reason for this conceptual `enrichment' is that the model seeks to
incorporate a global perspective in addition to the (dominant)
relative perspective of observations. When `going global' the
model must be very careful to not ``step on SR's toes", and this
is why global invariance is so important --- both for physical
quantities in the model, and the (hidden) background ontological
`limitation' of specific energy introduced in this section
(\ref{subsection:Reversal}).

\subsubsection{Mass dilation and clarifying the intended meaning of
the word ``real"}\label{subsubsection:mass dilation} The apparent
increase or dilation of mass in the measurements of high-speed
elementary particles is seen to arise by way of a mass `experiencing'
less real space, as compared to inertial conditions. By way of
relative motion the `weighting' of mass in the `real world' (left
available to it) is proportionally \emph{increased}. Note that from
the model's new perspective, it is the time loss that is ultimately
responsible for the dilation of mass; because the space lost is in
equal proportion to, and ensues from, the amount of time lost.

Two aspects of mass dilation, by way of space loss/contraction,
need to be distinguished. Firstly, a mass exists in a space that
is predominantly external to it; and its high-speed motion
`encounters' less of this space. Similarly, yet more importantly,
the `internal' space of a mass is also reduced --- effectively
increasing its density. It is this latter case that pertains to
the mass dilation of SR, or mass \emph{gain} from our new
perspective. It has been confirmed experimentally\footnote{Walter
Kaufmann was the first to confirm (1901-03) the velocity
dependence of electromagnetic mass by analyzing the ratio e/m
(where m is the mass and e is the [constant] charge) of cathode
rays. Paraphrased from Wikipedia: \emph{History of Special
Relativity}, 2011.}. As with length and time, the dilation/gain
effect (exists and) is quantified in a purely relative way.

Note that from a noumenal perspective, magnitudes at the
phenomenal level vary with respect to the noumenal's invariant
`nature'; but equal adjustments in the units of the Length and
Time dimensions ensure that (the observational/phenomenal value
of) the speed of light is invariable --- both spatially and
temporally.

The word ``real" in this section (\ref{subsection:Reversal})
retains its common usage, as in: (1) having objective existence,
(2) not imaginary, and (for our purposes) (3) being `phenomenally'
existent in the sense of being measurable. Note that relative
motion \emph{is} real, but it also has a further unmeasurable
effect in that it generates an `imaginary' component of latent
specific energy [say Im$(e)$], thereby reducing the maximum
potential specific energy available that can become `real'.

\subsubsection{A brief interim summary; \& replies to:
``Can everything be relative?"} In subsections
\ref{subsubsection:relative motion} through
\ref{subsubsection:mass dilation}, we have argued that: special
relativistic time dilation, length contraction effects, and mass
dilation effects can be traced back to (or grounded in) a hidden
supplementary (deeper) ontology. Subsequently, the Lorentz
transformations can be seen to describe \emph{both}: how spacetime
intervals remain invariant, so as to be compatible with
electromagnetism; as well as how \emph{uniform relative motion}
coexists with (and `bites' into) a maximum `available' specific
energy --- that remains invariant everywhere throughout the
universe, for all sequential `moments of observable time'. From
the observer's perspective, the time describing physical phenomena
is analog.

The Lorentz transformations effectively build a bridge between a
\emph{global} (i.e. universal) specific energy invariance, and
SR's (and EM's) \emph{local} invariance of two spacetime
intervals. Physics is arguably stymied by way of lacking an
appreciation of this `new' coexistence; preferring to pursue a
deeper understanding of both special and general relativity
through the study of the Minkowski metric.

Relativity implicitly answers the ontological question: ``Can
everything be relative?" in the affirmative, in line with its
stance on the formalistic representation of physical reality,
whereas the contents and sentiments of the preceding discussion
requires that we answer this question --- which is a borderline
oxymoron --- in the negative.

\subsubsection{Justifying a constant specific energy value throughout the universe}
\label{subsubsection:justify specific} The aim of this subsection
is to argue that reasonable justification for a homogeneous
background specific energy throughout all of space, at all `times'
--- upon which reality `draws-down' --- can be found within mechanical
classical mechanics (i.e. excluding electromagnetic effects). To
achieve this we examine the laws describing a range of classical
or everyday (macroscopic) mechanisms or phenomena.

Subsections \ref{Subsubsection:MM Differences} and
\ref{subsubsection:Prop/HAWT} argued for a qualitative distinction
between the physical laws of the macro- and microscopic realms.
Subsequently, four `force' unification is not endorsed. This
subsection pursues the (non-reductive) ramifications of this
distinction; particularly the unsuitability of `force' in
describing gravitational effects. Note that gravitation is the
only \emph{mechanical} phenomenon that (inevitably) exists on a
universal (or cosmological) scale. Also note that at the
microscopic level, i.e. up to the `level' of atoms and molecules,
the standard account of reality in terms of three (unified) forces
is fully embraced.

In non-chemical, non-thermal and non-electromagnetic
circumstances, macroscopic energy equations commonly have the form
$E\propto m$; e.g. $E=mgz$, $E=\frac{1}{2}m\omega^2x^2$ (linked to
$E=\frac{1}{2}Kx^2$), $E=\frac{1}{2}mv^2$, and $E=m \Gamma_e f$
(recall subsection \ref{subsubsection:Prop/HAWT}) concerning
(respectively): positional change in a constant gravitational
field, springs, kinetic energy, and wind turbines/`airscrews'. In
these `mechanical' systems, or aspects thereof, the establishment
of specific energy ($E/m=e$), from the kinetic or potential energy
expression, is easily obtained. Clearly, force plays a vital role
in the first two energy expressions, but not the latter two.

Regarding electromagnetism, the Planck relation --- between the
energy of a photon and its electromagnetic wave, $E=hf$
--- retains the prominence of energy; and regarding thermal
physics, the ideal gas law $pV=nRT$ retains the dimensionality of
energy, i.e. $[\frac{M L^2}{T^2}]$. Neither of these
expressions/laws is simply an integral of force over distance.
Energy (cf. force), or at least ``dimensions to that effect",
appears to be more general and useful at the macroscopic level of
physical formalism (i.e. representation).

No expressions from the mechanics of solids have been presented,
but the concepts of deformation and curvature have some `overlap'
with the formalism of general relativity. Accepting universal
`coordinated time' (see subsection \ref{subsubsection:universal
time}), the notions of a cosmological space continuum and space
curvature has validity --- albeit at an idealised level of
conceptualisation.

In fluid mechanics, we utilise the concept of a mass continuum, and
(necessarily) examine force per unit volume rather then force. By way
of utilising conservation of mass (i.e. continuity) and conservation of
momentum, we may obtain equations describing the motion throughout a
flow field. In the case of an inviscid flow we obtain Euler's equation.
If density ($\rho$) is constant (i.e. incompressible flow), then
integration of Euler's equation throughout an irrotational (i.e. zero
vorticity) flow, yields
$$\frac{\partial \phi}{\partial t}+\frac{v^2}{2}+\frac{p}{\rho}-gz=f(t)$$
where $\phi$ is velocity potential and $p$ is pressure \citep[
p.69]{Kuethe_76}. This is the same as (a special form of)
Bernoulli's equation $\frac{1}{2} \rho v^2+p-\rho g\Delta z=p_0$
\citep[ p.62]{Kuethe_76}, except for the inclusion of the unsteady
(flow) term and adjustment for the integration constant.
Importantly, the dimensions are those of \emph{specific energy}
$[\frac{L^2}{T^2}]$; and in steady inviscid incompressible flow,
the stagnation pressure ($p_0$) is \emph{constant throughout} an
irrotational flow.

It is now argued that the magnitude of our proposed (maximum)
latent/noumenal potential (specific) energy `source', at $c^2$, is
similar to the constant stagnation pressure (divided by constant
density) in (an idealised) fluid mechanics; in that it is constant
throughout (the `whole' of) space. Background space itself can be
seen as a continuum; but in regard to physical reality with
different local rates of time, it cannot be a systemic frame of
reference for observations. Nevertheless, this conceptualisation
permits the existence of a \emph{second} (and quite different)
type of gravitational `draw-down', that serves to indirectly
facilitate an explanation/model for the Pioneer anomaly ---
\emph{the} crucial step in permitting a \emph{real} Pioneer
anomaly.

In section \ref{Subsection:Model quantifies external} we shall see
that the model's value of (rotating space-warp) specific energy,
$\Delta e=\frac{1}{2}\Delta a^2 \Delta t^2$, is (also) constant
throughout space.

\subsubsection{A new reason for general covariance in General
Relativity's formalism}\label{subsubsection:new reason for GCoV}
Subsection \ref{subsubsection:justify specific} illustrated that
in fluid mechanics there is (often) a representational necessity
to work with specific energies. GR's principle of equivalence
means that a (test) body will `experience' the same acceleration
independent (or regardless) of its mass. Historically, general
relativity moved us away from an analysis based upon force;
although, in somewhat classical terms, we might (nowadays) think
of acceleration as a specific force ($F/m$).

Conventionally total energy is an \emph{addition} of energies
(from zero \emph{up}). From our reversed ontological perspective
we now (also) think of conventional \emph{macroscopic} (specific)
energy increases as actually a reduction in, or
drawing-\emph{down} upon, a (very large magnitude) invariant
specific energy `noumenal-source'\footnote{The reader may be
wondering whether the `mechanical' specific energy ($E/m$)
discussed herein also applies in the case of decidedly
electromagnetic quantities and processes, where ``the mass" in
question is not at all obvious. Since our concern herein only
involves the inertial/mass aspects of QM atomic/molecular systems,
electromagnetism shall not directly concern us; but it \emph{may}
be that electromagnetic \emph{quantities} (cf. their effects) need
to be treated quite differently at a deeper ontological level ---
especially if specific energy is a very important quantity. The
following may possibly be of relevance. \,\,\,Reducing the
units/dimensions of electromagnetic quantities to the fundamental
units/dimensions of (kg, m, sec, C) and (M, L, T, and Q)
respectively, it can be seen that the ratio $M/Q$ appears in the
dimensions of most electromagnetic quantities. Note that C is
coulombs and Q is (electrical) charge. For example, the inverse of
capacitance has the unit $[\rm{Farad}]^{-1}$ which can be
expressed (with some effort) in fundamental units as
$[\frac{\rm{kg.m^2}}{\rm{C^2.sec^2}}]$; which in fundamental
dimension notation is $[\frac{ML^2}{Q^2T^2}]$. The extra $[M/Q]$
ratio arises from physics giving primacy to \emph{mechanical}
energy, which has dimensions $[\frac{ML^2}{T^2}]$. Note that there
is no $Q$ dimension in work/energy, even if the system doing the
work is electrical and/or magnetic. Since the charge to (rest)
mass ratio of an electron $(q/m)_e$ is a fixed quantity and has
dimensions $[Q/M]$, we can use this to alter the dimensions of
(inverse) Farads to $[\frac{L^2}{QT^2}]$, which is now expressed
in a $Q, L, T$ system of units (without $M$). The $Q$ dimension is
in the denominator rather than the numerator (as $M$ is with
mechanical energy). Hence, it appears that the specific energy,
i.e. $[\frac{L^2}{T^2}]$, associated with this electromagnetic
quantity (capacitance$^{-1}$), in the new `units', involves
inverse capacitance multiplied by the (stored) electrical charge,
which is actually \emph{electrical potential}. In other words, in
this new non-mechanical ($Q, L, T$) unit/dimension system
especially suited to electrical quantities, specific energy
$[\frac{L^2}{T^2}]$ is simply (electrical) potential. In the
traditional system of units/dimensions, potential energy per unit
charge has the unit of `Volts' and dimensions:
$[\frac{ML^2}{QT^2}]$. The primacy of electrical potential in
electromagnetism makes it well suited to playing an important role
at a deeper ontological level, \emph{if}: specific energy is
important, and the use of the $M, Q, L, T$ unit/dimension system
for describing electromagnetic quantities is replaced at a deeper
level with the simpler (and conceivably more natural) $Q, L, T$
system.}.

Recall from subsection \ref{subsubsection:justify specific}, that
the physical representation of energy in macroscopic (mechanical)
mechanisms easily yields a specific energy. Further, mass itself
and its motion, are considered to also `draw-down' upon a
``background availability"; and at a macroscopic level this
circumstance is (now) conjectured to be associated with
conventional gravitational effects. From such a perspective, the
(standard) representation of various effects within general
relativity is complicated by the numerous ways in which such
draw-downs can occur; e.g. mass, momentum, and energy; as well as
temporal rate changes that accompany these effects. Subsequently,
it can be argued that only a theoretical approach that employs
general covariance is capable of coping with this `multiplicity'
of (draw-down) effects.

\subsubsection{Three gravitational `draw-downs' upon invariant specific
energy} \label{subsubsection:conceptual ramifications} SR's
relative motion, GR's gravitation, and the model's rotating
space-warps (RSWs) can all be understood as various ways of
``drawing-down" from a (hidden) uniform specific energy
background. Subsequently, the two types of non-Euclidean geometry
associated with GR and the new model's RSWs, although quite
different, can coexist --- without any conflict.

Firstly, we have the simple draw down of high-speed (relativistic)
relative motion, as discussed in subsection
\ref{subsubsection:relative motion} through
\ref{subsubsection:mass dilation}, giving rise to the
idiosyncrasies of special relativity --- and expressed by way of
Lorentz transformations. Herein, this has been additionally
understood as relative motion's (specific energy) introducing an
imaginary component of specific energy, i.e. a drawing-down from
the maximum background/noumenal specific energy
field/potential-hill; thus leaving less specific energy available
to `reality'.

Secondly, as mentioned in subsection \ref{subsubsection:new reason
for GCoV}, a plethora of effects result in (GR's) gravitational
effects. Further, by embracing the alternative ontology outlined
in this section, especially that universal (noumenal/global)
sequential evolution involves a hidden (beyond observation)
simultaneity; spacetime effects may then be re-interpreted as a
spatial separation change accompanied by a rate of time change.
Note that this alternative (and `neo-classical') conceptualisation
has no (direct) observational relevance. This interpretation
allows GR's gravitation to be visualised (solely) as space
curvature, in conjunction with both: a local time rate and a
(hidden/background) global time rate. This intrinsic curvature of
space\footnote{Relative to what \emph{would} be a naturally
existing flat/Euclidean space background, in the (idealised
circumstances of an) absence of: matter, momentum, and energy.},
as compared with extrinsic curvature, is seen to occur with the
assistance of imaginary numbers in the three space dimensions.
Conceivably, and neo-classically, these imaginary (number based)
spatial components are then related to a (specific) deformation
energy, which in turn equates to the (specific energy) `draw-down'
associated with (GR's) gravitation. The mathematical
quantification of the specific energy associated with deformation
into (imaginary number based) space shall not concern
us\footnote{It is assumed that an analogy to circumstances
involving the bending of a beam in ``mechanics of solids" can
achieve this, with an adjustment for the local time rate cf. the
fastest clock in the universe (recall subsection
\ref{subsubsection:fastest time}).}.

Thirdly, the model represents the specific energy of each rotating
space-warp (RSW) as $\Delta e=\frac{1}{2}\Delta a^2 \Delta t^2$.
This value is constant throughout (the whole of) space.
Additionally, it is \emph{process} based, occurring over a time
($\Delta t$). This duration is additionally a cyclic loop time
primarily determined by (lunar) \emph{motion}\footnote{Later we
shall more fully discuss how the mechanism involves the
self-interference of atoms/molecules.}; and subsequently,
relativistic effects upon this motion are accommodated. It is
quantified by way of a \emph{systemic} clock rate; most
conveniently a (solar system) barycentric time rate. In the case
of RSWs, the specific energy `draw-down' is simply
$\frac{1}{2}\Delta a^2 \Delta t^2$, which is quite distinct from
`draw-downs' associated with GR's gravitation --- although an
acceleration amplitude ($\Delta a$), i.e. \emph{gravitational}
space-warp magnitude, can also be envisaged in terms of curvature.

Importantly, the constancy of the model's specific energy appears
to be necessary, so as to allow GR's `gravitation' and the model's
new aspect (or supplement) of gravitation to be compatible and
\emph{coexist}. These two `gravitational' effects are quite
different; e.g. having their `basis' in different `levels' of
matter --- i.e. macroscopic mass in motion vs. atomic/molecular
mass in motion. Sections \ref{Section:general model} and
\ref{Section:Quantif Model} shall explain how the total
\emph{energy} associated with RSWs, incorporating a non-local
mass, equals the total virtual and unexpressed non-inertial energy
(at the atomic/molecular level) of a (suitable) moon-planet
system. Outstanding at this stage, is a discussion concerning the
model's reliance upon the principle of energy conservation (see
\mbox{section \ref{subsection:Noether}).}

\subsubsection{Summary and concluding remarks}\label{subsubsection:summary local
relativistic} The primary goal in this section has been to outline
how a second type of non-Euclidean geometry (i.e. gravitational
field) can arise, in addition to the macroscopic-based
gravitational effects of general relativity. This has involved the
introduction of a supplementary ontology involving a reversed
perspective upon (observational) reality. These two different (and
complementary) ontological perspectives are seen to be required
for a full description of reality --- if the Pioneer anomaly is a
real gravitationally based phenomenon. Respectively, one is
sub-global and physical, while the other is global and
`metaphysical'; i.e. they are phenomenal and noumenal
respectively.

This ``dual-ontology" allows us to simply posit that measurements
in spacetime do not exhaust `reality'. Measurements can only
`apprehend' the `real', i.e. non-imaginary, components of a
physical phenomena that in their (formalistic) entirety
conceivably also involve complex numbers. Aspects of the
supplementary ontology include: recognition of a special type of
hidden background/global time simultaneity, and a background
Euclidean `flat' space continuum (in idealised circumstances). The
latter exists (conceptually) in the absence of all mass, momentum
and energy (fields). Thus, we may (also) ascribe gravitational
effects to both a loss in \emph{measured} time rate cf. a
`universal' time rate, \emph{and} a change in measured spatial
separation cf. an (idealised) flat space separation.

A crucial aspect of the model (to come) is that a \emph{virtual}
QM energy is re-expressed as a (non-virtual) physically real
rotating curved/warped space (together with a non-local mass
distribution). In section \ref{subsection:Noether} we shall see,
somewhat by way of our new supplementary/complementary (and
reversed) perspective on `reality', that \emph{conservation of
energy}:
\begin{enumerate} \item{has applicability to the `gravitational'
(i.e. the space perturbation) aspect of the model;} \item{is
grandly systemic in that it is not necessarily restricted to one
`level' of `material' reality; and} \item{can involve a virtual
energy magnitude, and thus it is not merely restricted to directly
measured (i.e. `phenomenal' and real) quantities.}
\end{enumerate}
%********************************************************************************************
\subsection{Symmetry, Noether's Theorem, and energy conservation}
\label{subsection:Noether} Our model, requires we allow a
qualitative divide to exist between the micro- and macroscopic
realms (also see section \ref{Subsection:Macro/micro}), and permit
a non-reductive (i.e. not to a graviton particle) gravitational
phenomenon at the macroscopic level. It may or may not be that GR
is reducible to a graviton \emph{particle}, but with our proposed
supplementary space curvature/perturbation, gravitation (overall)
cannot be (solely) a ``gauge" theory, as the three microscopic
forces are. The ramifications of this non-particle reduction (and
non-solely-particle unification) approach are now explored in
relation to Noether's theorem. Gravitation is preferably seen as a
\emph{``geometric"} theory.

\subsubsection{Preliminary Remarks: macroscopic-microscopic `separation'}
A total macro-micro separation is not being espoused. Rather, we
are examining how the very different characteristics exhibited by
the quantum mechanical (QM) and non-QM realms, may coexist.
Particularly, when a physical phenomena spans both domains to
comprise an integrated system. We shall argue that energy is
common to both realms, and its conservation therefore spans both
domains; much in the manner that the emission and absorbtion of EM
radiation/photons involves both a quantum system and the external
world. This \emph{electromagnetic} micro-macro situation is
somewhat of a precedent for our \emph{inertial mass}-based model.

Restricting ourselves solely to the macroscopic level, the
instantiation of (gravito-quantum) rotating space-warps to explain
the Pioneer anomaly appears to violate energy conservation; but
(we shall see that) for the combined micro- \emph{and} macroscopic
system proposed herein, energy's conservation is maintained. In
fact, energy conservation (and geometric constraints) are
necessary in order to quantify the acceleration/gravitational
amplitude and mass ``carrying capacity" of a rotating space-warp
--- which in turn is indicative of a (virtual) non-inertial (mass
in `motion') circumstance at the QM level of matter (see
subsection \ref{subsubsection:preview mechanism} for a fuller
account of the model).

\subsubsection{Preliminary Remarks: invariance, energy and acceleration}
Concerning the benefits of homogeneity and/or invariance, two
remarks need to be made. Firstly, the model is able to cope with
the fact that energy is a frame dependent concept, whereas GR's
energy-momentum tensor is not; because (cycle-based) specific
energy ($\Delta e$) is homogeneous throughout (the whole of)
space. Note that the (global) frame is the whole universe.
Secondly, (amplitude) homogeneity makes coordinate variations
non-existent. Thus, by default we cannot deny that:
\emph{coordinate} acceleration and (systemic) physical
acceleration are the same (cf. GR); with these also equivalent to,
and indicative of, a `gravitational' phenomenon (in its broadest
sense/usage).

\subsubsection{Background to Noether's Theorem}\label{subsubsection:Noether}
Noether's theorem is considered to precisely illustrate a deep
connection between conservation laws and symmetries in nature.
Noether's theorem is restricted to systems which obey the
principle of least action \emph{and} have a Lagrangian and a
Hamiltonian, which then determine the field equations. It is an
important result, especially for symmetries in elementary particle
physics. The symmetries of Noether's theorem relate to the
invariance of the \emph{laws} of physics. The homogeneity and
isotropy of space, and the homogeneity of time, then imply
conservation of momentum (linear and angular) and energy,
respectively. Indeed: \mbox{``\ldots any symmetry} of the
Lagrangian corresponds to a conserved quantity (and \emph{vice
versa})."\footnote{\url{http://www.mathpages.com/home/kmath564/kmath564.htm}}
Noether's theorem is \emph{not} seen to apply to the model being
proposed herein, because the dual-level `system' being considered
cannot be described by a single Lagrangian and/or Hamiltonian; but
aspects associated with Noether's theorem \emph{are} conceptually
valuable to the new model's formulation and discussion.

\subsubsection{What remains in need of clarification}
Let us recall our hypothesised (supplementary) ontological stance
regarding noumenal/\emph{background} space and (hidden background)
sequential temporal evolution --- mainly developed in section
\ref{subsection:SR's ontology}. In the idealised circumstances of
an absence of mass, momentum and (mass- and radiation-based)
energy, (the) background space \emph{is} inherently homogeneous
and isotropic. Further, the phenomenal--noumenal
(pivot-pause-pivot-pause-...) sequential arrangement of
all-inclusive (i.e. cosmologically-`wide') `moments of reality'
--- occurring all together\footnote{With a beyond observation
process-interval between these coordinated and cosmologically all
inclusive (staccato-like) `moments' (recall subsections
\ref{subsubsection:non-locality} and \ref{subsubsection:universal
time}).} --- allows us to say that: (non-observational) universal
`time' \emph{is} homogeneous (i.e.
simultaneous)\footnote{Notwithstanding the fact that
\emph{observationally} clocks throughout the universe run at
different rates --- depending on their local circumstances.}.
Thus, conservation of energy, and conservation of linear and
angular momentum apply; and they are possibly valid
assumptions/principles for the model's real world application. We
now investigate the veracity of this possibility.

Two subtleties shall need to be addressed. Firstly, in the unique
circumstances of the model and a supplemented ontology, does the
presence of: mass, momentum, energy, and (hence) a non-uniform
(observational) time (rate), deny all of these conservation
laws/principles? Secondly, (the `quantity') energy, by way of its
`dimensional' make-up, also involves a mass `dimension', in
addition to the (spatial) length and time dimensions. Thus, if a
rotating space-warp has a total energy, how (then) is the mass
aspect of this energy to be understood?

\subsubsection{On the failure of local conservation of energy in
general relativity}\label{subsubsection:Failure of local}
Gravitational energy in GR cannot be conserved \emph{locally} as
it is in classical theories; although, in the presence of
asymptotically flat spacetime (S/T), global energy conservation
applies.

Somewhat in response to section \ref{Subsection:Tensions}, it
appears that it is the assumed existence of a \emph{local} nature
of gravitational potential that necessarily thwarts local
gravitational energy conservation. This is because a gravitational
`potential' is largely based upon (the antiquated notion of)
gravitational `force' applied over a distance, leading to an
energy in the sense of physical work. Logically, a local failure
of (this type of) energy conservation in GR is \emph{not} a
sufficient reason (as yet) to deny the applicability of energy
conservation to/in a \emph{global}/cosmological (curved space)
physical system.

In GR, the non-homogeneity (or non-simultaneity) of time gives
sufficient reason for the denial of conservation of energy. The
inappropriateness of this denial, in the case of our model, shall
(now) be discussed in subsection \ref{subsubsection:homo, iso}.

\subsubsection{Homogeneity, isotropy, and the model being
pursued}\label{subsubsection:homo, iso} Accepting the
dual-ontology outlined in section \ref{subsection:SR's ontology},
relied upon a distinction between formalism and ontology. The
former concerns: ``what we can say about reality", and the latter
concerns: ``what reality is" --- which possibly also includes
aspects behind the scenes, i.e. beyond (direct) observations. In
the \emph{idealised} situation of an absence of matter, momentum
and energy, and by way of embracing a dual-ontology, we may
embrace the (classical-like) symmetries of hidden/background space
--- (meaning/implying) that it is the same at every location
(homogeneous) and in every direction (isotropic).

By way of reality's underlying homogeneity of `time' (i.e. a
simultaneous digital cosmological temporal evolution), energy
conservation is implied --- in both real and idealised conditions.
By way of (noumenal) \emph{time's} (priority and) homogeneity,
energy (cf. momentum) conservation may be discussed ---
\emph{both} micro- and macroscopically. For the purposes of the
model, energy needs to span (and interweave) a physical phenomenon
involving both the microscopic and macroscopic realms
\emph{together}. In other words, energy conservation is seen to
`operate' across both/all systems through (a common systemic)
time. In the model, we shall see that energy and its linkage
between the two realms, is the major (guiding) principle embraced;
with the magnitude of this energy (per atom/molecule) ultimately
lying within limits imposed by ($\Delta t$ and $\hbar$ in)
Heisenberg's uncertainty principle (HUP).

Time isotropy does not exist at the level of the new model's
\emph{formalism}, because a total \emph{virtual} microscopic (QM)
energy is \emph{asymmetrically/irreversibly} re-expressed
externally as the (\emph{real}) energy of a rotating space-warp.
Further, to allow the space-warp's (real) existence at the
macroscopic level, i.e. the level of curved space, the model
forgoes the isotropy and homogeneity of idealised (background)
space. The correct implication (from this) is not that the model
violates conservation of linear and angular momentum; but rather
that a traditional reductive particle-based approach to
conservation of momentum is inappropriate (and inapplicable) to
the model's multi-realm (system) --- and its HUP-based
circumstances.

The model cannot be based upon an action principle in one
big/general system, as a reductive (and relativistic) approach
envisages, because Noether's theorem (only) applies to single
level `conservative' systems. Thus, Noether's theorem is not
applicable in a dual-realm (i.e. microscopic plus macroscopic)
physical system, that also employs a phenomena-to-noumena-like
ontological distinction. Only the conservation of (a non-vector,
scalar) energy, can `survive' this complexity. Indeed,
conservation of energy\footnote{Specifically (and additionally)
involving a cyclic rate of angular momentum (offset), which is yet
to be elucidated.} is the linchpin of the model, with energy's
`universality' and applicability to a field theory (see Section
\ref{Section:Quantif Model}) providing strong support.

\subsubsection{Briefly on dual-level, and non-local physical behaviour}
In the preface to ``Science and Ultimate Reality" \citep*[
xvii-xviii]{Barrow_04}, Freeman J. Dyson discusses an example of
(macroscopic) classical mechanics and quantum mechanics being both
required in an explanation. A description of Uranium 235 fission
by a slow neutron sits at the border of microscopic and
macroscopic physics, with a classical liquid drop
conceptualisation being very useful. Thus, (once again) the
model's approach is not unprecedented.

How is the mass dimension of energy, which has dimensions
$[\frac{ML^2}{T^2}]$, pertinent to the model? We shall see that:
at the QM (virtual energy) level, it arises (partly) by way of
Planck's constant, i.e. a minimum angular momentum quantity; and
at the macroscopic level, a (new) `non-local mass' ($m^*$)
quantity is (necessarily) introduced in addition to the
`invariant' specific energy ($\Delta e$) of rotating space-warps.
This new type of (non-local) mass relates to the fact that:
``\ldots QMs is `non-local' because the state of a quantum system
is spread throughout a region of space \citep[ p.20]{Davies_04}."
Similarly, by way of the virtual nature of the QM energy involved,
this non-local mass is always spread throughout a volume ($V$) of
space. In other words, it is never localised; but it can/does act
upon a localised mass, e.g. the Pioneer spacecraft. Indeed, we
shall see that: $m^* V=\rm{constant}$ (in subsection
\ref{subsubsection:enclosed volume}).

\subsubsection{Asymmetry in the model and two asymmetries in physical reality}
\label{subsubsection:asymmetry} Unlike Noether's theorem (recall the
beginning of subsection \ref{subsubsection:Noether}), the model herein
is considered to illustrate a deep connection between a conservation
law (i.e. energy) and an \emph{asymmetry} in nature.

In Section \ref{Section:general model} we shall see that the
virtual quantum mechanical (QM) energy arises from the inertial
mass (and spin) within an atom or molecule. By way of relative
angular momentum, a (virtual) non-inertial configuration is
attained within (a great many) \emph{atoms/molecules}. This arises
by way of: their macroscopic (celestial) motion (orbital path) in
curved spacetime; a small change/offset in \emph{geometric} phase;
and the dominant strength of electromagnetic forces within an
atom/molecule. This virtual QM energy is ultimately an expression
of a (different or newly recognised type of) asymmetry in nature.

What comprises the \emph{broad} basis for the existence of this
(virtual) QM (angular momentum rate) asymmetry? Gravitation (in
practice) involves only positive mass (i.e. an absence of
antimatter), and hence locally there is only positive curvature.
Further, the model exploits the unidirectionality (i.e. arrow) of
time. These two physical asymmetries pertain to two primary
aspects of reality. Indeed, physical dimensionality (standardly)
employs mass and time as two of its \emph{fundamental} dimensions.
Thus, asymmetry is deeply imbedded in reality, and it is
conceivable that in (positively) curved spacetime, an irreversible
inertial mass-based process --- that is asymmetrical in time ---
is indirectly behind the Pioneer anomaly.

\subsubsection{The model's unique circumstances}
Regarding the model's rotating space-warps\footnote{Later we see
that these are actually (in three spatial dimensions) rotating
axisymmetric warped space-`cylinders', of global size/scale or
`proportions' --- rather than a thin (two-dimensional) disk (also)
of global/universal extent.}, homogeneity of space and the
isotropy of space and time are clearly absent. We shall now
discuss the very restricted, and minor, sense in which this
occurs.

In the model, the energy of the asymmetry associated with the
inertial to non-inertial frame offset --- and a virtual to actual
physical (circumstance) offset --- equals the energy of a
supplementary (external) real gravitational/accelerational field
effect (a non-isotropic rotating space-warp\footnote{With its
associated non-local mass magnitude ($m^*$).}). Note that over the
time of a single rotation (this external field effect) is equally
distributed around an equilibrium field strength --- in that it
imparts (on average) no `overall' additional/extra curvature to
that already provided by general relativity. The constant
amplitude of the space-warp means that, over extended time spans
and \emph{on the whole}, an idealised `background space' in one
sense retains its inherent homogeneity and isotropy.

The existence of this rotating space-warp is ultimately based upon
the uncertainty principle of QMs, and quantum \emph{discreteness}
encountering the effects of (non-discrete or) \emph{continuous}
curved spacetime. Thus, the warp merely represents an uncertainty
principle based ``wiggle room" (perturbation) around an
equilibrium set of circumstances established by GR (i.e. standard
gravitation). Importantly, such a mechanism ensures the ongoing
stable existence of the \emph{mass} aspects of QM systems in
curved spacetime circumstances, especially angular momentum (see
Section \ref{Section:general model} for a more detailed
discussion).

Energy is the (common) `currency' of the model's micro- to
macroscopic mediation (or re-expression). The physical energy
being discussed is a virtual QM intrinsic angular momentum per
cycle time (i.e. a virtual spin energy). We shall see that because
the underlying quantum mechanical asymmetry is cyclic, and process
based, it reasserts itself after every completed spin-orbital
cycle; thus maintaining the continuous rotation of the
(effectively rigid\footnote{In the sense of not deforming from its
initial `shape'.}) space-warp. Note, that for the QM energy to be
virtual, it needs to be below the minimum energy level of any
atomic/molecular QM interaction; and that the not insignificant
magnitude of this virtual energy arises from it being (very nearly
the same magnitude and) shared by every atom/molecule throughout a
(suitable) spin-orbit coupled moon.

\subsubsection{Concluding remarks}
The previous discussion has shown that: Noether's theorem, which
is a part of the theoretical physicist's ``principle" based
approach to reality, has effectively over-reached regarding that
which \emph{cannot} exist in reality. To model a real Pioneer
anomaly, asymmetry and energy conservation (shall) play a crucial
role in linking together the microscopic (i.e. QM) and macroscopic
(i.e. gravitational) realms/domains. Without this approach, it is
hard to conceive how any meaningful physical linkage would
otherwise be attained. In the language of philosophers of science,
(herein) we are pursuing a (mainly) constructive theory (and
model), rather than a principle theory approach.
%**************************************************************************************
\subsection{Summary and concluding remarks regarding SR and GR}
This section provides an overview of conclusions reached regarding
our new stance concerning special and general relativity, as well
as an encapsulation of the model to be presented (and further
developed).

\subsubsection{Incompleteness cf. incorrectness}
Let it be said that in no way is either SR and/or GR's approach to
`spacetime' and physical phenomena seen to be \emph{incorrect}, as
regards relevant observational evidence. Our concern is whether
they provide a conceptually \emph{complete} theory of
``gravitation", in the widest sense of the word. By way of issues
surrounding QM non-locality, and the assumed existence of a real
Pioneer anomaly\footnote{Not to mention that: ``no one understands
why empty space should have [dark] energy \citep[
p.83]{Krauss_04}." Note that theory based upon vacuum energy
(seeking to explain dark energy) differs from measurements by
about 120 orders of magnitude, and thus this cannot be a valid
hypothesis.}, there is sufficient justification for hypothesising
an ontological extension to our (singular) understanding of what
space and time `are'; and how they relate to, and coexist with,
each other.

\subsubsection{Aspects of our new approach to \mbox{Special} Relativity and
Gravitation}\label{subsubsection:Eddington} In the spirit of (Sir)
Arthur Stanley Eddington we saw (he saw) (QMs and) GR ``\ldots as
fundamentally `epistemological' in character, meaning that they
provided insight into \emph{how} we see the world, rather than
\emph{what} the world is \citep*[ p.37]{Stanley_05}." Thus, in
line with certain philosophers of physics, general covariance
(GCoV) is regarded as simply a new mathematical technique, not an
expression of physical content.

In the spirit of Vladimir A. Fock, who sought to develop a new,
non-local, point of view of Einstein's theory \citep[
Preface]{Fock_59}, we recognise the need (in some cases) for a
global perspective upon gravitation. Further, herein we share
Fock's preference for viewing SR as a ``Theory of Invariance" and
GR as a ``Relativistic Theory of Gravitation" --- in order to
guide our supplementation of gravitation --- but his wish to have
privileged systems of coordinates (and `recast' Einstein's theory)
is not embraced.

\subsubsection{Benefits of a new supplementary conception of space and time}
To appease a (real) Pioneer anomaly, we are required to accept the
``dual-ontology" outlined in this Section. This dual-ontology involves
appreciating the `coexistence' of \emph{both} a relative
(observational), and a global (hidden background), conception of space
and time. The former is constrained to `working with' spacetime,
whereas the latter allows space and time to be treated as physically
separate notions. The latter conception, by way of a new appreciation
of universal sequential evolution (that cannot be directly appreciated
by way of observations), is consistent with QM non-locality.

This `other' global (hidden background) approach/perspective,
posits a type of hidden cosmological substratum; i.e. a space
continuum that is somewhat analogous to a fluid mechanical
molecular mass continuum. The nature of global (sequential)
evolution is such that an effective (hidden background) time
simultaneity exists; but observationally, clocks will run at
different rates depending on their local
circumstances\footnote{Note that a hidden variable background time
(itself) is \emph{not} being proposed; only a hidden cosmological
(sequential existence) go-pause-go-pause-... \emph{simultaneity}
is being proposed (recall subsection
\ref{subsubsection:non-locality}).}.

Further, we need to allow the standard notions of emptiness and
zero specific energy to coexist with their `opposites', i.e.
fullness and maximum specific energy (set at $c^2$). The latter is
an (everywhere) uniform and (`everywhen') unchanging background
potential (specific) energy field. We then need to hypothesise
that the magnitude of this `field' is `complex', in it has both
real and imaginary components. A ramification of this
supplementary conception, is that SR's Lorentz transformations can
be seen as arising from relative motion introducing an imaginary
component of specific energy, so as to reduce the availability of
space and time to observations. Observationally, this (\emph{loss}
of availability) \emph{appears} as time dilation, length
contraction, and mass dilation. We may then think of this as
involving a ``drawing-down" upon the uniform (hidden background)
specific energy field (or potential hill).

The `subtlety' of GR, is seen to arise from multiple draw-downs
coexisting (e.g. via mass and the motion of mass); hence the
(somewhat restrictive) requirement of general covariance in its
formalism. We conjectured (subsection \ref{subsubsection:SR's
ontological commit}) that: GR's need for general covariance (GCoV)
is a direct consequence of (further) adding gravitational effects
to the (already additional) conformal structure demanded by SR's
Lorentz symmetry. From our new (or complementary) perspective we
may alternatively recognise an intrinsic (cf. extrinsic) imaginary
(number based) space `dimension' into which space `curves' or
deforms. Although this new conceptualisation assists our
understanding, where observations are concerned, we must almost
always necessarily defer to standard GR. The Pioneer anomaly and
its explanation is the exception to this rule.

The consequence of this supplementary (noumenal cf. phenomenal)
conceptualisation is that a different type of `gravitational'
field, as compared to any describable within GR, can exist.
Subsection \ref{subsubsection:preview mechanism} briefly outlines
how this field arises from a different `level' of matter; i.e.
mass `within' (a great many) atoms/molecules --- subjected to
celestial closed loop/path orbital (and spin) motion --- cf.
macroscopic mass/matter (itself), its motion, and energy.

\subsubsection{Summary remarks and distinguishing the new approach to
`gravitation'}\label{subsubsection:distinguising Grav} A reduction
and unification agenda, Noether's theorem, GR's principle of GCoV,
SR's non-simultaneity, and the lack of local energy conservation
in GR, are seen to \emph{not} stand in the way of the model being
presented; if the reader is prepared to relax their standard
``concept package" of reality --- (so as) to embrace a conceivable
alternative. If one's standard ``concept package" is not relaxed,
then one can only have a high degree of scepticism regarding the
real (`gravitational') existence of the Pioneer anomaly. The model
requires the `universality' of energy conservation to be
applicable/restored (in its specific circumstances), and the
importance of an exclusively `principle'-based approach in
\emph{macroscopic} physics to be strongly downplayed.

The new approach being pursued, via the awkward Pioneer
observational evidence, involves/includes:
\begin{enumerate} \item{a macroscopic constructive theory, that arises
 in response to a microscopic asymmetry,}
\item{an energy basis, and a circulatory field effect,}
\item{continuum mechanics (involving space \& non-local quantum
mechanical mass --- separately),} \item{a richer ontology, and
acceptance of hidden variables --- albeit at the level of the full
system, and} \item{recognition of two quite distinct, yet
coexistent, physical realms; i.e. the micro- and macroscopic.}
\end{enumerate} This stands in stark contrast to the general
thrust (standard approach) of contemporary physics which
utilises/employs (respectively):
\begin{enumerate} \item{a principle based approach to theory (including an emphasis upon symmetry),}
\item{particle exchange as being responsible for force,}
\item{Lagrangian and Hamiltonian based formalisms,} \item{an
``operationalist" approach to the world, and} \item{a four force
unification agenda (3 micro+GR).}
\end{enumerate}
This write-up doesn't seek to discredit Special or General Relativity,
nor disparage the three force ``standard model"; rather, it seeks to
supplement and complement this approach --- which has failed, for more
than 25 years, to make \emph{substantial} `scientific' (cf.
mathematical) progress towards its objective of unifying the three
force quantum mechanical microscopic world and General Relativity.

\subsubsection{Previewing the mechanism that indirectly explains the Pioneer anomaly}
\label{subsubsection:preview mechanism} The ramifications of our
(noumenal plus phenomenal\footnote{In a scientific observational
sense.}) dual-ontological stance upon energy conservation was
examined in section \ref{subsection:Noether}. We saw that the
global approach, of the supplementary/complementary ontology, also
allows conservation of energy to regain its ascendency --- at
least with regard to the unique circumstances of the (new)
mechanism/model.

The model involves geodesic motion in curved spacetime influencing
the (spin-based) geometric phase of (all elementary fermion/matter
particles within) an atomic/molecular quantum mechanical system
(see Section \ref{Section:general model}). This geometric phase
influence is (effectively) common to (a great many) QM
\emph{systems} i.e. atoms and molecules within a `suitable'
celestial body.  This suitability requires the celestial body to
be a (predominantly solid-body) moon in spin-orbit coupled motion
around is host planet, which in turn orbits the Sun --- i.e. three
celestial bodies are involved in the model. Also involved are
inertial effects pertaining to QM intrinsic spin. The relative
(spin) geometric phase change `results' in a not insignificant
(non-inertial) QM intrinsic angular momentum (rate), or (spin)
energy. It is `non-inertial' because atoms/molecules are dominated
by the electromagnetic force, and because of the `internal'
spin-orbit coupling of atoms/molecules the geometric phase
change/offset is overridden/denied. A `\emph{rate} of angular
momentum' is involved, because self-interference of the
atoms/molecules per completed orbital/cyclic loop is a feature of
the mechanism to be proposed.

Only when this (`internal', relative, and non-inertial) energy
offset is virtual, i.e. existing below a decoherence `trigger' at
the first or a minimum QM energy level, does the situation result
in a non-local \emph{external} re-expression of the (same) energy
magnitude. The re-expression of the total virtual internal energy
($\Delta E$) as a real external energy is a one-way process. It
relies on energy being a scalar quantity, and having a
multi-faceted `nature'. Scalar energy is the common `currency' in
the `mediation' of this rather complicated physical phenomenon;
with `universal' energy conservation \emph{the} `governing'
principle.

The external re-expression of (virtual QM) energy expresses itself
as a neo-classical-like rotating and asymmetric perturbation of
space, that `piggy-backs' upon and distorts the pre-existing
gravitational field (as determined by GR). This perturbation takes
the from of a universal (or cosmological) size/scale rotating
space-warp (RSW) (introduced/discussed in Section
\ref{Section:PrelimModel}), whose energy --- upon completion of
one full rotation of the RSW --- is also $\Delta E$. Associated
with this (now non-virtual) energy is both: a non-local mass
distribution, and a specific energy distribution ($\Delta e$). The
latter, and an associated acceleration amplitude ($\Delta a$), is
the same everywhere (i.e. cosmologically) throughout the field. It
is the (distributed) non-local mass, at a point in the field, that
reduces as we move away from the RSW's source; and as such $\Delta
E$ is `dissipated' away from its source (atoms and molecules).

The representation and conceptualisation of this effect relies on
the dual-ontology previously mentioned. The existence of a hidden
background temporal coordination (or simultaneity), and a global
background space continuum, together with energy as drawing-down
from a maximum (reservoir) value, permits the representation of
the mechanism to be expressed in a `neo-classical' mathematical
formalism (that is separate from GR's formalism). Subsequently,
use of a metric shall \emph{not} be required.

The model is comprehensively quantified in \mbox{Section
\ref{Section:Quantif Model}.} Several of these RSWs (coexist and)
combine/superposition to produce the Pioneer anomaly. Preempting
this, (for each RSW) the (virtual) QM energy ($\Delta E$) is
proportional to a rotational rate [$\Delta t^{-1}$], multiplied by
the reduced Planck constant ($\hbar$) and the number of
atoms/molecules ($N_m$). The rotating space-warp's energy (also
$\Delta E$) is proportional to (half) the warp/wave amplitude
squared, i.e. acceleration squared. Later we shall see that
$\Delta E=\frac{1}{2}m_1^*\Delta a^2 \Delta t^2$ where $\Delta t$
is the duration of a single rotation, and $m_1^*$ is a non-local
mass encompassing a given (initial) volume.

In Section \ref{Section:general model} it shall become clear that
only a minimal geometric phase shift can lead to the
\emph{virtual} QM energy associated with the mechanism, and this
is dependent upon a number of things; one of which is the
existence of the \emph{weak} gravitational fields typical of our
solar system --- and almost certainly other similar solar systems.
Subsequently, the neo-classical approach employed herein is not
compromised by circumstances pertaining to very strong
gravitational fields.
%****************************************************************************************************************
%**************************************************************************************************************
\section{Outlining quantum \& general aspects of the model}
\label{Section:general model}In this Section the quantum
mechanical (QM) aspects of the model are primarily outlined. Each
section steps up the level of detail in which the model is
presented. Due to the length of this Section, the economical
reader may prefer to merely read the summary at the end of each
section --- \ref{Subsection:Macro/micro} to
\ref{Subsection:Condensate}. The main aim of this Section is
discussed in section \ref{subsubsection:Chiao's non-locality}.

\subsection{Revising and extending the relationships between the microscopic
and macroscopic `realms'} \label{Subsection:Macro/micro} Drawing
upon the alternative ontology presented in Section
\ref{Section:PhiloTheory}, this section
(\ref{Subsection:Macro/micro}) seeks to elaborate upon the
differences between the microscopic and macroscopic realms
previously discussed in subsection \ref{Subsubsection:MM
Differences}. It then highlights various ramifications ensuing
from this. The discussion is at times general and at other times
model specific. This is an important step in our attempt to
establish a second `means' (or manner) by which non-Euclidean
geometry is achieved. Preceding this we declare a major aim of
this paper/treatise.

\subsubsection{Main aim, and non-locality and locality at non-extreme
temperatures}\label{subsubsection:Chiao's non-locality} In very
general terms, the main aim of this Section and the paper is to
(indirectly) relate curved spacetime (rather than GR \emph{per
se}) to a quantum mechanical system's `state' at non-extreme
temperatures/energies. Of great influence in this regard has been
Raymond Chiao \citep{Chiao_04}. His, and our, primary concern is
the apparent tension between QMs' non-locality and GR's locality.
He addresses this issue in conjunction with Berry's geometric
phase (see section \ref{Subsection:Berry}). Note that `locality'
implies both: a particle basis, \emph{and} communication being
mitigated at (or below) the speed of light. The ramifications of
questioning a particle basis are investigated in this section,
i.e. \mbox{section \ref{Subsection:Macro/micro}.}

When examining an internal-QM to external-GR physical
relationship, Raymond Chiao and others examine either sub-atomic
particles or electromagnetic radiation's photons. In contrast, our
interest shall involve (wave-like) angular momentum aspects of the
internal `motion' within \emph{atoms and molecules}. Thus, the
Dirac equation in `flat' or curved spacetime shall not concern us.
Of interest shall be: maximum QM energy uncertainty, a minimum
difference in QM energy levels; and (quantitatively) Dirac's
constant (i.e. Planck's constant $\div ~2 \pi=h/2 \pi=\hbar$)
shall be important.

\subsubsection{Altering the perceived relationship between gravitation and
EM}\label{subsubsection:E/M vs Grav} Some close parallels are
\emph{assumed} to exist between gravitation and electromagnetism
(EM), e.g. the inverse square law. The dual (wave and particle)
nature of electromagnetic radiation/photons spans the micro-
\emph{and} macroscopic realms. Note that mechanistically, EM
exhibits quite distinct behaviour in the two realms: for example,
the role of EM in an electric motor is very distinct from its role
in an atom. To compliment the massless photon, the existence of a
(massless) graviton particle has been confidently postulated, but
there is a subtle difference between EM and gravitation.

In EM the force between two charges doesn't depend upon a third
charge, whereas with gravitational influences this is not the
case. If we deny the existence of a graviton particle and hence a
\emph{gauge theory} approach to gravitation, then a purely
particle based (gravitational) momentum to energy link flounders,
because graviton momentum can no longer be defined.
Unconventionally, we shall allow energy's manner of existence to
not be solely restricted to a \emph{local} particle (and force)
basis. As such, the \emph{energy} of a rotating space-warp can be
associated with both: the \emph{specific energy} of a rotating
space-warp/space-deformation, together with a (new)
\emph{non-local} distribution of \emph{mass} --- with the latter
elaborated upon in section \ref{Subsection:warp's mass}.

This is consistent with our previous denial of ``mass in SR"
always being secondary to energy, at the \emph{macroscopic} level;
which ensured the subsequent revival of the notion of relativistic
mass (see subsections \ref{subsubsection:relativistic mass} and
\ref{subsubsection:two timing}). In contrast,
\emph{microscopically} the relationship between the energy and the
momentum of a photon is simply mitigated by the speed of light.
Denying a (physical) reduction to the graviton (particle), in turn
denies both: that mass is \emph{always} secondary to energy, and
(the validity of) a gauge theory approach to gravitation. In place
of a particle emphasis, gravitation --- in the sense of a
gravitational acceleration (that can act upon a body) --- is given
a geometric emphasis; and subsequently, it is quite different to,
and distinct from, the other (three microscopic) `forces'.

\subsubsection{The size of `microscopic systems', and mass \& energy
by number}\label{subsubsection:by number} Our criteria for a
macro-micro (theoretical) dividing line is taken to be: whether
the size of a micro/QM object is determined by its QM wavelike
nature --- a `boundary' condition for the `microscopic' QM realm
one might say. For electrons and atoms, i.e. an elementary
particle and a QM bound system (respectively), this \emph{is} the
case\footnote{The wavelength of an electron `inside' an atom is
equal to the \emph{size} of an atom. The wavelength of the neutron
gives the nucleus its size. The wavelength of a quark inside a
proton is the same order of magnitude as the size of a proton
\citep[ p.138]{Rohlf_94}.}. Subsequently, the largest microscopic
systems are necessarily atoms and molecules\footnote{The molecules
may indeed be very very large.}.

\begin{quote} If in some cataclysm, all of scientific knowledge were to be
destroyed, and only one sentence passed on to the next generation
of creatures, what statement would contain the most information in
the fewest words? I believe it is \ldots that \emph{all
[macroscopic] things are made of atoms} \ldots Richard P. Feynman
\citep[ p.1]{Rohlf_94} \end{quote} We emphasise the somewhat
neglected idea that bulk mass (or large mass) is effectively a sum
of its constituent atoms/molecules. Loosely speaking, we have
``mass by number". Microscopically, mass is secondary to energy.
These two points together exhibit the possibility that an excess
``virtual energy by number" may arise. This notion was introduced
in subsection \ref{subsubsection:Large number}, and herein it
shall be applied with regard to an energy uncertainty upper limit
associated with the atoms/molecules `constituting' a celestial
three-body spin and orbital system.

Additionally, when discussing the macroscopic in terms of the
microscopic, we shall confine ourselves to the largest microscopic
systems: i.e. atoms/molecules. They represent the total angular
momentum of all smaller constituent particles (or lesser
elementary fermion wave aspects) --- as discussed in the following
quote.
\begin{quote} Formally [i.e. the equations involved], however,
there is no difference between the spin of a single particle and
the total angular momentum of any system regarded as a whole,
neglecting its internal structure. It is therefore evident that
the transformation properties of spinors apply equally to the
behaviour, with respect to rotations in space, of the wave
functions $\psi_{jm}$ of any particle or system of particles with
total angular momentum $j$, independent of whether orbital or spin
angular momentum is concerned \citep*[ p.200]{Landau_65}.
\end{quote}

\subsubsection{A mass emphasis as compared to an electric and magnetic
emphasis}\label{subsubsection:mass emphasis} It is the mass,
inertial, and mechanical aspects of an atom/molecule's angular
momentum that shall be our primary concern herein, even though
atomic/molecular angular momentum is dominated by electromagnetic
fields and the electrical charge of its constituent particles.

Importantly (re: the model being pursued), it shall be the
(indirect) effect of an \emph{external field}, gravitational cf.
electric or magnetic, that shall concern us; in particular, its
(non-negligible) affect upon spin-orbital angular momentum
coupling within an atom/molecule --- in certain `third'
(celestial) body circumstances.

Note that the (internal) position of the centre of mass of
atoms/molecules is effectively fixed cf. the distribution of
electrical charge; and (recalling subsection
\ref{Subsubsection:Equiv and Uncert}) if the correspondence
principle holds, then this centre of mass moves (according to
Ehrenfest's theorem) along a classical trajectory. Further, as a
property of atoms/molecules, the mass interaction of neighbouring
atoms/molecules is negligible cf. the electrical interactions of
(outer) electrons in atoms/molecules with their neighbours.

Regarding QMs and the behaviour of atoms/molecules in macroscopic
bodies (e.g. metals, and chemistry), it is just the outer electron
effects that are predominantly significant; whereas, the inertial
effects related to curved spacetime (proposed herein) involve
\emph{all} the mass of each atom/molecule --- in an `appropriate'
macroscopic (celestial `third') body.

Unlike the case with electromagnetic aspects of atoms/molecules,
each atom's/molecule's mass, and (as we shall see) virtual spin,
is treated as \emph{distinct} from its neighbouring
atoms/molecules. Thus, if the (energy) influence upon a collection
of neighbouring atom/molecules is the same --- e.g. via motion in
a gravitational field --- the total energy may then be described
by way of a simple \emph{addition} of the individual
atomic/molecular contributions. Loosely speaking, this is an
``energy by number" situation; i.e. energy is proportional to the
number of atoms/molecules involved.

\subsubsection{A point of order re: QMs, classical physics and
macroscopic systems}\label{subsubsection:Vlatko Vedral}
Elaborating upon comments made in a footnote of subsection
\ref{subsubsection:Forward_reach}, we note --- by way of
paraphrasing \citet[ p.40]{Vlatko_11} --- that the criteria for
treating a macroscopic system as purely classical depends upon the
way quantum systems interact with one another, rather than merely
depending upon the \emph{size} of the system itself. Furthermore:
\begin{quote} Larger things tend to be more susceptible to
decoherence than smaller ones, which justifies why physicists can
usually get away with regarding quantum mechanics as a theory of
the microworld. But in many cases, the information leakage can be
slowed or stopped, \ldots \,\,[In summary, the] division between
the quantum and classical worlds appears not to be fundamental
\ldots [and as such] quantum mechanics applies on all scales
\citep[ p.41,43]{Vlatko_11}.\end{quote} Two comments ensue:
firstly, information leakage in the model is `thwarted' by way of
the influence (we have proposed) being so small that it cannot
register its presence within the digital/quantum system itself;
effectively, it is a (proper) fraction of any minimal quantum
change. Secondly, there \emph{is} a clear division between solely
quantum-based and solely classical-based
\emph{descriptions}/models of the world; the challenge herein is
incorporating features of both descriptions (and worlds/domains)
into a single model.

Finally, experiments now leave very little room for the following
proposals to operate: (1) [Roger Penrose in the 1980s suggested]
``that gravity might cause QMs to give way to classical physics
for objects more massive than 20 micrograms, \ldots \,\,[or (2)]
\ldots that large numbers of particles spontaneously behave
classically \citep[ p.43]{Vlatko_11}."

\subsubsection{Quantum probability density and \mbox{acceleration/
gravitational} field undulations --- single particle case}
Celestial bodies in stable orbits move along geodesics, and are
\emph{not} acted upon by gravitational \emph{forces} (\emph{per
se}). It is worth mentioning that (via the time independent
Schr\"{o}dinger equation) for a (solitary) \emph{free} `particle'
travelling at a constant speed\footnote{Note that we are dealing
with the \emph{zero potential}; and `particle' also refers to
atoms and molecules.}, as atoms/molecules `slaved' to a moon
roughly/effectively do; (then) regarding the wavefunction, the
probability density has a \emph{constant amplitude} (for all x and
t)\footnote{The unobserved (single) particle is equally likely to
be found anywhere on the geodesic (i.e. a straight line in curved
spacetime) that forms a closed orbit. If this were the case, the
momentum (and energy) are known with complete precision.}. In our
case the ``free particle(s)" we consider are the constituent
atoms/molecules of a bulk mass in celestial motion.

This subsection's (hypothetical and idealised) examination of a
single free particle essentially disregards all but one
atom/molecule in a celestial body; and thus it is inappropriate
when/where \emph{real} quantum mechanical interactions upon or
between atoms/molecules occur. For a \emph{virtual} effect, common
to all atoms/molecules in a celestial body, the following
discussion could (conceivably) be valid.

Using a large number of wavefunctions to generate a group of
travelling waves, the wave `packet' \emph{spreads out} from its
(original) localised position. Similarly, and in addition to the
rotating space-warp's acceleration amplitude remaining constant
\emph{out} to `infinity', the space-warp's (non-local) mass at a
point in space reduces with the spherical volume enclosed around a
moon --- going to zero at (mathematical) `infinity' (see section
\ref{Subsection:warp's mass}). Thus, the nature of the
hypothesised rotating space-warp's associated (non-local) mass
\mbox{distribution} is not totally dissimilar to the quantum
mechanical (probability) wave function of a \emph{single} free
particle. We also note that interference effects are possible, as
are self-interference effects --- with the latter being especially
relevant to our model.

\subsubsection{De Broglie matter waves, decoherence, and reduction} By advocating a
standard two realm non-reductive approach, we also necessarily question
the nature of the assumed existence of \mbox{de Broglie} matter waves
for \emph{whole macroscopic} objects, e.g. a pitched baseball. We cite
a Paul Davies thought experiment, concerning a perceived quantum
violation of the equivalence principle, involving two perfectly elastic
balls; one made of rubber and one made of steel, dropped from the same
height.

\begin{quote} \ldots quantum mechanically, there will be the phenomenon of
tunneling, in which the two balls can penetrate into the
classically forbidden region \emph{above} their turning points.
The extra time spent by the balls in the classically forbidden
region due to tunneling will depend on their mass (and thus on
their composition) \citep[ p.268]{Chiao_04}. \end{quote} A
reductive assumption underlies the discussion of this concern.
Essentially by way of decoherence, we might alternatively conclude
that a discussion of quantum tunneling effects regarding
macroscopic objects (at room temperature) is invalid. This
alternative stance, and Paul Davies' perceived concern, supports
the contention that the microscopic and the macroscopic are quite
distinct realms/`worlds', with distinct behaviour\footnote{N.
Bohr, W. Heisenberg, and J. von Neumann all argued for a strong
distinction between the quantum realm and the classical world.}.
This is \emph{not} to say they cannot both be involved in a
phenomenon; rather, they may (herein) coexist together but not in
a ``unified" and reductive manner --- at least as regards the
`mass era' of the universe (cf. the hot `radiation era').

Decoherence, as an approach to reality is generally loyal to a
perceived reductive agenda; in that ``behind the scenes" a
macroscopic body is considered to still possess a matter wave
field associated with the \emph{whole} object cf. a macroscopic
body being essentially representative of the sum of its (QM)
parts. This unverified (whole body) conjecture is denied by our
non-reductive micro-macro phenomenal stance.

The alternative is that, there is a (dual) coexistence, with the
domain of independent matter waves only extending to the largest
QM systems, with these being atoms/molecules. Thus, in a
macroscopic body, and by way of decoherence we \emph{may} still
retain latent QM features; but there is now a huge multitude of
`individual' matter waves superpositioned ``behind the scenes".
Thus, the whole object's matter waves exist simply by way of the
sum/superposition of its parts --- which may in turn involve
further (and unknown) processes at the level of the whole object
(i.e. emergent processes).

Finally, regarding decoherence, we shall not need to employ
concepts such as the density matrix, or mixed states; nor shall
Schr\"{o}dinger's equation concern us --- the latter because we
are not concerned with \emph{finding} the allowed energy levels of
QM systems (especially atoms/molecules), nor determining the
positional probabilities of particles. Our emphasis lies solely
with the intrinsic angular momentum (spin vs. orbital) aspects of
microscopic (i.e. QM) matter, subjected to a common
\emph{external} effect; that then (under very special
circumstances) expresses itself (also) externally to the QM system
(or systems) concerned.

\subsubsection{A system comprising four levels of matter and
the system as a whole} Due to the different physical: phenomena,
laws, and principles involved, a `complete' system relevant to our
model involves five different levels of matter and size.

Note that the word ``macroscopic" is reserved for use in the sense
of ``not microscopic" --- thus applying at levels three, four and
five. These matter/size levels are:
\begin{enumerate} \item{Elementary fermion particles --- the
smallest \mbox{microscopic} level;} \item{atoms/molecules ---
conceived as a composite of (all) their constituent elementary
fermion (spin-1/2) particles --- the largest microscopic level;}
\item{the (composite) sum of all atoms/molecules, i.e. bulk
(lunar) mass --- the `mesoscopic' level\footnote{Although not the
conventional use of the word, we are avoiding the use of the word
``macroscopic" (here). Additionally, mesoscopic is used to
indicate that certain statistical properties have meaning at this
level.};} \item{three-body celestial motion involving a moon in
spin-orbit tidal lock orbiting a planet, with this moon-planet
pair orbiting a central star/sun --- the celestial (and bulk mass
in motion) level.} \item{Finally, there is the full system to
consider which incorporates \emph{all} four (lower) levels as well
as the background space-time\footnote{Generally, we shall use the
word ``spacetime" when referring to `everyday' or general
relativistic usage, but we shall use ``space-time" when the
model's non-local basis and (supplementary) noumenal/hidden
background ontology needs to be recognized/appreciated. Note that
some quotations (presented in this paper) use ``space-time" rather
than ``spacetime".} `stage' --- (effectively) the universal (and
cosmological) level.}\end{enumerate}

\subsubsection{The transition between the microscopic and macroscopic realms}
The transition between the microscopic and macroscopic realms
remains a contentious (and open) issue (recall subsection
\ref{subsubsection:decoherence}). By altering the
conventional/standard relationship of the microscopic to the
macroscopic world we attain a different understanding of the
demarkation and transition between the two realms.

Bohr's correspondence principle (CP) is seen to apply to
atoms/molecules, but its application is restricted to QMs that
obeys Schr\"{o}dinger's equation --- i.e. quantum (orbital)
mechanics, with the QM spin exempt/excluded. Regarding `orbital'
angular momentum, as in the case of the Rydberg atom where (the
orbital angular momentum's quantum number) $l\rightarrow\infty$;
the behaviour of the quantum mechanical system `reduces' to
classical physics in the limit of large quantum numbers. In
contrast, the spin of a quantum system never `goes to' infinity.
Subsequently, for QM spin the correspondence principle does
\emph{not} apply. This is one reason we are told to \emph{never}
conceptualise QM spin in a classical manner. While this is true
for QM systems on their own, it may not always be heuristically
useful --- especially for a QM system existing in curved
spacetime; (because) we shall argue that (in a rare type of
third-body system) \emph{externally} imposed effects of spacetime
curvature can \emph{attempt} to offset QM spin phase relative to
the phase demanded by EM spin-orbit coupling. This occurs in a
non-discrete (i.e. continuous/analog) manner --- albeit only in
the form of a tiny (sub-quanta) virtual effect.

A conceptual visualisation in the form of a relative over-spin, is
beneficial for our explanation --- albeit not physically
realistic. Also note that a ``virtual" effect is an effect that is
not actually occurring nor is it directly observable, but may
still be physically relevant --- (in our case) at the (external)
universal physical level. Our virtual effect is below the first
energy level --- and/or a minimum change in the (practical) energy
levels --- of a quantum atomic/molecular system, and thus it is
associated with (proper) \emph{fractions} of Planck's constant
($h$), where $h=2\pi\hbar$.

The use of $\hbar\rightarrow0$ to indicate the macroscopic limit,
so that $\hbar l$ is finite as $l\rightarrow \infty$, is
\emph{not} embraced unequivocally. Its use assumes that reduction
is always justified. It appeals to the relative smallness of
$\hbar$. Indeed if a (fermion \emph{spin}) offset effect is common
to a great number of atoms/molecules, the use of the
$\hbar\rightarrow 0$ macroscopic limit is inappropriate. Thus,
this use of $\hbar\rightarrow0$ (i.e. very very small) is only
appropriate for a single atom/molecule. A hypothetical intrinsic
angular momentum discrepancy/offset does not always have a
classical/macroscopic limit analogous to (`interior') orbital
angular momentum. Thus, any blanket denial of spin affecting a
macroscopic system, in the form of a (virtual) angular momentum
offset at the universal/systemic level, appears to be unjustified
--- albeit only in the special (vast number) additive and externally
`driven' circumstances of the model.

\subsubsection{Summary remarks for section \ref{Subsection:Macro/micro}}
This section has outlined how the model is able to restrict itself
to the study of (only) atoms and molecules: as a composite of all
their sub-system fermion elementary particle/wave constituents,
and independent of interactions with neighbouring atoms/molecules.
The model's emphasis (with regard to quantum mechanics) lies with
atomic/molecular inertial spin effects --- via atomic/molecular
geodesic motion in curved spacetime --- rather than the usually
dominant electromagnetic based effects. Thus, standard QM
approaches cannot be applied, e.g. involving Schr\"{o}dinger's
equation.

By way of (further) questioning both the physical/ontological and
theoretical aspects of a scientific reduction agenda, and emphasising
the differences in the micro- and macroscopic realms/domains, it was
argued that these realms coexist primarily in an independent manner
--- at least as far as our formal/representative understanding of
them is concerned. This standpoint influences issues such as: the
transition between the two realms; the superposition of
atomic/molecular effects; and highlights a subtle difference
between the classical limits of QM spin (i.e. intrinsic angular
momentum) vs./cf. orbital angular momentum --- in that QM spin has
no classical limit (nor a classical analogue). Thus, (in
exceptional circumstances) QM spin may be (additively) relevant to
a macroscopic body compromising a great many atoms and molecules;
the model herein shall exploit this case.
%**************************************************************************************************************
\subsection{Berry's geometric phase in \mbox{celestial
systems}}\label{Subsection:Berry}

\subsubsection{Curved space, time and geometric phase}
In Section \ref{Section:PhiloTheory} we examined what may be
termed an ``observational ontological oversight", arrived at by
way of our questioning that ``observations are reality" (without
remainder). Subsections \ref{subsubsection:conceptual
ramifications} and \ref{subsubsection:summary local relativistic}
utilised a `dual-ontology' to reach an alternative/supplementary
interpretation of curved `spacetime'. By placing all the
adjustment for the presence of: mass, momentum, and energy,
relative to an (idealised) empty universe Euclidean space and
time, into an imaginary (number based) space dimension(s) we may
effectively speak of curved space, whilst employing time
simultaneity --- so as to (more generally) appease QM
non-locality. Note that time rates \emph{are} variable (in the
presence of mass, momentum, and energy).

This `meta'-physical approach, involving ``hidden physicality"
(and hence hidden variables), facilitates this section's (i.e.
\ref{Subsection:Berry}'s) conceptualisation and quantitisation of
a relative and virtual (spin) geometric phase (offset) --- for the
motion of atoms/molecules \emph{along a geodesic} in curved
spacetime, cf. motion in `flat' space circumstances. Indeed the
use of the standard mathematical formalisms of both GR and QMs
implies that \emph{no} geometric phase offset, real or virtual,
should exist for motion upon a geodesic --- whatever the
circumstances. The phase offset described herein necessarily
applies only within the third-body of a three-body celestial
system --- relative to spin phases in the rest of the system.
Later we shall see that QM self-interference is involved (see
subsection \ref{subsubsection:A-B comparison}).

Note that \emph{observationally}, QMs, SR, and GR all hold as
before; but regarding the (virtual) geometric phase (GP) offset,
the additive/supplementary use of a curved space and `staccato'
noumenal/background time simultaneity idealisation is our only
option --- if we are to provide an explanation for a `real'
Pioneer anomaly incorporating non-local considerations.

\subsubsection{Background to Berry's geometric phase and anholonomy}
\label{subsubsection:GP background} A quantum system's
wavefunction may not return to its original phase after its
parameters cycle `slowly' around a circuit --- a circumstance
referred to as `anholonomy', or preferably `holonomy' by some
authors. If a stationary state for a QM system is maintained, as
it travels around a circuit, it may acquire a phase shift that is
of a geometric or topological origin
--- which is quite different to the usual dynamical phase changes
associated with Schr\"{o}dinger's equation\footnote{Note that
dynamical phase shift is proportional to time, whereas geometric
phase is path-based and never (directly) dependent upon time.}.
The QM system is said to: \begin{quote}``\ldots record its (path)
history in a deeply geometrical way, whose natural formulation
\ldots involves phase functions hidden in parameter-space regions
which the system has not visited \citep*{Berry_84}." \end{quote}

Note that geometric phase is always \emph{relative}. This is elaborated
upon in subsections \ref{subsubsection:Two faces} and
\ref{subsubsection:Two types}.

There are two types of anholonomies: topological phases (e.g. the
M\"{o}bius strip) and geometric phases --- ``geometric" because
phase shift depends only on the geometry of the circuit in the
parameter space [which is sometimes real physical space] \citep[
p.28]{Berry_88}.

In \citet*{CaiPW_90} the classical nature of Berry's phase for
photons, specifically the rotation of a linearly polarised beam
travelling along a single helically wound optical fibre (or, in
AmE, fiber), is shown to arise from the intrinsic topological
structure of Maxwell's theory --- if the Minkowski (flat)
spacetime is considered as a background. The phase is
(theoretically) developed in the context of fibre-bundle theory.

By way of contrast, we are examining the geometric phase of QM
\emph{intrinsic} angular momentum (i.e. spin) in the context of:
motion in (observational or phenomenal) curved spacetime
--- or equivalently and alternatively, motion in (noumenal time
and) curved space. Note that orbital angular momentum phase is not
affected. Of particular interest shall be:\begin{itemize}
\item{the path of atoms/molecules when `attached' to (i.e. are a
part of) celestial bodies,} \item{the inertial `commitment' of the
QM system,} \item{and the projection of QM spin by way of a
particular direction in space; with this direction being
externally established by way of the \emph{lunar spin axis} or the
plane of lunar orbital motion.}
\end{itemize} For large moons this orbital plane usually has very
low inclination relative to the planet's equatorial
plane\footnote{A planet's equatorial plane and a large moon's
orbital plane are usually very nearly parallel
--- with Neptune's captured moon Triton being an exception to this
(low relative inclination) rule.}, and the lunar spin axis is very
nearly perpendicular to the moon's orbital plane\footnote{One way
in which discrepancies can exist is via a (spin) rotation ``axial
precession" effect e.g. Earth's moon.}, i.e. the obliquities are
very small --- and effectively zero in some cases.

\subsubsection{Finer detail upon the use of Berry's geometric phase
in the model}\label{subsubsection:Berry background} We now turn to
the use of (Berry's) geometric phase in the model being
hypothesised/formulated.

\textbf{Firstly}, we recognise the \emph{global} and cyclic nature
of (relative) geometric phase change (or offset); which implies
the curvature (i.e. non-Euclidean geometry), and possibly
topology, of background space and time is important --- especially
its relationship to the `spin' of QM systems. \begin{quote} \ldots
only for a cyclic evolution the change in geometric phase $\beta$
can be defined invariantly. This makes it appear that $\beta$ is
global in the sense that it can be unambiguously defined only for
cyclic evolution \citep*[ p.1868]{Anandan_88}.
\end{quote} Three comments ensue: \begin{enumerate}[a)]
\item{The global (rather than local) origin of $\beta$ implies a
global phase `datum', relative to which $\beta$ can be
referenced\footnote{Certainly, for EM effects in a laboratory
curved spacetime is usually neglected.}.} \item{A cyclic process
conceivably establishes a (virtual) spin \emph{energy} --- in the
sense of a rate of angular momentum. With regard to the new
mechanism (under development), the external expression of energy
associated with an internal energy (of equal/common magnitude)
indicates the effect is non-cumulative; i.e. the phase offset and
virtual energy are not carried over to the next cycle.} \item{In a
two-body gravitational system the ``reduced mass" greatly
simplifies the orbital description, so that the two-body problem
can be solved as if it were a one-body problem. In any solely
two-body situation the (new) geometric phase change is seen to be
absent.}
\end{enumerate} \emph{Only} with a three-body (gravitationally
bound) celestial system\footnote{Asteroids occasionally have
satellites, e.g. Ida and its moon Dactyl; but a naturally
occurring stable four-body (gravitationally) bound system remains
unobserved. An artificial satellite in the form of a spacecraft,
orbiting our moon, effectively creates a four-body bound system.},
with its multiple types/sources of spacetime curvature, can (a
virtual) relative phase offset occur; with this offset being for a
moon's constituent atoms/molecules
--- and (equally for) all their constituent elementary
fermion/matter particles/waves --- relative to the rest of the
system. Note that the model \emph{also} requires celestial
spin-orbit resonance (i.e. phase-lock or synchronous motion) for a
moon in relation to its host planet. The importance of this is
discussed in subsection \ref{subsubsection:Two faces}. The host
planet (and Sun) remain unaffected, thus effectively forming a
`datum' of no atomic/molecular geometric (spin) phase change.
Later we see (subsection \ref{subsubsection:Celestial geometry
and}) that the (particular case of the) Sun-Earth-Moon system
fails to be geometrically suitable, and thus does not
produce/`coexist' with a rotating space-warp.

\textbf{Secondly}, the external `impetus' to an internal (virtual)
change in geometric phase (i.e. $\beta$) requires a relationship
between celestial kinematics and QM geometric phase be pursued.
\begin{quote} The essential difference between our [Anandan \&
Aharonov] approach and Berry's approach is that we regard $\beta$
as a geometric phase associated with the motion [i.e. kinematics]
of the state of the quantum system and not with the motion of the
Hamiltonian as Berry did \citep[ p.1864]{Anandan_88}. \end{quote}
In our case the motion of atoms/molecules within (and `slaved' to)
a celestial moon is examined. With the model requiring celestial
spin-orbit resonance, the lunar-constituent atoms/molecules
undergo both (classical) spin rotational motion \emph{and}
`linear'/geodesic motion (in curved spacetime) --- at the same
`time'.

It shall be seen that (pure/Euclidean) geometry also plays an
important role in determining geometric phase in the
model\footnote{We shall use the expression ``geometric phase",
rather than ``Berry's phase", to distinguish it from the latter
which is often associated with the motion of the Hamiltonian (in
parameter space).}; in conjunction with the `geometry' associated
with spacetime curvature. Finally, we note that the celestial
motion of atoms/molecules (within a large/bulk mass) is both:
adiabatic, and ``force-free" by way of being motion `upon' a
geodesic. Thus, within the model, dynamical phase concerns may be
(and are) neglected.

\textbf{Thirdly}, in certain EM two-body circumstances:
\begin{quote} \ldots geometric phase depends only on C [the closed
curve] and is independent of how the evolution around C takes
place \citep[ p.1866]{Anandan_88}. \end{quote} In our three-body
moon-planet-sun system, two distinct ``closed curves" can be
recognised: i.e. a planet around its sun, and a moon around its
host planet\footnote{Since Jupiter and Saturn's moons dominate the
model, the larger body's core is essentially the centre of
rotation.}. In both cases the orbital paths \emph{are} dictated by
gravity (and kinematics). In the model we consider one full
(sidereal) orbital cycle of a moon as the primary closed curve.
Later (section \ref{subsubsection:three-body phase}) we propose
that $\beta$ (the closed curve geometric phase change) is
dependent upon \emph{both} the moon's closed path and the planet's
angular rotation/progression around the Sun in this time
--- which is substantially less than a ($2\pi$ rad) planetary year.

Note that the value of $\beta$ is only \emph{indirectly} based
upon the mass and speed of the bodies concerned. Thus, once the
(celestial) orbital features are in place, it is only the paths
that are important; and the basis for our empirical
\emph{determination} of (quantum mechanical) geometric phase
change/offset can be considered solely ``geometric" --- i.e.
closed curves, arcs, lengths and angles. Loosely speaking,
celestial kinematics and dynamics establish a `path' geometry,
which then establishes the (spin) phase offset of lunar
atoms/molecules at the end of a closed (orbital) circuit relative
to the spin phase at the beginning of the (moon's) closed
loop/circuit. One again we note that \emph{orbital angular
momentum phase is unaffected}. That a geometric/topological
\emph{quantum mechanical} fermion wave phase effect should have a
purely geometric \emph{celestial} basis is initially surprising
but not unreasonable --- especially since this
(classical/Euclidean) path-based geometry is `coexistent' with
celestial (`phenomenal') curved spacetime\footnote{Or ``coexistent
with curved \emph{space}", from our alternative
noumenal/background perspective, i.e. the second of our `dual'
perspectives.}.

Outstanding, at this stage is that: it appears we may need to
accredit the background curved space continuum with some
additional (topological) fine detail (see subsection
\ref{subsubsection:foundation}).

\subsubsection{A different role for geometric phase in a `mechanistic'
system}\label{subsubsection:E/M and Berry's phase} Within
atoms/molecules, i.e. (microscopic) QM systems, the
electromagnetic force dominates any gravitational `forces' by a
very large amount; the ratio of the electrical to the
gravitational forces between a proton and an electron is about
$10^{40}$. Subsequently, any physical effects, that are solely
based upon mass cf. mass \emph{and} electric charge, are always
considered negligible. We are considering an exception to this
`rule'. Regarding geometric phase, the presence of curved
spacetime is (currently) seen to merely alter relationships that
are dominated by electromagnetic (EM) concerns.

Herein, we examine a curved spacetime effect upon a
\emph{non-electromagnetic} aspect of an atom or molecule --- hence
this is referred to as a ``mechanistic" approach. We shall take
geometric phase into a domain where its effects are considered to
be non-existent, or at best negligible; with this being: the
non-inertial status of the intrinsic angular momentum of an
autonomous (moon-based) bound QM system translating/propagating
along a geodesic in curved spacetime.

\subsubsection{Geometric phase \& celestial motion}
The effects of spacetime curvature upon the geometric phase of
different, although neighbouring, quantum mechanical (QM) systems
are generally considered to be \emph{all} the same; but on larger
scales, i.e. involving different celestial bodies, this local
idealisation is completely inappropriate. For example, a moon
orbits its host planet at a distance, and hence its path in curved
spacetime is profoundly different to the (curved spacetime) path
of the planet.

At laboratory scales curvature effects differing between QM systems are
imperceptible, and up to moon or planet size are very nearly identical
--- because the paths of neighbouring atoms/molecules are almost
\emph{parallel} in spacetime. In other words, throughout a given
celestial body (particularly moons), the geometric (spin) phase
offset (at the `level' of individual atoms/molecules) is
effectively the same --- essentially because any given moon's
radius is much smaller than the moon-planet semi-major axis.

\subsubsection{The model encapsulated}
The following brief overview of the new mechanism encapsulates much of
what shall be discussed in the rest of Section \ref{Section:general
model}.

For an atom/molecule (within an appropriately configured third
celestial body) moving in curved spacetime, a (relative) virtual
geometric phase offset (per loop duration/time) effectively
`equates' to a (virtual) over-spin of intrinsic angular momentum
per unit loop duration/time; with this being a spin \emph{energy}.
Further, this internal energy offset --- by way of being below the
first/lowest energy level\footnote{Or below a minimum change in
discrete energy levels.} --- is (effectively) shared by every
atom/molecule in an appropriate moon; and thus, upon summation
(throughout a moon) the virtual energy offset `produced' is
significant --- by way of an extremely small number
$(<\frac{1}{2}\hbar \Delta t^{-1})$ being multiplied by an
extremely large number of atoms/molecules.

The multifarious orientations of the elementary fermion particles
within atoms/molecules all `receive' the same \emph{externally
orientated} phase change effect. Thus, the \emph{virtual} phase
effect is effectively coherent. In other words, the magnitude of the
phase offset and the (common/shared) spin axes' orientations are
extrinsically based\footnote{Extrinsic (definition): not part of the
essential nature of something, and coming or operating from outside.}
--- with the latter only `being' possible in the case of a virtual
effect/offset. Note that only minor differences/variations in
phase change occur over the extended body, because the body's
radius is very small compared to the scale of the spacetime motion
and curvature/geometry concerned. We shall assume that all
(third-body) geometric phase changes are equal. Only at and above
the decoherence threshold, i.e. greater than a (fermion)
half-wavelength ($2\pi$) geometric phase offset, could the actual
(i.e. real) multifarious spin orientations be exposed. Below this
threshold, the virtual effect's common application (i.e. same
direction and magnitude for all atoms/molecules) facilitates the
existence of a new type of (externalised/environmental)
`condensate' behaviour --- (that is) not necessarily physically
associated with a constructive superposition of (virtual)
geometric phase offsets in the QM wavefunctions (see subsection
\ref{subsubsection:graviton's absence}).

The appeasement of this virtual energy imbalance, by way of a
\emph{systemic} (i.e. global) conservation of energy, necessitates
a (real) physical expression of energy \emph{external} to the
many\footnote{Of the order of $\sim 10^{50}$ atoms/molecules in a
large solar system moon.} QM systems. This external energy
expression imposes itself as a field upon the pre-existing
gravitational field. This unrecognised ability of nature acts to
always maintain the (internal) stability of atoms/molecules in the
presence of curved spacetime. Importantly, QMs and GR's curved
spacetime are \emph{both} involved in the generation of this new
`gravitational' field effect (i.e. rotating space-warps with an
associated non-local mass distribution), and yet QMs and GR
maintain their \emph{distinct} `character' --- rebuffing most
physicists' expectation that the next step in their relationship
is to be one of unification and reduction.

\subsubsection{Commenting upon two assumptions associated with Berry's
phase} \citet[ p.31]{Berry_88} discusses two assumptions implicit
in his derivation of the geometric phase. \begin{enumerate}
\item{The environmental parameters \emph{governing a system} can
be determined (at least in principle) to arbitrarily high
precision, and that} \item{the environment remains unaffected by
any phase changes it induces in the system.} \end{enumerate}

Regarding the first assumption, the environmental parameters herein
involve the geometric and kinematic characteristics of lunar and
planetary motion; and these \emph{are} known to high precision.

Regarding the second assumption, in contrast to EM instantiations
of Berry's phase, the environment --- in a different sense of the
word --- \emph{is} necessarily affected in the model's case; by an
amount equal to the \emph{virtual} energy offset it induces
`within' a bound QM system (or systems). It is the QM
atomic/molecular system that is unaffected --- as far as a
physically \emph{real} change is concerned. Interestingly, if a
physically real change is induced in the QM system, then
decoherence ensures the environment remains unaffected --- this is
further discussed in section \ref{Subsection:Spin, spinors}.

\subsubsection{Towards coherence and condensate behaviour, in the
graviton's absence}\label{subsubsection:graviton's absence}
Internal (virtual) QM \emph{coherent} behaviour and a (condensate)
external acceleration/gravitational field resultant effect is
being pursued herein; (but) with the photon particle mediating the
EM \emph{force}, and the graviton particle denied, things are
necessarily quite different.

The term ``condensate" is used here in the sense of `singular'
behaviour, but also in the sense of an effect related and common
to (i.e. shared by) a great many atoms/molecules ($N_{m}$). Thus,
an \emph{additive} approach to the quantification of virtual spin
energy is possible. Note that our phenomenon of interest lies
\emph{below} the lowest (i.e. first) energy level of a bound QM
system, whereas a Bose-Einstein condensate occurs \emph{at} the
first energy level; and as such the singular behaviour discussed
is \emph{not} that of a real physical (state of matter)
condensate. Nevertheless, in the model, quantum effects become
apparent on a macroscopic scale.

In QMs, geometric phase (and gauge theory) are usually discussed
in terms of such things as fibre bundles. Importantly, relative
electromagnetic based geometric phase effects are seen to extend
``all the way up" to Maxwell's equations. Thus, geometric phase
effects appear to be related to the \emph{long-range} effects of
classical electromagnetic fields, and hence they are also
conceivably steeped in curved spacetime's large-scale (geometric)
fields/influences. Subsequently, a back-reaction involving a
long-range field response, external to the QM systems concerned,
is not unreasonable.

This new long-range geometric phase-based effect requires us to
not disregard subtle properties of background space and time
\emph{in themselves}. Coherence and condensate-like behaviour are
further discussed in section \ref{Subsection:Condensate}.

\subsubsection{Summary remarks for section \ref{Subsection:Berry}}
Building upon the supplementary ontology discussed in Section
\ref{Section:PhiloTheory}, this subsection introduced the manner
in which (Berry's) geometric phase is used in the model. After a
brief introduction, the very restricted conditions of its
instantiation were outlined, representing a departure from
standard applications of Berry's phase --- which typically involve
neutrons and electromagnetic phenomena.

These restrictions involve: (1) application to the inertial mass
aspects of atoms and molecules within the (lunar) third-body of a
three-body celestial system, \emph{moving} `along' a geodesic in
curved spacetime; (2) a relative phase shift that is restricted to
intrinsic angular momentum (i.e. spin) phases, with orbital
angular momentum-based geometric phase remaining unaffected; and
(3) a dependence upon closed orbital loop motion (and
self-interference).

For relative geometric phase shifts less than half a fermion
wavelength (i.e. $< 2\pi$ rad), the phase `shift' represents a
virtual effect within an atom/molecule. Lunar motion, relative to
its host planet, determines the orientation of this (virtual)
phase shift for all (lunar) atoms/molecules; and since a moon's
diameter is small relative to its semi-major axis, the magnitude
of the phase shift is (also) effectively the same throughout an
appropriate moon. This permits an additive approach to spin
`energy' --- i.e. change of intrinsic angular momentum per closed
loop duration/time ($\Delta t$).

The general acceptance of a linkage between (standard) geometric
phase effects and long-range classical EM effects is seen to
support our hypothesis\footnote{By way of similarity, and hence
analogy.} that: curved spacetime based geometric phase effects can
externally `express' (a \emph{virtual} QM energy) into a (new type
of) long-range acceleration/gravitational field --- albeit only
for circumstances where a minimum (internal) QM energy `level' is
\emph{not} attained and (such that) geometric phase-based
decoherence is not `triggered'.
%***********************************************************************************************
\subsection{Spin, spinors, decoherence and Hannay angle}\label{Subsection:Spin, spinors}

\subsubsection{The model's use of QM spin and spinors} Note that herein
the QM mathematical device of a ``spinor" plays no quantitative
role in the model, but spinors are necessarily a part of the
discussion. With the physical energy involved in QM systems of
primary importance, the concept of ``spin" (i.e. intrinsic angular
momentum) is crucial to the discussion --- particularly in terms
of the physicality of a phase change. Note that a $4\pi$ spin
phase change of a fermion particle results in a return to its
initial state --- all other things being equal (\emph{ceteris
paribus}).

\subsubsection{Fermions, spin, spinors, Hannay \mbox{angle,} and decoherence}
\label{subsubsection:spinors} By utilising both microscopic
\emph{and} macroscopic phenomena (i.e. quantum mechanical and
non-QM gravitational/accelerational field phenomena) in the model,
a \emph{solely} QM approach is clearly not possible. In QMs,
physical quantities are quadratic in their wave functions. A $2
\pi$ change in the wave phase (of fermion particles) was
historically (pre-1967) seen to be merely associated with a sign
change of the wave function
--- with no associated experimental consequences; whereas nowadays
experimental consequences are well-appreciated. Generally, only
interference experiments are considered an appropriate vehicle for
the investigation of geometric phase effects. Our investigation
goes beyond this restriction.

If we restrict discussion of the phase change to $2 \pi$ `jumps',
this suits the reductive/dicrete particle approach of EM, but not
the continuous (i.e. analog) geometric phase changes proposed
herein --- that lie below a minimum energy level\footnote{That is,
below a level where the QM atomic/molecular system's state is
altered, so as to exhibit (local) particle-based consequences ---
that can be observed by experiments. Below such a level,
mathematical consequences, and wave-based interference
consequences cannot be ruled out; nor can our proposed
externalisation of `unresolved' (fractional quantum) energy be
ruled out.}, and are seen to arise from an external analog curved
spacetime effect. Subtleties of spinor sign change, related to the
case of a $2\pi$ rotation of one system \emph{relative} to another
system are briefly discussed in subsection
\ref{subsubsection:foundation}.

In (non-relativistic) QMs, spin states may be simply represented
by a two-component spinor (spin up and down); where ``spinors" are
wave functions of the intrinsic angular momentum of (usually)
elementary particles. Note that spinors are also associated with a
dilation effect, in addition to the rotational aspect.
Additionally, we note that a phase, for the purposes of QMs, is
not a state of matter but a complex number of unit modulus, an
element of group U(1). Herein, we cannot restrict ourselves to a
solely QM basis --- with its mathematical group theory
emphasis\footnote{Not to mention: Clifford algebra, Lie geometry,
etc.} --- but we \emph{can} restrict the quantification (and
conceptualisation) of the model to a solely \emph{classical} basis
--- e.g. energy, momentum, mass, and (something akin to) classical
spin. How this is achieved is outlined in the following paragraph.

The \emph{external} basis/origin of the QM (geometric) phase
change/offset proposed herein acts similarly upon both the up and
down spin components --- of all elementary fermion particles, and
(in equal measure) an atom/molecule as a whole. This simplicity,
regarding how the external effect acts, allows the spin
(geometric) phase offset to be (loosely) described and
conceptualised as a (classical) Hannay angle --- with this being
the mechanical/classical analogue of Berry's geometric phase.
Indeed the intrinsic spin of `quantum' particles should not be
considered classically, but the external origin of the internal
effect lends itself to this simple and useful
\emph{visualisation}. An example of a Hannay angle is the angular
rotation associated with Foucault's pendulum. We can imagine the
model's virtual (phase offset) effect as behaving like a simple
classical ``over-spin" (per loop/cycle). Note that the
\emph{attractive} (only) nature of mass-to-mass gravitational
interactions results in a phase asymmetry that is restricted to an
over-spin circumstance; whereas an under-spin circumstance (its
\emph{repulsive} opposite) cannot exist in a three-body (bound)
\emph{gravitational} system. This over-spin is clarified in
subsection \ref{subsubsection:What lies ahead}.

With our approach using `classical' energy as a bridging quantity
between the two realms, our classical rotation
conceptualisation/visualisation is a doubly useful explanatory
device. Thus, from an external perspective we are effectively
\emph{treating} the atoms/molecules ``mechanistically" --- not
that this is how they are described mathematically within a
(microscopic) QM \emph{system}, e.g. an atom or molecule.

Fermion wave functions have a $4 \pi$ [radian] periodicity with
respect to either: a physical \emph{rotation} (about an axis, and
within a coordinate system), or a rotation of the coordinate
system about the fermionic `system'; and unlike bosons it takes
two full rotations to return them to their original `orientation'.
Fermions that have undergone a $2 \pi$ [radian] rotation around an
axis experience a sign change in their wave function. Note that
this is a \emph{forced} rotation whereas celestial orbital motion
is an \emph{unforced} rotation. Thus, the relationship between
forced rotation and phase does \emph{not} simply `transpose' into
the model (see subsection \ref{subsubsection:three-body phase} for
the actual three-body relationship). In fact, (`unforced')
planetary and lunar orbital motion is generally considered to be
free of any interior QM phase changes; but for (third-body) lunar
motion an unforseen spin phase change/offset is seen to be
indirectly implied by the (inconclusively explained) Pioneer
observations.

This paper proposes that (relative) geometric phase changes can
arise within three-body `unforced' rotational systems. Planetary
and lunar orbits involve (straight line) geodesic motion in curved
spacetime; thus, this situation is quite unlike a rotation in
`flat' space. Nevertheless, it is proposed that the intrinsic
angular momentum of atoms/molecules constituting (i.e. comprising)
a moon are imposed upon by something analogous to a classical
rotation in flat space --- in that a geometric phase change/offset
can arise. Note that the orientation of the spin axis associated
with the \emph{virtual} geometric phase offset is actually
extrinsically determined. We further note that the quantitative
behaviour of the phase shift is not necessarily analogous to that
of a simple fermion rotation in flat spacetime. Decoherence is
seen to occur when $\beta \geq 2\pi$, i.e. when a sign change in
\emph{any} fermion wave function occurs\footnote{Note that with
atoms/molecules in a moon having varied spin orientations, the
first atoms/molecules to `trigger' the macroscopic decoherence
will be those with their spin axis most susceptible to the
external effect --- with this (conceivably) being (either) a spin
axis perpendicular to a moon's orbital plane or parallel to the
moon's spin axis.}. Note that this decoherence is \emph{not}
temperature related; it is (geometric) phase based. In short, an
externally `driven' phase effect is only ``found out" by the QM
system when it `alters' the QM system --- even if no `observables'
have changed.

Previously, we have indicated the model is based upon a situation
\emph{below} a maximum energy uncertainty or below a minimum
(discrete) change in energy levels (which quantitatively are a
function of the reduced Planck constant). We now make this
physically specific by using the relative $2 \pi$ phase advance as
the threshold/dividing line between decoherence and
non-decoherence of the system. Only for $0< \beta < 2 \pi$ can a
rotating space-warp coexist; (and) it is this scenario that we are
specifically examining\footnote{Possibly, for a \emph{single} atom
or molecule things may be different, especially if EM radiation
absorbtion or emission is involved. See the discussion on change
in the fine structure constant in subsection
\ref{subsubsection:Brief list}.}. For any \emph{single}
intra-atomic/molecular $\beta \geq 2\pi$ occurrence the `bulk'
moon is considered to loose (entirely) its ability to `behave' in
an (additive) QM condensate-like manner.

\subsubsection{Physicalising the spinor sign change arising from
a $2\pi$ relative rotation}\label{subsubsection:foundation} In
flat spacetime, a $2\pi$ ($6\pi$, $10\pi$, etc.) [radian] rotation
of either a physical fermion particle/system about \emph{any} axis
(within a coordinate system), or a rotation of the
\emph{coordinate system} (or an observer) itself around a
stationary (spin-1/2) fermion particle or fermionic system, leads
to the sign change of a spinor wave function, but not a change in
state --- with a $4\pi$ ($8\pi$, $12\pi$, etc.) rotation returning
the spinor to its initial sign. Beyond the quite abstract
mathematics associated with spinors and operators, the physical
interpretation of spinor sign change, arising from a physical
rotation, deserves examination.

\citet{Silverman_80} points out that for the spinor sign change to
be physically relevant we need to discuss a \emph{relative}
situation\footnote{If the \emph{whole} (i.e. entire) system is
rotated there can be no experimental consequences.}.
\begin{quote} Only a part of the [whole] system can be rotated; part
must remain fixed to provide a reference against which the
rotation is measured. [Additionally] \ldots experiments revealing
the $2\pi$ phase change of a spinor wave function resemble closely
well-known classical optical phenomena \citep[
p.116]{Silverman_80}. \end{quote}

To develop the model we need to make a distinction (and linkage)
between a classical ($2 \pi$) rotation/revolution and a quantum
mechanical ($2 \pi$) phase change. Our model involves: an inertial
mass-based emphasis, curved spacetime, a three-body celestial
system and a (fermionic system) atomic/molecular geometric phase
change; whereas, as a rule, QM fermion particle/system `rotation'
is dominated by external \emph{electromagnetic}
effects\footnote{With a (rotating) external magnetic field being
the most common means to induce a change, often involving neutron
particles in a beam, e.g. a Larmor precession.}. A $2\pi$
`rotation' in an electromagnetic force-dominated QM system is seen
to be \emph{physically} generated; thus, it is effectively a
`forced' rotation --- whereas celestial (geodesic) motion in
\emph{curved} spacetime is actually an `unforced' (orbital)
`rotation', if we embrace a purely geometric (i.e. curved
spacetime) approach to gravitation.

In both cases the change in a QM sub-system's spinor is
considered: physical\footnote{Note that the model makes real, by
way of an (external) rotating space-warp, the `condensate' QM
\emph{virtual} energy asymmetry.}, externally induced, and
relative. To simply examine a mathematical ($2\pi$) particle or
coordinate rotation upon a (mathematical) spinor overlooks all
this. Indeed, Maxwell's equations of electromagnetism and GR's
gravitation cover the two classes of external macroscopic physical
effects that can conceivably influence the geometric phase of a QM
particle, atom or molecule --- with an atom/molecule having both
electrical (plus magnetic) aspects and (inertial) mass aspects (in
its own right).

It should \emph{not} be readily thought that a spinor sign change
(arising from a $2\pi$ relative rotation) requires background
spacetime to have an additional and mysterious structure that
interacts with (i.e. topologically couples to) QM
spin/spinors\footnote{Even though in Section
\ref{Section:PhiloTheory} of this paper, in order to appease
non-locality, a noumenal/background/hidden `universal substratum'
and a different interpretation of SR have been propounded.}.
Background space is seen to merely support (or act as) a
(physically) relative `stationary' phase or `rotation' reference
datum, but it doesn't need to have a spin structure intrinsically
within itself.

Thus, we have reason to reject (principally by way of Ockham's
razor) the assertion of \citet{Gough_92}, who cites
\citet{Edmonds_73, Edmonds_76}, that a direct linkage between a
$2\pi$ rotation of \emph{coordinates}, in flat spacetime, and a QM
$2\pi$ phase advance is possible --- by way of a new `hidden
variable'/quantum number called ``tumble"\footnote{The idea of a
new quantum number ``tumble", with its apparent similarity to
spin, arose out of expressing the Dirac equation in (complex)
quaternion form \citep{Edmonds_73}. Edmonds explores ways to lift
its degeneracy.}. Although it should be noted that: ``\ldots,
conventional quantum mechanics requires lengthy arguments to
justify the sign change, involving the obscure [yet, highly
regarded] notion of fibre bundles \citep[ p.169]{Gough_92}."

We may downplay, by way of an internal to external distinction,
the concern that \citet[ p.122]{Silverman_80} raises, by way of
Byrne, regarding how to interpret the interference of spatially
separated and rotated\footnote{By way of some electromagnetic
process.} spin-$\frac{1}{2}$ particles in a beam; because recently
proposed interference free measurements may indeed be possible.
\begin{quote} \ldots for fermions, simultaneous observations of the
relative rotations of the spins in the two beams and of the phase
shift are incompatible [since observation of the relative rotation
of neutrons destroys the interference pattern]. [Thus, argues
Byrne:] \ldots the notion of relative rotation ceases to have
meaning as it corresponds to nothing that is measurable
\citep{Byrne_78}. \end{quote} Further, \citet[
p.122]{Silverman_80} notes that with regard to the interference
pattern: we cannot say that we can observe \emph{directly} the
sign reversal of a spinor wave function subjected to a $2\pi$
rotation.

To simply, but not exhaustively, understand the experiment (that
Silverman discusses) we may allow both: an externally generated
physical relative-rotation to act upon QM spin/spinors, and an
QM/internal phase change, to coexist (at least conceptually and
prior to measurement); in order to explain the interference
effects when the two beams are recombined. Clearly, the classical
conceptualisation of QM spin as a rotation remains forbidden, and
it is preferable to avoid implementing the `hidden' background
associated with the quantum number ``tumble".

Although the notion of tumble has its origin in Dirac's equation
(and hence electrical charge) \citep{Edmonds_73}, the
conceptualisation of a phase change (by way of relative motion) is
quite similar to the model's proposed geometric phase offset; with
both necessarily recognising (in different ways) the existence of
a `global' (or systemic) reference frame --- e.g. a barycentric
solar system reference frame. Hence, the additional notion of
tumble retains some appeal, especially in light of the fact that
the hypothesised non-local existence of the rotating space-warps
proposed herein lacks a truly `deep' physical
explanation\footnote{The author confesses to limitations regarding
his understanding of the subtleties of quantum mechanical
`reality'.}.

In summary, it can generally be said that quite elaborate
mathematics is needed to represent the `weirdness' of QM physical
`behaviour', and yet the simple notion of an externally induced
fermion $2\pi$ phase advance, leading to spinor sign change, is
easily appreciated. Externally induced (or forced) rotation and
internal phase change are well suited (if not `natural') partners.
The model's quite different conditions, i.e. exclusively examining
\emph{mass}-based aspects of atoms/molecules undergoing unforced
(closed loop) orbital motion, retains and exploits the simplicity
of this conceptual linkage; now (i.e. in the model) between lunar
celestial (orbital) `\emph{rotation}' --- within a
systemic/barycentric reference frame --- \emph{and} relative
geometric \emph{phase}. Finally, we recall that in the
mechanism/model macroscopic lunar spin-orbit resonance, also known
as `phase-lock', is additionally required.

\subsubsection{Closing and summary remarks for section
\ref{Subsection:Spin, spinors}} The two major issues addressed in
this section were: the `physicality' of a particular type of
(fermion) phase change, and the understanding of what that phase
change might also involve/implicate.

In the model, a classical quantity --- in the form of energy ---
shall be used to link a QM energy within atoms and molecules to a
new type of macroscopic (gravitational) field effect. This QM
(spin) energy is linked to intrinsic angular momentum, which in
turn is linked to a relative spin-based geometric phase offset
($\beta$). Since this phase offset arises by way of geodesic
motion and curved spacetime conditions \emph{external} to the
atoms/molecules (in question), a major simplification in the
treatment of $\beta$ is possible.

Our model is greatly simplified if we \emph{conceptualise} $\beta$
as an externally `driven' classical (Hannay) angular rotation ---
of both the spin up and spin down components. This is a completely
inappropriate representation of what is physically happening at
the QM level, and a rigorous mathematical representation would
preferably employ spinors, but the virtual nature of $\beta$
permits this simplification.

Instead of a rotation in an Earth-based laboratory leading to a
relative (QM) fermion phase change, our concern involves a
celestial (orbital) `rotation' --- acting in unison with a
celestial spin rotation. By establishing a (non-flat spacetime)
barycentric-based reference frame, it can be said that we are
examining the lunar (and planetary) rotations (i.e. orbits) of an
atom/molecule --- i.e. the quantum mechanical sub-systems of bulk
celestial matter. Only the \emph{lunar} (third body) motion is of
significance, in that its atoms and molecules experience a fermion
phase change/offset ($\beta$) relative to the rest of the system
--- i.e. Sun (first body) and the moon's host planet (second
body).

A sign change in any affected wavefunction is seen to trigger full
systemic decoherence, and thus a system characterised by a spinor
would seem necessary. Nevertheless, because our concern is merely
the magnitude of phase change, relative to a decoherence onset
value of $\beta=2 \pi$, and because $\beta$ shall be determined
(later) solely by external (celestial) geometric and kinematic
conditions, a rigorous internal representation of the
atomic/molecular circumstances is \emph{not} required ---
\emph{nor} could it (alone) represent the more elaborate
internal-external (and micro-macro) situation being studied.

Our approach is quite distinct from (standard) interference based
experiments (used) as a means of investigating a relative
geometric phase effect/offset; especially because the third body
in a celestial gravitational system is involved, and a
\emph{virtual} phase offset ($\beta<2 \pi$) (as well as
self-interference) are central to the model. Finally, we note that
the virtual nature of the spin phase offset permits it to be
associated with a quantum mechanical spin axis orientation that is
externally (i.e. extrinsically) determined --- with this
orientation being perpendicular to the plane of lunar orbital
motion, or parallel to the moon's spin axis.
%****************************************************************************************************
\subsection{A new type of precession --- a QM relative spin phase shift}
\label{subsection:Precession}
\subsubsection{Relating the model to other geometric phase based phenomena}
The non-Euclidean `geometry' associated with spacetime (S/T)
curvature is a vital qualitative ingredient behind the geometric
phase offset. By way of a comparison, geometric phase effects in
EM situations (e.g. the Aharonov-Bohm and the Aharonov-Casher
effects) are associated with ``\ldots a `vector-potential'
coupling [that] gives rise to a topological phase
\citep*{Aharonov_00}." We have effectively cut out the
`go-between' coupling, and allow the path of an atom/molecule in
curved spacetime to establish the geometric phase offset ---
albeit in quite special circumstances.

It is useful at this point to distinguish locally applicable
geometrical properties of \emph{the universe}, e.g. curved spacetime,
from (exclusively) global characteristics of the universe, e.g.
topological characteristics (such as shape).

It should also be noted that a gravitational analogue to the EM
Aharonov-Bohm effect is discussed in the literature. This analogue
to the EM situation is \emph{not} what is being pursed herein. The
situation being proposed is quite unique and different (recall
subsection \ref{subsubsection:E/M vs Grav}). The EM Aharonov-Bohm
and the Aharonov-Casher effects have gravitational analogues that
are associated with a classical time delay or lag
\citep{Reznik_95}. The model's new situation does \emph{not} have
an EM analogue; it is thoroughly independent of electromagnetism.
In other words, it is unique to ``geometric" curved spacetime.

\subsubsection{On the omission of global topological aspects from the model}
\label{subsubsection:omission} Although the model accredits the
geometric phase offset ($\beta$) solely to geodesic motion in
curved spacetime (recall subsection \ref{subsubsection:GP
background} on two types of anholonomies), and we shall (later)
quantify $\beta$ merely by way of celestial kinematics and
(Euclidean plane) geometry; (but) a topological aspect associated
with quantum spin (existing throughout the universe) cannot be
ruled out. Indeed, this topological aspect may (and probably will)
form part of a richer explanation/model than the one given herein
--- especially considering the rotating space-warps (to be further
discussed) are cosmological in their size/extent, and universal
conservation of energy and conservation of angular momentum
principles/laws are required by the model.

\subsubsection{Comparison to the Aharonov-Bohm effect}
\label{subsubsection:A-B comparison} The EM based Aharonov-Bohm
(A-B) effect is considered to be the \emph{only} type of
non-locality associated with fermion waves. Herein, a second
non-locality associated with fermion waves is being proposed,
involving the mass/inertial aspect of atoms/molecules. The
comparison is given in point form.
\begin{enumerate} \item{Importantly, we are (also) assuming
the self-interference of fermions, i.e. the constructive
interference of each elementary fermion wave/particle (within an
atomic/molecular fermion system) with itself after a full
loop/cycle\footnote{This is somewhat analogous to the physical
interpretation of self-interference given by \citet{Chiao_06}, who
refers to the closed path Feynman-Onsager quantitisation
rule/condition \citep{Donnelly_67}.}. Clearly, non-locality is
implicated/involved; additionally so, because no field lines are
crossed --- via the geodesic motion of lunar atoms/molecules.}
\item{Our interest lies with the (spin) inertia of fermions (which
have mass). Although Planck's (or rather the reduced Planck)
constant is vital to the (maximum) magnitude of the new effect
($\frac{1}{2}\hbar \Delta t^{-1}$), this energy magnitude acts
independently of any electromagnetic field effects.} \item{Whereas
the Aharonov-Bohm effect is observed for neutrons and electrons
(i.e. particles), we examine atoms/molecules. In particular,
ramifications of their (`internal') spin-orbit coupling.} \item{As
for the A-B effect, the parameter space remains the ordinary space
of the atom's motion, cf. the motion of the Hamiltonian; but this
space is now \emph{curved} space cf. Minkowski space.} \item{The
phase shift of interest in the model is virtual, cf. (in the A-B
effect) the real phase shift of particles acted upon by
topological macroscopic EM effects. The virtual phase shift arises
from the motion of a celestial spin-orbit `coupled' third-body in
curved space --- cf. an EM vector coupling (topological) effect in
Minkowski space --- but (herein) it is not determined by way of
the parallel transport of a vector.} \item{EM-based phase changes
associated with particle or radiation emission/absorption are
quantum-like (and gauge-based), whereas this motion and curvature
based phase change allows arbitrary analog (i.e. non-digital)
phase changes --- both virtual and `real', i.e. (respectively)
below and above (or rather `up to') a decoherence threshold.}
\end{enumerate}

\subsubsection{A new type of `precession'}\label{subsubsection:new precession}
The (electromagnetic) Aharonov-Bohm effect is considered to be
analogous to the general relativistic frame-dragging
effect\footnote{Also known as the Lense-Thirring effect.} of a
body orbiting a rotating mass. Loosely speaking, we can say
frame-dragging leads to a precession in the plane of (orbital)
rotation.

General relativity also involves an (orbital) geodetic effect upon
spinning celestial bodies. A spinning body's axis, when parallel
transported around a massive object's curved spacetime, (in one
complete revolution for argument's sake) does not end up pointing
in exactly the same direction as before. There is a `precession'
of the polar axis. Loosely speaking, one can say: ``\ldots the
gyroscope `leans over' into the slope of the space-time curvature
(Wikipedia: \emph{Gravity Probe B}, 2011)."

The new ``precession" being proposed herein is a function of
curved spacetime but it is quite different:
\begin{itemize} \item{Involving, a (relative) \emph{phase} of
intrinsic angular momentum --- and requiring atoms/molecules
`slaved' to a celestial body.} \item{It is a \emph{virtual}
effect, that is self-interference based and thus determined only
after a complete orbital loop/cycle; in a suitably configured
three-body spin-orbit coupled celestial system.} \item{This QM
``spin phase precession", relative to initial spin phase and an
unchanged orbital phase, does not arise from a (local) change in a
celestial body's \emph{orientation}. See section
\ref{Subsection:Spin-turning} for further discussion.}
\end{itemize} Loosely speaking, we may also visualise (i.e. imagine)
the (virtual) spin phase precession/offset as a QM ``over-spin",
although it is the relative virtual spin \emph{phase offset} that
is our primary focus.

\subsubsection{The three faces of one and the same (magnitude)
geometric phase offset}\label{subsubsection:Two faces} It helps to
recognise three different (conceptual) types of \emph{relative} phase
change (per loop/cycle), that are physically equivalent \emph{if}
moon-planet spin-orbit resonance/coupling is `active', \emph{and} a
cosmological substratum is recognised (recall section
\ref{subsection:SR's ontology}). Firstly, internal to an
atom/molecule we have: the QM phase associated with intrinsic angular
momentum (i.e. spin) relative to (the unchanging, non-topological)
orbital angular momentum phase. Secondly, we have an internal
(lunar-based) spin phase offset relative to a background global
reference `datum'; with this datum including the sun and host
planet's atoms/molecules
--- i.e. everything \emph{other} than the spin aspect of
atomic/molecular constituents within the third body of suitably
configured celestial systems. Thirdly, we have the spin phase
offset, after the closed loop motion, relative to its (own)
pre-loop/initial `value' --- by way of self-interference.

\subsubsection{The need for two types of spin-orbit `coupling': macro- and
microscopic}\label{subsubsection:Two types} Lunar orbital
resonance (i.e. macroscopic moon-planet spin-orbit
resonance/phase-lock) ensures an exact externally orientated (1:1)
spin-orbit relationship for all lunar atoms/molecules per ``closed
path loop" (i.e. one orbit). This externally imposed exact
phase-matching establishes a (resonance) `datum', against which
any unforseen externally driven atomic/molecular geometric phase
drift may be referenced. Subsequently, two different broad types
of relative phase offset are conceivable. The first is exclusively
internal/micro: i.e. spin vs. orbital, in conjunction with spin
(virtual) vs. spin (actual); and the second is lunar
(atoms/molecules) vs. an external reference `datum' --- which
includes the unaffected spin and orbital geometric phases of
planetary and solar atoms/molecules. In both cases the same
conceivable offset --- occurring upon the completion of a closed
path loop --- is being described; i.e. a lunar (atomic) spin phase
offset relative to either: an internal unchanging QM spin-orbital
phase configuration; and/or an external background space phase
`datum'.

In the model, any QM system (particularly atoms and molecules)
whose spin phase is offset by less than half a wavelength, must
`maintain' (and not lose) its (discrete) quantum spin-orbit
`configuration'; (and) thus, the energy associated with the
virtual geometric phase offset must be externalized into the
space-time `environment' --- if a systemic (i.e. universal)
conservation of energy principle applies. Actually, this is a
conservation of angular momentum over the lunar spin-orbital
duration/time $\Delta t$. Thus, the \emph{virtual} spin phase
offset `exists' relative to its own unchanging (actual) lunar
atomic/molecular spin phase datum (see subsection
\ref{subsubsection:What lies ahead}); with this being a
\emph{local} datum cf. a \emph{background} space phase datum.

\subsubsection{Global reference frames and lunar geodetic precession rate}
In relation to the Sun's tidal acceleration Kenneth Nordtvedt
recognises that: \begin{quote} \ldots de Sitter's `geodetically'
rotating inertial frames accompanying the Earth in its motion
around the Sun induce a precession of the lunar orbit which is
\emph{slightly greater} than the geodetic precession
rate\footnote{The Lense-Thirring effect is neglected in the
article [see p.1318].}, \ldots [because] the Moon is not a locally
isolated dynamical system, \ldots [Thus, the] Moon's dynamics in
the geodetically rotating inertial frame is intrinsically
different from its dynamics in a globally fixed
frame,\footnote{That is, a frame fixed with respect to distant
inertial frames. This is the barycentric frame (comprising the
Sun, planets, Earth, Moon, asteroids, etc.) employed in the
generation of ephemerides for the Moon and other solar system
bodies.} \ldots \citep*[ abstract]{Nordtvedt_96}.
\end{quote}

Somewhat similarly, in our model there is a tiny (virtual)
discrepancy in the form of a QM spin phase
`precession'\footnote{In the from of an intrinsic angular momentum
phase shift/offset per (closed loop) cycle time.}; arising from
the planet-moon system's orbital motion in the `barycentered'
solar system frame --- that would (always) be absent if only the
local (two-body orbital) dynamics was considered.

It is prudent to (once again) note that a GR-based understanding
of the motion of larger bodies in the solar system, especially the
Earth-Moon system, is \emph{not} (at all) being questioned. Indeed
the ``closed cycle/loop" referred to, regarding QM geometric
phase, is an \emph{orbital} loop that includes (non-Newtonian
mechanical) general relativistic-based effects.

\subsubsection{Summary remarks for section \ref{subsection:Precession}}
The section's main objective has been a discussion of the proposed
geometric phase (GP) offset, and how it is different to GP shifts in
(primarily) electromagnetic-based situations. Specifically, we make a
comparison with the Aharonov-Bohm (A-B) effect, because the A-B effect
(until now) has been considered to be the only phenomenon that involves
a non-locality associated with fermion waves.

We distinguish this (unique) GP offset from gravitational
analogues of the (EM) Aharonov-Bohm and Aharonov-Casher effects,
where the A-B effect is considered to be analogous to general
relativity's frame-dragging. The new GP offset has been portrayed
as (or likened to) a new type of (spin) precession, in that it is
(also) a function of motion in curved spacetime. Obviously, it is
quite distinct from geodetic precession in that it involves a
relative and \emph{virtual phase} offset that is figuratively
likened to a classical Hannay angle offset.

The special circumstances required for this new effect have been
discussed. Firstly, by way of applying to all (third-body)
elementary fermion particles, this GP offset is also seen as
applying --- in equal measure as that applied to each (and every)
`constituent' fermion particle --- to whole \emph{quantum
mechanical} systems (i.e. atoms or molecules). We note that
atoms/molecules have `internal' spin-orbit coupling. Secondly, we
require the parent body of the atoms/molecules (i.e. a moon) to be
in celestial spin-orbit resonance relative to its host planet.
These conditions ensure that a minimal phase offset can occur,
i.e. less than half a fermion wavelength ($2 \pi$), with this
circumstance being physically and quantitatively not
insignificant.

Finally, the relative nature of the geometric phase offset
($\beta$) was further discussed. It is seen as both: an internal
spin vs. orbital effect involving angular momentum \emph{phase};
and secondly, as a local vs. global/systemic effect involving spin
phase cf. a background (spin phase) datum. Additionally, we
introduced the idea that $\beta$ is also a virtual spin phase
offset vs. the actual spin phase as fixed by the EM dominated
`internal' spin-orbit coupling --- which acts to suppress (and
make virtual) the `would be' phase advance.
%*************************************************************************************************************
\subsection{Spin phase, orientation, and spin-turning}
\label{Subsection:Spin-turning} This section brings us to a core
aspect of Section \ref{Section:general model}.

\subsubsection{Spin phase and probability changes} Measurements yield
only two discrete values of (fermion) spin, i.e. $-\hbar/2$ or
$+\hbar/2$. In our case, the projection plane of a QM system's spin
is fixed by the plane of lunar motion. Relative phase
changes\footnote{Either a lunar (atomic/molecular) vs. planet/global
spin phase offset, or equivalently, a lunar atom/molecule's initial
spin phase vs. its (virtual) final phase.} [i.e. relative to a
systemic (phase) reference frame] are seen to alter the probability
of finding a lunar-based (and atomic/molecular-based) elementary
fermion particle in an spin-up or spin-down state. Loosely speaking,
a $2\pi$ fermion phase shift, for example, would alter a 100\%
spin-up probability to a 100\% spin-down probability
--- with this alteration also associated with the sign change of a
spinor. A \mbox{$\pi$ (radian)} phase shift/offset, i.e.
effectively 1/4 of a fermion wavelength, alters a 100\% spin-up
probability to a 50\% spin-up and 50\% spin-down
probability\footnote{Where probabilities are calculated from the
mathematical description of each state of a particle as a wave
function.}. Note that this discussion, and that of subsection
\ref{subsubsection:turning vs rotation}, draws upon the discussion
in a \citet*[ pp.97-99]{Bernstein_81} Scientific American article;
but we shall \emph{not} employ the notions of: base space, total
space and fibre bundles (preferring the model-specific stance
adopted in \mbox{subsection \ref{subsubsection:new precession}).}

\subsubsection{`Turning' as compared to `rotation'}
\label{subsubsection:turning vs rotation} Usually (although not in
our case), any change in spin probability, that is conceived (in
an idealised manner) as arising from an externally driven
(physical) rotation/precession of a QM system's spin `vector',
also implies a change in the spin orientation. Some authors prefer
to speak of a `turning' rather than a rotation to reinforce the
non-classicality of QM spin. In our case, this turning is
\emph{not} an axis orientation turning; rather, it is a phase
turning associated with a fixed/given spin vector's orientation
--- (classically visualised) much in the manner of the Earth's
(daily) turning on its axis. In the case of our model, we shall
see that maximum virtual energy occurs at a relative phase offset
occurs of $\pi$ radians, with the 0 and $2\pi$ radian offsets
being \emph{quantitatively} equivalent. Note that coherent
(virtual) phase offsets greater than $2\pi$ radians are not
considered because they are physically impossible; by way of
(actual) decoherence at $2\pi$ radians, which will `reveal', and
would also be incompatible with, the \emph{real}/actual
multifarious orientations of spin in the numerous atoms/molecules
of a bulk moon.

\subsubsection{Spin orientation and virtual spin phase offset
orientation}\label{subsubsection:Spin orientation} The
\emph{external} planar (orbital) rotation of a moon about its host
planet fixes both the projection angle of (internal)
atomic/molecular spin, and also the spin axis orientation (the
\emph{z} axis) for the \emph{virtual} phase `turning' associated
with motion in curved spacetime. This circumstance applies to
\emph{every} atom/molecule `in' a moon so long as it remains below
the decoherence limit.

In the solar system, planets and their moons orbit close to the
ecliptic plane; (and) the orbital orientations of large moons,
with the exception of Neptune's Triton, are closely aligned with a
planet's equatorial plane. Fortunately, we shall see later
(subsections \ref{subsubsection:3D external} and
\ref{subsubsection:three dimensions}) that the model's motion
retardation effect for (the Pioneer) spacecraft is the same
irrespective of the S/C's inclination to lunar orbital planes;
although for Earth-based measurements a correction for the angle
between spacecraft path and line-of-sight observation direction is
required (see Table \ref{Table:corrections} of Section
\ref{Section:Quantif Model}).

\subsubsection{Standard approaches to the interaction of spin with inertia
and gravity}\label{subsubsection:standard spin and gravity} Two
names that loom large in the field of gravity's interaction with
quantum mechanical systems are Bahram Mashhoon and Giorgio Papini.
Their emphasis is upon applications to particles (cf.
atoms/molecules) and: \begin{itemize} \item{utilises a
\emph{localised} approach to the formalism (cf. a non-local,
closed loop, Aharonov-Bohm like effect);} \item{involves the
(local) interaction of spin with: inertia, rotation, and gravity;
and} \item{concentrates upon motion that \emph{crosses}
gravitational field lines, i.e. non-geodesic motion.}
\end{itemize}

In broad terms, the interest for Papini, Mashhoon, and others lies in
the fact that: ``External gravitational fields induce phase factors
[and only phase factors] in the wave functions of particles \citep[
abstract]{Papini_07}." It was \citet*{Cai_89} who obtained a covariant
generalization of Berry's phase. That general covariance of GR can be
incorporated in the theory is a significant feat.

\begin{quote} We can finally conclude that the validity of
covariant wave equations in an inertial-gravitational context finds
support in experimental verifications of some of the effects they
predict [concerning neutrons and either: the Earth's rotation, or the
Earth's gravitational field], in tests of the general relativistic
deflection of light rays and also in the phase wrap-up in global
position system measurements \citep[ p.15]{Papini_07}. \end{quote} Our
main concern is to comment upon their proposed spin-rotation
coupling\footnote{For fermions, there exists at present only indirect
evidence for the existence of spin-rotation coupling
\citep{Mashhoon_06}.} and its extension to spin-rotation-gravity
coupling \citep[ section 3]{Mashhoon_00}.

In light of the fact that the model's non-local effect is
restricted to (`overall' atomic/molecular) intrinsic angular
momentum (spin), while leaving orbital angular momentum
unaffected, it is \emph{crucial} to note that \emph{locally}:
\begin{quote} In derivations based on the covariant Dirac
equation, the coupling of inertia and gravitation to spin is
identical to that for orbital angular momentum \citep[
p.16]{Papini_07}. \end{quote} Thus, a local approach to
establishing the model's new (virtual) spin geometric phase
offset, involving GR and QMs, is denied. Consequently, the offset
is (necessarily) systemic and non-local (as per subsection
\ref{subsubsection:A-B comparison}).

Clearly, the new mechanism must be different, and not included in
the broad scope of situations examined by this field. Were it not
for the Pioneer anomaly, the conceivable presence of this new
(non-local) curved spacetime/gravitational field effect upon
certain quantum systems, which is \emph{completely} `back-reacted'
into a gravitational field (in a different/new format), would have
remained far from consideration.

\subsubsection{Further reasons for denying a local approach to the phase shift/offset}
\label{subsubsection:unjustified spin precession} There are only
two theoretical options regarding the basis of the new phase
offset. Either it is solely related to an atom/molecule's local
physical motion, or an additional systemic and non-local effect
upon fermion spin phase is also present --- as is the case with
the Aharonov-Bohm effect. The former (i.e. local) scenario is
preferable, but for motion along a geodesic this subsection shall
present further reasons why it cannot be justified. Thus, by
default we need to embrace the latter scenario i.e. a
non-local/global effect arising from (closed loop) motion in
curved spacetime.

From a purely \emph{local} perspective the (shared by every
atom/molecule) QM virtual spin phase offset may be conceived as a
(virtual) angular offset/excess (or ``over-turning"); with this
being a measure of the curvature enclosed by the path of the moon
in curved spacetime. Note that in this (offset) case, parallel
transport of a vector \emph{is} (in some sense) necessarily
implied. We may then equate the common angular excess of all the
atoms/molecules in a moon to a (virtual) lunar bulk-mass
``over-spin" --- relative to a scenario involving uncurved (flat)
space. It was tempting (for the author) to then assume this
virtual ``over-spin" is actually an unrealised/thwarted (macro)
spin geodetic precession effect --- thwarted by way of tidal
forces that lead to celestial spin-orbit resonance. Thus, the
\emph{angular} offset could then be derived from a moon's path in
relativistic curved spacetime, so as to make unnecessary any
non-local basis (ultimately involving self-interference) `behind'
the \emph{phase} offset. But, there are problems with such an
expectation/approach.

If we idealise the moon-planet-sun three-body motion as: (all
bodies) lying in the same (ecliptic) plane (with the lunar and
planet spin axes orthogonal to this plane), then it is clear that
the over-spin does/could not arise from a precession of the
orientation of a moon's spin-axis. Thus, the notion of a geodetic
spin-axis precession is not relevant to the model. Furthermore,
the macroscopic (Hannay angle) ``over-spin" concept fails to be
analogous to the geodetic precession of a spinning gyroscope ---
as is examined in the Gravity Probe B experiment\footnote{An
alternative name for the Pioneer anomaly could be (the) Gravity
Probe C experiment --- even though two spacecraft were involved.}.
In other words, utilising GR's representation of spacetime
curvature and the concept of parallel transport of a vector, there
is no way to link curved spacetime with a \emph{spin}
over-rotation angle. It is just not in the `machinery' of general
relativity, (even) incorporating the parallel transport of a
vector, to do this. Furthermore, the heuristic and conceptual use
of a classical Hannay angle, regarding (an extrinsic/external
imposition upon) QM spin, would be inappropriately physicalised.

The failure of a solely local approach is further supported by the
inability of the QM spin (at a sub-system level) to be physically
understood as a classical rotation --- of any kind. Only a virtual
effect \emph{thwarted} by the dominant electromagnetic
intra-atomic `forces' is conceivable, and thus the thwarting of a
physically \emph{real} rotation (on any level) is inappropriate
and not possible. General relativity demands that the local frame
can always be configured as an inertial frame; in \mbox{section
\ref{Subsection:Inertial}} we see why this, in special celestial
`third-body' circumstances, is not (always) the case --- at least
virtually --- at the QM (atomic/molecular) systemic level.

\subsubsection{Ramifications of denying a local and reductive
approach (re: phase shift)} Subsection
\ref{subsubsection:unjustified spin precession} showed how a
solely reductive and local account of the model has fatal flaws.
Thus, regarding the model's physical associations, we are
restricted to only discussing a relative phase-shift (albeit
virtual). Any reference to spin \emph{rotation} angles lacks
physical validity, concerning \emph{both}: a moon (as a whole),
and at the quantum mechanical level of its constituent
atoms/molecules --- as well as the latter's constituent elementary
(fermion) particles.

The representation of a \emph{virtual} relative geometric phase
offset by way of either: a Hannay angular offset, an over-turning
angle, or over-spin angle, nevertheless remains a useful
heuristic/visual tool in the model. For atoms/molecules we are
restricted to speaking only of spinor or spin phase
changes\footnote{Mathematically we would need to deal with
`spinors', whereas for physical understanding the term `spin' is
preferred.}, but systemically/globally we allow ourselves to
envisage, by way of the conceptual tool of visualisation, a
spin-rotation or spin-turning based offset (as per
\mbox{subsection \ref{subsubsection:standard spin and gravity}).}

\subsubsection{Aspects of a (third-body) non-local virtual geometric
phase offset} It remains true that the virtual angular offset, or
rather virtual geometric phase offset, is (still) derivable from a
moon's path in relativistic curved spacetime; but herein it shall
simply be the (configurational) \emph{geometry} of the lunar and
planetary \emph{closed paths} that shall quantify the relative
phase shift (see subsection \ref{subsubsection:Celestial geometry
and}).

Non-locality in the explanation cannot be avoided. Our
(global/non-local) fermion self-interference scenario involves an
unappreciated and subtle three-body curved spacetime effect upon
the phase of a (given) spinor's wavefunction. We embrace a
phase-offset approach, rather than an orientation based approach.
For explanatory simplicity we have chosen to (somewhat
inappropriately) visualise this (externally `imposed' effect) as a
(relative) over-spin.

This spin phase offset exists in conjunction with what Michael
Berry termed: ``\ldots phase functions hidden in parameter-space
regions which the system has not visited \citep*{Berry_84}."
Regarding the model, this quoted phrase is best altered to:
``phase effects hidden in background space-time that the system
cannot recognise within itself, but systemically (i.e. globally)
are physically relevant."

The non-local basis of the spin phase offset is seen to extend the
geometric (two-body) `curious problem of spinor sign change' (and
$4\pi$ fermion wavelength) to a new scenario; this being an
appropriate third-body of a three-body celestial system. Note that
in both (the two- and three-body) cases the simple/pure geometric
relationship of (macroscopic physical) celestial (planetary
\emph{and} lunar) orbital rotation angles to (microscopic) phase
shift is determined within a plane, i.e. it is a relationship
describable in two dimensions.

\subsubsection{What lies ahead: the spin offset via a non-inertial
path cf. an inertial path}\label{subsubsection:What lies ahead} In
section \ref{Subsection:Inertial} the physical significance (and
ramifications) of the spin phase offset's existence is examined.
Up to this point we have simply examined the basis and motivation
for a phase offset. It shall be argued that a virtual (intrinsic
angular momentum) phase offset represents a phase offset between a
background global \emph{inertial} frame and a local frame that has
remained inertial (and unforced) in all ways \emph{except} for the
unique spin circumstances outlined herein.

Since the phase offset --- upon the completion of a geodesic
closed loop (orbital) motion and one complete lunar-celestial spin
cycle/loop --- is virtual, there is an absence of any actual (i.e.
real) (fermion) wave interference. Thus, at all times we retain
(cf. the implications of Berry's phase), a circumstance where
\mbox{(local)} quantum eigenstates are not made `multi-valued' ---
i.e. they (themselves) remain unaffected by the (externally
induced) virtual phase offset.

In other words, we shall argue that the system's spin phase lags
behind that which curved space (eventually) requires after the
lunar closed loop (and spin) motion(s). Thus, the atomic/molecular
system itself `under-turns' relative to a spin inertial frame ---
with this being the phase excess or over-turning we have been
discussing. It is the internal/QM spin-orbit coupling of an
atom/molecule that both: leads to a non-inertial circumstance; and
thwarts any internal (sub-$2\pi$) spin phase change, because
orbital phase is unaffected. Thus, a \emph{virtual} spin phase
offset has \emph{real} physical implications --- presuming that it
is recognised/appreciated by the global system it is a part of. In
QMs this situation can lead to an \emph{equal} and \emph{opposite}
compensating \emph{phase} shift in the environment \citep[
p.31]{Berry_88}. \emph{In our model the `reaction' in the
environment is based upon energy rather than phase.}

Importantly, in our case --- i.e. the model's hypothesised case
--- the virtual phase shift of intrinsic angular momentum also
allows an \emph{energy} value to be determined. The spin energy
deficiency needs to be ``made up" somewhere else in the larger
system (i.e. the universe). With no available means of internal
expression this energy, totalled up for all the atoms/molecules in
a (suitable) moon ($N_m$), gets expressed \emph{externally} as a
(single) rotating space-warp of amplitude $\Delta a$ --- rotating
in the \emph{same} direction and at the same angular velocity as a
moon's 1:1 spin-rotational (i.e. spin-orbit resonant)
motion\footnote{By way of energy conservation demands, the
`universal' system effectively `preempts' the offset dilemma and
`externalises' the energy involved --- over the course of the
cycle time $\Delta t$. Additional background/hidden temporal
effects are implied --- see \citet{Price_96} for an account of how
concerns arising from this situation may/might be appeased.}. Note
that decoherence negates (or invalidates) any external influence
arising from a greater \mbox{than $2\pi$ rad} geometric phase
offset/change (i.e. $\beta \geq 2\pi$ rad).

Finally, we note that this space-warp `rotation' --- introduced
and discussed in Section \ref{Section:PrelimModel} --- is quite
different to the \emph{internal/interior} virtual/non-inertial
phase shift or phase-`turning' per lunar spin-orbital cycle
discussed throughout (this) Section \ref{Section:general model}.
Initially, the mechanism is treated as a \emph{planar} phenomena,
but the space-warp's extension into three (physical) dimensional
space is required (see subsection \ref{subsubsection:3D
external}).

\subsubsection{Summary remarks for section \ref{Subsection:Spin-turning}}
This subsection has discussed further aspects of the model's
virtual and non-discrete geometric phase offset ($\beta$), and
finishes with a preview of what this offset is physically
representative of; i.e. a distinction between (or difference
involving) an inertial and a non-inertial situation.

Whilst the standard relationship of a change in phase to a change
in the probability of measuring a `system' as spin-up or spin-down
is retained, the external means by which the spin phase shift is
`generated' is quite \emph{different}; in that it does not arise
(solely) from an externally induced rotation of a spinor (or
particle's) spin vector orientation. This difference is best
understood as arising from the non-standard `influence' of a
gravitational (i.e. curved spacetime) field upon the
(wavefunction-based, and intrinsic angular momentum-based) phase
of (certain third-body) quantum mechanical sub-systems --- i.e.
the atoms and molecules of a moon. This interaction differs from
standard approaches in three ways: (1) atoms/molecules as
composite elementary fermion particle systems are examined cf.
merely particles; \mbox{(2) our} concern lies with geodesic motion
cf. effects arising from the crossing of gravitational field
lines; and (3) most importantly, the basis for the geometric phase
change/offset is (necessarily) global/systemic and non-local cf. a
localised approach that involves the parallel transport of a spin
vector's orientation. The unavoidability of this final distinction
has been discussed in some detail and reinforces the fact that:
the geometric phase change cannot be solely attributed to a local
motion and/or rotation, nor some form of local
spin-rotation-gravity coupling.

We also noted that: for this \emph{virtual} geometric phase shift,
the projection plane of each (and every) lunar
(intra-atomic/molecular) elementary fermion particle's spin axis
is fixed (externally) by either: the bulk mass moon's spin axis,
or the plane of lunar orbital motion around its host planet.
Finally, it was mentioned that a $\beta=\pi$ phase offset (i.e.
one quarter of a fermion wavelength) is associated with the
maximum (internally unresolved) \emph{virtual} spin energy
--- of every atom/molecule in a (solid) bulk lunar `mass'. This total
internally unresolved energy is expressed (externally) by way of a
`real' rotating space-warp, and a spherical non-local mass
distribution --- for the latter, see/await subsection
\ref{subsubsection:non-local inertial mass} for an introduction, and
section \ref{Subsection:warp's mass} for a comprehensive discussion.
%**************************************************************************************************************
\subsection{Further discussion upon the non-local geometrical (and
topological) spin phase offset}\label{Subsection:Inertial}

\subsubsection{The inertia of intrinsic spin} \citet*[ introduction]{Mashhoon_06}
note that: ``\ldots, in quantum theory the inertial properties of
a particle are determined by its inertial mass as well as spin."
We extend (the use of) `particle' (in this quote) to the bound
\emph{systems} of atoms and molecules. Our interest involves both
spin \emph{and} indirectly inertial mass --- the latter by way of
the \emph{number} of atoms/molecules being `equivalently'
influenced at the same time, \emph{and} (more importantly) the
role of mass in intrinsic angular \emph{momentum} (i.e. spin).

The interpretation of our model/mechanism is that the spin phase
offset indicates the local spin phase's lag (at the end of a loop
in curved space-time) relative to the spin phase required to
maintain an inertial spin frame --- with respect to a global or
systemic (inertial) spin frame. This systemic (spin) reference
frame implies there is a topological aspect to the geometric phase
(recall subsection \ref{subsubsection:omission}). As regards
geometric phase changes, the atoms/molecules effectively remain in
a stationary state throughout the cycle/loop --- with only phase
and probabilities experiencing a (virtual) alteration.

The \emph{orbital} characteristics of atoms/molecules are
\emph{not} (additionally) affected by third-body cyclic motion in
curved spacetime, i.e. there is no latent or hidden virtual effect
upon the phase of (lunar-based) atomic/molecular
orbital\footnote{The word ``orbital" has a somewhat loose
applicability to QM circumstances.} angular momentum.

Local atomic/molecular spin-orbit coupling dictates that the
spin's phase is necessarily `slaved' to the orbital
characteristics of an atom/molecule --- which is dominated by:
electromagnetic forces, Schr\"{o}dinger's equation, Planck's
constant, discreteness, etc.; as compared to minuscule (internal)
gravitational (attractive) `forces' which are $\sim10^{-40}$ times
weaker \emph{in the atom}. Subsequently, (within atoms/molecules)
any inertial `forces/effects' acting upon the elementary
fermion/matter particle's spin (alone), `driven' by (external)
gravitational curvature effects, have hitherto been assumed
negligible, or (alternatively) have simply not been
appreciated/anticipated.

Locally, the (relative) spin phase offset --- common to every
elementary fermion wave/particle and hence (equally) to an
atom/molecule `overall' --- is a background/hidden effect. Only a
system `greater' than just the atom/molecule can `register' its
presence --- and only then if it is below the ($2\pi$ phase shift)
decoherence limit. The under-turning of an atom's spin (i.e. its
non-inertial spin behaviour) fixes a (per loop) rate of angular
momentum offset --- i.e. a non-inertial spin \emph{energy}
difference (per atom/molecule).

Implicit in this discussion is the need to assume the existence of
a global (but \emph{not} absolute) reference frame where both
gravitational and quantum mechanical systems (and effects)
harmoniously coexist in nature. A systemic conservation of (spin)
energy is also embraced (recall section \ref{subsection:Noether}).
Also implicit is the model's use of noumenal/background/hidden
time simultaneity in conjunction with quantum mechanical
`non-locality' (as discussed in Section
\ref{Section:PhiloTheory}).

\subsubsection{Ramification of no graviton particle: a virtual
fermion condensate}\label{subsubsection:ramification no graviton}
Firstly, with a presumption of the non-existence of graviton
particles --- recall subsections \ref{subsubsection:graviton},
\ref{subsubsection:E/M vs Grav} and \ref{subsubsection:graviton's
absence} --- there is no gravitational momentum transfer (of
information) by particles; thus, only a sign change of a spinor
`activates' decoherence. In contrast to a charged
particle-to-photon coupling, the absence of a mass to graviton
coupling means an unlimited quantum coherence cannot be ruled out
--- i.e. a large number of quantum `particles' cooperating in a
single quantum `state'. This `state' is actually the energy
associated with the shared virtual phase difference --- with a
common (externally determined) projection angle of
atomic/molecular spin and a common/shared spin-axis orientation
facilitating this ``large number" effect (recall subsection
\ref{subsubsection:Spin orientation}).

The physical absence of charge interference, because charges are
decohered by their own electrical field, leads to a charge
superselection rule. Similarly, although by default, there is no
graviton interference. Thus, below the level of decoherence,
something akin to a (virtual) fermion `condensate' is possible.
Without a graviton particle a QM system, influenced by curved
spacetime effects, can remain isolated from its environment ---
but only so far as `real' (i.e. particle-based) physical
communication is concerned.

To simply believe a large mass is necessarily classical fails to
fully appreciate the \emph{qualitative} manner by which
decoherence \emph{is} achieved (recall subsection
\ref{subsubsection:Vlatko Vedral}), and overlooks the exceptional
circumstances, based upon a \emph{quantifiable} phase offset,
discussed herein.

The EM gauge and phase transformations that satisfy
Schr\"{o}dinger's equation and gauge invariance are only relevant
in so far as stipulating a minimum quanta of intrinsic angular
momentum (i.e. spin) (for fermions) of $\frac{1}{2}\hbar$. By way
of quantum discreteness and uncertainty the \emph{minimum}
(actual/real) energy associated with: a particle's, atom's, or
molecule's QM spin is $\frac{1}{2}\hbar \Delta t^{-1}$ (recall
subsection \ref{subsubsection:Large number}); which is actually a
\emph{maximum} possible virtual energy level (per atom/molecule)
in the new mechanism's spin-based inertial offset situation (see
subsection \ref{subsubsection:maximum virtual}). A prerequisite of
this energy (in the model/mechanism) is the coexistence of atomic
spin \emph{and} orbital behaviour, and their (interaction or)
coupling.

\subsubsection{Discussing relationships between spin and orbital
angular momentum}\label{subsubsection:relationships spin and} The
additional QM/microscopic spin phase advance needed to achieve
inertial status, is thwarted by the much stronger local EM
effects. Note that internal QM orbital angular momentum is
\emph{only} a function of EM effects, and it is (not surprisingly)
completely \emph{distinct} from macroscopic celestial orbital
angular momentum --- including GR's additional geodetic (orbital)
precession, and any ``frame-dragging" effects. With regard to
macroscopic spin angular momentum and QM spin this distinction is
even more pronounced.

Recalling subsection \ref{subsubsection:unjustified spin
precession}, it is wrong to consider the shared
(sub-half-wavelength) microscopic spin-phase offset, as
representing a virtual macroscopic ``over-spin" that is thwarted
by tidal effects. We also noted that even if a macroscopic
geodetic precession-like effect did exist, it could not be
`thwarted' --- so as to `establish' an unresolved non-inertial
energy. Additionally, with microscopic over-spins always below a
minimum (actual change) energy level, and decohered (i.e. lost) at
or above it, tidal effects can \emph{never} be said to physically
impede (quantised) lunar over-spin. All that remains physically
relevant is that, at a microscopic level, the many (virtual)
`over-turnings' remain active in the `awareness' of the global
system as a whole --- i.e. in the `environment'/surroundings of
the quantum mechanical (atomic/molecular) systems.

It is the unique ability of a global/systemic reference frame to
access how the spin probabilities of QM atoms/molecules are
counterfactually\footnote{`Counterfactual' (adjective) is defined
as: relating to or expressing what (is conceivable but) has not
happened or is not the case; or running contrary to the facts
(i.e. reality).} influenced, and the inability of any microscopic
state change or macroscopic motion to appease this change, that
necessitates the existence of the external rotating space-warp.
The total QM over-turning \emph{energy} (imbalance) cannot be
denied --- \emph{if} it is below the ($2\pi$ phase offset)
decoherence limit. A crucial feature of the model is the fact that
the counterfactual/virtual offset scenario \emph{is} the
\emph{inertial} frame circumstance. At, or above, the decoherence
limit this subtle situation, involving both macroscopic GR's
curved spacetime and microscopic QMs, is absent; and effectively
only macroscopic tidal effects\footnote{Tidal effects arise from
celestial bodies being non-rigid; notwithstanding the fact that
moons are (primarily) solid bodies.} act upon the celestial
third-body's spin --- as is currently envisaged by theorists.

Importantly, without sufficiently strong tidal effects, the
absence of planet-moon spin-orbit resonance would mean that the
finely tuned self-interference arising from: \begin{itemize}
\item{celestial (moon to planet) spin-orbit resonance,}
\item{motion of celestial mass `upon' a geodesic,} \item{the vast
majority of atoms/molecules in a moon being in a fixed location
relative to the moon's geocentre\footnote{In other words, large
moons are essentially (i.e. dominantly) ``solid" bodies.}, and}
\item{QM atomic/molecular spin-orbit coupling,} \end{itemize} is
not applicable, and thus the decoherence limit is (considered)
easily breached. Decoherence ensures that GR plus tidal effects
are sufficient to successfully account for most three-body
gravitational phenomena. We are examining the exception to the
rule, with celestial and QM spin-orbit resonance/coupling
(respectively) crucial to the new mechanism's existence.

With regard to the QMs of atoms/molecules, spin and orbital
angular momentum are linked in an \emph{interdependent} manner.
Classically, spin and orbital angular momentum are (usually) quite
\emph{independent} --- with the spin-orbit synchronisation of a
moon-planet system being an exceptional case. Thus, a
\emph{distinction} between how (external) curved space (cf. flat
space) influences an atom/molecule's intrinsic spin and orbital
momentum is not unduly surprising --- with our (moon-based) case's
specific `closed curve' (recall subsection
\ref{subsubsection:Berry background}) involving (orbital)
celestial motion along a geodesic, as well as celestial spin.

As a final encapsulating remark, it may be said that the
environment's `awareness' of each atom/molecule's non-inertial
(microscopic) spin status, throughout a (generally) solid moon,
does not entail non-geodesic (macroscopic) motion, but it does
entail a new (macroscopic) gravitational/accelerational phenomenon
in the form of a rotating space-warp\footnote{With an associated
non-local mass distribution.}.

\subsubsection{Uncertainty, angular momentum and energy in quantum
mechanics} \label{subsubsection:uncertainty, ang. mom., E}
Heisenberg's uncertainty principle tells us we may not know
simultaneously the exact values of two components of the angular
momentum; in our case orbital and spin angular momentum. Even
though the position of electrons in an atom (for example) may only
be known by way of probability, the value of its angular momentum
is an exact (or sharp) value. \begin{quote}[In an atom] the
angular momentum operator commutes with the Hamiltonian of the
system. By Heisenberg's uncertainty relation this means that the
angular momentum can assume a sharp value simultaneously with the
energy (eigenvalue of the Hamiltonian)\footnote{Sourced from
Wikipedia (modified 3 January 2012 at 21:31): \emph{Angular
momentum coupling}.}.\end{quote} This is supportive of the
existence of an exact value of virtual (intrinsic) angular
momentum in the model, and its change over cycle loop time $\Delta
t$ (i.e. virtual energy).

\subsubsection{A distinction between macroscopic spin and orbital
angular momentum facilitates the existence of a microscopic/QM
non-inertial situation} \label{subsubsection:macro spin and
orbital} Although macroscopic geodetic precession and the model's
proposed microscopic spin phase offset (or spin-turning) arise
from the \emph{same} motion in curved spacetime, these effects are
otherwise totally distinct. With external curved spacetime acting
upon the QM nature of atoms/molecules, any distinction between
(and independence of) spin and orbital angular momentum at the
classical level is easily carried through/`down' to the
microscopic level. Thus, there is no compulsion to have something
(local and reductive) like a geodetic spin-orbit coupling effect,
with an unknown extra celestial/macroscopic spin precession (cf.
NMs and GR) accompanying the extra macroscopic orbital precession
(of GR cf. NMs) for example. Macroscopically/celestially, spin and
orbital angular momentum are quite distinct, and subsequently, the
effects of geodesic celestial motion/momentum upon the (geometric)
phase associated with QM spin and orbital momentum in
atoms/molecules can conceivably retain this independence, i.e. the
effect can be `uncoupled'.

What remains true is that \emph{together}: the virtual microscopic
spin phase offset(s), which may be figuratively and collectively
conceived as a (virtual) lunar-\emph{spin} angular excess, and
general relativistic subtleties such as the geodetic effect(s)
etc.; act to physically establish what is inertial motion --- in
the sense of ``least/no action" applying. Subsequently, for ``mass
in motion", at the QM level, there is no good reason to deny an
inertial--to--non-inertial `frame' offset --- albeit virtual ---
concerning the spin phase and (thus) the \emph{mass} `within'
atoms/molecules. This offset (over the course of a cycle/loop) is
effectively a non-inertial intrinsic angular momentum (rate); it
arises from atomic/molecular spin-orbit coupling being quantised
and dominated by electromagnetic concerns (including electrical
\emph{charge}), rather than the (inertial) \emph{mass} of its
constituent sub-atomic (fermion) particles --- such that the
actual/real spin phase is necessarily coupled/slaved to (the
unaffected) orbital phase.

The confidence of Papini (subsection \ref{subsubsection:standard
spin and gravity}) in assigning \emph{all} orbital and spin
angular momentum the same coupling to inertia and gravitation is
(thus) cast into doubt, at least below the level of any change in
QM state and/or a sign change of a spinor. This concern is not
with regard to the (local) \emph{crossing} of gravitational field
lines, but rather with regard to the geometric phase offset effect
proposed herein --- arising from the closed loop \emph{geodesic}
motion/trajectory of atoms/molecules within a `third' celestial
body.

\subsubsection{Can dynamic curved spacetime considerations determine
the model's virtual spin phase offset?} \label{subsubsection:dynamic
vs geometry} In response to subsection \ref{subsubsection:macro spin
and orbital} it could be argued that: because general relativity
itself is seen to encompass ``all that is inertial", it follows
(somewhat trivially) that (even) a quantitative fibre bundle analysis
of (lunar atomic/molecular elementary fermion particle) QM spin in
curved spacetime, with its basis in GR's \emph{local} curvature,
cannot lead to an inertial vs. non-inertial offset at the QM level.
Clearly, this is not a `watertight' argument.

In subsection \ref{subsubsection:GP background} it was mentioned
that a relative geometric/Berry phase shift for photons has been
shown to arise from the intrinsic topological structure of
Maxwell's theory. Conceivably, the means by which the model's
geometric phase offsets \emph{could} be established is by way of
the (dynamic) geometry (and topology) of curved spacetime,
involving a metric tensor --- albeit in a \emph{three} body
system. One stumbling block for such an approach is its
complexity. Another possibility would be to use the (approximate)
classical/Newtonian gravitational potentials. Such top-down
reductive approaches cannot be ruled out, even though the closed
loop (local) path followed is a \emph{geodesic}, and regardless of
the fact that curvature is an inherently `non-local' phenomenon
--- in the sense of being distributed over a region of space.

For the sake of simplicity, and essentially by default, in Section
\ref{Section:Quantif Model} and initially in subsection
\ref{subsubsection:three-body phase}, we shall employ an
\emph{quasi-empirical} purely-geometric (and kinematical) basis to
quantify the (third-body) relative QM spin phase offsets; as
compared to a dynamic approach involving either: a metric tensor,
gravitational potential, or a Lagrangian and/or Hamiltonian. This
approach may or may not be the `best' quantitative approach, but
its use (and further development) in this paper achieves
consistency with (i.e. matches) all the various observational
evidence pertaining to the Pioneer anomaly. In agreement with the
third quotation in subsection \ref{subsubsection:Berry background}
we assume that: each geometric phase (offset) depends on a closed
(orbital) curve (C) and (its quantitative determination) is
independent of how the evolution around C takes place.

Notwithstanding our quasi-empirical approach, we nevertheless
achieve a relative superiority and comprehensiveness of
explanation (cf. other models arguing for a real non-systematic
based Pioneer anomaly), that \emph{quantitatively} is quite
conceivably only of provisional (or stepping-stone) status. On the
other hand, \emph{qualitatively}/conceptually there is little room
for the model to manoeuvre, and the approach is either:
fundamentally misguided, or has merit and can lead to further
progress. The latter is elucidated later in this paper
(particularly subsections \ref{subsubsection:flyby anomaly} to
\ref{subsubsection:Brief list}).

What does stand (somewhat) in the way of a \emph{dynamic
approach}, to the geometric phase offset, is the model's embrace
of non-local physical concerns/considerations, with this (embrace)
exhibited by both: the manner with which we argued that a
supplementary (to GR) type of acceleration/gravitational field can
exist, and the need to introduce the concept of ``non-local mass"
(at the macroscopic level). Further, it remains unclear as to
whether dynamic curved spacetime considerations could (ever)
determine the model's virtual spin phase offset (per
orbital/cycle-loop). For our purposes a solely
\emph{geometric-kinematical approach} (involving three celestial
bodies) is utilised, thus removing \emph{all} possible conflict
with non-locality (particularly with regard to QM spin
entanglement).

\subsubsection{Quantifying the geometric phase offset in the
third-body of a three-body gravitational system}
\label{subsubsection:three-body phase} The model's
`\emph{third-body}' (virtual and mechanistic) spin phase
offset\footnote{For atoms/molecules upon the completion of a lunar
orbit (closed circuit).} can be \emph{conceptually} appreciated as
a virtual (Hannay angle) over-spin; and physically as a
distinction between an inertial and a non-inertial frame. This
unorthodox phase offset requires the presence of curved spacetime,
but it goes beyond purely local and `force'-based considerations.

We seek to precisely determine the geometric phase offset for
specific three-body celestial motion circumstances (e.g.
Sun--Jupiter--Ganymede\footnote{Or more precisely: the (solar
system's) Barycentre--Jupiter--Ganymede.}). In the model, the
phase change's magnitude is a continuous (i.e. non-discrete)
quantity, cf. subsection \ref{subsubsection:foundation} where the
QMs based discussion was (implicitly) restricted to discrete
$2\pi$ (and $4\pi, 6\pi, 8\pi, \ldots$) phase changes.

For phenomena solely involving quantum mechanics `plus'
electromagnetism (QMs+EM) (with special relativistic effects
included), the conventional approach to spin phase change (and
spinor sign change) cannot be applied to the virtual spin
phase/Hannay angle (offset) conceptualised within our (QM+curved
S/T) model. With the Pioneer spacecraft observations dictating the
model's structure, a purely \emph{geometric approach} is the
simplest and most elegant option\footnote{These words may give the
sceptical reader some solace. ``\ldots for geometry, you know, is
the gate of science, and the gate is so low and small that one can
only enter it as a little child (William K. Clifford)."};
especially in the presumed absence of a graviton particle.
Fortunately, this (semi-empirical) closed-curve based approach
`bares fruit'\footnote{Recall that two conceivable (but unlikely)
alternative methods for establishing the geometric phase offsets
were discussed in subsection \ref{subsubsection:dynamic vs
geometry}.}.

In the model, the relative phase advance (of the inertial spin
frame) is restricted to atoms/molecules `experiencing' (lunar)
`third-body' spin-orbit resonant (or phase-locked) motion; with
the atoms/molecules making up the rest of the system (i.e. planets
and the sun) \emph{unaffected}. Distinctively, the ($2\pi$ rad
magnitude) `decoherence' spin phase advance --- of lunar
atoms/molecules over one spin-orbit cycle ($\Delta t$) --- needs
to play a further crucial role, as regards its relationship to the
geometry (and kinematics) of the (second-body) host planet's
orbital motion (over $\Delta t$).

By way of fitting the model to the observations: for \emph{one}
lunar spin-orbit cycle; \emph{and} a planetary orbital progression
angle (around the Sun), designated $\theta$, of
\mbox{$\tan^{-1}(8\pi)^{-1}\approx2.2785^o$}; the lunar phase
advance (applicable to intrinsic angular momentum) is seen to be
$2\pi$ radians. This specific and unique geometric configuration,
although highly unlikely in real moon-planet-sun circumstances,
becomes our reference scenario. It is used to determine (i.e.
quantify) the (virtual) lunar atomic/molecular (spin) geometric
phase offsets (over one spin-orbit cycle) --- of physically real
three-body celestial systems (e.g. Sun--Jupiter--Ganymede). This
is more fully examined in section \ref{Subsection:Model quantified
internal}.

The geometry of actual lunar and planetary orbits, arcs, angles,
etc., `supported' by the presence of spacetime's
curvature/geometry, is all that is needed to establish any active
third-body phase offset $\beta$ (as long as it is $<2\pi$
radians). The Earth's moon (Luna), whose origin is collision
based, is not a space-warp `generator' because its geometric phase
advance far exceeds the decoherence limit of $2\pi$ radians ---
where a spinor sign change occurs. Effectively, the spinor sign
change (associated with a $2\pi$ phase offset) in flat space has
been extended to a curved space scenario --- involving
\emph{three} bodies --- and hence the extra (planetary
progression) angle is required. Thus, our three-part reference
scenario is: \begin{itemize} \item{a $2\pi$ (geometric) phase
offset ($\beta_{\rm{ref}}=2 \pi$ rad),} \item{for one (closed
loop) lunar orbit and one lunar spin cycle [i.e. (in both cases) a
$2 \pi$ rad rotation],} \item{where the moon's host planet
advances \mbox{$\approx2.28^o$} [i.e. $\tan^{-1}(8\pi)^{-1}$]
around the sun in this time.}
\end{itemize}

This `yardstick' (three-part) relationship allows us to quantify
the `gravitational' geometric phase offset $\beta$ of a
third-body's atomic/molecular constituents. For cases where
$\theta\neq\tan^{-1}(8\pi)^{-1}$ over one lunar orbit then
$\beta\neq2\pi$. This is an important step in the eventual
determination of the total energy of each rotating space-warp, and
subsequently the Pioneer anomaly. In section
\ref{subsubsection:looking ahead} we outline the steps involved.

\subsubsection{Concluding and summary remarks for section
\ref{Subsection:Inertial}}\label{subsubsection:summary
topological} Of vital importance to the model is the fact that: in
quantum theory the inertial properties of a particle, atom, or
molecule are determined by its inertial mass \emph{as well as}
spin.

A core issue of this subsection is the physical relevance of the
virtual spin (geometric) phase offset upon the completion of a
closed loop/orbit; which then quantifies a rate of intrinsic
angular momentum, i.e. a virtual (spin) energy offset
--- within (and only within) an appropriate `third' (lunar)
celestial body. Fortuitously, the (lunar-based) atoms/molecules
(effectively) all experience the same virtual geometric phase offset
relative to the (pre-loop) initial phase or an unchanging/unaffected
phase. In the ``special circumstances" of our model, a frame offset
is present because each atom/molecule's \emph{actual} (local) spin
phase configuration (at the completion of a closed orbit) `has
become' non-inertial; with the virtual offset being indicative of the
departure from the inertial (spin) circumstance. This inertial vs.
non-inertial offset circumstance is dependent upon the local
existence of a \emph{spin status} that itself is influenced
(non-locally) by: the curved-spacetime-geodesic closed-path/orbital
\emph{motion} of (lunar-based) atoms and molecules, and all their
constituent elementary fermion particles.

This physical phenomenon has both local \emph{and} systemic/non-local
aspects. There is a systemic (or global) awareness of the (relative)
local asymmetry that is (then) collectively appeased externally (in
the environment) by way of both: a rotating space-warp, and an
accompanying non-local mass distribution. \emph{Individually},
neither quantum mechanics nor general relativity suggest that this
offset could exist, with the proposed phenomenon (necessarily)
involving features relevant to \emph{both} of these distinctly
different theories --- in particular QM spin, QM spin-orbit coupling,
non-local geometric phase effects, and curved spacetime.

The (aforementioned) `special circumstances' required include: (1) a
topological and geometric spin phase `precession' --- albeit virtual
--- that does not affect orbital (angular momentum) phase in any way;
and (2) a lack of decoherence triggered by the presence of gravitons,
which is distinct from the case with electromagnetism
--- where electrical charges are decohered by their own electrical
field. This is because a solely geometric approach to gravitation is
seen to deny the (very) existence of the graviton particle (recall
Section \ref{Section:PhiloTheory}). Thus, a significant quantum
coherence of the non-inertial offset effect and virtual fermion (and
atomic/molecular) condensate behaviour --- so vital to the model ---
cannot be ruled out (see section \ref{Subsection:Condensate}).

Further, the `fineness' of the spin phase offset (i.e. \mbox{$<2 \pi$
rad),} requires near exact self-interference of atoms/molecules after
closed loop motion. This is attained by way of: third-body celestial
(macroscopic) spin-orbit resonance\footnote{That is, lunar
synchronisation or `phase-lock' around its host planet (the
second-body).}, \emph{and} QM (EM dominated) spin-orbit coupling;
lunar geodesic motion; and an absence of atomic/molecular motion
relative to a lunar geocentre (i.e. the solid body content of a
moon).

The preliminary (empirical) quantification of the geometric phase
offset was also outlined. The determination of relative geometric
phase needs to go beyond a standard electromagnetic and QMs based
approach. Celestial orbital \emph{geometry}\footnote{This is in its
traditional (non-general relativistic) sense, involving: closed
curves, arcs, angles, etc.; i.e. plane Euclidean geometry as compared
to non-Euclidean geometry.} (alone) is sufficient to allow the
quantification of the spin phase offset in all actual
(\emph{three}-body) \mbox{moon-planet-sun} configurations. This is
achieved by way of recognising, and adjusting from, an idealised
(three-pronged) reference scenario that gives: a $2 \pi$ phase
offset, over the course of one lunar spin-orbit cycle, for a specific
planetary progression angle ($\theta$) around its (host)
Sun\footnote{Or more precisely, the solar system's barycentre or
centre of mass, which is (effectively) the three-body system's
`first-body' or (alternatively its) `central-body'.}. Later
(subsection \ref{Subsection:Model quantified internal}), we see that
the Earth's moon exceeds the $2 \pi$ decoherence limit, and thus it
doesn't have/generate an associated rotating space-warp.
%******************************************************************************************
\subsection{Energy, coherence and condensate behaviour in the model}
\label{Subsection:Condensate}

\subsubsection{Constructive superposition of virtual
phase}\label{subsubsection:constructive super} The geometric phase
effects upon neighbouring atoms/molecules in a celestial body,
arising indirectly from spacetime curvature, may be considered
equivalent because of their (effectively) parallel celestial
(orbital) motion. Certainly, the motion/path of lunar atoms is
very different to those of its `host' planet --- within a systemic
(barycentric) reference frame.

That the spin phase offset is common to all lunar atoms/molecules
means that effectively there is no geometric phase shift difference
between different lunar atoms/molecules\footnote{In other words, the
variation throughout a moon is negligible. A ``solid" body is also
implied, i.e. effectively negligible contributions from molten
material and water --- see subsection
\ref{subsubsection:non-solidity} for further discussion.}. Thus,
their respective phase shifts act in concert, i.e. coherently
yielding a type of constructive superposition --- albeit virtual.
Additionally, by denying the existence of gravitons, the coupling
between mass and gravitons is non-existent, and hence there exists no
limitation upon the coherence and superposition of (qualities and
quantities involving) spin and `mass'\footnote{In analogy to: ``If
the electromagnetic coupling between charge and photons is turned
off, then there exists no limitation on the coherence of
superpositions of charge \citep*[ p.1430]{Aharonov_67}."}. In our
case, this is indicative of a proportion of the (finite) minimum
intrinsic angular momentum of each and every elementary fermion
particle within each and every atom/molecule\footnote{This is
quantified in subsection \ref{subsubsection:maximum virtual}.}; that
in themselves are now appreciated to be not immune from an unforeseen
(global/systemic) inertial commitment/frame --- and from which they
`stand apart'.

Now, with the virtual phase shift representing a virtual (spin)
energy, constructive superposition allows a simple additive
summation of these individual energies to give the total energy
--- and this energy is then expressed externally, i.e. in the
environment, as a rotating space-warp.  Note that the rotation of the
space-warp is in the same sense/direction as the moon, so as to make
up for the system's `under-spin' relative to inertial (spin)
conditions. This scenario is seen to obey a global (i.e. universal or
systemic) conservation of energy principle, that encompasses both the
micro- and macroscopic realms (at the same `time' and in `unison')
--- i.e. QM and gravitational energies (respectively).

\subsubsection{A space-warp so as to disallow a global/non-local phase
change}\label{subsubsection:disallow} An external rotating
space-warp, with both an associated constant
acceleration/gravitational undulation amplitude ($\Delta a$) and
non-local `mass' distribution, effectively quantifies the energy
it takes to maintain/regain an `all-round' conserved energy
scenario. Additionally, the non-local mass is seen to exhibit
(something analogous to) a spreading of a quantum wave packet over
time (recall subsection \ref{subsubsection:Vlatko Vedral}). See
section \ref{Subsection:warp's mass} for how this (new concept of)
non-local mass, acting upon a given condensed/compact physical
mass in space, changes with radius away from a moon --- i.e. the
mass involved is a field based quantity, as is $\Delta a$.

For any QM system exhibiting `interior/internal' spin-orbit coupling,
`slaved' to a macroscopic third-body moving along a geodesic, the
departure of atomic/molecular-based spin from inertial conditions
(upon completion of a closed loop), due to external gravitational and
configuration conditions/circumstances, must remain an
\emph{external} gravitational condition. Rephrasing this, a systemic
or global curved space and motion \emph{input} effect produces a
systemic rotating curved space \emph{output} --- albeit a totally
different phenomenon. Remember, that with the `internal'/QM effects
of the celestial orbital motion essentially uniform throughout the
bulk matter of the lunar body, quantum mechanical coherence behaviour
exists. Thus, a new type of external `coordinated' behaviour is
conceivable --- in what is termed the ``environment" of the
(collective) QM sub-systems.

Non-locality is necessary (so as) to allow the internal and external
behaviours to coexist at (and over) the same (duration of) time
($\Delta t$).

\subsubsection{The benefit of a rotating space-warp}
It can be argued that the external physical effect/behaviour ensuing
from the lunar `constituent' atoms/molecules, allows these internal
QM sub-systems to be effectively (and continuously) in a stable
configuration --- that coexists with `unforced' movement along a
geodesic. The rotating space-warp acts to screen the state of each QM
sub-system (i.e. an atom or molecule) from the model's new non-local
(virtual) spin phase offset. Thus, in the presence of analog curved
space-time, `nature' \emph{always} ensures the on-going internal
stability of these QM systems --- particularly in the rare
circumstance of when a coherent effect exists \emph{below} both:  a
minimum change in energy levels, and a phase-based decoherence
threshold.

Although not physically \emph{measurable}, the total
(background/hidden) virtual QM energy \emph{quantifies} the energy of
the real (macroscopic) rotating space-warp --- by way of a global
principle of energy conservation.

\subsubsection{Inertial energy cf. fictitious force}
\label{subsubsection:inertial} The following discussion draws upon
the notion of inertial (or fictitious) forces\footnote{Fictitious
forces arise from the \emph{acceleration} of a non-inertial reference
frame itself. They are proportional to mass, and do not arise from a
(local) physical interaction.}, e.g. centrifugal force. We shall
(somewhat loosely) extend this notion to the energy discussed herein,
by way of there being a rate of `change' of (QM) momentum involved
(albeit a non-instantaneous change/process). In what follows, we
shall restrict the discussion of an inertial frame to the spin status
of quantum mechanical systems.

The term ``virtual energy" (as used previously) is now seen to be
indicative of the inability of a quantum mechanical system to
either: express or `match' the (global) inertial requirements of
spin energy. From a global/systemic perspective the energy is
real: representing an \emph{inertial energy}\footnote{In the sense
of inertial (or fictitious) \emph{force}.}; i.e. an `apparent'
energy resulting from the intrinsic angular \emph{momentum} of
atoms/molecules in a non-inertial frame of reference --- albeit
only upon the \emph{completion} of a closed circuit or cycle
(spanning a duration $\Delta t$). In the model, by way of a QM
(cyclic) rate of intrinsic angular momentum change, we establish a
relative (inertial to non-inertial frame) spin \emph{energy}
offset (i.e. discrepancy); with this energy magnitude now being
alternatively understood, and referred to, as an `inertial'
energy.

Unlike the case with ``inertial force" we are not primarily
dealing with macroscopic/classical ``mass in
motion"\footnote{Although we note that \emph{macroscopically}, the
motion of the atoms and molecules (slaved to a moon) is `along' a
(force-free) geodesic.}, nor `felt'\footnote{Unless we entertain
the notion that when this energy is expressed externally in an
atom/molecule's environment, it is `felt' there --- i.e.
`experienced' in the environment itself.} fictitious/inertial
forces, even though the energy is based upon an inertial to
non-inertial frame distinction. At the microscopic/QM level, the
non-inertial spin energy is not \emph{locally} `registered';
(firstly) because it is a non-local systemic effect (below a
`discrete' minimum energy level difference), and (secondly)
because there is no such thing as QM spin (rotational)
\emph{motion} in our understanding of: an elementary particle, a
subatomic particle, and an atom/molecule\footnote{Quoting Werner
Heisenberg: ``The atom of modern physics can be symbolized only
through a partial differential equation in an abstract space of
many dimensions. All its properties are inferential; no material
properties can be directly attributed to it. That is to say, any
picture of the atom that our imagination is able to invent is for
that very reason defective. An understanding of the atomic world
in that primary sensuous fashion \ldots is impossible".}.
Nevertheless, total virtual QM \emph{spin energy} is made
(externally) physically real, and this new term ``inertial energy"
nicely encapsulates a core conceptual feature of the new
mechanism.

One could possibly say the new mechanism exhibits
\mbox{``spinertia";} defined as the ability of a QM system, or a
collection of QM sub-systems, to export its/their non-inertial
(rate of intrinsic angular momentum) energy. This phenomenon
requires: QM self-interference, 3-body orbital (celestial) motion
in curved spacetime, a non-local (relative) geometric phase
change, as well as other special requirements outlined in
subsections \ref{subsubsection:relationships spin and},
\ref{subsubsection:three-body phase} and elsewhere.

\subsubsection{Decoherence, and the statistically additive nature
of total energy}\label{subsubsection:statistical induction} The
internal to external, and virtual to real, bifurcations of the
spin (inertial) energy previously discussed is supported by and/or
related to one central aspect of decoherence theory.

Standard QM decoherence theory speaks of the separation of a
quantum system into a subsystem (the relevant/distinguished part)
and its environment (the irrelevant/ignored part) \citep[
p.227]{Joos_03}. The (external space-time) environment plays a
special role in our new mechanism. Further, we note that:
\begin{quote} Decoherence follows from the irreversible coupling of the
observed system to the outside world reservoir. In this process, the
quantum superposition is turned into a statistical mixture, for which
all the information on the system can be described in classical
terms, so our usual perception of the world is recovered \citep[
introduction]{Davidovich_96}. \end{quote} The classical concept of
energy --- so central to QMs --- is fundamental to the model. To this
we add that within ``statistical physics" \emph{energy} plays a
primary and unique role, and total energy is attained
\emph{additively}.

Post-decoherence, QM effects are considered to be `delocalised'
rather then destroyed. Our proposed mechanism takes delocalisation
(of mass especially) to a conceptual limit, by allowing both the
constant acceleration/gravitational amplitude of the space-warp
and the (de-localised or) non-local mass component to extend to
the `end' of the universe --- albeit with the (non-local)
mass\footnote{As compared to gravitational mass (active or
passive) or inertial mass.} at a `point' in the field decreasing
with the volume of space enclosed around a (third-body) moon (i.e.
as $r^{-3}$).

Our denial of the graviton particle's existence removes the standard
particle-exchange means by which QM decoherence could be `activated'.
We shall consider the (neo-classical) space-warp as representative of
QM \emph{decoherence} only in the sense that it involves an internal
to external (physical) `information' exchange, and/or an internal to
external (physical) ``exchange" of energy `information' --- albeit an
exchange that occurs
`instantaneously'/non-locally\footnote{``Non-local" as in a process
that acts at the noumenal/hidden background `level', rather than at
the phenomenal particle-exchange level (recall subsection
\ref{subsubsection:non-locality}); and involving an unforeseen and
new/`foreign' (non-local) process that is yet to be fully understood
--- see section \ref{Subsection:warp's mass} for further
discussion.}, is ongoing, and is irreversible (i.e. one-way only).

\begin{quote} [Post-decoherence] (almost) all
information about quantum phases has migrated into correlations
with the environment and is thus no longer accessible in
observations of the subsystem alone \citep[ p.227]{Joos_03}.
\end{quote} Indeed, for the model's mechanism, its virtual phase
information is never directly (i.e. locally)
accessible\footnote{Further, virtual phase is not there (at the QM
level) to be measured, because (via non-local instantaneous
`causation') the external compensating rotating space-warp \emph{is}
present.}. This is further examined in subsection
\ref{subsubsection:quantum coher/entang} and it is not so much phase
that `migrates' into the environment, rather it is (some proportion
of the minimum `internally expressible') atomic/molecular spin energy
($\frac{1}{2}\hbar/\Delta t$) that is \emph{exported} (\emph{en
masse}) into the environment (or surroundings).

Note that all elementary fermion particles within an atom/molecule
share the same geometric phase offset; and it is this shared/common
phase offset that applies to the (composite) atom/molecule (i.e. as a
whole). This \emph{atomic/molecular} spin phase offset, over a cyclic
loop duration $\Delta t$, is indicative of a \emph{non-inertial} rate
of intrinsic angular momentum (i.e. spin energy).

\subsubsection{Quantum coherence, environmental energy, and quantum
entanglement}\label{subsubsection:quantum coher/entang} We shall now
further consider a rotating space-warp as representative of
\emph{quantum coherence} in some way; thus implying that the effects
of the internal (virtual) superposition are somehow `retained'. In
subsection \ref{subsubsection:statistical induction} we spoke of QM
effects being delocalised rather then destroyed.

The effective equivalence of the (virtual spin) phase offset for all
elementary fermion particles within an atom/molecule, and for all
atoms/molecules in a moon, means that the phase offset (relative to
that of an inertial frame) may be termed a `pure' phase offset rather
then a `mixed' phase offset. Further, the total \emph{energy}
associated with this phase offset --- which is proportional to the
number of atoms/molecules in a moon --- may be thought of as somewhat
like a pure quantum `state', albeit incapable of expression
internally (i.e. within a bound atomic/molecular QM system). Such a
pure coherent quantum state is well suited to a `singular' form of
expression, e.g. as an external QM `condensate'.

Consequently, we may think of the energy of a rotating space-warp,
with its associated non-local mass distribution, (loosely speaking)
as a single classical energy `state' in the (QM systems') environment
--- albeit a \emph{process} that is expressed over a finite cycle
time ($\Delta t$). Further, this (effectively) exact coherence
ensures that the total energy of the rotating space-warp is equal to
the simple additive \emph{sum} of the individual (atomic/molecular)
QM energies --- with these being (effectively) all equal.

One can go on to say there is a type of quantum \emph{entanglement}
between the (perfectly coherent) pure (offset) `state' of the
(atomic/molecular) QM sub-systems together, and the (singular) energy
process external to (i.e. in the environment of) this conglomerate
(lunar) QM ``super-system". It is the internal \emph{and} external
domains/realms\footnote{Quantum entanglement is usually discussed in
relation to two `objects' as compared to a distinction between micro-
and macroscopic realms/domains.} of the full/whole (conglomerate)
quantum system (i.e. a moon) that have to be described (non-locally)
with reference to one another, in the sense that these domains are
spatially separated\footnote{Thus, a dividing radius may exist to
demarcate where the external far-field begins. This is seen to be
loosely analogous to the EM wave situation where, at a particular
distance (from an aperture or slit), Fresnel diffraction (also known
as near-field diffraction) changes to Fraunhofer (far-field)
diffraction.}, and `instantaneously' coordinated in time
--- as far as observational physics is concerned.

Significantly, although the two domains/realms are physically
entwined, their formalisms remain distinct. Thus, we require both
quantum mechanical quantities (especially $\hbar$) and classical
quantities to express their (equal) energy magnitudes (see section
\ref{Subsection:Model quantified internal}). Indeed, the total
internal/virtual energy is our \emph{only} means for establishing the
environment's real energy magnitude --- involving both the
space-warp's acceleration magnitude, and the amount of non-local mass
(at a given radius) enclosed by a given volume.

Finally, for the new mechanism, let us additionally designate a (new)
quantum ``\emph{phase transition}", in the sense of a major
qualitative change in system behaviour. Transition is dependent upon
the amount of spin phase offset; with a solitary $>2\pi$ phase offset
(`interior' to any atom/molecule\footnote{Recalling that these are
the largest bound quantum mechanical systems.}) being enough to
induce a comprehensive (full lunar body) \emph{decoherence} `event'.
The different actual spin axis orientations within different
atoms/molecules (comprising a bulk lunar body) become apparent to the
global system, and the (externally orientated \emph{geometric}) phase
offsets are exposed as mixed or non-pure. Thus, `post'-decoherence,
i.e. $\beta>2 \pi$ over (a spin and) an orbital duration $\Delta t$,
a common virtual phase offset becomes meaningless and the lunar-based
rotating space-warps cannot occur. This is because the wiggle room
allowed by (the minimum quantum angular momentum) $\frac{1}{2}\hbar$
in Heisenberg's uncertainty principle has been breached, and
subsequently a moon reverts back to \mbox{non-QM} (i.e. exclusively
classical/macroscopic) behaviour
--- as is currently envisaged. At the celestial/macroscopic level
this is typical behaviour, as compared to the ``exception to the
rule" situation modelled herein.

\subsubsection{Concluding and summary remarks for section \ref{Subsection:Condensate}}
This subsection has built upon earlier work that involved: (1)
recognising a hidden (or noumenal) global/system background process
that observationally implies a form of time simultaneity, thus
facilitating non-locality (subsection
\ref{subsubsection:non-locality}); (2) accepting global conservation
of angular momentum and global conservation of energy (in
particular), (with the latter) involving spin energy in particular;
and (3) appreciating that a fractional analog geometric offset can
not be expressed by a `digital'/quantum ($\frac{1}{2} \hbar$) spin
effect.

For explanatory depth we have sought to extend certain QM concepts
such as: coherence, condensate, decoherence, and entanglement into
the model. With the geometric phase offset being (effectively) equal
throughout a (suitable) non-decohered moon, as is (subsequently) the
intrinsic angular momentum offset rate (per loop), the total
virtual/proper-fractional (quantum) energy involved can (simply) be
determined additively --- by way of multiplying the single
atomic/molecular energy value by the number of atoms/molecules in a
moon ($N_m$).

This (total) energy is effectively a non-inertial energy, somewhat
analogous to inertial/fictitious force effects. With a single
atomic/molecular $2 \pi$ phase offset triggering decoherence
throughout a moon, this ($2 \pi$) level of phase offset can be
associated with a new type of (quantum mechanical) `phase'
transition. Further, the same rate of spin phase (and rate of
intrinsic angular momentum) offset can be considered a coherent
effect throughout the lunar system --- albeit a virtual effect
`internal' to the moon, involving every single elementary fermion
particle nested within \emph{every single} (composite) atom and
molecule\footnote{That is, assuming the (non-rigid) moon is
completely comprised of `solid' material (cf. `fluid' material).}.

The (singular) resolution of this additive energy offset can be
understood by way of an analogy to quantum decoherence, in that we
make the universe a two-part system with the `relevant' part being
the lunar atoms/molecules, and the environment being all the
(cosmological) space that is external to these `sub-systems'. The
model suggests that: observations of these (atomic/molecular)
sub-systems alone will yield no sign of their individual
latent/virtual energies.

The external expression of the total lunar virtual energy as a
real energy, in the form of a rotating space-warp (RSW) and
non-local mass distribution, can be associated with a number of
things. Firstly, it appeases conservation of (universal spin)
energy; secondly, it illuminates the presence of non-local (and
instantaneous) spin entanglement; thirdly, we may argue that the
local (fermion-based) quantum coherence behaviour entails a
(wave-like) condensate behaviour (i.e. the RSW); and fourthly,
internal `information' is \emph{transferred} into the environment
in an \emph{irreversible} manner. This latter characteristic is
further supported by the need for a (classical) \emph{energy}
basis as a means to (or `currency' by which we can)
globally/systemically relate the internal atomic/molecular QMs to
an external neo-classical (new kind of
gravitational/accelerational field) effect.

The mechanism/effect being proposed involves both: a constant
acceleration/gravitational field amplitude ($\Delta a$), albeit
sinusoidally varying at a given `fixed' location in the field in
response to the space-warp's rotation around its source
region/`point'; and a variable non-local mass quantity [$m^*(r)$],
that accounts for dispersion of the energy at increasing distances
from its (lunar) source region --- see \mbox{Section
\ref{Section:Quantif Model}} for further discussion and
quantification.

We note that an \emph{external} curved spacetime dependent effect
upon (internal) atoms/molecules in (third-body) geodesic celestial
motion has resulted in a (curved space-time) effect also
\emph{external} to these QM sub-systems of a lunar bulk-mass. The
benefit of the presence of this ``insourcing--outsourcing" scenario
is that analog \emph{geometric} phase changes and digital QM energy
levels are never locally in conflict. In short, the environment
`deals with' (`fall-out' from) the non-flat spacetime based
effect/circumstance, and thus `nature' (\emph{in toto}) is inherently
\mbox{stable} in the presence of this potentially conflicting (analog
vs. digital) situation.
%************************************************************************************************
%************************************************************************************************
\section{General features and the quantification of the model}
\label{Section:Quantif Model} Having qualitatively justified the
model, in Sections \ref{Section:Status} to \ref{Section:general
model}, we now turn to the model's quantification. Firstly, we
briefly discuss errors and encapsulate the model's features; then
(beginning at section \ref{Subsection:Model quantified internal})
the `flesh' of quantification is applied to this framework.

\subsection{Discussion of errors and major features of the model}
\subsubsection{Looking ahead: steps to explaining the Pioneer
anomaly}\label{subsubsection:looking ahead} We have worked
`backwards' from the Pioneer observational evidence/constraints,
and assumed a `real' (non-heat based) anomaly. The model is able
to empirically \emph{match}, with the assistance of pure/classical
geometry:
\begin{itemize} \item{the physical characteristics of moon-planet-Sun
motions [e.g. lunar orbital period ($\Delta t$), lunar mass (or
rather number of atoms/molecules), semi-major axis (length),
planetary angular progression angle (per lunar orbit), etc.];}
\item{\emph{to the} geometric phase offsets ($\beta$) experienced
collectively (in a given moon) by all of its atoms and molecules
($N_{m}$), relative to the optimum\footnote{As in having the
maximum possible physical effect.} ($\pi$ rad) phase offset. This
offset is expressed as an energy efficiency factor $\eta$ (where
$0\leq\eta\leq1$).} \item{This efficiency factor then `quantifies'
the total (virtual) asymmetric/non-inertial internal energy
($\Delta E_w$), which is related to a time/process-based (maximum)
energy uncertainty within atoms/molecules; (and) via conservation
of energy, $\Delta E_w$ (the exact externalised energy) is
expressed as an external rotating space-warp and an
(initial/inception) non-local mass ($m^*_1$).} \item{The external
energy of each suitable moon is dependent upon: a (\emph{squared}
value of) acceleration/gravitational undulation \emph{amplitude}
($\Delta a$), a process time ($\Delta t$), and a
non-local\footnote{Non-local also in the sense of a wave's energy
distribution being distributed over a region of space.} mass
`continuity'/distribution quantification --- again by way of
employing (classical) plane geometry.} \item{Together, by way of
superposition, several of these rotating space-warps, of different
period ($\Delta t$) and amplitude ($\Delta a$), result in moving
`low mass' (celestial) bodies experiencing oscillatory/unsteady
speed perturbations --- around an equilibrium speed value governed
by general relativistic gravitation (recall section
\ref{Subsection:Shortfall}).} \item{This superposition of
individual unsteady motion (perturbation) effects, by way of the
coexistence of multiple rotating space-warps, `causes' the overall
shortfall (cf. predictions) in motion (over time) of the Pioneer
spacecraft --- i.e. the Pioneer (acceleration) anomaly
($a_{p}$)\footnote{Larger objects, such as moons and planets, are
not affected because their mass is above the cut-off masses of the
dominant (RSW `generating') moon-planet systems --- i.e.
Jupiter's: Io, Europa, Ganymede, and Callisto; Saturn's Titan; and
to a much lesser extent Neptune's Triton (see subsection
\ref{subsubsection:distrib eqn ramifi} and Table
\ref{Table:cut-off}).}.} \end{itemize} Although the approach used
is semi-classical, conceptually `vivid' and empirical, it fits all
the (awkward) observational evidence (see section
\ref{Subsection:Primary observational}). Additionally, it has
predictive capability (see section \ref{subsection:Extensions}),
and a significant further (dark energy related) ramification
(developed in Section \ref{section:Type1a}). Curved space-time,
independent of special relativity, plays a vital role. While such
a method is possibly anathema to mathematical physicists, this
approach is in all likelihood merely an intermediate step along a
path to fully understanding how the microscopic and macroscopic
domains, described by QMs and GR, interrelate in the physical
world --- to yield something new, unexpected, and progressive.

\subsubsection{Interim summary: Core issues of the model encapsulated}
\label{subsubsection:core} Arguably, \emph{the} core issue is
that: in the presence of curved spacetime, a macroscopic object
comprising innumerable quantum mechanical (atomic/molecular)
sub-systems can engage with its external surroundings to maintain
its on-going (`internal') stability. Thus, each sub-system is able
to ``carry on" (locally) as if the non-local effects of curved
spacetime upon its geometric phase, acting in concert with
background global space-time topological aspects, were absent.

The external expression of the (shared/common) internal (virtual)
spin phase offset\footnote{The (virtual) spin phase offset is
`relative' (in a broad sense of the word) to three things: i)
actual spin-orbit (coupled) phases; ii) a background datum; and
iii) itself, regarding: initial and (after $\Delta t$) `final'
cyclic-loop motion status.} involves both: an external
(gravito-quantum) rotating space-warp, quantified by a
\emph{specific} energy; and an external mass dilatory effect
(which is volume dependent). Together, these external effects
`match' (and substitute for) the total internal
``\emph{under-spin}" energy (per loop/cycle) --- with the latter
being a \emph{virtual} energy. Note that the phase offset is
associated with an angular momentum offset, and that the
under-spin is relative to an inertial circumstance/configuration.
Alternatively, we may say that an inertial frame \emph{over-spins}
relative to the actual QM spin phase --- which is held in
spin-orbit coupling by (the dominating) electromagnetic forces.

Ignoring special relativistic effects, the virtual internal and
\emph{real} external effects coexist in `universal\footnote{The
systemic (or global) perspective for rotating space-warps `takes
in' the whole universe.} real time'\footnote{Recalling section
\ref{subsection:SR's ontology}, this is a (hidden) background time
utilised by the Universe `in itself', especially for QM non-local
effects. It is not, and cannot be, indicated by direct
observational measurements involving EM radiation.}
(alternatively, a `systemic simultaneous pseudotime'\footnote{By
way of the American physicist John Cramer, we may alternatively
refer to this `other' time as a \emph{pseudotime}. In words
appearing upon John Gribbin's homepage (click on ``Quantum
mysteries", then ``Solving the mysteries"): ``Cramer's pseudotime
is a semantic device allowing us to stand outside of
(observational) time." As far as clocks in the everyday world are
concerned, any (pseudo-temporal) `gaps' between digital sequential
moments/frames (of observational/physical reality) are
instantaneous --- by way of them lying beyond observation-based
`apprehension'.}); with the space-warp rotating at the same rate,
and in the same rotational direction/sense, as a moon's spin. Note
that additionally a (`hidden background') global/systemic frame is
required to exist, and that the \emph{measured} effects of special
and general relativity must be upheld\footnote{They \emph{are}
upheld; (in part) because, on average and `overall', the general
relativistic gravitational field's strength is `unchanged' by the
space-warp's presence.}. All in all, conservation of the
\emph{full} system's energy is maintained by the field-based
`reaction' --- to each (moon-planet-Sun) system's (total) virtual
energy offset/shorfall.

The presence of both: shallow gravitational wells (pertaining to
moons and planets), and a spin-orbit coupled moon-planet system
(orbiting around a sun) are required --- with the moon being the
\emph{third} celestial body. Energy is seen as the linchpin that
relates a mechanical aspect of the micro/QM world to the macro
(`gravitational') world. The need to accept a (specific) ``energy
of a rotating space-warp" is unorthodox; further, the additional
acceleration/gravitational field associated with this energy is
supplementary to, and independent of, GR's spacetime curvature.
Thus, we are transcending the implicit restrictions of GR's
general principle of relativity, and the sole use of a metric
theory to explain `gravitational' influences --- in their widest
sense. This is further discussed in section
\ref{Subsection:EquivPr Comment}.

It is \emph{important} to note that locally the asymmetry-based or
offset-based virtual QM energy, although conceivable, does
\emph{not} physically/actually ``exist''. What does exist (in its
place) is the external rotating space-warp phenomenon.
Importantly, the internal virtual (and non-inertial) energy acts
to quantify the external energy magnitude. The importance of a
`universal' conservation of energy principle, to cope with this
situation, is paramount. Thus, at all times and in all
circumstances, the \emph{mass} based aspect (cf. electrical charge
based aspect) of QM systems moving in curved spacetime is
\emph{stable}. `All up', (discrete) QM systems obey SR, and can
always cope with the additional effects pertaining to: (analog) GR
and Heisenberg's (energy and time based) uncertainty principle
`demonstrated' herein --- at least in the universe's `mass era'.

The restriction of the energy magnitude to that: spanning/within a
finite time loop/cycle ($\Delta t$), \emph{and} involving an
angular momentum less than (half) Dirac's constant
($\frac{1}{2}\hbar$), assists (systemic) stability --- in that
things can only be minimally perturbed from an equilibrium
circumstance. By virtue of these (and other) circumstances, the
Pioneer anomaly has an extremely small magnitude ---
notwithstanding the vast number of atoms/molecules in each
(contributing) moon.

\subsubsection{Error discussion and aspects of the model's idealisation}
\label{subsubsection:Error and idealisation} This subsection
highlights how the Pioneer anomaly's error far exceeds the
\emph{numerical} error, and comfortably exceeds the total error,
of the/our \emph{idealised} model. Note that the Pioneer anomaly's
magnitude-to-error ratio is 8.74\,:\,1.33 (with both values
$\times ~10^{-8}~ \rm{cm~s^{-2}}$). Known to many more significant
figures are the physical quantities in the model's quantification:
Planck's (and Dirac's) constant, lunar masses, celestial body
orbital periods and distances. Further, the model's geometric
approach uses an \emph{exact} (celestial) angle:
$\tan^{-1}(8\pi)^{-1}$, and is free of adjustable parameters, i.e.
free of parametrization\footnote{Should the observational value of
the Pioneer anomaly be revised significantly, the model --- with
its $a_p$ value --- \emph{could not} be adjusted to fit this
altered $a_P$ value.}. Subsequently, the uncertainty or `error' in
the model's (eventual) quantification of total anomalous
acceleration ($a_p$) is well below the `observational' error
attributable to the Pioneer spacecraft's (measured) anomalous
acceleration ($a_P$).

Obviously, if the model is flawed, then this small error is
unjustified. Loosely speaking, the conceptual modelling `error'
(or rather inaccuracy) is (potentially) very much greater than the
model's numerical error.

The model, at this stage, is idealised and restricted; in point
form these (`simplifications') are: \begin{enumerate}[i)]
\item{Solar mass $>>$ planetary mass $>>$ lunar mass; noting that
the Earth's moon shall fail to be a (rotating) space-warp
`generator'.} \item{The relationship between lunar mass and its
number of constituent atoms/molecules is based upon Carbon 12 (and
any alteration in this relationship due to binding energy is
considered negligible).} \item{Large moons dominate the anomaly:
Jupiter's four Galilean moons and Saturn's Titan almost completely
quantify the anomaly.} \item{Moons are considered to be
(non-rigid) solid bodies, so as to not compromise the pure effect
of a common geometric phase offset. The minor ramification of a
(possible) variation from this idealisation is discussed in
subsection \ref{subsubsection:non-solidity}.} \item{Orbits are
considered to be circular. The influence of (small) planetary and
(very small) lunar orbital eccentricities is negligible (and
conceivably nonexistent) --- see subsections
\ref{subsubsection:Berry background} and
\ref{subsubsection:eccentricity}.} \item{In the global/systemic
reference frame, over a single lunar cycle (which is always $<17$
days), any change of the Pioneer spacecraft's (barycentric)
position, trajectory, and hence observational inclination angle,
has a negligible influence upon the anomalous acceleration (i.e.
\emph{rate} of speed shortfall).} \end{enumerate}

Assuming the idealised model is valid and (later)
adjustments/corrections to it (see Table \ref{Table:corrections}) are
also valid, the computational/numerical errors are very small
($<0.1\times 10^{-8}~ \rm{cm~s^{-2}}$). It is the veracity of the
model that is primarily at stake here, and further discussion of the
model's numerical errors is not pursued.

A possible modelling error also lies in corrections to the
idealised model. Later (in Table \ref{Table:corrections}) we see
that the adjustments/corrections to the idealised model for
Pioneer 10 are up to $0.56\times 10^{-8}~ \rm{cm~s^{-2}}$. Thus,
we get a (conservative) \emph{total error} magnitude of
\mbox{$0.66\times 10^{-8}~ \rm{cm~s^{-2}}$,} which is about half
the observational and experimental error of $1.33\times 10^{-8}~
\rm{cm~s^{-2}}$. If the idealised model's corrections are valid,
then a more refined (i.e. less idealised) model may well be
accurate to better than $0.10\times 10^{-8}~ \rm{cm~s^{-2}}$.

\subsubsection{The relationship between: QMs, GR, intrinsic
`spin' and conservation} For our model, the concept of intrinsic
spin has allowed an inter-linking between: (micro) QMs, (macro)
celestial motion, and `gravitation' (in its broadest sense);
whereas gauge field theoretical approaches have (to date) not
achieved \emph{this} type of (quantum mechanical to gravitational)
link, together with falsifiable predictions involving observable
physical phenomena. This linkage is achieved by way of a deeply
conceptual model involving: micro- and macroscopic spin-orbit
coupling\footnote{The latter involving the second and third bodies
of a (sun-planet-moon) three-body celestial system.}, a relative
(virtual) phase effect, (together with) a non-inertial vs.
inertial frame offset, conservation of (systemic) spin energy, and
a denial of the graviton particle's existence --- amongst other
things.

A non-\emph{special} relativistic formal representation of quantum
mechanical spin, in an atomic/molecular system, is independent of
(and additional to) Schr\"{o}dinger's equation\footnote{Note that
Dirac's equation incorporates \emph{special} relativistic effects,
but not general relativistic effects, and it provides a
description of elementary spin-$\frac{1}{2}$ particles.}.
Similarly, a virtual (lunar or) \emph{third-body}-based
(intrinsic) over-spin --- which itself is dependent upon: a common
(to all atoms/molecules) fermion spin phase offset, quantum
entanglement, and self-interference --- is independent of (and
beyond) \emph{general} relativity's formalism, but not necessarily
curved spacetime \emph{per se}.

Note that the (tiny) Lense-Thirring effect arises from a
\emph{central} body's spin, and similarly, the spin-down of a
binary pulsar (or dual-star) by way of gravitational wave `output'
is also a \emph{two-body} general relativistic effect. The new
mechanism features both: tidal effects of non-point-like
(moon/planet) masses, and requires a \emph{three-body} celestial
system; both of these features lie outside the `umbrella' of GR's
applicability.

It is as if (QM) intrinsic spin's failure to be a part of either
Schr\"{o}dinger's equation or GR's formalism (re: a third-body)
has paved the way for a micro-macro spin (conservation)
relationship to mitigate any effects arising from a (digital) QM
system existing in (analog) curved spacetime --- all the while
allowing Schr\"{o}dinger's (and Dirac's) equation and GR's
formalism to dominate their essentially separate domains. In
short, the general relationship between QMs and gravitation
appears to be one of (almost completely) detached coexistence,
rather then fully entwined (reductive) unification; and in order
to establish our (admittedly minor) physical link between
gravitation and QMs, a common ontological ground is required. The
dual-ontology pursued in this paper achieves this (see section
\ref{subsection:Reversal}). Also required is a reassertion of the
need for a global barycentric reference frame\footnote{In
conjunction with a systemic Barycentric Dynamical Time (TDB), with
this being how NASA treats the \emph{observations} of spacecraft
in the outer solar system --- cf. a (theory of) general relativity
based approach to S/C motion.}, and the resurrection of
conservation of angular momentum and energy principles (section
\ref{subsection:Noether}) that span both the micro- and
macroscopic realms (at the same time).

Interestingly, a global (classical-like) conservation of angular
momentum holds impressively in a (macroscopic only) moon-planet
system. The Earth's slowing spin rate leads to the Moon's orbital
radius (and period) increasing --- by way of celestial spin-orbit
coupling in conjunction with tidal effects and tidal locking.
\begin{quote} \ldots empirical observations, many of which actually
come from analysis of the LLR [Lunar laser ranging] data itself,
suggest that these conservation laws [energy, momentum, and
angular momentum] hold to high precision \citep[
A104-5]{Nordtvedt_99}. \end{quote} Note that this comment is made
in spite of the fact that for GR there is not a general law of
energy conservation, just a restricted version involving an
\mbox{$r\rightarrow \infty$} requirement (recall subsection
\ref{subsubsection:Failure of local}).
%***********************************************************************************************
\subsection{Extensions to, and peripheral issues influenced by, the
model}\label{subsection:Extensions} In this section a number of
the model's pragmatic features are briefly addressed, including:
its extension into three spatial dimensions, predictions, future
observations of interest, and ramifications of the model upon the
current understanding of various issues.

Note that any \emph{change} in a Sun-planet-moon's geometrical
(configuration) circumstances, that impacts upon the
offset/imbalance energy magnitude ($\Delta E_w$) and hence its
associated rotating space-warp, is propagated at the speed of
light. Over human and spacecraft lifetimes, the effect of changes
in planetary and lunar orbital characteristics upon the
model/mechanism (as presented herein) are negligible.

\subsubsection{Extending the model into three \mbox{spatial}
dimensions}\label{subsubsection:three spatial} We need to
distinguish three different aspects of the model's extension into
three spatial dimensions:
\begin{enumerate}[i)] \item{an internal quantum mechanical spin
aspect;} \item{an orthogonal extension to the (external) planar
rotating space-warp (discussed in Section
\ref{Section:PrelimModel});} \item{and motion of a spacecraft at
an angle to the plane of space-warp rotation.} \end{enumerate}

The orbital basis for a moon-planet offset in (overall\footnote{By
way of the same geometric phase offset applying to all elementary
fermion/matter particles within each atomic/molecular quantum
mechanical system.}) atomic/molecular geometric phase ($\beta$),
and subsequently virtual intrinsic angular momentum ($S_{z}$),
only considers a (planar) projection of spin. The
ensuing/associated rotating space-warp is initially idealised as a
two dimensional planar phenomenon. Due to the virtual nature of
the QM energy, corrections for the (internal) three dimensionality
of atoms/molecules, and their total spin magnitude, are \emph{not}
required.

The extension of the two-dimensional space-warp, we have been
considering (up to this point), into three spatial dimensions is
addressed in subsection \ref{subsubsection:3D external}.

Fortuitously, for (spacecraft) motion at an angle to the lunar
(and space-warp) plane of rotation, the motion shortfall along the
spacecraft's \emph{path} is unaffected\footnote{The Pioneer
spacecraft (S/C), whether moving parallel to the lunar orbital
plane, or at an angle to it, will `experience' the full strength
of the space-warp's cyclical acceleration amplitude. In subsection
\ref{subsubsection:3D external} we see that regardless of the
spacecraft's distance above or below the ecliptic plane, the
retardation effect (upon the S/C) is also the full effect.}; but a
minor correction for the \emph{geometric} inclination angle of the
spacecraft path to the  observational line-of-sight is required
(see subsection \ref{subsubsection:three dimensions} and Table
\ref{Table:corrections}). This correction employs the cosine of
the angle(s) between spacecraft path and line-of-sight
measurement. Two orthogonal angles are required: the first
(longitudinal) within the ecliptic plane, and the second
(latitudinal) involving inclination above or below the ecliptic
plane.

\subsubsection{Extending the planar rotating space-warp into three
dimensions}\label{subsubsection:3D external} 
\begin{figure}[h!]
\centerline{\includegraphics[height=10.0cm,
angle=0]{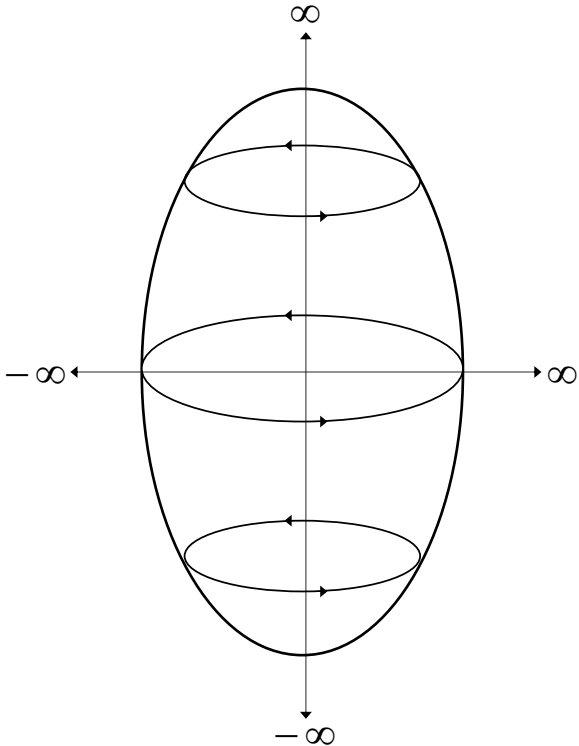}}
\begin{center}
\caption[Schematic diagram illustrating the cosmological extent of
the rotating space-warps in three spatial dimensions. Each and
every horizontal sectional disk extends to `infinity', with an
orthogonal depth (upwards and downwards) that also extends to
`infinity' --- i.e. the end of the universe.]{Schematic diagram
illustrating the cosmological extent of the rotating space-warps
in three spatial dimensions. Each and every horizontal sectional
disk extends to `infinity', with an orthogonal depth (upwards and
downwards) that also extends to `infinity' --- i.e. the end of the
universe.} \label{Fig:Ellipsoid}
\end{center}
\end{figure}

If we assume some
form of analogy with irrotational, non-viscous fluid flow (i.e.
potential flow), then by way of Kelvin's circulation theorem and
Helmholtz's theorems two things follow. Firstly, the strength of
the rotating space-warp, as indicated by the acceleration
amplitude, is constant along its length --- i.e. orthogonal to the
two-dimensional warp-plane. Secondly, because the space-warp does
not start or end on a boundary, it either extends to infinity or
forms a universal sized closed loop. Thus, the (infinitesimally
thin) planar rotating space-warp, implied by observations and
hypothesised in Section \ref{Section:PrelimModel}, is necessarily
a thickly rotating space-warp that effectively extends to
`infinity' --- in the warp plane and also orthogonal to this
plane\footnote{This extension to infinity, or the end of the
universe, might be mitigated by unforeseen/unknown effects.} (see
Figure \ref{Fig:Ellipsoid}). Energy dissipation/dispersion,
involving the dispersion of (non-local) \emph{mass} rather than
(constant/uniform) specific energy, is discussed in \mbox{section
\ref{Subsection:warp's mass}.}

%********************************** Fig:Ellipsoid *********************************
%********************************** end of Fig:Ellipsoid  *************************

The extension to a volume from a plane is not problematic because
we are dealing with a constant specific energy (at all points
throughout a plane) when quantifying the space-warp, i.e. $\Delta
e=\frac{1}{2}\Delta a^2 \Delta t^2$ (recall subsections
\ref{subsubsection:real_relative} and \ref{subsubsection:amplitude
to shortfall}). That the effect extends to `infinity' above and
below the lunar orbital plane was somewhat unexpected, but it
appears necessary if a (space) continuum mechanics approach is
employed. Compatibility with GR, via $\Delta e$'s (universal)
invariance, is maintained.

Interestingly, pursuing this conceptual analogy implies that
compressibility effects in fluid mechanics might be analogous to
special relativity's effects in space continuum mechanics, which
is compatible with our previous contention that `measured time' is
not necessarily all there is to `time'. The orthogonal extension
of the space-warp to `infinity' (or the end of the universe) is
consistent with our stance upon background time simultaneity
(recall section \ref{subsection:SR's ontology}) --- which was
introduced so as to `deal with' and more fully understand the
notions of QM (spin) entanglement and non-locality, both in
general and in the model.

\subsubsection{Briefly on different observations to improve our
understanding of $a_p$}\label{subsubsection:improve a_p} There are
a number of future space missions whose data may be beneficial to
an analysis of the Pioneer anomaly. \begin{enumerate}[i)] \item{A
dedicated mission to examine the Pioneer anomaly, e.g.
\citet*{Dittus_05}.} \item{The New Horizons mission\footnote{This
spacecraft lacks the navigational accuracy and precision of
Pioneer 10 and 11 due to thermal radiation effects, arising from
the proximity of the power supply to the spacecraft, but
nevertheless its motion allows certain hypotheses for the Pioneer
anomaly to be ruled \emph{out}.}, especially an investigation of
the ``Saturn jump" discussed in sub-section
\ref{subsubsection:Saturn jump} and \citet{Nieto_0710}. Is the
change in anomalous `acceleration' related to the Saturn encounter
itself, or is it simply a distance from the Sun effect? Assuming
the former, does the Pioneer anomaly begin at Saturn encounter or
do we have a sudden large increase from a much lower (non-zero)
value? Herein, the latter scenario is consistent with the model.}
\item{Further detailed analysis of past and future Earth
flybys\footnote{Occasionally referred to as an Earth `swingby'.}
by spacecraft\footnote{The gravitational field of the Earth is
much more accurately known, as compared to other planets such as
Mars.}, e.g. the ``Deep Impact" spacecraft in January 2008 and
2009. A clear confirmation of a real anomalous Earth flyby effect,
in preferably only some flyby cases, would possibly lend support
to a real (non-heat based) Pioneer anomaly. This view, and also
the existence of anomalous \emph{decreases} in K.E., is supported
by the papers of \citet{Anderson_08} and \citet*[ Section
3]{Nieto_09a} --- although \citet*[ p.174]{Turyshev_10a} are
sceptical. See subsection \ref{subsubsection:flyby anomaly} for
further discussion.} \item{Performance of the LISA Pathfinder
mission [a precursor to the Laser Interferometer Space Antenna
(LISA) mission]. The accurate positioning of the spacecraft (at
the Lagrange 1 point), to a few millionths of a millimetre in
space\footnote{Based upon an article by Will Gater in
\emph{Astronomy Now}, August 07, p.12.}, should be affected by the
anomalous acceleration ($a_p$).} \end{enumerate}

Finally, for a comprehensive overview of space missions (past,
present, and future) the \citet*{Lammerzahl_05} article is
recommended. Space-based (cf. Earth-based) missions investigating
various features of general relativity and issues fundamental to
physics are outlined and categorised. Interestingly, he
mentions\footnote{In relation to experiments examining possible
violations of Local Lorentz Invariance.} in the summary: ``\ldots
\,[observational] searches for anomalous couplings of
spin-particles with gravity." This type of `interaction' is
closely related to, although subtly different from, the mechanism
modelled herein.

\subsubsection{A conceivable extension of the model to explain the
Earth flyby anomaly}\label{subsubsection:flyby anomaly} There are
two points of view regarding the ``Earth flyby anomaly". Firstly,
a sceptical stance as favoured by P.G. Andreasian and S.G
Turyshev\footnote{As reported by Richard A. Lovett in Issue 21 of
the Australian popular science magazine ``Cosmos" (pp.78-81);
(titled) ``Magical mystery tour: The Pioneer anomaly".}, involving
an error in the computer code used to shift between Earth-bound
and space-based coordinate systems. This is done quickly in the
flyby and they believe flawed modelling results in a
velocity-dependent and (geocentric) latitude-dependent error.
Alternatively, \citet{Anderson_08} report that: \mbox{``\ldots the
Earth} flyby anomaly is a real effect inherent to the tracking of
spacecraft". The difference of opinion is stark.

The empirical relationship proposed by \citet{Anderson_08}
describing six Earth flyby events (for five spacecraft) is simple
and engaging. The remainder of this subsection outlines how the
rotating space-warps (RSWs), proposed herein to explain the
existence of a real Pioneer anomaly, are compatible with a real
Earth flyby anomaly. What is required is that the far-field
(roughly) ecliptic plane orientation of the RSWs be locally (i.e.
near field) \emph{distorted} so as to be aligned with the Earth's
equatorial plane. Considering the gravitational acceleration at
the Earth's surface ($9.81~\rm{m\,s^{-2}}$) is over ten billion
times stronger than the acceleration (amplitude) associated with
RSWs, local deformation/refraction of the plane of the RSWs in the
vicinity of a high mass body is conceivable. Also, by aligning
with the Earth's equatorial plane, the RSW is not inclined to the
orientation of the Earth's Lense-Thirring effect, and a ``least
action" circumstance prevails.

A distinctly local distortion of the RSW's orientation in the
vicinity of the Earth might also explain the apparent
distance/altitude dependence of the flyby anomaly, and (thus) why
an anomalistic effect might be acting over perhaps $\pm 10$ hours
of perigee --- as discussed in Section 1 of \citet{Anderson_10a}.

Appreciating that the Pioneer anomaly (as modelled herein) is
associated with both a speed \emph{and} distance shortfall per
cycle time of a RSW, and thus a distance travelled by a spacecraft
(S/C) approaching (and exiting an) Earth flyby encounter, we can
establish a length ratio ($\Delta L/L_\infty$). This dimensionless
ratio equates to the speed ratio\footnote{Note the difference in
the order of magnitude of these quantities: $\Delta V$ is measured
to the order of $[\rm{mm\,s^{-1}}]$, whereas $V_\infty$ is in the
order of $[\rm{km\,s^{-1}}]$.} $(\Delta V/V_\infty)$ at a standard
time rate, as utilised in Equation 2 of \citet{Anderson_08}. The
ratio $(\Delta V/V_\infty)$ is proportional to an (Earth-based)
proportionality constant and a \emph{difference} in (the cosine)
of geocentric inbound and outbound inclination angles.

This difference may be related to a spacecraft's inclination to
(both the RSWs and) the equatorial plane of the Earth (at
sufficiently low altitudes). For in-plane S/C motion the
\emph{observed} (Pioneer-like) shortfall is maximised. With
increasing angles to the plane of the RSWs (and Earth equatorial
plane) the speed shortfall, as measured by a line-of-sight
observation, is reduced
--- in proportion to the cosine of the (geometric) inclination
angle. Thus, for a flyby encounter, a reduction in geocentric
\emph{equatorial} latitude inclination (post- vs. pre-encounter)
leads to a greater (measured) shortfall and thus an apparent
anomalous \emph{reduction} in osculating velocity \emph{at}
encounter is observed\footnote{With $\Delta t \rightarrow
\rm{small}$ (for this process), this gives the impression that the
temporal duration is instantaneous.}. This change takes the pre-
and post-encounter change of (measured) far-field geodesic motion
circumstances into `account'. In the literature anomalous velocity
\emph{increases} are highlighted, indicating a reduction in the
\emph{measurable} (Pioneer-like) anomalous speed shortfall at
higher (post- vs. pre-encounter) inclination angles. Note that in
the far-field --- sufficiently far away from the Earth's
gravitational field --- a RSW's orientation refracts/reverts back
to its lunar-based plane.

In planetary flybys the ability of the rotating space-warp
hypothesis to provisionally accommodate: the sudden speed
variation `around' encounter; the existence of both anomalous
increases \emph{and} decreases in osculating velocity; and the
simple \emph{geometric} (directed) motion relationship presented
in \citet{Anderson_08} is a promising, but by no means (exhaustive
or) conclusive, account of the situation. A systematic effect may
still easily explain away the flyby anomaly, but the preceding
explanation, based upon a quirk relating to line-of-sight
observations rather than a (kinetic) energy conservation defying
change in speed, could satisfy both sceptics and
believers/accepters alike.

This application of our Pioneer anomaly based model appears to
fill the explanatory vacuum that currently exists with regard to
the Earth flyby anomaly.

\subsubsection{A brief outline of solar system-based predictions arising
from the model}\begin{enumerate}[i)] \item{A sharp increase in
$a_p$ (for Pioneer 11) post-Saturn encounter, cf. an (initial)
onset of the anomaly beyond Saturn --- recall subsection
\ref{subsubsection:Saturn jump}.} \item{In reply to oscillatory
perturbations in acceleration/gravitational field strength arising
from rotating space-warps, the LISA Pathfinder spacecraft should
experience both: a tiny (anomalous) reduction in speed, together
with some (tiny) quasi-stochastic changes in location around the
Lagrange point equilibrium value --- due to the underlying
(superpositioned effects of) variations in field strength and
hence LISA spacecraft speed.} \item{A comparison of paths
involving two asteroids of different mass, where one is
experiencing the Pioneer anomaly and the other (larger asteroid)
is of too great a mass to be affected. This relative comparison
would suffice to indicate if a ``cut-off" mass exists in support
of the model\footnote{Without the benefit of an Earth-pointed
antenna, it is not possible for the navigational tracking of
single asteroids to match the precision and accuracy of `in situ'
spacecraft navigational tracking. Onboard the spacecraft the
(second) `in situ' antenna (by way of a transponder) is in `phase
lock' with an Earth-based (transmitting and receiving) antenna,
which is referenced to a very accurate Hydrogen maser frequency
standard. This makes a world of difference to navigational
precision and accuracy.}. See subsection \ref{subsubsection:1862
Apollo} for further discussion.} \end{enumerate}

\subsubsection{A brief list of other (possible) ramifications
arising from the model}\label{subsubsection:Brief list} The
proposal of something in our solar system as noteworthy as a
(gravito-quantum) rotating space-warp --- that extends to infinity
or the end of the universe\footnote{Both in the plane of rotation
and orthogonal to this plane.} --- (unavoidably) has significant
effects upon numerous things, including the galactic and
cosmological motion of low mass bodies (and photons). The
space-warp's influence upon a variety of issues is now listed and
some brief comments are made. A fuller expos\'{e} shall not be
given herein. These implications remain underdeveloped and largely
unsubstantiated at this stage.
\begin{enumerate}[i)] \item{\textbf{Solar system formation}. Recall section
\ref{subsection:further concerns}.} \item{\textbf{Asteroid crater
size distribution}.} \item{\textbf{Near-Earth object (NEO) risk
assessment}.} \item{\textbf{Spiral galaxy density waves}. An
appreciation that there may be more than simply Newtonian
Mechanics (and dark matter) `at work' (gravitationally) in the
galaxy, opens the way for other energy expressions of/upon
gravitational fields. There is a distinct similarity between a
spiral density wave and a rotating space-warp\footnote{Jerry
Sellwood (Scientific American 21st Oct 1999, Ask the Experts) says
that: ``Fortunately, nearly everyone agrees that spiral density
patterns extract gravitational energy from the field of a galaxy."
This view is not inconsistent with the contention of
\citet{Prokhovnik_78} that: ``The gravitational property of matter
is associated with an energy field imbedded in a cosmological
substratum."} --- in terms of (both of them) being a
\emph{rotating} perturbation upon a stable gravitational field.}
\item{\textbf{Dark matter (DM)}. Once again going beyond Newtonian
Mechanics (NMs), it may be that (after stripping out DM effects)
the gravitational potential well of a galaxy is not as per NMs. A
galaxy comprises approximately 100 billion stars fairly evenly
distributed throughout. The extrapolation of 2-body NMs from the
solar system [with a dominant central Sun (comprising 99.86\% of
the total mass of the solar system)] to galaxies, may not be
valid. Hitherto unforeseen field interaction effects may
exist\footnote{Supported by a belief that NMs and GR are not the
`last word' in gravitation --- if (gravito-quantum) rotating
space-warps and non-local mass distributions are valid physical
phenomena.}, and thus dark matter might simply indicate a
misconceiving of galactic gravitational field strength. This is
speculative, but the lack of direct detection to date of (diffuse)
DM, in either Earth-based laboratories or the outer space of the
solar system, implies that DM may not necessarily `be' a missing
non-baryonic `particle'.} \item{\textbf{Cosmic microwave
background radiation (CMB radiation)}. Recall subsection
\ref{subsubsection:WMAP}. In particular, that a foreground effect
may arise from the anisotropic dipole \emph{motion} of the solar
system together with the existence of the model's (largest
possible scale) $360^o$-rotation symmetric rotating space-warps
(RSWs). Note that the latter, by way of Jupiter's Galilean moons,
are primarily aligned with the ecliptic plane of the solar system.
Furthermore, the wavelike nature of the RSWs (with amplitude
$\Delta a$ \emph{and} energy $\propto\Delta a^2$) could account
for the unexplained alignment of the quadrupole ($l=2$) \emph{and}
octopole ($l=3$) modes with each other and the ecliptic plane.}
\item{\textbf{Dark energy and cosmological accelerating
expansion}. In Section \ref{section:Type1a} a further ramification
is discussed, pertaining to electromagnetic radiation/photons,
rather than the retarding effect of oscillatory acceleration
perturbations ($\Delta a$) upon a mass in motion. Unlike the
anisotropic \emph{motion} based influence upon CMB radiation
previously discussed, this further effect is \emph{energy} based
and isotropic --- and pertains to EM radiation received from type
1a supernovae (standard candles). The (spherically symmetric)
\emph{non-local mass} (and hence energy) distribution
`surrounding' a GQ-RSW alters the energy (and hence frequency) of
approaching photons --- with the acceleration perturbations
playing no role whatsoever. The implication is that dark energy
may be a misinterpretation of the (redshift-based) observational
evidence, thus resurrecting the possibility that the expansion of
the universe is actually decelerating --- a far from outrageous
suggestion, albeit at odds with some aspects of the prevailing
``concordance model". Section \ref{section:Type1a} expands upon
this brief synopsis; it is exclusively devoted to this issue.}
\item{\textbf{Change in the fine structure \mbox{constant ($\Delta
\alpha$)}}. On a lesser note, an admittedly speculative and brief
conjecture regarding this open cosmological issue follows. Of
relevance may be the finding herein that spin geometric phase, but
not orbital phase, can be affected by the geodesic motion (of
atoms/molecules) in an external gravitational field ---
specifically, where the atoms/molecules are (a part of) the third
body of a rotating three-body (celestial) mass system. Under the
appropriate geometric conditions, this circumstance might disrupt
the (standard and cosmologically invariant) energy differences
pertaining to QM spin-orbital transitions `within' an
atom/molecule; thus affecting emission and absorption events. The
`fractional' quantum energy change involved would (then) give the
(observational) impression that the fine structure constant
(alpha) has changed from its standard laboratory
condition\footnote{This mechanistic approach to $\Delta \alpha$
regarding \emph{atoms and molecules} is indirectly supported by a
fairly recent null result \citep*{King_08} concerning the
cosmological evolution of another major dimensionless constant,
the proton-to-electron mass ratio ($\mu$). Admittedly, the status
of changes in dimensionless constants is made uncertain by way of
the complexities involved in processing the data.}. A good
background reference is \citet*{Murphy_03}\footnote{Their paper,
(and other) ``change in alpha" numerical results, are not
inconsistent with a mechanistic approach that conjectures the
(fractional quantum) ``wiggle room" is bounded by: $|\Delta
\alpha|\leq(\alpha/\pi)^2$ where $(\alpha/\pi)^2\approx
5.4\times10^{-6}$.}.}
\end{enumerate}

As can be seen from this list of implications, the influence of
(cosmological size) rotating space-warps --- each with an
attendant (or conjoint) non-local mass distribution --- is
conceivably quite far reaching and significant.

\subsubsection{Spacetime curvature energy \& the model's denial of
a graviton particle} In analogy with electromagnetic radiation
`waves': in GR, a time-varying mass-energy distribution leads to a
time-varying gravitational field, i.e. GR's gravitation waves.
Associated with this is a loss of angular momentum in binary
pulsars. The latter can occur with no mass actually lost from the
celestial bodies themselves, and thus we are dealing with a loss
of \emph{specific} orbital momentum and energy by way of specific
energy `radiated' into the gravitational field.

There is a clear distinction between saying that: \begin{quote}
The curvature of spacetime is itself a form of energy, which
produces its own gravitational field \citep[
p.229]{Harrison_00}.\end{quote} and the model's stance that: the
re-expression (or redistribution) of non-inertial (QM) energy, in
a universal system as a (macroscopic) curvature of space-time,
does \emph{not} then also produce its own gravitational field.
Herein the first of these two stances is discouraged.

The assumption of a graviton particle with an energy flux and
momentum flux, in analogy with quantum field theory and the photon
in EM, remains to be verified. In the absence of a graviton
particle, the transport of `momentum' within the field is
(technically) not possible. In subsections:
\ref{subsubsection:graviton}, \ref{subsubsection:E/M vs Grav},
\ref{subsubsection:graviton's absence},
\ref{subsubsection:ramification no graviton},
\ref{subsubsection:constructive super} and
\ref{subsubsection:statistical induction} we have argued that the
model actually depends upon the non-existence of the hypothetical
graviton (elementary particle) --- at least as far as it being a
decoherence `agent' in the model.

We are restricting the model's `physicality' to energy --- free of
any related `force'. In \mbox{subsection \ref{Subsection:warp's
mass}} the nature of the mass unit or `dimension', within the
physical quantities of energy and force, is more thoroughly
investigated. It shall become evident (later) that although a
rotating space-warp and its associated non-local mass distribution
has an energy representation, the model's mechanism does
\emph{not} have (nor need) an associated local (particle) momentum
aspect.

\subsubsection{A brief speculation concerning the origin of
inertial effects} The discussion in this subsection is decidedly
\emph{speculative}, but worthy of brief mention, due to the
ongoing debate that surrounds the basis of/for inertial effects.

That inertial effects require and relate to a global reference
frame is herein considered unavoidable. Additionally, macroscopic
mass is considered to be (in one sense) simply a summation of
atoms and molecules. Thus, with regard to inertia, the
\emph{quantity} of mass in QM bound systems shall not specifically
concern us; rather, it is the \emph{number} of atoms/molecules
(and their superposition) that is of major concern. The difference
is subtle, but it is conceptually important. The aim here is to
try and `reduce' the macroscopic concept of inertia to
non-macroscopic (i.e. QM) circumstances. Note that our concern is
\emph{not} with the ``origin of mass".

The content and discussion of \mbox{Section \ref{Section:general
model}} implies that `inertia', i.e. a resistance to a deviation
away from either rest or uniform motion, might have something to
do with a many `microscopic' body system (i.e. particles, atoms,
molecules) `settling' into some type of (minimum energy) global
phase coherence --- much in the manner that ocean waves, in the
absence of wind and other disturbances, settle into a (single
frequency and amplitude) ``ground swell"\footnote{At least upon
the ocean surface.}. Consequently, local macroscopic inertial
effects might be an indication of the global system's resistance
to any deviation away from a (long-term) `settled' systemic
phase-coherence (or phase-harmony). This situation is not
immediately obvious, because the phase status of QM (fermion)
`particles' are hidden from observational physics.

If this is the case, then only an indirect appreciation of inertia
has hitherto been recognised, hence its enduring mystery. This
(very raw) hypothesis relies upon the continual (and
non-local/instantaneous) outward `reach' of (non-observed QM)
particle mass effects in an `entangled' global system. It is
consistent with our stance on cosmological temporal
evolution/progression outlined in Section
\ref{Section:PhiloTheory}.

\subsubsection{Virtual phase and energy, imaginary numbers and
spatial dimensions} Our use of the `virtual' ties in with our
(complementary and noumenal) stance upon curved space involving
`imaginary' numbers in three spatial dimensions (discussed in
subsections \ref{subsubsection:summerview} and
\ref{subsubsection:conceptual ramifications}). There are strong
mathematical ties (or links) between complex numbers and geometry;
see for example work by Emeritus Professor David Hestenes on this
issue. Additionally, note that the standard notion of SR's and
GR's (`phenomenal') spacetime is unable to incorporate non-local
quantities.

\subsubsection{Some similarities in/of the model to magnetism}
\label{subsubsection:magnetism} That an external (lunar orbital or
spin axis) orientation, and the number of atoms/molecules, should
dictate the orientation and strength of the new effect
(respectively) has a number of similarities to magnetism.
\begin{enumerate} \item{Macroscopic magnetic properties arise by way of many
component/constituent atoms/molecules having the same magnetic
moments.} \item{As with paramagnetism, an \emph{external} field
ensures the direction of the macroscopic field.} \item{With
ferromagnetism, the magnetic field of a physical ``magnet" selects
a dominate/special direction in space. Analogous to this is the
planar nature of the space-warp's (initial) formal
representation.} \item{Magnetism involves angular momentum, both
directly and indirectly (i.e. only in the field). Quantum
mechanical (intrinsic) angular momentum is (arguably) the
foundation stone of the model.} \end{enumerate}

Finally, note that we are told to not accept either orbital, nor
spin, \emph{rotation} in an atom; and yet magnetic effects
necessarily pertain to the `rotational' motion of charged
particles, and hence magnetic dipole moments (in some sense at
least). A somewhat similar conceptual leeway is apparent (and
necessary) in our model (recall subsection \ref{subsubsection:new
precession}).
%*****************************************************************************************
\subsection{The physical model quantified: internal (virtual) energy}
\label{Subsection:Model quantified internal} The model has been
labouriously conceptualised, and its (potential) fecundity
displayed (section \ref{subsection:Extensions}). The emphasis now
shifts from a conceptual emphasis to the model's quantification,
with further conceptualisation included if/as necessary. We shall
derive $\Delta E_{w}$ --- the total energy offset magnitude,
or/and total non-inertial QM energy --- by way of quantum
mechanical (QM) considerations; then this (weighted) energy is
expressed both physically, and formally, as a rotating space-warp
(together/conjointly with an associated non-local mass
distribution). In equation form, we will see that (for each
individual rotating space-warp):
\begin{equation}\label{eq:2 by Energy}
\Delta E_w= \frac{1}{2}\hbar (\Delta t)^{-1}N_m \eta=\frac{1}{2}
m_{1}^* \Delta a_w^{2}\Delta t^{2} \end{equation} This (and
following) sections shall explain the various quantities in this
dual-equality. Only loop/cycle time ($\Delta t$) is common to the
very different internal (LHS) and external (RHS) forms of the
(equivalent) energy magnitude.

The nomenclature of $\Delta E$ is preferred over $E$, so as to
retain the form of energy uncertainty in Heisenberg's Uncertainty
Principle, and to signify both: an energy offset over the cyclic
duration time $\Delta t$; and secondly, an energy `transfer' (by
way of re-expressing the internal \emph{virtual} energy
externally\footnote{As a rotating space-warp, together with a
non-local mass distribution.}).

\subsubsection{A crucial difference between moons and atoms as
composite systems}\label{subsubsection:moons and atoms} An
atom/molecule as a whole is described by QMs, whereas a moon
\emph{as a whole} is not, even though it is comprised of
atoms/molecules (i.e. QM sub-systems) ---  recall subsection
\ref{subsubsection:by number}. An atom/molecule is a composite QM
system, (mainly) comprised of QM elementary fermion/matter
particles (parts). Regarding the model's virtual spin phase offset
we hypothesise that: unlike the case with a moon --- which is
(herein treated as) an additive sum of its (matter) parts --- an
atom/molecule as a whole `receives' the \emph{same} virtual spin
phase offset as does each (and every one) of its component
elementary fermion/matter particle parts.

\subsubsection{The total virtual non-inertial `spin' energy (for
a moon as a whole)}\label{subsubsection:maximum virtual} We have
argued that in suitable curved spacetime conditions, the
projection of atomic/molecular (spin) angular momentum in a plane
can become slightly non-inertial by way of a geometric phase
offset --- but only in a minimal/virtual manner, and only from a
systemic (or global) perspective. With orbital angular momentum
unaffected we have: \mbox{$\Delta J_{\rm{max}}=(\Delta
S_z)_{\rm{max}}=\frac{1}{2}\hbar$.}
\begin{quote} Electron's, and also some other fundamental particles
(protons, neutrons) have a spin whose magnitude is
$\frac{1}{2}\hbar$. This is found from experimental evidence, and
there are theoretical reasons showing that this spin is more
elementary than any other, even spin zero. The study of this
particular spin is therefore of special importance
\citep{Dirac_58}.\end{quote} In keeping with subsections
\ref{subsubsection:Large number} and \ref{subsubsection:three-body
phase}, and Section \ref{Section:general model} in general,
$(\Delta S_z)_{\rm{max}}$ corresponds to a (spin-based inertial
vs. non-inertial/actual `frame') geometric phase offset of
$\beta=\pi$, indicating an offset of one quarter of one fermion
wavelength ($4\pi$). For $\beta=0~\rm{or}~2\pi$, $\Delta S_z=0$.
Indeed, for $\beta\geq2\pi$ we have $\Delta S_z\rightarrow0$ by
way of decoherence, thus denying any (global presence of) virtual
QM energy.

With the angular momentum offset determined over a finite cyclic
process time ($\Delta t$) the \emph{maximum} virtual spin energy
offset, for a \emph{single} atom/molecule, is simply:
\begin{displaymath} \Delta E_{\rm{max}}=\frac{\Delta J_{\rm{max}}}
{\Delta t}=\frac{1}{2}\hbar (\Delta t)^{-1}
\end{displaymath} so that $\Delta E_{\rm{virtual}}\leq
\frac{1}{2}\hbar (\Delta t)^{-1}$. Compare this expression to one
form of Heisenberg's uncertainty principle: $\Delta
E_{\rm{real}}\Delta t \geq \frac{1}{2}\hbar$. The latter (HUP)
expression concerns the non-definite energy of a quantum
mechanical \emph{state}, whereas the former concerns a virtual
energy (i.e. a hidden systemic variable) --- of a definite/exact
magnitude.

Unlike the case with $\Delta p \Delta x \geq \frac{1}{2}\hbar$,
$\Delta t$ is \emph{not} an operator belonging to a particle; it
is an evolution parameter. Thus, in addition to (non-reversible)
measurement, the mechanism proposed herein is effectively a second
(and new type of) non-reversible time evolution of a quantum
mechanical ``system".

Note that the `mass' dimensional aspect of the non-inertial energy
offset has its basis in Planck's constant, which in turn is
indicative of the (`internal') `mechanics' of atoms/molecules ---
which herein (recall subsection \ref{subsubsection:by number}) are
considered to be the largest (bound) quantum mechanical system,
and the \emph{only} types of QM \emph{system} directly affected by
the new mechanism.

\begin{figure}[h!]
\centerline{\includegraphics[height=8.7cm, angle=0]{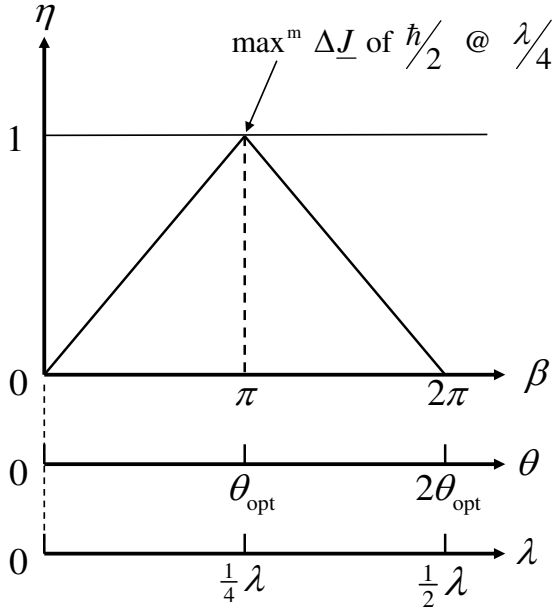}}
\begin{center}
\caption[Diagrammatic representation of the variation in the
proportion of (or efficiency with which) the smallest possible
\emph{quantum} intrinsic angular momentum ($\frac{1}{2}\hbar$)
offset (occurs).]{Diagrammatic representation of the variation in
the proportion of (or efficiency with which) the smallest possible
\emph{quantum} intrinsic angular momentum ($\frac{1}{2}\hbar$)
offset (occurs). Angular momentum magnitude is functionally
related to a QM geometric phase offset ($\beta$), which coexists
with a planetary progression angle ($\theta$) (also see Figure
\ref{Fig:Helical}). Full efficiency (i.e. $\eta=1$ and an angular
momentum offset of $\frac{1}{2}\hbar$) occurs at: $\beta=\pi$
radians, and $\theta=\theta_{\rm{opt}}=\frac{1}{2}\tan^{-1}({8
\pi})^{-1}$; with this value also representing a $\frac{1}{4}$ of
a (fermion) wavelength offset. For values where: $\theta>2
\theta_{\rm{opt}}$, $\beta>2\pi$ --- representing more than a
$\frac{1}{2}$ quantum wavelength offset --- quantum decoherence
occurs and thus $\eta=0$ (effectively).} \label{Fig:Triangle}
\end{center}
\end{figure}

For a non-optimal effect an efficiency factor $\eta$ is
introduced, where \mbox{$0\leq\eta\leq 1$}, such that $\Delta
J=\frac{1}{2}\hbar \eta=\frac{1}{2}\hbar_w$. We shall refer to
$\frac{1}{2}\hbar_w$ as the weighted intrinsic angular momentum
offset (see Figure \ref{Fig:Triangle}).

%********************************** Fig:Triangle ***********************************
%********************************** end of Fig:Triangle  *************************
%********************************** Fig:Helical **********************************
\begin{figure}[h!]
\centerline{\includegraphics[height=9.0cm, angle=0]{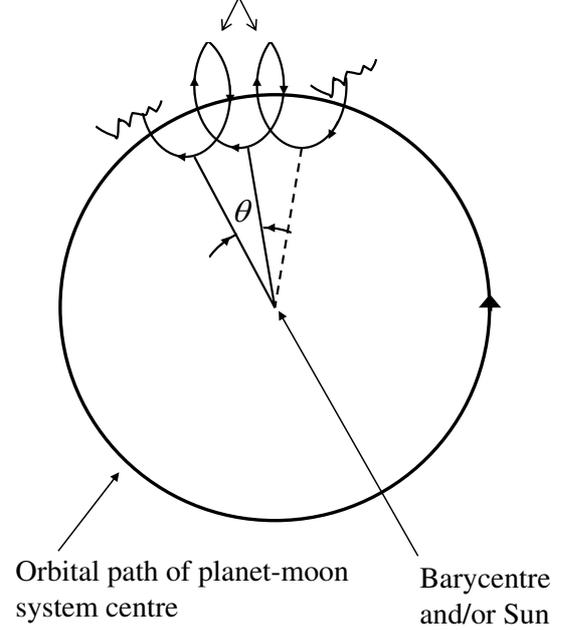}}
\begin{center}
\caption[Celestial (curved space) geometric constraints of the
model are indicated by this diagrammatic representation of
planetary progression angle ($\theta$), i.e. the angular
progression of a planet around the Sun --- over the duration of
one lunar spin-orbital cycle.]{Celestial (curved space) geometric
constraints of the model are indicated by this diagrammatic
representation of planetary progression angle ($\theta$), i.e. the
angular progression of a planet around the Sun --- over the
duration of one lunar spin-orbital cycle. Note that the view is
(downwards) from the north ecliptic pole. Theta i.e. $\theta$ (cf.
$\theta_{\rm{opt}}$) both: establishes (and coexists with) the
geometric phase offset ($\beta$), and the proportion (or
efficiency) of a minimum angular momentum ($\frac{1}{2} \hbar$)
that is (virtually) affected --- previously illustrated in Figure
\ref{Fig:Triangle}.} \label{Fig:Helical}
\end{center}
\end{figure}
%********************************** end of Fig:Helical  *************************

Further, with the offset applying (concurrently) to every (single)
atom/molecule in a moon ($N_m$), this gives an (additive based)
\emph{total} `weighted' energy of:
\begin{equation}\label{eq:internal energy}
\Delta E_w= \frac{1}{2}\hbar (\Delta t)^{-1}N_m \eta
\end{equation} This represents the total virtual energy
imbalance; it is locally hidden, but resident within the (wider)
system. The maximum (or optimum) available energy for a given
\emph{moon} is $\Delta E_o$ so that $\Delta E_w=\Delta E_o \eta$.

Note that with $\Delta E_w\propto N_m$ the largest moons of the solar
system (in general) have the largest energy (excess), and their
acceleration amplitudes dominate the response of the Pioneer
spacecraft. This response is due to undulations in the strength of
the acceleration/gravitational field, arising from a \emph{number} of
rotating space-warps, that are themselves centered upon a moon-planet
system\footnote{It is unclear as to whether the mechanism's `centre'
is: the moon's centre, or host planet's centre, or at the pair's
centre of mass. Most likely, it is at the moon's centre.}. Rotating
space-warp acceleration amplitudes associated with the large moons of
Uranus are too small to make any significant contribution to the
Pioneer anomaly; see subsection \ref{subsubsection:geometric acceler}
for further comment. Consequently, the remainder of this Section's
quantification excludes data based upon Uranus' large moons,
particularly Tables \ref{Table:angular progression},
\ref{Table:Efficiency}, \ref{Table:energy} and
\ref{Table:acceleration}.

%***************************** Table 1 and 2 ********************************
\begin{table*}[t]
\small \caption{Angular progression of planet, around the Sun, for
one lunar orbit. \label{Table:angular progression}}
\begin{center}
\begin{tabular}[t]{lcrrrrrrc} \hline
MOON$^a$              &\textit{Units}       &Luna$^{b}$ &Io$^{c}$
&Europa$^{c}$    &G'mede$^{c}$   &Callisto$^{c}$ &Titan$^{d}$
&Triton$^{e}$
\\\hline
Moon orbit frequency  &($10^{-6}{\rm s^{-1}}$) &0.42362 &6.5422     &3.2592          &1.6177           &0.69351         &0.72586   &1.9694$^f$ \\[0.5ex]
Moon's orbital period   &(days)               &27.3217     &1.76914   &3.55118        &7.15455         &16.68902       &15.94542     &5.87685  \\[0.5ex]
Host's orbital period   &(days)               &365.256       &4332       &4332            &4332             &4332            &10759         &60190 \\[0.5ex]
Orbital time ratio    &($\%$)            &7.480        &0.0408      &0.0820           &0.1652        &0.3852           &0.1482         &0.0098 \\[0.5ex]
Angular progression   &(deg)            &26.93        &0.1470     &0.2951          &0.5946           &1.3869          &0.5335        &0.0351 \\[0.2ex]\hline

\end{tabular}
\end{center}
\begin{center}
$^a$Lunar data is taken from NSSDC (National Space Science Data
Center web site), with lunar orbital time/period rounded off to 6
or 7 significant figures. $^b$Earth's moon. $^c$Large moon of
Jupiter. Note that `Ganymede' is abbreviated to G'mede. $^d$Large
moon of Saturn. $^e$Large moon of Neptune. $^f$Retrograde motion.
\end{center}
\end{table*}

\begin{table*}[t]
\small \caption{Geometric effectiveness/efficiency of various
Sun-planet-moon motions. \label{Table:Efficiency}}
\begin{center}
\begin{tabular}[t]{llcrrrrrrr} \hline
MOON &Symbol &\textit{Units} &Luna &Io    &Europa    &Ganymede
&Callisto &Titan &Triton
\\\hline

Angular progression &$\theta$ &(deg)  &26.93   &0.1470     &0.2951 &0.5946   &1.3869  &0.5335 &0.0351 \\[0.5ex]
Efficiency$^g$      &$\eta$ &(--)       &0.00    &0.1290 &0.2590
&0.5219   &0.7826          &0.4683 &0.0309
\\[0.5ex]\hline
\end{tabular}
\end{center}
\begin{center}
$^g$ Efficiency by way of optimum (planet) progression angle of
$\theta_{\rm{opt}}\approx1.13926$ deg
[\mbox{i.e.\,\,$\frac{1}{2}\tan^{-1}(8\pi)^{-1}$}]. A triangular
relationship is utilised to determine efficiency --- indicative of
a (quantum) wave-to-wave offset.
\end{center}
\end{table*}
%********************************************************************************

\subsubsection{The relationship between: phase, \mbox{angular} momentum, and efficiency}
A simple \emph{triangle function} is used to relate the
(continuous/analog) phase offset ($\beta$) to its (discrete
minimum and/or maximum \emph{virtual}) angular momentum
`counterpart' $\frac{1}{2}\hbar$ (see Figure \ref{Fig:Triangle}).
When \mbox{$\beta=\pi$} we have $\eta=1$; thus, when
\mbox{$\beta=\pi/2$} or $3\pi/2$ we have $\eta=1/2$, and when
\mbox{$\beta=0$} or $\beta\geq2\pi$ then $\eta=0$. This triangle
function is seen to indicate a wave-to-wave effect representing QM
(fermion) wave mechanics involving an inertial to non-inertial
intrinsic angular momentum (offset). Recall that for motion along
a geodesic, (QM `systemic') intrinsic angular momentum exists in
relation to an orbital angular momentum --- which itself remains
unchanged and thus also acts as a stable (i.e. unchanged)
reference frame or datum.

\subsubsection{A non-metric approach, but very much a geometric approach}
\label{subsubsection:Non-metric geometry} The model's external
re-expression of the energy `asymmetry' ($\Delta E_w$) is not
expressible by way of the exactitude of a (dynamic) metric theory,
and \emph{geometry} (necessarily) becomes the model's guiding
aspect or `principle'. A `frictionless' equation of motion is
fundamental to Newtonian and relativistic gravitation theory;
(but) herein the equation of motion merely defines the (geodesic)
path of celestial bodies, and hence the \emph{path} of their
constituent atoms and molecules.

The supplementary (i.e. perturbation) curvature, relative to the
pre-existing gravitational field (i.e. the field already
established by way of GR), causes a `dissipation' of a moving
body's kinetic energy; (or) in other words, the rotating
space-warp leads to a shortfall vs. predicted motion --- recall
section \ref{Subsection:Shortfall}. The field's cyclic nature can
still be represented by a field energy, but (recalling Section
\ref{Section:PhiloTheory}) it is a field that in its idealised
representation utilises a background (Eulerian) space continuum
(and formalism) rather than a spacetime continuum and GR's
formalism. Hence a secondary non-Euclidean geometry is (also)
conceivable, but it cannot be quantified by way of a metric
theory. Our (supplementary) continuum mechanics approach is
further discussed in \mbox{subsection \ref{subsubsection:two
non-Euclidean}.}

\subsubsection{Celestial geometry and $\frac{1}{2}\hbar$
efficiency}\label{subsubsection:Celestial geometry and} The
geometry of three-body celestial motion determines $\theta$, the
planetary angular progression angle (around the Sun, per lunar
orbit). This was discussed previously in subsection
\ref{subsubsection:three-body phase} and the results for major
solar system moon-planet systems appear in Table
\ref{Table:angular progression}.

By way of comparing this angle to the optimum planetary
progression angle ($\theta_{\rm{opt}}$) of
\mbox{$\frac{1}{2}\tan^{-1}(8\pi)^{-1} \approx1.14^o$} --- which
also represents a QM relative spin phase offset of $\pi$ radians
--- we may establish the efficiency of this planetary progression
angle, i.e. the proportion of $\Delta
J_{\rm{max}}=\frac{1}{2}\hbar$ `in play' systemically. The
efficiency values for different moons are presented in \mbox{Table
\ref{Table:Efficiency}.} Figures \ref{Fig:Triangle} and
\ref{Fig:Helical} provide visual assistance, (respectively)
clarifying the \emph{efficiency} value ($\eta$) and planetary
\emph{angular progression} per lunar orbit ($\theta$).

%************************************ Table 3 ***************************************
\begin{table*}[t]
\small \caption{Weighted QM `non-inertial' energy per orbit
(equaling space-warp energy) and associated values.
\label{Table:energy}}
\begin{center}
\begin{tabular}[t]{llcrrrrrr} \hline
MOON  &Symbol  &\textit{Units}   &Io    &Europa    &G'mede
 &Callisto &Titan &Triton \\\hline
Moon orbit frequency ($\Delta t^{-1}$)   &$f_{\rm m}$  &($10^{-6}{\rm \,s}$) &6.5422 &3.2592 &1.6177 &0.6935 &0.7259 &1.9694 \\[0.5ex]
Mass                   &M                   &($10^{21}{\rm \,kg}$)     &89.32     &48.00   &148.19     &107.59  &134.55  &21.4  \\[0.5ex]
Number of atoms/molecules$^h$  &$N_{m}$              &($10^{49}$)          &5.379    &2.891  &8.924      &6.479   &8.103 &1.289 \\[0.5ex]
Optimum Energy$^i$     &$\Delta E_{o}$            &($10^{9}{\rm \,J}$)     &18.555    &4.968   &7.612      &2.369 &3.101  &1.338 \\[0.5ex]
Efficiency &$\eta$  &(--)    &0.1290 &0.2590 &0.5219 &0.7826
&0.4683 &0.0309 \\[1.2ex]
Weighted Energy  &$\Delta E_{w}$  &($10^{9}{\rm \,J}$) &2.395
&1.287 &3.973    &1.854 &1.452 &0.041
\\[0.5ex]\hline
\end{tabular}
\end{center}
\begin{center}
$^h$ Based on carbon 12 molar mass and Avogadro's number per gram
mole ($6.0221415\times 10^{23}\rm{~mole^{-1}}$). $^i$ A full unit
($\eta=1$) of excess (virtual) spin energy ($\frac{1}{2} \hbar
\Delta t^{-1} N_{m}$) applies --- where $\hbar =
(2\pi)^{-1}(6.6260693\times 10^{-34})\rm{~J~s}$.
\end{center}
\end{table*}
%************************************************************************************

\subsubsection{On the effect of lunar and planetary orbital motion
eccentricity}\label{subsubsection:eccentricity} Note that in this
paper both lunar and planetary eccentricities are idealised as
zero, i.e. $e\rightarrow 0$ and orbits are thus treated as
circular. If \mbox{$e\neq0$} then the relationship between angular
progression and time is variable. It is not the case that variable
angular progression leads to variable efficiency ($\eta$), and
hence variable space-warp amplitude ($\Delta a_w$) and energy
($\Delta E_w$); because a variable rate of lunar and planetary
angular progression simply leads to a non-constant rate of angular
frequency/velocity for the rotating space-warp.

Examination into the possible existence of this second-order
temporal variation in $a_p(t)$ is complicated by the coexistence
of multiple RSW `signatures' in the $a_p(t)$ data, and
subsequently this minor issue shall not be pursued in this paper.
Importantly, the acceleration/gravitational field amplitudes and
the rate of speed shortfall (\emph{per cycle}) are
\emph{unaffected} by orbital eccentricities. Similarly, a very
minor second-order temporal variation of $a_p$ over Jupiter's
orbital time (11.86 years), due to the orbital eccentricity of
Jupiter (and its four large attendant Galilean moons) alone, could
also be present in the Pioneer data; but with just under 12 years
of data in the most accurate Pioneer analysis \citep[ Figure
14]{Anderson_02a}, the `noise' of the Pioneer
data\footnote{Arising from \emph{both} measurement noise, and the
inherent variation/wandering of $a_p(t)$ around the long-term mean
value ($\overline{a_p(t)}=a_p$), proposed/hypothesised by the
model (recall subsection
\ref{subsubsection:Velocity_vs_Doppler}).} means that this
(conceivable but) very minor effect is (also) beyond observational
recognition.

\subsubsection{Total (idealised) internal energy} Having determined
the efficiency of the internal (spin) energy offset, the total
virtual offset energy or weighted energy $\Delta E_w=
\frac{1}{2}\hbar (\Delta t)^{-1}N_m \eta$ (over time $\Delta t$)
pertaining to each moon may be found --- see Table
\ref{Table:energy}. Note that the number of atoms/molecules
($N_m$) in each moon is idealised so as to be based on carbon 12
molar mass, with Avogadro's number (per gram mole) (also) playing
a vital role. \emph{Significantly}, this idealisation has no
effect upon the all-important acceleration perturbation amplitudes
determined in section \ref{Subsection:Model quantifies external}
and presented in Table \ref{Table:acceleration}; whereas the
accuracy of non-local mass (and mass cut-off) values, established
in section \ref{Subsection:warp's mass} and presented in Table
\ref{Table:cut-off}, \emph{are} affected by the inaccuracy arising
from this all-inclusive idealisation\footnote{Not so much
Ganymede, Callisto and Titan, but more so the relatively denser
moons of Io and Europa.}. Fortunately, the low mass of the Pioneer
spacecraft --- much lower than any ``cut-off" mass --- makes this
inaccuracy ``of no consequence".

\subsection{The physical model qualified}

\subsubsection{Two types of equivalence}
\label{subsubsection:Violation of Eq.Pr} GR's curved spacetime
makes Newton's gravitational ``force" an inferior and approximate
approach to gravitation. Herein, we `decompose' the classical
quantity ``force" into its separate mass and acceleration
parts/components. From subsection \ref{subsubsection:field to S/C}
we recall that (model-based) acceleration (herein) refers to
either: a (monotonic) rate of change in (or rate of loss of)
translational velocity ($\delta a$), and/or a sinusoidal/cyclic
field curvature perturbation (with amplitude $\Delta a$) `at' the
spacecraft (over time $\Delta t$). In the model, mass and
acceleration have/play distinct and separate roles, and force
(\emph{per se}) plays no (physical) role whatsoever.

Standard gravitation (GR) employs two types of equivalence
\emph{together}: primarily, (inertial and gravitational)
\emph{acceleration}; and secondly, (inertial and passive
gravitational) \emph{mass}\footnote{With this equivalence dating
all the way back (at least) to Isaac Newton and experiments by
Galileo Galilei.} equivalence. The former, and historically more
recent equivalence, represents something of a ``dematerialisation
of gravitation", in the sense that the free-fall motion of a
(point) mass in a gravitational field is independent of the amount
of (passive gravitational) mass; i.e. we deal solely with
spacetime curvature, which can be conceptualised as a
gravitational acceleration acting upon a body. Section
\ref{Subsection:EquivPr Comment} further examines these
equivalence principles.

\subsubsection{Introducing non-local mass}
\label{subsubsection:non-local inertial mass}The non-local
probabilistic nature of a QM mass (preceding its observation) ---
for example, Schr\"{o}dinger's equation applied to an unobserved
free particle --- is clearly distinct from both:
\emph{condensed/compact} (macroscopic) matter, e.g. a planet, a
spacecraft, a comet, or a piece of chalk; and from gaseous and
plasma matter. The model requires a \emph{new} type of mass that
is similar and yet different to the case of an (unobserved) free
particle. We hypothesise that if the \emph{virtual} nature of the
model's total QM energy offset ($\Delta E_w$) --- via its basis in
energy uncertainty --- is externalised as an energy field, then
the mass aspect of this energy field is always ``non-localised";
in the sense that the probability of finding a particle with mass
anywhere in the field goes to \emph{zero}. This field `mass' is
(thus) effectively non-local. The model requires this external (to
atoms/molecules) distributed mass effect\footnote{Possibly,
``non-observable non-particle effect" is a preferable term.},
without any associated external particle; i.e. a \emph{non-local
mass} effect --- that will always be at a `subliminal' level as
far as direct physical observation is concerned.

Furthermore (re: the preceding paragraph's discussion), there is
\emph{insufficient} energy for a point mass to `materialise'
anywhere in this external field, i.e. anywhere in the (spherical)
volume associated with this non-particle-like (non-local) mass
($m^*$). Later (in section \ref{Subsection:warp's mass}) we shall
see that an invariant spherical volume to non-local mass
relationship is important. The model proposes that: the mass
aspect of the (virtual offset) energy is spread out \emph{evenly}
over a volume, thus making this mass aspect `quintessentially'
\emph{non-local}. The basis for this new type of mass is the new
relationship: $(\Delta E_{\rm virtual} \, \Delta t) \leq
\frac{1}{2}\hbar$ (recall subsection \ref{subsubsection:maximum
virtual}).

The extension of the non-local to local mass distinction (outlined
above) to `gravitational' field theorisation\footnote{So as to be
a supplementary or further feature of gravitational field energy.}
requires we accept that: (in certain unique circumstances)
microscopic/QM matter in `motion' can induce an (external) field
curvature/deformation, \emph{and} that (virtual QM) energy plays a
vital role in the model. To appease the dimensions of energy, and
to respect the requirements of a physical quantity\footnote{That
is, having a magnitude \emph{and} a unit/dimension(s).} existing
throughout a field, we unavoidably require this non-local mass to
have a magnitude, at all points in the field.

In a way, the non-local nature of this new type of QM-based mass
introduces a further stage in the \emph{dematerialisation of
gravitation}; in the sense that matter is involved, but in a
non-compact \emph{insubstantial} form. Non-local mass is not a
`substance' \emph{per se}, although in subsection
\ref{subsubsection:convolution} we explain the circumstances
whereby it does or does not influence the motion of (ordinary
celestial) matter --- be it: solid, liquid, gas or plasma.

\subsubsection{A second and different type of non-Euclidean
geometry}\label{subsubsection:two non-Euclidean} The model's
explanation requires a distinction between two different types of
curvature involving space and time. Firstly, there is standard GR,
where (active) mass, momentum and energy lead to spacetime
curvature, and the acceleration of a (passive gravitational) mass
is independent of its mass magnitude/amount. In spacetime we
describe physical \emph{events}, with these often occurring at a
specific time or between specific times, rather than over a period
or duration of time.

The (new) supplementary (i.e. cyclic perturbation) curvature,
arising from the coexistence of quantum mechanical systems and
celestial (geodesic) motion obeying GR, introduced by the model is
markedly different. There is no metric, nor is a new equation of
motion meaningful. To appease GR's invariance the (gravitational)
acceleration amplitude, associated with a rotating space-warp, is
\emph{constant} throughout space\footnote{Note that very slow
changes in amplitude over \emph{time} may occur and these are seen
to propagate at the speed of light, but the amplitude (in the
absence of change) is constant throughout \emph{space}. Such a
change only arises from changes in either lunar or planetary
orbital (and hence \emph{geometric}) characteristics, and/or the
\emph{number} of (lunar) atoms/molecules involved.}, which is in
stark contrast to GR's gravitational fields; whereas, to allow for
energy dispersion, the non-local mass at a point in the field
\emph{varies} (with enclosed spherical volume) throughout space.
Additionally, we are discussing an energy-based inherently
non-instantaneous physical \emph{process} cf. singular or multiple
`events'.

With the introduction of non-local or distributed mass (elaborated
upon in section \ref{Subsection:warp's mass}), there is no
violation of \emph{an equivalence principle} involving point-like
`condensed' inertial mass and (passive) gravitational mass (see
subsection \ref{subsubsection:convolution}). Neither is standard
(i.e. general relativistic) inertial acceleration to gravitational
acceleration equivalence violated\footnote{In GR this equivalence
is between \emph{uniformly} accelerated reference systems and
homogeneous gravitational fields. A \emph{sinusoidal} variation in
acceleration around a mean value is a very different
circumstance.}. We simply have an additional (and non-standard)
contribution to the overall `gravitational' field, in conjunction
with the spatial distribution of (a new) \emph{non-local} mass
`quantity' (denoted as) $m^*$ (see subsection
\ref{subsubsection:enclosed volume}).

\subsubsection{The Biot-Savart law, and extending continuum mechanics field
theory} The rotating space-warp proposed is considered to have
similarities to the magnetic (induction) vector field
$\underline{B}$ induced by a (steady) electrical current in a
wire, and the velocity field `induced by' a vortex
line/filament\footnote{Possibly, the phrase ``induced by" could be
replaced by ``coexisting with".}. Both of these physical phenomena
are described mathematically by the Biot-Savart law. A
(three-dimensional) circulatory aspect, which may be restricted to
a planar effect, is common in all cases. The rotating space-warp
is seen to be an external effect induced by an `internally'
inexpressible QM energy --- with this energy based upon a new
form/application of Heisenberg's uncertainty principle (and the
quantum or discrete nature of atomic/molecular angular momentum)
--- as discussed in subsection \ref{subsubsection:maximum
virtual}.

Recall that we are now discussing space curvature independently of
Special Relativity, in conjunction with a model that prefers a
\emph{space} continuum, cf. a spacetime continuum. Note that both
Mechanics of Solids and Fluid Mechanics utilise a \emph{mass}
continuum\footnote{From a microscopic/QM perspective, macroscopic
matter is also largely comprised of `empty' space.}. This is
clearly an unorthodox approach to `gravitation', or rather the
supplementation of gravitation, but it is actually (merely) a
simple extension of classical continuum mechanics into the
celestial realm\footnote{Recall that Newton unified terrestrial
and celestial (condensed matter) gravitation over 300 years ago.}
--- albeit now incorporating delocalised mass (i.e. non-local
mass).

\subsubsection{Extending the model's kinetic energy shortfall into
three dimensions}\label{subsubsection:planer} The derivation of
the kinetic energy `shortfall' of spacecraft motion (cf. currently
predicted motion) by way of the undulatory
acceleration/gravitational field(s) (outlined in section
\ref{Subsection:Shortfall}), was restricted to a planar
(space-warp) phenomenon --- albeit a distortion of a Euclidean
plane\footnote{If we neglect the presence of GR's gravitational
field.}, with intrinsic curvature assumed\footnote{The ontological
understanding of \emph{intrinsic} curvature deserves further
discussion. For our purposes we have envisaged an imaginary
(number based) space dimension, into which space effectively
deforms. This \emph{conceptualisation} is not inconsistent with
GR, and it is related to the discussion in section
\ref{subsection:Reversal}.}. The planar nature of the model lends
itself to a more visual (and geometric) understanding of curved
space.

The \emph{specific} energy of the undulation, over cycle time
$\Delta t$, is $\Delta e=\frac{1}{2}\Delta a^{2}\Delta t^{2}$. The
extension of this space curvature into three dimensions is quite
simple, resulting in a universe long axis about which an
`infinitely' wide space-warp rotates (recall subsection
\ref{subsubsection:3D external}). Thus, the specific energy
($\Delta e$) relationship is upheld. This three dimensionality of
the rotating space-warp is satisfying, because the Rayleigh
Theorem based quantification (beginning at subsection
\ref{subsubsection:Rayleigh Power}) was initially restricted to
simple \emph{one}-dimensional (in-line) motion, and then extended
by way of an (observationally demanded) hypothetical
conceptualization to a \emph{two}-dimensional rotating space-warp
(see subsection \ref{subsubsection:Space-warp}).

Recall that a rotating space-warp leads to the same (path-based)
loss of kinetic energy (per cycle) regardless of the (in-plane)
direction of motion of a celestial body or spacecraft. Radial and
circumferential motion in the solar system are equally affected.
For three-dimensional motion, i.e. motion inclined to the plane of
the space-warp's rotation, the same variation in $\Delta a$ over
duration $\Delta t$ is (or will be) `experienced'. Thus, speed
shortfall ($\delta v$ over duration $\Delta t$) along the path of
motion --- i.e. along the velocity vector --- is the same as for
(two-dimensional) in-plane motion.

Even though the kinetic energy `shortfall' of moving (low mass)
bodies is path-independent, there remains a need to make
corrections for \emph{line-of-sight} Doppler \emph{observations}
(geometrically) inclined to the Pioneer 10 path vector. This
correction is quantified in subsection \ref{subsubsection:three
dimensions}.

\subsubsection{General remarks concerning the \mbox{rotating}
space-warps (GQ-RSWs)}\label{subsubsection:general on GQ-RSWs} The
simplicity of these moon-planet based (gravito-quantum) rotating
space-warps is suitable for (i.e. not incompatible with) a field
based \emph{systemic} representation, that itself is (inevitably)
centered at the solar system barycentre. A systemic representation
is useful for determining speed shortfall cf. predicted speed (in
the absence of rotating space-warps).

We shall see that non-local mass, in the field, varies with
distance from each individual source (section
\ref{Subsection:warp's mass}), whereas acceleration amplitude is
fixed throughout the field --- although this (perturbation)
acceleration does vary cyclically/sinusoidally (over time at a
point or point mass) as the space-warp rotates.

In this ``constructive theory" (cf. principle theory) approach,
(non-local) mass and (RSW-based) acceleration are physically
distinct variables, with the concept of \emph{gravitational} force
of no significance. Energy is the model's primary physical
quantity and conservation of energy its linchpin. Further, we
shall see that the (initial) external field energy ($\Delta E_w$)
for an individual RSW cannot be completely localised (i.e.
spatially \emph{and} temporally) in the formalism; this is
because, even though the specific energy of each RSW is constant
throughout space\footnote{Except for a (relatively very small)
central near-field `hole'.}, the distribution (and nature) of
non-local mass is incompatible with a `point' mass based
determination of energy values.

In spite of the fact that the energy's microscopic QM basis
requires a closed (circumferential) loop (and hence a finite
process time) for its quantification, a systemic representation of
non-local mass ($m^*$) and acceleration undulation
amplitude\footnote{Either side of an equilibrium acceleration
value (that is) based upon general relativistic gravitation.} at
every point in the field (through time) is feasible in
principle\footnote{Knowing the timing of each phase of the
rotating space-warps' amplitude is required. Unfortunately, these
are not currently known, because only their superposition is
easily accessible to observations.}. Herein, the model deals
primarily with an average Pioneer anomaly value
($a_p$)\footnote{This average anomalous acceleration or speed
shortfall rate, acting upon all (celestially) bodies of
sufficiently low mass, physically coexists with the (on-going)
resultant `summation' of the acceleration/gravitational field
(perturbation) amplitudes of the various rotating space-warps.}
arising from several coexisting RSWs. A (classical) continuum
mechanics approach to the formalism of the (global) system is also
required to achieve the formalism's (in principle) localisation,
that (separately) involves constant acceleration (amplitude)
\emph{and} non-local mass at points throughout the field. Note
that the new/supplemenatry gravitational energy, like the
gravitational energy in GR, can \emph{never} be spatially or
temporally localised, because the quantification of $\Delta E_w$
is process based, i.e. requiring $\Delta t>>0$.

It shall become apparent (in subsection
\ref{subsubsection:convolution}) that the non-locality of what was
originally inertial mass in two different atomic/molecular QM
angular momentums (i.e. spin and orbital), and the coexistence of
rotating space-warps and (non-local) mass in the field, has lead
to the observational interpretation of an \emph{apparent}
violation of GR's (mass) equivalence principle --- in that `high
mass' bodies such as: Halley's comet, Vesta (asteroid), Europa
(moon), Pluto (dwarf planet), and the planet Saturn (for example)
do not respond to the rotating space-warps, whereas the (`low
mass') Pioneer spacecraft \emph{do} respond.

Actually, in the case of comet Halley this total absence of any
anomalous effect is not necessarily \emph{always} the case,
because in the unlikely event of comet Halley (with its current
mass of $\approx2.2\times10^{14}~\rm{kg}$) coming within
approximately: 2 AU of Jupiter, or 0.8 AU of Saturn, or 0.55 AU of
Neptune, there will be `some' influence --- only one
Sun-planet-moon system in the case of Saturn and Neptune ---
because the distance dependent mass cut-offs of the host planet's
moons are no longer less than the comet's total (`condensed')
mass. Note that these aforementioned distance values were
determined by way of interpolating the data (presented later) in
Table \ref{Table:cut-off} of subsection \ref{subsubsection:distrib
eqn ramifi}.
%***********************************************************************************
\subsection{The physical model quantified: external (real) energy}
\label{Subsection:Model quantifies external} In this section the
acceleration/gravitational amplitudes of the rotating space-warps
(RSWs) are determined, and the nature and coexistence of these
accelerations are discussed. Issues concerning (non-local) mass
shall be discussed in section \ref{Subsection:warp's mass}. Unless
otherwise specified a \emph{single} RSW is the basis of any
(rotating space-warp based) discussion in this section.

\subsubsection{Geometry's other role in the (semi-empirical) model}
\label{subsubsection:geometry's other role} In addition to
determining internal geometric phase offset ($\beta$) and virtual
angular momentum ($\frac{1}{2}\hbar_w$), geometry plays a further
important role in the model. This is unavoidable for two main
reasons. Firstly, the rotating space-warp needs to exhibit
dispersion with increasing distance away from the warp's
lunar-planetary source. Secondly, the energy of a rotating
space-warp comprises both: a \emph{non-spherical} (planar-based)
rotating space-warp of (constant) amplitude $\Delta a$, and a
presumably \emph{spherical} (non-local) mass distribution. We
shall need to appease, or at least understand, this (spherical vs.
non-spherical) situation which ostensibly involves a geometric
conflict.

The external expression of total internal virtual QM energy
($\Delta E_w$), as a (real) rotating space-warp and non-local mass
distribution, is a `hybrid' energy, involving mass and specific
energy ($\Delta e=\frac{1}{2}\Delta a^2 \Delta t^2$)
\emph{separately} --- although not independently.

We shall now argue that both: the magnitude of $\Delta a$, and the
non-local (initial) mass (associated with a given initial volume)
$m_1^*$, are dependent upon a reference (or `yardstick') radius
from a moon-planet system, and hence the volume enclosed within
this radius --- with this reference radius determined by geometric
circumstances. Additionally, cycle (or closed loop) time $\Delta
t$ acts as a reference (or `yardstick') time. The determination of
this (new) reference radius follows.

Geometry and differentiation give us the fact that: the derivative
of spherical surface area ($S$) with respect to radius is:
\mbox{${\rm d}S / {\rm d}r=8\pi r$}. If we set $r$ equal to lunar
orbital semi-major axis radius\footnote{We use the variable $r_o$
in preference to $a$ or $r_a$ to denote the length of the lunar
semi-major axis.} ($r_{o}$), the radius $8 \pi r_{o}$ describes a
unique radius. In the model this (geometrically unique) radius
sets the `initial' (non-local) mass value (for an initial volume
and surface area). In subsection \ref{subsubsection:acceleration}
we see that this radius also establishes the \emph{constant}
$\Delta a_w$ value associated with each moon. The (overall) energy
of a rotating space-warp, and the (initial) non-local mass
associated with it, is:
\begin{equation}\label{eq:squared by 2}\Delta E_w=\frac{1}{2} m_1^*
\Delta a_w^{2}\Delta t^{2} \end{equation} where $m_1^*$ represents
non-local mass at $8\pi r_o$ ($=r_1$ say), $\Delta a_w$ is a
weighted acceleration amplitude, and $\Delta E_w$ is the total
supplementary field energy over the course of a single lunar
loop/cycle. Note that use of $\Delta a$ previously (especially in
Section \ref{Section:PrelimModel}) is replaced by $\Delta a_w$
from here on --- see subsection \ref{subsubsection:acceleration}
for justification.

It can also be said that a (`defining' radius or) reference radius
$8 \pi r_{o}$ effectively represents a `dividing-radius' between
the near-field and far-field regions of the model, with our
interest dominated by the far-field\footnote{How to treat the
near-field (i.e. $r<8\pi r_o$) remains far from clear, and is not
considered herein. One possibility is as a hole in an infinite
field.}. Speaking of a rotating space-warp in terms analogous to a
fluid mechanical vortex: we have a space-warp external to an
(inner) \emph{\emph{tube}} rather than external to a linear (one
dimensional) filament.

The geometry-based assumptions presented in this section are
necessary if we are to make the model match the awkward
observational evidence. Once again the small angle
$\tan^{-1}(8\pi)^{-1}$ is important; this time in conjunction with
two sides of a right triangle: i.e. lunar semi-major axis length
$r_o$, and the (geometrically unique) reference length $8 \pi
r_o$.

%************************************ Table 4 *******************************************
\begin{table*}[t]
\small \caption{Weighted lunar space-warp acceleration amplitude and
associated values. \label{Table:acceleration}}
\begin{center}
\begin{tabular}[t]{llcrrrrrr} \hline
MOON  &Symbol  &\textit{Units}   &Io    &Europa    &G'mede
 &Callisto &Titan &Triton \\\hline

Mass                   &M                   &($10^{21}~{\rm kg}$)     &89.32     &48.00   &148.19   &107.59   &134.55  &21.4   \\[0.5ex]
Semi-major axis$^j$   &$r_o$               &($10^{9}~{\rm m}$)        &0.4218    &0.6711  &1.0704   &1.8827   &1.2219  &0.3548 \\[0.5ex]
Optimum acceleration  &$\Delta a_{o}$  &($10^{-8}{\rm \,cm~s^{-2}}$)  &39.821    &8.4537  &10.2590  &2.4076   &7.1485  &13.487 \\[0.5ex]
Efficiency         &$\eta$                           &(--)             &0.1290    &0.2590  &0.5219   &0.7826   &0.4683  &0.0309 \\[1.0ex]\hline
Weighted acceleration$^l$  &$\Delta a_{w}$  &($10^{-8}{\rm \,cm~s^{-2}}$) &5.139 &2.190   &5.354    &1.884    &3.348   &0.416$^k$ \\[0.5ex]\hline
\end{tabular}
\end{center}
\begin{center}
$^j$ Semi-major axis data, usually denoted as $a$, is taken from
the ``Planetary Satellites Mean Orbital Parameters" page of the
JPL Solar System Dynamics web site
(\url{http://ssd.jpl.nasa.gov}). Data for Titan (1221865 km) and
Triton (354759 km) exceed the significant figures in the table.
\,$^k$ Although Triton's retrograde motion `opposes' the other
major moons' rotations, its contribution to $\Delta a_{w}$ is
considered `positive'. \,$^l$ Note that square root of the
summation of squared (weighted) acceleration is $8.64\times
10^{-8}~{\rm cm\, s^{-2}}$ without Triton, and $8.65\times
10^{-8}~{\rm cm\, s^{-2}}$ with Triton's contribution; as compared
to the Pioneer anomaly's quoted magnitude of $a_{p} = 8.74\pm 1.33
\times 10^{-8}~{\rm cm\, s^{-2}}$.
\end{center}
\end{table*}
%****************************************************************************************
\subsubsection{The use of a weighted acceleration}
\label{subsubsection:acceleration}

Previously (section \ref{Subsection:Shortfall}) we used $\delta a$
to denote the spacecraft's speed change over time (in response to
a single space-warp) and $\Delta a$ to denote the warp amplitude.
The latter is now expressed as $\Delta a_w$, because it is a
weighted acceleration: $\Delta a_w=\Delta a_o \eta$ where $\Delta
a_o$ is the optimum acceleration amplitude, i.e. the acceleration
amplitude corresponding to $\beta=\pi$ where $\eta=1$.
Importantly, we recall (Equation \ref{eq:deltas_a}, subsection
\ref{subsubsection:amplitude to shortfall}) that $\Delta a=|\delta
a|$, which is now written as: $\Delta a_w=|\delta a|$.

Just as $\frac{1}{2}\hbar$ represents the (maximum) internal
uncertainty or `wiggle' room (of a single atom/molecule's
intrinsic angular momentum), (so) the amplitude $\Delta a_o$ is
seen to represent the corresponding (maximum) external
(acceleration/gravitational field) `wiggle' room of a single
rotating space-warp. Further, the same internal efficiency factor
($\eta$) is reproduced externally upon $\Delta a_o$, rather than
with regard to the total energy $\Delta E_w$. The empirical model
demands this `physical' connection between $\frac{1}{2}\hbar$ and
$\Delta a_o$ or $\frac{1}{2}\hbar_w$ and $\Delta a_w$ --- i.e.
(internal) intrinsic angular momentum and (external) acceleration
field amplitude. This is further discussed in subsection
\ref{subsubsection:external wiggle}.

\subsubsection{The geometric determination of weighted acceleration ($\Delta a_w$)}
\label{subsubsection:geometric acceler} We now turn our attention
to the (geometric) determination of $\Delta a_o$, and hence
$\Delta a_w$ where $\Delta a_w=\eta \Delta a_o$.

\begin{figure}[h!]
\centerline{\includegraphics[height=7.6cm,
angle=0]{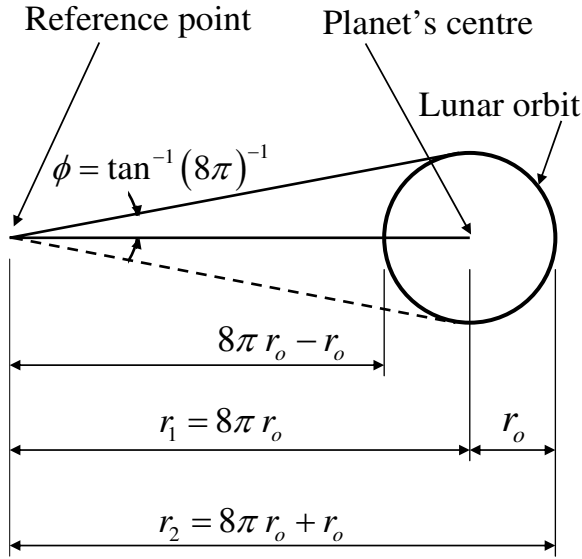}}
\begin{center}
\caption[Schematic diagram showing the model's (new) angular based
reference point which is used to determine rotating space-warp
(RSW) acceleration amplitudes.]{Schematic diagram showing the
model's (new) angular based reference point which is used to
determine rotating space-warp (RSW) acceleration amplitudes. This
reference point is $8\pi r_o$ distant from a planet's centre,
where $r_o$ is the lunar semi-major axis (length). Note that the
eccentricity of (large moon) lunar orbits is very small. The
variation in gravitational acceleration over a $2 r_o$ range is
used to determine the rotating space-warp acceleration amplitudes
($\Delta a$).} \label{Fig:EarthMoon}
\end{center}
\end{figure}

A right triangle comprising sides: $r_o$ and $8 \pi r_o$, and
angle: $\phi=2 \theta_{\rm{opt}}=\tan^{-1}(8\pi)^{-1}$ is used to
establish $\Delta a_o$. We utilise a reference point that is $8
\pi r_o$ distant from a moon-planet's orbital
centre\footnote{Dominating the model are the large moons of the
gas and ice giant planets, whose planet centres are very nearly
the moon-planet (reduced mass) orbital centres.}. Thus, the moon's
centre, during its orbital motion, subtends a (maximum) angle
$\phi$ relative to this external (geometrically established)
reference point --- see Figure \ref{Fig:EarthMoon}.

%********************************** Fig:EarthMoon *******************************
%********************************** end of Fig:EarthMoon  *************************

From the Newtonian gravitation approximation:
\begin{displaymath} \frac{F}{m}=a=\frac{G M}{r^2}
\end{displaymath} where $m$ is a test mass (e.g. a spacecraft) and
$M$ is lunar mass, we may determine (to good accuracy), \emph{at
the reference distance}, half the total variation (i.e.
$\frac{1}{2} \times 2\Delta a$) in acceleration due to the moon's
orbit. We define the maximum variation in gravitational
acceleration ($\Delta a_o$), around a mean value, as occurring
between $r_1=8 \pi r_o$ and \mbox{$r=8 \pi r_o \pm r_o$.}
Remembering that $r_o$ (cf. standard notation $r_a$ or $a$) is the
lunar semi-major axis, and letting $r_2=8 \pi r_o + r_o$ the
variation, or wiggle room, in acceleration is:
\begin{displaymath} \Delta a_o=G M
\left(\frac{1}{r_1^2}-\frac{1}{r_2^2}\right)\approx G M
\left(\frac{2 \,r_o}{r_1^3}\right) \end{displaymath} or
alternatively: \begin{equation} \Delta
a_o=1.18843\times10^{-4}\left(\frac{G
M}{r_o^2}\right)\approx\frac{2 G}{(8
\pi)^3}\left(\frac{M}{r_o^2}\right) \end{equation} Our
constructive empirical model asserts that this acceleration
magnitude has a correspondence to both: the maximum intrinsic
virtual angular momentum offset, i.e. $\frac{1}{2} \hbar$; and
also a one quarter QM (fermion) wavelength offset (i.e. $\pi$
rad). \mbox{Table \ref{Table:acceleration}} presents the various
$\Delta a_o$ and $\Delta a_w$ values of the moons dominating the
Pioneer anomaly. Note that the efficiency values are as per Table
\ref{Table:Efficiency}, and $G\approx6.67428\times10^{-11}\,{\rm
m^3\,kg^{-1}\,s^{-2}}$.

Regarding the moons of Uranus, the largest weighted acceleration
value is $0.132\times10^{-8}~{\rm cm~s^{-2}}$ for Titania, the
eighth largest moon of the solar system. All the other moons of
Uranus have values less than $0.10\times10^{-8}~{\rm cm~s^{-2}}$.
Thus, their quantitative contribution to the root sum
\emph{squared} based determination of the Pioneer anomaly ($a_p$)
is negligible.

\subsubsection{The model's acceleration as compared to MOND's acceleration}
\label{subsubsection:MOND} The `fixed' value of the space-warp's
optimum amplitude ($\Delta a_o$) should not be confused with the
additional \emph{fixed} acceleration
($a_0\approx1.2\times10^{-8}~{\rm cm~s^{-2}}$) associated with a
modified Newtonian dynamics (or MOND) based approach to spiral
galaxy equilibrium --- which presumes the non-existence of dark
matter \citep{Milgrom_06}. Acceleration drops off as $r^{-2}$ in
NMs, and as $r^{-1}$ in the MOND low acceleration regime of spiral
galaxies; whereas in our model the supplementary acceleration is
constant (i.e. it `drops off' as $r^0$).

Unlike MOND's force based approach to gravitation, or rather
gravitational modification\footnote{This includes modification of
a particle's inertia (under the action of a force) --- whatever
that physically entails.}, the new model's spatially
\emph{constant} (sinusoidal) acceleration (amplitude) allows it to
achieve a universality that is beyond GR's scope. This is because
the comprehensiveness of GR's approach to gravitation is
unavoidably local, whereas the model's local uniformity is
compatible with non-locality --- including quantum mechanical
(spin entanglement based) \emph{non}-locality.

\subsubsection{Overall effective acceleration value arising from multiple
space-warps}\label{subsubsection:multiple warps} Subsections
\ref{subsubsection:three spatial} and \ref{subsubsection:3D
external} discussed how spacecraft motion at a geometric angle to
the plane of the rotating space-warps (RSWs), as determined by
their lunar orbital planes, is quantitatively equivalent to a
spacecraft or body moving in the plane of the (various) lunar
orbits\footnote{This is fortuitous because Jupiter and Saturn's
equatorial planes are not parallel; additionally, Jupiter and
Saturn's orbital inclination (to the ecliptic plane) are
unequal.}. Thus, the weighted acceleration amplitudes of the
model's (major) space-warps ($\Delta a_w$) applies without
alteration, although a correction to the idealised model for the
geometric inclination of the line-of-sight observations to the
spacecraft path vector is required. This issue is further
discussed in subsection \ref{subsubsection:three dimensions}.

The manner in which these space-warps act \emph{together} upon a
\emph{moving} body's kinetic energy requires an explanation --- which
now follows.

We recall from subsection \ref{subsubsection:field to S/C} that
\mbox{$\frac{1}{2}\Delta a^2 \Delta t^2=\frac{1}{2} \delta v^2$,}
where $\delta v$ is the loss of a moving body's (spacecraft)
translational speed in one cycle (i.e. rotation of the warp) ---
albeit for a \emph{single} rotating space-warp.

In subsections \ref{subsubsection:a_p and mass comment},
\ref{subsubsection:Space-warp} and \ref{subsubsection:summations},
for multiple rotating space-warps, a root sum of squares (RSS)
approach was signalled. Recalling \mbox{Equation \ref{eq:SumA}:}
\begin{displaymath} a_p=\overline{a_p(t)}=\sqrt{\sum (\delta a_{\rm
proper})^{2}_i}=\sqrt{\sum (\Delta a_{\rm field})^{2}_i}
\end{displaymath} and substituting the various $\Delta a_w$ values
of \mbox{Table \ref{Table:acceleration}}, which equal the $\Delta
a_{\rm{field}}$ values --- whilst noting that the $\delta a_{\rm
proper}$ values are alternatively written (simply) as $\delta a$
--- we establish (for the idealised model) that:
\begin{displaymath} a_p=\sqrt{\sum (\Delta a_w)^{2}_i}=8.65\times
10^{-8}~{\rm cm~s^{-2}}
\end{displaymath} which is within the error bars of the quoted
value of the Pioneer anomaly, i.e. $a_p=8.74\pm 1.33 \times
10^{-8}~{\rm cm~s^{-2}}$. Bear in mind that: $1 \times 10^{-8}~{\rm
cm~s^{-2}}=1 \times 10^{-10}~{\rm m~s^{-2}}$.

Note that in the absence of the rotating space-warps from the five
other (primary) moons, Triton's RSW would necessarily (also)
retard the motion of (sufficiently low mass) bodies. Thus,
although Triton's retrograde motion (around Neptune) and
associated retrograde RSW oppose the direction of lunar prograde
RSWs, the root sum of squares approach (to quantifying $a_p$) is
seen to remain valid. See subsection \ref{subsubsection:Saturn
jump} for why Triton's largely negligible contribution may
actually be one of destructive interference.

The use of a (square) root (of) sum of squares (RSS) approach
implies the (constant amplitude) space-warps are uncorrelated ---
which is non-problematic. Somewhat problematically, there is also
the implication that the speed changes induced by the various
space-warps upon a moving body, are in some other sense orthogonal
to each other --- as is the case with motion in three
dimensions\footnote{For example, an overall (or `resultant')
velocity magnitude is the square root of the squared values of
speed in the x, y, and z directions --- so that
$v=\sqrt{v_x^2+v_y^2+v_z^2}$.}, or a (quantum mechanical) Hilbert
space. The issue of concern is that the various speed changes all
act along the \emph{same} velocity vector (or path vector) of a
given spacecraft.

We shall hypothesize that this (elegant) RSS approach is how the
superposition of the rotating space-warps is made physical for a
moving body (of sufficiently low mass). This assertion is
supported by:
\begin{enumerate} \item{the QM (energy) origins of the rotating space-warps
and their associated non-local mass,} \item{the agreement of the
model with the observational evidence (see section
\ref{Subsection:Primary observational} especially)\footnote{Noting
that temporal variation in $a_p \,$, around its very long-term
constant value, are `integral' to the model (recall subsection
\ref{subsubsection:Facade}); and that the resonance of Jupiter's
moons leads to only a minor smoothing effect (recall subsection
\ref{subsubsection:accel attenuation}).}.} \end{enumerate}

\subsubsection{On the hypothesised relationship \mbox{between}
$\frac{1}{2}\hbar$ and $\Delta a_o$}\label{subsubsection:external
wiggle} The model's proposed relationship between a
maximum/optimum virtual intrinsic angular momentum offset
($\frac{1}{2}\hbar$) and the magnitude of an optimum rotating
space-warp's amplitude ($\Delta a_o$), or between
(non-optimal/actual) $\frac{1}{2}\hbar_w$ and $\Delta a_w$ is now
examined. This subsection shall argue that it is a unique (and
\emph{new} type of) physical relationship.

This new relationship is the (non-local) backbone of the `new'
physics presented herein. It relates to the fact that in the
model: collective (virtual) QM \emph{spin} is (effectively)
entangled with the (macroscopic) RSW's
\emph{acceleration}/gravitational field amplitude. This (one-way)
non-local physical interrelationship is qualitatively
distinct/separate from the model's quantitative overarching
microscopic-to-macroscopic energy equality (Equation \ref{eq:2 by
Energy}), with the latter pertaining to a (one-way) energy
re-expression and (also) universal/systemic energy conservation.

The following discussion examines four primary physical
`quantities' and their units or (more specifically their)
dimensionality, with the latter expressed in terms of the
`fundamental' (mechanistic) dimensions\footnote{Units tend to be
specific like speed in metres per second, whereas the word
`dimensions' is better suited to non-specific and generalised
`units' such as: mass, length, and time.} of: mass, length, and
time\footnote{Distinguishing between vector and scalar quantities
is not of importance in this discussion.}.
\begin{enumerate} \item{Linear momentum ($p$), $\left[\rm{\frac{ML}{T}}\right]$}
\item{Angular momentum ($L$), $\left[\rm{\frac{ML^2}{T}}\right]$}
\item{Force ($F$), $\left[\rm{\frac{ML}{T^2}}\right]$}
\item{Energy ($E$) and Torque ($\tau$),
$\left[\rm{\frac{ML^2}{T^2}}\right]$} \end{enumerate}

\begin{figure}[h!]
\centerline{\includegraphics[height=7.0cm,
angle=0]{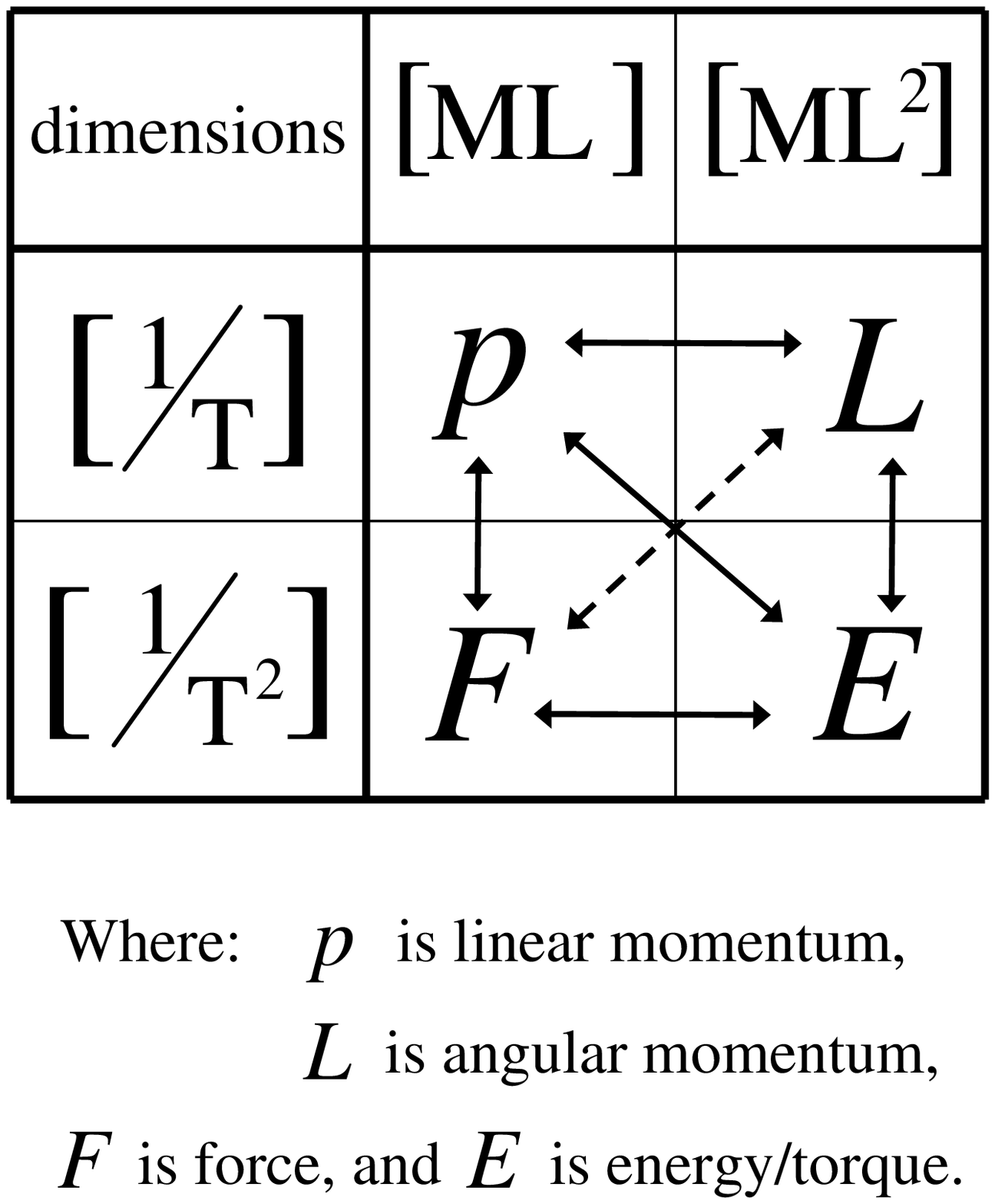}}
\begin{center}
\caption[Tabular schematic representation of the
\textit{dimensional} contributions of mass, length and time with
regard to the quantities: linear momentum, angular momentum, force
and energy.]{Tabular schematic representation of the
\textit{dimensional} contributions of mass, length and time with
regard to the quantities: linear momentum, angular momentum, force
and energy. For example, energy has dimensions of:
$[\frac{\rm{ML^2}}{\rm{T^2}}]$. This facilitates an understanding
of the dimensional relationships existing between these
quantities. A new relationship between (a QM offset of) angular
momentum ($L$) and Force ($F$) --- or rather acceleration ($F/m$)
--- is central to the model proposed in this paper. Unfortunately,
standard nomenclature dictates that `$L$' and `L' represent the
(vector) physical `quantity' of angular momentum, and the length
dimension (respectively); this subtle difference should be noted.}
\label{Fig:Dimension}
\end{center}
\end{figure}

In physics, especially electromagnetism and particle physics,
(instantaneous linear) particle momentum ($p$) is related to
particle energy ($E$) by way of (multiplying/dividing) the
particle's speed --- which has units/dimensions [$\frac{L}{T}$].
Relationships between the quantities: $p$ and $F$, and $L$ and $E$
involve a differentiation/intergration with respect to
\emph{time}, whereas a relationship between: $F$ and $E$ involves
an integration/differentiation with respect to \emph{distance}
(i.e. a length). Further, the relationship between $p$ and $L$
(and $F$ and Torque) utilises a cross product involving a radius
arm vector. Thus far we have discussed five permutations involving
pairs of the above four `primary' physical quantities ($p$, $L$,
$F$, and $E$) --- see \mbox{Figure \ref{Fig:Dimension}.} The use
of `$L$' and `L' in conventional nomenclature to denote angular
momentum (as a vector) and the length dimension (respectively) is
somewhat unfortunate and should be noted.

%********************************** Fig:Dimension *********************************
%********************************** end of Fig:Dimension  *************************

The only one of six permutations omitted from a conceivable
relationship, is a (direct) relationship between angular momentum
($L$) and force ($F$)\footnote{Admittedly, this discussion is fast
and loose, with (more specific) quantities such as impulse,
tension, and heat (for example) not mentioned; but our interest
lies in the sixth (and final) permutation --- involving any two of
these four primary physical quantities --- rather than further
instantiations of the permutations already recognised.}. In the
model we have needed to relate angular momentum, in the form of (a
proportion of) $\frac{1}{2}\hbar$ to an acceleration $\Delta a_w$,
with the latter's dimensionality, i.e. [$\frac{L}{T^2}$],
equivalent to a specific force ($F/m$). Thus, this is a new and
unique relationship; a systemic (i.e. global) relationship unique
to the model, that is implied (indirectly) by way of
``best-fitting" the observational evidence. This relationship
coexists with, facilitates, and supports the use of conservation
of (systemic/global) energy as the main guiding \emph{principle}
of the model.

Note that in the model, initial or starting non-local mass
($m_1^*$) is a fixed quantity, and thus the relationship between
angular momentum and a constant acceleration (amplitude) allows
the establishment of a pseudo-force term ($F=m_1^* \Delta a_w$)
for each moon --- not that it is a physically relevant quantity.
The primary quantities in the model's mechanism are: `initial'
non-local mass ($m_1^*$), constant acceleration amplitude ($\Delta
a_w$), total (virtual) angular momentum ($\frac{1}{2}\hbar_w
N_m$), and total (re-expressed) energy ($\Delta E_w$). Throw in
the predominantly linear momentum ($p$) of the (point mass-like)
Pioneer spacecraft, at least over relatively short time spans
($\Delta t$), and the broad scope of the physical quantities
comprising the model is evident.

\subsubsection{A comment upon dimensional variation and `reference'
quantities} We note that to change a quantity from a \emph{linear}
momentum ($p$) to an energy expression ($E$) involves multiplying
by a quantity with dimensions [$\frac{L}{T}$] (i.e. speed, e.g.
speed of light). By way of contrast, the physical link from a
(microscopic) angular momentum to a (macroscopic) pseudo force
(mass times an acceleration, $m^* \Delta a_w$) involves altering
the dimensions by [$\frac{1}{LT}$]. The former change is well
suited to the physics of particle motion, whereas the latter
`re-expression' --- also involving (only) one length and one time
dimension --- is unique to the new (non-graviton-particle,
non-local, systemic) model proposed herein.

Once again, we note that the model involves an unusual
re-expression of (systemic) atomic/molecular angular momentum
based energy, and has no place for a particle momentum based
`force' --- as embraced and prioritised by the standard model and
standard physics.

The coexistence of the physical re-expression of (virtual)
intrinsic angular momentum as a (constant amplitude) rotating
space-warp with a non-local mass component, in conjunction with
the three particle-based forces of the standard model does not
deny the model's implementation of an additional (hidden)
background Euclidean frame (as discussed in section
\ref{subsection:SR's ontology}). This (`idealised') frame
represents the flat space circumstances that exist in the
\emph{absence} of all matter and measurable energy effects.
Although not physically `realised', it is conceivable as well as
formally and conceptually useful. In a nutshell, we employ a
global `pseudo-frame' with an associated (global) pseudo-time, in
order to allow us to quantify the (equal) energy of both: a
(coherent and collective) QM fermion `wave' (phase) offset
(achieved over a given cycle time), and a macroscopic rotating
space-warp.

It can be argued that for the dimensional `multipliers' of
$[\frac{L}{T}]$ and [$\frac{1}{LT}$] to \emph{coexist}
non-problematically in nature, a systemic reference frame and a
systemic time unit must exist, and upon this stage (hidden from
observations) particle motion (also) occurs. It is ultimately the
observation of quantum non-locality, and the role of non-locality
in the model, that demands a (`noumenal') simultaneous
background/hidden time be instantiated --- both theoretically and
ontologically.

Our geometry based model has required the introduction of a
(geometrically significant) reference length ($8\pi r_o$) and a
(physically significant) reference cycle time ($\Delta t$), as
compared to the standard model's use of (what may be called) a
``reference speed" --- i.e. the \emph{speed} of light ($c$) which,
along with Planck's constant, is a \emph{universal} (i.e. global)
constant. The model also requires a reference \emph{angle}
$\tan^{-1}(8\pi)^{-1}$, which can also be thought of as a
`dimensionless' ratio of two lengths: $8 \pi r_o$ and (lunar
semi-major axis length) $r_o$ --- the non-hypotenuse sides of a
right triangle. Interestingly, the standard model has no
`universal' angle, i.e. an angle of particular importance; and it
is \emph{geometry}, with the assistance of energy conservation,
that has primarily allowed us to relate a (supplementary)
gravitational phenomenon to a quantum mechanical circumstance.

Recalling section \ref{Subsection:GR fortress}, hopefully the
reader now accepts that to simply assume ``measurements are
reality (without remainder)", and hence to embrace GR's general
covariance as indicative of a characteristic of
reality-in-itself\,\footnote{Or alternatively, ``all there is" to
reality.}, is conceivably a non-progressive stance
--- at least in the specific case of the Pioneer anomaly. To only
accept GR's stance on time necessarily leads to a complete denial
of the new model hypothesised. In the case of a \emph{real}
Pioneer anomaly, it appears that only by way of enriching Nature's
ontology, and appreciating the limitations of locality and
observational evidence, can a viable and progressive scientific
explanation be achieved.
%******************************** Table 5 new ****************************************
\begin{table*}[t]
\small \caption[Approximate corrections to the idealised Pioneer
acceleration ($a_p$) for spacecraft path inclination (i.e. angular
offset) relative to Doppler line-of-sight
observations.]{Approximate corrections to the idealised Pioneer
acceleration ($a_p$) for spacecraft path inclination (i.e. angular
offset) relative to Doppler line-of-sight observations. The
corrections are approximate because the angles for the
line-of-sight observations are idealised, in that they are
considered to be taken from the barycentre/heliocentre cf. the
Earth's surface (at 1 AU). We seek to \emph{correct} the
(one-dimensional) total (gravito-quantum rotating space-warp)
acceleration amplitude $\left[\sum( \Delta
a_w)^{2}_i\right]^{\frac{1}{2}}=8.649\times 10^{-8}~{\rm
cm\,s^{-2}}$, which [via $(\Delta a_w)_i=(\delta a)_i$] is also
idealised $a_p$, \emph{for} both solar ecliptic latitude and
longitude. Only the longitude is significant, arising from the
hyperbolic nature of the spacecraft's path out of the solar system
(recall Figure \ref{Fig:RevTang}). With the Pioneer 10 and 11 path
inclination to solar ecliptic latitude (as measured from the
heliocentre) being of the order of: less than 1 degree and about
1.5 degrees respectively, the cosines of these angular differences
are effectively negligible; subsequently, they are of no
significant consequence in the post-correction ``observable
$a_p$'' results given in the last column of the
table.}\label{Table:corrections}
\begin{center}
\begin{tabular}[t]{lccccccc} \hline
Spacecraft   &Date$^m$   &Radius$^n$    &SE Latitude$^n$    &SE
Longitude$^n$ &Angle$^o$ [$\varphi$] &$\cos \varphi$ &Observable
$a_p$
\\\hline

\textit{Units} &[D--M--Y] &AU &degrees &degrees &degrees &--- &$\times 10^{-8}{\rm \,cm~s^{-2}}$ \\[2.0ex]

Pioneer 10  &01 Jan 80   &20.54  &3.141  &59.67   &23   &0.9205    &\,\,7.96$^p$ \\[0.4ex]

Pioneer 10  &01 Jan 87   &40.01  &3.105  &70.99   &12   &0.9781    &8.46 \\[0.4ex]

Pioneer 10  &22 Jul 98   &70.51  &3.044  &76.45   &7   &0.9925     &8.58 \\[2ex]

Pioneer 11  &01 Jan 87   &22.39  &16.58  &254.93   &33   &0.8387   &7.25$^q$ \\[0.4ex]

Pioneer 11  &01 Oct 90   &31.70  &16.07  &265.66   &23   &0.9205   &7.96$^q$ \\[0.5ex]\hline

\end{tabular}
\end{center}
\begin{center}
$^m$ Time of day is 00:00 UT (i.e. Universal Time). $^n$
Information is derived from JPL's HORIZONS system, and is
consistent with $l_o$ and $b_o$ given in Table III of \citet[
p.45, Appendix]{Anderson_02a}. $^o$ Further approximation arises
because the (in plane) `inclination' angles were determined (by
the author, simply) by drawing the S/C path and the
Sun-to-spacecraft line and then measuring the angle of
inclination. This (approximate) approach gives angles that are
only accurate to the nearest whole degree. $^p$ This slightly
lower value is well within the noise of the observations --- as
evidenced by the variation in $a_P$ results for the three
intervals I, II and III comprising the 1987 to 1998 data set
\citep[ Table 1, p.22]{Anderson_02a}. $^q$ The model's low Pioneer
11 values are of concern but there are strong mitigating
circumstances (see discussion within subsection
\ref{subsubsection:Pioneer11}).
\end{center}
\end{table*}
%**********************************************************************************
\subsubsection{Extending the modelling of the Pioneer spacecraft to three
dimensions}\label{subsubsection:three dimensions} In this paper
(and generally) the word `dimension' is used  both as a
(non-specific/generalised) unit, and as an ontological (and
conceptual) aspect of reality. We now return to its (latter and)
more common meaning, especially regarding space.

When visualising a (macroscopic) rotating space-warp, it has been
useful to think in terms of an (idealised two [then actually
three] dimensional) perturbation upon either: an (idealised)
uncurved flat space, or the pre-existing general relativistic
gravitational field. We now quantitatively extend the model's
planar gravitational/accelerational (perturbation) influence to
spacecraft (S/C) motion inclined to a plane that is orthogonal to
the space-warp's axis of rotation. In short, the model is extended
to the motion of (passive `gravitational') mass in three spatial
dimensions.

Earlier (subsection \ref{subsubsection:3D external}) we were able
to extend the space-warp's (presence and) influence to having a
more than minimal thickness  --- i.e. we went beyond its idealised
planar instantiation. Regardless of how far the (moving) Pioneer
10 and 11 spacecraft (S/C) lie from the ecliptic plane, and
regardless of their velocity vector's inclination to the ecliptic
plane, the model suggests they will `receive' the rotating
space-warps \emph{full} influence\footnote{For example, the
Ulysses spacecraft.} (recall subsection \ref{subsubsection:three
spatial}). Similarly, within a given plane, e.g. the ecliptic
plane, there is no variation of anomalous speed shortfall arising
from the direction of motion; radial and circumferential motion
are equally affected.

The ubiquitous nature of the Pioneer anomaly (motion shortfall
rate\footnote{For `low mass' bodies, whose mass is below a ``mass
cut-off threshold", and recalling that the `shortfall rate' is
relative to a predicted motion (that assumes no
anomalous/additional physical effect).}) means there is no
correction required for S/C motion inclined to each specific lunar
orbital plane. But, this shortfall is attenuated if the
spacecraft's path/velocity vector is inclined (i.e. not aligned)
with the line-of-sight Doppler observations. This is inevitably
the case with Pioneer 10 and 11 and thus the observed Pioneer
anomaly will be different to the full path only based effect (see
\mbox{Table \ref{Table:corrections}).}

Since Jupiter's Galilean moons dominate the anomaly, and Jupiter's
equatorial plane is inclined at only $1.3^o$ to the ecliptic
plane, we shall (when loosely speaking) refer to the ecliptic
plane when using a single plane to `represent' a plane to which
the majority of various lunar orbits, and hence rotating
space-warps, are closely parallel.

\subsubsection{Lesser importance ascribed to the Pioneer 11
navigation experiment}\label{subsubsection:Pioneer11} There are
two (interconnected) reasons for assigning the Pioneer 11 results
a considerably lesser degree of confidence (and significance) than
the Pioneer 10 data; the Pioneer 10 data is discussed in
subsection \ref{subsubsection:corrected accel}. Firstly, by way of
quoting \citet[ p.23]{Anderson_02a} ``\ldots the [Pioneer 11] data
was relatively noisy, was from much closer to the Sun, and was
taken during a period of high solar activity. We also do not have
the same handle on spin-rate change effects as we did for Pioneer
10.'' Secondly and \textit{significantly}, the final value given
by \citet[ pp.39-40]{Anderson_02a} for the Pioneer anomaly ($a_P$)
omits the Pioneer 11 data completely, and \textit{only} uses the
Pioneer 10 data.

\citet[ p.37]{Anderson_02a} mention that: ``\ldots the [annual
variation] is particularly large in the out-of-ecliptic voyage of
Pioneer 11 \ldots"; whereas one can (just as easily) retort, by
way of \citet[ p.396]{Olsen_07}, that: ``\ldots the results for
Pioneer 11 show no clear annual variation of the anomalous
acceleration." These two opposing viewpoints, regarding Pioneer
11's `annual' variation in anomalous acceleration ($a_P$),
highlight a lack of consensus amongst different data analyses of
temporal variation in \mbox{Pioneer 11's} $a_P$ value. Such a
difference of opinion supports this subsection's main contention
that: considerably lesser importance should be (and indeed has
been) ascribed to the Pioneer 11 navigation experiment (cf. the
Pioneer 10 navigation experiment).

The recent paper by \citet{Turyshev_11a} relies heavily upon
results from Pioneer 11. Subsequently, their findings that the
Pioneer anomaly is: non path-based, and exhibits a reduction/decay
in its magnitude over time --- so as to be supportive of a
heat-based explanation --- is not seen as a definitive
rebuttal/refutation of the hypothesis (and model) of a real
Pioneer anomaly pursued throughout this paper.

\subsubsection{Corrected acceleration}
\label{subsubsection:corrected accel} By way of Table
\ref{Table:corrections} the model's average value for the
\emph{observed} anomalous acceleration, of Pioneer 10 between 1987
and 1998, is:
\begin{displaymath} (a_p)_{\rm{model}}=(a_{p10})_{\rm{model}}=8.52
\pm0.66 \times 10^{-8}~{\rm cm~s^{-2}}
\end{displaymath} which displays the modelling
error\footnote{Possibly ``uncertainty of the model'' is a better
expression.} argued for in subsection \ref{subsubsection:Error and
idealisation}. This magnitude is a sound quantitative match to the
``headline" Pioneer anomaly acceleration: \begin{displaymath}
(a_P)_{\rm{observation}}=8.74\pm1.33\times 10^{-8}~{\rm cm~s^{-2}}
\end{displaymath}

Note that the model's average value for the \emph{observed}
anomalous acceleration of Pioneer 11 between 1987 and late 1990
is\footnote{The orbit determination program of \citet[
p.18]{Toth_09a} generates a similar difference between the Pioneer
10 and 11 values: $9.03$ vs. $8.21 ~(\times~10^{-10}~{\rm m/s^2})$
respectively.}:
\begin{displaymath} (a_{p11})_{\rm{model}}=7.61\pm0.66
\times 10^{-8}~{\rm cm~s^{-2}} \end{displaymath}

This \emph{average} (additional/`anomalous') roughly sunward and
line-of-sight based `acceleration' value (for Pioneer 10) is most
appropriately understood as a \emph{measured} (cf. actual
path-based) total \emph{loss} of speed over a duration of time,
rather than as a (constant) force based acceleration. Actually,
the full acceleration ($8.65 \times 10^{-8}~{\rm cm~s^{-2}}$) is
directed against the \emph{path} of the Pioneer 10 (and 11)
spacecraft --- which happens to be a hyperbola whose direction is
predominantly radial, i.e. away from the solar system's
barycentre.

For non-radial spacecraft motion it is conceivable that the
spacecraft's rate of spin rotation might change with $a_p(t)$, as
implied by the observational evidence --- recall subsection
\ref{subsubsection:spin rate changes}.

\subsubsection{A possible minor correction for the partial non-solidity
of lunar bodies}\label{subsubsection:non-solidity} The analysis
(thus far) has assumed/idealised moons to be $100 \%$ \emph{solid}
bodies, although not necessarily \emph{rigid} bodies --- (the
latter) so as to allow tidal effects to achieve tidal locking
(i.e. moon-planet orbital resonance). The non-solid aspects of a
lunar body, e.g. molten core and/or mantle, vulcanism, and the
presence of water, have not been fully appreciated. For example,
Io and Ganymede are considered to have (partial) molten and liquid
cores respectively, whereas Europa, Callisto and Titan are
considered to have (or possibly have) an internal (relatively
thin) layer of liquid water\footnote{Assisted by the presence of
antifreezes, e.g. ammonia.} deep beneath their ice crusts.
Nevertheless, the vast majority of matter in these five large
moons central to the model \emph{is} solid (surely $> 95 \%$),
generally comprising rocky material (especially silicates) and
water ice.

Internal and/or surface motions of material (comprising
atoms/molecules) within or upon a moon, i.e. non-gravitational
lunar-based motions, will (conceivably) fail to keep the
(atomic/molecular) geometric phase shifts (per spin/orbital cycle)
to the minimal virtual level so crucial to the model --- recall
the numerous and tight constraints outlined in subsection
\ref{subsubsection:relationships spin and}. Thus, the number of
molecules/atoms ($N_m$) in a bulk lunar body contributing to the
energy of the (gravito-quantum) rotating space-warps will (in all
likelihood) be reduced from the model's ideal values given in
Table \ref{Table:energy} --- which assumed a completely ($100 \%$)
solid (and considerably rigid) bulk lunar body\footnote{Although,
if the external/extrinsic imposition upon `internal'
\emph{geometric} phase, by way of its virtual `nature', is
completely independent of quantum mechanical \emph{dynamical}
phase, then the macroscopic entanglement effect (associated with
it) may well remain at $100 \%$, and as such there would be no
correction required for (partial) non-solidity.}.

Fortunately, because the amplitude of the various rotating
space-warps ($\Delta a_w$) is dependent upon total lunar mass
($M$) rather than the total number of atoms/molecules comprising a
`solid' body (recall subsection \ref{subsubsection:geometric
acceler}), any correction arising from lunar non-solidity is
restricted to only altering values of $m_1^*$ and $\Delta E_w$.
Corrections to the (Pioneer anomalous) acceleration value $a_p$
--- based upon a path to observational line-of-sight angle (see
Table \ref{Table:corrections} and subsection
\ref{subsubsection:corrected accel}) --- are also independent of
$N_m$. The model's pre-correction `acceleration' value (or rather
speed shortfall rate) $a_p=\left[\sum( \Delta
a_w)^{2}\right]^{\frac{1}{2}}=8.65\times 10^{-8}~{\rm cm~s^{-2}}$
--- as quoted in subsection \ref{subsubsection:multiple warps} and
determined from weighted acceleration values in Table
\ref{Table:acceleration} --- is effectively a fixed (or immutable)
quantity.

Unlike many other scientific models applied to anomalous
observations in need of an explanation, the model has no
`flexible' parameters that can be easily adjusted or `tweaked'. In
other words, the model's \emph{primary} quantity $(a_p)$, i.e. the
full anomalous acceleration magnitude acting along the \emph{path}
vector of a moving body (not corrected for the observational
line-of-sight)\footnote{That is, $a_p$ as given is subsection
\ref{subsubsection:multiple warps}, as compared to the (Pioneer 10
based) `observational' $(a_p)_{\rm{model}}$ value given in
subsection \ref{subsubsection:corrected accel}.}, is essentially
closed to (theoretical or) model-based parameter adjustments.
Further, we note that $N_m$ is the (celestial physical) quantity
whose magnitude is known with (by far) the least accuracy. With
$a_p$ not directly dependent upon `participating' $N_m$, its
`component' $\Delta a_w$ ($=\delta a$) values are determined (in
the model) with better accuracy than either $\Delta E_w$, $m_1^*$,
or the various $m^*(r)$ distribution values discussed in section
\ref{Subsection:warp's mass} and given in Table
\ref{Table:cut-off}.
%**********************************************************************************************
\subsection{Non-local mass distribution}\label{Subsection:warp's mass}
This section continues the quantification of the macroscopic
physical model, involving a number of rotating space-warps (RSWs)
in superposition, and the non-local mass distribution associated
with each particular RSW. The following discussion of non-local
mass needs to clearly distinguish mass from (ordinary/non-exotic)
matter, in the sense of \emph{matter} being a tangible or
observable physical `substance' that: occupies space, has rest
mass, and is comprised of ``building blocks" (or is itself a
building block) --- e.g. a moon, a lake, a grain of sand, an atom,
an electron, or a quark.

\subsubsection{Background to non-local mass} To distinguish non-local
mass from standard (baryonic) mass we indicate it as: $m^*$. The
non-particle-like nature of the model's non-local mass
effect\footnote{I am reluctant to say non-local mass has a
(classical) `wave-like' nature because `wave' implies a
\emph{transport} of energy, whereas we only require the wave
characteristic of being a de-localised phenomena, i.e. one spread out
in space.}, means that it is associated with both: conditions
\emph{at a point} in a (space) field; and also a \emph{distribution}
effect throughout space $[m^*(r)]$. We shall see that $m^*(r)$ exists
at every point upon a spherical surface that encloses a volume (of
specified radius $r$). As is the case with time, the `essence' of
$m^*$ shall not be of major concern herein; rather, our major concern
is with the model's quantification of $m^*$, particularly its spatial
variation.

Note that $m^*$ has its basis in the mass `dimension' of Dirac's
constant $\hbar$, which appears in the (total) virtual QM energy
expression (Equation \ref{eq:2 by Energy}, section
\ref{Subsection:Model quantified internal}).

The geometric basis of the model ensures that non-local mass is
best understood in association with the total volume
\emph{enclosed} by the `establishment' (or reference) distance
from the source ($r_1=8\pi r_o$). With its volume dependence,
$m^*$ is somewhat like the density of standard matter --- which
has dimensions: $[\rm{\frac{M}{L^3}}]$. In the spirit of QMs, the
introduction of $m^*$ means that `mass' is exhibiting a
(local--nonlocal) duality --- rather than (i.e. as distinct from)
a wave--particle duality\footnote{Possibly, a (non-reductionist)
local--nonlocal duality is more general than the wave--particle
duality. Such a duality or complementarity has been discussed in
relation to the Aharonov--Bohm effect by \citet[
Abstract]{Aharonov_00} --- albeit with a different emphasis `in
mind'.}.

The distribution of non-local mass (at any point) in the field is
assumed to be continuous, and change very gradually between
neighbouring regions. Of special importance is the change in the
magnitude of non-local mass as one moves away from its `source' (or
initial/establisment) surface and volume, and thus, there is a need
for a (non-local) mass distribution function. This is addressed from
subsection \ref{subsubsection:enclosed volume} onwards.

\subsubsection{Comments regarding the governing energy expression}
\label{subsubsection:governing energy} In the expression for the
total supplementary field energy of a rotating space-warp (and its
conjoint mass): \begin{displaymath}\Delta E_w=\frac{1}{2}m_1^*\Delta
a^2_w\Delta t^2\end{displaymath} (i.e. Equation \ref{eq:squared by
2}, section \ref{Subsection:Model quantifies external}) the mass and
acceleration components are separate (recall subsection
\ref{subsubsection:Violation of Eq.Pr}). The total scalar energy
$\Delta E_w$ can be understood as simply a `mass' multiplied by
specific energy. That is:
\begin{displaymath}\Delta E_w=m_1^* \Delta e_w\end{displaymath}

The specific energy ($\Delta e_w$) is comprised of (a sinusoidal)
acceleration amplitude and cyclic duration ($\Delta t$) and it is
essentially always fixed, because of the dynamically `stable'
moon-planet-Sun celestial systems in our solar system\footnote{At
least over human time scales, with these being of the order of
tens, hundreds, or thousands of years. The proportion of non-solid
matter in a moon also remains essentially fixed (recall subsection
\ref{subsubsection:non-solidity}).}. Note that $m_1^*$ is actually
non-local mass \emph{at} the (initial) `establishment' radius,
i.e. the reference radius, $r_1=8\pi r_o$; further, we signify
$V_1$ as the spherical volume enclosed within this radius. Thus,
upon establishment of the space-warp, all quantities: $\Delta
E_w$, $m_1^*$, and \mbox{$\Delta e_w=\frac{1}{2}\Delta a^2_w
\Delta t^2$} may be considered \emph{fixed}, i.e. (long-term)
constant values.

\subsubsection{Total energy constancy, and the conceivable dispersion of non-local
mass}\label{subsubsection:total energy constancy} The external
counterpart of the mass aspect of the total excess (non-inertial) QM
(intrinsic angular momentum based) energy ($\Delta E_w$) is $m_1^*$.
The (total energy) equality: $\Delta E_w=m_1^* \Delta e_w$ is assumed
to (only) apply at \emph{every `point'} on the initial (spherical)
\emph{surface} and not throughout the (spatial) field exterior to the
initial radius\footnote{The conditions interior to $r=r_1$ shall not
concern us in this subsection, nor have they been of any significant
importance (to the model) elsewhere in this paper.}, i.e. $r>r_1$.
Further, the magnitude of $m^*_1$ is associated with an initial
(enclosed) volume ($V_1$).

Note that if we (were to) propose an equality, involving a surface
energy, such that (at $r=r_1$):
\begin{displaymath} \Delta E_w=(E_w)_{\rm{surface}}=\gamma_1^* S_1
\end{displaymath} then $\gamma_1^*$ has dimensions $[\frac{M}{T^2}]$.
Interestingly, both spring stiffness ($K$) and fluid mechanical
surface tension ($\gamma$) also have dimensions $[\frac{M}{T^2}]$.
This notion is not pursued.

Physical circumstances pertaining to the space-warp's energy
description when $r>r_1$ are quite distinct from the ($r=r_1$)
establishment conditions. Due to the very slow evolution of
moon-planet-Sun celestial systems, and the constancy of the
various $\Delta a_w$ values, the various (lunar) specific energies
$\Delta e_w$ can fortunately be treated as a fixed quantity at
\emph{every} point \emph{throughout} their respective rotating
space-warp fields. Thus, for $r>r_1$ our concern lies with how
$m^*$ varies with $r$, or rather (enclosed) spherical volume. We
denote \emph{variable} (and distributed) $m^*$ as $m^*(r)$ which
is abbreviated to $m_r^*$; with $m^*(r_1)$ written as $m^*_1$.

For a given space-warp the magnitude of the \emph{total} field
\emph{energy} $\Delta E_w$ is independent of position ($r$) in the
field, even though (in the formalism) a specific radius ($r=r_1$)
is implicitly associated with $m^*_1$. This is because non-local
mass --- which is unlike standard (inertial, passive and active
gravitational) mass --- requires an initial reference radius
(recall section \ref{Subsection:Model quantifies external}).
Nevertheless, rather than a \emph{total} field energy, we can
still conceive of a (non-local) field energy at \emph{points} in
the field: (say) $E^*_{\rm{field}}(r)=m_r^* \Delta e_w$, which
varies with $m_r^*$. We also recollect that this energy --- which
pertains to a \emph{single} rotating space-warp and non-local mass
distribution --- is process based, i.e. occurring over $\Delta t$.
Thus, equating this energy $E^*_{\rm{field}}(r)$ [or rather
$(E^*_r)_{\rm{component}}$, using the nomenclature of Section
\ref{section:Type1a}] to the notion of gravitational energy (at a
point in the field) is nonsensical; as is (the case with)
gravitational energy at a point in the field (at a given time) in
general relativity. In stark contrast to this, we have the well
defined \emph{specific} energy of a ($360^o$) rotating space-warp
--- in two and three dimensions
--- i.e. $\Delta e=\frac{1}{2}\Delta a_w^2 \Delta t^2$, which is
invariant throughout `space' and effectively fixed/constant
throughout (cosmologically recent) `time'.

\subsubsection{Global or universal aspects of the wave-like non-local mass
distribution}\label{subsubsection:Global or universal} Just as
$\Delta e_w$ fills the whole of space (i.e. the universe), the
mechanism's non-local mass ($m_r^*$) also fills the universe. For
points $r>r_1$ (i.e. in the far-field) the associated (spherical)
enclosed volume and surface area increase in magnitude relative to
the initial reference conditions (where $r=r_1$). In analogy with
classical mechanics, we hypothesise the notion of \emph{energy
dissipation}, and hence (non-local) \emph{mass dispersion}, as we
examine conditions at points progressively further away from the
source, or rather `core', of the non-local mass distribution. This
core is the (initial and) minimum volume (and surface area) of the
mass component of the virtual QM energy's externalisation.

With $\Delta e_w$ constant, a far field (pseudo)-energy at a point
in the field [$(E^*_r)_{\rm{component}}$] is seen to vary in
proportion to $m_r^*$. The different physical `natures' of
non-local mass and acceleration/gravitational (field) warp
amplitude in the model means that this far-field energy cannot be
localised. What can be asserted with confidence is that
\emph{total} (QM-based) scalar energy $\Delta E_w$ is
fixed/unchanged/constant (i.e. invariant) regardless of the radius
$r$, surface area $S$, or enclosed volume $V$ considered. This
\emph{scale} independence of total (non-local mass and space-warp)
energy arises from its (closed loop) QM virtual energy basis, and
its real exterior expression of universal extent or size --- in
that the space-warp extends to infinity in the plane, and its
rotation axis also extends to infinity. It is (only) non-local
\emph{mass} that varies with (the) scale (considered) --- thus
incorporating all scales from an minimum/`initial' reference
(sub-universal) system, all the way up to the full (universal)
system (where $r^3 \rightarrow \infty$ and $m_r^*\rightarrow 0$).
This variation in \emph{non-local} mass magnitude is necessarily
subject to another \emph{geometric} constraint of the model ---
see discussion in subsections \ref{subsubsection:enclosed volume}
and \ref{subsubsection:distrib eqn ramifi}.

Note that these (spatially based) changes (in non-local mass) can
only be apprehended by way of the existence of a systemic (i.e.
global and/or universal) space continuum or space
substratum\footnote{In the sense discussed in subsection
\ref{subsubsection:substantivalism}.}. \emph{Variation} in $m_r^*$
[or alternatively $m^*(r)$] by way of a variation in the enclosed
(spherical) \emph{volume} (and indirectly surface area) is our
prime concern; we note/recall that $m_r^*$ (itself) describes
conditions on the surface of the (enclosed) volume. The magnitude
and distribution of $m_r^*$ has been constrained (and indirectly
implied) by observation evidence (recall section
\ref{subsection:further concerns}) concerning small comets of
approximately 1 km diameter (with mass of $\sim2.6\times10^{11}$
kg) apparently being (of the order of) the largest size/mass of
bodies `anomalously' influenced.

\subsubsection{Quantifying the variation of non-local mass with the enclosed volume}
\label{subsubsection:enclosed volume} The inception, i.e.
establishment, of a (single/specific) rotating space-warp and
non-local mass distribution involves both: a reference (or
minimal) \emph{enclosed} volume $V_1$, and $m_1^*$ --- which is
non-local mass \emph{at a distance} $r=r_1=8\pi r_o$ from the
space-warp's central point\footnote{The central point of the
space-warp generator/emitter is probably the moon's centre, but
possibly it is the centre of the moon-planet system. This
conceptual inexactitude has a negligible quantitative effect upon
the model (as presented herein).}, where $r_o=r_a$ i.e. lunar
semi-major axis.

Note that a rotating space-warp (slice in two dimensions) has a
point-like centre of rotation (i.e. an origin), but there is no
`point' source (\emph{per se}) for the space-warp's (associated)
non-local mass. The term `source', rather than `origin', shall
designate the finite: radius ($r_1$), volume ($V_1$), and surface
area associated with the initialising/establishment of non-local
mass. The relevance of this distinction is lessened when large
radii from the centre of rotation are considered.

The model has proposed/argued that: although the non-local mass
$m_1^*$ in the expression for $\Delta E_w$ is a fixed quantity,
the value of the non-local mass \emph{at a point} in the far-field
($m_r^*$) diminishes as $r^{-3}$ (i.e. $m_r^*$ reduces with
increased volume enclosed) for radii $r\geq8 \pi r_{o}$ (i.e.
$r\geq r_1$) away from the source of a rotating space-warp. This
is based upon observational evidence and theoretical compatibility
with GR, the latter because the specific energy ($\Delta e_w$) of
the rotating space-warp is the \emph{same} throughout the field
--- at least for stable lunar orbital conditions, and also because
(cosmologically very short) time scales ($\Delta t$), of the order
of weeks, are involved. Recall that the quantity $\Delta E_w$
--- i.e. the virtual QM energy requiring external `re-expression' ---
only applies to non-local mass (and rotating space-warp)
\emph{establishment} --- in that it contains the `initialising'
non-local mass value ($m^*_1$).

Herein we have been able to treat specific energy $\Delta e_w$ as
a constant. Slow changes in specific energy $\Delta e_w$, i.e.
changes in either orbital time and/or acceleration amplitude,
propagate (and spread) into the field at the speed of light. In
contrast, the apparent simultaneity associated with quantum
non-locality means that, in `\emph{measurable time}', non-local
mass `propagates' (and spreads) into the field/universe
instantaneously. In other words, non-local mass is coexistent
throughout the field at any given (`noumenal') `\emph{systemic
time}'\footnote{Except for a small central area (or tube in three
dimensions) of radius $8\pi r_o$.}. In accord with section
\ref{subsection:SR's ontology} we may say that systemic time is a
theoretical second time, for global (hidden) processes occurring
between the moments grasped by measurements
--- such that some processes (especially entanglement)
\emph{appear}, from an observational/`phenomenal' perspective, to
involve instantaneous `interactions' (and ``action at a
distance").

We have seen that the \emph{dissipation} of total (non-local mass
and gravitational/accelerational) field \emph{energy} lies solely
in the (spatial) \emph{dispersion} of the non-local mass term
($m_r^*$). Subsection \ref{subsubsection:distrib eqn ramifi} shall
argue that at greater distances/volumes from a source there is a
similar reduction in the (passive) mass of bodies ($m_p$) that the
space-warp's conjoint/associated $m_r^*$ can
affect\footnote{``Passive mass" or ``passive `gravitational' mass"
is now being used in the sense of a mass (that also) responds to
the new acceleration/gravitational fields proposed herein --- in
addition to standard gravitation.}. We shall see that
\emph{non-local} mass $m_r^*$ is markedly distinct/different from
inertial mass $m_i$ and gravitational mass (both active and
passive) $m_g$, in that the equality $m_i=m_g$ is not extended so
as to include $m_r^*$. In general, the non-local mass field value
does not equal a body's (local) mass value, i.e. $m_r^* \neq m_i$
and $m_r^* \neq m_g$.

The model proposes a (non-local mass) \emph{distribution function},
with dimensions [$M L^3$], of the form:
\begin{displaymath} m^*(r)\, V(r)={\rm{constant}}=m^*(8\pi
r_o)\, V(8\pi r_o) \end{displaymath}or more succinctly:
\begin{equation} \label{eq:distribution} m_r^*\, V_r=
{\rm{constant}}=m_1^*\, V_1 \end{equation} where: $m^*(r)$ or
$m_r^*$ is the non-local mass value at (spherical) radius $r$, and
$V(r)$ or $V_r$ is the \emph{volume enclosed} within $r$. Once
again we note that this distribution function is restricted to
$r\geq8 \pi r_o$, i.e. $r \geq r_1$.

%************************ Table 6 ******************************************
\begin{table*}[t]
\small \caption{Mass cut-offs at various distances for large
(effective) solar system moons. \label{Table:cut-off}}
\begin{center}
\begin{tabular}[t]{lcrrrrrrrrrrr} \hline
MOON   &\textit{Units}   &Io    &Europa    &Ganymede
 &Callisto &Titan &Triton \\\hline

Weighted Energy ($\Delta E_{w}$)  &($10^{9}{\rm \,J}$)                     &2.395    &1.287   &3.973   &1.854   &1.452    &0.041\\[0.5ex]
Moon orbit frequency ($f_{\rm m}$) &($10^{-6}~{\rm s^{-1}}$)                &6.5422   &3.2592  &1.6177  &0.69351 &0.72586  &1.9694 \\[0.5ex]
Weighted acceleration$^r$ ($\Delta a_{w}$)  &($10^{-10}{\rm \,m\,s^{-2}}$) &5.139    &2.190   &5.354   &1.884   &3.348    &0.416 \\[0.5ex]
Mass cut-off at \mbox{$8\pi r_o$} ($m^*_1$)  &($10^{15}{\rm \,kg}$)        &692      &507     &64.6    &44.7    &12.2     &165    \\[1.5ex]

Semi-major axis ($r_o$)       &($10^{9}{\rm \,m}$)     &0.4218    &0.6711  &1.0704  &1.8827  &1.2219  &0.3548 \\[0.5ex]
Transition radius (\mbox{$8\pi r_o=r_1$})  &(AU)$^s$   &0.0709    &0.1127  &0.1798  &0.3163  &0.2053  &0.0596  \\[0.5ex]
No. of transition radii in 1\,AU           &(--)       &14.11     &8.87    &5.56    &3.16    &4.87    &16.78 \\[1.5ex]

Mass cut-off at 1\,AU  &($10^{12}{\rm \,kg}$)                   &276    &817    &422     &1590    &118   &39.2 \\[0.5ex]
Mass cut-off at 2.5\,AU$^t$  &($10^{12}{\rm \,kg}$)             &17.7   &52.3   &27.0    &101.7   &7.56  &2.51 \\[0.5ex]
Mass cut-off at 10\,AU$^t$  &($10^{9}{\rm \,kg}$)               &276    &817    &422     &1590    &118   &39.2 \\[0.5ex]
Mass cut-off at 100\,AU  &($10^{9}{\rm \,kg}$)                  &0.276  &0.817  &0.422   &1.590   &0.118 &0.039 \\[0.5ex]
Mass cut-off at 100\,000\,AU$^u$  &(${\rm kg}$)                 &0.276  &0.817  &0.422   &1.590   &0.118 &0.039 \\[0.7ex]\hline
\end{tabular}
\end{center}
\begin{center}
$^r$ Non-corrected values. $^s$ Astronomical Unit ($149.598\times
10^{9}~\rm{m}$). $^t$ Compare to comets of various diameters
(comet density assumed equal to $0.5\times 10^{3}{\rm \,kg \,
m^{-3}}$)\,: \mbox{1\,km $\rightarrow$ $0.26 \times 10^{12}{\rm
\,kg}$}\,, \mbox{3\,km $\rightarrow$ $7.1 \times 10^{12}{\rm
\,kg}$,} 5\,km $\rightarrow$ $33 \times 10^{12}{\rm \,kg}$. \\
$^u$ 100\,000 AU is a distance equal to 1.58 \,light years or
approximately half a parsec.
\end{center}
\end{table*}
%*****************************************************************************

\subsubsection{Distribution equation ramifications and values for
the different moons}\label{subsubsection:distrib eqn ramifi} We
can compare and contrast this non-local-mass distribution
function/equation in \emph{space} to one form of the continuity
equation in fluid mechanics for a steady flow: \mbox{$\rho_n A_n
u_n=\rm{Constant}$}\footnote{Where: $n=1,2,3, ...$\,, \,$\rho$ is
fluid density, $A$ is flow area, and $u$ is speed. Usually,
continuity equations are expressed as a differential equation, and
thus are a (stronger) \emph{local} form of conservation laws ---
paraphrased from Wikipedia: \emph{Continuity equation}, 2010-12.}.
This is a conservation of mass equation associated with (mass)
motion in \emph{time}, and has dimensions [$\frac{M}{T}$]. It
follows that in one sense Equation \ref{eq:distribution} --- with
its dimensions/units of $[M L^3]$ --- can conceivably be described
as a \emph{spatial non-local-mass continuity equation}.

In Table \ref{Table:cut-off} the variation of non-local mass with
distance from its source is detailed. Each of these non-local
masses ($m^*_r$) are referred to as a ``cut-off mass", because if
the mass of a moving body ($m_p\,$ say) is above this (lunar
specific) mass, i.e. $m_p>m_r^*$, then the body cannot respond to
the model's (rotating space-warp based) additional perturbation of
acceleration/gravitational field strength --- see subsections
\ref{subsubsection:at-a-point} and \ref{subsubsection:convolution}
for further discussion. Clearly, the Pioneer S/C with a mass of
approximately\footnote{The Pioneer spacecraft launch mass
comprised a total module ``dry weight" of 223\,kg and an
additional 36\,kg of hydrazine propellant for the three thruster
pairs \citep[ pp.2-3]{Anderson_02a}.} 259\,kg lies well below the
various cut-off masses of the various (`active') moon-planet-Sun
systems --- regardless of the relative distance between the moon
and spacecraft. By contrast, the large mass of the planet Uranus
for example, at $8.68\times10^{25}$\,kg, is always immune from any
influence. Thus, the \emph{apparent} presence of a violation of
the weak principle of equivalence with regard to (local inertial
and/or gravitational) mass in a gravitational field, as associated
with the Pioneer spacecraft anomaly, is (i.e. has become)
non-problematic.

\subsubsection{Relating non-local mass to local `bulk' physical
mass (at a point)}\label{subsubsection:at-a-point} Obviously,
matter-comprised bodies are not point masses; they have
(three-dimensional) extension in (three dimensional) space. In
this subsection we enquire: how do the model's distributed
``at-a-point" non-local mass field values [$m^*(r)$ or $m^*_r$]
--- that extend throughout space --- relate to a finite
size macroscopic physical body ($m_p$)? We seek to address this
issue fully aware of the ambiguity surrounding just what non-local
mass ($m^*$, or $m^*_r$ at a distance $r$ from its source)
actually is.

In (classical) Newtonian gravitation we can usually assume that
the extended mass of a macroscopic body acts as if it is (totally)
located at a \emph{central point}. This idealisation relies on two
things: firstly, the spherical nature of (macroscopic)
gravitational influence, and the fact that the field values of
interest lie well beyond the surface of the `extended'
(gravitational source) mass itself; and secondly, that a passive
(gravitational) mass in a gravitational field is treated as a
point mass. In the model, quite different reasons shall result in
a similar (point-mass) circumstance. Clearly, we need to go beyond
classical physics (alone) to elucidate the nature of an
interaction/relationship concerning $m^*_r$ and the motion of an
(isolated in space) macroscopic physical body ($m_p$).

In physics, the correspondence principle states that: the behavior of
systems described by the theory of quantum mechanics reproduces
classical physics in the limit of large quantum numbers. In
subsection \ref{Subsubsection:Equiv and Uncert} we mentioned that:
``whenever the correspondence principle holds, the centre of mass of
a quantum wave packet (for either a single particle or for an entire
quantum object) moves according to Ehrenfest's theorem along a
classical trajectory."

Obviously, any collection of neighbouring atoms/molecules in a
bulk-matter body has a centre of mass. A bulk/macroscopic body may
alternatively be conceived (of) from a quantum mechanical
(wave-like) `perspective'; as such, each and every atom/molecule
(only) has a quantum wave packet associated with it. Importantly,
these numerous wave packets are distributed (over the ``whole of
space"). This (corresponding and complementary) conception of a
macroscopic body is analogous to the inherently \emph{distributed}
(throughout the universe) `nature' of non-local mass.
Subsequently, we note that, just as non-local mass (in the model)
can have \emph{both} a point-like value in a field, as well as a
distribution of values throughout the field; so the mass of a
macroscopic body (e.g. a spacecraft) may be conceived of in both a
(distributed/non-local) wave-like manner as well as in a
corresponding (local) particle-like manner --- with both
conceptions `appreciating' the usefulness/relevance of a
(point-like) centre of mass.

Taking this (dual) correspondence one step further, we hypothesise
that: as far as the interaction between $m^*_r$ and $m_p$ and the
model's quantification is concerned, the point-like `existence' of
non-local mass in a field has a `correspondence' to bulk/macroscopic
matter existing effectively at a \emph{point} in the field, and thus
we assume that $m^*_r$ interacts with the macroscopic matter $m_p$
``as if" the $m_p$ existed solely/totaly at its centre of mass.
Importantly, we (further) conjecture that $m^*_r$ is
\emph{concurrently} interacting with (the QM `face' of)
macroscopic/bulk matter $m_p$ in a (corresponding) quantum mechanical
wave-like manner, and as such the `at-a-\emph{point}' (in space)
relationship between $m^*_r$ and (total) $m_p$ is not an idealisation
--- as is the case in classical gravitational physics --- but a true
representation of their (field point-to-`particle')
interrelationship. Note that conceptually `we' --- meaning people
and scientists --- have a strong tendancy/bias towards a
particle-like (cf. a wave-like) perspective on things.

Planets, moons, comets, asteroids, and spacecraft are all
(considered) sufficiently compact/condensed (so as) to obey this
($m^*_r$ to $m_p$) interactive feature, but not so a (whole)
galaxy (for example). In subsection
\ref{subsubsection:convolution} we build upon this
hypothesis/conjecture and further describe how the $m^*_r$
\emph{upon} $m_p$ `interaction' is perceived and treated in the
model.

\subsubsection{How non-local mass influences physical objects with
(passive) mass}\label{subsubsection:convolution} If the passive
(condensed) mass ($m_p$) of (for example) a spacecraft
($m_{\rm{s/c}}$) or celestial body ($m_{\rm{body}}$), is
\emph{less} than the non-local mass ($m_r^*$) --- at a given
distance from a \emph{particular} rotating space-warp `generator'
--- then that mass \emph{is} considered to be influenced by the
new cyclical acceleration field variation, otherwise it is not.
Thus, the Pioneer anomalous acceleration affects spacecraft
(because $m_{\rm{s/c}}<< m_r^*$), but not planets, moons, and
`large' asteroids, where $m_{\rm{body}}> m_r^*$ --- at all times
and distances\footnote{Thus, one can create a demarcation between
large and small asteroids (and comets) based upon this criteria.
Interestingly, small comets ($<1$ km diameter) are unusually rare
(recall section \ref{subsection:further concerns}).}. See
\mbox{Table \ref{Table:cut-off}} for the different `cut-off'
masses (associated with each particular moon-planet-Sun system)
and typical comet size to comet mass relationships; (and) for
comparative purposes see/recall Table \ref{Table:acceleration} for
``large moon" masses.

The physics behind this `interactive' relationship between (each
specific) $m_r^*$ value and a body's $m_p$ is indeed hypothetical.
The physical nature of this relationship is considered to be
somewhat analogous to the mathematical (and physical) concept of
\emph{convolution}, in the sense of: ``the mathematical technique
for determining a system output given an input signal and the
system impulse response".

In the analogy: $m_r^*$ (and a space-warp undulation/sinusoid of
amplitude $\Delta a_w$) are the input signals; the system impulse
response is determined by a body's $m_p$; and the system output is
the additionally perturbed (or non-perturbed) motion of the
spacecraft\footnote{Like the photoelectric effect, this is an
``all or nothing" system output.}. Although usually associated
with the frequency response of a physical observing instrument
(that possesses inertia), \citet[ p.24]{Bracewell_00} notes that:
\begin{quote} Later we show that the appearance of convolution is
coterminous\footnote{Meaning: being the same in extent; coextensive
in range or scope.} with linearity plus time or space invariance, and
also with \emph{sinusoidal} response to sinusoidal stimulus.
\end{quote} With the (multiple) sinusoidal stimuli/inputs being the
multiple/various oscillatory RSWs (of amplitude $\Delta a_w$) over
their respective cycle times $\Delta t$, and the response being
the spacecraft's various speed ($\Delta v$) variations/sinusoids
and speed losses ($\delta v$) per cycle. In the model (only) five
moon-planet systems dominate the Pioneer acceleration anomaly; Io,
Europa, Ganymede, Callisto with Jupiter; and Saturn's Titan. The
Earth-Moon system, with its collision based heritage, is (for
geometric reasons) not a space-warp `generator' --- as indicated
in Tables \ref{Table:angular progression} and
\ref{Table:Efficiency}.

The physical nature of this (proposed) $m_r^*$ to $m_p$ interaction
appears to implicate some kind of inherently quantum mechanical
wave-like aspect of $m_p$ in the interaction (recall subsection
\ref{subsubsection:at-a-point}), that can only (at this stage) be
figuratively appreciated and formally represented in a rudimentary
manner. The conceptual intractability (or `weirdness') of QMs
provides little assistance in elaborating upon this coarse
description.

How the model's non-local mass might `interact' with the Higgs field
(and mechanism) is not discussed.

\subsubsection{Commenting upon local and global spatial quantities
in the new model}\label{subsubsection:local global spatial} The
constancy of the (non-local) mass-volume product and distribution
relationship of Equation \ref{eq:distribution} in subsection
\ref{subsubsection:enclosed volume}, with dimensions $[\rm{M
L^3}]$, ensures this new \emph{mass-spatial} product/relationship
describes a \emph{global} quantity. In contrast to this, the
non-local mass at a point in the field ($m_r^*$) is a \emph{local}
quantity, albeit of a distributed non-local `pedigree'. By its
very nature, the density of standard mass, which has dimensions
$[\rm{M/L^3}]$, is also a (spatially) distributed
quantity\footnote{Indeed, density is (arguably) `undefined' at a
point.}. Density is usually treated as a non-global (i.e.
body-specific) quantity --- although it (along with total baryonic
mass) also has one extremely global application, i.e. its
application to the universe as a whole.

The constant quantity $m_r^*V_r$, where $V_r$ is an \emph{enclosed}
spherical volume (of radius $r$), is: new, unnamed, and without a
representing symbol. Tentatively, we could call this quantity a
``volumetric non-local mass distribution constant", in that it
establishes the volumetric distribution of the non-local mass
component --- of the externally re-expressed (QM spin offset) energy
$\Delta E_w$. Dimensionally, this new quantity is peripherally
`related' to both: first moment of area $[L^3]$, and (mass) moment of
inertia $[M L^2]$ --- both of which involve \emph{rotation} in some
sense, with the latter also involving a \emph{distribution} of
matter.

Interestingly, the hypothesised existence of (each) space-warp
involves a rotation; further, each (gravito-quantum) rotating
space-warp is most readily idealised as an
(\emph{acceleration}/gravitation based) phenomenon, with a tubular
core\footnote{Planar in two dimensions, tubular in three
dimensions.}, which exists in \emph{conjunction} with a spherical
non-local \emph{mass} distribution that has a spherical ($r=r_1$)
core region\footnote{This tubular core (around the space-warp's axis
of rotation) vs. spherical core distinction further discourages the
possible implementation of a local energy ($E^*_{\rm{field}}$)
throughout space (as mentioned in subsections
\ref{subsubsection:total energy constancy} and
\ref{subsubsection:Global or universal}).}. Thus, this new quantity's
constancy, and its (unique) dimensionality of $[M L^3]$, might be
related to the geometric conflict between (a pure) spherical volume
and a rotating tubular-centered volume discussed in subsection
\ref{subsubsection:geometry's other role}.

Finally, it appears that any further rotation of a moon-planet-sun
system, e.g. within a galaxy (and so forth), is not problematic to
the model --- in that $m^*_r \,V_r=\rm constant$ (Equation
\ref{eq:distribution}) applies regardless of this motion. On a
more speculative note, there is a possibility that the constancy
of a distributed mass times volume product/relationship could have
application to the (formation and) evolution of spiral galaxies,
and/or the very early formation of the solar system.

\subsubsection{1862 Apollo: an asteroid exhibiting intermittent
anomalous behaviour?}\label{subsubsection:1862 Apollo} The mass
and orbital characteristics of the asteroid ``1862 Apollo" makes
it an ideal `exhibiting body' of, and `test body' for examining,
the model's proposed (distance dependent) non-local mass cut-offs
[$m^*(r)$, or alternatively $m^*_r$] associated with the four
Galilean moons of Jupiter. Saturn's greater distance ensures that
the rotating space-warp, associated with its large moon Titan,
fails `at all times' to influence this asteroid's motion, i.e.
$m^*(r)<m_{\rm{1862}}$ for all possible distances ($r$) between
1862 Apollo and Saturn-Titan.

By way of various sources/websites\footnote{Including: IAU Minor
Planet Center, JPL Small-Body Database Browser, Johnston's
Archive, and Wikipedia.} 1862 Apollo has a mass of
$\approx5.1\times 10^{12} \rm kg$ (denoted $m_{1862}$), a mean
diameter of approximately 1.7\,km\footnote{Associated with a mean
density of $\approx2.0 \, \rm{g/cm^3}$, i.e. $\approx2.0\times
10^{3} \, \rm{kg/m^3}$.}, an aphelion of 2.294AU, a perihelion of
0.647AU, an orbital inclination of $6.354^o$, and an orbital
period of 651.4 days. It is a Near Earth Asteroid (NEA) that
crosses the orbits of Mars, Earth and Venus. These physical and
orbital characteristics ensure that `Apollo' can receive either:
most, some, or none of the retardation effect stemming from the
individual (lunar-`driven') rotating space-warps (RSWs) ---
depending on the location of Jupiter (with its four Galilean
moons) and the asteroid itself, and the separation distance
between them. With Jupiter having an aphelion of 5.458AU and a
perihelion of 4.950AU, it turns out that the values of
$m^*(r)=m_{\rm{1862}}$ for Io, Europa, Ganymede and Callisto all
lie at distances \emph{between} their closest possible approach of
2.656AU, and a maximum possible separation distance of 7.752AU.

By way of using the mass cut-off values at 2.5AU (or/and 10AU) as
presented in Table \ref{Table:cut-off}, we may extrapolate (or
interpolate) to determine the distance ($r$) where
$m^*(r)=m_{\rm{1862}}$, i.e. where the (non-local) mass cut-off
value equals the asteroid's mass. These values, in ascending order
of distance, are: \mbox{Io 3.78AU,} Ganymede 4.36AU, Europa
5.43AU, and Callisto 6.78AU\footnote{Note that the value for
Saturn's large moon Titan is a mere 2.85AU.}. Thus, none, one,
two, three, or all four RSWs associated with Jupiter's Galilean
moons can be actively `causing' the (type of) motion retardation
(cf. predicted motion) that also affects the Pioneer 10 and 11
spacecraft\footnote{Albeit without the contributions arising from
`distant' Titan and Triton.}.

Assuming that the true mass and density values of asteroid 1862
Apollo are near their currently estimated values, the distance
dependent activation and deactivation of the RSW-based motion
retardation mechanism (upon the asteroid) then becomes an `in
principle' key prediction of the model. In practice, the
observation of this (four `geared') engaging and disengaging
behaviour will be extremely difficult to measure, due to the
`smallness' of the effect and the asteroid's position in the inner
solar system where radiation pressure effects are not
insignificant\footnote{Note that if the model's non-local mass
cut-off values were out by a factor of two (for example), due to
some minor mathematical oversight, the choice of the asteroid 1862
Apollo would still be of some use (but no longer ideal).}.

Our ability to directly observe asteroid 1862 Apollo from Earth
over an extended period of time (albeit intermittently) is a
positive; but the tiny motion variation effects\footnote{As the
strength of each supplementary gravitational field, associated
with each individual RSW, undulates the asteroid's speed responds
sinusoidally around an equilibrium speed --- assuming
$m^*(r)>m_{\rm{1862}}$. Note that in practice only a superposition
of these individual effects actually exists.} \emph{and} the
cumulative (per cycle) motion retardation effect involved appear
to be beyond the combination of our best models and best
observational techniques --- both now and in the near/foreseeable
future. Methods involving: an accurate radiation pressure model,
Doppler effects, and angular triangulation can be envisaged, but
they are `futuristic'. Conceivably, this would involve (for
example) either: (1) a transmitter/transponder (of some kind) upon
the asteroid itself, in conjunction with observing equipment upon:
Earth, Mars and a `large' (unaffected) asteroid; or (2) an
observing device upon the (unfortunately rotating\footnote{We also
note that the YORP (Yarkovsky-O'Keefe-Radzievskii-Paddack) effect
upon this asteroid has been successfully modelled.}) asteroid
itself, utilising the accurately known locations of three or more
planets\footnote{If this motion variation effect were measurable
with a high degree of accuracy, and the proposed model is valid,
then the mass and average density of `small' asteroids could
conceivably be determined with significantly greater accuracy than
is currently the case.}.

A more qualitative approach involving comparing the errors
accumulated over time in the predicted paths of asteroids (of
various masses), e.g. (re: Apollo asteroids) the more massive 1620
Geographos and/or the less massive 1566 Icarus, appears to be the
best option \emph{currently} available. Furthermore, a somewhat
step-like discrepancy in predictive accuracy involving the orbit
determination of asteroids whose mass is either side of an
`intermittent zone' --- spanning $\approx1.0\times 10^{12} \rm kg$
to $\approx2.7\times 10^{13} \rm kg$ (roughly corresponding to a
mean diameter range of 1.0\,km to 3.0\,km) --- would indirectly
support the existence and quantification of the non-local mass
cut-offs proposed by the model. Subsequently, such a result would
(also) provide good support for the model ``as a whole". Note that
`large' asteroids (i.e. $>3~$km in diameter) will not display the
additional/anomalous `deceleration' and thus they should have a
lesser discrepancy in their orbit determination --- all other
things being equal.
%**********************************************************************************
\subsection{The Pioneer anomaly, GR, and the three Equivalence Principles}
\label{Subsection:EquivPr Comment} This section investigates the
new model argued for herein in relation to two principles
associated with the establishment of general relativity: the
principle of equivalence and the general principle of relativity.
Such an analysis serves to clarify the veracity of the
gravitational supplementation that has been proposed.

\subsubsection{Introduction} Currently, general relativity (GR)
is generally regarded as the only theory (and ``final word") on
gravitation. If we assume that all systematic causes of the
Pioneer anomaly are ruled out, and embrace the rotating
space-warps (RSWs) --- or more formally the `gravito-quantum'
rotating space-warps (GQ-RSWs) --- implied by the Pioneer
anomalous observations, then a standard
appreciation/conceptualization of gravitation is necessarily
incomplete.

In this section (i.e. \ref{Subsection:EquivPr Comment}) we more
fully examine the relationship of the Pioneer anomaly to different
forms of the equivalence principle (EP)\footnote{Weak, strong, and
the Einstein equivalence principle.}. Of particular interest is
the apparent violation of the weak EP, also known as the principle
of universality of free fall, with regard to the motion of low
mass bodies\footnote{Below the cut-off mass at a given position
[i.e. $m^*(r)$ or alternatively $m^*_r$].} (e.g. the Pioneer
spacecraft). The ramifications of a `real' Pioneer anomaly upon
our understanding of gravitation in its widest sense is also
investigated, with an emphasis given towards understanding how our
new model is \emph{not} ruled out by GR's general principle of
relativity (GPrR).

Previously, it was argued that: the Earth flyby anomaly is
(merely) an observational artifact (subsection
\ref{subsubsection:flyby anomaly}) that arises indirectly from the
(gravito-quantum) RSWs associated with (and instigating) the
Pioneer anomaly. Consequently, our discussion of equivalence
principle violation, within the solar system\footnote{Subsection
\ref{subsubsection:Brief list} raised the issue of a
measured/apparent `change' in the fine structure constant. This
implies a violation of the Lorentz Position Invariance aspect of
Einstein's equivalence principle (EEP). Due to the provisional
nature of this observation, and the great distances involved, this
issue has not been pursued.}, can be solely restricted to the
(idiosyncratic) `Pioneer anomaly' case.

\subsubsection{General remarks regarding a violation of the Equivalence
Principle} This subsection provides a preliminary guide as to how
a violation of the equivalence principle (EP) is handled by
respected authors. We briefly explore this issue in terms of the
Pioneer anomaly specifically, and also the general conceivability
of an EP violation.

Firstly, the type of violation is unexpected. \begin{quote}
Neither case [Pioneer anomaly, nor Earth flyby anomaly] matches
expectations for an EP violation; for example, the directions of
the anomalous accelerations do not match [the authors] Equation
(1) \citep*[ p.2449]{Williams_01}. \end{quote}

Secondly, the violation itself is based upon an extremely minor
additional acceleration/gravitational field occurring in the solar
system's \emph{weak} gravitational field. Such a minor
violation\footnote{At least over fairly short (celestial) time
intervals, i.e. of the order of days, weeks, or months --- and
even years.}, especially one associated with quantum mechanical
energy in `partnership' with a curved spacetime field, is not a
cause for alarm --- and indeed something like it has been
`heralded' by Clifford Will. \begin{quote} \ldots there is
mounting theoretical evidence to suggest that EEP [Einstein
equivalence principle] is \emph{likely} to be violated at some
level, whether by quantum gravity effects, by effects arising from
string theory, or by hitherto undetected interactions. Roughly
speaking, in addition to the pure Einsteinian gravitational
interaction, which respects EEP, theories such as string theory
predict other interactions which do not \citep*[ p.23]{Will_06}.
\end{quote}

Thirdly, our approach of considering the GQ-RSWs as
additional/supplementary to general relativity's (approach to)
gravitation is not without some loose `ideological' support.
\begin{quote} \ldots the only theories of gravity that have a hope
of being viable are metric theories, or possibly theories that are
metric apart from very weak short range non-metric couplings (as
in string theory). \ldots In principle, however, there could exist
other gravitational fields besides the metric, such as scalar
fields, vector fields, and so on \citep*[ p.26]{Will_06}.
\end{quote} In our case/model the situation is somewhat different, with
the proposed very weak (non-metric-based) `coupling' being
(non-local and) long range, and taking the form of two scalar
fields: a constant amplitude ($\Delta a$) sinusoidal/oscillatory
field (also representing non-Euclidean geometry), and a variable
non-local mass [$m^*(r)$] scalar field.

Thus, to simply dismiss the Pioneer anomaly on the grounds of it
indicating a violation of the EP is too quick and inappropriate;
although there appear to be (\emph{prima facie}) sound reasons for
this dismissal, not the least of which is how (multiple
instantiations of) these scalar fields (particularly $\Delta a$)
coexist with standard general relativistic spacetime curvature. To
fully appreciate how a real (i.e. non-systematic based) Pioneer
anomaly may be contravening GR's equivalence principle (in an
acceptable manner), we need to reconsider (more deeply) what our
supplement to (GR's) gravitation really entails and its
relationship to the various guises of the EP. The latter is
pursued in subsections \ref{subsubsection:weak and strong} and
\ref{subsubsection:EEP exclusive}, the former from
\mbox{subsection \ref{subsubsection:GR supplprelim} onwards.}

\subsubsection{Pioneer anomaly as regards the weak and strong
Equivalence Principles}\label{subsubsection:weak and strong} For
the weak EP (WEP), \citet{Jammer_00} distinguishes between a
kinematic version, the principle of universality of free fall,
where at a given location all bodies fall with the same
acceleration; and a dynamic version where the gravitational
acceleration of (or acting upon) a body is independent of its
structure (i.e. composition and the amount of mass).

The long temporal and (long) distance propagational requirements,
as well as the very accurate measurements necessary to recognise
the very tiny (low mass only) Pioneer anomaly, are exceptional.
Comparatively \emph{short} temporal/duration torsion balance and
terrestrial `free fall' experiments involve masses
($m_{\rm{body}}$) on the \emph{same} side of all of the (lunar
specific) cut-off mass values [$m^*(r)$]; and gravitational
redshift experiments by way of the M\"{o}ssbauer effect are
unsuited to recognising this (Pioneer) EP violation. Similarly, we
note that the Earth-Moon (strong EP) ongoing `experiment' has (for
both the Earth and Moon) $m_{\rm{body}}>>[m^*(r)]_{\rm max}$.

The strong EP (SEP) assumes the complete physical equivalence of a
gravitational field and a corresponding acceleration of the
reference system, which then implies that there is no physical
difference between inertial motion without a gravitational field,
and free fall in a gravitational field. Regarding the `felt'
aspect of the SEP\footnote{As compared to the acceleration of a
\emph{reference system} in GR, which then allows the
scientifically rigorous (quantitative) theory of general
relativity to be formulated.}, we extend Einstein's
lift/elevator-based thought/gedanken experiment, where
$\underline{a} \equiv \underline{g}$, to a sinusoidal/undulatory
scenario with amplitude \mbox{$\Delta a \equiv \Delta g$;}
recalling that the latter is associated with a speed/motion
undulation of amplitude $\Delta v$ (recall section
\ref{Subsection:Shortfall}) --- in the case of low mass bodies
i.e. \mbox{$m_{\rm{body}}<[m^*(r)]_{\rm minimum}$.}

Interestingly, within the Pioneer spacecraft (itself) a `felt'
(temporal sinusoidal) acceleration variation\footnote{Using
nomenclature consistent with subsection
\ref{subsubsection:CSHPM}.} [$a_{\rm{proper}}=a=a(t)=-\Delta a
\cos(\omega t)$], whether delivered by way of a RSW [$\Delta g
\sin(\omega t)=\Delta a \sin(\omega t)$] or conceivably by a
mimicking variation in thrust [$-\Delta a_t \cos(\omega t)$],
causes a purely oscillatory/sinusoidal variation in motion
[$v_{\rm{proper}}=-\Delta v \sin(\omega t)=\Delta v \sin(\omega
t-\pi)$] around a steady mean speed that is pendulum-like or
swing-like in nature\footnote{We note that $\omega$ is the
rotating space-warp's angular frequency, and that the (proper)
velocity variations are 180 degrees out of phase with
acceleration/gravitational field variations (recall Figure
\ref{Fig:VelocityAccel}). For a spacecraft radially exiting the
solar system at negligibly relativistic speeds, we can
figuratively conceive (i.e. imagine) the forward looking
\emph{space} curvature `encountered' as acting like an inclined
ramp. The undulatory variation in acceleration/gravitational field
strength can be imagined as a smoothly varying cyclic change in
the angular slope of such a (curved space) `ramp'.}.
\emph{Locally}, i.e. within the (windowless) spacecraft, there
would be no physiological sense of an unmodelled monotonic motion
`shortfall' over extended intervals of time, i.e. (no sense of) an
additional/supplementary `constant' (Pioneer anomaly-like)
deceleration\footnote{Regardless of the spacecraft's direction of
motion: circumferential, radially inward, or radially outward,
etc.}. A number of provisos (concerning this statement) need to be
appreciated.

We recall (from section \ref{Subsection:Shortfall}) that the
motion shortfall is relative to \emph{predicted} motion. We also
note that the \emph{Pioneer} anomalous acceleration arises from
(the superposition of) a \emph{number} of these (coexistent)
rotating space-warp based sinusoids.

Further, it is only in the global/barycentric frame, and by way of
having/using a theoretical model to \emph{predict} a spacecraft's
trajectory and motion, that a small redistribution of S/C
translational/propagational kinetic energy (down from $100\%$)
into undulatory/oscillatory kinetic energy has physical validity.
Predictions of S/C motion that omit this feature, inevitably
notice a motion shortfall (over time) that is `declared' the
Pioneer anomaly. Importantly, `locally' we could (still) in
principle transform away gravitation by choosing a particular
reference frame, with the additional terms being `understood' as
inertial force terms. Thus, (in this sense) the strong equivalence
principle, associated with Einstein's (windowless) lift thought
experiment (or a windowless moving spacecraft), is \emph{retained}
when extended to a pure undulatory/oscillatory gravitational field
--- i.e. a first order cyclic perturbation (of \emph{constant}
amplitude) superimposed upon a (locally\footnote{Both spatially
and temporally --- in relation to one's location `onboard' the
spacecraft.}) \emph{steady} gravitational field.

We note that a global/barycentric perspective is necessary, in
addition to a local `felt force' perspective, to appreciate the
(counter to predicted motion) Pioneer anomaly\footnote{Adding
windows to the spacecraft (or elevator) would in principle (i.e.
conceivably) `allow' anomalous (i.e. unpredicted/unmodelled)
barycentric position and/or speed variations to be
\emph{indirectly} ascertained.}. By comparison, GR is/was
constructed/established upon a \emph{local} principle of
equivalence (and `felt-force', or lack thereof) perspective. This
need for the coexistence of two (or dual) perspectives, i.e. local
and systemic as regards this new (supplementary) gravitational
effect, is symptomatic of the new model's approach.

It is not unreasonable to surmise/conclude that: both the
kinematic and dynamic versions of the weak EP `can' be violated
(in some sense), although in practice this violation is
observationally both: exceptional (or even unique), and
exceedingly small in magnitude. A comparison of planetary motion
and Pioneer spacecraft motion confirms a violation (of some kind)
--- albeit requiring a global/barycentric perspective, as well as
a local Earth-based observer. It becomes a question of (semantics
and) one's own conceptual perspective as to whether one considers
the \emph{acceleration} (equivalence to gravitation) \emph{aspect}
of the strong equivalence principle (SEP) to have been violated by
the Pioneer S/C. If we \emph{restrict} ourselves to a local
physiological perspective, as Einstein did with his windowless
lift/elevator experiment, then the SEP is \emph{not} violated ---
even in our undulatory/cyclic gravitational field circumstance
(i.e. subject to the effects of RSWs). The SEP \emph{is} violated
when it subsumes the existence of a mass-based violation of the
WEP. Further, the proposed existence of a non-local mass field
$[m^*(r)]$ (or fields) --- related to the non-local implications
of (lunar-based) quantum mechanical systems --- is supplementary
to the scope of the assumed `equivalence' of inertial mass and
passive gravitational mass: $m_i \equiv m_p\,(=m_g)$.

`Large-mass' solar system bodies such as: planets, moons, large
comets, and large asteroids do not respond to the additional
acceleration/gravitational field associated with GQ-RSWs, whereas
the Pioneer 10 and 11 spacecraft (and other `low mass' bodies) do
respond; for the latter we can `figuratively' write that:
$\underline{a}_{\rm{body}} \equiv \underline{g}_{\rm{total}}$.
Consequently, by way of \emph{not} indicating the existence of the
supplementary additional acceleration/gravitational field, it is
(surprisingly) these large-mass celestial bodies that are all in
\emph{violation} of the `SEP', when the SEP is extended to first
order field perturbations --- albeit a very minor violation
(hitherto unappreciated). In other words, the motion of
larger/more-massive bodies does not `completely' exhibit/indicate
the (entire) non-Euclidean geometry of the solar system, such that
(for them): $\underline{a}_{\rm{body}} \neq
\underline{g}_{\rm{total}}$.

\subsubsection{The Einstein Equivalence Principle and
exclusivity of the metric in GR}\label{subsubsection:EEP
exclusive} Extending the SEP, Einstein made the Einstein
equivalence principle (EEP) part of the bedrock of his general
theory of relativity (GR). It acts (somewhat) as a correspondence
principle between SR and GR, by ensuring that locally SR's Lorentz
invariance and position invariance are enforced. Clearly, the
model's supplementary acceleration/gravitational field ---
associated with non-Euclidean geometry --- needs to be in denial
of the `exclusivity' of the metric tensor(s) arising from GR's
field equations, in the sense that we need to question the
following `exclusive' statements: \begin{enumerate} \item{If SEP
is strictly valid, the (components of the) metric \emph{alone}
determines the effect of gravity.} \item{Matter and
non-gravitational fields respond \emph{only} to the spacetime
metric.} \item{The gravitational field is \emph{entirely}
describable by a universal coupling of the (macroscopic)
mass-energy contents of the `world' to the metric tensor.}
\item{Test bodies follow geodesics of \emph{the} metric.}
\end{enumerate}

As mentioned previously, there is nothing wrong with GR; rather
our concern is necessarily with the scope or domain of application
inherent in GR's conceptualisation and formalism, and hence the
idea that GR is the `final word' on gravitation --- such that GR
`completes' gravitational theorisation. Supporting this concern we
recognise the exclusions and idealised circumstances associated
with GR outlined in subsection \ref{subsubsection:Beyond GR}, and
the various restrictions to gravitational theorisation discussed
in subsection \ref{subsubsection:logic of GCoV}. We reiterate a
few pertinent examples. Firstly, tidal effects, which arise from
non-uniformity in gravitational fields, illustrate why there is a
need to restrict the acceleration/gravitational/curvature aspect
of the SEP to a \emph{local} effect; this is in stark contrast to
the cosmological scale amplitude uniformity ($\Delta a$)
associated with the unsteady/undulatory field of the model's
GQ-RSWs. Secondly, effectively \emph{only macroscopic} bodies are
considered, with all (`isolated' cases of) microscopic/quantum
mechanical contributions to gravity (including energy effects)
effectively assumed negligible; note that this negligibility
includes the additive sum of a QM angular momentum rate (i.e.
energy), externalised as GQ-RSWs, in our model. Thirdly, in
subsection \ref{subsubsection:Scope} we briefly discussed how the
non-locality of QM entanglement\footnote{Also known as ``the
quantum non-local connection"; (and) which herein involves a
`transmutation' of QM spin \emph{energy} into `gravitational'
energy cf. (simply involving) an equal spin to spin (i.e. angular
momentum) relationship.}, and a finite communication rate (i.e.
light speed) in SR and GR, sit together uneasily
\citep{Albert_09}\footnote{Contemporary physics `accepts' a
situation whereby: non-local correlations do occur, but because
they cannot be used to transmit information, they do not violate
causality. Our `uneasiness' stems from a lack of explanatory
depth, in that an (additional) process/mechanism to account for
the ongoing `sustenance' of these correlations is not considered
absolutely necessary when explaining entanglement.}.

Furthering our concern, Einstein, in creating and developing GR,
(as late as 1916) sought to \emph{generalise} the relativity of
motion in special relativity (SR) to accelerated
motion\footnote{Additionally, he also sought, and eventually
abandoned (as late as mid 1918), a `Machian' relativity of
inertia, thus resigning himself to a physics in which spacetime
has independent existence and physical qualities
\citep{Hoefer_94}.}. The resultant General Theory of Relativity is
astounding, but this (aforementioned) `elegant' objective was not
entirely fulfilled; nevertheless its `footprint' remains in GR's
(approach to its) formalism.

Furthermore, recalling subsection \ref{subsubsection:Non-rel
aspect of g}, acceleration and rotation (in at least one sense)
resist a purely relativistic understanding, even though
\emph{observations} of accelerated motion are relative. We note
that (coincidentally and fortuitously) rotation \emph{and}
acceleration/gravitational field undulations together, in the form
of (gravito-quantum) rotating space-warps (GQ-RSWs), are central
to the model\footnote{Recall that (each): lunar, atomic/molecular,
self-interference, and geometric phase based, supplementary
rotating space-warp [of undulatory acceleration amplitude ($\Delta
a$)], coexists with a non-local mass distribution [$m^*(r)$].
Together, they express the total non-local externalised
(`fractional' and inexpressible) quantum mechanical energy
--- associated with the many atoms/molecules within each:
geometrically suitable, spin-orbit coupled, (large) moon of the
solar system. The contribution of the various RSWs (with
perturbation amplitude $\Delta a_w$) to non-Euclidean geometry is
necessarily a superposition of \emph{secondary} `gravitational'
effects, in that they `piggy-back' upon, and require the
\emph{pre-existence} of, standard general relativistic (spacetime)
non-Euclidean geometry throughout the solar system.}.

Finally, but on a lesser note, it can be argued that the
`extension' of SR to GR has a `downside', in that if a weakness
and/or alternative interpretation exists in our physical
understanding of (all that is within) the Theory of Special
Relativity, then this flaw will somehow `express' (or even
compound) itself in GR. Our concern lies with the `extension' of
SR's (very successful use of) Minkowski/flat \emph{spacetime} into
GR's (equally successful use of) curved spacetime. Specifically,
the latent presumption that only four-dimensional spacetime is
associated with gravitation (in its broadest sense); this
(ontological feature) is contrary to what our model requires, i.e.
a perspective where space and time (need to) retain some degree of
independent existence. \emph{Three} features, associated with
`diminishing returns' or restrictions as we move from SR to GR,
loosely support this concern. Firstly, \emph{global} Lorentz
covariance goes to \emph{local} Lorentz covariance. Secondly, GR
requires (more complicated) covariant derivatives and metric
tensors cf. the partial differentials of non-relativistic
macroscopic physics. Thirdly, GR's formalism, particularly the
(non-linear) Einstein field equations, have very few
simple/elegant solutions. These features of localisation and
complication, can be interpreted as merely signifying
\emph{restrictions} necessary for the establishment of a `general'
theory of relativity, but additionally/alternatively these
restrictions may be interpreted as providing indirect support for
our model's supplementation of gravitation, with its
`supplementary' use of (independent) space and time cf. spacetime
--- as argued for in Section \ref{Section:PhiloTheory}.

This subsection has sought to illustrate how general relativity,
by way of its `make-up', is actually \emph{incapable} of
describing the very minor additional gravitational phenomenon
associated with the Pioneer anomaly. We have needed to cast doubt
upon the necessity and exclusively of a (spacetime) metric-based
approach, partly by way of an exceptional circumstance in our new
model; i.e. a constant amplitude (rotating) `warped'
acceleration/gravitational field --- albeit undulatory/sinusoidal
at a `point' in space, or rather upon a fixed direction radius
`arm' (centered at a lunar GQ-RSW `generator'). Further signs of a
restrictiveness within GR's approach, especially regarding its
domain of application, include: a non-relative aspect pertaining
to (the motion-based concepts of) rotation and acceleration; the
existence of QM \emph{non-local} relationships/behaviour; and the
fact that GR is necessarily a \emph{relativistic} theory of (and
approach to) gravitation --- requiring special relativity and
Minkowski spacetime as a `foundation'. Aspects of GR, such as
localisation (regarding the Einstein equivalence principle), and
the inherent difficulty in solving the Einstein field equations
are loosely supportive of these concerns; as is the failed
agenda/goal to `relativise' accelerated motion and/or inertia (in
the final form of the theory). Citing these `imperfections',
within and peripheral to GR, indirectly acts to support our new
model; this \emph{modus operandi} (i.e. method of operating) is
furthered in subsections \ref{subsubsection:GR supplprelim} and
\ref{subsubsection:G Pr of Rel}.

\subsubsection{Discussing the new model's need to supplement General
Relativity}\label{subsubsection:GR supplprelim}  A real Pioneer
anomaly cuts to the core of gravitational conceptualisation,
although it does \emph{not} imply that general relativity is in
error. Instead, GR is seen as an incomplete account of
`gravitation' --- in the broadest sense of the word --- in need of
a quantum mechanical (and Heisenberg uncertainty principle-based)
supplementation; with quantum mechanical entanglement and
non-locality, in conjunction with geometric phase, also playing a
vital role. It remains the case that (i.e. the new model respects
that): \begin{enumerate} \item{The only theories of gravity that
can fully embody the EEP (by default) are those that satisfy the
postulates of metric theories of gravity.} \item{In local Lorentz
frames, the non-gravitational laws of physics (describing
observations of physical phenomena) are those of special
relativity.} \item{The strong equivalence principle (SEP) implies
that gravitational acceleration is (in one sense) an entirely
geometrical (natural) phenomenon.}\end{enumerate} The model's
sinusoidal gravitation/acceleration field supplement (of amplitude
$\Delta a$), in conjunction with non-local mass [$m^*(r)$], is
\emph{inconsistent} with both the EEP and the exclusivity of GR's
metric, even though removing gravitational effects locally regains
SR\footnote{Such that the theoretical/formal requirements of local
Lorentz invariance and local position invariance are
\emph{observationally} upheld, and such that SR remains a limiting
case of both GR and `gravitation' in its widest sense --- with the
latter additionally involving the RSW supplementation.}. Our aim
here is not to pursue a drawn-out scholarly review of GR; rather,
we have sought to highlight how GR's non-Euclidean
(spacetime-based) curvature/geometry needs to coexist with the
RSW-based \emph{supplementary} non-Euclidean curvature/geometry.
Recalling section \ref{subsection:SR's ontology}, this supplement
needs to (additionally cf. alternatively) treat space and time as
distinct `features' of reality --- albeit requiring an additional
conceptual level of `reality', separate from (and `prior' to)
observational/phenomenal time and reality. This ontological
feature is introduced so as to be able to `deal with': QM
non-locality and entanglement, (an externalisation of)
non-inertial QM energy, and the new gravitational effect,
\emph{concurrently}.

The primary difference, between GR's standard gravitation and our
supplementary gravitational effect, lies in there being two
distinct \emph{source `classes'} and two distinct theoretical
representations of non-Euclidean geometry. In GR the sources are
formally represented by the stress-energy tensor, whose components
involve: (macroscopic) energy density and flux, momentum density
and flux (as well as shear stress and pressure). In our model a
physically inexpressible sub-quantum mechanical-based
energy\footnote{Arising (in part) from a conflict involving the
\emph{discrete} energy levels associated with atoms and molecules,
and \emph{analog} geometric phase offsets arising from the closed
loop motion of these QM systems in analog curved spacetime.}, in
the form of a (quantifiable) \emph{process}-based per cycle rate
of (intrinsic) angular momentum (shared by a great many
atoms/molecules), is necessarily `re-expressed' as (macroscopic)
non-Euclidean geometry (i.e. as a GQ-RSW)\footnote{A loosely
analogous situation would be the `expression' of ripples on a pond
when some additional energy is imparted to the water/medium.} with
an accompanying non-local mass distribution --- so as to ensure
that global conservation of energy is maintained `through' time.
This conservation of energy involves both: microscopic \emph{and}
macroscopic aspects of reality (with seamless functional
coordination); as well as a form of functional separation of
energies between these two domains/realms\footnote{Note that
electromagnetism is unique amongst physics' `four forces', in that
it can be bifurcated in at least two special ways:
macroscopic/classical level phenomena vs. microscopic/QM
phenomena, and wave vs. particle aspects.}, so as to ensure a
(micro-to-macro) re-expression/`transfer' of (a given/equal
quantity of) energy\footnote{Note that an \emph{agenda} that seeks
to `unify' gravitation/general relativity and quantum mechanics
--- regardless of their generally disparate phenomena and
disparate mathematical descriptions --- into a ``theory of
everything" or ``unified field theory" does not need to
`subscribe' to the `dual-conception' stance alluded to herein.}.
Note that the `macroscopic' in our model unavoidably requires a
systematic/global (i.e. non-purely-relativistic) perspective.

In a manner of speaking, the inertial \emph{force} terms of GR
(cf. SR), need to be supplemented by, and \emph{coexist} with, a
(systemically/globally relevant) inertial \emph{energy} offset
(established) at the microscopic level (recall subsection
\ref{subsubsection:inertial}). This supports our assertion of
\emph{two} quite distinct types/sources of curvature/non-Euclidean
geometry, with two quite distinct types/methods of (theoretical)
representation. Differences between GR and our new approach
involve: the field types, the role of energy conservation, and the
nature of space and time in theory formulation\footnote{Einstein's
SR drew upon Poincar\'{e}'s light signal based operationalist
methodology, so as to ensure that the speed of light between
different inertial frames, and Maxwell's equations of
Electromagnetism, remain invariant. This required the
implementation of Minkowski spacetime in the formalism. Our
concern herein is with the further presumption, as based upon
(fully valid) scientific \emph{observational} methodology, that
spacetime is ontologically real, and \emph{completely} replaces
the prior/earlier conceptualisation of space and time. As we have
discussed previously (particularly in sections
\ref{subsection:SR's ontology} and \ref{subsection:Reversal}), our
model requires a noumenal supplementation to this purely
observational/phenomenal (spacetime) approach; in that outside of
observations (and relativistic theoretical formulations), space
and time as distinct `entities' needs to maintain theoretical
validity --- especially/crucially when quantum mechanical energy
and QM non-locality and entanglement are involved.}. Unlike GR,
the new model makes use of: QM non-locality, QM self-interference,
(hidden/background) `noumenal' systemic/global aspects, and
three-body orbital motion --- with the latter being described by
`phenomenal'/observational based `relativity'. Furthermore,
non-locality appears to necessitate the use of a
\emph{non-observationally} perceptible global space substratum (or
continuum) coexisting with an `idealised' background (hidden) time
simultaneity (recall Section \ref{Section:PhiloTheory}); this
suits the `weirder' aspects of QMs, but it is incompatible with
the foundation ``principles" and ontological assumptions upon
which general relativity was constructed and \emph{is} built.

For the devotee of general relativity's perfection and
exclusivity, who is incapable of appreciating the possibility of a
`holistic' supplementation\footnote{Holistic, in the sense of
needing to appreciate the universal/cosmological system as a whole
at different given `moments' (i.e. from a non-relativistic
noumenal perspective), rather than only from a `sum' of the parts
perspective --- involving for example: the metric at different
points, different spacetime intervals between two events, and
clocks `running' at different rates.} that coexists with a
`relativistic' approach to gravitation, the model's new approach
and ontological stance will necessarily remain `anathema' to them;
as will the idea that the Pioneer anomaly is physically real and
indicative of a new physical phenomenon --- as compared to an
undetected/unappreciated mundane systematic effect. Note that in
achieving our new model, we have employed an eliminative
methodology guided by the highly constraining observational
evidence (outlined in section \ref{Subsection:Primary
observational}) --- subsequent to \emph{assuming} the
(non-systematic based) `reality' of the Pioneer anomaly. This
eliminative methodology involved proceeding much in the
manner/spirit of Sir Arthur Conan Doyle's fictional detective
Sherlock Holmes, in that we have sought to adhere to his creed
that: ``\ldots when you have eliminated the impossible, whatever
remains, however improbable, must be the truth." In our case,
Doyle's `truth' is replaced by the model's proposed/hypothesised
`new physics'.

\subsubsection{Post-production chinks in the general
principle of relativity's armour}\label{subsubsection:G Pr of Rel}
The general principle of relativity (GPrR) is the requirement that
the equations describing the laws of nature/physics have an
equivalent form in all systems of reference (and observation) ---
i.e. inertial and accelerated frames. \begin{quote} The great power
possessed by the general principle of relativity lies in the
comprehensive limitation which is imposed upon the laws of
nature\footnote{A view endorsed by philosopher of science Karl
Popper, who believed (\emph{The Logic of Scientific Discovery}, 1972,
Addendum to Section 40): the more laws of nature prohibit, the more
they say.} \ldots \citep{Einstein_20} \end{quote}

On the proviso that: the $2 \Delta a$ range of sinusoidal
acceleration/gravitational field strength lies within the `wiggle
room' allowed by Heisenberg's uncertainty principle, the
\emph{compatibility} of our new and supplementary model with GR's
GPrR is (trivially) assured because the undulatory field
strength/amplitude ($\Delta a$) of the RSWs is \emph{constant}
throughout the universe. We note that the variable non-local mass
[$m^*(r)$] accommodates energy dissipation with increasing
distance from the source of a given GQ-RSW. Thus, [excluding the
source-centered $m^*(r)$ distribution] all systems of (far-field)
reference are (necessarily) equivalent with respect to the
formulation of laws concerning (the model's supplementary)
gravitation/acceleration/curvature --- in the form of a number of
gravito-quantum rotating space-warps (GQ-RSWs).

Our model accepts that there is no (justifiable) preferred
reference frame for describing: SR, GR's
gravitation\footnote{``Just as one could formulate Newtonian
mechanics or special relativity in generally covariant
coordinates, so it is possible to formulate general relativity in
a preferred coordinate system. Einstein does not argue that it is
not possible to do so, but merely that such a formulation imposes
a formal structure without physical relevance \citep*[
p.106]{Lehner_05}."}, and the \emph{laws} of physics; and that the
notion of an (observationally `active') aether/medium is ``without
merit". Nevertheless, the cosmic microwave background
radiation\footnote{First detected in 1964 by American radio
astronomers Arno Penzias and Robert Wilson.} (CMB radiation)
appears to ``fit the bill" of a privileged universal frame (of
sorts) in our expanding universe, in that it permits a local
determination of global/absolute speed and/or direction of travel
in space --- i.e. with respect to the rest frame of the CMB
radiation --- e.g. Earth dipole motion/velocity. A
thought/gedanken experiment involving two spacecraft (in inertial
motion) fitted with equipment that can ascertain the spacecraft's
speed and direction relative to this CMB `rest' frame, is
indirectly also ascertaining a common third `frame' (cf. observer)
which is both: at `rest' and of cosmological extent. Such a
circumstance, while not denying the special principle of
relativity, is not (at all) in the spirit of special relativity as
originally formulated. Similarly, quantum mechanical non-local
behaviour\footnote{Observationally well supported by Alain Aspect
and his colleagues (including P. Grangier, G. Roger, and J.
Dalibard) at Orsay, Paris in 1981-1982 with their `Bell test
experiments' (post-dating an `initial' Bell-test of S. J. Freedman
and J. F. Clauser in 1972). These experiments confirmed (John S.)
Bell's theorem (or Bell's inequality), first published in 1964 and
later refined [e.g. to the CSHS (Clauser-Horne-Shimony-Holt)
inequality in 1969]. John Bell drew his inspiration from the
Einstein-Podolsky-Rosen (EPR) paradox, originally published in
1935.} `goes against the grain' of special relativity as
originally conceived.

Drawing upon \citet{Lehner_05}, the general principle of
relativity (GPrR) is meant to be a principle about the real world.
As such, general covariance, which seems to be (only) a formal
property of a theory, cannot alone imply GPrR. Possibly it can if:
``Nature's laws are merely statements about temporal-spatial
coincidences; \ldots"; but as John Norton has pointed out, this
claim is far from trivial. Furthermore: \begin{quote} [Such a
claim] is not even plausible without the fundamental
reinterpretation of space-time properties that general relativity
produced. Newton's [rotating] bucket [argument] is --- on the face
of it --- a striking counterexample to this claim \citep[
p.105]{Lehner_05}."
\end{quote} Subsequently, Lehner regards the mature formulation of
Einstein's principle of general relativity (\emph{circa} 1921) that: ``all
states of motion are equivalent in principle, before a specific
distribution of mass in the universe is specified \citep[
p.106]{Lehner_05}", as having its (real world) basis upon/within the
principle of equivalence (cf. Mach's principle or general covariance) ---
which requires that there is no physical difference between inertial motion
without a gravitational field and free fall in a gravitational field.
Additionally, Einstein believed that:
\begin{quote} If this equivalence is to be a fundamental principle
rather than a coincidence in physical phenomena, there can be no
structure like Newtonian absolute space that distinguishes a
priori which motions are inertial and which are not \citep[
p.106]{Lehner_05}." \end{quote} A statement we endorse, as far as
regards \emph{Newtonian} absolute space, but shall reject in
response to our additional and idealised noumenal (beyond/prior to
observation) perspective --- hypothesised, discussed, and argued
for in Section \ref{Section:PhiloTheory}.

\subsubsection{Summarising aspects of GR's principle approach and further
discussion}\label{subsubsection:GRposthistorical} A major
objective of this section (\ref{Subsection:EquivPr Comment}) has
been to show how and why our model can supplement and coexist with
GR. Interestingly, the new model is not in conflict with much of
GR's underlying \emph{conceptual} structure and (formative)
\emph{principles}. For example:
\begin{enumerate} \item{We endorse special relativity's stance
that there can be no structure like a (physically independent/`inert')
Newtonian absolute space and no (same \emph{rate} for all observers)
absolute time.} \item{Special relativity maintains its role as a
limiting case, in the absence of gravitational effects.} \item{The
acceleration aspect of the strong equivalence principle is maintained,
so as to not preclude its extension to first order
(undulatory/sinusoidal) variations upon a (pre-existing `steady')
gravitational field.} \item{Furthermore, gravitational acceleration
remains conceptually `reducible' to, and concomitant with, a
curvature-based (and non-Euclidean geometry-based) `situation'.}
\end{enumerate}

A clear omission of our new model is that the discussion of
curvature (i.e. non-Euclidean geometry) is only conceptual; no
formal representation of the (supplementary) \emph{curvature}
itself is given. The model's major distinction (cf. GR) --- as
demanded by the awkward observational constraints and a
macroscopic re-expression of (an externally imposed\footnote{By
way of three-body celestial motion in smoothly (cf. discontinuous)
curved spacetime.}) many atoms/molecules-based `fractional' QM
energy `offset' --- lies in the introduction of a new (variable,
and source-based) non-local mass scalar field [$m^*(r)$]. This
field, in conjunction with the RSW constant ($\Delta a$) amplitude
undulations, has the ability to influence masses in celestial
motion in a manner that lies beyond GR's theoretical `coverage'
--- both formally and (in some regards) conceptually.

The model's nonconformity with the (empirical rule of) inertial
mass to (passive) gravitational mass equivalence\footnote{Based
upon (macroscopic) experiments dating back to the 17th century.},
suggests that we (need to) further (conceptually) `detach'
gravitational acceleration from \emph{passive} (gravitational)
mass. This move (further) undermines the notion of gravitational
`force' [with dimensions ($\frac{ML}{T^2}$)] as a primary
`gravitational' concept --- a feature initiated by GR's
curvature-based understanding of gravitation. Inevitably, this
[$m^*(r)$ based] exception/violation to a general/all mass-type
equivalence --- that is generally considered to be restricted to
$m_i \equiv m_g$, but now (newly) incorporates $m^* \neq m_g$ ---
means that the principle of equivalence no longer attains
universal physical \emph{principle} status. Recalling the
discussion in subsection \ref{subsubsection:G Pr of Rel}, this
violation subsequently denies the (without exception)
generalisation of SR's relativity principle; although the general
principle of relativity (GPrR) still applies when we restrict
applications to solely macroscopic physics --- but only regarding
the type of theoretical \emph{formalisms} that may be established,
cf. the GPrR having watertight `real world' \emph{physical}
validity. Actually, the Theory of General Relativity can be
applied almost universally, except in the rare/exceptional and
probably unique (macro- \emph{and} microscopic/QM combined)
circumstances exhibited by our modelling and explanation of the
Pioneer anomaly by way of a QM energy \emph{source} --- which is
an ``exception to the rule" of general relativity's gravitational
`law'.

Importantly, our stance regarding the principles of: equivalence,
general relativity, and general covariance mimics (and is
compatible with) that of Einstein (\emph{circa} 1920 and
beyond\footnote{As discussed by \citet[ p.105]{Lehner_05}.}), in
that these principles are seen to have played a historical and
heuristic role. They are no longer seen to be an essential part of
the `basis' of the theory, although (in this author's opinion)
their formative `footprint' remains. This legacy allows doubt to
be cast upon the \emph{completeness} of a general relativistic
approach to gravitation, and allows concern to be raised regarding
GR's (scope or) domain of application. There is nothing like a
gravitational ``completeness theorem" for general relativity's
theoretical method and scope, merely a firm belief in its
comprehensiveness.

\emph{Circa} 1920 \mbox{`` \ldots Einstein} [now] postulates the
metric field as the fundamental entity of his theory \citep[
p.105]{Lehner_05}." This is understandable, in that having
completed the theory's formulation, and appreciating its
observational confirmation, it is the metric tensor (at each
point) --- i.e. the solution(s) of the Einstein field equations
--- that inevitably becomes the new and more appropriate focus of a
relativistic theory of gravitation\footnote{In a manner of
speaking, the `metric' in GR does the `heavy lifting' of the
theory. Aspects/features of \emph{the metric} include: that the
curved spacetime geometry around a star (for example) is described
by a metric tensor at every point; it generates the connections
which are used to construct the geodesic equation of motion and
the Riemann curvature tensor; the Ricci tensor and scalar
curvature depend on the metric in a complicated non-linear manner;
it is used to raise and lower indices; (and finally) in the weak
field approximation (applicable in our solar system) the metric
tensor is closely related to `gravitational potential'.}. Such a
\emph{change} in attitude/focus, although
theoretically/epistemologically neutral\footnote{In that
subsequent to GR's formalism having been successfully established
in 1915 (the Einstein field equations etc.), the nature of the
formalism has remained fixed/unmodified --- notwithstanding the
contentious issue of whether or not to include a cosmological
constant term.}, nevertheless involves two major \emph{implicit}
assumptions that have needed to be disassembled.

Firstly, our new/supplementary model's very existence has demanded
that we find fault with the assumed `exclusivity' of GR's metric,
as well as the finality/completeness associated with the Einstein
field equations (see subsections \ref{subsubsection:EEP exclusive}
to \ref{subsubsection:G Pr of Rel}). This exclusivity is based
upon both: the need for a generally covariant approach, and the
physical validity of the general principle of relativity. We
argued that the latter requirement has not been achieved, and note
that general covariance --- a formal (rather then a physical)
requirement/constraint for a relativistic \emph{theory} of
gravitational phenomena --- is insufficient on its own to warrant
this exclusivity.

A second implicit set of assumptions that needed to be
disassembled concerns certain ontological assumptions/beliefs,
related to the (long past and completed) theoretical
\emph{formative} process that led to GR, which have become `built
into' GR's relativistic approach to gravitation. In Section
\ref{Section:PhiloTheory} we discussed how a model for a `real'
Pioneer anomaly has no choice but to revisit the `standard'
\emph{ontological} commitments/assumptions of special relativity
(that are extended `into' GR); especially the veracity/reality of
the independent existence of Minkowski spacetime and its
associated physical qualities. This specific concern was dependent
upon our taking issue with the `standard' (and
\emph{solitary}/exclusive) interpretation given to space and time
in special relativity. In Section \ref{Section:PhiloTheory} we
proposed a \emph{alternative}/complementary (less simple, more
subtle) interpretation that is consistent with \emph{both}: the
observational evidence relevant to the Special Theory of
Relativity, and non-local behaviour in Quantum Mechanics. Only
this secondary/supplementary interpretation could have facilitated
the construction of the model presented in this paper, so that
this model can now stand alongside Relativity Theory --- upon
which we have performed a partial \emph{conceptual}
`reconstruction'.

\subsubsection{Brief summary and closing remarks regarding section
\ref{Subsection:EquivPr Comment}} Our investigation of
the Pioneer spacecraft's `apparent' violation of (at least one form of) the
Equivalence Principle has required a fairly extensive investigation of
general relativity's conceptual foundations. Combining this investigation
with an appreciation of the new model's content, allows us to reach an
understanding of how the motion of the Pioneer spacecraft is not in
defiance of general relativity (i.e. a relativistic theory of gravitation).

This section (\ref{Subsection:EquivPr Comment}) has argued that
the `footprint' left upon the Theory of General Relativity by
both: Special Relativity's use of Minkowski spacetime, and the
somewhat antiquated notions of: the equivalence principle (in its
various forms) and the general principle of relativity --- by way
of their use in the formative/development stage of general
relativity (GR) --- have acted to restrict (in a good way) GR's
theoretical formalism. Unfortunately, if the conceptual argument
used to establish a \emph{complete} and proper physical
generalisation of (special) relativity is open to doubt, or indeed
invalid, then the formative `footprint' left by these conceptual
`building blocks' can also limit/restrict (in a negative manner)
the scope/`domain' of General Relativity's application. We have
argued that this is the case. This less than perfect situation is
indirectly supportive of our model, with its stated aim of
establishing a viable `best' explanation/hypothesis for the case
of the Pioneer anomaly being `real' cf. a systematic effect.

It has become apparent that the formal content of our (new and
provisional) model is mutually exclusive from/to GR; the model
provides a supplementation of gravitational theory that in no way
disparages GR, nor finds fault with any of its formal/mathematical
content. Whereas GR is applicable to all forms of macroscopic mass
and energy, our new model is something of a `boutique' model with
only a few applications. The model's underlying mechanism is
unique, and involves a very specific set of microscopic \emph{and}
macroscopic circumstances coexisting in Nature. The model's
incorporation of QM non-local behaviour/phenomena, and an
alternative ontological stance regarding space and time
`underlies' this segregation of Gravitational (approaches and)
descriptions. Thus, our (boutique) model's complementary approach
and quantification of `gravitation' (in its widest/new sense) is
neither denied nor in direct conflict with GR's `unique' (and
broad-ranging or `general') approach to gravitation.

In Section \ref{Section:Summary} an overview summary and final
discussion is given, drawing upon all Sections of the paper, but
prior to that (in Section \ref{section:Type1a}) a further and
somewhat surprising ramification/application of the new model is
presented.
%*****************************************************************************************
\section{Rotating space-warps, non-local mass distributions and
type 1a supernovae results}\label{section:Type1a} This Section
brings together a number of aspects of the preceding model to
argue that the current interpretation of type 1a supernovae
observations, implying \mbox{\emph{accelerated}} cosmological
expansion, may be misguided --- subsequent to gravito-quantum
rotating space-warps (GQ-RSWs), and (particularly) their conjoint
non-local mass distributions (NMDs), achieving recognition.
Although a not insubstantial level of idealisation is used in this
informal investigation, the main aspects of the discussion are not
compromised.

Recently, the ``standard candle" supernovae method has been
supplemented and corroborated by a ``standard ruler" method,
involving baryon acoustic oscillations (BAOs)\footnote{Baryon
acoustic oscillations (BAOs) correspond to a preferred length
scale imprinted in the distribution of photons and baryons by the
propagation of sound waves in the relativistic plasma of the early
Universe \citep[ Introduction]{Blake_11}.} and (statistical
features of) the large scale clustering of galaxies
\citep*{Blake_11}.

\subsection{Preliminaries} Previously (subsections
\ref{subsubsection:local global spatial} and
\ref{subsubsection:planer}), we have argued that \emph{below} a
certain mass cut-off threshold, GQ-RSWs (or more succinctly, RSWs)
influence the \emph{motion} of low mass moving bodies equally
regardless of the body's direction of motion; i.e. the influence
(from the body's standpoint) is \emph{isotropic} --- although the
line-of-sight \textit{measurement} of motion shortfalls, relative
to predictions ignoring this influence, is not isotropic. We
cannot simply extend this influence to photons because the speed
of photons/light (in vacuum) is `physically' invariable (or
constant). Nevertheless, by way of the change in the non-local
mass value with radial distance from a RSW source, a new and
interesting (isotropic) effect upon the redshifted spectroscopic
lines of distant type 1a supernovae (\mbox{SNe 1a}) may be
hypothesised. Interestingly, this further
effect\footnote{Previously (subsections \ref{subsubsection:WMAP}
and \ref{subsubsection:Brief list}), we discussed how RSWs moving
with the solar system might anisotropically affect CMB radiation
measurement.}, involving photon propagation (and/or EM radiation
energy propagation), is \textit{obscured} by way of its isotropic
nature; in that the observer moves \textit{with} the solar system
and the RSWs and non-local mass distributions (NMDs) it harbours.

\subsection{Proposing a fourth type of `redshift' (of cosmological extent)}
\label{subsection:4th redshift} We hypothesise that a fourth (and
new) type of general\footnote{Pertaining to the equal shifting of
spectroscopic lines regardless of their frequency or wavelength.}
photon \emph{frequency shift} exists in addition to: (1) a Doppler
motion based redshift, (2) a scale factor (cosmological)
redshift\footnote{Some cosmologists prefer to interpret
cosmological redshift as a (kinematic) Doppler shift
\citep*{Bunn_09,Davis_10}. We shall refrain from this
interpretation, citing concern with the downplaying of a wavelike
aspect of electromagnetic radiation's propagation, as evidenced in
Young's double slit experiment.  This wavelike aspect is
considered to be (unavoidably) affected by space (distance between
two points) expansion --- in an expanding universe. If we restrict
Planck's relation ($E=hf$) to emission and absorption events, then
concern with regard to photon energy loss during its propagation
is not an issue. The Doppler interpretation of `cosmological'
redshift is not being denied; rather it is considered a secondary
interpretation that may (possibly) be of benefit to our
understanding of general relativity.}, and (3) a gravitational
redshift\footnote{The latter involves photons changing position in
the gravitational field of an uncharged, non-rotating, spherically
symmetric mass --- as confirmed by the Pound-Rebka experiment.}.
Whereas the three standard redshifts can be understood by way of
laws involving a transformation between reference frames, the
manner of frame transformation for this proposed (fourth type of)
redshift/blueshift is subtly different.

Conceivably, (4) an isotropic \emph{blueshift} arises for photons
as they travel \emph{towards} the source of RSWs and NMDs in
\emph{our solar system}, which is also the location of an Earth
based observer. We hypothesise that this occurs by way of the
greater (local) value of non-local mass (and non-local energy),
associated with the NMDs, closer to their source. This frequency
blueshift is similar to, but quite distinct from, the
gravitational blueshift experienced by photons as they move
(`downhill') into a stronger gravitational field. The new
blueshift, as measured upon Earth, pertains to far distant
objects, and thus it is a blueshift of \emph{cosmological} extent.
As such, its contribution to the electromagnetic-based frequency
shifts of the Pioneer spacecraft's observational Doppler data is
negligible.

Note that (`now', i.e. present time $\pm$ 100 years) the stability
of Sun-planet-moon motion in the solar system ensures that the
specific energy of each RSW ($\Delta e_w$) is (effectively)
constant throughout the universe. Subsequently, our primary
concern in this Section is the effect of NMDs upon the propagation
of electromagnetic waves/radiation (and photons).

Quantitatively, we \emph{assume} (essentially by default) that
this perturbation is observationally indicated by way of a
relation similar to the energy to frequency proportionality in the
Planck relation\footnote{There is no other simple way to link the
energy of electromagnetic radiation to both: a frequency
change/offset, and a QM mass-energy system. In the case of the
latter, we are referring to both: the atomic/molecular basis of
(conjoint) RSW and NMD energy, and the necessary externalisation
of this (non-inertial) energy.} (or the Planck-Einstein equation):
$E=h\nu$. As such, there will be a (hitherto unappreciated)
distance dependent blueshift offset ($\delta \nu$) in the
frequency of observed photons (cf. expectations) --- regardless of
the magnitude of their wavelength --- due to the presence of (all)
the solar system's (RSWs and) NMDs. Additionally, we shall define
a path dependent \emph{change} in this blueshift (offset):
$(\delta \nu)_{\rm{final}}-(\delta \nu)_{\rm{initial}}=\Delta
(\delta \nu)$.

An order of magnitude (`ball-park') investigation shows that
Planck's constant ($h$) cannot be the proportionality constant in
this new `negative redshift' (i.e. blueshift) phenomenon. The
introduction of a new (mechanism-specific) proportionality
constant, denoted $h^*$, is proposed; it has dimensions $[ML^2/T]$
or units $[\rm{kg \, m^2 \, s^{-1}}]$, as is the case with
Planck's constant, angular momentum, and the action (of a path or
trajectory). This proportionality constant relates the total (RSW
and) NMD based energy \emph{gain} to a frequency blueshift. It is
unlikely that a quantum mechanical `universal' constant
(especially relevant to atoms and molecules in emission and
absorption events) would be applicable at the
macroscopic/cosmological level, i.e. external to (microscopic) QM
systems. Possibly, $h^*$ is somewhat like a \emph{least} action
constant that applies to photon/`light' propagation in the
presence of a variable/dispersive non-local energy field.

The effects of the dispersion (with increasing radius from a RSW's
core) of the non-local mass field, and hence energy field, is more
fully discussed and algebraically formulated from section
\ref{subsection:volumetric dispersion} onwards, but prior to this
a case for the validity of this hypothesis is built.

\subsection{Dark Energy transition redshift \& the time from RSW
initiation}\label{subsection:RSW initiation} Support for this
hypothesis comes from consistency between the transition redshift of
dark energy\footnote{That is, when a deceleration in the expansion of
the universe `gives way' to an accelerating expansion.} ($z_t$), and
the light travel time (also known as lookback time) `back' to the
\emph{establishment} of lunar spin-orbit resonance in the early solar
system's history.

Good quality recent values of $z_t$ are: $0.43\pm0.07$ \citep[
Section 3.1, p.32]{Riess_07} and $0.46\pm0.13$ \citep[
Abstract]{Riess_04}. Note that the underlying physics of $z_t$ is
generally considered unknown \citep*{Rapetti_07}; and that
alternatives to dark energy, such as grey dust and/or the
evolution of supernovae properties, appear to be without merit
\citep{Wood-Vasey_07}.

Taking the age of the solar system as 4.567 billion
years\footnote{Wikipedia: \textit{Age of the Earth}, 2010-12.},
and that the time to lunar spin-orbit resonance (i.e. synchronous
rotation) of the Galilean moons is 2.5 \textit{million} years
\citep*[ p.560]{Peale_99}; RSW initiation or establishment is of
the order of 4.56 billion `lookback' years. By way of Edward (Ned)
Wright's or Siobhan Morgan's (internet based) ``Cosmic Calculator"
one can generate a `best' estimate of the redshift of this
lookback time. Taking: $H_0=71.5$ $\rm{(km/s)/Mpc}$,
$\Omega_M=0.265$, and assuming a flat universe, a $z_t$ value of
0.44 is associated with a universe age of 13.64 billion years and
a lookback time of 4.55 billion years. Thus, it is not beyond
conception that RSW (and non-local mass distribution)
establishment might be the `cause' of (or reason behind) a
perceived universal expansion inflexion point (in time) ---
dividing a decelerating universal expansion from a more recently
(perceived) accelerating one.

\subsection{A new energy, and questioning a pivotal cosmological
assumption}\label{Subsection:question pivotal} As discussed
previously (subsections \ref{subsubsection:core} and
\ref{subsubsection:summary topological}), the \emph{energy}
`driving' the GQ-RSWs (or just RSWs) and NMDs arises from an
external (collective) expression of non-inertial (frame-based)
atomic/molecular energy\footnote{Effectively analogous to an
`inertial' \emph{force} (or more appropriately to avoid confusion,
a `fictitious' or pseudo force), but concerning energy. Herein,
the expression ``virtual energy" is preferred, in a bid to (fully)
avoid confusion.}, common to a great many atoms/molecules, that
occurs below a minimum (internally) expressible energy value.
Generally, regarding the expansion of the universe, and in the
absence of (Doppler) peculiar motion and gravitational redshift
effects, it is assumed that:
\begin{quote} \ldots there is a direct one to one relationship
between observed redshift and comoving coordinate \citep[
p.107]{Davis_04}.
\end{quote} It is this assumption, underlying the case for dark
energy, that we seek to question. In other words, we are
questioning the validity of the currently conceived relationship
between redshift ($z$) and (time-dependent) cosmic scale factor(s)
at observation and emission, i.e:
$$1+z=\frac{a(t_o)}{a(t_e)}=\frac{a_{\rm{now}}}{a_{\rm{then}}}=
\frac{\nu_e}{\nu_o}=\frac{\lambda_o}{\lambda_e}$$

Note that comoving distance is (cosmological) proper distance
($D$) multiplied by this ratio of scale factors, and that in a
`flat' universe, line-or-sight comoving distance ($D_C$) equals
transverse comoving distance ($D_M$). Further, comoving distance
factors out the expansion of the universe whereas proper distance
does not; it equals proper distance at the present time.

\subsection{De-prioritising Dark Energy}
\label{subsection:de-prioritizing} The accelerating
expansion of the universe, as implied by the type 1a supernovae
(and baryon acoustic oscillations) results, is but one of
\emph{three} pillars upon which the theoretical concept of dark
energy rests. The second pillar arises by way of a need to
reconcile the cosmic microwave background (CMB) anisotropies, that
indicate a flat/uncurved geometry of (cosmological) space, with
the total amount of matter (baryonic and dark) in the
universe\footnote{Between 25 and 30 percent of the critical
density.} --- implied by measurements of: the CMB, large-scale
structure, and gravitational lensing. Lastly, we have the
late-time integrated Sachs-Wolfe (ISW) effect, closely related to
the Rees-Sciama effect \citep*{Granett_08}. With temperature
variations of ``about a millionth of a
Kelvin\footnote{\url{http://physicsworld.com/cws/article/news/35368}
(Jon Cartwright, Aug 8, 2008).}", the meticulous research
associated with validating the ISW effect as a ``direct signal of
dark energy" is cutting edge but it is not beyond reasonable
doubt.

The ontological alternative discussed in Section
\ref{Section:PhiloTheory}, regarding space and time, took issue
with the assumed extension of General Relativity to the universe
``as a whole", in the sense that $k\neq0$ is denied\footnote{By
way of ontological ``boundary conditions", i.e. additional
features of the universe (as a whole) that go beyond a purely
reductive (to general relativity) approach.} --- although this in
no way denies the validity of (observations based upon) the
Friedmann-Lema\^itre-Robertson-Walker (FLRW) metric with $k=0$.
Accepting this stance removes the logical implication that a
`flat' universe and sub-critical density necessarily implies dark
energy.

Subsequently, the pivotal experimental evidence behind the
presumed existence of dark energy is the type 1a supernovae (and
BAOs) data, and any potential reinterpretation of the SNe 1a data
is worth considering and potentially significant.

Furthermore, tests of the gravitational inverse-square law below the
dark-energy length scale \citep*{Kapner_07} yield a ``result [that]
is a setback in the search for the gravitational effects of dark
energy, which cosmologists believe should begin to appear at this
length
scale\footnote{\url{http://physicsworld.com/cws/article/news/26826}
(Hamish Johnston, Jan 17, 2007).}." Finally, \citet*{Frieman_08} in
their review paper --- \textit{Dark Energy and the Accelerating
Universe} --- highlight two other problems: (1) ``the coincidence
problem", where $\Omega_{DE}\sim\Omega_M$ [pp.404-405]; and (2) the
smallness of the energy density of the quantum vacuum [p.386].

\subsection{Volumetric `dispersion' of the `non-local mass \& RSW' energy}
\label{subsection:volumetric dispersion} Subsections
\ref{subsubsection:geometry's other role} and
\ref{subsubsection:governing energy} gave the total energy of a
single rotating space-warp (RSW) that coexists with non-local mass
as: $$\Delta E_w=\frac{1}{2} m_1^* \Delta a_w^{2}\Delta
t^{2}=m_1^* \Delta e_w$$ where: $m_1^*$ represents non-local mass
at (the inception radius) $8\pi r_o=r_1$, $\Delta a_w$ is the
(weighted, i.e. actual) acceleration's sinusoidal amplitude,
$\Delta t$ is lunar spin-orbit duration, and $\Delta E_w$ is the
total supplementary field energy over the course of a single lunar
loop/cycle. With our solar system quite stable (over the
long-term), $\Delta e_w$ (i.e. the weighted specific energy) is
(essentially) a constant and thus:
$$\Delta E_w \propto m_1^*$$

Secondly, we recall the \textit{non-local mass} distribution
function of subsection \ref{subsubsection:enclosed volume}
(Equation \ref{eq:distribution}), which was alternatively
conceived of as a spatial non-local mass continuity equation (in
subsection \ref{subsubsection:distrib eqn ramifi}):
$$m_r^* V_r={\rm{constant}}=m_1^* V_1$$ where: $m_1^*$ and $V_1$
are (respectively) the \textit{fixed} non-local mass, and enclosed
volume, at inception. Note that $m_r^*$ is both: the maximum
(inertial or passive gravitational) `compact' mass value at radius
$r$ (that can be affected by a NMD), and the non-local mass value
at radius $r$ in the field's distribution, with $V_r$ the (total)
\emph{volume} enclosed within this radius\footnote{The `$r$'
subscript simplifies the algebraic notation; its use mimics an
elementary form of mass continuity equation. Recall (subsection
\ref{subsubsection:enclosed volume}) that: $m^*_r\equiv m^*(r)$
and $V_r\equiv V(r)$.}. This non-local mass distribution
relationship describes a type of (non-local) mass dispersion as we
move away from the (single case) RWS and NMD inception radius,
such that: $m^*_r \propto 1/V_r$ and
$$m^*_r \propto (1/r)^3$$

Although the energy of a (conjoint) RSW and NMD is `defined' at
inception (i.e. at $r=r_1$), the dispersion of non-local mass
magnitude (for $r_1<r< \infty$) allows us to (equally) conceive of
the dispersion of this RSW and NMD \textit{energy} for $r>r_1$ and
$V_r>V_1$.

Let us propose and \textit{define} a \emph{total} energy (at
cosmological distance/radius\footnote{Small differences in
distance ($r$) between the different Sun-planet-moon (RSW and NMD
`generating') systems, by way of their `sources' lying in
different parts of the solar system (i.e. near Jupiter, Saturn or
Neptune), are negligible when \emph{cosmological} distances are
being considered.} $r$ from an Earth-based observer) applicable to
\emph{all} `conjoint' RSWs and NMDs:
\begin{equation}\label{eq:SumD}E^*_r=\sum (m^*_r \, \Delta
e_w)_i\end{equation} $E^*(r)$, or alternatively $E^*_r$ in our
abbreviated notation/nomenclature, is the total energy (or work)
`capacity' of the RSWs and NMDs (collectively) at
\emph{cosmological} (`comoving') distance $r$. It comprises a
summation of non-local mass carrying capacity terms multiplied by
their respective wave-like $\frac{1}{2} \Delta a_w^2 \Delta t^2$
(specific energy) space-warp terms --- both of which can act upon
a (localised) `compact' mass. Note that the `$i$' subscript in
Equation \ref{eq:SumD} refers to the index of summation applicable
to the different RSW and NMD systems (involving for example:
Sun--Jupiter--Ganymede, Sun--Saturn--Titan, etc.). At the
inception (i.e. establishment) radius ($r=r_1$) of each RSW and
NMD we have: $\Delta E_w=E^*_1=m^*_1 \Delta e_w$. Thus, the
dispersion of the field energy (and/or the `work' capacity of a
particular RSW and NMD), as $r$ \textit{increases} from $r_1$, is
reduced such that each: $(E^*_r)_{\rm component} \propto 1/V_r$ or
$(E^*_r)_{\rm component} \propto (1/r^3)$.

To proceed without undue complication, it is appropriate to this
Section's goal that we idealise circumstances by way of three
reasonable approximations. Firstly, we assume solar system
Sun-planet-moon orbital stability over most of the solar system's
history. The $(\Delta e_w)_i=(\frac{1}{2}\Delta a_w^2\,\Delta
t^2)_i$ (component) terms are effectively constant at the present
time ($\pm$ thousands of years). We extend this treatment of $e_w$
values as \emph{constants} to billions of years. Secondly, with
the $e_w$ values being of similar magnitude (see Table
\ref{Table:acceleration} in section \ref{Subsection:Model
quantifies external}) we assume all $e_w$ values are \emph{equal}.
Thus, $E^*_r\propto \sum m^*_r$, where $\sum
m^*_r\equiv(m^*_r)_{\rm total}$. Note that we are using $E^*_r$
rather then writing $(E^*_r)_{\rm total}$. Thirdly, with the
$m^*_r$ values being of similar magnitude (see \mbox{Table
\ref{Table:cut-off}} in section \ref{Subsection:warp's mass}) we
assume all $m^*_r$ values are (also) \emph{equal}. Thus, we have
$E^*_r\propto \overline{m^*_r}$, where $\overline{m^*_r}$ is a
mean NMD of the various $m^*_r$ distributions; and (similarly to
an individual/component $m^*_r$) we obtain: $$E^*_r \propto
(1/r^3)$$ A subtlety worth recalling/noting is that $E^*_r$ is a
total energy (that is) based upon the entire enclosed
(cosmological scale) volume at a distance $r$; but by way of the
non-locality (or global nature) of non-local mass ($m^*$), this
energy magnitude applies to \textit{all} locations on/at the
spherical \textit{surface} of the NMD's enclosed/spanned volume
$V_r$.

In section \ref{subsection:Type1a} we begin to more fully examine
how the presence of (constant acceleration/gravitational amplitude
$\Delta a_w$) RSWs with their (conjoint) $m^*_r$ distributions may
influence, i.e. do work on or alter the energy of `inbound'
electromagnetic radiation (and photons). Note that this paper has
previously concentrated upon the influence of the RSWs'
acceleration undulations upon the motion of \emph{low mass
bodies}, whereas the wave-like propagation of photons as EM
radiation (at $c$), and their \emph{zero} (rest) \emph{mass},
means that the propagation speed of photons cannot be influenced
by RSWs in the manner that compact (low mass) bodies are ---
although in subsection \ref{subsubsection:WMAP} we argued for the
possibility that RSWs may have left a `signature' upon CMB
radiation data (as measured by the Wilkinson Microwave Anisotropy
Probe i.e. WMAP).

\subsection{Circumventing the ambiguity of `light' propagation
distance} \label{subsection:distance trouble} 
\begin{figure}[h!]
\centerline{\includegraphics[height=10.5cm,
angle=0]{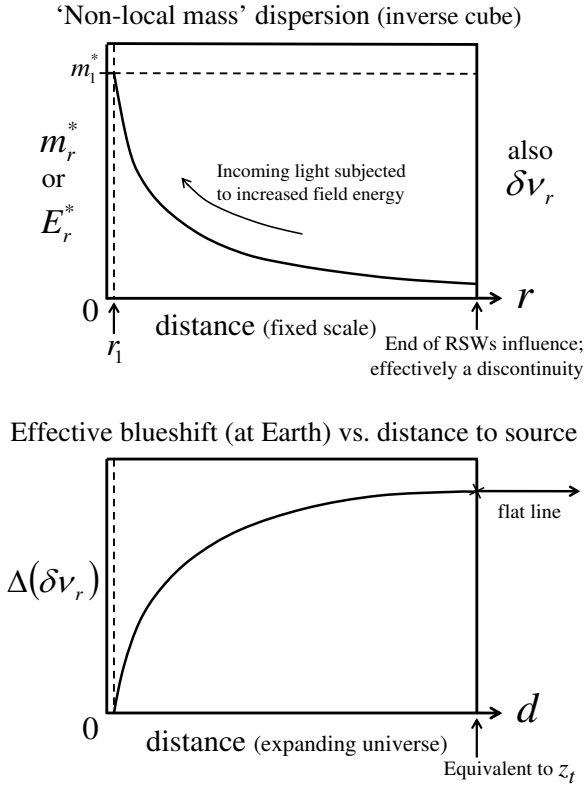}}
\begin{center}
\caption[The top schematic diagram represents the dispersion of
both: a single non-local mass distribution ($m^*_r$), and
total/collective non-local energy ($E^*_r$), with increasing
distance ($r$) away from the `source' of the rotating space-warps
(RSWs) and their conjoint non-local mass distributions (NMDs).
This inverse cube-like relationship also represents a distance
dependent blueshift frequency offset ($\delta \nu_r$). The lower
schematic diagram illustrates the presently \emph{observable}
difference in this blueshift offset $\Delta (\delta \nu_r)$, as
experienced by EM radiation `during' its propagation towards the
solar system.]{The top schematic diagram represents the dispersion
of both: a single non-local mass distribution ($m^*_r$), and
total/collective non-local energy ($E^*_r$), with increasing
distance ($r$) away from the `source' of the rotating space-warps
(RSWs) and their conjoint non-local mass distributions (NMDs). We
hypothesise that this inverse cube-like relationship also
represents a distance dependent blueshift frequency offset
($\delta \nu_r$), relative to a case without RSWs and NMDs
present. The lower schematic diagram illustrates the presently
\emph{observable} difference in this blueshift offset [$\Delta
(\delta \nu_r)$], as experienced by light/electromagnetic
radiation `during' its propagation (towards the solar system)
through the (non-local mass based) $E^*_r$ energy field (arising
from multiple RSWs and NMDs). Beyond the time of RSW and NMD
initiation, i.e. $>4.56$ billion years ago --- which is
effectively the (dark energy) transition redshift
($z_t\approx0.44$) --- there can be no further increase in this
value.} \label{Fig:Dispersion}
\end{center}
\end{figure}

Section
\ref{subsection:volumetric dispersion} discussed collective/total
RSW and NMD based energy changes for increasing distance
\emph{from} their source `region' (i.e. the solar system);
whereas, the arrival of photons at the Earth's surface involves
propagation \textit{towards} the solar system --- i.e.
\textit{decreasing} distance. The arriving photons have had to
climb into the dispersed total non-local mass [$(m^*_r)_{\rm
total}]$ and total energy $E^*_r$ field of the RSWs and NMDs so as
to reach the observer. In other words: a propagating
photon/electromagnetic radiation \emph{en route} to the solar
system (inadvertently) `experiences' a radius/distance dependent
increasing scalar field. Note that a photon's overall/total
$m^*_r$ (`starting') value is set at emission/absorption, with
$m^*_1$ (at $r_1$) being the maximum achievable value of any
$m^*_r$ component. The greater the distance travelled, the greater
the difference/increase (is) between the $m^*_r$ and $m^*_1$
values, see Figure \ref{Fig:Dispersion}. Beyond the dark energy
transition (and RSW and NMD initiation) redshift distance $z_t$
(defined and discussed in section \ref{subsection:RSW
initiation}), the difference between these two values flattens
out, i.e. no longer changes. Note firstly, that the discontinuity
mentioned in Figure \ref{Fig:Dispersion} (top diagram, maximum
distance) refers to the extent of the RSWs' and NMDs' influence
upon electromagnetic radiation (as observed `now'); and secondly,
that the current (i.e. `now') field of the (non-local,
global/systemic) RSWs and NMDs actually fills the \emph{entire}
universe, regardless of its `size', with even lower (individual)
$m^*_r$ and (collective) $E^*_r$ values achievable (in the future)
in a further expanded universe.

%********************************** Fig:Dispersion ********************************
%********************************** end of Fig:Dispersion  *************************

A major issue (until now unmentioned) is that the distance ($r$)
from the various rotating space-warp (RSW) source radii has been
treated as independent of the universe's expansion --- i.e. it has
been effectively treated as a comoving distance --- whereas the
distance travelled by a photon is dependent upon the (metric)
expansion of space. Thus, our discussion of distance travelled is
inherently ambiguous. Fortunately, the inverse \emph{cubic}
dispersion of each $(m^*_r)_i$ and $(m^*_r)_{\rm total}$ with
increasing $r$ means that: for small cosmological distances,
cosmological expansion can be neglected. Additionally, for larger
$r$, i.e. between (say) 2.0 and 4.56 billion light (travel time)
years, the relative error in the $m^*_r$ values, and hence
$E^*_r$, is very small --- even though a non-expanding distance
scale is oversimplified.

For our approximate (i.e. idealised) purposes the expansion of the
universe, although obviously significant for cosmological redshift
$\Delta\nu$ (or $\Delta \lambda$), is fairly insignificant for the
(RSW and NMD based) $\delta \nu$ (or $\delta \lambda$) offset
values proposed in this Section. We shall proceed without
correcting $E^*_r$, $\delta \nu$, or (the various individual)
$m^*_r$ (values) for the expansion of the universe. Additionally,
it should be noted that we have not considered solar system
(peculiar) motion, with respect to the type 1a supernovae
(`standard candles'), over the past (approximately) 4.56 billion
years.

Finally, the non-locality of RSWs and their conjoint NMDs
(implying `instantaneous' field entanglement) makes comoving
distance --- i.e. the distance between two points measured along a
path defined at the \textit{present} cosmological time ---
equivalent to the RSW and NMD distance measure discussed
throughout this paper (i.e. $D_C=r$), but only \emph{`now'}/at the
present cosmological time --- a feature that suits the
non-locality inherent to the model. As regards this/our idealised
investigation, the distinction between comoving distance and
proper distance (the distance scale that expands with the
universe) is not a pressing concern.

\subsection{An alternative interpretation of the type
1a supernovae data} \label{subsection:Type1a} The top schematic
diagram of Figure \ref{Fig:Dispersion} represents the spatial
distribution of field strength for both: a single non-local mass
distribution ($m^*_r$), and total/collective non-local (scalar)
energy ($E^*_r$) --- arising from (one or multiple, respectively)
rotating space-warp(s) and (respectively) its/their conjoint
non-local mass distribution(s) --- that have influenced photons
`received' (in the present epoch\footnote{With ``present epoch"
being used in a cosmological sense, such that any given time in
the past 50 years or so can be effectively considered `now'.}). In
section \ref{subsection:volumetric dispersion}, Equation
\ref{eq:SumD} defined the total energy (or work) `capacity' of
multiple (coexisting) RSWs and NMDs as $E^*_r=\sum (m^*_r \,
\Delta e_w)_i$. Subsequently, let us propose and define
$$E^*_r=h^* (\delta \nu_r)$$ or (alternatively) written more conventionally
$$E^*(r)=h^*[\delta \nu(r)]$$ as being indicative of the
(radius/distance dependent) frequency \emph{shift} associated with
the \emph{existence} of the new (total) scalar energy field. Note
that each $\delta \nu_r$ [i.e. $\delta \nu$ at $r$] value represents
the \emph{offset} in cosmological `redshift' due to the presence of
RSWs and NMDs cf. the case without the presence of this scalar field
effect. We also note that the use of `$\delta$' does \emph{not}
represent a (mathematical) infinitesimal change in the value of a
variable (as is the case) in infinitesimal calculus.

We use this (unconventional) `$\delta$' nomenclature/notation for
two reasons. Firstly, to indicate an offset from idealised (no
$E^*$ field) conditions; in this sense it is similar to $\delta$
as used in Engineering Mechanics to represent a deflection (e.g.
of a beam from equilibrium conditions). Secondly, it represents an
anomalous/second type of change in addition to a standard
change/decrease (of frequency with distance, $\Delta
\nu=\nu_e-\nu_o$), in the same manner that \emph{two different}
speed changes were required earlier (e.g. in subsection
\ref{subsubsection:Interpret power}) --- one for the speed
oscillation amplitude ($\Delta v$) induced by a rotating
space-warp, and the other for the associated monotonic speed loss
($\delta v$) of a moving body (cf. predicted speed without the
RSWs' presence).

Actually, neither $E^*_r$ nor $\delta \nu_r$ can be directly
measured; only a \emph{change}/increase in their values is
observationally relevant. Thus (via final and initial values):
$$E^*_1-E^*_r=h^*(\delta \nu_1-\delta \nu_r)\quad {\rm{or~alt^y}} \quad
\Delta E^*_r=h^*[\Delta (\delta \nu_r)]$$ represents the
difference/increase that exists between initial photon
emission/absorption and (final) reception/observation. This
blueshift effect is supplementary to the standard cosmological
expansion effect (i.e. a `cosmic' scale factor based
\emph{redshift} effect, \mbox{$\Delta \nu=-(\nu_o-\nu_e$).} The
spatial distribution of this difference [$\Delta (\delta \nu_r)$],
stripped of the proportionality constant ($h^*$), is shown in the
bottom schematic diagram of \mbox{Figure \ref{Fig:Dispersion}.}
This quantity is physically `real' in that it represents a
\emph{quantifiable} blueshift of EM radiation arising during its
propagation \emph{toward} our solar system, as realised by
measurements at the present `time'.

The variation of (i.e. change in) the proposed (frequency)
blueshift offset [$\Delta(\delta \nu_r)$] with distance, and
(recalling section \ref{subsection:RSW initiation}) the
\mbox{$z_t=0.44$} transition (for $H_0=71.5$ $\rm{(km/s)/Mpc}$,
$\Omega_M=0.265$, and assuming a flat universe), concur with the
graphical supernovae representation of \citet*[ Figure
7]{Davis_07}, which plots redshift vs. `distance'
modulus\footnote{Distance modulus is the apparent magnitude (of an
astronomical object) minus the absolute magnitude, i.e.
\mbox{$\mu=m-M$.}} ($\mu$).

\emph{Physically}, this (\emph{non-local} mass based) frequency
blueshift effect appears to be restricted to (i.e. only active
upon) the wave-like nature of EM radiation. In other words, as
light propagates toward the observer this hitherto unforseen
blueshift (energy increase) occurs. Note that nothing
discontinuous is seen to occur suddenly upon measurement. Emission
and absorption events and their associated spectra, as well as the
generation of continuous spectra, and also measurement events, are
all governed by standard quantum electrodynamics.

\subsection{Ramifications of the NMD-based supplementary blueshift}
Standard cosmological redshifts arise by way of comparing distant
spectroscopic lines with the wavelengths of reference
spectroscopic lines --- as determined/calibrated in Earth-based
laboratories. The measured brightness of a type 1a supernova (SNe
1a) indicates the distance (now) of the object, whereas the
redshift illustrates the intervening expansion history of the
universe\footnote{Distances need to be determined in conjunction
with some particular model of spacetime; usually this is a
(standard) Lambda-Cold Dark Matter ($\rm \Lambda$CDM) cosmology.}.
Including this supplementary blueshifting `process' makes for
greater cosmic scale factor redshifts than is currently
appreciated; i.e. the currently \emph{perceived} relationship
between measured redshift and cosmic scale factor is
`significantly' altered, as is the redshift-magnitude relation of
type 1a supernovae, with the most significant effects at
\emph{low} redshifts.

Upon accepting the presence of NMDs, the (once
surprising\footnote{The type 1a supernovae observations/results
were a major surprise to physicists at the time of their
`discovery'. Today, physicists (in general) are very accepting of
this observation, although decidedly `less so' the interpretation
given to it.}) dimness of type 1a supernovae need not be
attributed to a recent acceleration of the universe's expansion,
and the hypothetical existence of dark energy. Correcting for the
new blueshift effect, means that objects such as SNe 1a are
actually more distant cf. distance (expectations) based upon
(recalling section \ref{Subsection:question pivotal}) an
\emph{assumed} ``\ldots direct one to one relationship between
observed redshift and comoving coordinate." Furthermore, the
(possible existence of) foreground `contamination' in the CMB
radiation data --- which possibly arises indirectly from the
existence of RSWs (see subsection \ref{subsubsection:WMAP}) --- is
supportive of the idea that a re-interpretation of the universe's
expansion history may be required.

This re-interpretation potentially allows a return to (the more
logically simpler status of) a \emph{decelerating} expansion.
Certainly, issues such as the ``age problem" of the universe will
resurface; but the ramifications of the existence of rotating
space-warps, with their associated non-local mass (and energy)
distributions, deserve investigation. As such, the scenario
proposed in this Section, albeit involving a not insubstantial
degree of idealisation, conceivably explains why there has been a
remarkable lack of progress in characterising the `essence' of
dark energy, and the physics underlying the transition redshift of
dark energy ($z_t$). Supporting evidence for this conjecture is
presented in section \ref{subsection:Mimic}.

\subsection{How the RSWs' non-local mass distributions act to mimic
effects attributed to dark energy}\label{subsection:Mimic}
Detailed investigation has revealed that \textit{dark} energy
density (i.e. energy\,$\div$\,volume) remains constant with time
\citep*{Wang_04}. Indeed, Max Tegmark told
\mbox{PhysicsWeb}\footnote{\url{http://physicsworld.com/cws/article/news/19736}
(Belle Dum\'e, Jun 24, 2004).}: ``I'm struck by the fact that the
dark energy seems so `vanilla'." Furthermore, the cosmological
\textit{constant} term in gravitational theory is considered to
act like a constant (positive) vacuum energy density. The
implication here is that: $\Delta E_{\rm{dark}}\propto \Delta V_u$
where $V_u$ is the volume of the universe; with temporal changes
in $V_u$ (i.e. $\Delta V_u$) dependent upon changes in cosmic
scale factor [$a(t)=a=a_{\rm scale}$ (say), in order to avoid
confusion with acceleration amplitude]. Thus:
$$\frac{\Delta E_{\rm{dark}}}{(\Delta a_{\rm scale})^3} \approx \rm{constant}$$

For a photon \emph{approaching} the solar system, section
\ref{subsection:volumetric dispersion} argued that: $E^*_r \propto
r^{-3}$, which then implies: $\Delta E^{*}_{r} \propto (\Delta
r)^{-3}$. Alternatively, if we replace the model's solar system
(origin) based distance ($r$) with a distance ($R$) that
\emph{increases} with photon propagation duration, and whose
`origin' lies well beyond any photon emission points, then we have
(roughly) \mbox{$\Delta E^{*}_{R} \propto (\Delta R)^3$.}
Subsequently, noting that $R$ is a (towards Earth) distance and
\emph{not} cosmic scale factor:
$$\frac{\Delta E^{*}_{R}}{(\Delta R)^3} \sim \rm{constant}$$ Over
the past two billion years --- the period of time where the change
in dark energy is most noticeable --- both $a_{\rm scale}$ and $R$
evolve (approximately) linearly with time (i.e. the \emph{same}
time). Thus, $\Delta a_{\rm scale} \propto \Delta R$ is feasible
for recent (cosmological) times and low redshifts ($z<0.16$). With
a common \emph{cubic} distance relationship, or volume-like
relationship, present in the denominator of both the two (change
in) energy based ratios (with numerators $\Delta E_{\rm{dark}}$
and $\Delta E^{*}_{R}$), the similarity of their (through time)
variation is greatly enhanced. Indeed it is their definitive
shared attribute.

For this reason, at low redshifts with $R^3$ proportional to a
volume $V_R$ we have: $$\frac{\Delta E^{*}_{R}}{\Delta V_R} \sim
\rm{constant}$$ which mimics the near constancy of dark energy
density through time as the universe undergoes expansion (devoid
of the model's NMDs), such that: $$\frac{\Delta
E_{\rm{dark}}}{\Delta V_u} \approx \rm{constant}$$ These two
constant (energy density) ratios, describe very different and
unrelated physical scenarios. The very similar proportionality
displayed by both (dimensionless) $\Delta a_{\rm scale}$ and
(`dimensional') $\Delta R$, as well as $\Delta V_u$ and $\Delta
V_R$, makes \emph{these} two (temporal change based) ratios
approximately equal. We reiterate that $V_R$ is a (photon-point)
motion based volume, whereas $V_u$ is an expanding scale (line
segment) based volume.

If we accept the model's alternative energy `density'
relationship, then the `plain vanilla' dark energy, that Max
Tegmark refers to, may be regarded as symptomatic of a (i.e. our)
completely different mechanism; a mechanism that is largely
independent of (i.e. only marginally influenced by) the universe's
expansion. This mechanism (similarly) requires the existence of an
unforseen energy ($E^*_R$), or more appropriately $E^*_r$ (as
defined in section \ref{subsection:volumetric dispersion}).
Subsequently, dark \emph{energy} ($E_{\rm{dark}}$) may no longer
be in need of `dispensation'.

\subsection{Summary and closing remarks}\label{subsection:altered}
In this Section we conjectured that the model's various non-local
mass distributions (NMDs), i.e. the various $m^*_r$ distributions,
and the overall/total energy distribution ($E^*_r$) --- associated
with the superposition of (multiple) rotating space-warps --- can
influence the energy (and frequency) of electromagnetic radiation.
We hypothesised --- in relation to the observation (in our solar
system) of photons arriving from (standard candle) type 1a
supernovae --- that: Earth bound EM radiation (during its journey)
has been \emph{blueshifted}, by way of encountering the
(non-local) energy distribution/dispersion of the (multiple)
GQ-RSWs and NMDs scalar fields --- such that: \mbox{$\Delta
E^*_r=h^*[\Delta(\delta \nu_r)]$} where $h^*$ is a (new)
proportionality constant with dimensions $[ML^2/T]$. Further,
section \ref{subsection:distance trouble} established that the
quantification of this `supplementary' blueshift
effect/\emph{offset} ($\delta \nu$), as measured by its
\emph{change} between (distant) emission/absorption and
reception/measurement at Earth locations in the present epoch
[$\Delta (\delta \nu)$], is only slightly dependent upon the
expansion of the universe; as compared to the (cosmological)
cosmic scale factor redshift, which is solely determined by the
expansion of the universe. In other words, this (lesser and)
hitherto unrecognised cosmological blueshift is almost totally
dependent upon an Earth \emph{approaching} photon (and/or EM
radiation) gaining energy as it climbs into the (distance, or
rather volume dependent) scalar energy field ($E^*_r$) of the
solar system `centered/based' RSWs and NMDs.

Support for this conjecture/hypothesis lies in the remarkable
`coincidence' that the dark energy transition redshift
($z_t\sim0.44$) --- i.e. from decelerating universal expansion to
an accelerated expansion --- matches the establishment `time' of
lunar spin-orbit resonance in the solar system's very early
history (i.e. $\approx4.55$ billion `lookback' years ago), see
section \ref{subsection:RSW initiation}. This synchronisation
feature is arguably \emph{the} major prerequisite required for
`generating' the model's RSW and NMD scalar fields. Other features
supportive of dark energy were critically evaluated in section
\ref{subsection:de-prioritizing}.

An important finding of this Section was that the (through time)
manner of the EM radiation's (frequency based) energy change is
very similar to, and effectively mimics, the presence of an
apparent constant `dark' energy \emph{density} and accelerated
expansion of the universe. This second (and currently accepted)
interpretation is implied by the redshift-magnitude relation of
type 1a supernovae (and observations of baryon acoustic
oscillations) --- in the assumed \emph{absence} of RSWs and their
`conjointly existing' NMDs. Further, by way of the overall/total
energy distribution (exhibiting dispersion) that surrounds
(multiple) RSWs and NMDs in our solar system, we need to
adjust/correct the measured (and assumed) value of the universe's
expansion-based cosmological redshift in order to compensate for
this new cosmological blueshift effect. Upon implementing this
adjustment (it turns out that) the universe is quite conceivably
(completely) free of dark energy and undergoing a
gravitation-based \emph{deceleration} in its expansion.

Concerns raised by this (suggested) return to what was once
conventional gravitational cosmology have been briefly addressed,
but not with any rigour; e.g. ``the age problem", and the status
of the late-time integrated Sachs-Wolfe effect. The objective of
this Section has been to broadly outline, with the assistance of a
not insubstantial degree of idealisation, a conceivable
\emph{alternative} to dark energy. This was achieved by way of
drawing upon content within the model and explanation of a `real'
Pioneer anomaly presented throughout this paper --- with this
model being molded and governed by the awkward observational
evidence associated with the Pioneer anomaly (and other peripheral
issues such as the Earth flyby anomaly).
%*****************************************************************************************************************
% \url{http://www.philosophynow.org/issue82/Hawking_contra_Philosophy}
% Christopher Norris is Professor of Philosophy at Cardiff University
%
% \begin{enumerate}[a)] \item{} \item{} \item{} \item{} \end{enumerate}
%  cf. \begin{itemize}  which gives `dots'
%*****************************************************************************************************************
\section{Summary, Discussion and Conclusions}
\label{Section:Summary}Due to the length and
broad/multidisciplinary scope of this paper, rather than merely
present a conclusion, this (final) Section contains both: a
summary of the major features and findings of the model/mechanism,
as well as a fairly protracted discussion of its ramifications. As
such, this concluding Section is interspersed with a diverse range
of constitutive conclusions. Sections \ref{subsection:End_Results}
to \ref{subsection:End_Equalities} are largely model-specific;
they outline the model's: primary conceptual features, major
quantitative results and equations, as well as a number of
predictions and applications. A somewhat different approach is
evident in sections \ref{subsection:End_Conclusion} and
\ref{Subsection:Big picture}, in that they are largely devoted to
discussing and encapsulating the wider and widest ramifications
(respectively) of the new model/mechanism. In section
\ref{Subsection:In closing} a brief final retrospective commentary
is delivered. In the interests of being productive and
progressive, some new information has been introduced throughout
this Section in order to achieve this treatise's ultimate aims. An
underlying theme of this Section (particularly prominent in
section \ref{subsection:End_Discussion}) is the philosophy of
science and philosophy of physics based
apologetics\footnote{Reasoned arguments or writings in
justification (or the defense) of something.} given to appease the
sceptical reader's hostility towards both a real and
non-systematic based Pioneer anomaly and the unorthodox
explanation (of it) `reasonably' pursued herein.

\subsection{The model's major results, predictions, applications
\& features}\label{subsection:End_Results} This paper has
developed a model for what many scientists believe is the
impossible or highly unlikely case of a `real' (non-systematic
based) Pioneer anomaly, i.e. one where thermal-radiation/heat
based effects and other (external, on-board, and computational)
systematics are present but play only a minor role. The new model
explains the four principal observational features/constraints of
the Pioneer anomaly; these being: (1) an mean anomalous (inward)
deceleration ($a_P$) of $8.74 \pm 1.33 \times 10^{-10}~{\rm
m~s^{-2}}$  (in the outer solar system and beyond); (2) the
(quasi-stochastic) temporal variation of the anomaly around this
long-term mean/average value; (3) an apparent violation of general
relativity's weak equivalence principle (WEP)
--- in that the anomaly affects the motion of spacecraft and `low'
mass celestial bodies but not the motion of `high' (or large) mass
celestial bodies such as planets and moons; and (4) the lower value
of the (Pioneer 11 based) anomaly --- on approach to Saturn
encounter.

A major qualitative feature of the model is that the anomalous
`deceleration' --- which is generally referred to (simply) as an
\emph{acceleration magnitude} --- acts in the (opposite) direction
of the translational \emph{path} of a (low mass) body, as compared
to being Sun-directed or Earth-directed. Furthermore, the
anomalous acceleration itself --- although not its line-of-sight
measurement --- is independent of a body's speed\footnote{Noting
that no celestial body in the solar system exists at rest (i.e. in
the absence of some kind of motion) relative to the solar system's
barycentre.} and direction of motion, i.e. it is velocity
independent. The energy source `driving' the anomalous (Pioneer)
acceleration is solar system based, involving (amongst many other
things) three-body Sun-planet-moon celestial systems (see
Introduction) --- with each moon needing be in a 1:1 or
synchronous spin-orbit resonance `around' its host planet. For an
Earth-based observer the mean \emph{measured} anomaly, although
essentially constant in the far outer solar system --- by way of
the line-of-sight observations approaching a maximum constant
value (asymptote) at large distances from the barycentre --- and
always present to some degree, is \emph{not} position independent;
especially when a spacecraft lies between Jupiter and Saturn, as
was the case for Pioneer 11 on its approach to Saturn encounter.
For each Sun-planet-moon system the model establishes a ``cut-off
mass", i.e. the maximum (passive gravitational\footnote{Where the
word ``gravitational" is used here in a wider sense than normal,
referring to both: standard (general relativistic) gravitational
fields, \emph{and} the supplementary acceleration/gravitational
field type proposed herein.}) mass of a body that can be
influenced. These (numerous and different) cut-off masses (all)
reduce with increasing distance away from their `source' region,
by way of an inverse $r$ cubed relationship\footnote{Where `$r$'
is the separation distance of a body from a given (lunar-based)
source region.} --- i.e. an inverse spherical volume relationship.
A rough (``rule of thumb") guide is that solar system bodies (of
average density) with a diameter of approximately 1 to 2
kilometres lie around these cut-off masses; \mbox{Table
\ref{Table:cut-off}} in section \ref{Subsection:warp's mass} gives
the actual (moon specific and) distance dependent cut-off
\emph{mass} values.

The model, whose basis and major features were encapsulated in the
Introduction/Section \ref{Section:introduction}, proposes two new
physical field mechanisms associated with each (participating)
``gravito-quantum" Sun-planet-moon system --- of which
(quantitatively) there are five dominant contributors involving
Jupiter's four Galilean moons and Saturn's Titan. Apart from a
very minor contribution from Neptune's Triton, effects
attributable to all other Sun-planet-moon systems are either
non-existent (e.g. Earth's moon) or negligible (e.g. Uranus'
largest moon Titania\footnote{The eighth largest moon in the solar
system.}). These five major superpositioned (`circumferentially'
sinusoidal-like) three dimensional rotating space-deformations or
rotating space-warps (RSWs) --- existing (in a `planar' disk
slice) as opposing perturbations `above and below' (and `upon')
the (equilibrium) spacetime curvature produced by general
relativistic gravitation\footnote{There is no change in the
overall/net spacetime curvature arising from the inclusion of
rotating space-warps (recall section \ref{Subsection:Shackles}).
For a diagrammatic representation see Figures \ref{Fig:FlatDisk}
to \ref{Fig:R3FrontElev} in Section \ref{Section:PrelimModel} and
Figure \ref{Fig:Ellipsoid} in Section \ref{Section:Quantif
Model}.} --- account for the path-based Pioneer anomaly.
Quantitatively, a \emph{root sum squared} (RSS) value is required,
with this (RSS) proper deceleration (superposition) value ($a_p$)
--- cf. the (Pioneer S/C) observation based value ($a_P$) ---
`formalised' in section \ref{subsection:End_Equalities}. This RSS
(anomalous `acceleration magnitude') value is related to (both)
the (sinusoidally varying) constant acceleration/curvature
\emph{field} amplitudes [$(\Delta g)_i$] and the ensuing
(effectively equal) proper acceleration amplitude (components) of
a \emph{moving body} [$(\Delta a)_i$] --- by way of these field
effects/sinusoids producing (component) sinusoidal/oscillatory
\emph{speed} variations around the moving body's equilibrium
(speed) value. Note that the `$i$' subscript (as used) here
indicates both: the index of summation of the RSS components, and
that multiple instantiations of the particular variable are being
referred to. Importantly, each unsteady/oscillatory motion/speed
component also produces a component (rate of) \emph{speed
shortfall} [$(\delta a)_i$] --- and a (component) translational
kinetic energy loss (rate) --- \emph{relative} to (speed)
predictions/expectations that \emph{omit} the presence of the
supplementary gravitational/accelerational mechanism, i.e. where
the \emph{predicted} kinetic energy (K.E.) is assumed
(effectively) steady/translational\footnote{The kinetic energy
associated with a spacecraft's spin/rotational motion may be
safely ignored.}. Additionally, the model (necessarily)
proposes/introduces a separate (i.e. second) field mechanism
feature --- involving the magnitude and distribution of a new
scalar physical quantity\footnote{A \emph{physical quantity} is
expressed as the product of a numerical value (i.e. a number)
\emph{and} a physical unit.} called ``\emph{non-local mass}"
($m^*$) --- that pertains to both the appeasement of the apparent
violation of the weak equivalence principle, and the dispersion of
energy away from each supplementary field's source
region\footnote{Note that this source `region' is centered upon a
point on the moon's \emph{spin axis} in the case of the
(cylindrical-like or parallel planar-like) rotating space-warps,
whereas it is centered upon a moon's core (or \emph{central
point}) in the case of the (spherical-like) non-local mass
distributions.}.

The model's \textbf{quantitative results} are divided into three
categories. Firstly, the (outer solar system and beyond) long-term
average constant (idealised maximum/asymptotic) `Pioneer' anomaly
value ($a_p$) is: $8.65 \pm 0.66 \times 10^{-10}~{\rm m~s^{-2}}$.
This is a path-based result, and (subsequently) the model's
proposed average line-of-sight-measurement value of the anomaly
associated with the Pioneer 10 spacecraft (01 Jan 1987 to 22 July
1998) is $8.52 \pm 0.66 \times 10^{-10}~{\rm m~s^{-2}}$. An
average line-of-sight value of $7.61 \pm 0.66 \times 10^{-10}~{\rm
m~s^{-2}}$ is associated with the Pioneer 11 spacecraft (01 Jan
1987 to 01 Oct 1990). Note that for a number of reasons including:
data arc duration/length, general spacecraft performance (e.g. gas
leaks), magnitude of solar radiation pressure, and level of solar
activity \citep[ p.34]{Anderson_02a}, \emph{the Pioneer 10 data
is/has been considered (significantly) superior to the Pioneer 11
data}.

Secondly, in Section \ref{Section:PrelimModel} the model is able
to account for aspects of the temporal variation in the Pioneer 10
data. Noting that $10\rm~{mHz}\equiv0.652 ~\rm{mm/s}$:
\begin{enumerate}[a)] \item{The model-based variation in speed,
around an equilibrium speed, of $16$ to $19 \rm~mHz$ due to the
RSWs, combined with the (root mean square) raw measurement
noise/residual of ``a few$\rm~mHz$", sum (i.e. add up) to match
the overall observation-based (post-fit) residual/noise of $20$ to
$25 \rm~mHz$.} \item{The Earth-based \emph{diurnal} residual
confirms the Pioneers' exceptional navigational precision and
accuracy. Section \ref{Subsection:Approx annual} argued that
although a ``true-annual" residual exists, its amplitude is not
significant, and the (more appropriately labelled) $\sim$annual
residual (C. Markwardt's nomenclature) has been incorrectly
interpreted as Earth-based.} \item{The Callisto-Titan beat
frequency amplitude of $0.733 \rm~mm/s$ (i.e. $11.24 \rm~mHz$) and
temporal duration of 357.9 days --- corrected to 356.1 days from
the perspective of the Pioneer 10 spacecraft --- matches the
amplitude ($\approx0.7\rm~mm/s $) and duration of the $\sim$annual
oscillatory term (i.e. $355\pm2$ days, or $0.0177\pm0.0001$
rad/day angular velocity).} \item{Noting that Io-Europa-Ganymede
obey a 4:2:1 (spin and) orbital resonance, the (unusual) 200-day
correlation time sensitivity of the $\sim$annual residual can be
attributed to this duration being almost exactly four times the
50.1 day (lesser known) \mbox{7-to-3} Ganymede-Callisto orbital
resonance duration.} \end{enumerate} Thirdly, the
initial/establishment values of non-local mass [$m^*(r_1)$ or more
simply/succinctly $m^*_1$], specific to each Sun-planet-moon (or
barycentre-planet-moon) system, and their spatial distribution
[$m^*(r)$ or rather $m^*_r$] around their (lunar) source region
are discussed in section \ref{Subsection:warp's mass} and
quantitatively displayed in Table \ref{Table:cut-off}. Note that
this subscript based nomenclature is tailored to equalities
involving the continuity of non-local mass.

Primary \textbf{predictions} specific to the model's
conceptualisation and quantification are:
\begin{enumerate}[a)] \item{That the accuracy of celestial body
and spacecraft ``orbit determination", e.g. future interplanetary
missions and near-Earth objects (NEOs) --- as well as the proposed
Laser Interferometer Space Antenna (LISA) mission --- will benefit
from adjusting for the (model's) Pioneer anomaly-like influence.}
\item{For Pioneer 11 `between' Jupiter and Saturn the amplitude of
temporal variation in the anomalous acceleration should be muted,
in proportion to the lessening of the anomalous acceleration
magnitude.} \item{Asteroid 1862 Apollo (1.7 km in diameter with a
mass $\approx5.1\times 10^{12} \rm~kg$) is an ideal asteroid for
examining the model's distance dependent non-local mass cut-off
values --- i.e. the maximum (passive gravitational) mass that can
be influenced (subsection \ref{subsubsection:1862 Apollo}). This
is because its orbital path spans the transition zone of four of
the five dominant Sun-planet-moon contributors --- with
Saturn-Titan being out of `range' (for an asteroid of this mass).}
\end{enumerate} It should be noted that due to the very small
magnitude of the (Pioneer) anomaly, verification or falsification
of these predictions is not easily (nor inexpensively) achieved;
it will take time, effort, and resources.

Major \textbf{applications} of the multiple (and superpositioned)
instantiations of the two proposed (supplementary) field
mechanisms include: \begin{enumerate}[a)] \item{Section
\ref{section:Type1a}'s proposal that a foreground effect related
to the non-local mass distributions (NMDs) leads to a
supplementary cosmological blueshift of incoming electromagnetic
radiation (and \emph{massless} photons); for example, the `light'
from ``standard candle" type 1a supernovae, and ``standard ruler"
baryon acoustic oscillations (re: galaxy clustering). This effect
arguably dispenses with the need to enact the existence of
(mysterious) \emph{dark energy}.} \item{Subsection
\ref{subsubsection:flyby anomaly} gives a qualitative account of
how the \emph{Earth flyby anomaly} can be attributed to the
existence of rotating space-warps (RSWs), if the RSWs are
`locally' refracted by the Earth's ($\sim$10 billion times
stronger) gravitational field so that their planes of rotation are
(all) parallel with the Earth's equatorial plane. This qualitative
account complements (and validates) the quantitative modelling and
empirical prediction formula of the Earth flyby anomaly proposed
by \citep{Anderson_08}.} \item{Due to the large moons of the solar
system achieving spin-orbit resonance very early in the solar
system's history, explanatory accounts of \emph{solar system
formation} --- at least as regards asteroids and comets of less
than approximately 1 or 2 km diameter
(respectively)\footnote{Although larger/higher mass bodies can be
affected `on occasion' --- for example Halley's comet at $\approx
2.2 \times 10^{14}{\rm \,kg}$ (as discussed in subsection
\ref{subsubsection:general on GQ-RSWs})
--- this range of size values applies to a more `continual'
anomalous influence. For a comet of (say) 1 km diameter with a
(typical comet) density of $0.5\times 10^{3}{\rm \,kg \, m^{-3}}$
and (thus a) mass of $0.26 \times 10^{12}{\rm \,kg}$, the rotating
space-warp based anomalous speed/motion retardation effects are
largely `continuous' --- applying at distances of less than
approximately: 18 AU from Jupiter, 7.5 AU from Saturn, and 5 AU
from Neptune (via Table \ref{Table:cut-off}). For a 0.5 km
diameter comet (cf. 1 km), or a (typical density) 0.3 km diameter
asteroid --- noting that asteroids have an average density of
about $2.5\times 10^{3}{\rm\,kg \, m^{-3}}$ or $2.5{\rm \,g/cm^3}$
--- these distances increase eight-fold, and thus the speed
retardation/loss effect is made effectively (or at least
increasingly) distance-independent.} --- will be similarly
affected by the overlooked/supplementary (anomalous)
speed-retardation effect, as the Pioneer spacecraft have been in
the present era. Over one million years the extremely small speed
retardation rate can produce a significant $\leq$\,27.3~km/s
reduction in speed\footnote{This speed reduction arguably cleanses
the solar system of a great deal of (relatively) low mass
material, by way of this matter losing kinetic energy and
spiralling into the Sun. By way of comparison the mean orbital
velocities of Venus, Jupiter, and Neptune are: 35.0, 13.1, and 5.4
km/s.}.}\end{enumerate}

Other possible applications include: (d) that an explanation of
the anomalous \emph{alignment of the cosmic microwave background
radiation (anisotropy) with the ecliptic plane} may have its basis
in the new model; noting that the planes of space-warp rotation,
associated with Jupiter's equatorial plane and its four Galilean
moons, are closely aligned with ($<2^o$ inclination to) the solar
system's (Earth based) ecliptic plane (see subsection
\ref{subsubsection:WMAP}). Further, by way of having (indirectly)
argued that other (non-general relativistic) gravitational
`sources' can exist, it may be the case that: (e) \emph{galaxies},
i.e. systems of billions of roughly equally massive stars --- (at
least) when compared to circumstances in our solar system --- that
are fairly evenly (and non-centrally) distributed throughout,
exhibit more than simply Newtonian gravitation; i.e. galaxies
might exhibit an additional non-Euclidean geometry contribution. A
myriad of field interaction effects (for example) could possibly
be involved.

\subsection{Physical and conceptual supplements inherent to the
model}\label{subsection:End_Discussion} In this paper the Pioneer
anomaly is treated as a real (i.e. non-systematic based) ``gateway
observation" that suggests a need for new physics, with the Earth
flyby anomaly being supportive of this stance. The `boutique',
unique and very specific new physics model proposed herein,
although completely independent of general relativistic
gravitational sources, needs to (and does) coexist (harmoniously)
with general relativity. General relativity (GR) is neither
considered wrong (in any way) nor in need of modification; and on
the whole (i.e. ``all in all"), general relativity `rules' (i.e.
dominates) Gravitation. An ``exception to the rule" solitary new
source of gravitation is proposed --- i.e. a new source of
non-Euclidean space-time geometry is proposed. This supplementary
source type is based upon (a many atomed/moleculed) internally
inexpressible fractional quantum mechanical (spin-based)
\emph{energy} being expressed externally, so as to appease
systemic (i.e. universal) energy conservation --- see the
Introduction for a more elaborate account/overview.

Basically, this additional source-type of non-Euclidean (spatial)
geometry, in addition to GR's (mass, momentum, and energy based)
non-Euclidean geometry/spacetime curvature, affects a (low-mass)
body's motion in a manner that `parallels' both: solar radiation
pressure, and the Poynting-Robertson and Yarkovsky effects. This
(remark) is true in the sense that the Pioneer anomaly is a
supplementary/additional effect upon celestial motion, not
necessarily equally applicable to all sizes/scales of matter, and
it acts so as to merely alter or perturb a body's motion in a
comparatively minor manner --- at least over
(celestially/cosmologically) brief time scales. The
boutique/unique nature of the model relates (in part) to the
unique fact that (with regard to quantum mechanics) intrinsic
angular momentum (i.e. \emph{spin}) \emph{has neither a classical
limit nor a classical analogue}. We also note that the development
of GR preceded/predated quantum mechanics (QMs) and that (to date)
a unification of GR and QMs has not been convincingly validated;
as such, the `new' physical model hypothesised is \emph{not}
inconsistent with the physics of \emph{today}.

The primary conceptual move supporting the new model/mechanism is
the implementation of an energy (`transfer'/re-expression) based
mechanism that spans (microscopic) systems described by quantum
mechanics and (macroscopic) systems `described' by space-time
curvature, so as to resolve a unique/solitary conflict scenario
pertaining to \emph{discrete} quantum mechanical atomic/molecular
systems moving (along a geodesic) in \emph{analog} curved
spacetime. The implementation of a (model-specific) energy
conservation principle is dependent upon (and consequential to) a
proposed supplementation of our understanding of time ---
specifically a background/hidden stance (or `take') on the
sequencing of events on a universal scale --- that takes issue
with the conventional (ubiquitous non-simultaneity)
\emph{interpretation} of light signal-based measurements;
notwithstanding the fact that `different' clocks run at different
rates. This (temporal) interpretive supplementation is implied by
quantum entanglement and quantum non-locality. Support for this
(energy and time based) conceptual supplementation is buttressed
by an appreciation of the following \emph{four} (largely
philosophy of science and philosophy of physics based)
\emph{issues/discussions}.

Firstly, one's attitude to the Pioneer anomaly is crucial. Are we
in a period of (what Thomas Kuhn calls) ``normal science", or not
(by way of the existence of a number of disconcerting anomalies)?
The former (of these two stances) will tend to: dwell on the
strengths of a `research programme' cf. concentrate upon and
investigate its weaknesses; dismiss or seek to explain away the
(creditably established) Pioneer and Earth flyby anomalies; and
bypass or downplay the need to question/re-access core
concepts/notions such as mass, time, and energy --- much less
enter into an interconnected re-conceptualization of aspects of
these fundamental physical `quantities'\footnote{Even though
physicists admit that there are deficiencies concerning/around the
understanding of these concepts.}. In broad terms, `normal
science' undergoes incremental advances, and a big risk factor is
an associated sense of confidence that (on the whole) ``all is
well", such that `opportunities' may not be fully
examined/explored. Recently, \citet*[ p.13]{Hawking_10}
confidently proclaimed/concluded that: ``\ldots \emph{philosophy
is dead} [italics added]." Sympathising with the latter of the two
schools of thought given at the beginning of this paragraph, one
hears a faint echo of the confident proclamation attributed to
Lord Kelvin (William Thomson) and his contemporaries --- at the
turn of the (19th to 20th) century\footnote{``There is nothing new
to be discovered in physics now. All that remains is more and more
precise measurement."}. Indeed, philosophical thinking can
(easily) appreciate two \emph{hidden assumptions} in this `morbid'
stance/quote: (1) that we \emph{are} in a period of ``normal
science", and (2) the primacy given to the role of mathematics
(over conceptual `wrestling'/evaluation); as well as appreciating
a type of subtle elitism peculiar to physicists
--- who (by their behaviour) display no pressing need to draw upon
philosophical expertise. The downside risk here is one of
physicists tending to become an island unto themselves, albeit
with `selected' visitors welcome, confident that the
\emph{mathematical} nature of (current) physical models/theories
(largely) contains the seeds of its future
development\footnote{For example, 11 dimensional M-theory with its
dearth of clear-cut experimentally falsifiable predictions.}.
Philosophers can clearly highlight the ``straw-man" nature of the
argument given to arrive at Stephen Hawking and Leonard Mlodinow's
(moribund) proclamation/conclusion (see
\url{http://www.philosophynow.org/issue82/Hawking_contra_Philosophy}\footnote{The
article's author Christopher Norris (at the time of writing) was
Professor of Philosophy at Cardiff University.}); and as a
`society' they (and philosophers of science and physics) give a
decidedly greater weighting/emphasis to seeking a deeper
understanding of what (the `weirdness' of) quantum entanglement
and non-locality might entail --- as compared to (arguably) a
latent \emph{bias} within physics to downplay this
awkward/non-conformist aspect of scientifically verified reality.

Secondly, minor imperfections regarding special and (particularly)
general relativity (SR and GR) have been presented, involving: the
intractability of (total) systemic energy (and mass); the
global/systemic (cf. relative) nature of rotation and
acceleration; the physical implications of GR's (mathematical
technique of) general covariance; the reliance of GR's formulation
upon SR, and the scaffolding role played by the general principle
of relativity in establishing the Einstein field equations; and
(lastly) lingering doubts regarding SR's clock problem/twin
paradox. These issues allow us to (at least) question the totality
of GR's physical (and conceptual) scope. To this we might add two
questions: (1) Can everything be relative? and in response to the
question: ``Why is there something rather than nothing?" (2) Can
something and nothing exist independently of the concept of (a
physical) everything? These two questions, and relativity's
imperfections, strongly suggest the need (on some level, in some
manner) for a `universal' (i.e. global or systemic) perspective.
By way of quantum entanglement our new model uncovers and then
exploits/utilises (such) a universal/global physical perspective.

Thirdly, armed with an open mind to physical reconceptualisation and
an appreciation of SR's and GR's imperfections\footnote{This in no
way should detract from the reverence deservedly bestowed upon the
scientist/physicist (and philosopher) Albert Einstein.}, the model's
(necessary) next step was to question an agenda of reductionism to a
unified theory of everything, such that a fairly clear divide between
microscopic and macroscopic realms/domains is maintained cf. overcome
--- which additionally casts into doubt the existence of the graviton
particle. Assisting this `divided' stance was: QMs and GR having
so little in common, both physically (re: scale and `nature') and
in terms of their laws/mathematics; and (also) the fact that the
(\emph{linear and rotational} based) classical vortex theory of
propellers and wind turbines (see subsection
\ref{subsubsection:Prop/HAWT}) requires a power (or at least
energy) based governing expression (recall Equation
\ref{eq:Vortex}: $E_p=m \Gamma_{e} f$) --- cf. microscopic physics
where a \emph{force} and/or energy basis is sufficient.
Subsequently, a ``\emph{pairism}" of `dual' concepts is
embraced\footnote{With ``pairism" obeying ``dualism" only in the
sense of: the division of something conceptually into two opposed
or contrasted aspects, or the state of being so divided. The new
term `pairism' is preferred because `dualism' has a lot of
additional conceptual baggage associated with it --- more so in
philosophy than in physics. Furthermore, the expression ``duality"
has quite different meanings in these two distinct academic
fields, and markedly different applications.}, and this structural
tool is then used to delineate different facets of
physical-reality/ontology. `Pairs' particularly relevant to the
model are:

macroscopic vs. microscopic (domains and laws);

analog vs. digital (i.e. continuous vs. discrete);

Euclidean vs. non-Euclidean (re: geometry);

local vs. non-local (re: physical `interactions');

internal vs. external (re: a physical system);

relative vs. systemic/global (physical perspective);

local vs. global (re: physical representation);

dynamical phase cf. geometric phase (re: QMs);

force (at a given time) cf. a process based energy;

real cf. virtual (re: physical energy);

spacetime cf. (distinct) space and time\footnote{Written as
``space-time" herein.};

mathematical cf. conceptual (re: theory advance);

symmetry cf. asymmetry (re: theoretical physics);

linear vs. rotational (re: physical circumstances);

reductionist cf. non-reductionist (physics agenda);

unification cf. complementarity (re: guiding aim);

\mbox{the phenomenal/observational vs. the `noumenal';} with this
last `pair' having been discussed in some detail in the
Introduction. Excluding the first three pairs, the second/latter
of these paired concepts are given at least equal priority in the
model/mechanism presented herein. In standard physics, with regard
to its: overall perspective, self reflectivity, and ultimate
objectives --- i.e. ``the big picture" --- these (second) notions
are generally considered of equal or \emph{lesser} importance.
Herein, our (non-reductive) dualist/`pairist' leanings
(necessarily) \emph{equally} recognises both sides of such
`divides'.

Fourthly (and finally), rather than downplay the `weirdness' of
quantum entanglement and non-locality, we recognise that
relativity's light speed limit (conceivably) places physics at
risk of being in something like a (scientific) ``double bind" or a
``Catch-22" (in its common idiomatic ``no-win situation"
usage)\footnote{``Gregory Bateson and Lawrence S. Bale describe
double binds that have arisen in science that have caused
decades-long delays of progress in science because science had
defined something as outside of its scope (or `not science')
\ldots" --- extracted from Wikipedia: \textit{Double Bind},
Dec-Jan 2011-12.}. Combining this dilemma with the different (and
`occasionally' incompatible) treatment of time in quantum
mechanics cf. special and general relativity \citep{Albert_09},
Section \ref{Section:PhiloTheory} argued for the introduction of a
supplementary/additional ontological scenario involving a (hidden)
universal/systemic `noumenal' background `scanning process' that
occurs `between' incrementally different staccato phenomenal (or
able to be observed) temporal moments\footnote{That is, separate
(digital) instantiations of a universal-wide reality, with each
`phenomenal' discrete (incrementally different) moment separated
by a (noumenal process-based) `pause' --- that is beyond direct
observational/phenomenal appreciation.} (see \mbox{section
\ref{subsection:SR's ontology}).} From a scientific observer's
(restricted) local and relative phenomenal/measurement-based
perspective, time (only) appears continuous/analog, whereas at a
deeper level of understanding it involves a digital `phenomenal'
aspect. Importantly, by proposing a (non-relative) relationship
between: a body's motion/speed, light/EM radiation as an observing
`agent', and the background universe itself --- with the latter
two not being independent of each other (cf. the case implicit in
SR and GR\footnote{Particularly SR's Principle of the
(\emph{observed}) invariance of the speed of light (in a vacuum),
regardless of an inertial observer's or light source's state of
(relative) motion/speed.}) --- a (noumenal/universal perspective)
supplementary interpretation of the Lorentz transformations and
SR's need for spacetime interval invariance was proposed. Six
ramifications ensuing from: the aforementioned considerations
(albeit a brief discussion\footnote{Encapsulating Section
\ref{Section:PhiloTheory} for the purposes of this brief summary
unfortunately results in some degree of ambiguity.}), and the
noumenal--phenomenal (physical) \emph{complementarity} arising
from these (temporal and interpretive) supplementations, are now
presented.
\begin{enumerate}[a)] \item{A universe-encompassing
(background/hidden) event/`time' simultaneity, albeit with
\emph{different} rates of \emph{measured} time duration between
events --- implies a (deformable/curvable) \emph{space}
substratum, that in the (idealised) absence of all: mass, motion,
energy, etc. would be Euclidean (i.e. a `flat' space).} \item{The
`latent' all-encompassing (`beyond' direct measurement)
reality/event simultaneity allows \emph{energy conservation} to be
applicable in certain `universal' situations, and it
supports/strengthens the case for a well coordinated background
inertial frame, particularly when non-local QM spin entanglement
is involved --- as is the case in our model/mechanism.} \item{From
a noumenal/global perspective special relativistic time dilation,
and subsequently length contraction and mass dilation, are
alternatively/complementarily attributed to a
\emph{loss}\footnote{This loss is achieved by way of an
accumulation of numerous incremental time losses (relative to
maximum \emph{availability}) --- see subsection
\ref{subsubsection:relative motion}. Further, this situation is
dependent upon `time passing' as systemic/universal (extremely
small/short finite duration) discrete `moments'.} in
available-to-measurement specific energy --- by way of the
specific energy associated with relative motion
($\frac{1}{2}v^2$). Residual unease pertaining to the clock/twin
paradox is vanquished, and maximum relative specific K.E. is
bounded by a $c^2$ limit.} \item{Perceiving a standard/phenomenal
specific energy ($e=E/m$) increase (from zero \emph{up}) as
alternatively a ``drawing-\emph{down}" from a maximum `potential
reservoir' value of $e=c^2$, that (in idealized conditions) would
exist homogeneously throughout the universe if it were
`completely' empty. This \emph{reversed}/`opposed' perspective
(further) builds upon the use of `paired' or complementary
perspectives.} \item{GR's need for a tensor based generally
covariant mathematical formulation is seen to be a consequence of
(further) adding gravitational effects to the (already existing)
effects of (special relativistic) high-speed relative motion, and
as such (from a noumenal and reversed perspective) a double
``draw-down" (from the spatially homogeneous $e=c^2$ ``maximum
potential reservoir") is involved.} \item{Finally,
noumenal--phenomenal ontological complementarity will have an
impact upon: (1) how quantum mechanics can/should be interpreted;
and (2) the belief that randomness, and hence
indeterminism/uncertainty, are deeply inherent to
Nature\footnote{To what extent (``behind the scenes") noumenal
based \emph{non-local hidden variables and processes} will impact
upon these issues is (itself) an open and debatable issue.}.}
\end{enumerate} Note that the model does not
establish these ramifications for the reason of discrediting
special and general relativity. As previously mentioned, the
theoretical structures of special and general relativity are
neither: wrong, nor in need of modification; rather, by way of the
phenomenon of quantum entanglement, some of their ontological
assumptions (and foundations) have been questioned\footnote{With
this questioning, and the ensuing ontological supplementation,
having (minimally consequential) ``time of an event" based
\emph{numerical} implications/ramifications.} and GR's
scope/domain of application is (merely) considered to be in need
of a (solitary) additional/supplementary physical circumstance
(that is itself beyond GR's ken). Finally, we also note that
although GR's non-Euclidean geometry is expressed in terms of a
metric tensor of (phenomenal) spacetime, the new model's
supplementary accleration/gravitational field is necessarily
expressed in terms of (distinct) curved space and (a
systemic/universal) `time' --- by way of (i.e. in response to)
utilising/adopting a noumenal perspective.

\subsection{Major mathematical equalities and relationships of
the model}\label{subsection:End_Equalities}The model/mechanism
(developed throughout this paper) benefits from circumstantial
\textbf{simplifications} such as: its field \emph{perturbation}
basis; the geodesic (i.e. unforced) motion of (the lunar based)
atoms/molecules involved in the mechanism; the prevalence of
geometry (in a number of guises\footnote{These include:
non-Euclidean geometry; (quantum mechanical) geometric phase; and
classical geometry --- (the latter) in connection with celestial
orbits and kinematics.}); and the use of a simple scalar
(total/aggregate) energy --- that is proportional to the total
\emph{number} of atoms/molecules ($N_m$) within a lunar non-rigid
(predominantly solid) body (i.e. within a moon). Indeed,
(quantifiable) energy is the physical linking (or bridging) means
whereby the aggregate of (each and every) atom's or molecule's
virtual (and `localised') quantum mechanical based (spin) energy
is re-expressed collectively/singularly and \emph{externally} as a
real (distributed) macroscopic (non-Euclidean geometry based)
\emph{rotating} acceleration/gravitational field. At a `fixed'
point in space this field induces a \emph{sinusoidally} varying
field amplitude/strength. Each of these ``rotating space-warps"
exists conjointly with a non-local mass field. In tandem with this
energy conservation `bridging' effect, and recalling the model's
five distinct sizes of particles and systems (outlined in the
Introduction), we note that the model's `physicality' stretches
from the (common and virtual) spin phase offsets of (lunar
atomic/molecular residing) \emph{elementary fermion particles}
through to space-time curvature(s) (and non-local mass
distributions) that span/encompass \emph{the universe} as a whole
(i.e. in its entirety).

This (aforementioned) macroscopic re-expression of total
fictitious/non-inertial QM spin-based (virtual) \emph{energy}: (1)
addresses a unique discrepancy that can arise when a
\emph{digital} \emph{system} is moving in \emph{analog} curved
spacetime; (2) maintains universal/systemic energy conservation;
and (3) is time-irreversible; so as to ensure (ongoing)
universal/systemic \emph{stability} (without fail). The
\emph{irreversibility} of this energy re-expression allows the
model/mechanism to be (effectively) regarded as a second type of
non-reversible/irreversible time evolution of a QM `system' --- in
addition to the irreversible process of a (quantum mechanical)
measurement. Note that the large scale nature of the
(superpositioned) rotating space-warp (RSW) based `gravitational'
field undulations (of amplitude $\Delta g$)
--- whose ultimate basis is spatial curvature (cf. spacetime
curvature in GR) --- ensures that the sinusoidal field strength
(amplitudes) \emph{at}, and proper accelerations acting upon, a
\emph{moving} spacecraft ($\Delta a$) effectively equal the
$\Delta g_{\rm{field}}$ values (see subsection
\ref{subsubsection:intra-undulation}).

The \textbf{compatibility} of each (RSW based) $\Delta a$
\textbf{with special and general relativity} is ensured by way of:
(1) each supplementary acceleration/gravitational field's rotating
sinusoidal perturbation having a (universally) \emph{position
invariant} constant `amplitude' --- i.e. (at any given `universal
moment') the maximum possible amplitude ($\Delta a$) is the same
everywhere --- although changes to $\Delta a$ will propagate
`outwards' at the speed of light; and (2) by way of the
secondary/supplementary $\Delta a$ field amplitude being envisaged
as a different (and independent) type of energy
`draw-down'\footnote{As compared to GR's `draw-down' situation ---
mentioned near the end of section \ref{subsection:End_Discussion}
[bullet points d) and e)\,].} --- such that a second class/type of
non-Euclidean geometry (involving space-time cf. spacetime) can
coexist with GR's gravitation (`in its own right')\footnote{This
ensures that the aforementioned virtual energy to real energy
`re-expression' does not create new energy.}.

The three major equalities of this paper are firstly, Equation
\ref{eq:SumA} (noting that $\Delta g_{\rm{field}}\approx\Delta
a_{\rm{field}}$ at a body):
$$a_p=\overline{a_p(t)}=\sqrt{\sum (\delta a_{\rm
proper})^{2}_i}=\sqrt{\sum (\Delta a_{\rm field})^{2}_i}$$
representing the average superposition-based overall/resultant
effect of all the fixed and constant amplitude (i.e. time
invariant and position invariant amplitude)
gravitational/accelerational space-warp fields \emph{upon}
spacecraft (and low mass body) acceleration; i.e. the (model's)
path-based constant \emph{anomalous Pioneer `deceleration'} value
--- \emph{relative} to predictions that do not include the
supplementary (RSW) field type. Each individual (field-based)
$\Delta a$ `equals' and corresponds to its respective (and
consequent) spacecraft based $\delta a$ (speed \emph{shortfall}
rate) value (see section \ref{Subsection:Shortfall}). Further, we
note that each $\delta a$ is equal to (and `derived' from) a
(single cycle) anomalous speed loss ($\delta v$) `divided by' its
(respective) sinusoidal period/duration ($\Delta t$).

Secondly, we have Equation \ref{eq:2 by Energy} --- rewritten here
with some minor nomenclature modifications --- which is applicable
to a \emph{single} (barycentre/Sun-planet-moon) `gravito-quantum'
rotating space-warp system:
$$\Delta E^{\,\rm{Vi}}_{\,\rm qm}=\left[\frac{1}{2}\hbar\,(\Delta
t)^{-1}\,\eta\right]N_m=\Delta E^{\,\rm{Re}}_{\,\rm
gr}=\frac{1}{2}\Delta a^{2}\Delta t^{2}\,m_{1}^*$$ It describes
the (irreversible) external \emph{re-expression} of: a
\emph{non-inertial} additive-based ($\times \,N_m$
atoms/molecules) \emph{virtual} fractional QM spin (exact) energy
(i.e. $\Delta E^{\,\rm{Vi}}_{\,\rm qm}$), (conjointly) as the
product of: (1) a \emph{real}/actual acceleration/gravitational
field perturbation-based specific energy --- with this specific
energy proportional to perturbation/`wave' \emph{amplitude}
($\Delta a=\Delta a_w=\Delta a_o \, \eta$) \emph{squared} --- and
(2) an initial (spherical surface-based) distributed non-local
mass value [$m^*_1$, i.e. $m^*(r)$ at $r=r_1$]. We note three
things: (a) the (quantum-based) efficiency factor:
\mbox{$0\leq\eta\leq1$}; (b) that with a \emph{process} based
$\Delta t$ term on both sides of the equality, some form of
simultaneity is required --- implicating the presence of quantum
entanglement and non-locality; and finally, (c) that with constant
$\hbar$, and $\Delta t$, $N_m$ and $\eta$ effectively fixed (over
human time scales i.e. decades), the $\Delta a$ values are also
`effectively' fixed.

Thirdly, Equation \ref{eq:distribution}: \mbox{$m_r^*\,
V_r={\rm{constant}}=m_1^*\, V_1\,$} describes $m^*$ ``continuity",
via the product of (QM entanglement based) non-local mass and
(enclosed) spherical volume, relative to establishment/inception
conditions (at the ``reference radius" $r_{\rm 1}$). This
relationship\footnote{Note the contrast with density ($\rho$)
where $\rho=m/V$.} ensures that the (non-local) field energy
[$\Delta E^{\,\rm{Re}}_{\,\rm gr}(r)$], at any given point/radius
in the `far'-field (where $r>r_1$), diminishes (i.e. undergoes
dispersion) away from the (lunar) source region. Note that the
initial non-local mass value ($m^*_1$) exhibits the idiosyncrasy
of (concurrently) being \emph{both}: a contributor to systemic
\emph{total} `gravitational' energy (in $\Delta
E^{\,\rm{Re}}_{\,\rm gr}$), and a local value (i.e. a point-like
instantiation value) that is \emph{homogeneously distributed}
over/upon a (lunar-centered) spherical surface (of radius $r_1$).

The model has at least four similarities or parallels to magnetism
(see subsection \ref{subsubsection:magnetism}). In particular, we
note that a (further) \emph{simplifying} aspect of the model is
the \emph{common} (spin-based) geometric phase offset and (common)
spin axis orientation applicable to \emph{every} constituent
fermion `particle' within (the atoms and molecules constituting) a
particular moon. This situation is dependent upon
\emph{externally} imposed (barycentre/Sun-planet-moon) orbital and
lunar spin characteristics, and is `bookended' by the appeasement
of any ensuing \emph{virtual} rate of intrinsic angular momentum
(i.e. spin energy) imbalance as a single (real) \emph{external}
rotating space-warp (RSW) and conjoint (external) non-local mass
distribution. \emph{Importantly}, this re-expression process is
demanded/facilitated by (the `rigidity' or inflexibility of)
\emph{quantised} electromagnetic spin-orbit coupling within atomic
and molecular systems --- when confronted by a (relative)
geometric phase offset that is $<2\pi$ radians (i.e. less than
half a fermion wavelength).

Furthermore, with \emph{geometry} --- both pure/classical geometry
(via barycentre/Sun-planet-moon celestial orbits) and a (quantum
mechanical) geometric phase offset angle --- playing a vital role
in the model, the introduction of a fixed ``universal constant"
(reference) \emph{angle}: $\phi=\tan^{-1}(8\pi)^{-1}
\approx2.28^o$ was necessary --- in order to `best-fit' the
(Pioneer spacecraft based) experimental/observational data to a
model. With regard to each (and every) particular Sun-planet-moon
(gravito-quantum) system, this angle is essential in determining
both: the efficiency factor (subsection
\ref{subsubsection:Celestial geometry and}) and the reference
radius $r_1=8\pi r_o$ (subsection \ref{subsubsection:geometric
acceler}) mentioned in relation to Equation \ref{eq:distribution}
(above)\footnote{Where $r_o$ is lunar (elliptical) semi-major axis
(length), usually denoted (in standard nomenclature) as $a$ or
$r_a$.}, as well as the optimum and weighted acceleration
amplitudes, i.e. $\Delta a_o$ and $\Delta a_w$ respectively (see
subsections \ref{subsubsection:acceleration} and
\ref{subsubsection:geometric acceler}). This
(fixed/universal-constant) reference angle is also used to
determine the different Sun-planet-moon (atomic/molecular, and
fermion based) (relative) geometric (spin) phase offsets
[$(\beta)_i$] --- common to all atoms/molecules, and their
constituent elementary fermion particles, within a specific (and
`geometrically' appropriate) moon. Each (lunar based) spin phase
offset ($\beta$) determines (and corresponds to) its particular
efficiency factor ($\eta$). This (`final' or post-orbit)
virtual/fractional spin-phase offset --- \emph{relative} to
initial spin phase (i.e. at lunar orbital loop commencement) ---
is indicative of an (`over-spin' or) `\emph{precession}' of the
background inertial (spin) frame --- \emph{relative to} the actual
unchanged (electromagnetic force constrained/dominated)
atomic/molecular spin-orbit configuration. Ultimately, this
relative (and virtual/latent) spin frame precession arises from
(lunar/third-body) geodesic motion in curved spacetime (over
closed loop/orbital duration $\Delta t$). For a pi radian (i.e.
one quarter fermion wavelength) phase offset $\eta=1$, whereas
$\eta=0$ for both a zero/null radian ($\beta=0$) and
$\beta\geq2\pi$ radian (relative) geometric spin phase offsets. A
triangle function (interpolation) is seen to apply between these
three (phase offset vs. efficiency factor `coordinate') values
(recall Figure \ref{Fig:Triangle}).

A central physical relationship of the model --- applying to each
individual/single Sun-planet-moon system --- is the
inter-relationship existing between (virtual/fractional) quantum
mechanical intrinsic angular momentum (i.e. spin):
$\frac{1}{2}\hbar \eta=\frac{1}{2}\hbar_w$ and (ensuing, real)
RSW-based perturbation amplitude $\Delta a_o \eta=\Delta a_w$.
Subsection \ref{subsubsection:external wiggle} argued that this
angular momentum to (gravitational) acceleration relationship is a
new and unique physical relationship. Facilitating our
understanding of the physical `connection' existing between these
two physical `quantities', we employ the \emph{conceptual} notion
of (a relative) \emph{twist}, in the sense of both a (spin)
\emph{turning} and a (space) \emph{warping} respectively, as a
conceptual bridging device --- noting that at a \emph{physical}
level the first of these (two) conceptualisations is particularly
inappropriate. Thus, a \emph{virtual} (QM spin phase based)
\emph{turning} (or precession) relative to an actual
electromagnetic force dictated/constrained QM spin (and orbital)
configuration --- and achieved over a (lunar celestial spin and)
orbit-based process time $\Delta t$ --- is physically
linked/related to a space \emph{warping}
(acceleration/gravitational) perturbation that rotates (also) with
period $\Delta t$, and in the same sense/direction as lunar spin.
Note that the `gravitational' equilibrium (i.e. unperturbed)
condition or state, and a moving body's unperturbed
\emph{kinematic} condition/state, are determined by standard
gravitational theorisation and the (conventional) gravitational
sources it encompasses. This physical `twist' relationship may be
(additionally) considered as a second type/face of spin
entanglement --- albeit involving a (spin) \emph{energy
transmutation} into (a new type of) `gravitational' field energy
--- with both `sides' of this relationship dependent upon their
own (distinct) macroscopic `rotational' \emph{process} (of
equivalent duration/period $\Delta t$).

Non-local mass ($m^*$) is considered to be a second type/class of
mass, primarily because its associated acceleration/gravitational
field (i.e. a rotating space-warp) does \emph{not} affect
\emph{all masses} in an equivalent manner (cf. GR). Non-local mass
varies with distance ($r$) from its source region, diminishing in
an inverse-cubed manner (i.e. inverse of the volume enclosed); and
subsequently its effect upon a passive (gravitational) mass
($m_p$) is not distance independent. More precisely, this mass
`interaction' is an ``all or nothing effect"; ``all" if
$m^*(r)\geq m_p$ and ``nothing" if $m^*(r)<m_p$ (at the body, i.e.
at the $m_p$), as is similarly (and generally) the case with the
\emph{energy} based (and EM radiation interacting with matter
based) photoelectric effect. This weak equivalence principle
defying feature of the new model/mechanism also utilises a
point-mass idealisation as regards the `interaction' of $m^*$ (at
$r$) with a compact/condensed $m_p$ (see subsection
\ref{subsubsection:at-a-point}), albeit for quite different
reasons to those applicable in traditional gravitation theory. By
its very nature, non-local mass is devoid of both (tangible)
matter/`material' and hence \emph{local} active gravitational
mass\footnote{Whether or not \emph{non-local} mass (itself) is an
\emph{active} gravitational mass has not been pursued. Being a
secondary effect arising from Sun-planet-moon motion it is (most)
likely that any (active gravitational) contribution will be very
minor, if not entirely negligible; the latter either by way of its
cosmological extent or simply via its very different `nature'.}.

\subsection{A crucial distinction and the model's broader
implications}\label{subsection:End_Conclusion} In this section,
further implications and ramifications peripheral to the model
(itself) are presented. The discussion takes the form of a series of
brief remarks.

Assuming a real and non-systematic based Pioneer anomaly has given
us no choice other than: (1) to considerably re-examine the
(physically) foundational concepts of time, mass, and energy; and
(2) establish a model that is independent of general relativity
(GR) but nevertheless (is) physically compatible (and coexistent)
with GR. We addressed the misconception that GR's success, and/or
its principle-based approach to a theory of gravitation --- i.e.
principles/principle of: equivalence, general relativity, and
general covariance --- denies \emph{all} other sources (and forms)
of non-Euclidean geometry (see section \ref{Subsection:EquivPr
Comment} in particular).

Crucial to this new understanding has been the fundamental
distinction drawn by Sir Arthur S. Eddington (and reprised by Matthew
Stanley, see subsection \ref{subsubsection:Eddington}) --- that
quantum mechanics (QMs) and GR provide insight into \emph{how} we see
the world (i.e. its epistemological characterisation), rather than
\emph{what} the world [entirely] is (i.e. its ontological
characterisation). Thus, physical observations and their
mathematical/theoretical modelling and understanding do not
necessarily provide an unambiguous picture of `what' the world is,
free of subtlety and/or oversights.

Eddington's fundamental conceptual (cf. empirical) distinction
facilitated the introduction of the (non-reductive)
paired/complementary phenomenal \emph{and} noumenal perspectives
proposed herein; particularly its implementation in achieving a
supplementary/complementary interpretation of SR's Lorentz
transformations by way of appreciating that: the `world'/universe,
an observer's motion/speed, EM radiation/light itself, and
observations utilising EM radiation are interwoven --- when
considered/`viewed' from a noumenal perspective. As such, SR's and
GR's use of ``spacetime" and general covariance need not
necessarily be understood as simply ``how \emph{and} what the
world is." Although there is a reduction in explanatory
simplicity, this new approach is not in defiance of Occam's razor
because there is a (more than) compensatory increase in
explanatory capability.

Further implications and ramifications of these related
(philosophical) epistemological--ontological and
(pragmatical/physical) phenomenal--noumenal dichotomies fills out
the remainder of this section and parts of section
\ref{Subsection:Big picture}.

The \emph{unavoidability} of ``observer dependence"
--- regarding measurements made at extremely low energies or
momentums, or over extremely short distances or durations ---
exhibited in microscopic QMs is (herein considered) similarly
present in macroscopic physics, but it is exhibited in an
altogether different way; i.e. involving an observer's
(relativistic) high speed motion (and substantial specific kinetic
energy). This macroscopic observer dependence is downplayed and
masked by SR's phenomenal (spacetime) perspective, whereas it has
been unmasked by our new/complementary noumenal (distinct space
and time, i.e. space-time) and global/systemic perspective (see
section \ref{subsection:Reversal}).

Within a many-body barycentric solar system (for example), the
tactile/sensory basis and \emph{historically} problematic
hypothesis of a gravitational \emph{force} --- acting upon a
moving object --- was superseded by general relativity's curved
spacetime based gravitation. Note that in a \emph{many-bodied
barycentric system}, the use of `gravitational
\emph{accelerations}' (at different points in space) --- that
apply to all bodies irrespective of their (passive gravitational)
mass --- has retained its pragmatic validity, e.g. regarding an
interplanetary spacecraft's ``orbit determination". The model's
(unique/specific) implementation \mbox{of: (1) non}-local mass
($m^*$), which effectively involves a \emph{dematerialisation} of
the concept of `mass' (cf. matter); and (2) a (general) scepticism
regarding (both) the graviton particle's existence and the
physical reduction of gravitation to a (boson) particle ``exchange
force" basis; is \emph{arguably} the second-stage of a two-stage
process of ``gravitational dematerialisation", if we
consider/designate GR's (pro-curvature) ``de-forcing"
innovation/advancement as the first-stage of this
dematerialisation of ``gravitation" --- in the broadest imaginable
sense of the word.

In response to: (1) the question ``Can everything be relative?"
which distinguishes a (systemic) whole from (relationships
involving) its constituent parts; and (2) the observation that
spacetime on a cosmological scale is uncurved (i.e. `flat' such
that in GR $k=0$); as well as (3) Section \ref{section:Type1a}'s
querying of dark energy; it was proposed (subsection
\ref{subsubsection:GR incomplete indi}) that whole universe/global
curvature ($k=\pm1$ in GR) may not (actually) be physically
'realisable'. As such, the assumption that (\emph{relativistic})
gravitation has a `whole universe' (or cosmological)
\emph{physical} application --- as compared to a theoretical
implication --- is quite conceivably unjustified. Pursuant to this
stance, the model proposed (subsection
\ref{subsubsection:substantivalism}) that in the \emph{idealised}
scenario of a complete absence of any mass, momentum and physical
field energy in the universe, a homogeneous space
`substratum'/continuum exhibiting Euclidean geometry would
`exist'; i.e. an uncurved empty Euclidean space `container'. In
(actual) reality, this idealised background uniformity/homogeneity
is \emph{locally} (but not globally) curved/deformed by its
contents (up to the scale of its largest substructures). The
(root) motivation and basis for this change in stance has been an
embracing of quantum entanglement and quantum non-locality,
together with an openness to (and `demarginalisation' of) their
ontological implications.

\subsection{Big picture `meditations' arising from the model
proposed herein}\label{Subsection:Big picture}In this subsection,
discussion of the implications of the new model is extended to
`big picture' topics\footnote{In the sense of crucial features,
facts and issues that contribute to mankind's understanding and
explanation of the physical world/universe (that we `find'
ourselves in).}.

The Nobel laureate Daniel Kahneman, an Israeli-American
psychologist noted for his work on judgment and decision-making,
states (in an interview\footnote{Liz Else, New Scientist, Vol.
212, No. 2839, pp. 34-35, ``Nobel psychologist reveals the error
of our ways".}) --- regarding forming an impression on the basis
of the information you have --- that:
\begin{quote} \ldots we can't live in a state of perpetual doubt
--- so we make up the best story possible and live as if the story
were true. \dots we like the stories to be good stories.
\end{quote} When dealing with a complicated problem,
\emph{correctness} differs depending upon the task and its
perceived solution. Scientific instrumentation and machinery
either work (as planned) or they don't; similarly, an elaborate
computer program either `runs' or it doesn't, and a mathematical
proof or a logical deduction is either valid or invalid. The
investigation of a complex scientific circumstance or problem is
(predominantly) not as straightforward as in these cases, e.g.
star formation. Scientific `correctness' (or truth) in the
presence of \emph{conceptual} complexity or intricacy, especially
if it is combined with limited/restricted knowledge and/or
ambiguous information, is often provisional with its maturity
requiring: time, effort, insight, and new or (at least) refined
information. It appears that the lack of a `timely' (and generally
favoured) explanation for the well credentialed Pioneer anomaly
has led to a dispelling of lingering doubt by virtue of embracing
a thermal-radiation/heat based explanation, which \emph{is} a
viable (and `good') story; although this scenario leaves the well
credentialed Earth flyby anomaly (isolated and) in need of a
similarly `good' explanation. Herein, we have looked beyond the
(psychological) weight of ongoing doubt to both: query the
correctness of this favoured explanation, and construct/formulate
an alternative model that is unavoidably/necessarily conceptually
(cf. mathematically) intricate\footnote{This approach appears to
be anathema to some theoretical physicists, particularly when the
complex conceptual discussion is in a language other than their
first language.}.

This emphasis upon conceptualisation has necessitated the copious
use of both bracketed words and the (forward) slash (or virgule)
punctuation character throughout this document, primarily to add
detail and/or clarify the meaning of a word, phrase, or sentence
--- cf. the slash's (other) ``choice between two words" application.
Additionally, various punctation marks, especially the: comma,
semi colon, and em dash (i.e. ---), as well as scare quotes (i.e.
`...'), are liberally employed in order to minimise ambiguity.
Note that scare quotes (herein) are used to alert the reader to
the non-standard or special use of a word or phrase.

Philosophers of science warn about being overly demanding or
prematurely dismissive of a recently formulated theory/model that
matches the evidence, especially one that is conceptually
`different'. In other words, the exactitude demanded of well
established theoretical models in physics, as exhibited by
mathematical/computational reasoning, should not be equally (or
blindly) applied to a recently formulated
theory/model\footnote{The (``intricate defensive") strategy of
continually ``raising the bar" of acceptance for a new
theory/approach, is (ultimately) symptomatic of preferring the
\emph{status quo}.}. Deficiencies will surely exist, but these do
not logically entail that the young model is entirely `wrong' or
unworthy of ongoing consideration.

To simply proclaim that general relativity, because of its
success, denies the Pioneer anomaly is unjustified and hence
unscientific (although understandable); whereas a conceptual
enriching and supplementation of physical concepts, driven by a
creditably established and striking observation, (although
possibly flawed) \emph{is} scientific --- especially if the
model/mechanism proposed can be falsified, and encompasses all the
Pioneer anomaly's awkward observational constraints. Herein, the
latter were not easily `fitted', e.g. by way of (mere) theory
modification; and a valid (and progressive) model's fit to the
evidence is necessarily complicated/difficult, in so much that
physical reality/ontology and fundamental physical concepts have
required (innovative) modification. The physical model/mechanism
established herein supports the saying/proverb that: ``truth is
stranger than fiction\footnote{The genesis of this saying is
(commonly) attributed to: Lord Byron's poem \emph{Don Juan}, Canto
XIV (1823).}", in the sense that (regarding the explanation for
the Pioneer and Earth flyby anomalies proposed herein):
\begin{quote}`\emph{a posteriori} scientific knowledge' driven by
a striking new scientific result\footnote{That is, a new physical
explanation deduced/derived from a new and anomalous scientific
experimental finding.} is stranger than and beyond any `\emph{a
priori}' reality or theory conceivable by means of human reason
and imagination alone\footnote{Inclusive of mathematical deduction
and hypotheses, and drawing upon all (pre-discovery) scientific
knowledge.}. \end{quote} This point of view is supported by the
model's use of virtual (relative) geometric phase and the model's
\emph{introduction} of: non-local mass; (real) constant amplitude
sinusoidal perturbations upon the pre-existing relativistic
gravitational field; and a latent ontological
`phenomenal--noumenal' complementarity; with the latter somewhat
debasing the centrality of the `observer' (and physicists) as
regards achieving progress towards a further/deeper understanding
of physical reality --- e.g. quantum entanglement (and causation).

The reversal of (an observer's) \emph{locational} perspective
arising from Nicolaus Copernicus' seminal (16th century) celestial
motion proposal has recently been revisited with the embrace of
the `multiverse'. A number of cosmologists (and theoretical
physicists) now believe that we only observe (figuratively
speaking) ``the tip of the (cosmological) iceberg", with the rest
(i.e. other universes) hidden --- in the sense of \emph{never}
being directly amenable/available to observations. It has been
proposed herein that something similar is occurring with respect
to `time' (recall subsection \ref{subsubsection:universal
time})\footnote{As was the case in Section
\ref{subsection:End_Discussion}, ``time" --- for want of a better
word --- is (occasionally) being used in an extended and `beyond'
(direct) measurement manner.}. Interrelated to this
\emph{temporal} aspect --- but occurring in the classical
Copernican ``\emph{reversal} of perspective" sense --- is an
energetics/energy aspect, that pertains to a body's specific
energy (see subsection \ref{subsubsection:Copernican}). Recalling
section \ref{subsection:End_Discussion}, this reversed (specific
energy) perspective is dependent upon a description of quantum
entanglement based situations/phenomena requiring the
prioritisation of a (complementary) noumenal perspective over the
standard (amenable to scientific measurements/observations)
`phenomenal' perspective.

The uniqueness of the model/mechanism hypothesised herein (implies
and) is illuminated by a (heuristic) two-by-two
categorisation/bifurcation of the laws of nature into four main
categories/classes. The first bifurcation involves a distinction
pertaining to laws describing (quantum mechanical) microscopic
`reality' as compared to macroscopic reality, with the largest
molecules demarcating the extent of the microscopic realm/domain
(recall subsections \ref{Subsubsection:MM Differences} and
\ref{subsubsection:Prop/HAWT}). Note that electromagnetism has
distinct laws in each domain, and that a total physical
micro-macro separation is not being espoused. The second
bifurcation involves a distinction between local (i.e. speed of
light limited) and non-local based processes/laws. Clearly, all
(known) physical laws, and almost all (known) physical phenomena
--- except for quantum entanglement and quantum non-locality ---
have a local basis cf. a non-local basis; with often neglected
(engineering based) fluid mechanics and the mechanics of solids
lying in the local macroscopic class of physical laws. The (unique
and boutique) model/mechanism proposed herein involves a
(repeating) cyclic \emph{process} that spans the \emph{non-local}
microscopic-macroscopic `divide' or bifurcation (previously
articulated). By way of the exemplar of aircraft propellers and
horizontal-axis wind turbines, subsection
\ref{subsubsection:Prop/HAWT} sought to show that a force (and/or
energy) basis to local macroscopic physics, as is the case with
local microscopic physics, is not a fail-safe (core and/or
working) assumption. When a vortex/circulation based theoretical
(potential flow) approach is used to describe this linear
\emph{and} rotational mechanism (in its three-dimensional
entirety), the `physical quantity' of power (in addition to
momentum, pressure, force and energy) is crucial/necessary.
Consequently, and notwithstanding the unquestionable validity of
the (three force) unification of: the strong force, the weak
force, and electromagnetism, a four \emph{force} unification (i.e.
theory of \emph{everything}) unification agenda is quite
conceivably misguided (recall sections \ref{Subsection:Tensions}
and \ref{Subsection:GR fortress}). This claim was borne out in the
model by the virtual offset of (geometric phase and) intrinsic
angular momentum (i.e. QM spin), applicable to each and every
elementary fermion particle `comprising' an atom or molecule,
being \emph{equivalent} to the spin offset of the atom/molecule in
its (systemic) entirety, whereas the total QM intrinsic angular
momentum (offset) associated with an entire moon is determined by
way of an additive \emph{summation} of the (equal) spin (offset)
values of its multitudinous `constituent' atoms and molecules.

A second `story' that physicists currently seek to (and do freely)
tell --- in addition to a preferred thermal-radiation/heat (or
thermal reaction force) based explanatory story/account of the
Pioneer anomaly --- relates to the future direction of physics and
its ultimate goals. This (`big picture') challenge involves
dealing with what is colloquially referred to as ``a final
theory". A projected solution, and generally considered `good
story', is the often mentioned/discussed: string theory, and the
more (recent and) elaborate M-theory --- requiring 11 cf. 10
spatial dimensions and incorporating supersymmetry --- espoused by
a majority of theoretical physicists. If this paper's explanatory
story and findings (both `physical' and conceptual) are correct,
it follows that this (almost unassailable) unification (and
reductive) agenda will (continue to) suffer stagnation as regards
its experimental confirmation and (to a lesser extent) its
predictive ability. In hindsight, the pursuit of a reductive
four-force unification goal, although understandable by way of
historical unification successes, is seen (in its current form) as
somewhat blinkered and too prescriptive. Additionally, dark matter
and dark energy are not integrally (corralled and) woven into this
agenda, and the ``make up" of both these entities has (to date)
not been assuredly determined. In light of this current ``dark
daze" and dubious ``theory of everything" progress, to suggest a
more generalist approach to physical reality (in its widest sense)
--- particularly involving/incorporating the stronger
elements/aspects of philosophical practice --- is not
unreasonable.

Bearing in mind that ``certainty is [sometimes] the enemy of
progress", the phenomenal--noumenal distinction (and its
complementarity) proposed herein is at risk of encountering
resistance and delay (within physics); arguably/possibly similar
to that which befell an appreciation of the usefulness of the
number ``zero" (and the subtle concept of nothing) in some
cultures in the past \citep*{Kaplan_99,Seife_00}. Zero, in the
sense of ``the absence of a number", somewhat parallels the
conceivable existence of a `noumenal physicality' that is not
susceptible to \emph{direct} physical observation. By way of
quantum probability and entanglement, and (indirectly via) Bell's
inequality, the existence of non-local (hidden) variables (and
processes) has been entertained for many years. Furthermore,
nothingnes\footnote{See (for example) the Stanford Encyclopedia of
Philosophy, \url{http://plato.stanford.edu/entries/nothingness/}},
abstract objects, (Kant's version and other `takes' on) the
phenomenal--noumenal distinction, and continental philosophy in
general (cf. analytic philosophy) are examples of ongoing active
areas of philosophical investigation; and it is not unreasonable
to presume that the elaborate explanation of the Pioneer anomaly
proposed herein has partially tilled the ground of this
\emph{potentially} rich field of indirect (analytical)
`scientific' investigation.

\subsection{A closing conditional comment}\label{Subsection:In closing}
Lastly, in this ``summary, discussion and conclusions" Section
(i.e. Section \ref{Section:Summary}), some retrospective remarks
are made and then a conditional scenario is put forward.

It was totally unexpected at the outset of this research project
that a new scientific model designed specifically to account for
the Pioneer anomaly's (solar system based) awkward observational
constraints --- by way of introducing two separate cosmological
scale/size mechanisms\footnote{With the two (three dimensional)
scalar fields associated with these mechanisms both extending to
`infinity'.}, i.e. (gravito-quantum) rotating space-warps (RSWs)
and non-local mass distributions (NMDs) --- could conceivably also
resolve or at least enlighten (via two distinct cosmological
foreground effects) both the cosmic microwave background (CMB)
anomalies and the dark energy issue (by way of the RSWs and NMDs
respectively). The model, within which energy and geometry feature
prominently (cf. force and position coordinates) also produces a
viable explanation of the (unresolved) Earth flyby anomaly, whilst
imposing (only) very minor changes to the solar system's
historical behaviour (i.e. its early history and subsequent
evolution) --- because only `low' mass bodies are
anomalously/additionally affected. Furthermore, the model's
conceptual structure casts doubt upon (both) a theory of
everything unification agenda and the veracity of the dark matter
hypothesis; the latter (hypothesis) in the sense that it assumes
gravitational theorisation is `complete', and thus any observed
transgression cannot be gravitationally-real
--- e.g. ``flat galactic rotation curves".

The broad scope of the model proposed in this treatise
--- which involves reconsideration (to some extent) of: time,
mass, quantum entanglement, QM (relative) geometric phase, and
physical `reality' (in general) --- means that the model's very
existence raises a (big picture) concern. The nature of this
concern is that: \emph{if} the new model is viable/correct (i.e.
``on the right track") \emph{then} to ignore its progressive
implications and ramifications --- e.g. by way of antipathy
towards deep (philosophically inclined) reasoning, or by way of
preferring a systematic based explanation of the Pioneer anomaly
--- does more than simply maintain the \emph{status quo}. The
downside risk of overlooking the model's ability to be a catalyst
for change/action, is that an (ulterior) form of (scientific)
conceptual stagnation might be unnecessarily perpetuated, in so
much that ongoing and future deliberations (specifically) aimed at
establishing the conceptual and physical \emph{crux} of (some or
all of) these (six) aforementioned issues/anomalies ---
particularly the (seemingly unrelated) CMB anomalies and the (what
is) dark energy issue --- would effectively be misappropriating
time, effort, and resources; notwithstanding: the new information
to be (surely) gleaned by these efforts, the erudition and skill
of the participating scientists, and the complexity involved in
performing the research.

Putting to one side this possible counterfactual and
meta-scientific discussion, this research has been primarily a
comprehensive investigation of direct and peripheral evidence
pertaining to the Pioneer anomaly.
%*******************************************************************************************
\bibliographystyle{aa}
\bibliography{P8}

\end{document}